\documentclass[11pt]{article}

\usepackage[margin=1in]{geometry}
\usepackage{amsmath,amssymb,amsthm}\usepackage{bbm}
\usepackage{enumitem}
\usepackage{booktabs}
\usepackage{xcolor}
\usepackage{graphicx}
\usepackage{hyperref}
\usepackage{algorithm}
\usepackage{algpseudocode}

\newcommand{\paperTitle}{Matrix Spencer: Eight Standard Deviations Suffice and an Almost-Linear Time Algorithm for Dense Input}
\newcommand{\paperAuthor}{Zhao Song\thanks{Independent Researcher. Email: \texttt{magic.linuxkde@gmail.com}.} \and Lichen Zhang\thanks{Massachusetts Institute of Technology. Email: \texttt{lichenz@mit.edu}.}}

\DeclareMathOperator{\tr}{tr}
\DeclareMathOperator{\Cov}{Cov}
\newcommand{\E}{\mathbb E}
\newcommand{\R}{\mathbb R}
\newcommand{\V}{\mathbb V}
\renewcommand{\d}{\mathrm{d}}
\newcommand{\Tmat}{\mathcal{T}_{\mathrm{mat}}}

\renewcommand{\epsilon}{\varepsilon}

\hypersetup{colorlinks=true,linkcolor=blue,citecolor=blue,urlcolor=blue,hypertexnames=false,pdftitle={\paperTitle},pdfauthor={Zhao Song and Lichen Zhang}}
\allowdisplaybreaks

\theoremstyle{plain}
\newtheorem{theorem}{Theorem}[section]
\newtheorem{lemma}[theorem]{Lemma}
\newtheorem{proposition}[theorem]{Proposition}
\newtheorem{corollary}[theorem]{Corollary}
\newtheorem{fact}[theorem]{Fact}
\theoremstyle{definition}
\newtheorem{definition}[theorem]{Definition}
\theoremstyle{remark}
\newtheorem{remark}[theorem]{Remark}

\begin{document}

\title{\paperTitle}
\author{\paperAuthor}
\date{}
\maketitle

\begin{abstract}
The Matrix Spencer conjecture asserts that for all symmetric matrices $A_1,\ldots,A_n\in\mathbb{R}^{n\times n}$ with $\|A_i\|\le1$ there are signs $\varepsilon_1,\ldots,\varepsilon_n\in\{-1,1\}$ with $\|\sum_{i=1}^n\varepsilon_iA_i\|=O(\sqrt n)$. We prove it: a signing of discrepancy below $8\sqrt n$ always exists. We also give a randomized algorithm that finds a signing of discrepancy below $12\sqrt n$ with failure probability at most $p$. The algorithm uses $n^{3+o(1)}\operatorname{polylog}(1/p)$ arithmetic operations in the real-arithmetic model. This matches the size $n^3$ of the dense input up to subpolynomial factors. In the other direction, we prove that for every $n$ there are collection of symmetric matrices such that every signing has discrepancy at least $(2-o(1))\sqrt{n}$. We conjecture the answer to the matrix Spencer constant is 2.

We present three different proofs of the matrix Spencer conjecture. The key to every proof is a hereditary small-ball estimate. This is a lower bound on the Gaussian measure of the spectral body $\{x\in \mathbb{R}^n:\|\sum_ix_iA_i\|\le R\}$ that holds for every subfamily of the matrices. The other ingredient turns that Gaussian measure into a partial signing. We give three approaches to obtain such a signing. The first one covers the cube by partially signed faces through Gaussian concentration with a constant $7\cdot10^9$. The second proof replaces the covering by a projection lemma with explicit parameters for a constant $156{,}000$. The third proof turns Gaussian measure into signs by a lossless coding, with no union bound. It proves the estimate at the right radius with smooth spectral barriers and certified coefficients. It gives a constant below $7.88$. For algorithms, the main idea is to project Gaussian points onto a smoothed spectral body. The $n^{3+o(1)}$ time algorithm tracks the Gibbs matrix of that body across coordinate-descent steps with sketched increments and random refreshes.

After the completion of our first proof, a preprint by Akbas and Sra resolved the conjecture by a hereditary small-ball estimate. Their estimate depends quadratically on the logarithm of the aspect ratio. In contrast, ours already depends on it linearly. A concurrent work of Kathuria obtained the optimal square-root dependence in the rectangular case through a random walk with a free-probability potential.  Both works leave the constant unspecified or large, and both find a signing in unspecified polynomial time. We prove explicit constants 8 and 12, a lower bound on the constant $2-o(1)$ and an $n^{3+o(1)}$ time algorithm.

\end{abstract}

\tableofcontents
\newpage

\section{Introduction}\label{sec:introduction}

Spencer's ``six standard deviations suffice'' theorem~\cite{spencer85} states
that every family of $n$ sets on an $n$-element ground set admits a
$\{-1,1\}$-coloring of discrepancy at most $6\sqrt n$, removing the
$\sqrt{\log n}$ loss of independent random signs.  The proof is a
partial-coloring argument with an entropy count and gives no algorithm.
Rothvoss~\cite{rot17} showed that
projecting a Gaussian point onto the intersection of the cube with any
symmetric convex body of Gaussian measure $e^{-\Omega(n)}$ yields a partial
coloring.  Hence the only question left about a body is its Gaussian
measure.

The Matrix Spencer conjecture is the noncommutative form of Spencer's
theorem~\cite{spencer85}.  For symmetric matrices $A_1,\ldots,A_n\in\R^{n\times n}$ with
$\|A_i\|\le1$ it asks for signs $\varepsilon_1,\ldots,\varepsilon_n\in\{-1,1\}$
with $\|\sum_{i=1}^n\varepsilon_iA_i\|=O(\sqrt n)$.  Diagonal matrices recover
the set-system case, and random signs give only the matrix-concentration
bound $O(\sqrt{n\log n})$ (Tropp~\cite[Chapter~4]{t15}).  The conjecture
therefore asks whether the $\sqrt{\log n}$ can be removed for matrices as
Spencer removed it for sets.  Before this work the conjecture was known only
under rank, block-diagonal, or Frobenius-norm restrictions, which we recall in
Section~\ref{subsec:previous_work}.  Akbas and Sra~\cite{as26} and Kathuria~\cite{k26}
resolved the conjecture independently and concurrently, after we completed our
first proof (Section~\ref{sec:first_existence}).

We present three suites of results: existence, algorithm and lower bound.   A signing of discrepancy below $8\sqrt n$ always exists.  We give a
randomized algorithm that finds a signing of discrepancy below $12\sqrt n$ with
failure probability at most $p$ in $n^{3+o(1)}\operatorname{polylog}(1/p)$
arithmetic operations in the real-arithmetic model.  This running time
matches the input size up to subpolynomial factors. We also show that for every $n$ there exists a collection of $n$ symmetric matrices that any signing has discrepancy $(2-o(1))\sqrt{n}$. In Section~\ref{sec:overview}, we discuss in detail how we obtain the
result.

We state our main theorems informally in Section~\ref{subsec:intro_our_results}:
Theorem~\ref{thm:matrix_spencer} on existence, Theorem~\ref{thm:lower_bound_informal}
on the lower bound, and Theorems~\ref{thm:dense_runtime_informal}
and~\ref{thm:bit_complexity_informal} on the algorithm, in arithmetic and in bit
operations.  Tables~\ref{tab:constants} and~\ref{tab:algorithms} in that
subsection collect the three constants and the exponents of the four
algorithms.  We say at the end of that subsection which statements count
arithmetic operations and which count bit operations.  In
Section~\ref{subsec:previous_work} we recall the constructive-discrepancy line
from Spencer to Rothvoss and the partial results on the conjecture under rank,
block-diagonal, or Frobenius-norm restrictions.  There we also place our work
beside the concurrent proofs of Akbas and Sra~\cite{as26} and
Kathuria~\cite{k26}, and close with the organization of the paper.

\subsection{Our results}\label{subsec:intro_our_results}
We state our main results as follows.
\begin{theorem}[Matrix Spencer, informal version of
Theorems~\ref{thm:first_main}, \ref{thm:matrix_spencer_bound}, and~\ref{thm:existence_constant_sub10}]
\label{thm:matrix_spencer}
There is a universal constant $C<8$ such that, for all symmetric matrices
$A_1,\ldots,A_n\in\R^{n\times n}$ with $\|A_i\|\le1$, there is a vector
$\varepsilon\in\{-1,1\}^n$ satisfying
$\|\sum_{i=1}^n\varepsilon_iA_i\|\le C\sqrt n$.
Three proofs give three constants. Theorem~\ref{thm:first_main} gives
$C=7\cdot10^9$,  Theorem~\ref{thm:matrix_spencer_bound} gives $C=156000$, and
Theorem~\ref{thm:existence_constant_sub10} gives $C<7.88$.
\end{theorem}

Section~\ref{sec:first_existence} is our first proof.
In Section~\ref{sec:comparison} we place it beside the concurrent work.
In the rest of the paper we sharpen the constant and make the proof algorithmic.
We bound the constant from below in Section~\ref{sec:lb_lower_bound}, and
Table~\ref{tab:constants} places the lower bound beside our smallest
existence constant and our smallest algorithmic constant.

\begin{theorem}[Lower bound, informal version of Corollary~\ref{cor:lb_constant},
Proposition~\ref{prop:lb_eight}, and Theorems~\ref{thm:lb_sixteen} and~\ref{thm:lb_field}]
\label{thm:lower_bound_informal}
For every $n$ there are symmetric matrices $A_1,\ldots,A_n\in\R^{n\times n}$
with $\|A_i\|\le1$ such that every $\varepsilon\in\{-1,1\}^n$ satisfies
$\|\sum_{i=1}^n\varepsilon_iA_i\|\ge(2-o(1))\sqrt n$.  An explicit $o(1)$ is
available: for $n\ge4$ the bound holds with $o(1)=\delta_n$, where
$\delta_n=O(n^{-1/5}(\log n)^{3/5})$ (Corollary~\ref{cor:lb_constant}(iv)).
Asymptotically, with constants that are not explicit, the $o(1)$ can be
taken $O(n^{-2/7}(\log n)^{6/7})$ (Proposition~\ref{prop:lb_rate_two_sevenths}).
\end{theorem}

\begin{table}[t]
\centering
\begin{tabular}{@{}lllp{0.44\textwidth}@{}}
\toprule
Constant & Value & Statement & Qualifier \\
\midrule
Lower bound & $2-o(1)$ & Corollary~\ref{cor:lb_constant} &
$o(1)=O(n^{-1/5}(\log n)^{3/5})$.  At finite $n$ the constant is at least
$1.77$ at $n=10^9$, for example (Table~\ref{tab:lb_delta}) \\
Existence & $<7.88$ & Theorem~\ref{thm:existence_constant_sub10} &
every $n$, existence \\
Algorithm & $<11.51$ & Theorem~\ref{thm:algorithm_constant_13} &
$n^{3+o(1)}\operatorname{polylog}(1/p)$ arithmetic operations, failure
probability at most $p$ \\
\bottomrule
\end{tabular}
\caption{The three constants of this paper.  The first row bounds the
constant of the conjecture from below and holds for every $n$, with an
explicit $o(1)$.  The other two rows are the smallest existence constant and
the smallest algorithmic constant that we prove in this paper.}
\label{tab:constants}
\end{table}

\begin{theorem}[Algorithm, informal version of
Theorem~\ref{thm:algorithm_constant_13}]
\label{thm:dense_runtime_informal}
For dense input matrices and every $p\in(0,1/2)$, a signing of discrepancy
below $11.51\sqrt n$ can be found with failure probability at most $p$
using
\[
n^{3+o(1)}\operatorname{polylog}(1/p)
\]
arithmetic operations in the real-arithmetic model.
\end{theorem}

\begin{remark}
\label{rem:dense_runtime_informal}
The algorithm of Theorem~\ref{thm:algorithm_constant_13} either outputs a
signing or reports failure, with failure probability at most $p$.  It is
designed for large $n$.  For $n\le10^{25}$ it runs capped independent sign
trials with a norm certificate below $10.89\sqrt n$.  It returns the accepted
signing, or reports failure if the cap is reached.  This step appears in
Algorithm~\ref{alg:main_cubic_small_constant}, and the certified trials are
those of Lemma~\ref{lem:algorithm_certified_implementation}, evaluated at
$n\le10^{25}$ in the proof of Lemma~\ref{lem:cubic_global_complete}.  It is
built on the
algorithm of Theorem~\ref{thm:cubic_dense_runtime}, which finds a signing of
discrepancy at most $156000\sqrt n$ within the same running time.  Three
algorithms lead up to that one, Theorems~\ref{thm:dense_runtime},
\ref{thm:faster_dense_runtime}, and~\ref{thm:accelerated_dense_runtime}, and
Table~\ref{tab:algorithms} lists their exponents.
\end{remark}

The dense input has $n^3$ entries, so
$n^{3+o(1)}$ is the time to read it, up to subpolynomial factors.  Every
algorithm runs the partial-coloring loop of Section~\ref{subsec:overview_constant}.
Each projection there is a bisection over one Lagrange multiplier, and the
four algorithms differ only in the solver for a fixed multiplier
(Section~\ref{subsec:overview_ladder}).  The last of them uses only two facts from fast
matrix multiplication, $\alpha_{\mathrm{dual}}>1/4$ and $\omega<5/2$, where
$\alpha_{\mathrm{dual}}:=\sup\{b:\omega(1,b,1)=2\}$ is the dual
matrix-multiplication exponent.  These facts are theorems of Vassilevska
Williams,
Xu, Xu, and Zhou~\cite[Section~3.4]{vxxz24} and of Alman, Duan, Vassilevska
Williams, Xu, Xu, and Zhou~\cite[Table~1]{almanetal25asymmetry}.  We collect the four bounds in Table~\ref{tab:algorithms}.

\begin{table}[t]
\centering
\begin{tabular}{@{}llll@{}}
\toprule
Section & Fixed-multiplier solver & Exponent & Numerical value \\
\midrule
\ref{sec:algorithm} & full Gibbs matrix, accelerated blocks & $2+\omega/2$ & $<3.19$ \\
\ref{sec:faster_algorithm} & selected-coordinate Gibbs oracle & $1+\omega(1,1/2,1)$ & $<3.05$ \\
\ref{sec:accelerated_block_algorithm} & accelerated block oracle & $3+\gamma_\star$ & $\le3.02$ \\
\ref{sec:cubic_time} & recursive Gibbs tracking & $3+o(1)$ & $3$, unconditionally \\
\bottomrule
\end{tabular}
\caption{The four algorithms.  Each finds a signing of discrepancy at most
$156000\sqrt n$ with failure probability at most $p$.  An exponent $e$ means
$n^{e+o(1)}\operatorname{polylog}(1/p)$ arithmetic operations in the
real-arithmetic model of Section~\ref{sec:preliminaries}, so every entry hides
$n^{o(1)}$ and $\operatorname{polylog}(1/p)$ factors.
Here $\gamma_\star:=\min_{0\le b\le1/2}\max\{\omega(1,b,1)+b-5/2,\ 1/2-b\}\le0.02$,
and $\gamma_\star=0$ whenever $\omega(1,1/2,1)=2$.
In Section~\ref{sec:small_algorithm_constant} we lower the constant of the last
algorithm below $11.51$ at the same exponent.
We run that small-constant algorithm on
fixed-point words of $O(\log n)$ bits in Section~\ref{sec:bit_complexity}.  The numerical values use
$\omega<2.38$, $\omega(1,0.4,1)<2.01$, and $\omega(1,1/2,1)<2.05$
from Alman, Duan, Vassilevska Williams, Xu, Xu, and
Zhou~\cite[Table~1]{almanetal25asymmetry}.  The last row uses only
$\alpha_{\mathrm{dual}}>1/4$ and $\omega<5/2$.}
\label{tab:algorithms}
\end{table}

\begin{theorem}[Bit complexity, informal version of Theorem~\ref{thm:bit_complexity}]
\label{thm:bit_complexity_informal}
The small-constant algorithm of Section~\ref{sec:small_algorithm_constant}, a cubic-time algorithm with constant
below $11.51$, can be run on fixed-point words of $O(\log n)$ bits.  Assume the known
bounds $\alpha_{\mathrm{dual}}>1/4$ and $\omega<5/2$, where $\alpha_{\mathrm{dual}}:=\sup\{b:\omega(1,b,1)=2\}$
is the dual matrix-multiplication exponent in the asymptotic-rank sense of
Definition~\ref{def:bit_rank_exponent}.  For every rational $p\in(0,1/2)$ with
$p\ge2^{-n}$, it then finds a signing of discrepancy below $11.51\sqrt n$ with failure probability at most $p$ using
$n^{3+o(1)}\operatorname{polylog}(1/p)+\widetilde O(n^3L_{\mathrm{in}})$ bit operations on inputs
whose entries are rationals of bit length at most $L_{\mathrm{in}}$.
\end{theorem}

We discuss dense input and the real-arithmetic model in Section~\ref{sec:model}. Note that
Theorem~\ref{thm:dense_runtime_informal} and the four
theorems of Remark~\ref{rem:dense_runtime_informal} count arithmetic
operations, not bit operations.  Theorem~\ref{thm:bit_complexity_informal}
covers the small-constant cubic-time algorithm only.  The four algorithms of
Table~\ref{tab:algorithms} remain real-arithmetic statements.

\subsection{Previous work}\label{subsec:previous_work}

\paragraph{Constructive discrepancy.}
Spencer's bound~\cite{spencer85} is proved by partial coloring with an entropy
argument, which gives no algorithm.  Bansal~\cite{bansal10} first obtained the bound
algorithmically by a random walk whose steps are guided by semidefinite
programs.  Lovett and Meka~\cite{lovettmeka15} obtained it algorithmically by
the edge walk, a
Brownian motion inside the cube that sticks to the constraint hyperplanes it
meets.  Rothvoss~\cite{rot17} showed that projecting a Gaussian point onto
the intersection of the cube with a symmetric convex body of Gaussian measure
$e^{-\Omega(n)}$ yields a partial coloring with a constant fraction of the
coordinates frozen.  This is the step that Lemma~\ref{lem:refreshed_projection}
analyzes.

\paragraph{Matrix Spencer.}
{\setlength{\emergencystretch}{2em}Zouzias~\cite{zouzias12} extended the hyperbolic-cosine method to
matrices, an early step toward the noncommutative setting.  Two works from
2022 went below the matrix-concentration bound under hypotheses.  Hopkins,
Raghavendra, and Shetty~\cite{hrs22} proved the conjectured bound when
$\|A_i\|_F\le n^{1/4}$, which includes matrices of rank at most $\sqrt n$,
using ideas from quantum communication.  Dadush, Jiang, and
Reis~\cite{djr22} developed a mirror-descent partial-coloring framework.
They proved the low-rank and block-diagonal cases, and reduced the general
problem to a geometric covering question.  Bansal, Jiang, and Meka~\cite{bjm23} then
handled rank at most $n/\log^3n$ by combining partial coloring with the
refined noncommutative Khintchine inequality of Bandeira, Boedihardjo, and
van Handel~\cite{bbh23} for correlated Gaussian matrices.\par}

\paragraph{Concurrent work.}
After we completed our first proof (Section~\ref{sec:first_existence}),
Akbas and Sra~\cite{as26} resolved the general conjecture by a hereditary
small-ball estimate.  Their estimate is proved through an interpolation
from a diagonal model under a matrix-weighted Poincar\'e inequality.  Their
radius depends quadratically on the logarithm of the aspect ratio.  In
contrast, our Proposition~\ref{prop:first_small_ball} already depends on it
linearly.  Their algorithm is the partial-coloring theorem of Reis and
Rothvoss~\cite{reisrothvoss22} and of Dadush, Jiang, and Reis~\cite{djr22}.
It runs in polynomial time with no explicit constant.
Bansal and Kook~\cite{bk26} subsequently simplified the proof of Akbas and Sra.
Kathuria~\cite{k26}, independently and concurrently, proved the
conjecture by a random walk.  The potential of the walk comes from free
probability and has a semidefinite formulation.  The bound has the optimal
square-root dependence on the logarithm of the aspect ratio in the
rectangular case.  The constant is $10^8$, and the running time is
polynomial.  Neither work gives an explicit constant below $10^8$ or a
running time beyond polynomial with an unspecified exponent.
Section~\ref{sec:comparison} compares the three papers part by part.

\paragraph{Organization.}
We give an informal overview in Section~\ref{sec:overview}, first of the proofs and
of the constant, then of the four algorithms, finally of the lower bound of the constant.
In Section~\ref{sec:comparison} we
compare each part of the paper with the concurrent works of Akbas and
Sra~\cite{as26} and of Kathuria~\cite{k26}.
Section~\ref{sec:preliminaries} introduces notation and the
computational model and collects the matrix-analytic tools that we use in
Section~\ref{sec:small_ball_proof}.  Section~\ref{sec:first_existence} is our first proof of the conjecture,
with the constant $7\cdot10^9$.
We prove the hereditary Gaussian small-ball
estimate with explicit constants in Section~\ref{sec:small_ball_proof}.  Section~\ref{sec:proof} combines it
with a refreshed Gaussian projection step to prove the discrepancy bound
$156000\sqrt n$.  We then prove the
existence bound $7.88\sqrt n$ in Section~\ref{sec:existence_constant_sub10}.
Section~\ref{sec:algorithm} gives our $n^{2+\omega/2+o(1)}$ algorithm.
Section~\ref{sec:faster_algorithm} gives the $n^{1+\omega(1,1/2,1)+o(1)}$
algorithm based on a selected-coordinate Gibbs oracle.
We give the accelerated block
algorithm with exponent $3+\gamma_\star$ in Section~\ref{sec:accelerated_block_algorithm}.  Section~\ref{sec:cubic_time}
gives the $n^{3+o(1)}$ algorithm, which tracks the Gibbs matrix recursively.
Section~\ref{sec:small_algorithm_constant} improves the constant of the
cubic-time algorithm below $11.51$.
We prove in Section~\ref{sec:bit_complexity} that this small-constant algorithm,
which drives the solver of Section~\ref{sec:cubic_time}, keeps its exponent
and its constant on fixed-point words of $O(\log n)$ bits. Finally we prove the a lower bound of $2-o(1)$ on the constant of matrix Spencer in Section~\ref{sec:lb_lower_bound}.

\section{Technical overview}
\label{sec:overview}

We treat the discrepancy bound, the running time, and the lower bound in
turn.  In Section~\ref{subsec:overview_constant} we take the discrepancy
bound in three steps, our first proof of the conjecture, its refinement to
$156000$, and the constant below $8$.  Every proof has the same two
ingredients, a hereditary small-ball estimate and a conversion of Gaussian
measure into a partial signing, and we say where each constant comes from.
From the three proofs we obtain Theorems~\ref{thm:first_main},
\ref{thm:matrix_spencer_bound}, and~\ref{thm:existence_constant_sub10}, and
we end the subsection with the constant that the algorithms carry
(Theorem~\ref{thm:algorithm_constant_13}).  The running time is the subject
of Section~\ref{subsec:overview_ladder}.  We bring it down in four steps
from $n^{2+\omega/2}$ to $n^{3+o(1)}$, one algorithm per step, each removing
one obstacle left by the one before it.  The four steps are
Theorems~\ref{thm:dense_runtime}, \ref{thm:faster_dense_runtime},
\ref{thm:accelerated_dense_runtime}, and~\ref{thm:cubic_dense_runtime}.  In
a fifth step, not about the exponent, we run the small-constant version of
the last algorithm on words of $O(\log n)$ bits
(Theorem~\ref{thm:bit_complexity}).  We close that subsection with what the
four algorithms share and with open questions.  Last, in
Section~\ref{subsec:overview_lower_bound} we describe the lower bound of
Section~\ref{sec:lb_lower_bound}, which shows that the lower constant of the
matrix Spencer problem is at least $2$ in the limit.  There we give the hard
distribution, the one-signing-at-a-time argument behind
Theorem~\ref{thm:lb_existence} with its explicit finite-$n$ rate
(Corollary~\ref{cor:lb_constant}), and the explicit families at $n=8$,
$n=16$, and constant $\sqrt2$.

\subsection{Proving the conjecture and sharpening the constant}
\label{subsec:overview_constant}

Every proof of the conjecture in this paper has the same two ingredients.
Ingredient one is a hereditary small-ball estimate.  This is a lower bound of
the form $e^{-cm}$, with an explicit $c$, on the Gaussian measure of the
spectral body $\{x\in\R^m:\|\sum_{i=1}^mx_iA_i\|\le R\}$ of every subfamily
of $m$ matrices, at a radius $R$ that is $\sqrt m$ times a factor that varies
slowly with $m$.  Ingredient two turns that Gaussian measure into a partial
signing.  Our three proofs share the form of the first ingredient.  They
differ in how we prove it, with headroom, with every constant tracked, or by a
different mechanism, and in how we do the second.  All three proofs sign in rounds of partial coloring, and they differ in how
a round turns Gaussian measure into a partial signing: by a covering of the
cube by partially signed faces (Section~\ref{sec:first_existence}), by a
Gaussian projection refreshed in every round (Section~\ref{sec:proof}), and
by a lossless Gaussian-to-sign coding that pays no union bound
(Section~\ref{sec:existence_constant_sub10}).
Section~\ref{sec:first_existence} is the skeleton, and it gives the constant
$7\mathbin{\cdot}10^9$ (Theorem~\ref{thm:first_main}).
Sections~\ref{sec:small_ball_proof} and~\ref{sec:proof} keep its architecture,
and there we track every constant and replace one of its two steps, together
they give $156000$ (Theorem~\ref{thm:matrix_spencer_bound}).
We refine both ingredients in Section~\ref{sec:existence_constant_sub10}, by a
different mechanism for each, and obtain $7.879493715947<7.8795$
(Theorem~\ref{thm:existence_constant_sub10}).  Last,
Section~\ref{sec:small_algorithm_constant} gives the constant that the
algorithms carry.

Proving the conjecture and reaching a constant below $8$ are separate
challenges.  The first needs one estimate inside a frame that does not care
about constants.  Akbas and Sra~\cite[Theorem~1.3]{as26} prove an estimate of
the same shape, and we make the comparison in Section~\ref{sec:comparison}.
The second needs us to locate every loss that the framework tolerates and pay for
it exactly.  We take the three proofs in turn, and for each we say what its
two ingredients are and where its constants come from.

\paragraph{The skeleton: the partial-coloring frame.}
For symmetric contractions
$A_1,\ldots,A_n$, a uniformly random sign vector gives
$\|\sum_i\varepsilon_iA_i\|=O(\sqrt{n\log n})$ by matrix concentration
(Tropp~\cite[Chapter~4]{t15}), and the conjecture asks for the $\sqrt{\log n}$
to be removed.  In Section~\ref{sec:first_existence} we choose the signs in the partial-coloring framework of Rothvoss~\cite{rot17} and of Dadush, Jiang, and
Reis~\cite{djr22}.  Keep a fractional coloring $y\in[-1,1]^n$, starting at
$y=0$.  Call a coordinate active while $|y_i|<1$, let $S:=\{i:|y_i|<1\}$ be
the active set, and let $m:=|S|$.  A coordinate that reaches $\pm1$ is a sign
and is never touched again.  One round looks at the spectral body of the
active family, the set of increments $x\in\R^S$ with $\|\sum_{i\in S}x_iA_i\|$
at most a radius of order $\sqrt m$.  It picks a Gaussian point of $\R^S$ and
moves $y$ by a bounded dilate of a point of the body, chosen so that $y$ stays
in the cube and a fixed fraction of the active coordinates lands exactly on
$\pm1$.  Then it repeats on the smaller active set.  By the triangle
inequality, the norm of the final signed sum is at most the sum over rounds of
the dilation times the radius of that round's body.  The radius is $\sqrt m$
times a factor that varies slowly with $m$, and $m$ shrinks by a fixed factor
per round, so the sum is a geometric series with total $O(\sqrt n)$.  Two
things could go wrong.  The Gaussian point might be nowhere near the spectral
body, since a convex set in $\R^m$ can have Gaussian measure far below
$e^{-m}$.  And even if the body is large, nothing so far forces any coordinate
onto $\pm1$.  The two ingredients answer these two failures in turn.  A
small-ball estimate says that the spectral body has Gaussian measure at least
$e^{-cm}$ for an explicit $c$, uniformly over active sets.  A covering
statement says that a body of that measure reaches, from every point of the
cube, a point of the cube with a fixed fraction of its coordinates equal to
$\pm1$.

\paragraph{Ingredient one in the skeleton: the hereditary small-ball estimate.}
The first ingredient is Proposition~\ref{prop:first_small_ball}.  For
$1\le m\le N$ and symmetric contractions $A_1,\ldots,A_m\in\R^{N\times N}$,
put $\ell:=\max\{1,\log(N/m)\}$.  Then
\[
\gamma_m\{x\in\R^m:\|\textstyle\sum_{i=1}^mx_iA_i\|\le10^6\sqrt m\,\ell\}
\ge e^{-m/1000},
\]
where $\gamma_m$ is the standard Gaussian measure on $\R^m$.  The estimate is
hereditary because it holds for every subfamily, with the same exponent.  The
framework needs exactly this, since the active set is different in every round and
depends on the increments chosen before.  Only the radius may depend on the
subfamily, through the aspect ratio $N/m$, and it does so only through the
factor $\ell$, which equals $1$ while $m\ge N/e$.  Matrix concentration does
not give this as it bounds the typical norm of $\sum_ix_iA_i$ by
$O(\sqrt{m\log N})$ and says nothing about the small probability of the
atypical event that the norm is $O(\sqrt m\,\ell)$.  To reach the
measure we interpolate from a model in which it can be computed.  Rescale
the matrices by the target radius, $B_i:=A_i/r$ with $r:=10^6\sqrt m\,\ell$.
Replace the indicator of the body $\{\|\sum_i\xi_iB_i\|\le1\}$ by
$e^{-\Phi(X)}$ with $\Phi(X):=\tr[\phi(X)]$ for a barrier $\phi$ on $(-1,1)$
that is even, nonnegative, convex and blows up at $\pm1$, so that the measure
of the body is at least $\E[e^{-\Phi(X_1)}]$ with $X_1:=\sum_i\xi_iB_i$
(Eq.~\eqref{eq:first_14}).  The barrier is
$\phi=\kappa_0\phi_0+(\kappa_1-\kappa_0)\phi_{1,\tau}$ with
$\phi_0(u)=-\log(1-u^2)$ and the tail series
$\phi_{1,\tau}(u)=\sum_{k\ge\tau}u^{2k}/k$.  Split each $B_i$ into its
diagonal part $D_i$ and its off-diagonal part $O_i$, put
$X_t:=\sum_i\xi_i(D_i+tO_i)$ and $U(t):=\log\E[e^{-\Phi(X_t)}]$.  Then
$U'(0)=0$, and $U''(t)$ is a nonnegative variance, which we discard, minus
the Hessian of the barrier along $\sum_i\xi_iO_i$ averaged under the tilted
Gaussian (Eq.~\eqref{eq:first_13}).  Integrating $U''$ leaves two quantities
to bound, the interpolation loss and the diagonal endpoint $U(0)$.  We bound
the Hessian by a spectral $0$--$1$ split of the eigenvalue pairs
(Eq.~\eqref{eq:first_18}), whose edge part is absorbed by the curvature of the
barrier and whose regular part is small because the rescaled matrices have a
tiny variance proxy.  The second is the matrix-weighted Poincar\'e inequality of
Eq.~\eqref{eq:first_2}, which bounds a weighted variance by $a^{-1}$ times the
weighted Dirichlet energy when the weight has curvature at least $a$ times
itself (Eq.~\eqref{eq:first_1}).  We apply it with $a=1/2$ to the tilted
density times the resolvents $(I+\sigma X_t)^{-1}$, $\sigma=\pm1$, and their
tensor squares.  Iterating it gives resolvent moments of order
$\tau$, which bound the interpolation loss (Eq.~\eqref{eq:first_34}).  At the
diagonal endpoint, \v Sid\'ak's strip bound of Eq.~\eqref{eq:first_7} keeps
every diagonal entry of $X_0$ inside $[-1/2,1/2]$, and a moment bound on those
entries controls the barrier there (Eq.~\eqref{eq:first_36}).

\paragraph{Where $10^6$ and $1/1000$ come from.}
We fix the constants at the start (Eqs.~\eqref{eq:first_10}
and~\eqref{eq:first_11}):
\[
C:=10^6,\qquad r:=C\sqrt m\,\ell,\qquad \rho:=\frac{1}{C^2\ell^2},\qquad
\tau:=\lceil8\ell\rceil,\qquad w:=\frac{1}{100},
\]
and the barrier weights
\[
\kappa_0:=\frac{10^{-8}m}{N\rho},\qquad
\kappa_1:=100(1+\kappa_0+\kappa_0^{-1}).
\]
Here $C$ is the radius coefficient, $\rho$ is the variance proxy of the
rescaled matrices, since $\sum_iB_i^2\preceq(m/r^2)I=\rho I$ and
$\rho\le10^{-12}$ throughout, $\tau$ is the order of the tail series and of
the resolvent moments, $w/\tau$ is the width of the edge zone, and
$\kappa_0,\kappa_1$ are the barrier weights.  The letter $\rho$ is reused in
Section~\ref{subsec:overview_ladder} for a projection tolerance.  Two choices
decide the shape of the result.  The tail term of the loss,
$(100\tau^2\rho)^\tau<2^{-20\tau}$, stands against a prefactor
$\kappa_1\tau N/m$ that grows like $e^{2\ell}$ times a polynomial in $\ell$,
so the moment order must be proportional to $\ell$.  With $\tau$ proportional
to $\ell$, the curvature check needs $\rho$ proportional to $1/\ell^2$, hence
the radius is $C\sqrt m\,\ell$, linear in $\ell$.
This linearity is the point we compare in
Section~\ref{sec:comparison} with the square of Akbas and Sra.
The check that pins the radius is the regular part of the split
(Eq.~\eqref{eq:first_23}): we need
$\tau^2\rho/w^2\le81\mathbin{\cdot}10^{-8}<1/4$, so that the regular part
costs at most $1/4$ of the unit Gaussian curvature per tensor factor and
half of it survives after both factors, which is the $a=1/2$ above.  The
exponent comes out of the final result, Eq.~\eqref{eq:first_37},
$-U(1)/m\le50\kappa_0N\rho/m+6\kappa_1\tau(N/m)(100\tau^2\rho)^\tau
+4(N/m)e^{-1/(8\rho)}$.  The regular term equals $50\mathbin{\cdot}10^{-8}$ by
our choice of $\kappa_0$, the tail term is below $10^{-6}$, and the strip
term (what we need to keep the diagonal entries of $X_0$ inside
$[-1/2,1/2]$) is below $10^{-6}$.  The total is below
$3\mathbin{\cdot}10^{-6}$, and the proposition claims $1/1000$.  The
instructive point is that we choose the constants of Section~\ref{sec:first_existence}
rather loosely just for the argument to go through.  The exponent we prove is more than three
hundred times smaller than the exponent we claim, and $C=10^6$ is far above
what the curvature check alone needs, since $81\mathbin{\cdot}10^{-8}$ sits far
below $1/4$.  We tune nothing.  Section~\ref{sec:small_ball_proof} is what
happens when we run the same argument with the headroom removed and constants tracked.

\paragraph{Ingredient two in the skeleton: covering the cube by partially signed faces.}
Proposition~\ref{prop:first_covering} turns measure into signs.  Suppose
$m\ge1000$ and a symmetric closed convex $K\subseteq\R^m$ has
$\gamma_m(K)\ge e^{-m/1000}$, then every point of $[-1,1]^m$ lies within $8K$
of a point of $[-1,1]^m$ having at least $\lceil m/100\rceil$ coordinates
equal to $\pm1$.  Applied with $K$ the spectral body of the active family, it
says that from the current fractional coloring we can move by a point of
$8K$ and freeze one percent of the active coordinates.  To prove it we
compare two distances.  Fix $y\in(-1,1)^m$ and the box $Q_y:=([-1,1]^m-y)/8$,
which lies inside $Q:=[-1/4,1/4]^m$.  For $c\in Q_y$ the point $y+8c$ lies in
$[-1,1]^m$, and its coordinate $i$ equals $\pm1$ exactly when the box
constraint of coordinate $i$ is tight at $c$.  Project a Gaussian point $g$
onto $K\cap Q_y$, call the projection $c$ and the set of tight coordinates $I$,
and take $y+8c$ as the new coloring.  If $|I|\le m/100$, then
$d(g,K\cap Q_y)\le d(g,K\cap H(I))$, where $H(I)$ is the subspace on which
the coordinates in $I$ vanish.  But a Gaussian point is close to every
low-codimension section of a body of measure at least $e^{-m/1000}$
(Eqs.~\eqref{eq:first_6} and~\eqref{eq:first_40}) and far from the small cube
$Q$: a suitable $g$ is within $(13/25)\sqrt m$ of every such section and
farther than $(16/25)\sqrt m$ from $Q$, so more than $m/100$ coordinates are
tight.  Given $g$ the covering is a deterministic statement about $K$, valid
for every $y$. For one set $I$ with $|I|\le m/100$,
$\E[d(G,K\cap H(I))]\le\sqrt{|I|+2m/1000+1}<(3/25)\sqrt m$
(Eq.~\eqref{eq:first_41}).  The bound is needed for every $I$ at once, and
there are at most $e^{m\,\mathsf h(1/100)}$ such sets, with $\mathsf h$ the binary entropy.  Here
$\mathsf h(1/100)<3/50$, so we pay for the union bound by a deviation
$(2/5)\sqrt m$ of the Lipschitz distance, which fails with probability at most
$e^{-m/50}$, the exponent being $(2/5)^2/2-3/50=1/50$, this gives union bound the
$13/25=3/25+2/5$ (Eq.~\eqref{eq:first_42}). For the cube, $\E[\|G\|_2]\ge\sqrt{m-1}$ and $Q$ has
half-width $1/4$, so $\E[d(G,Q)]\ge(37/50)\sqrt m$ for $m\ge1000$, and the
lower tail at deviation $\sqrt m/10$ has probability at most $e^{-m/200}$,
outside which $d(G,Q)>(16/25)\sqrt m$ (Eq.~\eqref{eq:first_43}).  The cube
must be small enough that $16/25$ beats $13/25$, and half-width $1/4$ does
it. The step from that box back to the unit cube is the dilation $8$.  The
one percent is what keeps $\mathsf h(1/100)$ small enough for the union bound
to be absorbed by the deviation $2/5$ as a larger freezing fraction raises the
entropy, hence the $13/25$, hence the dilation.  The condition $m\ge1000$
makes $e^{-m/50}+e^{-m/200}<1$, so a point $g$ satisfying both bounds exists.

\paragraph{Assembling $7\mathbin{\cdot}10^9$.}
Theorem~\ref{thm:first_main} runs the loop of the framework with the two
ingredients.  At stage $j$, with active set $S_j$, $m_j:=|S_j|$ and
$\ell_j:=\max\{1,\log(n/m_j)\}$, the body is
$K_j:=\{x\in\R^{S_j}:\|\sum_{i\in S_j}x_iA_i\|\le10^6\sqrt{m_j}\,\ell_j\}$.
Proposition~\ref{prop:first_small_ball} with $N=n$ gives
$\gamma_{m_j}(K_j)\ge e^{-m_j/1000}$, and
Proposition~\ref{prop:first_covering} then moves $y$ by a point of $8K_j$ and
freezes at least one percent of the active coordinates, so
$m_{j+1}\le(99/100)m_j$ and the increment has norm at most
$8\mathbin{\cdot}10^6\sqrt{m_j}\,\ell_j$ (Eq.~\eqref{eq:first_44}).  Summing
the stage costs, we obtain $6.4\mathbin{\cdot}10^9\sqrt n$, through the elementary
bound $\max\{1,u\}e^{-u/2}\le2e^{-u/4}$ at $u=\log(n/m_j)$ and the geometric
ratio $q^{1/4}\le399/400$ for $q=99/100$.  Rounding the fewer than $1000$
coordinates left when the loop stops costs at most $1000\sqrt n$, and the
total is $(6.4\mathbin{\cdot}10^9+1000)\sqrt n<7\mathbin{\cdot}10^9\sqrt n$.
The constant is a product of three factors: the dilation $8$, the radius
coefficient $10^6$, and a round sum of $2\mathbin{\cdot}400=800$, the $2$ from
the elementary bound and the $400$ from the geometric series.  Every later
improvement of the constant acts on one of these three factors, on the radius
coefficient through the small-ball proof, on the dilation through the sign
step, and on the round sum through the freezing fraction and the summation.

\paragraph{Both ingredients with the constants tracked.}
Sections~\ref{sec:small_ball_proof} and~\ref{sec:proof} keep the
two-ingredient architecture of Section~\ref{sec:first_existence} and change
three things, acting on the radius coefficient, on the sign step, and on the
round sum.  We prove the hereditary small-ball estimate in
Section~\ref{sec:small_ball_proof} with every constant tracked, at a smaller
radius coefficient.  In Section~\ref{sec:proof} we turn the measure into signs
by repeated partial coloring, a projection lemma with explicit parameters
refreshed in every round, and sum the increments exactly.  Together they
lower the constant from $7\mathbin{\cdot}10^9$ to $156000$
(Theorem~\ref{thm:matrix_spencer_bound}).

\paragraph{Ingredient one with constants tracked: the six-regime small ball.}
In Section~\ref{sec:small_ball_proof} we write the spectral body of the active
set $S$ as
$K_S:=\{x\in\R^S:\|\sum_{i\in S}x_iA_i\|\le c_{\rm rad}(\ell_{\rm asp})\sqrt m\,\ell_{\rm asp}\}$
with $m:=|S|$ and $\ell_{\rm asp}:=\max\{1,\log(n/m)\}$, the $\ell$ of
Section~\ref{sec:first_existence}, and the radius coefficient
$c_{\rm rad}(\ell_{\rm asp})$ replaces $10^6$.  The
partial-coloring step needs $\gamma_m(K_S):=\Pr[\xi\in K_S]\ge e^{-cm}$ for a
standard Gaussian vector $\xi\sim\mathcal N(0,I_m)$, with a small explicit
$c$, uniformly over every active set $S$.  After rescaling by the target
radius, define
$\mathcal{A}(\xi):=(c_{\rm rad}(\ell_{\rm asp})\sqrt m\,\ell_{\rm asp})^{-1}\sum_{i\in S}\xi_iA_i$,
so that $\gamma_m(K_S)=\Pr[\|\mathcal{A}(\xi)\|\le1]$.  The proof is that of
Proposition~\ref{prop:first_small_ball} with every constant tracked and four
of its choices changed.  The moment order is
$\tau(\ell_{\rm asp}):=\lceil256\ell_{\rm asp}/17\rceil$ in place of
$\lceil8\ell\rceil$, so that $\tau\le16\ell_{\rm asp}$.  We keep the edge
cutoff of the spectral split, fixed at $w=1/100$ in Eq.~\eqref{eq:first_10},
free as $\widehat\kappa_3=b^2/(1+b)$ with curvature coefficient
$\widehat c_0=(1-2b)/2$ in place of $1/4$, and we tune $b$ in each regime: in
the first regime $b=0.40084$ and the cutoff is $0.1147$.  A wider edge zone
matters because the curvature check pinning the radius reads
$\tau^2\rho\le w^2/4$ for the variance proxy
$\rho=1/(c_{\rm rad}\ell_{\rm asp})^2$, so a larger cutoff admits a smaller
radius. What stops $b$ short of $1/2$ is that $\widehat c_0$ then vanishes and
the barrier weights are paid for in the exponent through the term $27D/A$.  We replace the trace-moment bound $N(2s^2w)^s$ of Eq.~\eqref{eq:first_8}
by Harg\'e's convex-domination theorem with a direct Gaussian
integration by parts, giving $N(2q-1)!!\,\nu^q\le N(2q\nu)^q$. We certify the
exponent by rational Taylor enclosures, every coefficient carried at its
value where the earlier proof rounded each one up to $3$ at $\rho\le10^{-12}$.
One step we keep unchanged on purpose: we average the independent-direction
factors of every word in the moment bound through the completely positive
map $U\mapsto\sum_iR_iUR_i$, at cost $4\rho$ on every Schatten class, in both
proofs.  This is what keeps the radius linear in $\ell_{\rm asp}$.
The comparison with the
cube that Akbas and Sra carry is in Section~\ref{sec:comparison}.
We obtain $\gamma_m(K_S)\ge\exp(-0.003712m)$ for all $1\le m\le n$
(Lemma~\ref{lem:scale_small_ball}).  The coefficient
$c_{\rm rad}(\ell_{\rm asp})$ takes six values, from $279$ at
$\tau(\ell_{\rm asp})=16$ to $185$ for $\tau(\ell_{\rm asp})\ge44$
(Table~\ref{tab:six_regimes}), because the most efficient barrier parameters
change as the moment order grows with the aspect ratio. We record where $279$
comes from in Remark~\ref{rem:smallball_reading_the_ledger}.  In every row
$c_{\rm rad}$ is the least integer passing the curvature check
$(\tau_0/(c_{\rm rad}\ell_0))^2<\widehat\kappa_3^2/4$, which in the first
regime reads $16^2/279^2<\widehat\kappa_3^2/4$ with $\widehat\kappa_3=0.1147$,
so the radius is pinned by curvature, not chosen for headroom.  On the binding
cell $[1,17/16]$ the three visible terms of the exponent term, $27D/A$, the
interpolation integral and the diagonal endpoint, are $0.50$, $0.39$ and
$0.11$ of the total. The diagonal endpoint, which we will remove altogether in
Section~\ref{sec:existence_constant_sub10}, costs below
$0.2$ percent from the third row on.  The check is the direct descendant of
$\tau^2\rho/w^2\le81\mathbin{\cdot}10^{-8}<1/4$.  With $\tau\le9\ell$ and
$w=1/100$ that check is passed by every
$C>2\mathbin{\cdot}9\mathbin{\cdot}100=1800$, and we took $C=10^6$ in
Section~\ref{sec:first_existence}.

\paragraph{Ingredient two with constants tracked: repeated partial coloring.}
Lemma~\ref{lem:refreshed_projection} turns the measure bound for $K_S$ into a
partial coloring by the comparison of two distances that proves
Proposition~\ref{prop:first_covering}, with three changes.  Suppose
$\gamma_m(K_S)\ge\exp(-0.00372m)$ and
$m\ge N_0:=3\mathbin{\cdot}10^8$, then one trial produces $x\in K_S$ with
$y+2x/\varepsilon_{\mathrm{box}}\in[-1,1]^m$ and at least $\delta m$
coordinates of the new coloring equal to $\pm1$, with probability greater than
$0.12$, for $\delta=0.0281672$ and $\varepsilon_{\mathrm{box}}=0.4235$.  The
gap between $0.003712$ and $0.00372$ is the headroom we later spend on smoothing
the body.  A successful round adds at most
$(2c_{\rm rad}(\ell_{\rm asp})/\varepsilon_{\mathrm{box}})e^{-u/2}\ell_{\rm asp}\sqrt n$
to the operator norm, $u:=\log(n/m)$
(Eq.~\eqref{eq:explicit_round_increment}).  Here, we have three modifications.
First, we draw the Gaussian point afresh in every round and discard and repeat
a failed trial, where Proposition~\ref{prop:first_covering} fixed one
point $g$ surviving two events of failure probabilities $e^{-m/50}$ and
$e^{-m/200}$.  Second, we compute the distance to the cube exactly, through $\psi(t):=\E[(|G|-t)_+^2]$ for a standard normal $G$, and we use the two facts about Gaussian measure of Fact~\ref{fact:gaussian_section_transport}, the section inequality and the distance bound, as Section~\ref{sec:first_existence} does at Eqs.~\eqref{eq:first_6} and~\eqref{eq:first_40}.  Third, the box is the cube shrunk by $\varepsilon_{\mathrm{box}}/2$
about $y$, inside $[-\varepsilon_{\mathrm{box}},\varepsilon_{\mathrm{box}}]^m$,
so the dilation is $2/\varepsilon_{\mathrm{box}}\approx4.72$ in place of $8$.
The constants are forced by one inequality.  The proof must hold for every
coordinate set $I$ of size at most $\delta m$ at once, a union bound over
$e^{\mathsf h(\delta)m}$ sets with $\mathsf h$ the binary entropy, paid for by an entropy
reserve $\sqrt{2\mathsf h(\delta)}\sqrt m$ in the deviation.  We write the normalized separation at $m=N_0$
in the letter of Section~\ref{sec:proof}, which is not the radius $r$ above:
\[
r:=\sqrt{\psi(\varepsilon_{\mathrm{box}})-1/N_0}
-\sqrt{\delta+2(0.00372)+1/N_0}
-\sqrt{2\mathsf h(\delta)},
\]
whose three terms are about $0.6953$, $0.1887$ and $0.5066$, so that
$10^6r>77.2$.  The two distances are separated by less than one ten-thousandth
of $\sqrt m$.  The entropy reserve is the largest of the three terms and grows
with $\delta$, the section term grows with $\delta$ too, and the cube term
shrinks as $\varepsilon_{\mathrm{box}}$ grows, so $\delta=0.0281672$ and
$\varepsilon_{\mathrm{box}}=0.4235$ are where the inequality closes.  Since
$\psi(t)<1$ for every $t>0$, no box makes $\sqrt\delta+\sqrt{2\mathsf h(\delta)}\ge1$
work, and this sum already exceeds $1$ at $\delta=0.075$: this shape of
argument cannot certify a freeze above a few percent, no matter how large the
Gaussian measure of the body is.  We spend the remaining separation on
concentration.  The cube and section deviations are
$3\mathbin{\cdot}10^{-5}\sqrt m$ each, and at $m=N_0$ the two failure
probabilities sum to $e^{-0.135}+e^{-4558}<0.874$, which is where both the
success probability $0.12$ and the threshold $N_0$ come from.  For comparison,
Proposition~\ref{prop:first_covering} spent $(2/5+1/10)\sqrt m$ on its two
deviations and kept the two distances $(16/25-13/25)\sqrt m$ apart.  All of
the headroom of Section~\ref{sec:first_existence} has gone into the larger box and
the larger $\delta$.

\begin{figure}[htbp]
\setlength{\abovecaptionskip}{4pt}
\centering
\includegraphics[width=\linewidth]{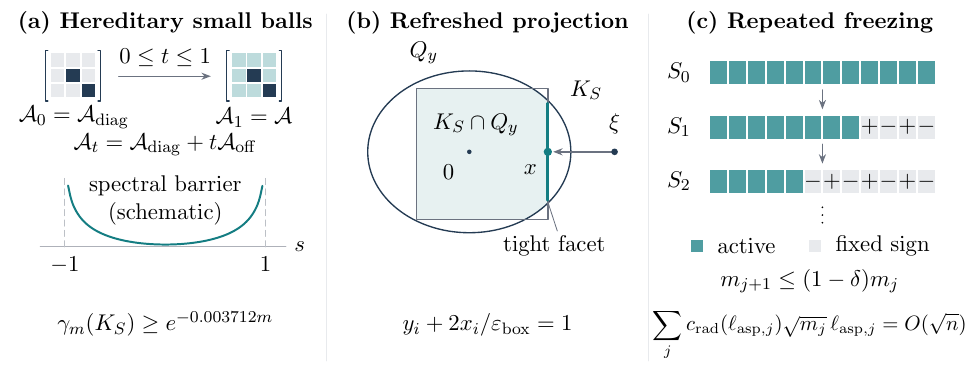}\par
\nointerlineskip
\caption{The partial-coloring mechanism
(Lemmas~\ref{lem:scale_small_ball} and~\ref{lem:refreshed_projection}).
Panel (a) shows the normalized interpolation $\mathcal{A}_t$ of
Definition~\ref{def:smallball_interpolation_data}, which gives Gaussian mass
for every active set $S$.  Panel (b) shows a successful projection trial,
on which tight facets of $Q_y$ become fixed signs after affine rescaling.
Panel (c) shows the active sets shrinking geometrically until terminal
rounding applies.  Here $m_j:=|S_j|$ and
$\ell_{{\rm asp},j}:=\max\{1,\log(n/m_j)\}$.  The barrier, geometry, and
coordinate counts are schematic.}
\label{fig:partial-coloring-mechanism}
\end{figure}

\paragraph{Summing the increments to $156000$.}
Each round freezes at least $\delta m$ coordinates, so the successive values
of $u$ are separated by at least $\lambda:=-\log(1-\delta)$ but need not lie
on a grid.  While $u<1$, we have $c_{\rm rad}=279$ and the increment
$(2\mathbin{\cdot}279/\varepsilon_{\mathrm{box}})e^{-u/2}\sqrt n$ is monotone
in $u$.  At most $35$ such rounds occur, since $34\lambda<1<35\lambda$, and we
sum them by an exact geometric series to at most $36550.1\sqrt n$.  For $u\ge1$ the
summand $c_{\rm rad}(u)ue^{-u/2}$ jumps at the five regime endpoints, and the
separated-sum estimate (Lemma~\ref{lem:bv_packing}) bounds every
$\lambda$-separated sum by the sum of the suprema over consecutive bins of
length $\lambda$, evaluated by rational Taylor enclosures.  These rounds add at
most $118159.2\sqrt n$.  All projection rounds together contribute less than
$36550.1+118159.2=154709.3$ times $\sqrt n$.  Below $N_0$, we round
independently with the fractional coloring as mean, and by the matrix
Laplace-transform bound of Tropp~\cite[Chapter~3]{t15} this contributes at most
$\sqrt{2m\log(4n)}\le6.5\sqrt n$ with probability at least $1/2$, replacing
the deterministic rounding that cost $1000\sqrt n$ before.  We discard and
repeat failed trials.  Hence a signing of
discrepancy at most $156000\sqrt n$ exists
(Theorem~\ref{thm:matrix_spencer_bound}), and the algorithms of
Sections~\ref{sec:algorithm}--\ref{sec:cubic_time} retain this bound.
Figure~\ref{fig:partial-coloring-mechanism} shows how the small-ball estimate,
projection, and summation fit together.

\paragraph{Where the factor between $7\mathbin{\cdot}10^9$ and $156000$ comes from.}
Both constants are a dilation times a radius coefficient times a round sum.
The radius coefficient falls from $10^6$ to $279$ in the first regime and to
$185$ at large aspect ratio, a factor above $3500$, because we now pass the
curvature check with the least integer and tune the cutoff, the moment bound
and the enclosures rather than take them with headroom.  The dilation falls
from $8$ to $2/\varepsilon_{\mathrm{box}}\approx4.72$, a factor of about
$1.7$.  The round sum falls because the covering froze one percent per round,
about a hundred rounds per unit of $\log(n/m)$, where the refreshed lemma
freezes $\delta=0.0281672$, about $35$ rounds per unit, and because we replace
the bound $2\mathbin{\cdot}400$ on the geometric series by the exact bin
sum.  Of the factor of about $4.5\mathbin{\cdot}10^4$ between the two
constants, the radius accounts for more than $3500$, the dilation for about
$1.7$, and the round sum for the remainder, a factor below ten.  There is also
a trade in the exponent, and it explains why the radius could fall so far.
Section~\ref{sec:first_existence} demands measure at least $e^{-m/1000}$ and
proves $e^{-3\mathbin{\cdot}10^{-6}m}$.  Section~\ref{sec:proof} is content
with $e^{-0.00372m}$, a weaker demand, and Section~\ref{sec:small_ball_proof}
delivers $e^{-0.003712m}$, just inside it, at a radius smaller by the factor
$10^6/279>3500$.  The headroom that Section~\ref{sec:first_existence} left in its
exponent is what we have spent on the radius.

\paragraph{Why the framework stops in the hundreds of thousands.}
Two losses are built into the partial-coloring framework as we use it above, and
no choice of its parameters removes them.  The first is in the step from
Gaussian measure to signs.  Both the covering of
Proposition~\ref{prop:first_covering} and the projection of
Lemma~\ref{lem:refreshed_projection} prove that many coordinates freeze by
ruling out every small set of tight coordinates at once, a union bound that
the measure bound survives only if $\delta$ is small, which makes the rounds
many, and only if the box is small compared with the cube, which makes the
dilation large.  The second loss is in the radius.  The radius coefficient
$c_{\rm rad}$, between $185$ and $279$, is the radius at which
Lemma~\ref{lem:scale_small_ball} proves its small-ball exponent, and a proof
that reached the same exponent at a smaller radius would lower every
increment in proportion.  We can read off how much headroom there is from the
first row of the signing below $7.8795$ (Section~\ref{subsec:sub81_first_row}),
which we certify at radius $\sqrt{n/\eta_0}$ with $\eta_0=0.065$, below
$4\sqrt n$.  The radius of Section~\ref{sec:small_ball_proof} is not what the
body needs but what the split barrier and the diagonal endpoint can prove.
In Section~\ref{sec:existence_constant_sub10} we replace both steps.  We
convert Gaussian measure into signs by a counting argument that pays no union
bound.  To prove the small-ball estimate we integrate along the
amplitude of the matrices, under smooth barriers and with a covariance-tilted
reference Gaussian, at radii $\sqrt{w_kn/\eta_k}$, where $w_kn$ bounds the
active count and $\eta_k$ is a variance parameter, between $0.065$ and $0.19$
in the eleven prefix phases of the signing below $7.8795$.  The cost is that
the counting argument proves that a partial signing exists and computes
nothing, so the constant it gives is an existence constant, and the
algorithms of Sections~\ref{sec:algorithm}--\ref{sec:cubic_time} keep the
framework.

\paragraph{Ingredient two refined: lossless Gaussian-to-sign coding.}
Lemma~\ref{lem:1070_lossless_coding} converts measure into signs without the
union bound.  Let $K\subseteq\R^m$ be symmetric and convex with
$\gamma_m(K)\ge e^{-Em}$, let $0<q<1$, and suppose $\log2-E>h((1-q)/2)$, then
there is $y\in\{-1,0,1\}^m$ with more than $(1-q)m$ nonzero coordinates such
that $s_\star y\in K$, where $s_\star:=0.67448$.  The choice $E=q^2/2$
satisfies the hypothesis.  As a signing step, $y$ signs more than
$(1-q)m$ of the active coordinates outright, and the signed increment lies in
$K/s_\star$, a dilation of $1/s_\star\approx1.48$ in place of $8$ or $4.72$.
The proof is a count, and it never rounds.  For $Z$ with an explicit even
density supported on $[-2s_\star,2s_\star]$ and an independent uniform sign
$\sigma$, the law of $Z+s_\star\sigma$ gives every symmetric convex set at
least its Gaussian measure (Lemma~\ref{lem:1070_peaked_density}).  Averaging
over $z=(Z_1,\ldots,Z_m)$, we find that the expected number of sign vectors
$\sigma\in\{-1,1\}^m$ with $z+s_\star\sigma\in K$ is at least
$e^{m(\log2-E)}$, so some translate $z$ has at least this many.  Kleitman's
cube-diameter theorem~\cite{kleitman66} (see also~\cite[Theorem~1.1]{hkp20})
bounds a family of Hamming diameter $D$ by $e^{mh(D/(2m))}$, and the
hypothesis puts the family above that bound at $D=(1-q)m$, so two members
$\sigma,\sigma'$ differ on more than $(1-q)m$ coordinates.  Their
half-difference $y:=(\sigma-\sigma')/2$ is the partial signing, and
$s_\star y$ is the midpoint of $z+s_\star\sigma$ and $-(z+s_\star\sigma')$,
both in $K$ by symmetry, so it lies in $K$ by convexity.  The entropy the
argument needs is a reserve checked separately for every row, by
Lemma~\ref{lem:sub10_certificate} for the stage rows of
Definition~\ref{def:sub10_parameters} and by the exact entropy tests of
Section~\ref{subsec:sub81_data} for the signing below $7.8795$.  We spend it
once, so a stage signs most of its active coordinates instead of the $2.8$
percent that the projection lemma guarantees.  In the signing below $7.8795$
each of the eleven prefix phases signs more than $77$ percent.  No union
bound, no fractional coloring and no box appear.  The lemma does not compute
the signs, and this is why the algorithms of
Sections~\ref{sec:algorithm}--\ref{sec:cubic_time} cannot use it.

\paragraph{Ingredient one refined: the covariance tilt, the radial interpolation, and smooth barriers.}
With the coding lemma in place, the constant is decided by the radius at which
we can prove the small-ball exponent.  We interpolated from a diagonal model
in Section~\ref{sec:first_existence} and paid twice, for the interpolation and
for the diagonal endpoint.  In Section~\ref{sec:existence_constant_sub10} we
interpolate along the amplitude of the matrices instead, and the endpoint
costs nothing.  Let $Y:=\sum_{i=1}^m\xi_iB_i$ be the rescaled series, with
$\sum_{i=1}^mB_i^2\preceq\eta I$, and let $\Gamma_{i,j}:=\tr[B_iB_j]$ be its
Gram matrix.  The barrier $\phi$ is an even power series whose quadratic term
$\kappa_0u^2$ we move into the reference Gaussian
$\nu:=\mathcal N(0,(I+2\kappa_0\Gamma)^{-1})$.  What remains,
$\psi:=\phi-\kappa_0u^2\ge0$, tilts $\nu$ along the path
$\d\mu_t:=Z_t^{-1}e^{-\tr[\psi(tY)]}\d\nu$ with
$Z_t:=\E_\nu[e^{-\tr[\psi(tY)]}]$, $0\le t\le1$.  At $t=0$ the potential
vanishes and $Z_0=1$ exactly, so there is no endpoint to bound, and at $t=1$
we have $Z_1\le\nu[\|Y\|<1]$.  The quantity we integrate between them is a
first-order identity, not a second derivative,
$-\frac{\d}{\d t}\log Z_t=\frac2t\,\E_{\mu_t}[\tr[g(t^2Y^2)]]$ with
$g(u^2)=u\psi'(u)/2$ (Lemma~\ref{lem:fixed_reference_radial}).  We bound its
ordinary moments with the covariance retained, as
$N^{-1}\E_{\mu_t}[\tr[(tY)^{2j}]]\le\operatorname{df}(j)(\eta t^2)^j\omega_j$
with $\omega_j:=\tr[\Gamma(I+2\kappa_0\Gamma)^{-j}]/(N\eta)\le1$
(Lemma~\ref{lem:1070_covariance_power}), by Harg\'e's convex domination and
the sharp Gaussian noncommutative Khintchine inequality.  All of this is a
statement about $\nu$, and the H\"older determinant transfer
(Lemma~\ref{lem:1070_transfer}) returns it to the standard Gaussian
$\gamma_m$ that the coding lemma needs: with
$Z_\Gamma:=\det(I+2\kappa_0\Gamma)^{-1/2}$ and any $H>1$,
$(Z_\Gamma\,\nu(\mathcal K))^H\le\gamma_m(\mathcal K)^{H-1}\det(I+2H\kappa_0\Gamma)^{-1/2}$.
The two determinants reduce, eigenvalue by eigenvalue of $\Gamma$, to the
scalar profile $L_H(x):=(H\log(1+x)-\log(1+Hx))/((H-1)x)$, which, together
with the rational terms that the $\omega_j$ produce, is capped on the whole
half-line by a certified constant $B_{\rm cap}$.  The exponent that
results is
\[
e=\kappa\eta B_{\rm cap}+bw/2+HC/(H-1)
\]
(Eq.~\eqref{eq:sub10_exponent}), with $\kappa$ the quadratic coefficient
$\kappa_0$, $C$ the part of the radial bound not attached to any $\omega_j$
(a different $C$ from the radius coefficient $C=10^6$ of
Section~\ref{sec:first_existence}), and $b$ a rank correction, zero in every
row of the signing below $7.8795$.  One clause of that lemma carries the
hereditary property: replacing $\eta$ and $w$ by $\alpha\eta$ and $\alpha w$
for $0<\alpha\le1$ scales every term of the exponent by $\alpha$, so a row
certified for families of size $wn$ at radius $\sqrt{wn/\eta}$ gives every
subfamily of size $m\le wn$ Gaussian measure at least $\exp(-me/w)$ at the
same radius.  The radial path needs a barrier whose curvature we can certify
at every pair of spectral values, and this is the second replacement.  In
place of the spectral $0$--$1$ split of Lemma~\ref{lem:scale_small_ball}, a
hard cutoff, we use smooth barriers, even power series with certified rational
coefficients (Definition~\ref{def:sub10_parameters}).  We certify curvature
by an all-pairs inequality on the divided difference $\phi'[u,v]$ at every
pair $-1<u,v<1$ (Eq.~\eqref{eq:1070_curvature_condition}), which is what the
weighted Poincar\'e inequality of Lemma~\ref{lem:1070_weighted_poincare}
consumes.  On each cell we make every polynomial comparison in the Bernstein
basis, where nonnegative coefficients certify the whole cell, and we bisect a
failed cell, never accept it (Definition~\ref{def:rational_certificate_rules}).
We prove the estimate of stage $k$ at radius $\sqrt{w_kn/\eta_k}$, and the
retention, cost and transfer parameters $q_k,c_k,H_k,z_k,B_{{\rm cap},k}$
certify each stage (Lemma~\ref{lem:sub10_certificate}).  The spectral barriers,
allocations and degrees are retained by the centered rows of
Definition~\ref{def:slack_rows}.  The signing below $7.8795$ uses rows of its
own: the first row and the ten residual rows each use a one-slot barrier
(Section~\ref{subsec:sub81_barriers}), the first retaining covariance credit
in its ordinary and derivative moments and the residual rows using
common-envelope trace powers, and the four suffix phases use four explicit
reflected rows of Definition~\ref{def:slack_rows}.  Every scalar inequality
that a stage needs is an all-domain certificate with exact rational data.  The
coefficient tables are part of the proof, and the text is the argument that
they suffice.

\paragraph{The signing below $7.8795$.}
As the active set shrinks, the aspect ratio $n/m$ grows and the best barrier
changes with it, exactly as the six regimes of Table~\ref{tab:six_regimes}
change with $\tau(\ell_{\rm asp})$.  In the sharpened proof
(Section~\ref{subsec:sub81_signing}) we sign in eleven prefix phases, four
suffix phases and a uniform tail.  Its target is $C_0:=7.8795$, and its coding
amplitude for the final signing is $s_+:=0.674489750196$
(Lemma~\ref{lem:slack_refined_coding}).  Every phase's estimate is hereditary.
Phase one is the first row of Lemma~\ref{lem:unit_first_row}, which also
imposes one shared trace constraint through the function $f_3$ of
Section~\ref{subsec:sub81_first_row}, and phases two through eleven are the
ten residual rows of Lemma~\ref{lem:sub81_residual_rows}.  Phase $i$ signs at
norm $L_i\sqrt n$ with $L_i^2s_+^2\eta_i\ge w_{i-1}$, where $w_in$ bounds the
active count after phase $i$: we certify the row at radius
$\sqrt{w_{i-1}n/\eta_i}$, the coding lemma signs at $1/s_+$ times that radius,
and $L_i$ is $\sqrt{w_{i-1}/(s_+^2\eta_i)}$ rounded up to twelve decimal
places (Section~\ref{subsec:sub81_data}).  With $w_0=1$ and $\eta_1=\eta_0=0.065$ the first phase alone costs
about $5.82$ of the $7.88$, against the
$(2\mathbin{\cdot}279/\varepsilon_{\mathrm{box}})\sqrt n$ of the first round
of Section~\ref{sec:proof}.  We certify the residual rows at $\eta_i$
larger than $\eta_1$, because they assume trace caps on the family they
receive, and their retentions $q_i$ are all below $0.23$, so the active
fraction falls geometrically (Section~\ref{subsec:sub81_data}).  Two devices
make the phases add.  The first acts after phase $i$ has produced the signed
sum $M_i$, when we replace every remaining matrix $B$ by
$C_i\sqrt n\,H_i^{-1/2}BH_i^{-1/2}$ with $H_i:=C_{i-1}\sqrt n\,I-|M_i|$ and
$C_i:=C_{i-1}-L_i$.  The residual family stays a family of contractions, and a
signing of it of norm $F\sqrt n$ with $F\le C_i$ lifts back to a signing of
the current one of norm $(L_i+F)\sqrt n$
(Lemmas~\ref{lem:slack_transform} and~\ref{lem:slack_schedule_completion}).
The second is that the trace caps a residual row assumes are supplied by the
phase before it.  Each of the first ten phases asks the signed increment to
satisfy six trace-moment constraints, and phase eleven asks for one.  By
log-concavity each costs a fixed factor $e^{-1}$ of Gaussian mass
(Lemma~\ref{lem:unit_log_moment}), Gaussian correlation lets us impose them
together with the norm body, and the entropy reserve of the coding lemma
absorbs the charges, which are below $8$, $7$ and $2$ in the three kinds of
phase.  The tail is the one place where Section~\ref{sec:small_ball_proof}
enters, through Lemma~\ref{lem:1070_infinite_tail}, the per-row clause of
Lemma~\ref{lem:scale_small_ball} on the row $\tau\ge44$ of
Table~\ref{tab:six_regimes}, and the coding lemma continues the signing to the end.  In the proof of
Theorem~\ref{thm:existence_constant_sub10} we assemble the three parts by
Lemma~\ref{lem:slack_schedule_completion} into its bound,
$B=7.879493715947<C_0$, with a positive reserve.  The target $C_0$ defines the
slack matrices $H_i$ and is not assumed, so the bound has to come in under
it, and it does.  Hence there is a signing of discrepancy at most
$7.879493715947\sqrt n<7.8795\sqrt n$
(Theorem~\ref{thm:existence_constant_sub10}), stated for symmetric $N\times N$
contractions with $1\le N\le n$, with every scalar input listed in
Section~\ref{subsec:sub81_data}.  For $n\le2^{43}$ random signs already give a
signing below $7.82\sqrt n$.

\paragraph{The algorithmic constant.}
We recover in Section~\ref{sec:small_algorithm_constant} a small constant for
the cubic-time algorithm without the coding lemma, with two kinds of
projection.
The near-integral step projects a uniform point of the cube onto the smoothed
body inside a shrunken box.  A width bound for uniform targets in terms of the
small-ball exponent (Lemma~\ref{lem:cubic_refined_width}) shows that the
output is nearly integral in a uniform-weighted sense
(Lemma~\ref{lem:cubic_refined_projection}), so this step needs no count of
tight facets.  We round the remaining fractional mass one binary digit at a time by a
deterministic binary repair at a fixed number of levels, four
auxiliary repair levels above a certified suffix signing
(Lemma~\ref{lem:cubic_refined_complete}) and fourteen in the complete
algorithm (Lemma~\ref{lem:cubic_global_complete}).  Each digit is fixed by a
helper signing drawn with a fresh random orientation so that the repair is
unbiased (Lemma~\ref{lem:algorithm_21_rounding}).  At depth zero the helper is
a certified suffix signing, which projects toward a point sampled from a
Euclidean ball and does count tight facets, more than $m/4$ per accepted trial
(Lemma~\ref{lem:algorithm_centered_box}), using the constant $0.7994$ of that
lemma and the $\delta:=1/4$ of Definition~\ref{def:cubic_refined_suffix}, a
different $\delta$ from the $\delta=0.0281672$ of Section~\ref{sec:proof}.  See
also Lemma~\ref{lem:algorithm_certified_implementation}.  In the suffix signing
we reuse the seven original small-ball estimates of
Section~\ref{sec:existence_constant_sub10}
(Definition~\ref{def:1070:stage-parameters}) and the centered rows of
Definition~\ref{def:slack_rows}, not the stage parameters of
Definition~\ref{def:sub10_parameters}, on capacity-matched suffix schedules of
hereditary radii (Definition~\ref{def:cubic_refined_suffix}) with centered
small-ball rows at the root (Definition~\ref{def:cubic_centered_root}).  For
$n>N_*=10^{30}$, Lemma~\ref{lem:cubic_refined_suffix} bounds the discrepancy
of the suffix at capacity $w_i=(19/20)^i$ by $C_i^{\rm suf}\sqrt n$.  We run
all projections through the certified multiplier wrapper and recursive solver
of Section~\ref{sec:cubic_time}, so the running time stays
$n^{3+o(1)}\operatorname{polylog}(1/p)$ arithmetic operations in the
real-arithmetic model.  The output has discrepancy at most
$11.503394001765\sqrt n<11.504\sqrt n$ with failure probability at most $p$
(Lemma~\ref{lem:cubic_global_complete},
Theorem~\ref{thm:algorithm_constant_13}).  For $n\le N_{\rm root}=10^{25}$ the
main procedure runs the capped independent-sign trials of
Algorithm~\ref{alg:main_cubic_small_constant}.  The algorithm computes
Euclidean projections only, and the polar bodies and coordinate sections of
the analysis never appear in it.
\subsection{From polynomial time to \texorpdfstring{$n^{3+o(1)}$}{n\string^(3+o(1))}}
\label{subsec:overview_ladder}

Our four algorithms take the round's projection down from a convex program
with one spectral constraint.  That program is solvable in polynomial time but far
above $n^3$.  The algorithms bring it down to $n^{3+o(1)}$, and each removes
one obstacle left by the previous one.  The first replaces the spectral
constraint by a smooth one and batches all derivatives through one Gibbs
matrix, and its cost is a dense diagonalization at every block query.  The
second forms the Gibbs matrix densely only at snapshots and sketches its
increment between them, and its exponent is $1+\omega(1,1/2,1)$, which is $3$
only under an open conjecture.  The third shares one sketch across a block of
coordinates and narrows the gap to $\gamma_\star\le0.018320$, and its loss is
that the increment is sketched afresh at every query.  The fourth tracks the
Gibbs matrix across coordinate steps, so that the variance of an update is
paid for by the descent energy of the step.  It reaches $n^{3+o(1)}$ using
only $\omega<5/2$ and $\alpha_{\mathrm{dual}}>1/4$.  A fifth step, not about
the exponent, runs the small-constant version of the fourth algorithm, which
is Section~\ref{sec:small_algorithm_constant}, on words of $O(\log n)$ bits.  In
the four figures of this subsection the total cost covers all partial-coloring
rounds, and we suppress logarithmic accuracy and confidence factors.

\paragraph{The $n^{2+\omega/2}$ algorithm.}
The trajectory needs a point of $K_S\cap Q_y$ within a small distance $\rho$
of the exact projection, not the projection itself.  Two ideas survive in all
four algorithms.  The first is to smooth the spectral constraint, and the
second is to batch all requested derivatives through one matrix function.
With $R:=c_{\rm rad}(\ell_{\rm asp})\sqrt m\,\ell_{\rm asp}$ and
$B(x):=\sum_{i\in S}x_iA_i$, the smooth body is
\[
D:=\{x\in\R^S:h(x)\le1-s\},
\qquad
h(x):=\frac1a\log\bigl(\tr[\exp(aB(x)/R)]+\tr[\exp(-aB(x)/R)]\bigr),
\]
with $a:=2\mathbin{\cdot}10^6\log(2n)$ and $s:=5\mathbin{\cdot}10^{-7}$.  It
satisfies $(1-2s)K_S\subseteq D\subseteq(1-s)K_S$ and
$\gamma_m(D)\ge(1-10^{-6})^m\gamma_m(K_S)\ge\exp(-0.00372m)$, so the
projection lemma applies to $D$ unchanged.  With one Lagrange multiplier
$\lambda\ge0$, a different $\lambda$ from the separation $-\log(1-\delta)$ of
Section~\ref{sec:proof}, we reduce the projection of $\xi$ onto $D\cap Q_y$ to
minimizing the strongly convex $\frac12\|x-\xi\|_2^2+\lambda h(x)$ over the
box.  We find the multiplier $\lambda$
by bisection, which inexact minimization does not derail.  A radial
rescaling toward the origin repairs slight infeasibility, and we snap
coordinates within $2\rho$ of a facet to it, the slack $s$ keeping the snapped
point in $K_S$ (Theorem~\ref{thm:dense_runtime}).  For a block of $\ell$
coordinates the block gradient of $h$ is $\widetilde O(\ell)$-Lipschitz, so
accelerated random block-coordinate descent needs $\widetilde O(m/\sqrt\ell)$
block queries.  One query diagonalizes $B(x)$ in $n^{\omega+o(1)}$ operations
and forms the signed Gibbs matrix $W_x$, the matrix with
$\partial_ih(x)=R^{-1}\tr[W_xA_i]$.  After that, the $\ell$ requested
derivatives are Frobenius products costing $O(\ell n^2)$ in total.  A
projection costs $\widetilde O((m/\sqrt\ell)(n^{\omega+o(1)}+\ell n^2)+mn^2)$,
the balance $\ell\asymp n^{\omega-2}$ gives $mn^{1+\omega/2+o(1)}$ per round,
and $\sum_tm_t=O(n)$ over the rounds gives $n^{2+\omega/2+o(1)}$.  We use
nothing cheaper than the full Gibbs matrix.

\begin{figure}[htbp]
\setlength{\abovecaptionskip}{4pt}
\centering
\includegraphics[width=\linewidth]{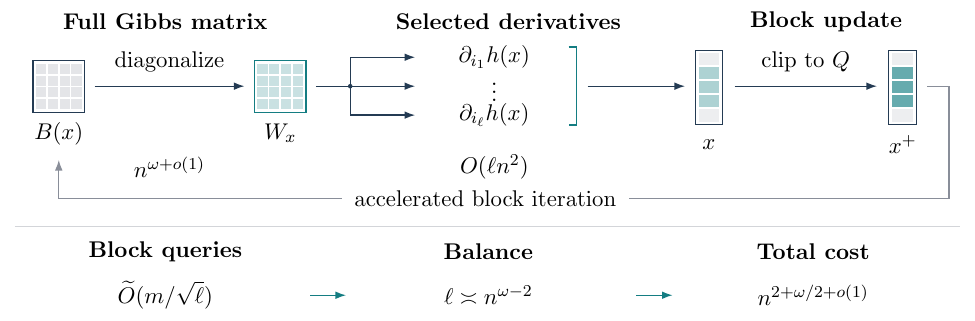}\par
\nointerlineskip
\caption{Full-Gibbs block-coordinate method
(Theorem~\ref{thm:dense_runtime}). One approximate diagonalization of
$B(x):=\sum_{i\in S}x_iA_i$ supplies the signed Gibbs matrix $W_x$ for all
$\ell$ spectral derivatives in the selected block $I$. The columns
schematize selected-block clipping in the coordinate box $Q=Q_y$ within the
accelerated recurrence.  The choice $\ell\asymp n^{\omega-2}$ balances
diagonalization and Frobenius products. Block-query counts are per
fixed-multiplier solve in the dominant rounds.}
\label{fig:full-gibbs-block}
\end{figure}

In Figure~\ref{fig:full-gibbs-block} we show how one Gibbs matrix supplies a
block of derivatives.

\paragraph{The selected-coordinate Gibbs oracle.}
{\setlength{\emergencystretch}{2em}\hbadness=10000 Every query above
constructs a full Gibbs matrix, although a coordinate method consumes one
derivative per step.  Put $C(x):=\operatorname{diag}(B(x)/R,-B(x)/R)$, which
stacks the two sides of the spectral constraint.  Put
$\Phi(x):=\tr[\exp(aC(x)-abI)]$ with $b:=1-s$.  This is a different $b$ from
the exponent parameter of $\omega(1,b,1)$ below.  The function satisfies
$h(x)\le b$ exactly when $\Phi(x)\le1$.  Minimize
$F_\nu(x):=\frac12\|x-\xi\|_2^2+\nu\Phi(x)$ over the box, with minimizer
$x_\nu$.  At a snapshot $w$ we form the Gibbs matrix $P_x:=\exp(aC(x)-abI)$ densely,
to inverse-polynomial accuracy.  The oracle estimates only the
increment $P_x-P_w$.  A Gaussian range finder of rank
$\widetilde\Theta(\sqrt m)$ captures its leading part, and
$\widetilde\Theta(\sqrt m)$ fresh Gaussian probes estimate the deflated
residual.  The generalized matrix Pinsker
inequality~\cite[Theorem~1.3]{bansaletal25panoply} and the Bregman divergence
of $F_\nu$ convert the trace norm of the residual into the average-coordinate
variance bound
\[
\E[|\widehat g_i-\partial_iF_\nu(x)|^2\mid x,w]
\le\frac\eta m(\Delta_x+\Delta_w),
\qquad
\Delta_x:=F_\nu(x)-F_\nu(x_\nu),
\]
for a small universal constant $\eta$.  The factor $1/m$ is the analytic gain.
A loopless variance-reduced coordinate method reduces the objective gap by a
constant factor in $\widetilde O(m)$ selected-coordinate queries with a
bounded expected number of dense snapshot refreshes.  The oracle is finite and
certified.  With squared polynomial actions we replace the exponential by
positive-semidefinite surrogates on a guarded spectral interval.  Fresh
one-sided trace guards keep the iterate on the level set where the variance
bound holds.  We accept candidates and multiplier brackets only from
outward-rounded intervals.  A trial that exhausts its cap returns
\textsc{Fail}, and we charge its probability to the budget $p$.  One query applies
the surrogates to $\widetilde O(\sqrt m)$ vectors and reconstructs two dense
matrices from rank-$\widetilde O(\sqrt m)$ factors.  It therefore costs
$\Tmat(n,n,\lceil\sqrt m\rceil)+\Tmat(n,\lceil\sqrt m\rceil,n)+n^2$ up to
subpolynomial factors.  The bound $\sum_tm_t=O(n)$ with the monotonicity of
$\Tmat$ gives $n^{1+\omega(1,1/2,1)+o(1)}<n^{3.042776+o(1)}$
(Theorem~\ref{thm:faster_dense_runtime}).  The speedup comes from a cheaper
certified Gibbs oracle, not from a new acceleration theorem.\par}

\begin{figure}[htbp]
\setlength{\abovecaptionskip}{4pt}
\centering
\includegraphics[width=\linewidth]{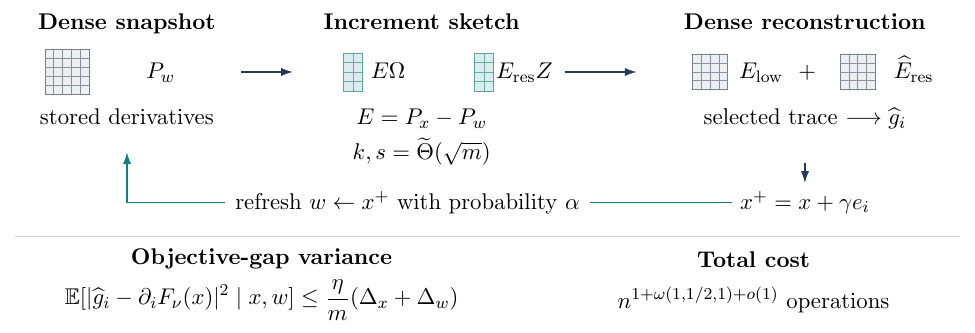}\par
\nointerlineskip
\caption{Selected-coordinate Gibbs oracle in the ideal exact-action analysis
(Theorem~\ref{thm:faster_dense_runtime}). Gaussian range and residual probes
$\Omega,Z$ sketch $E:=P_x-P_w$ with widths $2k$ and $s$.  Here $\Omega$ is
the Gaussian range finder of width $2k$, and $Z:=[z_1,\ldots,z_s]$ collects
the $s$ residual probes.  This $s$ is not the slack $s$ of the smooth body.
The dense residual reconstruction is
$\widehat E_{\rm res}:=E_{\rm res}ZZ^\top/s$.  Its selected trace combines
with $E_{\rm low}$ and the stored snapshot derivative.  The refresh rate is
$\alpha:=1/(m\overline L)$, with $\overline L$ the coordinate curvature
including the trust-step correction, and
$\Delta_z:=F_\nu(z)-F_\nu(x_\nu)$. The snapshot stays fixed between
refreshes.  The variance bound is conditional on $x,w$ and averages over
the uniformly selected coordinate and oracle randomness on the guarded
level.}
\label{fig:selected-coordinate-oracle}
\end{figure}

Figure~\ref{fig:selected-coordinate-oracle} shows the snapshot and increment
calculations used for a selected derivative.

\paragraph{The accelerated block algorithm.}
The exponent $1+\omega(1,1/2,1)$ equals $3$ only if the dual exponent
$\alpha_{\mathrm{dual}}:=\sup\{b:\omega(1,b,1)=2\}$ is at least $1/2$, which
is open.  We narrow the gap in Section~\ref{sec:accelerated_block_algorithm}.  One
sketch of width $\widetilde O(\sqrt{m/\ell})$ serves a block of
$\ell=n^{1-2b}$ coordinates.  Catalyst acceleration~\cite{linetal18catalyst}
with regularization $\kappa=\ell$ brings the block queries to
$\widetilde O(m/\sqrt\ell)$ at the cost of $\widetilde O(\sqrt\ell)$ dense
proximal subproblems.  A safe-level invariant keeps the extrapolated centers
where the sketch variance bound holds.  Balancing
$(n/\sqrt\ell)\Tmat(n,n,\lceil\sqrt{n/\ell}\rceil)$ against $n^3\sqrt\ell$,
we obtain $3+\gamma_\star$, $\gamma_\star\le0.018320$ at $b_\star=0.48168\ldots$
(Theorem~\ref{thm:accelerated_dense_runtime}).  The loss $\gamma_\star$ is the
cost of sketching $P_x-P_w$ afresh at every query.

\begin{figure}[htbp]
\setlength{\abovecaptionskip}{4pt}
\centering
\includegraphics[width=\linewidth]{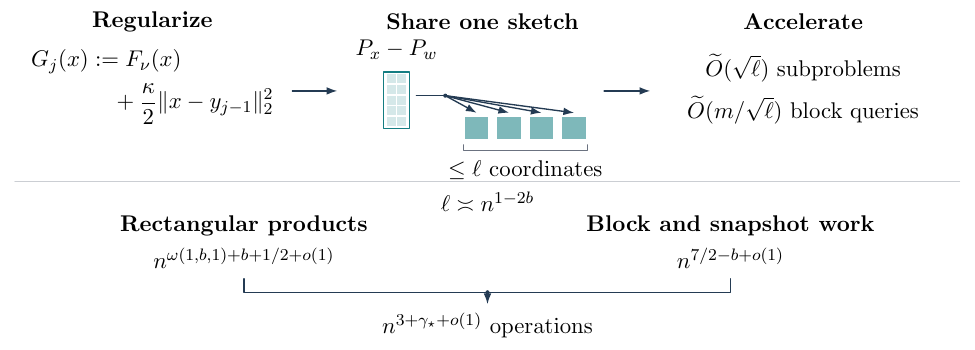}\par
\nointerlineskip
\caption{Accelerated block method
(Theorem~\ref{thm:accelerated_dense_runtime}). Here $\kappa:=\ell$,
$y_{j-1}$ is the Catalyst~\cite{linetal18catalyst} center, and $w$ is the
snapshot. A fresh range sketch and residual probes of the ideal difference
$P_x-P_w$, with widths $\widetilde O(\sqrt{m/\ell})$, serve the whole
block. The upper counts are per fixed-multiplier solve in the accelerated
regime. The safe-level invariant keeps accepted iterates on the guarded
Gibbs trace level.  The block size is
$\ell:=\max\{2,\lceil n^{1-2b}\rceil\}$.  Choosing $1/3\le b\le1/2$
balances the lower costs over all partial-coloring rounds.  The loss
$\gamma_\star$ minimizes the larger exponent minus three.}
\label{fig:accelerated-blocks}
\end{figure}

We show in Figure~\ref{fig:accelerated-blocks} the tradeoff between shared
block sketches and dense proximal solves.

\paragraph{Recursive Gibbs tracking.}
In Section~\ref{sec:cubic_time} we instead track the Gibbs matrix across
coordinate steps.  There we maintain a dense $\widehat P_t$ and, after the step
$x_{t+1}=x_t+d_te_{i_t}$, add an unbiased range-finder-and-probe estimate
$\widehat\Delta_t$ of $P_{x_{t+1}}-P_{x_t}$ whose variance is proportional to
$d_t^2$ (Lemma~\ref{lem:cubic_local_increment}).  The squared step length is
exactly what the negative step-energy term of coordinate descent pays for.  An
independent refresh $\widehat P_{t+1}:=P_{x_{t+1}}$ with probability $\pi$
keeps the error from accumulating.  With $\mathcal E_t$ the tracking error in
the gradient seminorm and $L$ the coordinate curvature, the coupled potential
$V_t:=\Delta(x_t)+\mathcal E_t/(\pi L)$ contracts,
\[
\E_t[V_{t+1}]\le(1-\alpha)V_t-\frac L8\E_t[d_t^2],
\qquad
\alpha:=\frac1{mL}
\]
(Lemma~\ref{lem:cubic_lyapunov}).  We build the guard around the level
$H:=64U$, $U:=F_\nu(0)$.  A phase starts from a certified point $z$ with
$F_\nu(z)\le H/16$.  We stop the ideal process at the first proposal above
$5H/16$, which happens with probability at most $1/20$.  Every proposal gets a
fresh certificate accurate to $H/48$, and we reject it above $H/3$.  We cap
the refreshes, for a universal constant $C$ and a per-phase failure
probability $\delta_{\mathrm{ph}}$, at
$M_{\mathrm{ref}}=\lceil C(T\pi+\log(1/\delta_{\mathrm{ph}}))\rceil$.  A phase
of $T=\lceil16/\alpha\rceil$ queries then halves the objective gap with a
universal positive probability (Lemma~\ref{lem:cubic_tracked_phase}).  Then
$J$ halving phases with repetition and certified comparison return an
$\varepsilon_{\mathrm{opt}}$-accurate solution.  Here
$\varepsilon_{\mathrm{opt}}$ is the accuracy parameter we introduce in
Section~\ref{sec:faster_algorithm}.  The solution comes with the interval
certificates the multiplier wrapper requires
(Lemma~\ref{lem:cubic_recursive_multiplier}).  With $\pi$ of order $n^{-1/2}$
and sketch width $n^{1/4+o(1)}$, a nonrefresh query costs $n^{2+o(1)}$ because
$\alpha_{\mathrm{dual}}>1/4$.  A refresh costs $n^{\omega+o(1)}$ at rate
$n^{-1/2}$, and $\sum_tm_t=O(n)$ gives
$(n^3+n^{\omega+1/2}+n^\omega)n^{o(1)}=n^{3+o(1)}$ because $\omega<5/2$
(Lemma~\ref{lem:cubic_global_runtime}, Theorem~\ref{thm:cubic_dense_runtime}).

\begin{figure}[htbp]
\setlength{\abovecaptionskip}{4pt}
\centering
\includegraphics[width=\linewidth]{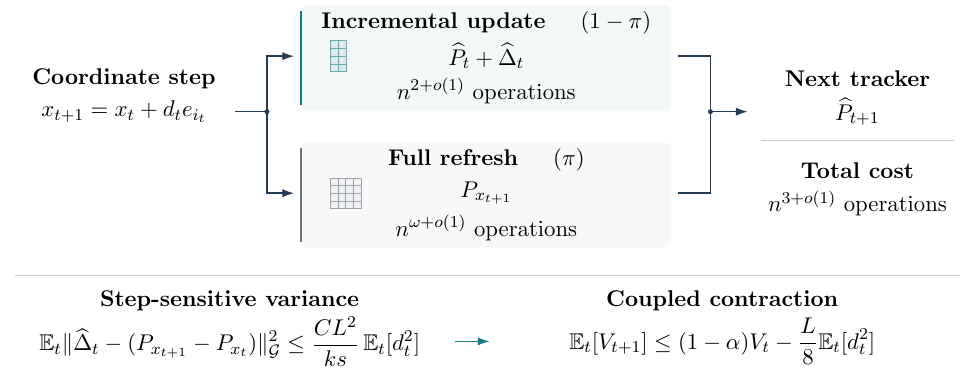}\par
\nointerlineskip
\caption{Recursive Gibbs tracking in the ideal exact-action analysis
(Lemmas~\ref{lem:cubic_local_increment} and~\ref{lem:cubic_lyapunov}).
The step-sensitive variance is absorbed by the descent energy.
Here $\Delta(x):=\Delta_x$,
$\mathcal E_t:=\|\widehat P_t-P_{x_t}\|_{\mathcal G}^2$, and
$V_t:=\Delta(x_t)+\mathcal E_t/(\pi L)$.  The seminorm
$\mathcal G$ is the coordinate-gradient seminorm, and $L$ is the coordinate
curvature.  The remaining parameters are $\alpha:=1/(mL)$ and
$\pi=\pi_m:=\min\{1,\max\{n^{-1/2},4\alpha\}\}$.
The range-finder rank and residual-probe count are
$k=s=\lceil C\pi_m^{-1/2}\rceil=O(n^{1/4})$, with $C$ large enough that
$ks\ge K_0/\pi_m$. Conditional expectations are given the history before
choosing $i_t$.
The displayed costs suppress
$\operatorname{polylog}(1/p)$ confidence factors and use
$\alpha_{\mathrm{dual}}>1/4$ and $\omega<5/2$
(Lemma~\ref{lem:cubic_global_runtime}).}
\label{fig:recursive-gibbs-tracking}
\end{figure}

Figure~\ref{fig:recursive-gibbs-tracking} shows the two tracker updates and
the variance estimate that links them to coordinate descent.

\paragraph{Fixed-point words.}
The real-arithmetic analysis never controls three things that a finite word
forces us to control.  They are decisions taken on rounded quantities, random
objects that cannot be drawn exactly, and the size of the tracked matrix,
which the gradient seminorm does not see.  In Section~\ref{sec:bit_complexity}
we specify a finite version of the small-constant algorithm of
Section~\ref{sec:small_algorithm_constant} as a list of replacements of its
primitives.  We prove one lemma per primitive and assemble them
(Theorem~\ref{thm:bit_complexity}).  Every stored number, apart from a
constant number of $2W$-bit scalars of the solve, is a fixed-point word of
$W:=I+f$ bits.  We form products in temporaries of $O(W)$ bits and round them
once.  Here $f$ is the number of fractional bits and $I$ the number of integer
bits.  This $I$ is a bit count, not the identity matrix.  Both are given by
explicit formulas in $\log_2(1/\varepsilon_{\mathrm{opt}})$,
$\log_2(H_{\max}/\underline\nu)$, $\log_2(2n)$, the top spectral scale
$\lambda_{\mathrm{top}}$, the spectral bound $M_{\mathrm{sp}}$, and the target
cap $c_B=10^{108}$ of that section.  We evaluate the formulas at one
projection that dominates every projection the algorithm performs, so that
$W=O(\log n)$ independent of $p$, about $10^4$ bits at $n=10^{30}$.  The
projection engine is the wrapper of Section~\ref{sec:faster_algorithm} driving
the solver of Section~\ref{sec:cubic_time}, made finite.  The list runs from
the input rounding to $b_{\mathrm{in}}:=2\lceil\log_2n\rceil+10$ bits to the
three fair-sign fallbacks, and the bridge to the real-arithmetic proof is a
few explicit constants we fix there.  The total is
$n^{3+o(1)}\operatorname{polylog}(1/p)+\widetilde O(n^3L_{\mathrm{in}})$ bit
operations on inputs of bit length $L_{\mathrm{in}}$
(Theorem~\ref{thm:bit_complexity_informal}), with $\alpha_{\mathrm{dual}}>1/4$
and $\omega<5/2$ taken in the asymptotic-rank sense of
Definition~\ref{def:bit_rank_exponent}, because the fast products need exact
bilinear algorithms with rational coefficients.

\paragraph{What is shared and what is not.}
The four algorithms follow the same pipeline.  There are $O(\log n)$ rounds
with active dimensions $m_t$ and $\sum_tm_t=O(n)$.  Per round there is a
bounded expected number of projections, each trial succeeding with probability
greater than $0.1$, and per projection there are polylogarithmically many
multiplier tests.  The cost per fixed-multiplier solve is what we compute in each of
Sections~\ref{sec:algorithm}--\ref{sec:cubic_time}.  The repetitions
that amplify the constant success probability to $1-p$ and the union bound
over all randomized calls give the factor $\operatorname{polylog}(1/p)$.  We
prove the outer partial-coloring trajectory once, and likewise the smoothed body $D$ with its constants
$a$ and $s$.  Here the second constant is the smoothing parameter of
Section~\ref{sec:algorithm}, not the probe count $s$ of
Section~\ref{sec:cubic_time}.  We also prove once the radial repair and snapping rule, the
terminal rounding with its $6.5\sqrt n$, and the discrepancy calculation.  All of these proofs are in Sections~\ref{sec:proof}
and~\ref{sec:algorithm}.  The
algorithms of Sections~\ref{sec:faster_algorithm}--\ref{sec:cubic_time}
further share the certified multiplier wrapper, the finite
positive-semidefinite surrogates, and the guard-and-certificate discipline of
Section~\ref{sec:faster_algorithm}.  They differ in one place, the solver of
the smooth projection.  We run accelerated
block-coordinate descent on a dense Gibbs matrix in Section~\ref{sec:algorithm}.
In Sections~\ref{sec:faster_algorithm}--\ref{sec:cubic_time} we change only the
fixed-multiplier part of the solver, each section with its own variance bound,
level-set invariant, and cost count.  The bit-complexity analysis of
Section~\ref{sec:bit_complexity} keeps the wrapper and the guards and replaces
the polynomial surrogate by a scaled Taylor power
(Lemma~\ref{lem:bit_taylor}).  There is no general theorem behind the four,
and each section proves its own.

\paragraph{Open questions.}
What is the smallest constant this route can give?  The existence bound
$7.879493715947$ of Section~\ref{sec:existence_constant_sub10} is still larger
than Spencer's $6$ in the diagonal case~\cite{spencer85}.  For the
projection-based algorithms, a smaller radius coefficient
$c_{\rm rad}(\ell_{\rm asp})$ at the same exponent, or a projection lemma with
a larger $\delta$, would lower the discrepancy bound.  Can the running time be
made proportional to the number of nonzero input entries, given that a dense
Gibbs matrix is what all four solvers maintain in some form?  Is there a
deterministic polynomial-time algorithm with an explicit constant, all four
algorithms here being randomized in the Gaussian draw and in their solvers?
What is the bit complexity of the $n^{2+\omega/2}$ algorithm, whose
diagonalization step we analyze only in real arithmetic?

\begingroup\setlength{\emergencystretch}{2em}There is a gap of roughly $3.6$ between the existence constant of
Section~\ref{sec:existence_constant_sub10}, which is $7.879493715947$, and the
algorithmic constant of Section~\ref{sec:small_algorithm_constant}, which is
$11.503394001765$.  We do not know how to close it.  Is there a fine-grained
lower bound, constants $c,c'>0$ such that every algorithm finding a signing of
discrepancy at most $c\sqrt n$ needs $\Omega(n^{3+c'})$ time in the worst
case?\par\endgroup

\subsection{The lower bound}
\label{subsec:overview_lower_bound}

Our upper bounds give the constants $7\cdot10^9$, $156000$ and
$7.879493715947$.  In Section~\ref{sec:lb_lower_bound} we bound the same
constant from below.  Let $D_n^{\mathrm{mat}}$ be the supremum of the discrepancy
$\min_{\epsilon\in\{-1,1\}^n}\|\sum_{i=1}^n\epsilon_iA_i\|$ over all families
of $n$ real symmetric matrices with $\|A_i\|\le1$ in dimension at most $n$
(Definition~\ref{def:lb_dmat}).  There we prove that the lower constant
is at least $2$ in the limit, by an existence argument with an explicit
finite-$n$ rate, and give two explicit families at $n=8$ and $n=16$ with
exact discrepancies and an explicit family at constant $\sqrt2$.  For even $n\ge4$ we have
$D_n^{\mathrm{mat}}\ge(2-\delta_n)\sqrt n+1$, and for odd $n\ge5$ we have
$D_n^{\mathrm{mat}}\ge(2-\delta_{n-1})\sqrt n$, where $\delta_n\to0$ with
$\delta_n=O(n^{-1/5}(\log n)^{3/5})$ (Corollary~\ref{cor:lb_constant}).
With Theorem~\ref{thm:existence_constant_sub10}, the least universal
constant lies in $[2,\,7.879493715947]$ (Remark~\ref{rem:lb_scope}).  Two
published inputs enter the existence theorem and its explicit rate, and we
quote them where they enter, Talagrand's convex distance inequality
(Theorem~\ref{thm:lb_talagrand}) and the norm bound of Bandeira and van
Handel (Theorem~\ref{thm:lb_bvh}).  Every other step of those proofs is on
the page.  The asymptotic Proposition~\ref{prop:lb_rate_two_sevenths} adds
one imported limit theorem (Theorem~\ref{thm:lb_ss}) and nothing else.

\paragraph{The hard distribution.}
The hard instance is a distribution, not a family.  Let $n\ge4$ be even.
Take the complete graph $K_n$ on the vertex set $\{\infty\}\cup\mathbb Z_{n-1}$
and split its $N:=n(n-1)/2$ edges into the $n-1$ perfect matchings
$\mathcal M_r$, $r\in\mathbb Z_{n-1}$, where $\mathcal M_r$ pairs $\infty$
with $r$ and pairs $r+j$ with $r-j$ for $1\le j\le(n-2)/2$
(Definition~\ref{def:lb_matchings}).  Every edge of $K_n$ lies in exactly
one matching (Lemma~\ref{lem:lb_model}).  Draw one independent uniform sign
$x_e$ per edge.  For $r\le n-1$ let $A_r$ carry $x_e$ at the two positions
of each edge $e\in\mathcal M_r$ and $0$ elsewhere, and let $A_n:=I_n$.  Each
$A_r$ is a symmetric signed permutation matrix with $A_r^2=I_n$, so
$\|A_r\|=1$ exactly.  Fix a signing $\epsilon=(\epsilon',\epsilon_n)$ and
put $W_{\epsilon'}:=\sum_{r=1}^{n-1}\epsilon'_rA_r$.  Since every edge lies
in exactly one matching, $W_{\epsilon'}$ has zero diagonal and carries at
the edge $e$ the sign $x_e$ times the sign of the matching containing $e$,
and this change of variables preserves the uniform measure on $\{-1,1\}^N$.
So for every fixed signing, $W_{\epsilon'}$ is a Rademacher Wigner matrix,
real symmetric with zero diagonal and entries above the diagonal independent
and uniform in $\{-1,1\}$, and
$\sum_{i=1}^n\epsilon_iA_i=W_{\epsilon'}+\epsilon_nI_n$
(Lemma~\ref{lem:lb_model}).  The matrices $W_{\epsilon'}$ for different
signings are dependent, so we use one signing at a time and a union
bound.

\paragraph{One signing at a time.}
Fix $\epsilon'$, write $W$ for $W_{\epsilon'}$ and $\widehat W:=W/\sqrt n$,
and put $a:=2-\delta$.  The goal is $\lambda_{\max}(W)>a\sqrt n$ with
probability so close to one that a union bound over the $2^{n-1}$ signings
$\epsilon'$ still leaves a realization of $x$ (Lemma~\ref{lem:lb_union}).
We read the largest eigenvalue through the hinge functional
$F:=\frac1n\sum_i(\lambda_i(\widehat W)-a)_+$, where $u_+:=\max(u,0)$, which
vanishes exactly when $\lambda_{\max}(\widehat W)\le a$
(Lemma~\ref{lem:lb_edgetail}).  As a function of the $N$ edge signs, $F$ is
convex and $(\sqrt2/n)$-Lipschitz (Lemma~\ref{lem:lb_hinge}).  Talagrand's
convex distance inequality gives two-sided tails around a median for convex
Lipschitz functions of uniform signs (Lemma~\ref{lem:lb_tails}), and for $F$
with median $m_F$ the lower tail reads $\Pr[F\le m_F-s]\le2e^{-n^2s^2/32}$.
Hence $\Pr[\lambda_{\max}(W)\le a\sqrt n]\le\Pr[F=0]\le2e^{-n^2m_F^2/32}$
once $m_F>0$.  This is the $\exp(-cn^2)$ speed we need for the union bound.  The
functional $F$ has Lipschitz constant $\sqrt2/n$ over $N=n(n-1)/2$
coordinates while its median is of order one in $\delta$.  The sign
$\epsilon_n$ of the identity is not in the union.  Applying the same bound
to $-\epsilon'$, we obtain $\lambda_{\min}(W_{\epsilon'})<-a\sqrt n$, and the
identity adds one in whichever direction $\epsilon_n$ points, so every
signing has $\|\sum_{i=1}^n\epsilon_iA_i\|>(2-\delta)\sqrt n+1$.  In
general $\|-S\|=\|S\|$ pairs the $2^n$ signings, so one sign is always
free.  We now need a lower bound on the median $m_F$.  The upper tail
puts the median at most $\sqrt{32\pi}/n$ below the mean
(Lemma~\ref{lem:lb_median}), and we bound the mean from below by the moment
method.  The even moment $\E[\tr W^{2p}]$ counts the closed walks of length
$2p$ in $K_n$ without loops in which every edge is traversed an even number
of times (Lemma~\ref{lem:lb_moments}).  The walks that trace a tree on
$p+1$ distinct vertices, one for each Dyck path of semilength $p$ and each
ordered list of $p+1$ vertices, give the Catalan floor
$\E[\frac1n\tr\widehat W^{2p}]\ge C_p\prod_{j=1}^p(1-j/n)$, with $C_p$ the
Catalan number.  A chord comparison of $u\mapsto u^{2p}$ on $[a,M]$ turns a
lower bound on the $2p$-th moment, truncated at a level $M$, into a lower
bound on $\E[F]$ (Lemma~\ref{lem:lb_chord}).  The truncation is where the
quoted bound of Bandeira and van Handel enters.  It controls the norm from
above only, $\E[\|W\|]\le\sqrt n\,(2+\eta_0)$ for the explicit $\eta_0$ of
Theorem~\ref{thm:lb_existence}, and with the upper tail of
Lemma~\ref{lem:lb_tails} for $\|W\|$ it bounds the part of the moment above
$M$ (Lemma~\ref{lem:lb_tailmoment}).  Proposition~\ref{prop:lb_edgemean}
assembles these into $\E[F]\ge L$ with every constant written.  We use no
semicircle law and no limit theorem.  The displayed hypothesis of
Theorem~\ref{thm:lb_existence}, Eq.~\eqref{eq:lb_star}, is equivalent to the condition
$2^{n-1}\cdot2e^{-n^2(L-\sqrt{32\pi}/n)^2/32}<1$, which closes the union
bound.

\paragraph{The rate and the explicit families.}
The admissible $\delta$ form the open interval $(\delta_n,2)$ or the empty
set (Corollary~\ref{cor:lb_constant}), and
$\delta_n=O(n^{-1/5}(\log n)^{3/5})$ (Proposition~\ref{prop:lb_rate}).
In Table~\ref{tab:lb_delta} we print upper bounds on $\delta_n$, every entry
admissible or rounded up from an admissible value: $2-\delta_n\ge1.0461$ at
$n=10^5$, $1.2952$ at $n=10^6$, $1.7790$ at $n=10^9$ and $1.9381$ at
$n=10^{12}$, and the bound is vacuous at $n=10^3$.  The theorem beats the
trivial floor (constant $1$) from about $n=10^5$ and beats the explicit
constant $\sqrt2$ of Theorem~\ref{thm:lb_field} from about $n=10^7$
(Remark~\ref{rem:lb_reading}).  If we replaced the mean of $F$ by its
semicircle value, a heuristic and not a theorem, the same concentration and
union bound would give about $4.98\,n^{-1/5}$, and the explicit theorem is
within the factor $(\log n)^{3/5}$ of that.  About seventy percent of the
loss over this heuristic, at every $n$ of the table, is the finite-$p$
moment method.  The explicit exponent $1/5$ comes from Talagrand's inequality with the
global Lipschitz constant of the hinge functional.  Its gradient on the
upper half of the space improves the exponent to $2/7$ in the asymptotic
register, at the cost of one imported limit theorem
(Proposition~\ref{prop:lb_rate_two_sevenths}), and whether the rate can
be of order $n^{-1/3}$ is open (Remark~\ref{rem:lb_scope}).  Theorem~\ref{thm:lb_existence} names a
distribution and certifies a sign pattern by existence.  The explicit
families of Section~\ref{sec:lb_lower_bound} are exact at their $n$, and
the theorem does not improve them at any enumerable $n$.  At $n=8$, eight
three-letter Kronecker words in the letters $I,X,Z,J$ have
$\|\sum_{i=1}^8\epsilon_iA_i\|=4+\sqrt2$ for every signing, so
$D_8^{\mathrm{mat}}\ge4+\sqrt2=1.914213562\ldots\times\sqrt8$
(Proposition~\ref{prop:lb_eight}).  At $n=16$, sixteen four-letter words
have discrepancy $1+\sqrt{z^*}$ with $z^*$ the largest root of an explicit
quartic, so $D_{16}^{\mathrm{mat}}/4\ge1.962541779\ldots$, and the proof is
an exact enumeration certificate over the $32768$ signings with
$\epsilon_1=+1$, run twice (Theorem~\ref{thm:lb_sixteen}).  For every
$n=2^k$, the family $\{P_{x,\,x^3+x}:x\in\mathrm{GF}(2^k)\}$ of
Theorem~\ref{thm:lb_field} has $\tr S^2=n^2$ and $\frac1n\tr S^4=2n^2$ for
every signed sum $S=\sum_x\epsilon_xP_{x,x^3+x}$, hence
$\|S\|\ge\sqrt{2n}=1.414213\ldots\times\sqrt n$ and
$D_n^{\mathrm{mat}}\ge\sqrt2\,\sqrt n$.  At $n=10^3$ the explicit families
give the bound, and the theorem passes $\sqrt2$ only from about $n=10^7$.
An explicit family at constant $2-o(1)$ is not known
(Remark~\ref{rem:lb_scope}).

\section*{Acknowledgment \& AI Disclosure}
\addcontentsline{toc}{section}{Acknowledgment \& AI Disclosure}

On July 28, 2026, the authors started prompting AI models for a proof of the
matrix Spencer conjecture.  The authors obtained the first proof
(Section~\ref{sec:first_existence}) on August 20, 2026. On August 25, 2026, the authors obtained an $n^{3.5}$-time SDP algorithm for the result of Section~\ref{sec:first_existence}. After that, the
authors aimed at improving the running time to almost- or nearly-linear in
the input size, and at achieving as small a constant as possible.  The hope is
to get a constant below 6, so that one can claim ``six standard deviations
suffice'' as in Spencer's original work. The paper was uploaded to arXiv
since the authors believe that the current techniques could not achieve
further significant improvements on the constant.  The AI tools used in the
preparation of
this paper are ChatGPT Pro 5.6, Codex Sol 5.6, Claude Fable 5 and Claude Fable
5.1.  The proof of the existence of such a signing comes mainly from AI.  The
majority of the ideas for improving the constants down to below 8 come
from the authors.

The authors developed the main ideas for the efficient algorithms.  In
particular, the $n^{2+\omega/2}$-time algorithm was developed by the authors.
The algorithms with running times $n^{1+\omega(1,1/2,1)}$ and $n^{3.018}$ were
designed by the authors prompting the LLMs with ideas from~\cite{cls19}.  The
final $n^{3+o(1)}$-time algorithm comes mainly from the authors as well.

The bit complexity of the small-constant cubic-time algorithm is analyzed by
LLMs.  The authors thank Victor Reis for very insightful discussions regarding the lower bound; these insights were passed to Claude Fable 5.1 for the final lower-bound proof. The proofs have been rewritten by the authors for necessary technical details, and
have been carefully verified by the authors, who take full responsibility for
their correctness.

The authors have tried to simplify the presentation.  However, because of the
number of results, and because the proofs have to track and optimize every
constant explicitly, the length could not be reduced significantly without
losing technical clarity. A detailed technical overview is therefore given in
Section~\ref{sec:overview} to explain the main ideas of the approach.  Readers
interested in a short, self-contained proof are encouraged to read
Section~\ref{sec:first_existence}. The authors would also like to note that the current techniques also lead to the square-root dependence on the logarithm of the aspect ratio that Kathuria obtained, but it does not help with improving the constant.  Since the paper already exceeds 300 pages, the authors chose not to include that result and its proof.

The authors want to make two remarks to close this disclosure.  First, the timeline above is recorded for
clarity, not to establish precedence.  In the age of advanced LLMs for
mathematics, the authors believe that the purpose of such a project is to
map a problem's full landscape end to end, with constants, lower bounds, algorithms and bit complexity. Second, although
the authors' prompting may have helped the LLMs solve these problems, for
instance in obtaining the $n^{3+o(1)}$-time algorithm and the constant below
$8$, the authors regard the LLMs as the main workhorse behind the results.

\section{Comparison with the concurrent works}
\label{sec:comparison}
We compare our result with two independent and concurrent works by~\cite{as26} and~\cite{k26} that also resolve the matrix Spencer conjecture.

\paragraph{The major differences.}
Both concurrent works prove the conjecture independently of us, and
each has something we do not.  Akbas and Sra prove a
non-hereditary small-ball estimate with explicit constants by rank-one
pinchings, their Theorem~1.4.  Kathuria proves the rectangular case with the
optimal square-root dependence on the logarithm of the aspect ratio.  On our side there are 4 the differences.  The constants: we
prove $7\cdot10^9$, $156000$ and $7.879493715947$, and our cubic-time
algorithm carries a constant below $11.504$, where Akbas and Sra leave the
constant unspecified and Kathuria states $10^8$.  The algorithms: four
algorithms with the running-time exponents $2+\omega/2$, $1+\omega(1,1/2,1)$,
$3+\gamma_\star$ and $3+o(1)$, and a bit-complexity analysis, where both
concurrent works give randomized polynomial-time algorithms without a stated
exponent.  The mechanisms: three sign steps, of which the covering of
Section~\ref{sec:first_existence} and the lossless coding of
Section~\ref{sec:existence_constant_sub10} appear in neither work, and a
small-ball proof in Section~\ref{sec:existence_constant_sub10} that
integrates along the amplitude of the matrices with the reference Gaussian
fixed.  The lower bound: in Section~\ref{sec:lb_lower_bound} we prove that the
constant is at least $2-o(1)$, with explicit families at $n=8$ and $n=16$
whose discrepancies we compute exactly.  Neither
concurrent work bounds the constant from below.  Sections~\ref{sec:small_ball_proof} and~\ref{sec:proof} and the paper of
Akbas and Sra follow the same route, modulo constants.  With Kathuria the
common ground is the statement proved.  In the paragraphs below we take the comparison part by part.

\paragraph{Akbas and Sra: an estimate of the same type, a different sign step.}
Akbas and Sra~\cite{as26} prove the conjecture independently and
concurrently.  Their preprint of August 28, 2026 appeared after
we completed the proof of Section~\ref{sec:first_existence}.  Their proof and our Sections~\ref{sec:small_ball_proof} and~\ref{sec:proof}
follow the same route, modulo constants.  Their proof and our
Section~\ref{sec:first_existence} have small-ball estimates of the same type
and different sign steps.  Their first step is a
Gaussian small-ball estimate~\cite[Theorem~1.3]{as26} of the same type as
our Proposition~\ref{prop:first_small_ball}: it holds for every active set of
every size, the property that both papers call hereditary.  In the notation of
Section~\ref{sec:overview}, it states that for every active set $S$ of size
$m$, their $n$, in ambient dimension $N$, their $d$, the spectral body of
radius $\kappa\sqrt m\,(1+\log(2N/m))^2$ has Gaussian measure at least
$e^{-c_*m}$.  This holds for any $c_*>0$ and a constant $\kappa=\kappa(c_*)$
that is not computed.  Its proof and our proof of
Proposition~\ref{prop:first_small_ball} run through the same steps.  These steps are the split
log-barrier tilt, the interpolation from the diagonal model through the second
derivative of the log-partition function, the spectral split, and the
matrix-weighted Poincar\'e inequality that this split feeds.  The spectral
split separates an edge component absorbed by the curvature of the barrier
from a regular component controlled by the variance proxy.  Their second
step~\cite[Theorem~1.2]{as26} feeds this estimate, at $c_*=\log2$, into a
partial-coloring theorem cited from Reis and Rothvoss~\cite{reisrothvoss22}
and from Dadush, Jiang, and Reis~\cite{djr22}.  That theorem freezes at least
half of the active coordinates per round inside an unspecified dilate of the
body.  There are four differences relative to
Section~\ref{sec:first_existence}.
The first difference is the aspect ratio.  Their radius is quadratic in the
logarithm of the aspect ratio, against a linear dependence in
Proposition~\ref{prop:first_small_ball}.  In Section~\ref{sec:first_existence},
and in Section~\ref{sec:small_ball_proof} after it, we average the
independent-direction factors of the moment bound through a completely
positive map, the map $U\mapsto\sum_iR_iUR_i$ of
Section~\ref{subsec:overview_constant}.  The moment bound therefore carries
the square of the moment order per word, $N(8\tau^2\rho)^\tau$ in
Section~\ref{sec:first_existence} and $N(4\tau^2\rho)^\tau$ in
Section~\ref{sec:small_ball_proof}, against $d(Cq^3\eta)^q$ in their text.
Here $q$, $\eta$ and $d$ stand for the moment order $\tau$, the variance proxy
$\rho$ and the ambient dimension $N$.  With $\tau$ proportional to $\ell$, our
curvature check and moment term are uniform in the aspect ratio.
The second difference is the constants.  Their radius constant $\kappa(c_*)$
is not computed, and their exponent is any positive constant, used at $\log2$.
The loss bound of their partition function is closed with universal
constants.  In contrast, we fix $C=10^6$,
moment order $\tau=\lceil8\ell\rceil$, edge cutoff $w=1/100$ and exponent
$1/1000$ at the outset of Section~\ref{sec:first_existence}.
The third difference is the sign step, where the two proofs differ.  Both
descend from the idea of Rothvoss~\cite{rot17}, a Gaussian
point projected onto the spectral body cut by a box.  Akbas and Sra take the
step from the cited theorem.  Each round draws a fresh Gaussian point, and its
projection freezes at least half of the active coordinates.  The dilation of
the body is a constant that the theorem does not name.  The theorem is
algorithmic, so a signing can be found in randomized polynomial time.
In Section~\ref{sec:first_existence} we prove our own step,
Proposition~\ref{prop:first_covering}, in the different form of a
deterministic covering of the cube.  Every point of the cube lies within $8K$
of a point with at least one percent of its coordinates equal to $\pm1$.  In the
proof of the covering we draw one Gaussian point.  Off an event of probability
$e^{-m/50}+e^{-m/200}$, the point lies within $\frac{13}{25}\sqrt m$ of every
coordinate section of $K$ cut by at most one percent of the coordinates, by a
union bound over the $e^{m\,\mathsf h(1/100)}$ coordinate sets.  The same
point also lies farther
than $\frac{16}{25}\sqrt m$ from the small cube $[-1/4,1/4]^m$.  Its
projection onto $K$ cut by the shifted box is then forced to have more than
one percent of its coordinates tight.  We then apply the covering to the
active coordinates round after round with the dilation $8$ and the freeze
$1/100$ fixed.  In contrast, their theorem freezes half at an unspecified
dilation.
Our proof of the covering is constructive as well, with one Gaussian point and
one Euclidean projection per round, but we state existence only in
Section~\ref{sec:first_existence}.
The fourth difference is that their Section~2 proves a separate and
non-hereditary estimate~\cite[Theorem~1.4]{as26} by rank-one pinchings.  A
Gaussian series whose matrix variance $\sum_iA_i^2$ is at most the identity
stays in the ball of radius $4$ with probability at least $e^{-cd}$ in ambient
dimension $d$, where $c\approx2.615$.  Their Section~3 invokes from their
Section~2 only the matrix-weighted Poincar\'e criterion, their Lemma~2.3.
Nothing in our paper corresponds to their Theorem~1.4.

\paragraph{Kathuria: a different route.}
In independent and concurrent work, Kathuria~\cite{k26} proves the
conjecture by a different route.  His route is not a small-ball estimate
inside a partial-coloring loop.  It is a random walk on the cube whose
potential comes from free probability and has a semidefinite formulation.
The potential is a soft spectral edge of the discrepancy matrix perturbed by
an operator-valued free semicircular.  For $n$ matrices of dimension
$N\ge n$ his randomized polynomial-time algorithm finds a signing of norm at
most $10^8\sqrt{n(1+\log(2N/n))}$, and of norm at most $10^8\sqrt n$ for
$N\le n$.  The square root is optimal already for diagonal matrices.  Among
the hereditary small-ball radii, for an active set of size $m$ in dimension
$N$, that of Akbas and Sra~\cite[Theorem~1.3]{as26} carries
$(1+\log(2N/m))^2$.  Those of our Proposition~\ref{prop:first_small_ball} and
Lemma~\ref{lem:scale_small_ball} carry $\max\{1,\log(N/m)\}$.  We state our
theorems for $n\times n$ matrices, or for $N\times N$
matrices with $N\le n$.  The rectangular regime $N>n$ enters them only
through the hereditary estimates.

\paragraph{The refinement to $156000$.}
Sections~\ref{sec:small_ball_proof} and~\ref{sec:proof} are
Section~\ref{sec:first_existence} with the two-ingredient architecture kept,
every constant tracked, and the second ingredient replaced.  We obtain
$156000$ in Theorem~\ref{thm:matrix_spencer_bound}.
These two sections and the paper of Akbas and Sra follow the same route in
both ingredients, modulo constants.  Our sign step is a projection step of the same kind as
their cited theorem, and we prove it here with explicit parameters.
Lemma~\ref{lem:scale_small_ball} re-derives
Proposition~\ref{prop:first_small_ball} with the same barrier,
interpolation, Poincar\'e inequality and completely positive averaging.  It
changes four choices, the moment order, the edge cutoff, the trace-moment
bound and the certification of the exponent.  We list these in
Section~\ref{subsec:overview_constant}.  In the six regimes of
Table~\ref{tab:six_regimes} we re-run the proof on six ranges of the moment
order under the single exponent $0.003712$.  Across them
$c_{\rm rad}(\ell_{\rm asp})$ runs from $279$ to $185$ in place of $10^6$.
Our Lemma~\ref{lem:scale_small_ball} and Theorem~1.3 of Akbas and Sra follow
the same route, with the same steps in the same order.  In the lemma we track every constant, and its radius is
linear in $\ell$ as in Proposition~\ref{prop:first_small_ball}.  Every
constant that their text leaves as a universal $C$ is a number here.  Our
edge cutoff is $0.1147$ in the first regime against their $1/64$, and it is
$1/100$ in Section~\ref{sec:first_existence}.  The three loss terms at the
diagonal endpoint coincide, with identical constants.
In Section~\ref{sec:proof} we replace the covering of
Proposition~\ref{prop:first_covering}, a fixed Gaussian point surviving a
union bound, by the refreshed projection of
Lemma~\ref{lem:refreshed_projection}, a random Gaussian trial that succeeds
with probability above $0.12$.  The refreshed projection has a freeze of
$\delta=0.0281672$ per round and dilation
$2/\varepsilon_{\mathrm{box}}\approx4.72$ in place of one percent and $8$.
The rounds are therefore about $35$ per unit of $\log(n/m)$, against about a
hundred in Section~\ref{sec:first_existence} and $1/\log2\approx1.44$ in
their paper.  We also replace the geometric series of Theorem~\ref{thm:first_main}
by the separated-sum estimate of Lemma~\ref{lem:bv_packing} over
the six regimes.  In place of the cleanup at $1000\sqrt n$ we use a random
rounding at $6.5\sqrt n$.  All three constants,
$7\cdot10^9$, $156000$ and theirs, are a dilation times a radius coefficient
times a round sum.  Their text leaves the dilation and the radius
coefficient unspecified, and we account
for the factor between $7\cdot10^9$ and $156000$ in Section~\ref{subsec:overview_constant}.

\paragraph{The constant below $8$.}
In Section~\ref{sec:existence_constant_sub10} we keep the framework of
Section~\ref{sec:first_existence}, a small-ball estimate feeding a partial
signing round by round.  We rebuild both ingredients so that the two losses
the framework tolerates are paid exactly.  These losses are the union bound that
turns Gaussian measure into signs and the radius at which we prove the small-ball
estimate.  We obtain $7.879493715947<7.8795$,
Theorem~\ref{thm:existence_constant_sub10}.  Apart from the one import we name
below, that section and the paper of Akbas and Sra have only standard tools in common.  These
tools are the barrier as a soft indicator of the ball, the Daleckii--Krein
formula, the weighted Poincar\'e criterion, resolvent means, H\"older and
Harg\'e for moments, and Brascamp--Lieb.  Its small-ball mechanism is
different from that of Section~\ref{sec:first_existence} and from theirs.
We integrate the partition function along the amplitude of the matrices
with the reference Gaussian fixed.  Its initial value is therefore exactly
$1$, and no diagonal endpoint needs bounding.  What we integrate is a
first-order identity rather than a second derivative.  The reference
Gaussian carries the quadratic part of the barrier in its covariance, and a
H\"older determinant transfer pays for the change of measure.  We certify
curvature by an all-pairs inequality with a continuous allocation rather
than a hard spectral cutoff.  Covariance-power moments bound the
noncommutative products.  Our sign step is the lossless coding of
Lemma~\ref{lem:1070_lossless_coding}.  This is a counting argument through
Kleitman's cube-diameter theorem~\cite{kleitman66} with no union bound and
no fractional coloring.  It signs more than $77$ percent of the active
coordinates in each of the eleven prefix phases.  We certify the eleven prefix phases
and the four suffix phases.
The small-ball estimate of Section~\ref{sec:small_ball_proof} enters
Section~\ref{sec:existence_constant_sub10} in one place.  That estimate and
their Section~3 follow the same route, and it is of the same type as
Proposition~\ref{prop:first_small_ball}.
Lemma~\ref{lem:1070_infinite_tail}, the small-ball bound for
$\log(N/m)>731/256$ that continues the signing past the last certified row, is
the per-row clause of Lemma~\ref{lem:scale_small_ball} on the row
$\tau\ge44$ of Table~\ref{tab:six_regimes}, which we import with its radius
$185\sqrt m\,\log(N/m)$ and exponent $0.000051714$ unchanged.  Its
share of the total
$7.879493715947$ is $0.000000090723$.

\paragraph{The algorithms and the bit complexity.}
Section~\ref{sec:first_existence} is an existence proof.
Section~\ref{sec:proof} is algorithmic in its structure, one Gaussian trial
and one projection per round, and our four algorithms implement that
per-round structure with explicit running times.  The algorithmic content of Akbas and Sra is the cited
partial-coloring theorem.  No constant, running-time exponent or failure
probability is stated.  Kathuria's is a randomized polynomial-time algorithm
whose semidefinite programs are solved by the ellipsoid method.  Its
operation count is in a real-arithmetic model that also charges one operation
for each nonnegative square root and each exact symmetric
eigendecomposition~\cite[Definition~14.1]{k26}, and our model of
Section~\ref{sec:model} has neither primitive.  It has the constant $10^8$ and no exponent stated.  Our
four algorithms of Sections~\ref{sec:algorithm}--\ref{sec:cubic_time} have
exponents $2+\omega/2$, $1+\omega(1,1/2,1)$, $3+\gamma_\star$ and $3+o(1)$.
The cubic-time algorithm of Section~\ref{sec:small_algorithm_constant} has
its constant below $11.504$.  We carry out the bit-complexity analysis of
Section~\ref{sec:bit_complexity} for that algorithm.  None of these has a
counterpart in either concurrent work.  In
Section~\ref{sec:small_algorithm_constant} the shared loop is only the
bottom layer, the suffix signing, whose cost at full capacity is
below $110.42\sqrt n$.  This layer is what corresponds to the constant $C$ of
Theorem~1.2 of Akbas and Sra.  Everything above it, the smoothed body, the
projection engine, the binary repair, the registry of small-ball rows and
the certificates, is not in their paper.  Kathuria's algorithm, a random
walk driven by semidefinite programs, has a different architecture
altogether.  The hereditary tail of
Section~\ref{sec:small_algorithm_constant} is again
Lemma~\ref{lem:1070_infinite_tail}.

\paragraph{What is shared and what is not.}
Our Sections~\ref{sec:small_ball_proof} and~\ref{sec:proof} and
the paper of Akbas and Sra follow the same route, modulo constants, arrived
at independently on both sides.  The route is the barrier
tilt, the interpolation from the diagonal model, the matrix-weighted
Poincar\'e inequality, and a projection step inside the partial-coloring
framework in which a hereditary small-ball estimate is turned into signs.
Our small-ball estimate of Section~\ref{sec:first_existence} and theirs are
of the same type.  The radius of that estimate is linear rather than quadratic in the
logarithm of the aspect ratio.  We prove the sign step of
Section~\ref{sec:first_existence}, the covering of the cube by partially
signed faces, in a different form from their cited theorem,
with its constants fixed.
Our paper and Kathuria's have the statement proved in common, not the
route. Three works differ on the constants obtained, and our work in particular gives an $n^{3+o(1)}$ time algorithm with bit complexity analysis.~\cite{as26} in addition prove the non-hereditary Theorem 1.4, and~\cite{k26} obtains optimal square-root dependence on the logarithmic of the aspect ratio in the rectangular case.
\section{Preliminaries}
\label{sec:preliminaries}

We fix our notation in Section~\ref{subsec:preli_notation}: matrix and Schatten norms, the Loewner order, the functional calculus, Kronecker products, Gaussian measure, and Fr\'echet derivatives. Our conventions for constants, failure probabilities, and tolerances close that subsection. In Section~\ref{sec:model} we specify the dense input and the real-arithmetic model in which we count the running times of Sections~\ref{sec:algorithm}--\ref{sec:cubic_time}. There we also define the matrix-multiplication exponents $\omega(a,b,c)$ and $\alpha_{\mathrm{dual}}$ and record the bounds on them that we use. In the remaining subsections we collect the tools shared by the later sections. We begin with the tensor Gram inequalities, Fact~\ref{fact:tensor_gram_inequalities} in Section~\ref{subsec:preli_tensor_gram}, through which we bound the noncommutative products of the small-ball argument of Section~\ref{sec:small_ball_proof}. We prove Fact~\ref{fact:gaussian_trace_moments}, the moment bound for a Gaussian matrix series, in Section~\ref{subsec:preli_gaussian_trace_moments}, and we reduce the polynomial moment bounds of Sections~\ref{sec:small_ball_proof} and~\ref{sec:existence_constant_sub10} to it. Our projection step in Section~\ref{sec:proof} rests on two facts about Gaussian sections and Gaussian distances, which we prove as the two parts of Fact~\ref{fact:gaussian_section_transport} in Section~\ref{subsec:preli_gaussian_sections}. Our small-ball arguments integrate against even convex tilts of the Gaussian measure. In Section~\ref{subsec:preli_tilts_series} we bound the covariance of such a tilt (Fact~\ref{fact:even_convex_tilts}) and the Kronecker second moment of a matrix series (Fact~\ref{fact:kronecker_cauchy_schwarz}). We also bound the Schatten norms of a completely positive averaging map in Fact~\ref{fact:cp_schatten_shift}. We close that subsection with Fact~\ref{fact:matrix_hoeffding}, the matrix Hoeffding bound for biased signs, which serves as the tail bound of our cleanup rounding in Sections~\ref{sec:proof} and~\ref{sec:algorithm}. We introduce the first divided difference, Definition~\ref{def:first_divided_difference}, in Section~\ref{subsec:preli_divided_differences} and prove two Hessian formulas there. We express the spectral Hessian of a trace function through it in Fact~\ref{fact:spectral_hessian}, also at coincident eigenvalues. In Fact~\ref{fact:trace_exp_hessian} we bound the second derivative of the trace exponential. We record the basic positive-semidefinite inequalities in Section~\ref{subsec:preli_psd} as Fact~\ref{fact:basic_psd_inequalities}: congruence preserves the Loewner order, and $(A\pm B)^2\preceq2A^2+2B^2$ for symmetric $A$ and $B$. For the matrix relative entropy of positive definite matrices with different traces, we state the generalized Pinsker inequality, Fact~\ref{fact:generalized_pinsker}, in Section~\ref{subsec:preli_relative_entropy}. We use it in Sections~\ref{sec:faster_algorithm}, \ref{sec:accelerated_block_algorithm}, \ref{sec:cubic_time}, and~\ref{sec:bit_complexity} to compare Gibbs matrices. Finally, our later certificates enclose square roots, logarithms, exponentials, and $\pi$ by the rational rules of Definition~\ref{def:rational_enclosures} in Section~\ref{subsec:preli_rational_enclosures}. We fix the precision parameters $d$ and $K_{\mathrm{series}}$ at each use.

\subsection{Notation and conventions}\label{subsec:preli_notation}

We use $n$ to denote the number of input matrices and $N$ to denote their matrix dimension. We use $m=|S|$ to denote the number of active matrices in a partial-coloring step. In proofs that pad the matrices to order $n$, the effective matrix dimension is $N=n$.

For an integer $r\ge1$, $[r]:=\{1,\ldots,r\}$. If $S$ is a finite set, $\R^S$ is the Euclidean space of real vectors indexed by $S$. We use $\R^{N \times N}$ to denote the space of real $N\times N$ matrices. We use $I_N$ and $0_N$ to denote the identity and zero matrices, and we suppress their subscripts when the dimension is clear.

For $A\in \R^{N \times N}$, $A^\top$ is its transpose, $\tr[A]$ its trace, and $|A|:=(A^\top A)^{1/2}$. The unadorned norm $\|A\|$ is the operator norm. For $1\le q<\infty$, the Schatten norms are
\[\|A\|_{S_q}:=(\tr[|A|^q])^{1/q},\qquad \|A\|_{S_\infty}:=\|A\|,\]
and we use $\|A\|_F:=\|A\|_{S_2}$ to denote the Frobenius norm and $\|A\|_1:=\|A\|_{S_1}=\tr[|A|]$ to denote the trace norm. A \emph{symmetric contraction} is a symmetric matrix $A$ with $\|A\|\le1$. The Frobenius inner product is $\langle A,B\rangle_F:=\tr[A^\top B]$. For symmetric matrices $A$ and $B$, $A\preceq B$ means that $B-A$ is positive semidefinite.

If $X=U\operatorname{diag}(\lambda_1,\ldots,\lambda_N)U^\top$ is symmetric and $f$ is scalar-valued on its spectrum, then $f(X):=U\operatorname{diag}(f(\lambda_1),\ldots,f(\lambda_N))U^\top$. We use $\operatorname{diag}(A_1,\ldots,A_k)$ to denote a block-diagonal matrix, and we set $A\oplus B:=\operatorname{diag}(A,B)$. We use $\operatorname{vec}(A)$ to denote column-wise vectorization and $A\otimes B$ to denote the Kronecker product. Thus $\operatorname{vec}(AXB)=(B^\top\otimes A)\operatorname{vec}(X)$ whenever the product is defined.

For vectors, $\langle x,y\rangle:=x^\top y$, $\|x\|_2:=\sqrt{\langle x,x\rangle}$, and $\|x\|_1:=\sum_i|x_i|$. For a nonempty closed set $K\subseteq\R^m$,
\[d(x,K):=\inf_{z\in K}\|x-z\|_2.\]
For $c\ge0$, $cK:=\{cx:x\in K\}$, and $K$ is symmetric when $K=-K$. For open sets $U$ and $V$, $U\Subset V$ means that the closure of $U$ is compact and contained in $V$. We use $\mathbbm{1}[\mathcal E]$ to denote the indicator of an event $\mathcal E$, and we set $a_+:=\max\{a,0\}$.

Unless stated otherwise, $\xi:=(\xi_1,\ldots,\xi_m)\sim \mathcal N(0,I_m)$ is a standard Gaussian vector and
\[\gamma_m(K):=\Pr[\xi\in K].\]
For a probability law $\mu$, $\E_\mu[X]$, $\Pr_\mu[\mathcal E]$, $\V_\mu[X]$, and $\Cov_\mu[X]$ are the expectation, probability, variance, and covariance under $\mu$, so that
\[\V_\mu[X]:=\E_\mu[(X-\E_\mu[X])^2],\qquad \Cov_\mu[X]:=\E_\mu[(X-\E_\mu[X])(X-\E_\mu[X])^\top].\]
All logarithms are natural.

For a differentiable map $F$, $D^jF(X)[H_1,\ldots,H_j]$ is its $j$th Fr\'echet derivative. The symbols $\partial_i$, $\partial_{i_1,i_2}$, and $\nabla^j$ denote coordinate derivatives. For $F:\R^m\to \R^{N \times N}$,
\[\|\nabla^jF(x)\|_{\mathrm{HS}}^2:=\sum_{i_1=1}^m\cdots\sum_{i_j=1}^m\|\partial_{i_1}\cdots\partial_{i_j}F(x)\|_F^2,\]
with the convention $\|\nabla^0F(x)\|_{\mathrm{HS}}:=\|F(x)\|_F$.

\paragraph{Constants.}
A symbol $C>0$ or $c>0$ without a subscript denotes a universal constant, $C$ large and $c$ small, whose value may change from line to line. The letter $p\in(0,1/2)$ is reserved for the failure probability of a randomized algorithm. A subscripted $p_0$, $p_{\mathrm{clean}}$, or $p_{\mathrm{s}}$ is the failure budget of the subroutine it names. The letter $\varepsilon$ is reserved for the sign vector $\varepsilon\in\{-1,1\}^n$ and its coordinates. The box scale of the partial-coloring step is $\varepsilon_{\mathrm{box}}:=0.4235$. Every tolerance or accuracy carries a subscript naming its role or level ($\varepsilon_{\mathrm{opt}}$, $\varepsilon_{\mathrm{op}}$, $\varepsilon_{\mathrm{ph},j}$, \ldots). The symbol $\varepsilon_i$ with a bare index is always a sign coordinate. A subscripted constant keeps, within the section that introduces it, the value fixed there. Every explicit numerical constant, such as $156000$, $0.003712$, or $N_0=3\mathbin{\cdot}10^8$, denotes the exact number written.

\subsection{Computational model}
\label{sec:model}

The input to every algorithm in this paper is dense: the $n$ symmetric matrices $A_1,\ldots,A_n\in\R^{n\times n}$ are given entry by entry, as $n^3$ real numbers. Here $n^3$ is the input size against which the running times of Sections~\ref{sec:algorithm}--\ref{sec:cubic_time} are measured. The algorithms use no structure of the matrices beyond the standing hypotheses (symmetric, $\|A_i\|\le1$). The phrase ``for dense input matrices'' in a theorem statement refers to this comparison with the input size, not to a hypothesis on the $A_i$. Running times are counted in the real-arithmetic model: each addition, subtraction, multiplication, division, or comparison of two real numbers costs one arithmetic operation. Each draw of a random real number that is uniform on $[0,1]$ or standard Gaussian also costs one arithmetic operation. The numbers themselves are exact, so the count does not depend on their bit length. Every running-time bound in Sections~\ref{sec:algorithm}--\ref{sec:cubic_time} is a count of arithmetic operations in this model, not a bit-complexity bound. Bit complexity is the subject of Section~\ref{sec:bit_complexity}, which runs the small-constant cubic-time algorithm of Section~\ref{sec:small_algorithm_constant} on fixed-point words of $O(\log n)$ bits.

Matrix products are charged at their asymptotic cost. For positive integers $r,s,t$, define
\[\begin{aligned}
\Tmat(r,s,t):=\min\{T:\ &\text{every $r\times s$ matrix can be multiplied by every $s\times t$ matrix}\\
&\text{using at most $T$ arithmetic operations}\}.
\end{aligned}
\]
For $a,b,c\ge0$, define the rectangular matrix-multiplication exponent by
\[
\omega(a,b,c):=\inf\{\tau\ge0:\Tmat(\lceil N_{\mathrm{mult}}^a\rceil,\lceil N_{\mathrm{mult}}^b\rceil,\lceil N_{\mathrm{mult}}^c\rceil)\le N_{\mathrm{mult}}^{\tau+o(1)}\text{ as }N_{\mathrm{mult}}\to\infty\}.
\]
Thus $\omega:=\omega(1,1,1)$ is the matrix-multiplication exponent. Transposition and permutation of the matrix-multiplication tensor imply that $\omega(a,b,c)$ is invariant under permutations of $(a,b,c)$. The dual matrix-multiplication exponent is
\[
\alpha_{\mathrm{dual}}:=\sup\{b\in[0,1]:\omega(1,b,1)=2\},
\]
the largest aspect ratio at which an $N_{\mathrm{mult}}\times N_{\mathrm{mult}}^b$ by $N_{\mathrm{mult}}^b\times N_{\mathrm{mult}}$ product still costs $N_{\mathrm{mult}}^{2+o(1)}$. Vassilevska Williams, Xu, Xu, and Zhou~\cite[Section~3.4]{vxxz24} prove $\alpha_{\mathrm{dual}}>0.321334$. Alman, Duan, Vassilevska Williams, Xu, Xu, and Zhou~\cite[Table~1]{almanetal25asymmetry} prove $\omega<2.371339$. The unconditional cubic-time bound of Section~\ref{sec:cubic_time} and the bit-complexity analysis of Section~\ref{sec:bit_complexity} use only $\alpha_{\mathrm{dual}}>1/4$ and $\omega<5/2$. Here the bit-complexity analysis is that of the small-constant algorithm of Section~\ref{sec:small_algorithm_constant}, and for Section~\ref{sec:bit_complexity} these exponents are taken in the asymptotic-rank sense of Definition~\ref{def:bit_rank_exponent}.

Finally, $\widetilde O(\cdot)$ suppresses factors polylogarithmic in $n$ and in the accuracy and failure parameters. Because $\omega(a,b,c)$ is an infimum, a running time stated as $n^{\tau+o(1)}$ with $\tau$ an expression in $\omega(\cdot,\cdot,\cdot)$ carries an unavoidable $n^{o(1)}$ factor. Where such a bound is written $\widetilde O(n^{\tau})$, this factor is included in the notation.

\subsection{Tensor Gram inequalities}\label{subsec:preli_tensor_gram}

In the small-ball arguments of Sections~\ref{sec:first_existence} and~\ref{sec:small_ball_proof} we bound their noncommutative products through the following Gram forms.

\begin{fact}[Tensor Gram inequalities]
\label{fact:tensor_gram_inequalities}
Let $A_1,\ldots,A_m,B_1,\ldots,B_m\in\R^{N\times N}$ be symmetric.
Let $\mathcal H$ be a finite-dimensional Hilbert space, and let
$z_1,\ldots,z_m\in\R^N\otimes\mathcal H$. Then
\begin{align*}
|\sum_{i_1=1}^m\sum_{i_2=1}^m\langle z_{i_1},(A_{i_2}A_{i_1}\otimes I_{\mathcal H})z_{i_2}\rangle|
&\le\sum_{i_1=1}^m\sum_{i_2=1}^m\tr[A_{i_1}A_{i_2}]\langle z_{i_1},z_{i_2}\rangle,\\
|\sum_{i_1=1}^m\sum_{i_2=1}^m\langle z_{i_1},(A_{i_2}B_{i_1}\otimes I_{\mathcal H})z_{i_2}\rangle|
&\le
(\sum_{i_1=1}^m\sum_{i_2=1}^m\tr[A_{i_1}A_{i_2}]\langle z_{i_1},z_{i_2}\rangle)^{1/2}
(\sum_{i_1=1}^m\sum_{i_2=1}^m\tr[B_{i_1}B_{i_2}]\langle z_{i_1},z_{i_2}\rangle)^{1/2}.
\end{align*}
The two Gram forms on the right-hand sides are nonnegative. Moreover, if
$U_1,\ldots,U_m$ are symmetric and $\sum_{i=1}^mU_i^2\preceq sI$, then
\[
|\sum_{i_1=1}^m\sum_{i_2=1}^m\langle z_{i_1},(U_{i_2}U_{i_1}\otimes I_{\mathcal H})z_{i_2}\rangle|
\le s\sum_{i=1}^m\|z_i\|_2^2.
\]
\end{fact}

\begin{proof}
We choose an orthonormal basis of $\mathcal H$ and write the coordinates of $z_i\in\R^N\otimes\mathcal H$ as $(z_i)_{a'h}$ with $a'\in[N]$ and $h\in[\dim\mathcal H]$. Define the four-index arrays
\[
T^A_{aba'h}:=\sum_{i=1}^m(A_i)_{ab}(z_i)_{a'h},
\qquad
T^B_{aba'h}:=\sum_{i=1}^m(B_i)_{ab}(z_i)_{a'h},
\]
and equip such arrays with the inner product $\langle S,T\rangle:=\sum_{a,b,a',h}S_{aba'h}T_{aba'h}$ and the norm $\|T\|_2:=\langle T,T\rangle^{1/2}$. Since each $A_i$ is symmetric, $\tr[A_{i_1}A_{i_2}]=\sum_{a,b}(A_{i_1})_{ab}(A_{i_2})_{ab}$. Expanding $\langle z_{i_1},z_{i_2}\rangle=\sum_{a',h}(z_{i_1})_{a'h}(z_{i_2})_{a'h}$, we then obtain
\[
\sum_{i_1=1}^m\sum_{i_2=1}^m\tr[A_{i_1}A_{i_2}]\langle z_{i_1},z_{i_2}\rangle=\|T^A\|_2^2,
\]
and likewise the second Gram form equals $\|T^B\|_2^2$. Both are nonnegative. Let $\sigma$ be the cyclic shift of the three $[N]$-indices, $(\sigma T)_{pqrh}:=T_{rpqh}$. It permutes the entries of an array and is therefore an isometry for $\|\cdot\|_2$. Writing out the first left-hand side coordinatewise, we get
\[
\begin{aligned}
\sum_{i_1=1}^m\sum_{i_2=1}^m\langle z_{i_1},(A_{i_2}A_{i_1}\otimes I_{\mathcal H})z_{i_2}\rangle
&=\sum_{a,e,b,h}\Bigl(\sum_{i_1}(A_{i_1})_{eb}(z_{i_1})_{ah}\Bigr)\Bigl(\sum_{i_2}(A_{i_2})_{ae}(z_{i_2})_{bh}\Bigr)\\
&=\sum_{a,e,b,h}T^A_{ebah}T^A_{aebh}
=\langle T^A,\sigma T^A\rangle,
\end{aligned}
\]
and the same computation with $B_{i_1}$ in place of $A_{i_1}$ identifies the second left-hand side as $\langle T^B,\sigma T^A\rangle$. We apply Cauchy--Schwarz and the isometry of $\sigma$ to get $|\langle T^A,\sigma T^A\rangle|\le\|T^A\|_2^2$ and $|\langle T^B,\sigma T^A\rangle|\le\|T^B\|_2\|T^A\|_2$, which are the first two inequalities.

For the last inequality, we apply $2|\langle x,y\rangle|\le\|x\|_2^2+\|y\|_2^2$ to $x=(U_{i_2}\otimes I_{\mathcal H})z_{i_1}$ and $y=(U_{i_1}\otimes I_{\mathcal H})z_{i_2}$ and obtain
\begin{align*}
|\sum_{i_1=1}^m\sum_{i_2=1}^m\langle z_{i_1},(U_{i_2}U_{i_1}\otimes I_{\mathcal H})z_{i_2}\rangle|
&\le\frac12\sum_{i_1=1}^m\sum_{i_2=1}^m
(\|(U_{i_2}\otimes I_{\mathcal H})z_{i_1}\|_2^2
+\|(U_{i_1}\otimes I_{\mathcal H})z_{i_2}\|_2^2)\\
&=\sum_{i_1=1}^m\langle z_{i_1},
(\sum_{i_2=1}^mU_{i_2}^2\otimes I_{\mathcal H})z_{i_1}\rangle
\le s\sum_{i_1=1}^m\|z_{i_1}\|_2^2.
\end{align*}
Here the equality interchanges $i_1$ and $i_2$ in the second of the two sums. The last inequality is $\sum_{i_2=1}^mU_{i_2}^2\preceq sI$.
\end{proof}

\subsection{Gaussian trace moments}\label{subsec:preli_gaussian_trace_moments}

We reduce the polynomial moment bounds of Sections~\ref{sec:small_ball_proof} and~\ref{sec:existence_constant_sub10} to the following moment bound for a Gaussian matrix series. It is the scalar-proxy case of the matrix Khintchine inequality with the sharp double-factorial constant (Buchholz~\cite[Theorem~5]{b01}). See also Tropp~\cite[Eq.~(4.17)]{t12} and~\cite{t15}. In Section~\ref{sec:first_existence} we derive a variant, Eq.~\eqref{eq:first_8}, with the constant $N(2s^2w)^s$ for every even measure satisfying the Poincar\'e inequality there.

\begin{fact}[Gaussian trace moments]
\label{fact:gaussian_trace_moments}
Let $R_1,\ldots,R_m\in\R^{N\times N}$ be symmetric with $\sum_{i=1}^mR_i^2\preceq\nu I_N$ for some $\nu\ge0$. Let $h_1,\ldots,h_m$ be independent standard Gaussian variables, and put $G:=\sum_{i=1}^mh_iR_i$. Then, for every integer $s\ge1$,
\[
\E[\tr[|G|^{2s}]]=\E[\tr[G^{2s}]]\le N(2s-1)!!\,\nu^s\le N(2s\nu)^s.
\]
\end{fact}

\begin{proof}
The equality holds because $G$ is symmetric, so $|G|^{2s}=G^{2s}$. We apply Gaussian integration by parts, $\E[h_if(h)]=\E[\partial_if(h)]$ for polynomial $f$, to $\E[\tr[G^{2s}]]=\sum_{i=1}^m\E[h_i\tr[R_iG^{2s-1}]]$, and obtain
\[
\E[\tr[G^{2s}]]
=\sum_{i=1}^m\sum_{k=0}^{2s-2}
\E[\tr[R_iG^kR_iG^{2s-2-k}]].
\]
Diagonalize $G=U\operatorname{diag}(\theta_1,\ldots,\theta_N)U^\top$ and put $\widetilde R_i:=U^\top R_iU$, so that
\[
\tr[R_iG^kR_iG^{2s-2-k}]=\sum_{u=1}^N\sum_{v=1}^N(\widetilde R_i)_{uv}^2\theta_u^k\theta_v^{2s-2-k}.
\]
For $s\ge2$, the weighted arithmetic--geometric mean inequality gives
\[
|\theta_u|^k|\theta_v|^{2s-2-k}\le\frac{k}{2s-2}|\theta_u|^{2s-2}+\frac{2s-2-k}{2s-2}|\theta_v|^{2s-2},
\]
and summing over $0\le k\le2s-2$ gives the weight $(2s-1)/2$ on each of the two terms. Since $(\widetilde R_i)_{uv}^2$ is symmetric in $u,v$ and $\sum_v(\widetilde R_i)_{uv}^2=(\widetilde R_i^2)_{uu}$, the sum over $k$ is at most $(2s-1)\sum_u(\widetilde R_i^2)_{uu}|\theta_u|^{2s-2}=(2s-1)\tr[R_i^2|G|^{2s-2}]$. For $s=1$ the sum over $k$ is $\tr[R_i^2]$, which is the same bound with $|G|^0:=I_N$. We sum over $i$ and use $\sum_{i=1}^mR_i^2\preceq\nu I_N$ together with $|G|^{2s-2}\succeq0$ to obtain
\[
\E[\tr[G^{2s}]]\le(2s-1)\nu\,\E[\tr[|G|^{2s-2}]].
\]
By induction from $\E[\tr[G^0]]=N$ we obtain $\E[\tr[G^{2s}]]\le N(2s-1)!!\,\nu^s$. The inequality $(2s-1)!!\le(2s)^s$ gives the last bound.
\end{proof}

\subsection{Gaussian sections and distances}\label{subsec:preli_gaussian_sections}

Our projection step in Section~\ref{sec:proof} rests on two facts about Gaussian measure. A central section of a symmetric convex set is at least as likely as the set, and a set of Gaussian measure $q$ is within distance $\sqrt{2\log(1/q)}$ of a Gaussian point on average. Part~(i) of the fact is the section inequality of the Gaussian partial-coloring framework of Rothvoss~\cite{rot17} and of Dadush, Jiang, and Reis~\cite{djr22}. Section~\ref{sec:first_existence} uses both parts at Eqs.~\eqref{eq:first_6} and~\eqref{eq:first_40}.

\begin{fact}[Gaussian sections and Gaussian distances]
\label{fact:gaussian_section_transport}
Let $H\subseteq\R^m$ be a linear subspace. Let $\gamma_H$ be its standard Gaussian measure, and let $\xi_H$ be a standard Gaussian vector on $H$.
\begin{enumerate}[label=(\roman*)]
\item If $C\subseteq\R^m$ is symmetric and convex, then $\gamma_H(C\cap H)\ge\gamma_m(C)$.
\item If $C\subseteq H$ is closed with $\gamma_H(C)=q>0$, then $\E[d(\xi_H,C)]\le\sqrt{2\log(1/q)}$.
\end{enumerate}
\end{fact}

\begin{proof}
For (i), we write $\R^m=H\oplus H^\perp$, so that $\gamma_m=\gamma_H\otimes\gamma_{H^\perp}$. For $w\in H^\perp$, let $s(w):=\gamma_H((C-w)\cap H)$ be the Gaussian measure of the section of $C$ by the translate $w+H$. Fubini gives $\gamma_m(C)=\int_{H^\perp}s(w)\,\d\gamma_{H^\perp}(w)$. The function $s$ is even, because $C=-C$ and $\gamma_H$ is symmetric. It is log-concave by Pr\'ekopa's theorem~\cite{prekopa73}, because it is the marginal over $v\in H$ of the log-concave function $(v,w)\mapsto\mathbbm{1}[v+w\in C]e^{-\|v\|_2^2/2}$ on $H\times H^\perp$, up to the normalizing constant of $\gamma_H$. An even log-concave function is maximized at the origin, since $s(0)\ge s(w)^{1/2}s(-w)^{1/2}=s(w)$. Hence we obtain $\gamma_m(C)\le s(0)=\gamma_H(C\cap H)$.

For (ii), let $\nu$ be $\gamma_H$ conditioned on $C$, so that $\d\nu=q^{-1}\mathbbm{1}[C]\,\d\gamma_H$. The relative entropy of $\nu$ with respect to $\gamma_H$ is then $\int\log(q^{-1})\,\d\nu=\log(1/q)$. We apply Talagrand's transportation inequality~\cite{talagrand96} for the standard Gaussian measure to bound the quadratic Wasserstein distance by $W_2(\nu,\gamma_H)\le\sqrt{2\log(1/q)}$. In any coupling $(\xi_H,Y)$ of $\gamma_H$ and $\nu$ the point $Y$ lies in $C$ almost surely, so $d(\xi_H,C)\le\|\xi_H-Y\|_2$. We take expectations, apply $\E[\|\xi_H-Y\|_2]\le(\E[\|\xi_H-Y\|_2^2])^{1/2}$, and pass to the infimum over couplings to get $\E[d(\xi_H,C)]\le W_2(\nu,\gamma_H)\le\sqrt{2\log(1/q)}$.
\end{proof}

\subsection{Even convex tilts and matrix series}\label{subsec:preli_tilts_series}

Our small-ball arguments in Sections~\ref{sec:first_existence}, \ref{sec:small_ball_proof}, and~\ref{sec:existence_constant_sub10} integrate against even convex tilts of the Gaussian measure. The next four facts give the covariance and convex-domination bounds of such a tilt and the Kronecker second moment of a matrix series. They also give the Schatten norms of a completely positive averaging map. The last fact is the tail bound of the cleanup rounding in Sections~\ref{sec:proof} and~\ref{sec:algorithm}.

\begin{fact}[Even convex Gaussian tilts]
\label{fact:even_convex_tilts}
Let $F:\R^m\to(-\infty,\infty]$ be even and convex, and finite on a set of positive Lebesgue measure. Put $Z:=\int e^{-F}\,\d\gamma_m$ and $\d\mu:=Z^{-1}e^{-F}\,\d\gamma_m$. Then $0<Z<\infty$, the measure $\mu$ has mean zero, and $\Cov_\mu[\xi]\preceq I_m$. Moreover, $\E_\mu[g(\xi)]\le\E_{\gamma_m}[g(\xi)]$ for every $\gamma_m$-integrable convex $g:\R^m\to\R$.
\end{fact}

\begin{proof}
Since $F$ is even and convex, $F\ge F(0)>-\infty$, so $Z\le e^{-F(0)}$ is finite, and $Z>0$ because $F$ is finite on a set of positive measure. The density of $\mu$ is even, so $\mu$ has mean zero. Suppose first that $F$ is finite and smooth. The potential $\|x\|_2^2/2+F(x)$ of $\mu$ has Hessian at least $I_m$, so the Brascamp--Lieb inequality~\cite{brascamplieb76} gives $\V_\mu[\langle\theta,\xi\rangle]\le\|\theta\|_2^2$ for every $\theta\in\R^m$, which is $\Cov_\mu[\xi]\preceq I_m$. In Section~\ref{sec:first_existence} we prove the Brascamp--Lieb inequality at Eq.~\eqref{eq:first_5}. We apply Harg\'e's theorem~\cite{harge04} to the even log-concave density $e^{-F}$ with respect to $\gamma_m$ and obtain $\E_\mu[g(\xi)]\le\E_{\gamma_m}[g(\xi)]$ for convex $g$. For general $F$, the Lipschitz envelopes $x\mapsto\inf_y[F(y)+k\|x-y\|_2]$, mollified with an even kernel at scale $k^{-2}$, are finite, smooth, even and convex potentials $F_k$. They converge to $F$ outside the boundary of its domain, a Lebesgue null set, and $F_k(0)\to F(0)$. The densities $e^{-F_k}\le e^{-F_k(0)}$ are therefore uniformly bounded for large $k$, and dominated convergence passes both conclusions to $\mu$.
\end{proof}

\begin{fact}[Kronecker second moment of a matrix series]
\label{fact:kronecker_cauchy_schwarz}
Let $A_1,\ldots,A_m\in\R^{N\times N}$ be symmetric, and let $\xi$ be a random vector in $\R^m$ with $\E[\xi\xi^\top]\preceq I_m$. Put $X:=\sum_{i=1}^m\xi_iA_i$. Then
\[
\|\E[X\otimes X]\|\le\|\sum_{i=1}^mA_i^2\|.
\]
For centered $\xi$, the hypothesis is $\Cov[\xi]\preceq I_m$. Moreover, $\|\sum_{k=1}^mN_k\otimes N_k\|\le\|\sum_{k=1}^mN_k^2\|$ for all symmetric $N_1,\ldots,N_m\in\R^{N\times N}$.
\end{fact}

\begin{proof}
Write $\E[\xi\xi^\top]=WW^\top$ with $W\in\R^{m\times m}$, and put $N_k:=\sum_{i=1}^mW_{ik}A_i$. Then $\E[X\otimes X]=\sum_{i,j}\E[\xi_i\xi_j]A_i\otimes A_j=\sum_{k=1}^mN_k\otimes N_k$. Similarly $\sum_kN_k^2=\sum_{i,j}(WW^\top)_{ij}A_iA_j=\mathbf A^\top(WW^\top\otimes I_N)\mathbf A$, where $\mathbf A\in\R^{mN\times N}$ stacks $A_1,\ldots,A_m$. Since $WW^\top\otimes I_N\preceq I_{mN}$, part~(a) of Fact~\ref{fact:basic_psd_inequalities} gives $\sum_kN_k^2\preceq\mathbf A^\top\mathbf A=\sum_iA_i^2$. It remains to prove the operator Cauchy--Schwarz inequality. For $z=\operatorname{vec}(Z)$ with $Z\in\R^{N\times N}$, symmetry of $N_k$ gives $\langle z,(N_k\otimes N_k)z\rangle=\tr[Z^\top N_kZN_k]=\langle N_kZ,ZN_k\rangle_F$. Moreover, $\sum_k\|N_kZ\|_F^2=\tr[Z^\top(\sum_kN_k^2)Z]\le\|\sum_kN_k^2\|\|Z\|_F^2$, and the same bound holds for $ZN_k$. Applying Cauchy--Schwarz twice and using these two bounds, we then bound $|\langle z,(\sum_kN_k\otimes N_k)z\rangle|$ by $\|\sum_kN_k^2\|\|z\|_2^2$. The matrix $\sum_kN_k\otimes N_k$ is symmetric, so this proves the operator bound.
\end{proof}

\begin{fact}[Schatten shift for completely positive maps]
\label{fact:cp_schatten_shift}
Let $\Phi$ be a completely positive linear map on $\R^{N\times N}$, self-adjoint for the trace pairing $\tr[\Phi(U)^\top V]=\tr[U^\top\Phi(V)]$. Here completely positive means that $\Phi$, applied blockwise to a positive semidefinite block matrix with $N\times N$ blocks, returns a positive semidefinite matrix. Put $S:=\Phi(I_N)$ and let $1<\ell\le\infty$. Then, for every $U\in\R^{N\times N}$,
\begin{equation}
\|\Phi(U)\|_{S_p}\le\|S\|_{S_\ell}\|U\|_{S_q},
\qquad\frac1p-\frac1q=\frac1\ell,
\qquad\frac{\ell}{\ell-1}\le q\le\infty,
\label{eq:cp_schatten_shift}
\end{equation}
with $1/\infty:=0$ and $\ell/(\ell-1):=1$ at $\ell=\infty$. At $\ell=\infty$ the bound reads $\|\Phi(U)\|_{S_p}\le\|S\|\,\|U\|_{S_p}$ for $1\le p\le\infty$. The map $U\mapsto\sum_{i=1}^mN_iUN_i$ with symmetric $N_i$ satisfies the hypotheses with $S=\sum_{i=1}^mN_i^2$, and so does $U\mapsto\E[XUX]$ for a random symmetric $X$ with $\E[\|X\|^2]<\infty$, with $S=\E[X^2]$.
\end{fact}

\begin{proof}
Let $\|U\|\le1$. The block matrix with diagonal blocks $I_N$ and off-diagonal blocks $U,U^\top$ is positive semidefinite, so its blockwise image, with diagonal blocks $S$ and off-diagonal blocks $\Phi(U),\Phi(U^\top)$, is positive semidefinite. In particular, $\Phi(U^\top)=\Phi(U)^\top$. For $\delta>0$, we add $\delta I_{2N}$ and take the Schur complement of $S+\delta I_N$ to get $\Phi(U)=(S+\delta I_N)^{1/2}K_\delta(S+\delta I_N)^{1/2}$ with $\|K_\delta\|\le1$. Schatten H\"older gives $\|\Phi(U)\|_{S_\ell}\le\|(S+\delta I_N)^{1/2}\|_{S_{2\ell}}^2=\|S+\delta I_N\|_{S_\ell}$, and $\delta\to0$ gives $\|\Phi\|_{S_\infty\to S_\ell}\le\|S\|_{S_\ell}$. At $\ell=\infty$ this is the Russo--Dye theorem, which bounds a positive map by its value at the identity. For the trace norm, we use duality and self-adjointness to get $\|\Phi(U)\|_{S_1}=\sup_{\|V\|\le1}\tr[U^\top\Phi(V)]\le\|U\|_{S_{\ell/(\ell-1)}}\|S\|_{S_\ell}$ by H\"older. Riesz--Thorin interpolation for Schatten classes, as in Simon~\cite{simon05} or Pisier and Xu~\cite{pisierxu03}, between $\Phi:S_\infty\to S_\ell$ and $\Phi:S_{\ell/(\ell-1)}\to S_1$ gives Eq.~\eqref{eq:cp_schatten_shift}. Section~\ref{sec:first_existence} uses the case $\ell=\infty$ for the averaging map $\sum_iR_iHR_i$ in the proof of Eq.~\eqref{eq:first_30}. For that map, Cauchy--Schwarz on unit vectors, $|x^\top\Phi(U)y|\le\|U\|(x^\top Sx)^{1/2}(y^\top Sy)^{1/2}$, gives the endpoint $p=\infty$ of this case directly, in place of the block argument. For the two examples, $\sum_iN_iUN_i$ and each realization $XUX$ act blockwise as congruences, which preserve positive semidefiniteness, and $\tr[(N_iUN_i)^\top V]=\tr[U^\top N_iVN_i]$ gives self-adjointness.
\end{proof}

\begin{fact}[Matrix Hoeffding bound for biased signs]
\label{fact:matrix_hoeffding}
Let $A_1,\ldots,A_m\in\R^{N\times N}$ be symmetric with $\sum_{i=1}^mA_i^2\preceq\sigma^2I_N$ for some $\sigma>0$. Let $\varepsilon_1,\ldots,\varepsilon_m\in\{-1,1\}$ be independent with $\E[\varepsilon_i]=y_i$. Then, for every $t\ge0$,
\[
\Pr[\|\sum_{i=1}^m(\varepsilon_i-y_i)A_i\|\ge t]\le2N\exp(-\frac{t^2}{2\sigma^2}).
\]
In particular, $\|\sum_{i=1}^m(\varepsilon_i-y_i)A_i\|\le\sqrt{2\sigma^2\log(4N)}$ with probability at least $1/2$.
\end{fact}

\begin{proof}
Put $X_i:=(\varepsilon_i-y_i)A_i$. The scalar $\varepsilon_i-y_i$ is centered with a range of length $2$, so Hoeffding's lemma gives $\E[e^{\theta(\varepsilon_i-y_i)}]\le e^{\theta^2/2}$ for every real $\theta$. We apply this to each eigenvalue of $A_i$ by the functional calculus and get $\E[e^{\theta X_i}]\preceq e^{\theta^2A_i^2/2}$. The matrix Laplace-transform method of Tropp~\cite[Chapter~3]{t15} then bounds $\Pr[\lambda_{\max}(\sum_iX_i)\ge t]$ by $\inf_{\theta>0}e^{-\theta t}\tr[\exp(\frac{\theta^2}2\sum_iA_i^2)]\le N\inf_{\theta>0}e^{-\theta t+\theta^2\sigma^2/2}=Ne^{-t^2/(2\sigma^2)}$. We combine the same bound for $-\sum_iX_i$ with a union bound to get the tail bound. The case $y=0$ is Tropp's matrix Rademacher inequality~\cite[Theorem~4.1]{t12}. At $t=\sqrt{2\sigma^2\log(4N)}$ the right-hand side equals $1/2$.
\end{proof}

\subsection{Divided differences}\label{subsec:preli_divided_differences}

The spectral Hessian of a trace function depends on a divided difference of its derivative, including when eigenvalues coincide.

\begin{definition}[First divided difference]\label{def:first_divided_difference}
Let $J\subseteq\R$ be an interval. Let $f\in C^2(J)$. For $x,y\in J$, define
\[
f'[x,y]:=
\begin{cases}
\dfrac{f'(x)-f'(y)}{x-y}, & x\neq y,\\
f''(x), & x=y.
\end{cases}
\]
We call $f'[x,y]$ the first divided difference of $f'$.
\end{definition}

\begin{fact}[Spectral Hessian of a trace function]
\label{fact:spectral_hessian}
Let $J\subseteq\R$ be an open interval, let $f\in C^2(J)$, and let $X=U\operatorname{diag}(\lambda_1,\ldots,\lambda_N)U^\top$ be symmetric with every $\lambda_a\in J$. Then $X\mapsto\tr[f(X)]$ is twice Fr\'echet differentiable on the symmetric matrices with spectrum in $J$, and for every symmetric $Y$,
\[
D^2\tr[f(X)][Y,Y]
=\sum_{a=1}^N\sum_{b=1}^Nf'[\lambda_a,\lambda_b]\,|(U^\top YU)_{ab}|^2,
\]
with the divided difference $f'[\cdot,\cdot]$ of Definition~\ref{def:first_divided_difference}. Moreover, for all $u,v\in J$,
\[
f'[u,v]=\int_0^1f''((1-h)u+hv)\,\d h.
\]
\end{fact}

\begin{proof}
For $g\in C^1(J)$, the Daleckii--Krein formula (Bhatia~\cite[Theorem~V.3.3]{bhatia97}) gives
\[
Dg(X)[Y]=U\bigl(g^{[1]}(\lambda_a,\lambda_b)\,(U^\top YU)_{ab}\bigr)_{a,b=1}^NU^\top .
\]
Here $g^{[1]}(x,y):=(g(x)-g(y))/(x-y)$ for $x\ne y$ and $g^{[1]}(x,x):=g'(x)$. With $g=f$, we take the trace of this formula and get $D\tr[f(X)][Y]=\tr[f'(X)Y]$. With $g=f'$, so that $g^{[1]}=f'[\cdot,\cdot]$, we differentiate once more and obtain
\[
D^2\tr[f(X)][Y,Y]=\tr[Df'(X)[Y]\,Y]=\sum_{a=1}^N\sum_{b=1}^Nf'[\lambda_a,\lambda_b](U^\top YU)_{ab}(U^\top YU)_{ba},
\]
and $U^\top YU$ is symmetric. For $u\ne v$, the fundamental theorem of calculus gives $\int_0^1f''((1-h)u+hv)\,\d h=(f'(v)-f'(u))/(v-u)$, and for $u=v$ the integral is $f''(u)$. Section~\ref{sec:first_existence} uses this formula at Eq.~\eqref{eq:first_18}.
\end{proof}

\begin{fact}[Second derivative of the trace exponential]
\label{fact:trace_exp_hessian}
Let $Z,H\in\R^{N\times N}$ be symmetric and put $M_t:=Z+tH$ for $t\in\R$. Then
\[
\frac{\d^2}{\d t^2}\tr[\exp(M_t)]
=\int_0^1\tr[He^{(1-s)M_t}He^{sM_t}]\,\d s
\le\tr[H^2e^{M_t}]
\le\|H\|^2\tr[\exp(M_t)],
\]
and the integrand is nonnegative.
\end{fact}

\begin{proof}
Duhamel's formula and cyclicity of the trace give
\[
\frac{\d}{\d t}e^{M_t}=\int_0^1e^{(1-s)M_t}He^{sM_t}\,\d s,
\qquad
\frac{\d}{\d t}\tr[e^{M_t}]=\tr[He^{M_t}],
\]
and we differentiate once more to get the integral formula. Diagonalize $M_t=U\Lambda U^\top$ with $\Lambda:=\operatorname{diag}(\lambda_1,\ldots,\lambda_N)$, and put $\widetilde H:=U^\top HU$. The integrand equals $\sum_{j,l}|\widetilde H_{jl}|^2e^{(1-s)\lambda_j+s\lambda_l}$, which is nonnegative. In the eigenbasis this is the case $f=\exp$ of Fact~\ref{fact:spectral_hessian}, since $\exp'[u,v]=\int_0^1e^{(1-s)u+sv}\,\d s$. We use the weighted arithmetic--geometric mean inequality $e^{(1-s)\lambda_j+s\lambda_l}\le(1-s)e^{\lambda_j}+se^{\lambda_l}$ and the symmetry of $|\widetilde H_{jl}|^2$ in $j,l$ to bound the integrand by
\[
\sum_{j,l}|\widetilde H_{jl}|^2e^{\lambda_j}=\sum_j(\widetilde H^2)_{jj}e^{\lambda_j}=\tr[H^2e^{M_t}].
\]
Integrating over $s$ gives the middle bound. The row sums satisfy $(\widetilde H^2)_{jj}=\sum_l|\widetilde H_{jl}|^2\le\|H\|^2$, which gives the last bound.
\end{proof}

\subsection{Positive-semidefinite inequalities}\label{subsec:preli_psd}

\begin{fact}[Basic PSD inequalities]
\label{fact:basic_psd_inequalities}
Let all matrices below have compatible dimensions.
\begin{itemize}
\item[(a)] \hypertarget{fact:basic_psd_inequalities_a}{}
If $X$ and $Y$ are symmetric with $X\preceq Y$, then
$R^\top X R\preceq R^\top YR$ for every matrix $R$.  Consequently, if
$0\preceq P\preceq I$ and $M$ is symmetric, then
$MPM\preceq M^2$.  Loewner inequalities may also be added termwise.
\item[(b)] \hypertarget{fact:basic_psd_inequalities_b}{}
For symmetric matrices $A$ and $B$,
\[
(A+B)^2\preceq 2A^2+2B^2,
\qquad
(A-B)^2\preceq 2A^2+2B^2.
\]
\end{itemize}
\end{fact}

\begin{proof}
For part~\hyperlink{fact:basic_psd_inequalities_a}{(a)},
$R^\top(Y-X)R\succeq0$ because $Y-X\succeq0$. The bound $MPM\preceq M^2$ is the case $R=M$, $X=P$, $Y=I$. Termwise addition is the case of sums of positive semidefinite differences.
For part~\hyperlink{fact:basic_psd_inequalities_b}{(b)}, we expand the positive
semidefinite matrices $(A-B)^2$ and $(A+B)^2$ to obtain, respectively,
$AB+BA\preceq A^2+B^2$ and $-AB-BA\preceq A^2+B^2$. Substituting these into $(A+B)^2=A^2+B^2+AB+BA$ and $(A-B)^2=A^2+B^2-AB-BA$, we prove the two
claims.
\end{proof}

\subsection{Relative entropy}\label{subsec:preli_relative_entropy}

In Sections~\ref{sec:faster_algorithm}, \ref{sec:accelerated_block_algorithm}, \ref{sec:cubic_time}, and~\ref{sec:bit_complexity} we compare Gibbs matrices of different traces through the following relative entropy and its Pinsker inequality.

\begin{fact}[Generalized Pinsker inequality]
\label{fact:generalized_pinsker}
For positive definite matrices $P,P'\in\R^{N\times N}$ of possibly different traces define
\[
\mathcal D(P\|P'):=\tr[P(\log P-\log P')-P+P'].
\]
Then
\[
\|P-P'\|_1^2\le4\max\{\tr[P],\tr[P']\}\mathcal D(P\|P').
\]
\end{fact}

\begin{proof}
This is Theorem~1.3 of Bansal, George, Ghosh, Sikora, and Zheng~\cite{bansaletal25panoply}. There the relative entropy is $\tr[P(\log P-\log P')]$, and the terms $-P+P'$ appear as the correction $-\tr[P-P']$ on the right-hand side.
\end{proof}

\subsection{Rational enclosures}\label{subsec:preli_rational_enclosures}

Our later certificates enclose square roots, logarithms, exponentials, and $\pi$ by the following rational rules, and we fix the precision parameters $d$ and $K_{\mathrm{series}}$ at each use.

\begin{definition}[Rational enclosures]
\label{def:rational_enclosures}
Enclose $\sqrt x$ for rational $x\ge0$ by $[a/2^d,(a+1)/2^d]$, choosing the integer $a\ge0$ by the exact comparisons $a^2\le 2^{2d}x<(a+1)^2$.

\begin{samepage}
For logarithms, reduce the argument to $y\in[1,2]$ by powers of two and put $z:=(y-1)/(y+1)$. Define
\begin{align}
\ell_{K_{\mathrm{series}}}(z)&:=2\sum_{j=0}^{K_{\mathrm{series}}}\frac{z^{2j+1}}{2j+1},
\label{eq:1070_log_partial}\\
0\le\log y-\ell_{K_{\mathrm{series}}}(z)&\le
\frac{2z^{2K_{\mathrm{series}}+3}}{(2K_{\mathrm{series}}+3)(1-z^2)}.
\label{eq:1070_log_remainder}
\end{align}
The second line bounds the remaining positive series by a geometric series. The same formula at $z=1/3$ encloses $\log2$.
\end{samepage}

For $0\le y<K_{\mathrm{series}}+2$, define $e_{K_{\mathrm{series}}}(y):=\sum_{j=0}^{K_{\mathrm{series}}}y^j/j!$. The exponential remainder satisfies
\begin{equation}
0\le e^y-e_{K_{\mathrm{series}}}(y)\le
\frac{y^{K_{\mathrm{series}}+1}}{(K_{\mathrm{series}}+1)!}\frac1{1-y/(K_{\mathrm{series}}+2)}.
\end{equation}
Consecutive terms of the omitted tail have ratio at most $y/(K_{\mathrm{series}}+2)<1$, so the tail is at most the geometric series with that ratio. For a larger positive argument use $e^x=(e^{x/2^r})^{2^r}$ with $x/2^r\le1$. For a negative argument take reciprocals. Finally, the identity $\pi=16\arctan(1/5)-4\arctan(1/239)$ and consecutive alternating partial sums of $\arctan y=\sum_{j=0}^\infty(-1)^jy^{2j+1}/(2j+1)$ give rational bounds for $\pi$. Propagate rational endpoints by interval arithmetic, reversing their order for reciprocals. Compare polynomial coefficients and terminating decimals exactly.
\end{definition}

\section{A hereditary Gaussian small ball and a signing by partially signed faces}
\label{sec:first_existence}

In this section, we give our first proof of the conjecture with its constant $7\cdot10^9$. Our proof has three parts, one per subsection. We collect the Gaussian tools in Section~\ref{subsec:first_small_balls}: a weighted variance inequality for matrix weights, the Gaussian Poincar\'e and Brascamp--Lieb inequalities, log-concave sections, \v{S}id\'ak's strip bound, and trace moments. With them we prove Proposition~\ref{prop:first_small_ball}, the hereditary small-ball bound: the spectral body of $m$ contractions of size $N$, at radius $10^6\sqrt m\,\max\{1,\log(N/m)\}$, has Gaussian measure at least $e^{-m/1000}$. The second part is Section~\ref{subsec:first_covering}, where we prove Proposition~\ref{prop:first_covering}, the covering of the cube by partially signed faces, from Gaussian concentration and the section bound. Here $m \ge 1000$, and $K \subseteq \R^m$ is symmetric, closed, and convex with Gaussian measure at least $e^{-m/1000}$. Every point of the cube then lies within $8K$ of a cube point with at least $m/100$ coordinates equal to $\pm1$. The third part is Section~\ref{subsec:first_signing}, where we combine the two into Theorem~\ref{thm:first_main}, the discrepancy bound $7\cdot10^9\sqrt n$. We apply the covering to the active coordinates stage by stage, the hereditary small-ball bound stays available after each restriction, and the increments are bounded by a geometric series. In Section~\ref{sec:overview} we describe the mechanism and the refinements of the constant, and in Section~\ref{sec:comparison} we compare the proof with the concurrent work.

\subsection{Gaussian small balls}\label{subsec:first_small_balls}

We use the notation of Section~\ref{subsec:preli_notation}, and we cite the facts of Section~\ref{sec:preliminaries} where they apply.

\paragraph{Weighted variance and Gaussian sections.}

Let $D \subseteq \R^d$ be open and convex. Let $M : D \to \R^{r \times r}$ be a smooth, integrable, symmetric positive definite matrix weight. Define
\[
\mathcal K_{ij} := (\partial_i M) M^{-1} (\partial_j M) - \partial_{ij} M.
\]
If a number $a > 0$ satisfies
\begin{equation}\label{eq:first_1}
\sum_{i,j} x_i^\top \mathcal K_{ij} x_j \ge a \sum_i x_i^\top M x_i \quad (x_1, \dots, x_d \in \R^r),
\end{equation}
then
\begin{equation}\label{eq:first_2}
\int_D f^\top M f - (\int_D M f)^\top (\int_D M)^{-1} (\int_D M f) \le a^{-1} \int_D \sum_i (\partial_i f)^\top M \partial_i f.
\end{equation}
Here and below, we apply a weighted inequality to fields with finite energy. A field $f$ has finite energy if it is locally integrable with weak first derivatives $\partial_i f$ and both $\int_D f^\top M f$ and $\int_D \sum_i (\partial_i f)^\top M \partial_i f$ are finite, that is, if $f$ has finite weighted Sobolev norm $(\int_D f^\top M f + \int_D \sum_i (\partial_i f)^\top M \partial_i f)^{1/2}$.

For completeness, we first consider a bounded smooth convex subdomain $D' \Subset D$. On $L^2(D', M)$ put
\[
\mathcal L u := -M^{-1} \sum_i \partial_i (M \partial_i u), \qquad \langle u, v \rangle_M := \int_{D'} u^\top M v.
\]
The Neumann condition $\partial_\nu u = 0$ gives $\langle v, \mathcal L u \rangle_M = \int_{D'} \sum_i (\partial_i v)^\top M \partial_i u$. Differentiating this identity and integrating by parts once more, we obtain
\begin{equation}\label{eq:first_3}
\begin{split}
\|\mathcal L u\|_M^2 = {} & \int_{D'} \sum_{i,j} (\partial_{ij} u)^\top M \partial_{ij} u + \int_{D'} \sum_{i,j} (\partial_i u)^\top \mathcal K_{ij} \partial_j u \\
& + \int_{\partial D'} \sum_{p,q} \mathrm{II}_{pq} (\partial_{e_p} u)^\top M \partial_{e_q} u.
\end{split}
\end{equation}
The vectors $e_p$ form an orthonormal tangential frame, and $\mathrm{II}$ is the second fundamental form with outward-normal convention. This convention is $\mathrm{II}_{pq} := \langle \partial_{e_p} \nu, e_q \rangle$, so that $\mathrm{II}$ is positive semidefinite on a convex domain. To check the order of the matrix factors, we expand $\mathcal L u = -\sum_i (\partial_i^2 u + M^{-1} (\partial_i M) \partial_i u)$. The coefficient left after the first-derivative terms cancel is $-M \partial_i (M^{-1} \partial_j M) = \mathcal K_{ij}$. The boundary term comes from tangential differentiation of $\partial_\nu u = 0$, and it is nonnegative by convexity.

Let $g := f - (\int_{D'} M)^{-1} \int_{D'} M f$, and solve $\mathcal L u = g$ with Neumann boundary condition. On this bounded domain, $M$ and $M^{-1}$ are bounded on the closure of $D'$, so the Dirichlet form is coercive modulo constant vectors by the Poincar\'e inequality on $D'$. The compatibility condition $\int_{D'} M g = 0$ holds by the definition of $g$. Hence the weak solution exists by the Hilbert-space variational argument (Lax--Milgram~\cite[Section~6.2.1]{evans10}). For smooth $f$, the solution is smooth up to the boundary by elliptic regularity for the conormal problem, due to Agmon, Douglis, and Nirenberg~\cite{adn64}. Hence we can apply Eq.~\eqref{eq:first_3} to it. We extend the resulting inequality to finite-energy $f$ by density in the energy norm.
Equations~\eqref{eq:first_1} and \eqref{eq:first_3} give $a \int \sum_i (\partial_i u)^\top M \partial_i u \le \|g\|_M^2$. We then apply Cauchy--Schwarz in the Green identity and obtain
\[
\|g\|_M^2 \le (\int \sum_i (\partial_i f)^\top M \partial_i f)^{1/2} \|g\|_M / \sqrt a.
\]
This proves Eq.~\eqref{eq:first_2} on $D'$. We prove it on $D$ by an increasing smooth convex exhaustion. The left side is $\inf_c \int (f - c)^\top M (f - c)$, and integrability makes both the minimizing constants and the integrals converge.

In the scalar case $M = e^{-U}$, the curvature is $\mathcal K_{ij} = M \partial_{ij} U$. Thus a probability measure $\mu$ with density proportional to $e^{-U}$ on a convex domain and $D^2 U \succeq I$ satisfies
\begin{equation}\label{eq:first_4}
\V_\mu[f] \le \E_\mu \|\nabla f\|_2^2.
\end{equation}
The same proof gives the sharper formula of Brascamp and Lieb~\cite{brascamplieb76},
\begin{equation}\label{eq:first_5}
\V_\mu[f] \le \E_\mu [\nabla f^\top (D^2 U)^{-1} \nabla f].
\end{equation}
Indeed, in the Green identity we use Cauchy--Schwarz with the two weights $(D^2 U)^{-1}$ and $D^2 U$. Equation~\eqref{eq:first_3} bounds the second factor by $\|f - \E_\mu f\|_{L^2(\mu)}$. By smooth approximation and exhaustion we also prove Eq.~\eqref{eq:first_4} for a Gaussian restricted to a convex set. More generally, we prove it for a Gaussian multiplied by $e^{-\psi}$, where $\psi$ is a convex potential allowed to equal $+\infty$ and finite on a set of positive measure. Equation~\eqref{eq:first_5} holds as well when $\psi$ is smooth on the interior of its domain.

We will also need the log-concavity of Gaussian sections and a strip bound, Eq.~\eqref{eq:first_7} below. By Pr\'ekopa's theorem~\cite{prekopa73}, marginals of log-concave functions are log-concave: for every convex $U : \R^d \to (-\infty, +\infty]$ and every splitting $z = (v, w)$ of the coordinates of $\R^d$, the section integral $s(w) := \int e^{-\|v\|_2^2/2 - U(v, w)} \, dv$ is log-concave in $w$. If the integrand is even, the marginal is even and has its maximum at zero. In particular, for every symmetric convex set $C \subseteq \R^d$ and linear subspace $H$,
\begin{equation}\label{eq:first_6}
\gamma_H(C \cap H) \ge \gamma_d(C).
\end{equation}
This is Fact~\ref{fact:gaussian_section_transport}(i), proved there from the same theorem: we decompose $\R^d = H \oplus H^\perp$ and integrate the section measure, which is at most its value at the origin.

For a centered Gaussian vector $(x_1, \dots, x_q)$ and positive $a_j$, the same marginal argument gives \v{S}id\'ak's inequality~\cite[Theorem~1]{sidak67},
\begin{equation}\label{eq:first_7}
\Pr[|x_j| \le a_j \text{ for all } j] \ge \prod_{j=1}^q \Pr[|x_j| \le a_j].
\end{equation}
We argue by induction on $q$. If $x_q$ has zero variance, the claim reduces to the inductive hypothesis. Otherwise, conditional on $x_q = t$, the other coordinates have the law of $Lv + ty$ for a standard Gaussian vector $v$, a fixed matrix $L$, and a fixed vector $y$. This representation covers singular covariance matrices. The probability $p(t)$ that they lie in the closed rectangle $\{|x_j| \le a_j \text{ for all } j < q\}$ is, up to a constant factor, the section integral of the closed symmetric convex set $\{(v, t) : |(Lv + ty)_j| \le a_j \text{ for all } j < q\}$. By Pr\'ekopa's theorem it is even and log-concave in $t$, hence nonincreasing in $|t|$. Thus $\E[p(x_q) \mid |x_q| \le a_q] \ge \E p(x_q)$, which supplies the inductive step.

\paragraph{Trace moments.}

Suppose $\mu$ is even, has finite polynomial moments, and satisfies Eq.~\eqref{eq:first_4}. Let $G(\xi) := \sum_i \xi_i R_i$ with symmetric matrices $R_i$ such that $\sum_i R_i^2 \preceq w I_N$. Evenness and the scalar inequality imply $\E_\mu \xi = 0$ and $\Cov_\mu[\xi] \preceq I$ (compare Fact~\ref{fact:even_convex_tilts}, which gives the same two bounds for even convex tilts of $\gamma_m$). Therefore $\E_\mu G^2 \preceq w I_N$. For every integer $s \ge 1$ we claim
\begin{equation}\label{eq:first_8}
\E_\mu \tr[G^{2s}] \le N (2 s^2 w)^s .
\end{equation}
Put $M_s := \E_\mu \tr[G^{2s}]$ and $M_0 := N$. Applying Eq.~\eqref{eq:first_4} to the entries of $G^s$, polynomials in $\xi$ that have finite energy because $\mu$ has finite polynomial moments, we obtain
\begin{equation}\label{eq:first_9}
M_s \le \|\E_\mu G^s\|_F^2 + \E_\mu \sum_i \|\partial_i G^s\|_F^2 .
\end{equation}
For any random vector $b$, Cauchy--Schwarz against a fixed unit vector gives $\|\E_\mu G b\|_2^2 \le w \E_\mu \|b\|_2^2$. We apply this to the columns of $G^{s-1}$ and obtain $\|\E_\mu G^s\|_F^2 \le w M_{s-1}$.

The product rule and Cauchy--Schwarz give
\[
\sum_i \|\partial_i G^s\|_F^2 \le s \sum_{a=0}^{s-1} \sum_i \|G^a R_i G^{s-1-a}\|_F^2 \le s^2 w \tr[G^{2s-2}] .
\]
For the second inequality suppose $s \ge 2$. We diagonalize $G = \Gamma \Theta \Gamma^\top$ with $\Theta = \operatorname{diag}(\theta_1, \dots, \theta_N)$, and replace each $R_i$ by $\Gamma^\top R_i \Gamma$, which preserves $\sum_i R_i^2 \preceq w I_N$. Weighted arithmetic--geometric mean gives
\[
|\theta_u|^{2a} |\theta_v|^{2(s-1-a)} \le \frac{a}{s-1} |\theta_u|^{2s-2} + \frac{s-1-a}{s-1} |\theta_v|^{2s-2} .
\]
We multiply by $\sum_i |(R_i)_{uv}|^2$ and sum over $u, v$. The row and column sums of these weights are at most $w$, so each of the $s$ terms is bounded by $w \tr[G^{2s-2}]$. For $s = 1$ the assertion is simply $\sum_i \tr[R_i^2] \le N w$. Equation~\eqref{eq:first_9} now yields $M_s \le (s^2 + 1) w M_{s-1}$. Iterating and using $s^2 + 1 \le 2 s^2$, we prove Eq.~\eqref{eq:first_8}.

\paragraph{The small-ball estimate.}

The barrier interpolation gives a uniform measure bound for every subfamily.

\begin{proposition}[Hereditary small-ball bound]\label{prop:first_small_ball}
Let $1 \le m \le N$, and let $A_1, \dots, A_m \in \R^{N \times N}$ be symmetric contractions. Put $\ell_{\mathrm{asp}} := \max\{1, \log(N/m)\}$. Then
\[
\gamma_m\{ x : \| \sum_{i=1}^m x_i A_i \| \le 10^6 \sqrt m\, \ell_{\mathrm{asp}} \} \ge e^{-m/1000} .
\]
\end{proposition}

\begin{proof}
\hypertarget{first-sb-part-1}{}\emph{1. The barrier and the interpolation.} Put
\begin{equation}\label{eq:first_10}
\begin{split}
& C := 10^6, \qquad r_{\mathrm{sb}} := C \sqrt m\, \ell_{\mathrm{asp}}, \qquad B_i := A_i / r_{\mathrm{sb}}, \\
& \rho := 1/(C^2 \ell_{\mathrm{asp}}^2), \qquad \tau := \lceil 8 \ell_{\mathrm{asp}} \rceil, \qquad w := 1/100 .
\end{split}
\end{equation}
Then $\sum_i B_i^2 \preceq \rho I_N$, $8\ell_{\mathrm{asp}} \le \tau \le 9\ell_{\mathrm{asp}}$, and $\rho \le 10^{-12}$. Fix the positive barrier parameters
\begin{equation}\label{eq:first_11}
\kappa_0 := 10^{-8} m/(N \rho), \qquad \kappa_1 := 100(1 + \kappa_0 + \kappa_0^{-1}) .
\end{equation}
In particular $\kappa_1 \ge \kappa_0$, $\kappa_1 > 8$, and $\kappa_0(\kappa_1 - 8) = 100 + 92\kappa_0 + 100\kappa_0^2 > 100$. In Step 3 we use the weaker bound $\kappa_0(\kappa_1 - 8) > 16$. For $|u| < 1$ define
\[
\phi_0(u) := -\log(1 - u^2), \qquad \phi_{1,\tau}(u) := \sum_{k=\tau}^{\infty} u^{2k}/k, \qquad \phi := \kappa_0 \phi_0 + (\kappa_1 - \kappa_0) \phi_{1,\tau} .
\]
We extend $\phi$ by $+\infty$ outside $(-1, 1)$, and define $\Phi(X) := \tr[\phi(X)]$ for symmetric $X$ by spectral calculus. These functions are even, nonnegative, and convex.

We choose an orthonormal basis. Let $D_i$ be the diagonal part of $B_i$ and $O_i := B_i - D_i$. For $0 \le t \le 1$ set
\[
X_t := \sum_i \xi_i (D_i + t O_i), \qquad V := \sum_i \xi_i O_i, \qquad \mathcal Z_t := \E_{\gamma_m} e^{-\Phi(X_t)}, \qquad U(t) := \log \mathcal Z_t .
\]
The coefficient variances satisfy
\begin{equation}\label{eq:first_12}
\sum_i (D_i + t O_i)^2 \preceq \rho I_N, \qquad \sum_i O_i^2 \preceq 4 \rho I_N .
\end{equation}
Indeed $D_i + t O_i = t B_i + (1 - t) D_i$, and convexity of the square in the Loewner order bounds its square by $t B_i^2 + (1 - t) D_i^2$. Also $\sum_i D_i^2 \preceq \operatorname{diag}(\sum_i B_i^2) \preceq \rho I_N$, and $(B_i - D_i)^2 \preceq 2 B_i^2 + 2 D_i^2$.

A neighborhood of $\xi = 0$ lies in $\{\|X_t\| < 1\}$, so $\mathcal Z_t > 0$. Define $d\mu_t := \mathcal Z_t^{-1} e^{-\Phi(X_t)}\, d\gamma_m$. The measure is even. Its density is proportional to $e^{-\|\xi\|_2^2/2 - \Phi(X_t)}$ on the open convex set $\{\|X_t\| < 1\}$, and this potential is smooth there with Hessian at least $I$. Hence Eq.~\eqref{eq:first_4} applies to $\mu_t$. Differentiation gives
\begin{equation}\label{eq:first_13}
U''(t) = \V_{\mu_t}[D\Phi(X_t)[V]] - \E_{\mu_t} D^2 \Phi(X_t)[V, V], \qquad U'(0) = 0 .
\end{equation}
The last identity holds because $X_0$ and $\phi'(X_0)$ are diagonal and $V$ has zero diagonal. To justify differentiation up to the spectral boundary, we use
\[
\phi(u) = -\kappa_1 \log(1 - u^2) - (\kappa_1 - \kappa_0) \sum_{k=1}^{\tau - 1} u^{2k}/k .
\]
With $\Delta_+ := \max\{0, 1 - \|X_t\|\}$, this gives $e^{-\Phi(X_t)} \le c_\Phi \Delta_+^{\kappa_1}$, where $c_\Phi$ depends only on the fixed parameters and $N$. On $\{\|X_t\| < 1\}$ we have $|D\Phi(X_t)[V]| \le 2\kappa_1 \sqrt N \|V\|_F / \Delta_+$ and $0 \le D^2\Phi(X_t)[V, V] \le 4\kappa_1 \|V\|_F^2 / \Delta_+^2$. Here we use that $\phi'$ is odd with $0 \le \phi' \le \kappa_1 \phi_0'$ on $[0, 1)$, that $0 \le \phi'' \le \kappa_1 \phi_0''$, and that $\phi_0'(u) \le 2/(1 - u)$ and $\phi_0''(u) \le 4/(1 - u)^2$ on $[0, 1)$. Hence the first two $t$-derivatives of $e^{-\Phi(X_t)}$ are at most a constant times $(1 + \|V\|_F^2) \Delta_+^{\kappa_1 - 2}$ there, with a constant independent of $t \in [0, 1]$. Since $\kappa_1 > 8$, we have $\kappa_1 > 2$, and $\Delta_+ \le 1$. The majorant is therefore at most a constant times $1 + \|V\|_F^2$, a quadratic polynomial in $\xi$, which is Gaussian integrable. At $t_0$ with $\|X_{t_0}\| = 1$ we have $\Delta_+ \le \|V\| \, |t - t_0|$, so the extension by zero is twice continuously differentiable in $t$. We prove Eq.~\eqref{eq:first_13} by dominated convergence against this majorant. Finally,
\begin{equation}\label{eq:first_14}
\gamma_m\{ \| \sum_i \xi_i A_i \| < r_{\mathrm{sb}} \} \ge \mathcal Z_1,
\end{equation}
since $0 \le e^{-\Phi(X_1)} \le \mathbbm{1}[\|X_1\| < 1]$. The closed ball of the proposition contains this open ball.

\hypertarget{first-sb-part-2}{}\emph{2. Scalar and spectral curvature.} For a scalar function $f$ of class $C^2$ on an interval, we use the divided difference $f'[u, v]$ of Definition~\ref{def:first_divided_difference}. We need
\begin{equation}\label{eq:first_15}
\phi_0'[u, v] = \sum_{\sigma = \pm 1} [(1 + \sigma u)(1 + \sigma v)]^{-1},
\end{equation}
\begin{equation}\label{eq:first_16}
\phi_{1,\tau}'[u, v] \ge \frac14 \sum_{\sigma = \pm 1} \frac{\mathbbm{1}[\min(1 + \sigma u, 1 + \sigma v) \le w/\tau]}{(1 + \sigma u)(1 + \sigma v)},
\end{equation}
\begin{equation}\label{eq:first_17}
\phi_{1,\tau}'[u, v] \le 2\tau \max(|u|, |v|)^{2\tau - 2} \phi_0'[u, v] .
\end{equation}
Differentiating $-\log(1 - s^2)$, we obtain the first identity. For the lower bound we consider the edge $+1$, which is the term $\sigma = -1$ of Eq.~\eqref{eq:first_16}. We assume $u \ge v$ and put $\delta_u := 1 - u$, $\delta_v := 1 - v$, with $\delta_u \le w/\tau$. If $\delta_v \le 1/(4\tau)$, then $s^{2\tau - 2} \ge 1/2$ on $[v, u]$, and
\[
\phi_{1,\tau}''(s) = \frac{2 s^{2\tau - 2}((2\tau - 1) - (2\tau - 3) s^2)}{(1 - s^2)^2} \ge \frac{1}{2(1 - s)^2} .
\]
Integration gives $\phi_{1,\tau}'[u, v] \ge 1/(2\delta_u\delta_v)$. If $\delta_v > 1/(4\tau)$, the formula $\phi_{1,\tau}'(s) = 2 s^{2\tau - 1}/(1 - s^2)$ gives $\phi_{1,\tau}'(u) \ge (1 - 2w)/\delta_u$ and $\phi_{1,\tau}'(v) \le 2/\delta_v$. Since $\delta_u/\delta_v < 4w$, their divided difference is at least
$(1-10w)/(\delta_u\delta_v) > 1/(2\delta_u\delta_v)$. Since $\phi_{1,\tau}'$ is odd, $\phi_{1,\tau}'[u, v] = \phi_{1,\tau}'[-v, -u]$. We apply the one-sided estimate to the pair $(-v, -u)$ and obtain $\phi_{1,\tau}'[u, v] \ge 1/(2\delta_u^-\delta_v^-)$. Here $\delta_u^- := 1 + u$ and $\delta_v^- := 1 + v$. The bound $1/(2\delta_u^-\delta_v^-)$ is the term $\sigma = +1$ of Eq.~\eqref{eq:first_16}, and it holds whenever $\delta_v^- \le w/\tau$. The divided difference is therefore at least the larger of $1/(2\delta_u\delta_v)$ and $1/(2\delta_u^-\delta_v^-)$. Hence it is at least one quarter of $1/(\delta_u\delta_v) + 1/(\delta_u^-\delta_v^-)$, which is Eq.~\eqref{eq:first_16}. We omit a term whose indicator there vanishes. Continuity handles $u = v$. For the upper bound, $\phi''_{1,\tau}(s) \le 2\tau |s|^{2\tau-2} \phi''_0(s)$, and we integrate along $[u,v]$ to prove Eq.~\eqref{eq:first_17}. We use Eq.~\eqref{eq:first_17} in Step 6.

For symmetric $X$ with eigenvalues $\theta_p$ in $(-1,1)$ and symmetric $H$, spectral differentiation gives
\begin{equation}\label{eq:first_18}
D^2 \tr[f(X)][H,H] = \sum_{p,q} f'[\theta_p,\theta_q] |H_{pq}|^2
\end{equation}
in an eigenbasis of $X$. This is Fact~\ref{fact:spectral_hessian} on the interval $(-1, 1)$. Since $\phi'' \ge 0$ and the same fact gives $\phi'[u, v] = \int_0^1 \phi''((1 - h)u + hv) \, dh \ge 0$, the trace potentials of our barriers are convex on $\{\|X\| < 1\}$, as used above.

Fix $t$, and put $X := X_t$ and $R_i := D_i + tO_i$. These $R_i$ replace the generic family of the paragraph on trace moments. Put $Y_\sigma := I + \sigma X$. Define $C_i^\sigma := Y_\sigma^{-1/2}(\sigma R_i) Y_\sigma^{-1/2}$. Let $P_\sigma$ be the spectral projection of $Y_\sigma$ onto $(0, w/\tau]$, and $\bar P_\sigma := I - P_\sigma$. We split
\[
C_{i,1}^\sigma := \bar P_\sigma C_i^\sigma \bar P_\sigma, \qquad C_{i,0}^\sigma := C_i^\sigma - C_{i,1}^\sigma.
\]
For vectors $x_i$ in any finite-dimensional Hilbert space put $f_{\beta,\sigma}(x) := \sum_{i,j} \tr[C_{i,\beta}^\sigma C_{j,\beta}^\sigma] \langle x_i, x_j \rangle$ for $\beta = 0,1$. Equations~\eqref{eq:first_15}--\eqref{eq:first_18} give
\begin{equation}\label{eq:first_19}
\sum_{i,j} D^2\Phi(X)[R_i,R_j] \langle x_i, x_j \rangle \ge \sum_{\sigma=\pm 1} ( (\kappa_1/4) f_{0,\sigma}(x) + \kappa_0 f_{1,\sigma}(x) ).
\end{equation}
Indeed the $0$ component consists precisely of entries touching an edge eigenspace. Its coefficient is at least $\kappa_0 + (\kappa_1 - \kappa_0)/4 \ge \kappa_1/4$. The other entries retain coefficient $\kappa_0$. We expand the $x_i$ in an orthonormal basis to obtain the vector-valued form. Furthermore $\bar P_\sigma Y_\sigma^{-1} \bar P_\sigma \preceq (\tau/w) \bar P_\sigma$, so Eq.~\eqref{eq:first_12} implies
\begin{equation}\label{eq:first_20}
\sum_i (C_{i,1}^\sigma)^2 \preceq (\tau^2 \rho / w^2) I.
\end{equation}
To prove this, we place the bound $(\tau/w)\bar P_\sigma$ in the middle of each square, use $R_i \bar P_\sigma R_i \preceq R_i^2$, and place the same bound in the two outside factors.

\hypertarget{first-sb-part-3}{}\emph{3. Tensor products and resolvent variance.} We restate the tensor estimate of Fact~\ref{fact:tensor_gram_inequalities} in the notation of this proof. For symmetric families $U_i, T_i$, a finite-dimensional Hilbert space $\mathcal H$, and vectors $x_i \in \R^N \otimes \mathcal H$, define $G_U(x) := \sum_{i,j} \tr[U_i U_j] \langle x_i, x_j \rangle$. Then $G_U(x) \ge 0$ and
\begin{equation}\label{eq:first_21}
\lvert \sum_{i,j} \langle x_i, (U_j T_i \otimes I) x_j \rangle \rvert \le \sqrt{G_U(x) G_T(x)}.
\end{equation}
If $\sum_i U_i^2 \preceq sI$, the same fact also gives
\begin{equation}\label{eq:first_22}
\lvert \sum_{i,j} \langle x_i, (U_j U_i \otimes I) x_j \rangle \rvert \le s \sum_i \|x_i\|_2^2 .
\end{equation}

On $\Omega_t := \{\xi : \|X_t\| < 1\}$ we consider the one-factor weights $e^{-\|\xi\|_2^2/2 - \Phi(X)} Y_\sigma^{-1}$ and the two-factor weights $e^{-\|\xi\|_2^2/2 - \Phi(X)} (Y_{\sigma_2}^{-1} \otimes Y_{\sigma_1}^{-1})$. For either weight $M$, we normalize its curvature by $\widetilde{\mathcal K}_{ij} := M^{-1/2} \mathcal K_{ij} M^{-1/2}$. The derivative identity $\partial_i Y_\sigma^{-1} = -Y_\sigma^{-1} (\sigma R_i) Y_\sigma^{-1}$ gives, for two factors,
\begin{equation}\label{eq:first_23}
\widetilde{\mathcal K}_{ij} = \delta_{ij} I + D^2\Phi(X)[R_i,R_j] I - C_j^{\sigma_2} C_i^{\sigma_2} \otimes I - I \otimes C_j^{\sigma_1} C_i^{\sigma_1}.
\end{equation}
For one factor there is just one of the negative products. This computation uses the inverse derivative twice. Mixed derivatives on different tensor factors cancel between $(\partial_i M) M^{-1} (\partial_j M)$ and $\partial_{ij} M$.

We split each $C_i^\sigma$ into its $0$ and $1$ components. Equations~\eqref{eq:first_21}, \eqref{eq:first_22}, and \eqref{eq:first_20} bound each negative factor by
\[
f_{0,\sigma} + 2\sqrt{f_{0,\sigma} f_{1,\sigma}} + (\tau^2 \rho / w^2) \sum_i \|x_i\|_2^2.
\]
For the second tensor factor, we apply the same bounds after swapping the two factors, which leaves every Gram form unchanged. Young's inequality and $\kappa_0(\kappa_1 - 8) > 16$ give
\[
f_{0,\sigma} + 2\sqrt{f_{0,\sigma} f_{1,\sigma}} \le (\kappa_1/8) f_{0,\sigma} + \frac{8}{\kappa_1 - 8} f_{1,\sigma} \le (\kappa_1/8) f_{0,\sigma} + (\kappa_0/2) f_{1,\sigma}.
\]
By Eq.~\eqref{eq:first_19}, this uses at most one half of the barrier Hessian for each factor. When $\sigma_1 = \sigma_2$, the two halves together use exactly the $\sigma$-part of Eq.~\eqref{eq:first_19}. For one factor there is one negative product. It uses one half of the barrier Hessian, and the curvature bound below holds with headroom. Also
\[
\tau^2 \rho / w^2 \le 81 \cdot 10^{-8} < 1/4.
\]
It follows from Eq.~\eqref{eq:first_23} that both normalized curvatures are at least $I/2$ as block quadratic forms. The weights are integrable: on $\{\|X_t\| < 1\}$ each inverse factor costs at most $\Delta_+^{-1}$, and the density vanishes as $\Delta_+^{\kappa_1}$. Thus we can apply Eq.~\eqref{eq:first_2} with constant $2$. The fields to which we apply it below are polynomials in $\xi$, and the weights are bounded by a constant times the Gaussian density. The fields therefore have finite energy. After we normalize by $\mathcal Z_t$, it says that for either $Q = Y_\sigma^{-1}$ or $Q = Y_{\sigma_2}^{-1} \otimes Y_{\sigma_1}^{-1}$,
\begin{equation}\label{eq:first_24}
\E_{\mu_t}[f^\top Q f] - (\E_{\mu_t}[Qf])^\top (\E_{\mu_t} Q)^{-1} \E_{\mu_t}[Qf] \le 2 \sum_i \E_{\mu_t}[(\partial_i f)^\top Q \partial_i f].
\end{equation}

\hypertarget{first-sb-part-4}{}\emph{4. Removing the weighted mean.} For one factor put $Z_\sigma := \E_{\mu_t} Y_\sigma^{-1}$. Evenness gives $\E_{\mu_t} Y_\sigma = I$ and hence $Z_\sigma \succeq I$. The last assertion also follows by averaging the positive block matrix with blocks $Y_\sigma, I, I, Y_\sigma^{-1}$ and taking its Schur complement. We apply Eq.~\eqref{eq:first_24} to $f = (Y_\sigma - I)x$. Expanding its weighted variance, we obtain
\[
I - Z_\sigma^{-1} \preceq 2 \sum_i R_i Z_\sigma R_i.
\]
Therefore $J_\sigma := \|Z_\sigma\|$ satisfies $1 - J_\sigma^{-1} \le 2\rho J_\sigma$. To select the smaller root, we replace $X$ continuously by $\lambda X$ for $0 \le \lambda \le 1$, with the same barrier parameters and the corresponding normalized tilted measure. The variance bound becomes $\lambda^2 \rho \le \rho$, so all curvature conditions remain valid and $1 - J_\sigma(\lambda)^{-1} \le 2\lambda^2\rho J_\sigma(\lambda)$ for every $\lambda$. The density times the resolvent is continuous after extension by zero and bounded by a constant times the Gaussian density. This uses $\kappa_1 > 8$, and in particular $\kappa_1 > 1$. Hence the normalizer and $J_\sigma(\lambda)$ are continuous in $\lambda$, the normalizer is positive, and $J_\sigma(0) = 1$. The two roots of $2\lambda^2\rho J^2 - J + 1 = 0$ stay separated, and the larger one tends to $+\infty$ as $\lambda \to 0$, so the continuous path from $J_\sigma(0) = 1$ stays on the smaller branch. Therefore
\begin{equation}\label{eq:first_25}
I \preceq Z_\sigma \preceq \frac{2}{1 + \sqrt{1 - 8\rho}} I \preceq (1 + 4\rho) I.
\end{equation}
The last inequality follows by squaring, using $\rho \le 10^{-12}$.

For two factors define
\[
Q := Y_{\sigma_2}^{-1} \otimes Y_{\sigma_1}^{-1}, \qquad Q_0 := \E_{\mu_t} Q, \qquad Q^{-1} = Y_{\sigma_2} \otimes Y_{\sigma_1}, \qquad R_0 := \E_{\mu_t} Q^{-1}.
\]
Since $\mu_t$ is even, the cross terms of $Q^{-1} = (I + \sigma_2 X) \otimes (I + \sigma_1 X)$ have mean zero, and Fact~\ref{fact:kronecker_cauchy_schwarz} with $\Cov_{\mu_t}[\xi] \preceq I$, which follows from evenness and Eq.~\eqref{eq:first_4} as in the paragraph on trace moments, and with $\sum_i R_i^2 \preceq \rho I$ from Eq.~\eqref{eq:first_12} gives
\begin{equation}\label{eq:first_26}
R_0 = I + \sigma_1 \sigma_2 \E_{\mu_t}[X \otimes X], \qquad \|R_0 - I\| \le \rho.
\end{equation}

We apply Eq.~\eqref{eq:first_24} to $f = (Q^{-1} - I)x$. Its weighted variance is $x^\top (R_0 - Q_0^{-1}) x$. Moreover,
\[
\partial_i Q^{-1} = \sigma_2 R_i \otimes Y_{\sigma_1} + Y_{\sigma_2} \otimes \sigma_1 R_i .
\]
The quadratic inequality for a sum, $Y_\sigma \preceq 2I$, and Eq.~\eqref{eq:first_25} give
\[
\sum_i \E_{\mu_t}[(\partial_i Q^{-1}) Q (\partial_i Q^{-1})] \preceq 8(1 + 4\rho)\rho I .
\]
There is a factor 2 from the quadratic inequality and another from $Y_\sigma \preceq 2I$ for each tensor term. The remaining sum is $\sum_i R_i Z_\sigma R_i \preceq (1 + 4\rho)\rho I$. Consequently $0 \preceq R_0 - Q_0^{-1} \preceq 16(1 + 4\rho)\rho I \preceq 17\rho I$. Together with Eq.~\eqref{eq:first_26}, this gives
\begin{equation}\label{eq:first_27}
\|Q_0^{-1} - I\| \le 18\rho, \qquad \|Q_0 - I\| \le 19\rho, \qquad Q_0^{-1} \preceq (1 + \rho) I .
\end{equation}
Here $18\rho/(1 - 18\rho) \le 19\rho$, since $\rho \le 10^{-12}$.

For a matrix field $U$ put $\mathcal E(U) := \E_{\mu_t}[(\operatorname{vec} U)^\top Q \operatorname{vec} U]$. Let $u := \operatorname{vec} U$, $b := \E_{\mu_t} Q u$, and $v := \E_{\mu_t}(Q - I)u$, so $b = \E_{\mu_t} u + v$. Taking expectations in the identity $(Q - I)Q^{-1}(Q - I) = Q - 2I + Q^{-1}$ gives $Q_0 - 2I + R_0$. This expectation is positive semidefinite, and by Eqs.~\eqref{eq:first_26} and \eqref{eq:first_27} it is at most $(19\rho + \rho) I = 20\rho I$. Weighted Cauchy--Schwarz gives $\|v\|_2^2 \le 20\rho \mathcal E(U)$, and hence
\[
b^\top Q_0^{-1} b \le 2(1 + \rho) \E_{\mu_t} \|U\|_F^2 + 40(1 + \rho)\rho \mathcal E(U) .
\]
We substitute into Eq.~\eqref{eq:first_24} and divide by $1 - 40(1 + \rho)\rho$ to prove
\begin{equation}\label{eq:first_28}
\mathcal E(U) \le 3 \E_{\mu_t} \|U\|_F^2 + 3 \sum_i \mathcal E(\partial_i U) .
\end{equation}
Both resulting coefficients are less than 3 at $\rho \le 10^{-12}$. This holds for every choice of the two signs in $Q$.

\hypertarget{first-sb-part-5}{}\emph{5. Polynomial resolvent moments.} For $H(\xi) := X^{\tau - 1} V$, we iterate Eq.~\eqref{eq:first_28} until all derivatives vanish. Derivative tensors use ordered indices, and $\|\nabla^j H\|_{\mathrm{HS}}$ is the norm of Section~\ref{subsec:preli_notation}. The iteration gives
\begin{equation}\label{eq:first_29}
\mathcal E(H) \le \sum_{j=0}^{\tau} 3^{j+1} \E_{\mu_t} \|\nabla^j H\|_{\mathrm{HS}}^2 .
\end{equation}
We now prove the bound
\begin{equation}\label{eq:first_30}
\E_{\mu_t} \|\nabla^j H\|_{\mathrm{HS}}^2 \le N (8\tau^2 \rho)^{\tau} \qquad (0 \le j \le \tau).
\end{equation}
Let $g_1, \dots, g_j$ be independent standard Gaussian directions, independent of $\xi$. Contraction of the ordered derivative tensor gives
\[
\|\nabla^j H(\xi)\|_{\mathrm{HS}}^2 = \E_g \|D^j H(\xi)[g_1, \dots, g_j]\|_F^2 .
\]
The product rule expands the derivative into at most $\tau^j$ words. Each has $\tau$ symmetric factors, with $j$ factors evaluated at distinct $g_a$ and $s := \tau - j$ factors at the common vector $\xi$. Every coefficient family has variance at most $4\rho I$ by Eq.~\eqref{eq:first_12}.

An independent factor $G = \sum_i g_i R_i$, with $R_i$ either coefficient family $D_i + tO_i$ or $O_i$, acts, after averaging, by $\mathcal T(H) := \sum_i R_i H R_i$. This map is completely positive and self-adjoint with $\mathcal T(I) = \sum_i R_i^2 \preceq 4\rho I$ by Eq.~\eqref{eq:first_12}, so Fact~\ref{fact:cp_schatten_shift}, in its operator-norm case $\|\mathcal T(H)\|_{S_p} \le \|\mathcal T(I)\| \, \|H\|_{S_p}$, gives it norm at most $4\rho$ on every Schatten class $S_p$, $1 \le p \le \infty$.

For a word $W = H_1 \cdots H_\tau$, form $B_0 := I$, $B_a := H_a B_{a-1} H_a$. Then $\|W\|_F^2 = \tr B_\tau$, and every $B_a$ is positive semidefinite. Conditional on $\xi$, we average over the independent directions in this recursion and track the Schatten norms of the averaged matrices $\E_g B_a$, not of $B_a$. An independent factor $H_a = G$ enters as $\E_g B_a = \mathcal T(\E_g B_{a-1})$, because its direction appears in no other factor of the word. It therefore costs at most $4\rho$ without changing the Schatten exponent. A common-$\xi$ factor enters as $\E_g B_a = H_a (\E_g B_{a-1}) H_a$. At the $a$th common-$\xi$ factor we use Schatten H\"older in the form
\[
\|HUH\|_{S_{s/a}} \le \|H\|_{S_{2s}}^2 \|U\|_{S_{s/(a-1)}} ,
\]
where $s/0 := \infty$. Starting at $S_\infty$ and ending at $S_1$, we obtain, for $s \ge 1$, the following bound. Write $G_1(\xi), \dots, G_s(\xi)$ for the common-$\xi$ factors of $W$ in their order. Each is $X$ or $V$. Then
\[
\E_g \|W\|_F^2 \le (4\rho)^j \prod_{a=1}^{s} \|G_a(\xi)\|_{S_{2s}}^2 .
\]
By scalar H\"older and Eq.~\eqref{eq:first_8} under $\mu_t$, with variance bound $4\rho$, we bound the expectation of this product by $N(8s^2\rho)^s$. Here we use that the measure is even and satisfies Eq.~\eqref{eq:first_4} by Step 1. For $s = 0$, the $\tau$ averaged maps followed by $\tr B \le N\|B\|$ give $N(4\rho)^\tau$. In both cases, we apply Cauchy--Schwarz to the at most $\tau^j$ words and obtain
\[
\E_{\mu_t} \|\nabla^j H\|_{\mathrm{HS}}^2 \le N(4\rho)^{\tau} \tau^{2j} (2s^2)^s \le N(8\tau^2\rho)^{\tau} ,
\]
with $0^0 := 1$. The last step uses $j + s = \tau$ and $s \le \tau$. This proves Eq.~\eqref{eq:first_30}. Since $\sum_{j=0}^{\tau} 3^{j+1} \le 3 \cdot 4^\tau \le 12^\tau$, Eq.~\eqref{eq:first_29} implies
\begin{equation}\label{eq:first_31}
\mathcal E(X^{\tau - 1} V) \le N(96\tau^2\rho)^{\tau} .
\end{equation}
For the linear field $V$, the covariance bound gives $\E_{\mu_t} \|V\|_F^2 \le 4N\rho$. For its constant derivatives Eq.~\eqref{eq:first_28} gives $\sum_i \mathcal E(O_i) \le 3 \sum_i \|O_i\|_F^2 \le 12N\rho$. We apply the same inequality once more and obtain
\begin{equation}\label{eq:first_32}
\mathcal E(V) \le 48N\rho .
\end{equation}

\hypertarget{first-sb-part-6}{}\emph{6. The interpolation loss.} For equal signs in $Q$, Eqs.~\eqref{eq:first_15} and \eqref{eq:first_18} say that the logarithmic Hessian is the sum of its two resolvent quadratic forms at $V$. For the tail we use Eq.~\eqref{eq:first_17}, followed by
\[
\max(|\theta_p|, |\theta_q|)^{2\tau - 2} \le |\theta_p|^{2\tau - 2} + |\theta_q|^{2\tau - 2} .
\]
The symmetry of $V$ makes the two resulting sums equal. Their common sum is the resolvent quadratic form at $X^{\tau - 1} V$. Equations~\eqref{eq:first_31} and \eqref{eq:first_32} therefore give
\begin{equation}\label{eq:first_33}
\E_{\mu_t} D^2 \Phi(X_t)[V, V] \le 96\kappa_0 N\rho + 8\kappa_1 \tau N(100\tau^2\rho)^{\tau} .
\end{equation}
The coefficient 8 is 4 from the tail comparison times the two signs. Replacing 96 inside the power by 100 only increases the bound. We discard the nonnegative variance in Eq.~\eqref{eq:first_13} and integrate with weight $1 - t$. Since $U'(0) = 0$, we obtain
\begin{equation}\label{eq:first_34}
U(1) \ge U(0) - 48\kappa_0 N\rho - 4\kappa_1 \tau N(100\tau^2\rho)^{\tau} .
\end{equation}

\hypertarget{first-sb-part-7}{}\emph{7. The diagonal endpoint.} Let $x_a$ be the diagonal entries of $X_0$ and $E := \{|x_a| \le 1/2 \text{ for all } a \in [N]\}$. Each $x_a$ has Gaussian variance at most $\rho$. The scalar exponential moment formula and Markov's inequality give $\Pr[|x_a| > 1/2] \le 2e^{-1/(8\rho)}$. The strip bound Eq.~\eqref{eq:first_7} and $\log(1 - x) \ge -2x$ for $0 \le x \le 1/2$ imply
\begin{equation}\label{eq:first_35}
\log \Pr[E] \ge -4N e^{-1/(8\rho)} .
\end{equation}
The Gaussian conditioned on $E$ is even and is restricted to a convex set, so it satisfies Eq.~\eqref{eq:first_4}. Consequently its second moment of $x_a$ is at most $\rho$, and Eq.~\eqref{eq:first_8} in dimension one gives $\E[x_a^{2\tau} \mid E] \le (2\tau^2\rho)^{\tau}$. On $|x| \le 1/2$ the two barrier series satisfy
\[
\phi_0(x) \le 2x^2, \qquad \phi_{1,\tau}(x) \le 2x^{2\tau}/\tau .
\]
Jensen's inequality conditional on $E$ now gives
\begin{equation}\label{eq:first_36}
U(0) \ge -4N e^{-1/(8\rho)} - 2\kappa_0 N\rho - (2\kappa_1 N/\tau)(2\tau^2\rho)^{\tau}.
\end{equation}
Indeed $\mathcal Z_0 \ge \Pr[E]\E[e^{-\Phi(X_0)} \mid E]$, and the conditional expectation is at least $e^{-\E[\Phi(X_0)\mid E]}$.

\hypertarget{first-sb-part-8}{}\emph{8. Substitution of the parameters.} We combine Eqs.~\eqref{eq:first_34} and \eqref{eq:first_36}, and increase positive coefficients, using that the last term of Eq.~\eqref{eq:first_36} is at most $2\kappa_1\tau N(100\tau^2\rho)^{\tau}$ in absolute value. This gives
\begin{equation}\label{eq:first_37}
-U(1)/m \le 50\kappa_0 N\rho/m + 6\kappa_1\tau(N/m)(100\tau^2\rho)^{\tau} + 4(N/m)e^{-1/(8\rho)}.
\end{equation}
The first term is $50\cdot 10^{-8}$. Put $u := \log(N/m)$, so $0 \le u \le \ell_{\mathrm{asp}}$. Equations~\eqref{eq:first_10} and \eqref{eq:first_11} give
\[
\kappa_0 = 10^4\ell_{\mathrm{asp}}^2 e^{-u}, \qquad \kappa_1 e^{u} = 100e^{u} + 10^6\ell_{\mathrm{asp}}^2 + 10^{-2}e^{2u}/\ell_{\mathrm{asp}}^2 < 2\cdot 10^6\ell_{\mathrm{asp}}^2 e^{2\ell_{\mathrm{asp}}}.
\]
Also $100\tau^2\rho \le 8100/10^{12} < 2^{-20}$ and $\tau \ge 8\ell_{\mathrm{asp}}$. Using $\tau \le 9\ell_{\mathrm{asp}}$, $\ell_{\mathrm{asp}}^3 \le e^{3\ell_{\mathrm{asp}}}$, and $e < 4$, the second term in Eq.~\eqref{eq:first_37} is less than
\[
10^9\ell_{\mathrm{asp}}^3 e^{2\ell_{\mathrm{asp}}}2^{-160\ell_{\mathrm{asp}}} \le 10^9\, 2^{-150\ell_{\mathrm{asp}}} < 10^{-6}.
\]
For the last strict inequality it suffices that $\ell_{\mathrm{asp}} \ge 1$ and $2^{50} > 10^{15}$. Finally $1/(8\rho) = 10^{12}\ell_{\mathrm{asp}}^2/8 \ge 100\ell_{\mathrm{asp}}$, so the third term is at most $4e^{-99\ell_{\mathrm{asp}}} < 4\cdot 2^{-99} < 10^{-6}$. Thus the right side of Eq.~\eqref{eq:first_37} is less than $3\cdot 10^{-6} < 1/1000$. By Equation~\eqref{eq:first_14}, we obtain the claimed bound.
\end{proof}

\subsection{Covering the cube by its faces}\label{subsec:first_covering}

From a Gaussian measure bound we obtain a covering of the cube by translates of faces on which a fixed proportion of the coordinates are signs. Define
\[
\mathcal F_m := \{x \in [-1,1]^m : |\{i : |x_i| = 1\}| \ge \lceil m/100\rceil\}.
\]

\begin{proposition}[Covering by partially signed faces]\label{prop:first_covering}
Let $m \ge 1000$, and let $K \subseteq \R^m$ be symmetric, closed, and convex. If $\gamma_m(K) \ge e^{-m/1000}$, then
\[
[-1,1]^m \subseteq \mathcal F_m + 8K.
\]
Equivalently, since $K = -K$, for every $y \in [-1,1]^m$ there is $x \in \mathcal F_m$ such that $x - y \in 8K$.
\end{proposition}

\begin{proof}
\hypertarget{first-cov-part-1}{}\emph{1. Gaussian concentration and distances.} We give the concentration argument needed for the covering. For smooth positive $h$ define the Gaussian averaging operators
\[
P_t h(x) := \E[h(e^{-t}x + \sqrt{1 - e^{-2t}}\,G)], \qquad G \sim \gamma_m.
\]
Differentiation and Gaussian integration by parts give
\[
-\frac{d}{dt}\E[(P_t h)\log(P_t h)] = \E[\|\nabla P_t h\|_2^2/(P_t h)].
\]
As $t \to \infty$, $P_t h$ tends to $\E h$. Also $\nabla P_t h = e^{-t}P_t(\nabla h)$, and Cauchy--Schwarz gives $\|P_t\nabla h\|_2^2 \le (P_t h)P_t(\|\nabla h\|_2^2/h)$. Integrating in $t$ and using the invariance of Gaussian measure, we obtain
\begin{equation}\label{eq:first_38}
\E[h\log h] - \E h\log\E h \le \frac{1}{2}\E[\|\nabla h\|_2^2/h].
\end{equation}
We first make the computation for smooth $h$ that is bounded, bounded away from zero, and has bounded derivatives. For a 1-Lipschitz $f$ and real $s$ we apply Eq.~\eqref{eq:first_38} to $h := e^{s f_j}$, where $f_j$ is $f$ truncated to $[-j, j]$ and then mollified at scale $1/j$. Each $f_j$ is smooth, bounded, 1-Lipschitz, and has bounded derivatives. Moreover $f_j \to f$ pointwise, and $|f_j(x)| \le |f(0)| + \|x\|_2 + 1$ gives Gaussian-integrable domination of the exponential moments.

If $f$ is 1-Lipschitz, we apply Eq.~\eqref{eq:first_38} to $h = e^{sf}$, first for the approximants $f_j$ and then in the limit $j \to \infty$, and put $\Lambda(s) := \log\E e^{sf}$. Then $s\Lambda'(s) - \Lambda(s) \le s^2/2$. Since $\Lambda(s)/s \to \E f$ as $s \to 0$, we integrate $(\Lambda(s)/s)' \le 1/2$ over $(0, s]$ and obtain $\E e^{s(f - \E f)} \le e^{s^2/2}$. We use the same inequality for $-f$ and optimize Markov's inequality to get
\begin{equation}\label{eq:first_39}
\Pr[f \ge \E f + t] \le e^{-t^2/2}, \qquad \Pr[f \le \E f - t] \le e^{-t^2/2} \quad (t \ge 0).
\end{equation}
Lipschitz functions have Gaussian exponential moments, so the approximations preserve these inequalities. The same approximation extends Eq.~\eqref{eq:first_4} to 1-Lipschitz functions, whose variance is therefore at most one. For a closed set $\mathcal A$ of Gaussian measure $q > 0$, Fact~\ref{fact:gaussian_section_transport}(ii) gives
\begin{equation}\label{eq:first_40}
\E d(G,\mathcal A) \le \sqrt{2\log(1/q)}.
\end{equation}
The bound also follows from the lower tail in Eq.~\eqref{eq:first_39}, applied to the 1-Lipschitz function $d(\cdot, \mathcal A)$ at $t = \E d(G, \mathcal A)$, since $\Pr[d(G, \mathcal A) = 0] = q$. Equations~\eqref{eq:first_39} and~\eqref{eq:first_40} apply in every linear subspace with its own standard Gaussian measure.

\hypertarget{first-cov-part-2}{}\emph{2. Simultaneous proximity to coordinate sections.} For $I \subseteq [m]$ put $H(I) := \{x : x_i = 0 \text{ for every } i \in I\}$. Equation~\eqref{eq:first_6} gives $\gamma_{H(I)}(K \cap H(I)) \ge e^{-m/1000}$. Let $G_H$ be the orthogonal projection of $G$ onto $H(I)$. Equation~\eqref{eq:first_40} gives $\E d(G_H, K \cap H(I)) \le \sqrt{2m/1000}$. By Gaussian Poincar\'e, Eq.~\eqref{eq:first_4}, the variance of this distance is at most one. Consequently, by Pythagoras and Cauchy--Schwarz, we obtain
\begin{equation}\label{eq:first_41}
\E d(G, K \cap H(I)) \le \sqrt{|I| + 2m/1000 + 1}.
\end{equation}
Indeed the squared distance is the sum of the squared distance within $H(I)$ and $\sum_{i \in I} G_i^2$, whose expectation is $|I|$. For $|I| \le m/100$ and $m \ge 1000$, the right side is at most $\sqrt{13m/1000} < (3/25)\sqrt{m}$.

The number of these coordinate sets is at most $e^{m\,\mathsf h(1/100)}$, where $\mathsf h(p) := -p\log p - (1-p)\log(1-p)$. To check the counting bound, we expand $1 = (p + (1-p))^m$ and keep terms $k \le pm$. Each weight $p^k(1-p)^{m-k}$ in this range is at least $p^{pm}(1-p)^{m-pm} = e^{-m\,\mathsf h(p)}$. Furthermore $\mathsf h(1/100) < 3/50$: we use $\log 100 < 5$ and $\log(100/99) < 1/99$. The first scalar inequality follows, for example, from $\sum_{j=0}^{6} 5^j/j! > 100$, and the second from $\log(1+x) < x$. From the upper tail in Eq.~\eqref{eq:first_39}, with deviation $(2/5)\sqrt{m}$, and a union bound we therefore see that, except on a set of probability at most $e^{-m/50}$,
\begin{equation}\label{eq:first_42}
d(G, K \cap H(I)) < \frac{13}{25}\sqrt{m} \quad \text{for every } I \text{ with } |I| \le m/100.
\end{equation}
The exponent is $(2/5)^2/2 - 3/50 = 1/50$.

\hypertarget{first-cov-part-3}{}\emph{3. Distance from the small cube.} Put $Q := [-1/4,1/4]^m$. The triangle inequality gives $d(G,Q) \ge \|G\|_2 - \sqrt{m}/4$. The Gaussian Poincar\'e inequality gives $\V[\|G\|_2] \le 1$, so $\E\|G\|_2 \ge \sqrt{m-1}$. Since $m \ge 1000$, we have
\[
\E d(G,Q) \ge (99/100 - 1/4)\sqrt{m} = (37/50)\sqrt{m}.
\]
From the lower tail in Eq.~\eqref{eq:first_39}, with deviation $\sqrt{m}/10$, we obtain that, except on a set of probability at most $e^{-m/200}$,
\begin{equation}\label{eq:first_43}
d(G,Q) > \frac{16}{25}\sqrt{m}.
\end{equation}
The sum $e^{-m/50} + e^{-m/200}$ is less than one for $m \ge 1000$. We choose a point $g$ satisfying both Eqs.~\eqref{eq:first_42} and \eqref{eq:first_43}.

\hypertarget{first-cov-part-4}{}\emph{4. Projection onto the shifted box.} First let $y \in (-1,1)^m$, and define
\[
Q_y := \prod_{i=1}^{m}[-(1+y_i)/8, (1-y_i)/8].
\]
The box contains zero in its interior and is contained in $Q$. Let $c$ be the Euclidean projection of $g$ onto $K \cap Q_y$. Since $\gamma_m(K) > 0$, the set $K$ is nonempty, and symmetry and convexity give $0 \in K$. Thus the intersection is nonempty and compact, so $c$ exists. Let $I$ be the set of coordinates whose box inequality is tight at $c$. At most one of the two inequalities is tight in each coordinate. Suppose $|I| \le m/100$ and discard every nontight box inequality. The point $c$ remains the projection onto the intersection of $K$ with the retained halfspaces. Otherwise a sufficiently short segment from $c$ toward a closer point of that intersection would satisfy all the discarded inequalities, which have positive slack at $c$. Such a segment would contradict the definition of $c$. This enlarged set contains $K \cap H(I)$, because zero satisfies every retained coordinate inequality. Hence
\[
d(g, K \cap Q_y) \le d(g, K \cap H(I)) < \frac{13}{25}\sqrt{m}.
\]
But $Q_y \subseteq Q$ and Eq.~\eqref{eq:first_43} give $d(g, K \cap Q_y) \ge d(g, Q) > (16/25)\sqrt{m}$, a contradiction. Thus more than $m/100$ coordinates are tight, hence at least $\lceil m/100\rceil$ of them. The point $x := y + 8c$ belongs to $[-1,1]^m$, and each tight coordinate becomes a sign. Therefore $x \in \mathcal F_m$ and $x - y \in 8K$.

For $y$ on the boundary of the cube, we approximate it by interior points. Their corresponding points $x$ have a convergent subsequence in the compact set $\mathcal F_m$, which is a finite union of closed faces of the cube. Closedness of $K$ preserves $x - y \in 8K$ in the limit. Symmetry of $K$ then gives the asserted covering.
\end{proof}

\subsection{Existence of a signing}\label{subsec:first_signing}

We apply the covering successively to the active coordinates. The small-ball bound is hereditary, so it remains available after each restriction.

\begin{theorem}[Matrix Spencer discrepancy]\label{thm:first_main}
For every integer $n \ge 1$ and symmetric contractions $A_1, \ldots, A_n \in \R^{n \times n}$, there exists $\varepsilon \in \{-1,1\}^n$ such that
\[
\| \sum_{i=1}^n \varepsilon_i A_i \| \le 7 \cdot 10^9 \sqrt{n}.
\]
\end{theorem}

\begin{proof}
We start with $y^{(0)} := 0$. At stage $j$ define $S_j := \{ i : |y_i^{(j)}| < 1 \}$ and $m_j := |S_j|$. If $m_j < 1000$, we stop. Otherwise put
\[
\ell_{\mathrm{asp},j} := \max\{1, \log(n/m_j)\}, \qquad K_j := \{ x \in \R^{S_j} : \| \sum_{i \in S_j} x_i A_i \| \le 10^6 \sqrt{m_j}\, \ell_{\mathrm{asp},j} \}.
\]
This is a symmetric closed convex set. Proposition~\ref{prop:first_small_ball} gives $\gamma_{m_j}(K_j) \ge e^{-m_j/1000}$. Proposition~\ref{prop:first_covering} supplies a point on a partially signed face of the active cube whose difference from $y^{(j)}|_{S_j}$ belongs to $8K_j$. We use that point as the new active coloring and leave all other coordinates fixed. Then
\begin{equation}\label{eq:first_44}
m_{j+1} \le \frac{99}{100} m_j, \qquad \| \sum_i ( y_i^{(j+1)} - y_i^{(j)} ) A_i \| \le 8 \cdot 10^6 \sqrt{m_j}\, \ell_{\mathrm{asp},j}.
\end{equation}
This process stops after finitely many stages.

For a stage that runs put $u_j := \log(n/m_j)$ and $q := 99/100$. Equation~\eqref{eq:first_44} implies $m_j \le q^j n$. For every $u \ge 0$,
\[
\max\{1, u\} e^{-u/2} \le 2 e^{-u/4}.
\]
Indeed $e^{-u/4} \le 1$, and the maximum of $u e^{-u/4}$ is $4/e < 2$ at $u = 4$. Thus we bound the total discrepancy added in all stages by
\[
16 \cdot 10^6 \sqrt{n} \sum_{j=0}^{\infty} q^{j/4} = \frac{16 \cdot 10^6}{1 - q^{1/4}} \sqrt{n}.
\]
Concavity of $x^{1/4}$ gives $q^{1/4} \le 1 - (1-q)/4 = 399/400$. The total is therefore at most $6.4 \cdot 10^9 \sqrt{n}$.

Fewer than $1000$ coordinates remain. We round each to a nearest sign, and keep every previously fixed sign. Each rounding changes a coordinate by at most one. By the triangle inequality we therefore bound the additional discrepancy by the number of remaining coordinates, which is less than $1000 \le 1000\sqrt{n}$. This also handles $n < 1000$, when no stage runs. The resulting complete signing satisfies
\[
\| \sum_{i=1}^n \varepsilon_i A_i \| \le (6.4 \cdot 10^9 + 1000)\sqrt{n} < 7 \cdot 10^9 \sqrt{n}.
\]
\end{proof}

\section{A hereditary Gaussian small ball with a smaller radius constant}
\label{sec:small_ball_proof}

In this section, we prove the hereditary small-ball estimate consumed by Section~\ref{sec:proof}, Lemma~\ref{lem:scale_small_ball} below.

We already have a hereditary small-ball estimate in Section~\ref{sec:first_existence}, Proposition~\ref{prop:first_small_ball}, and Lemma~\ref{lem:scale_small_ball} has the same hypotheses and the same shape. Both take $1\le m\le N$ and symmetric contractions $A_1,\ldots,A_m\in\R^{N\times N}$, both bound the Gaussian measure $\gamma_m$ of the spectral body, which we write as a probability over the standard Gaussian vector $\xi$ in the lemma, and both hold for every subfamily with an exponent that does not depend on the subfamily. Both radii are $\sqrt m$ times the aspect-ratio factor $\max\{1,\log(N/m)\}$, which we write $\ell$ there and $\ell_{\rm asp}$ here, times a coefficient. The radius coefficient falls from $10^6$ to $c_{\rm rad}(\ell_{\rm asp})$, which takes the six values of Table~\ref{tab:six_regimes}, from $279$ for $\ell_{\rm asp}\le17/16$ down to $185$ for $\ell_{\rm asp}>731/256$. We select the row by the moment order $\tau(\ell_{\rm asp})=\lceil256\ell_{\rm asp}/17\rceil$, and we track every constant of the proof in each row. The exponent moves the other way. Proposition~\ref{prop:first_small_ball} gives measure at least $e^{-m/1000}$, and the uniform clause of the lemma gives at least $\exp(-0.003712m)$. Since $0.003712>1/1000$, the second bound is the smaller, and a larger exponent is the weaker probability guarantee. The per-row clause of the lemma is sharper than its uniform clause. Its exponents $10^{-9}E_{\rm tab}$ run from $0.003711824$ in the first row to $0.000051714$ in the last, so only the last row has an exponent below $1/1000$. Every row entry is at most $3711824$, which is why the single exponent $0.003712$ covers all six rows. The improvement is in the radius, not in the uniform exponent. Section~\ref{sec:proof} consumes the lemma through these two constants. The exponent enters only through the threshold of Lemma~\ref{lem:refreshed_projection}, which asks for $\gamma_m(K)\ge\exp(-0.00372m)$, and $0.003712<0.00372$. The radius coefficient enters the final bound directly, since the norm of the signed sum is at most the sum over rounds of the dilation times the radius of that round's body, and that sum is a geometric series.

\begin{table}[!ht]
\centering
\caption{The six small-ball regimes. A row applies when its range of moment orders contains $\tau(\ell_{\rm asp})=\lceil256\ell_{\rm asp}/17\rceil$. The second column lists the values $\ell_{\rm asp}\ge1$ for which this holds. Every terminating decimal denotes the corresponding exact rational number. The last two columns are an upper bound for $10^9E$ and a lower bound for $10^9\Delta_{\mathrm{curv}}$, the normalized small-ball exponent and the curvature margin of the row, both defined in the proof of Lemma~\ref{lem:scale_small_ball}. The columns $b$, $D$, and $A$ are the parameters fixed in Step~2 of the proof of Lemma~\ref{lem:explicit_small_ball} and, for the remaining rows, in Section~\ref{subsec:smallball_regimes}.}
\label{tab:six_regimes}
\begin{tabular}{ccrrrrrr}
\toprule
$\tau$ & $\ell_{\rm asp}$ & $c_{\rm rad}$ & $b$ & $D$ & $A$ & $10^9E$ & $10^9\Delta_{\mathrm{curv}}$\\
\midrule
$16$       & $[1,17/16]$ & $279$ & $0.400840$ & $323$    & $4.7\mathbin{\cdot}10^6$   & $3711824$ & $118.0$\\
$17$--$19$ & $(17/16,323/256]$ & $266$ & $0.412210$ & $365$    & $5.6\mathbin{\cdot}10^6$   & $3517582$ & $1156.8$\\
$20$--$26$ & $(323/256,221/128]$ & $240$ & $0.435510$ & $497$    & $11.3\mathbin{\cdot}10^6$  & $2376271$ & $2086.7$\\
$27$--$30$ & $(221/128,255/128]$ & $200$ & $0.481310$ & $1713$   & $31\mathbin{\cdot}10^6$    & $2980959$ & $629.4$\\
$31$--$43$ & $(255/128,731/256]$ & $190$ & $0.494840$ & $6210$   & $122.5\mathbin{\cdot}10^6$ & $2730038$ & $834.3$\\
$\tau\ge44$  & $(731/256,\infty)$ & $185$ & $0.499855$ & $221000$ & $228\mathbin{\cdot}10^9$   & $51714$   & $168.4$\\
\bottomrule
\end{tabular}
\end{table}

In the partial-coloring argument we need a Gaussian measure bound for every active subfamily. While we may let the radius depend on the aspect ratio, the exponential loss must stay uniform across subfamilies.

\begin{lemma}[Six-regime hereditary small ball, Table~\ref{tab:six_regimes}]
\label{lem:scale_small_ball}
Let $1\le m\le N$, let $A_1,\ldots,A_m\in \R^{N \times N}$ be symmetric
contractions, and put
\[
\ell_{\rm asp}=\max\{1,\log(N/m)\},\qquad
c_{\rm cut}:=\frac{256}{17},\qquad
\tau(\ell_{\rm asp}):=\lceil c_{\rm cut} \ell_{\rm asp}\rceil.
\]
Let $c_{\rm rad}(\ell_{\rm asp})$ be the entry $c_{\rm rad}$ of the row of Table~\ref{tab:six_regimes} whose range of moment orders contains $\tau(\ell_{\rm asp})$. Then
\[
\Pr[\|\sum_{i=1}^m \xi_iA_i\|
\le c_{\rm rad}(\ell_{\rm asp})\sqrt m\ell_{\rm asp}]
\ge\exp(-0.003712m).
\]
Moreover, the same probability is at least $\exp(-10^{-9}E_{\rm tab}\,m)$, where $E_{\rm tab}$ is the entry $10^9E$ of the same row.
\end{lemma}

In the rest of the section we prove Lemma~\ref{lem:scale_small_ball}. We build the small-ball estimate with universal constants in Sections~\ref{subsec:smallball_interpolation}--\ref{subsec:smallball_moments} and fix the numbers in Sections~\ref{subsec:smallball_explicit} and~\ref{subsec:smallball_regimes}. We start in Section~\ref{subsec:smallball_interpolation} with the barrier potential, the normalization of the radius to one, and the diagonal-to-full interpolation of the matrix series. Lemma~\ref{lem:smallball_interpolation} there reduces the small-ball probability to the diagonal endpoint and an integrated lower bound on the curvature of the log-partition function $a(t)$. Section~\ref{subsec:smallball_divided_differences} proves Lemma~\ref{lem:smallball_divided_difference}, the two divided-difference bounds for the tail barrier $\phi_{1,\tau}$: a lower bound at the spectral edge and an upper bound in the interior. From these scalar bounds, Section~\ref{subsec:smallball_zero_one} derives the spectral $0$--$1$ decomposition and Lemma~\ref{lem:smallball_zero_one}. That lemma bounds the Hessian of $\Phi$ from below by an edge component and a regular component, and it controls the variance proxy of the regular components. The next three subsections supply the Poincar\'e machinery. Section~\ref{subsec:smallball_poincare} proves Lemma~\ref{lem:matrix_weighted_poincare}, a curvature criterion under which a matrix weight satisfies a Poincar\'e inequality. We apply that criterion in Section~\ref{subsec:smallball_resolvent_poincare} to the one- and two-slot small-ball weights and prove Lemma~\ref{lem:resolvent_poincare}. The $0$--$1$ estimates absorb the negative curvature of the resolvent factors into the barrier Hessian there. Section~\ref{subsec:smallball_weighted_resolvent} turns these inequalities into Lemma~\ref{lem:smallball_weighted_poincare}, the defective Poincar\'e estimate with the constant $c_2<3$ that the moment argument iterates. Section~\ref{subsec:smallball_moments} controls the noncommutative polynomials that repeated differentiation produces, uniformly along the interpolation path. There we prove Lemma~\ref{lem:smallball_tail_hessian_comparison}, the resolvent moment bounds and the upper bound on the interpolation Hessian that feeds the curvature of $a(t)$. In Section~\ref{subsec:smallball_explicit} we track every constant for one parameter set and prove Lemma~\ref{lem:explicit_small_ball}, the radius $279$ with the exponent of the first row of Table~\ref{tab:six_regimes}. We repeat that computation for each row of Table~\ref{tab:six_regimes} in Section~\ref{subsec:smallball_regimes}. Checks at the first cell of each row, a monotonicity argument along the row, and exact rational evaluation of the endpoint exponents in Table~\ref{tab:six_regime_endpoint} complete the proof of Lemma~\ref{lem:scale_small_ball}.

\subsection{Normalization and the interpolating partition function}
\label{subsec:smallball_interpolation}
We first rescale the matrix series so that its quadratic variance is small. By a
diagonal-to-full interpolation we convert the small-ball probability into estimates for
a log-partition function.

\begin{definition}[Barrier potential]
\label{def:smallball_potential}
For an integer $\tau\ge2$ and parameters $\kappa_{1}\ge\kappa_{0}>0$, define on $(-1,1)$
\[
\phi_0(u):=-\log(1-u^2),
\qquad
\phi_{1,\tau}(u):=\sum_{k=\tau}^{\infty}\frac{u^{2k}}{k},
\qquad
\phi(u):=\kappa_{0}\phi_0(u)+(\kappa_{1}-\kappa_{0})\phi_{1,\tau}(u).
\]
Extend $\phi$ by $+\infty$ outside $(-1,1)$, and, for every symmetric matrix $X$,
define
\[
\Phi(X):=\tr[\phi(X)].
\]
\end{definition}

\begin{lemma}[Boundary decay of the barrier]
\label{lem:smallball_boundary_decay}
Let $\phi$ and $\Phi$ be as in Definition~\ref{def:smallball_potential}. For every symmetric matrix $X$ with $\|X\|<1$, put $\Delta:=1-\|X\|$. Then
\[
e^{-\Phi(X)}\le2^{\kappa_1}\exp[(\kappa_1-\kappa_0)\sum_{k=1}^{\tau-1}1/k]\,\Delta^{\kappa_1}.
\]
\end{lemma}

\begin{proof}
For $|u|<1$, Definition~\ref{def:smallball_potential} gives
\[
\phi(u)=-\kappa_1\log(1-u^2)
-(\kappa_1-\kappa_0)\sum_{k=1}^{\tau-1}\frac{u^{2k}}{k}.
\]
Also $\phi(u)\ge0$, since its power-series coefficients are positive. Choose an eigenvalue $\lambda$ of $X$ with $|\lambda|=\|X\|$. Then
\begin{align*}
e^{-\Phi(X)}
&\le e^{-\phi(\lambda)}\\
&\le(1-\lambda^2)^{\kappa_1}
\exp[(\kappa_1-\kappa_0)\sum_{k=1}^{\tau-1}1/k]\\
&\le2^{\kappa_1}\exp[(\kappa_1-\kappa_0)\sum_{k=1}^{\tau-1}1/k]\Delta^{\kappa_1}.
\end{align*}
The first step uses $\phi\ge0$ for all the other eigenvalues. The second uses $\kappa_1\ge\kappa_0$ and $\lambda^{2k}\le1$. The last uses $1-\lambda^2=(1-|\lambda|)(1+|\lambda|)\le2\Delta$.
\end{proof}

We normalize the target radius to one so that the small-ball event matches the domain of the barrier potential.

\begin{definition}[Standing normalization]
\label{def:smallball_normalization}
For integers $1\le m\le N$, symmetric contractions
$A_1,\ldots,A_m\in\R^{N\times N}$, and a parameter $c_{\rm rad}>0$, define
\[
\ell_{\rm asp}:=\max\{1,\log(N/m)\},\qquad
r_{\rm sb}:=c_{\rm rad}\sqrt m\ell_{\rm asp},\qquad
\widehat{A}_i:=r_{\rm sb}^{-1}A_i,\qquad
\rho:=(c_{\rm rad}^2\ell_{\rm asp}^2)^{-1}.
\]
Here $\ell_{\rm asp}$ denotes the logarithmic aspect ratio floored at one,
$c_{\rm rad}>0$ is the radius multiplier, and $r_{\rm sb}$ is the target small-ball radius.
The matrices $\widehat A_i$ are the rescaled coefficient matrices, and $\rho$ is their scalar variance bound.
Unless explicitly redefined, these symbols retain this meaning throughout
Section~\ref{sec:small_ball_proof}.
\end{definition}

The diagonal matrix series has a scalar Gaussian description. We connect this endpoint to the full series by turning on the off-diagonal entries continuously.

\begin{definition}[Diagonal-to-full interpolation]
\label{def:smallball_interpolation_data}
Let $A_1,\ldots,A_m$, $\widehat{A}_1,\ldots,\widehat{A}_m$, $\ell_{\rm asp}$, $r_{\rm sb}$, and $\rho$ be defined as in Definition~\ref{def:smallball_normalization}. Let $\xi_1,\ldots,\xi_m$ be independent standard Gaussian random variables.
Fix an integer $\tau\ge2$ and parameters $\kappa_{1}\ge\kappa_{0}>0$, and let $\Phi,\phi$ be as in
Definition~\ref{def:smallball_potential}. Define
$\mathcal{A}:=\mathcal{A}(\xi):=\sum_{i=1}^m \xi_i\widehat{A}_i$. Fix an orthonormal basis and write
$\mathcal{A}=\mathcal{A}_{\mathrm{diag}}+\mathcal{A}_{\mathrm{off}}$, where $\mathcal{A}_{\mathrm{diag}}$ and $\mathcal{A}_{\mathrm{off}}$
are respectively the diagonal and off-diagonal parts of $\mathcal{A}$ in this basis. Likewise, $\widehat{A}_{i,\mathrm{diag}}$ and $\widehat{A}_{i,\mathrm{off}}$ denote the diagonal and off-diagonal parts of $\widehat{A}_i$ in the same basis.
For $t\in[0,1]$, define
\[
\begin{aligned}
\mathcal{A}_t&:=\mathcal{A}_{\mathrm{diag}}+t\mathcal{A}_{\mathrm{off}}, & 
\widehat{A}_i(t)&:=\widehat{A}_{i,\mathrm{diag}}+t\widehat{A}_{i,\mathrm{off}},\\
a_t&:=\E[e^{-\Phi(\mathcal{A}_t)}], & a(t)&:=\log a_t.
\end{aligned}
\]
\end{definition}

\begin{lemma}[Variance proxies]
\label{lem:smallball_variance_proxies}
In the setting of Definition~\ref{def:smallball_interpolation_data}, for every $t\in[0,1]$,
\[
\sum_{i=1}^m\widehat{A}_i(t)^2\preceq\rho I_N,
\qquad
\sum_{i=1}^m\widehat{A}_{i,\mathrm{diag}}^2\preceq\rho I_N,
\qquad
\sum_{i=1}^m\widehat{A}_{i,\mathrm{off}}^2\preceq4\rho I_N.
\]
\end{lemma}

\begin{proof}
Since $A_i^2\preceq I_N$, the normalized matrices satisfy
$\sum_{i=1}^m\widehat{A}_i^2\preceq\rho I_N$.
Let $\mathcal S_t$ be the Schur multiplier that fixes diagonal entries and multiplies
off-diagonal entries by $t$. It is unital and completely positive, because its symbol $(1-t)I+tJ$, with $J$ the all-ones matrix, is positive semidefinite with unit diagonal for $t\in[0,1]$. By Kadison's
inequality we obtain
\[
\sum_{i=1}^m\widehat{A}_i(t)^2
\preceq\sum_{i=1}^m\mathcal S_t(\widehat{A}_i^2)
=\mathcal S_t(\sum_{i=1}^m\widehat{A}_i^2)
\preceq\rho I_N.
\]
The case $t=0$ is the second bound, since $\widehat{A}_i(0)=\widehat{A}_{i,\mathrm{diag}}$, and Fact~\ref{fact:basic_psd_inequalities}(b) with $\widehat{A}_{i,\mathrm{off}}=\widehat{A}_i-\widehat{A}_{i,\mathrm{diag}}$ gives the third.
\end{proof}

With the interpolation argument we reduce the probability estimate to the diagonal endpoint and an integrated lower bound on the curvature of the log-partition function.

\begin{lemma}[Normalized interpolation]
\label{lem:smallball_interpolation}
Let $A_i$, $\xi_i$, $\widehat{A}_i$, $\mathcal{A}$, $\mathcal{A}_{\mathrm{diag}}$, $\mathcal{A}_{\mathrm{off}}$, $\mathcal{A}_t$, $\widehat{A}_i(t)$, $a_t$, and $a(t)$ be defined as in Definition~\ref{def:smallball_interpolation_data}. Assume moreover that $\kappa_{1}\ge2$. Then the following statements hold.
\begin{enumerate}[label=(\roman*)]
\item
For every $t\in[0,1]$, we have $\mathcal{A}_t=\sum_{i=1}^m \xi_i\widehat{A}_i(t)$ and $a_t>0$.
\item
For $t\in[0,1]$, define $\d\mu_t:=a_t^{-1}e^{-\Phi(\mathcal{A}_t)}\d\gamma_m$. Then
$a'(t)=-\E_{\mu_t}[D\Phi(\mathcal{A}_t)[\mathcal{A}_{\mathrm{off}}]]$.
\item
For every $t\in[0,1]$,
$a''(t)=\V_{\mu_t}[D\Phi(\mathcal{A}_t)[\mathcal{A}_{\mathrm{off}}]]-\E_{\mu_t}[D^2\Phi(\mathcal{A}_t)[\mathcal{A}_{\mathrm{off}},\mathcal{A}_{\mathrm{off}}]]$.
\item
The normalized matrices satisfy
$\sum_{i=1}^m \widehat{A}_i^2\preceq\rho I_N$, and the initial derivative satisfies $a'(0)=0$.
\item
The Gaussian small-ball probability satisfies
$\Pr[\|\sum_{i=1}^m \xi_iA_i\|<r_{\rm sb}]\ge a_1$.
\end{enumerate}
\end{lemma}

\begin{proof}
Items (i), (iii), and (v) and the identity $a'(0)=0$ of (iv) are part~\hyperlink{first-sb-part-1}{1} of the proof of Proposition~\ref{prop:first_small_ball}, which we read with $\mathcal{A}_t$ for $X_t$, $\mathcal{A}_{\mathrm{off}}$ for $V$, $\widehat{A}_i(t)$ for $D_i+tO_i$, $a_t$ for $\mathcal Z_t$, and $a(t)$ for $U(t)$.
There Eq.~\eqref{eq:first_13} gives (iii) and $a'(0)=0$, and Eq.~\eqref{eq:first_14} gives (v).
The dominated convergence there uses $(1-\|\mathcal{A}_t\|)^{\kappa_{1}-2}\le1$, which needs only $\kappa_{1}\ge2$.
The bound $\sum_{i=1}^m\widehat{A}_i^2\preceq\rho I_N$ of (iv) is Lemma~\ref{lem:smallball_variance_proxies} at $t=1$.

\emph{Proof of (ii).}
By (i), the measure $\mu_t$ is well defined. Put $G_t:=D\Phi(\mathcal{A}_t)[\mathcal{A}_{\mathrm{off}}]$ and $\Delta_t:=1-\|\mathcal{A}_t\|$. Here and in the rest of the section, we write $c_{\kappa_{1},\kappa_{0},\tau}$ for a constant depending only on $\kappa_{1},\kappa_{0},\tau$, whose value may change from line to line. Lemma~\ref{lem:smallball_boundary_decay} gives, on $\{\|\mathcal{A}_t\|<1\}$,
$e^{-\Phi(\mathcal{A}_t)}\le c_{\kappa_{1},\kappa_{0},\tau}\Delta_t^{\kappa_{1}}$. After we extend by zero, the integrand of $a_t$ against $\gamma_m$ and its first $t$-derivative are bounded, uniformly for $t\in[0,1]$, by $c\,(1+\|\xi\|_2)^2$. Here $c$ depends on $\kappa_{1},\kappa_{0},\tau,N,\rho$. This bound is $\gamma_m$-integrable, so we may apply dominated convergence without a moving cutoff. Since $\mathcal{A}_t'=\mathcal{A}_{\mathrm{off}}$, by the chain rule we obtain
\[
a_t'=-\int G_t e^{-\Phi(\mathcal{A}_t)}\d\gamma_m
=-a_t\E_{\mu_t}[G_t].
\]
Consequently,
\[
a'(t)=\frac{a_t'}{a_t}
=-\E_{\mu_t}[D\Phi(\mathcal{A}_t)[\mathcal{A}_{\mathrm{off}}]],
\]
which is (ii).
\end{proof}

\subsection{Divided-difference estimates}\label{subsec:smallball_divided_differences}
We quantify the scalar barrier near the spectral edge and in
the interior by divided-difference estimates, in the notation of Definition~\ref{def:first_divided_difference}.

\begin{lemma}[Divided differences]
\label{lem:smallball_divided_difference}
Let $0<w<1/20$, let $\tau\ge2$ be an integer, and let $\phi_0$ and $\phi_{1,\tau}$ be as in Definition~\ref{def:smallball_potential}. For every $u,v\in(-1,1)$, the following statements hold.
\begin{enumerate}[label=(\roman*)]
\item
\[
\phi_{1,\tau}'[u,v]
\ge\frac14\sum_{\sigma\in\{-1,1\}}
\frac{\mathbbm{1}[\min\{1+\sigma u,1+\sigma v\}\le w/\tau]}
{(1+\sigma u)(1+\sigma v)}.
\]
\item
\[
\phi_{1,\tau}'[u,v]
\le 2\tau\max\{|u|,|v|\}^{2\tau-2}\phi_0'[u,v].
\]
\end{enumerate}
\end{lemma}

\begin{proof}
Both bounds are Eqs.~\eqref{eq:first_16} and~\eqref{eq:first_17}, proved in part~\hyperlink{first-sb-part-2}{2} of the proof of Proposition~\ref{prop:first_small_ball} for the same $\phi_0$ and $\phi_{1,\tau}$.
That proof fixes $w=1/100$ but requires of $w$ only $1-10w>1/2$, so it holds for every $0<w<1/20$.
\end{proof}

\subsection{Spectral Hessian and the 0--1 decomposition}\label{subsec:smallball_zero_one}
From the scalar estimates we obtain spectral Hessian and $1$-component variance bounds.

\begin{definition}[Spectral 0--1 decomposition]
\label{def:smallball_zero_one_decomposition}
Fix an arbitrary $\kappa_3\in(0,1/20)$, retained throughout the general estimates below.
Fix $t\in[0,1]$, let $\widehat{A}_i(t)$ be as in Definition~\ref{def:smallball_interpolation_data}, and let $X$ be symmetric with $\|X\|<1$. For $\sigma\in\{-1,1\}$, define
\[
Y_\sigma:=I+\sigma X,
\qquad
C^\sigma(Y):=Y_\sigma^{-1/2}(\sigma Y)Y_\sigma^{-1/2}.
\]
Let $P_\sigma$ be the spectral projection of $Y_\sigma$ onto $[0,\kappa_3/\tau]$,
put $Q_\sigma:=I-P_\sigma$, and define
\[
C^\sigma_{1}(Y):=Q_\sigma C^\sigma(Y)Q_\sigma,
\qquad
C^\sigma_{0}(Y):=C^\sigma(Y)-C^\sigma_{1}(Y).
\]
Here index $0$ denotes the edge-touching component, and index $1$
denotes the regular component supported on $Q_\sigma$.
For the coefficient matrices at time $t$, define
\[
C_i^\sigma:=C^\sigma(\widehat{A}_i(t)),
\qquad
C_{i,1}^\sigma:=Q_\sigma C_i^\sigma Q_\sigma,
\qquad
C_{i,0}^\sigma:=C_i^\sigma-C_{i,1}^\sigma.
\]
\end{definition}

We apply the tensor Gram inequalities separately to the edge and regular components through the following quadratic forms.

\begin{definition}[Spectral quadratic forms]
\label{def:smallball_zero_one_forms}
Let $C_{i,0}^\sigma$ and $C_{i,1}^\sigma$ be defined in Definition~\ref{def:smallball_zero_one_decomposition}.
For $z_1,\ldots,z_m$ in a finite-dimensional Hilbert space, define
\begin{align*}
f_{0,\sigma}(z)&:=
\sum_{i_1=1}^m\sum_{i_2=1}^m\tr[C_{i_1,0}^\sigma
C_{i_2,0}^\sigma]\langle z_{i_1},z_{i_2}\rangle,\\
f_{1,\sigma}(z)&:=
\sum_{i_1=1}^m\sum_{i_2=1}^m\tr[C_{i_1,1}^\sigma
C_{i_2,1}^\sigma]\langle z_{i_1},z_{i_2}\rangle.
\end{align*}
\end{definition}

The barrier supplies strong curvature on the edge component, while the regular component retains a bounded quadratic variance. We use these two controls in the curvature calculation for the resolvent weights.

\begin{lemma}[Spectral 0--1 estimates]
\label{lem:smallball_zero_one}
Let the notation be as in Definitions~\ref{def:smallball_zero_one_decomposition}
and~\ref{def:smallball_zero_one_forms}.
Then the following statements hold.
\begin{enumerate}[label=(\roman*)]
\item
For every symmetric $Y$,
\[
D^2\Phi(X)[Y,Y]\ge
\sum_{\sigma\in\{-1,1\}}
(\frac{\kappa_{1}}{4}\|C^\sigma_{0}(Y)\|_F^2
+\kappa_{0}\|C^\sigma_{1}(Y)\|_F^2).
\]
\item
The $1$-components satisfy
\[
\sum_{i=1}^m(C_{i,1}^\sigma)^2
\preceq \kappa_3^{-2}\tau^2\rho I_N.
\]
\item
The forms $f_{0,\sigma}$ and $f_{1,\sigma}$ are nonnegative,
and
\[
\sum_{i_1=1}^m\sum_{i_2=1}^mD^2\Phi(X)[\widehat{A}_{i_1}(t),\widehat{A}_{i_2}(t)]\langle z_{i_1},z_{i_2}\rangle
\ge\sum_{\sigma\in\{-1,1\}}
(\frac{\kappa_{1}}{4} f_{0,\sigma}(z)+\kappa_{0} f_{1,\sigma}(z)).
\]
\end{enumerate}
\end{lemma}

\begin{proof}
Items (i) and (iii) are Eq.~\eqref{eq:first_19} in part~\hyperlink{first-sb-part-2}{2} of the proof of Proposition~\ref{prop:first_small_ball}, with $\kappa_3$ for $w$, $\widehat{A}_i(t)$ for $R_i$, and $Q_\sigma$ for $\bar P_\sigma$.
Item (i) is its entrywise form through Fact~\ref{fact:spectral_hessian} and Lemma~\ref{lem:smallball_divided_difference}(i) at $w=\kappa_3$, and the argument there uses nothing about $X$ beyond $\|X\|<1$.
Fact~\ref{fact:tensor_gram_inequalities} gives the nonnegativity in (iii).
Item (ii) is Eq.~\eqref{eq:first_20} there, whose input $\sum_{i=1}^m\widehat{A}_i(t)^2\preceq\rho I_N$ is Lemma~\ref{lem:smallball_variance_proxies}.
\end{proof}

\subsection{The matrix-weighted Poincar\'e estimate}
\label{subsec:smallball_poincare}
For convex matrix weights we need a Poincar\'e inequality that respects their
noncommutative geometry. Positive curvature turns weighted variance into a
Dirichlet-energy bound. For a positive definite weight $A$ on an open set $\Gamma\subseteq\R^{q_1}$, we say that a smooth field $f:\Gamma\to\R^{q_2}$ has finite energy if $\int f^\top Af$ and $\int\sum_{i=1}^{q_1}(\partial_if)^\top A(\partial_if)$ are finite.

\begin{lemma}[Matrix-weighted Poincar\'e criterion]
\label{lem:matrix_weighted_poincare}
Let $\Gamma\subseteq\R^{q_1}$ be open and convex, let $q_3>0$, and let $A:\Gamma\to \R^{q_2 \times q_2}$ be
smooth, symmetric positive definite, and integrable. Define
\[
\mathcal K_{i_1,i_2}(A):=(\partial_{i_1}A)A^{-1}(\partial_{i_2}A)-\partial_{i_1,i_2}A.
\]
If, for every $x_1,\ldots,x_{q_1}\in\R^{q_2}$,
\[
\sum_{i_1=1}^{q_1}\sum_{i_2=1}^{q_1}x_{i_1}^{\top}\mathcal K_{i_1,i_2}(A)x_{i_2}
\ge q_3\sum_{i=1}^{q_1} x_i^\top Ax_i,
\]
then every smooth $f:\Gamma\to\R^{q_2}$ of finite energy satisfies
\[\int f^\top Af-(\int Af)^\top(\int A)^{-1}(\int Af)\le q_3^{-1}\int\sum_{i=1}^{q_1}(\partial_if)^\top A(\partial_if).\]
\end{lemma}

\begin{proof}
This is Eq.~\eqref{eq:first_2} of Section~\ref{sec:first_existence} under Eq.~\eqref{eq:first_1}, with $\Gamma$ for $D$, $q_1$ for $d$, $A$ for $M$, $q_2$ for $r$, and $q_3$ for $a$. The hypotheses agree clause by clause. The domain is open and convex. The weight is smooth, symmetric positive definite, and integrable. The blocks $\mathcal K_{i_1,i_2}(A)$ are the blocks $\mathcal K_{ij}$ of that section. The finite-energy convention we stated before the lemma is the one of Section~\ref{sec:first_existence}. The conclusion there holds for every finite-energy field, in particular for the smooth ones. The proof given there applies verbatim. It uses the Neumann problem for $\mathcal Lu=-A^{-1}\sum_i\partial_i(A\partial_iu)$ on a smooth bounded convex subdomain, the Green identity, the identity Eq.~\eqref{eq:first_3} with its boundary term nonnegative by convexity, and an increasing smooth convex exhaustion of $\Gamma$.
\end{proof}

\subsection{One- and two-slot resolvent Poincar\'e inequalities}\label{subsec:smallball_resolvent_poincare}
From the matrix-weighted criterion we get Poincar\'e inequalities for the weights $e^{-\Phi(X)-\|\xi\|_2^2/2}Y_\sigma^{-1}$ and $e^{-\Phi(X)-\|\xi\|_2^2/2}Y_{\sigma_2}^{-1}\otimes Y_{\sigma_1}^{-1}$ on the small-ball domain. We call one Kronecker factor of the weight a slot, so these are the one- and two-slot inequalities.

\begin{definition}[Weighted resolvent energy]
\label{def:weighted_resolvent_quadratic_forms}
In the setting of Definition~\ref{def:smallball_interpolation_data}, fix $t\in[0,1]$, put $X:=\mathcal{A}_t(\xi)$ and $Y_\sigma:=I+\sigma X$ for $\sigma\in\{-1,1\}$, and let $\mu_t$ be defined as in Lemma~\ref{lem:smallball_interpolation}(ii). For $\sigma_1,\sigma_2\in\{-1,1\}$ and every matrix field $U=U(\xi)$ for which the following expectation is finite, define
\[
\mathcal E_{\sigma_1,\sigma_2}(U):=\E_{\mu_t}[\operatorname{vec}(U)^\top(Y_{\sigma_2}^{-1}\otimes Y_{\sigma_1}^{-1})\operatorname{vec}(U)].
\]
\end{definition}

Each resolvent factor contributes a negative term to the normalized curvature of the matrix weight. With the spectral $0$--$1$ estimates we absorb these terms into the barrier Hessian.

\begin{lemma}[One- and two-slot resolvent Poincar\'e inequalities]
\label{lem:resolvent_poincare}
Fix $t\in[0,1]$, put $X:=\mathcal{A}_t(\xi)$ and $Y_\sigma:=I+\sigma X$, and let $\mu_t$ be defined as in Lemma~\ref{lem:smallball_interpolation}(ii). Let $\kappa_3$ be as in Definition~\ref{def:smallball_zero_one_decomposition}. Assume $\kappa_{1}>8$, $\kappa_{0}(\kappa_{1}-8)>16$, and $\tau^2\rho\le\kappa_3^2/4$. Define
\[
c_1(\kappa_0,\kappa_1,\kappa_3,\tau,\rho):=\frac{1}{1-2\tau^2\rho/\kappa_3^2}.
\]
In the following statements, $c_1$ denotes this value.
\begin{enumerate}[label=(\roman*)]
\item
For every $\sigma\in\{-1,1\}$ and every smooth vector field $v:\R^m\to\R^N$ of finite energy, define $S_\sigma:=\E_{\mu_t}[Y_\sigma^{-1}]$ and $b_\sigma(v):=\E_{\mu_t}[Y_\sigma^{-1}v]$. Then
\[
\E_{\mu_t}[v^\top Y_\sigma^{-1}v]-b_\sigma(v)^\top S_\sigma^{-1}b_\sigma(v)
\le c_1\sum_{i=1}^m\E_{\mu_t}[(\partial_i v)^\top Y_\sigma^{-1}(\partial_i v)].
\]
\item
For every $\sigma_1,\sigma_2\in\{-1,1\}$ and every smooth vector field $f:\R^m\to\R^{N^2}$ of finite energy, define
\[
\mathcal Y_{\sigma_1,\sigma_2}:=Y_{\sigma_2}^{-1}\otimes Y_{\sigma_1}^{-1},\qquad
S_{\sigma_1,\sigma_2}:=\E_{\mu_t}[\mathcal Y_{\sigma_1,\sigma_2}],\qquad
b_{\sigma_1,\sigma_2}(f):=\E_{\mu_t}[\mathcal Y_{\sigma_1,\sigma_2}f].
\]
Then
\[
\E_{\mu_t}[f^\top \mathcal Y_{\sigma_1,\sigma_2}f]-b_{\sigma_1,\sigma_2}(f)^\top S_{\sigma_1,\sigma_2}^{-1}b_{\sigma_1,\sigma_2}(f)
\le c_1\sum_{i=1}^m\E_{\mu_t}[(\partial_i f)^\top \mathcal Y_{\sigma_1,\sigma_2}(\partial_i f)].
\]
\end{enumerate}
\end{lemma}

\begin{proof}
We apply Lemma~\ref{lem:matrix_weighted_poincare} to the one- and two-slot small-ball weights
\[
A_\sigma(\xi):=e^{-\|\xi\|_2^2/2-\Phi(X)}Y_\sigma^{-1},
\qquad
A_{\sigma_1,\sigma_2}(\xi):=e^{-\|\xi\|_2^2/2-\Phi(X)}
(Y_{\sigma_2}^{-1}\otimes Y_{\sigma_1}^{-1}),
\qquad \sigma,\sigma_1,\sigma_2\in\{-1,1\},
\]
with the normalized curvature blocks
\[
\widetilde{\mathcal K}^\sigma_{i_1,i_2}:=A_\sigma^{-1/2}\mathcal K_{i_1,i_2}(A_\sigma)A_\sigma^{-1/2},
\qquad
\widetilde {\mathcal K}_{i_1,i_2}^{\sigma_1,\sigma_2}:=A_{\sigma_1,\sigma_2}^{-1/2}\mathcal K_{i_1,i_2}(A_{\sigma_1,\sigma_2})A_{\sigma_1,\sigma_2}^{-1/2}.
\]

\emph{Step 1: curvature of the two-slot weight.}
The identity for $\widetilde {\mathcal K}_{i_1,i_2}^{\sigma_1,\sigma_2}$ is Eq.~\eqref{eq:first_23} in part~\hyperlink{first-sb-part-3}{3} of the proof of Proposition~\ref{prop:first_small_ball}, which we read with $\widehat{A}_i(t)$ for $R_i$, $\kappa_3$ for $w$, $Q_\sigma$ for $\bar P_\sigma$, and $z_i$ for $x_i$.
For $z_1,\ldots,z_m\in\R^N\otimes\R^N$, that part bounds each negative slot by $f_{0,\sigma}(z)+2\sqrt{f_{0,\sigma}(z)f_{1,\sigma}(z)}+\kappa_3^{-2}\tau^2\rho\sum_{i=1}^m\|z_i\|_2^2$, which here follows from Fact~\ref{fact:tensor_gram_inequalities} and Lemma~\ref{lem:smallball_zero_one}(ii).
Its Young step uses only $\kappa_{1}>8$ and $\kappa_{0}(\kappa_{1}-8)>16$, and absorbs the first two terms of each slot into one half of the barrier Hessian form of Lemma~\ref{lem:smallball_zero_one}(iii).
The two slots together give
\[
\sum_{i_1=1}^m\sum_{i_2=1}^m\langle z_{i_1},\widetilde {\mathcal K}_{i_1,i_2}^{\sigma_1,\sigma_2}z_{i_2}\rangle
\ge(1-2\kappa_3^{-2}\tau^2\rho)\sum_{i=1}^m\|z_i\|_2^2.
\]
That part rounds this curvature to $1/2$ at its fixed parameters, while here we keep the explicit value.
Hence $\lambda:=1-2\kappa_3^{-2}\tau^2\rho$ satisfies $\lambda\ge1/2$ by the hypothesis $\tau^2\rho\le\kappa_3^2/4$, and $[\widetilde {\mathcal K}_{i_1,i_2}^{\sigma_1,\sigma_2}]_{i_1,i_2}\succeq\lambda I$.

\emph{Step 2: curvature of the one-slot weight.}
The identity for $\widetilde {\mathcal K}_{i_1,i_2}^{\sigma}$ has a single negative product, so from the same estimates we obtain
\[
[\widetilde {\mathcal K}_{i_1,i_2}^{\sigma}]_{i_1,i_2}\succeq(1-\kappa_3^{-2}\tau^2\rho)I\succeq\lambda I,
\]
where the second step follows from $\kappa_3^{-2}\tau^2\rho\ge0$.

\emph{Step 3: the domain and integrability.}
We apply the curvature criterion on
\[
\Omega_t:=\{\xi\in\R^m:\|\mathcal{A}_t(\xi)\|<1\}.
\]
This set is open, convex, and contains the origin. The weights are smooth
and positive definite on $\Omega_t$. If
$\Delta(\xi):=1-\|\mathcal{A}_t(\xi)\|$, then Lemma~\ref{lem:smallball_boundary_decay} gives $e^{-\Phi(\mathcal{A}_t(\xi))}\le c_{\kappa_{1},\kappa_{0},\tau}\Delta(\xi)^{\kappa_{1}}$.
Since each resolvent has norm at most $\Delta(\xi)^{-1}$, the one-slot and
two-slot weights are bounded by a Gaussian density times, respectively,
$c_{\kappa_{1},\kappa_{0},\tau}\Delta^{\kappa_{1}-1}$ and $c_{\kappa_{1},\kappa_{0},\tau}\Delta^{\kappa_{1}-2}$. Thus they and their
products with polynomial test fields are integrable because $\kappa_{1}>8$.
We apply Lemma~\ref{lem:matrix_weighted_poincare} with $q_1=m$, $\Gamma=\Omega_t$, and $q_3=\lambda=1-2\kappa_3^{-2}\tau^2\rho$. For (i), we take $q_2=N$ and $A=A_\sigma$. For (ii), we take $q_2=N^2$ and $A=A_{\sigma_1,\sigma_2}$. Since
\[
\d\mu_t(\xi)=\frac{e^{-\|\xi\|_2^2/2-\Phi(X)}}{(2\pi)^{m/2}a_t}\d\xi,
\]
we divide each resulting inequality by $(2\pi)^{m/2}a_t$ and obtain the corresponding assertion with
\[c_1(\kappa_0,\kappa_1,\kappa_3,\tau,\rho)=q_3^{-1}=\lambda^{-1}.\]
\end{proof}

\subsection{The weighted resolvent estimate}\label{subsec:smallball_weighted_resolvent}
The one- and two-slot inequalities yield the resolvent estimate that we use in the moment argument.

\begin{lemma}[Weighted resolvent estimate]
\label{lem:smallball_weighted_poincare}
Fix $t\in[0,1]$, and put $X:=\mathcal{A}_t(\xi)$ and $Y_\sigma:=I+\sigma X$. Let $\mathcal E_{\sigma_1,\sigma_2}$ be defined as in Definition~\ref{def:weighted_resolvent_quadratic_forms}. Let $0<\kappa_3<1/20$ be as in Definition~\ref{def:smallball_zero_one_decomposition}, and let $\tau\ge2$ be an integer.
Assume $\kappa_1>8$, $\kappa_0(\kappa_1-8)>16$, and $\tau^2\rho\le\kappa_3^2/4$. Let $c_1=c_1(\kappa_0,\kappa_1,\kappa_3,\tau,\rho)$ be as defined in Lemma~\ref{lem:resolvent_poincare}, and define
\[
\eta:=\bigl(1+8c_1(1+2c_1\rho)\bigr)\rho,
\qquad
c_2(c_1,\rho):=
\frac{2(1+\rho)}{1-2(1+\rho)(\rho+\eta/(1-\eta))}.
\]
The denominators are positive, and $c_2(c_1,\rho)<3$ throughout the admissible parameter range. Hereafter $c_2$ denotes this value. For every $\sigma_1,\sigma_2\in\{-1,1\}$ and every smooth matrix field $U$ of finite energy,
\[
\mathcal E_{\sigma_1,\sigma_2}(U)
\le c_2\E_{\mu_t}[\|U\|_F^2]+c_2\sum_{i=1}^m\mathcal E_{\sigma_1,\sigma_2}(\partial_iU).
\]
\end{lemma}

\begin{proof}
We retain the notation $X:=\mathcal{A}_t(\xi)$. The assumptions $0<\kappa_3<1/20$ and $\tau\ge2$ give
\[
\rho\le\frac{\kappa_3^2}{4\tau^2}<\frac1{6400},\qquad
c_1=\frac{1}{1-2\tau^2\rho/\kappa_3^2},\qquad 1\le c_1\le2.
\]
The first bound is the hypothesis on $\tau^2\rho$. The second uses $\kappa_3^2<1/400$ and $\tau^2\ge4$. The formula for $c_1$ is its definition in Lemma~\ref{lem:resolvent_poincare}, and $c_1\le2$ follows from $\tau^2\rho\le\kappa_3^2/4$. In particular,
\[
\eta\le17\rho+64\rho^2<\frac1{200}.
\]
The first step uses $c_1\le2$, and the second uses $\rho<1/6400$. It follows that
\[
2(1+\rho)(\rho+\eta/(1-\eta))
<2(1+1/6400)(1/6400+1/199)<1/50.
\]
The first step uses $\eta/(1-\eta)<1/199$ and $\rho<1/6400$. The second is a rational comparison. Thus both denominators in the definition of $c_2$ are positive, and
\[
c_2(c_1,\rho)<\frac{2(1+1/6400)}{1-1/50}<3.
\]
The first step uses the preceding denominator bound, and the second is a rational comparison. We retain the parameter-dependent coefficients below, using the one-slot and two-slot estimates of Lemma~\ref{lem:resolvent_poincare}(i)--(ii) with coefficient $c_1$.

\emph{Step 1: the projected mean of a one-slot weight.}
Put
\[
S_\sigma:=\E_{\mu_t}[Y_\sigma^{-1}].
\]
Evenness gives $\E_{\mu_t}[X]=0$, hence
$\E_{\mu_t}[Y_\sigma]=I$. Operator convexity of inversion yields
$S_\sigma\succeq I$. For $z\in\R^N$, we apply the one-slot inequality to
\[
v_{\sigma,z}:=(Y_\sigma-I)z.
\]
Since $Y_\sigma^{-1}v_{\sigma,z}=(I-Y_\sigma^{-1})z$,
$\partial_iv_{\sigma,z}=\sigma \widehat{A}_i(t)z$, and
$\E_{\mu_t}[Y_\sigma]=I$, we expand the weighted variance directly and obtain
\[
I-S_\sigma^{-1}\preceq c_1\sum_{i=1}^m\widehat{A}_i(t)S_\sigma \widehat{A}_i(t).
\]
If $L_\sigma:=\|S_\sigma\|$, then
\[
1-L_\sigma^{-1}\le c_1\rho L_\sigma.
\]
To exclude the large root, for $u\in[0,1]$ we define
\[
X_u:=uX,\qquad Y_{\sigma,u}:=I+\sigma X_u,
\qquad
a_{t,u}:=\E[e^{-\Phi(X_u)}],\qquad
\d\mu_{t,u}:=a_{t,u}^{-1}e^{-\Phi(X_u)}\,\d\gamma_m,
\]
\[
S_{\sigma,u}:=\E_{\mu_{t,u}}[Y_{\sigma,u}^{-1}],
\qquad L_\sigma(u):=\|S_{\sigma,u}\|.
\]
The function $c_1(\kappa_0,\kappa_1,\kappa_3,\tau,\rho)$ is nondecreasing in $\rho$ on the admissible range. Since $u^2\rho\le\rho$, Lemma~\ref{lem:resolvent_poincare} applies to the coefficients $u\widehat{A}_i(t)$ with a coefficient no larger than the fixed value $c_1$ at $\rho$. The same argument therefore gives
\[
1-L_\sigma(u)^{-1}\le c_1u^2\rho L_\sigma(u).
\]
Put $\Delta_u:=1-\|X_u\|$ on $\{\|X_u\|<1\}$. By Lemma~\ref{lem:smallball_boundary_decay}, $e^{-\Phi(X_u)}\le2^{\kappa_1}\exp[(\kappa_1-\kappa_0)\sum_{k=1}^{\tau-1}1/k]\,\Delta_u^{\kappa_1}$.
Consequently, the integrand in the numerator defining $S_{\sigma,u}$ satisfies
\[
\|e^{-\Phi(X_u)}Y_{\sigma,u}^{-1}\|
\le2^{\kappa_1}\exp[(\kappa_1-\kappa_0)\sum_{k=1}^{\tau-1}1/k]\Delta_u^{\kappa_1-1}
\le2^{\kappa_1}\exp[(\kappa_1-\kappa_0)\sum_{k=1}^{\tau-1}1/k].
\]
The first step uses $\|Y_{\sigma,u}^{-1}\|\le\Delta_u^{-1}$, and the second uses $0<\Delta_u\le1$ and $\kappa_1>8$. We extend the numerator and denominator integrands by zero outside $\{\|X_u\|<1\}$. These extensions are continuous at the boundary, and the denominator integrand is bounded by $1$. Dominated convergence, together with $a_{t,u}>0$, shows that $L_\sigma(u)$ is continuous. Since $L_\sigma(0)=1$, this path stays on the smaller branch of
the quadratic. Consequently,
\begin{equation}
I\preceq S_\sigma\preceq
\frac{2}{1+\sqrt{1-4c_1\rho}}I
\preceq(1+2c_1\rho)I.
\label{eq:smallball_one_slot_resolvent}
\end{equation}
The last step uses $2/(1+\sqrt{1-4x})\le1+2x$, valid for $0\le x\le(\sqrt2-1)/2$. Squared out, the inequality is $4x^2+4x-1\le0$. We apply the inequality with $x=c_1\rho\le2\rho\le\kappa_3^2/8<1/3200$, since $\tau\ge2$ and $0<\kappa_3<1/20$.

\emph{Step 2: the two-slot expectation.}
For the two-slot weight, put
\[
\mathcal Y:=Y_{\sigma_2}^{-1}\otimes Y_{\sigma_1}^{-1},
\qquad
S:=\E_{\mu_t}[\mathcal Y],
\qquad
\mathcal R:=\mathcal Y^{-1}=Y_{\sigma_2}\otimes Y_{\sigma_1},
\qquad
T:=\E_{\mu_t}[\mathcal R].
\]
The potential $\xi\mapsto\Phi(\mathcal{A}_t(\xi))$ is even, convex, and finite near the origin, so Fact~\ref{fact:even_convex_tilts} gives $\Cov_{\mu_t}[\xi]\preceq I$. Fact~\ref{fact:kronecker_cauchy_schwarz} with $\sum_{i=1}^m\widehat A_i(t)^2\preceq\rho I$ then gives $\|\E_{\mu_t}[X\otimes X]\|\le\rho$. Therefore
\begin{equation}
T=I+\sigma_1\sigma_2\E_{\mu_t}[X\otimes X],
\qquad
\|T-I\|\le\rho.
\label{eq:smallball_T_close}
\end{equation}

\emph{Step 3: closeness of $S$ to the identity.}
For $z\in\R^{N^2}$, we apply the two-slot inequality to
$\Psi_z:=(\mathcal R-I)z$. Its weighted variance is
$z^\top(T-S^{-1})z$. Moreover,
\[
\partial_i\mathcal R
=\sigma_2 \widehat{A}_i(t)\otimes Y_{\sigma_1}
+Y_{\sigma_2}\otimes \sigma_1 \widehat{A}_i(t).
\]
By $(P+Q)^\top\mathcal Y(P+Q)\preceq2P^\top\mathcal YP+2Q^\top\mathcal YQ$, $Y_\sigma\preceq2I$, $\sum_{i=1}^m\widehat{A}_i(t)^2\preceq\rho I$, and Eq.~\eqref{eq:smallball_one_slot_resolvent},
\begin{align*}
\sum_{i=1}^m\E_{\mu_t}[(\partial_i\mathcal R)\mathcal Y(\partial_i\mathcal R)]
&\preceq4(\sum_{i=1}^m\widehat{A}_i(t)S_{\sigma_2}\widehat{A}_i(t))\otimes I\\
&\quad+4I\otimes(\sum_{i=1}^m\widehat{A}_i(t)S_{\sigma_1}\widehat{A}_i(t))\\
&\preceq8(1+2c_1\rho)\rho I.
\end{align*}
The first step uses a factor $2$ from the quadratic-form bound and a factor $2$ from $Y_\sigma\preceq2I$ in each tensor term. The second uses $S_\sigma\preceq(1+2c_1\rho)I$ and $\sum_i\widehat{A}_i(t)^2\preceq\rho I$. The two-slot estimate with coefficient $c_1$ therefore gives
\[
0\preceq T-S^{-1}\preceq8c_1(1+2c_1\rho)\rho I.
\]
Together with Eq.~\eqref{eq:smallball_T_close}, this implies
\[
\|S^{-1}-I\|\le\rho+8c_1(1+2c_1\rho)\rho=\eta.
\]
The first step is the triangle inequality, and the second is the definition of $\eta$. Since $\eta<1$, we have $S^{-1}\succeq(1-\eta)I$. Hence $\|S\|\le(1-\eta)^{-1}$. Consequently,
\[
\|S-I\|=\|S(I-S^{-1})\|\le\|S\|\|I-S^{-1}\|\le\frac{\eta}{1-\eta}.
\]
The first step is the identity $S(I-S^{-1})=S-I$. The second is submultiplicativity, and the last uses the two preceding bounds. Also $S^{-1}\preceq T\preceq(1+\rho)I$. We have proved
\begin{equation}
\|S^{-1}-I\|\le\eta,\qquad
\|S-I\|\le\frac{\eta}{1-\eta},\qquad
S^{-1}\preceq(1+\rho)I.
\label{eq:smallball_two_slot_resolvent}
\end{equation}

\emph{Step 4: the projected mean and the conclusion.}
Now let $f:=\operatorname{vec}(U)$ and put
\[
b:=\E_{\mu_t}[\mathcal Yf],\qquad
w:=\E_{\mu_t}[(\mathcal Y-I)f],\qquad
\text{so that}\quad b=\E_{\mu_t}[f]+w.
\]
The identity
\[
(\mathcal Y-I)\mathcal Y^{-1}(\mathcal Y-I)
=\mathcal Y-2I+\mathcal Y^{-1}
\]
and Eq.~\eqref{eq:smallball_T_close} and Eq.~\eqref{eq:smallball_two_slot_resolvent} give
\[
0\preceq\E_{\mu_t}[(\mathcal Y-I)\mathcal Y^{-1}(\mathcal Y-I)]
=S-2I+T
\preceq(\rho+\eta/(1-\eta))I.
\]
The first step uses positive semidefiniteness of the integrand. The second follows from the displayed identity. The last uses $S-I\preceq\eta(1-\eta)^{-1}I$ and $T-I\preceq\rho I$.
Weighted Cauchy--Schwarz and ordinary Cauchy--Schwarz give, respectively,
\[
\|w\|_2^2\le(\rho+\eta/(1-\eta))\mathcal E_{\sigma_1,\sigma_2}(U),\qquad
\|\E_{\mu_t}[f]\|_2^2\le\E_{\mu_t}[\|U\|_F^2].
\]
Therefore
\begin{align*}
b^\top S^{-1}b
&\le(1+\rho)\|b\|_2^2\\
&\le2(1+\rho)(\|\E_{\mu_t}[f]\|_2^2+\|w\|_2^2)\\
&\le2(1+\rho)\E_{\mu_t}[\|U\|_F^2]
+2(1+\rho)(\rho+\eta/(1-\eta))\mathcal E_{\sigma_1,\sigma_2}(U).
\end{align*}
The first step uses $S^{-1}\preceq(1+\rho)I$. The second uses $b=\E_{\mu_t}[f]+w$ and $\|x+y\|_2^2\le2\|x\|_2^2+2\|y\|_2^2$. The last uses the two preceding Cauchy--Schwarz bounds.
Substituting this projected-mean bound into Lemma~\ref{lem:resolvent_poincare}(ii), we obtain
\begin{align*}
\mathcal E_{\sigma_1,\sigma_2}(U)
&\le2(1+\rho)\E_{\mu_t}[\|U\|_F^2]
+c_1\sum_{i=1}^m\mathcal E_{\sigma_1,\sigma_2}(\partial_iU)\\
&\quad+2(1+\rho)(\rho+\eta/(1-\eta))\mathcal E_{\sigma_1,\sigma_2}(U).
\end{align*}
The denominator $1-2(1+\rho)(\rho+\eta/(1-\eta))$ is positive, as we checked at the start of the proof. Moving the last term to the left and dividing, we obtain
\begin{align*}
\mathcal E_{\sigma_1,\sigma_2}(U)
&\le\frac{2(1+\rho)\E_{\mu_t}[\|U\|_F^2]+c_1\sum_{i=1}^m\mathcal E_{\sigma_1,\sigma_2}(\partial_iU)}
{1-2(1+\rho)(\rho+\eta/(1-\eta))}\\
&\le c_2(c_1,\rho)\E_{\mu_t}[\|U\|_F^2]
+c_2(c_1,\rho)\sum_{i=1}^m\mathcal E_{\sigma_1,\sigma_2}(\partial_iU).
\end{align*}
The first step is this rearrangement. The last uses the displayed definition of $c_2(c_1,\rho)$ and $c_1\le2\le2(1+\rho)$.
\end{proof}

\subsection{Polynomial resolvent moments and the interpolation Hessian}
\label{subsec:smallball_moments}
Repeated differentiation produces noncommutative polynomials in Gaussian matrix
series. We control these polynomials uniformly over the interpolation path
by Schatten interpolation and Gaussian moment bounds.

\begin{lemma}[Polynomial resolvent and interpolation bounds]
\label{lem:smallball_tail_hessian_comparison}
In the notation of Definitions~\ref{def:smallball_interpolation_data}, \ref{def:smallball_zero_one_decomposition}, and~\ref{def:weighted_resolvent_quadratic_forms}, and with $c_2$ as in Lemma~\ref{lem:smallball_weighted_poincare}, define
\[
c_3:=4c_2(1+c_2),\qquad c_4:=2c_3.
\]
These constants are uniformly bounded over the admissible parameter range, and the following statements hold.
\begin{enumerate}[label=(\roman*)]
\item
For symmetric $X,V$ with $\|X\|<1$,
\[
D^2\tr[\phi_{1,\tau}(X)][V,V]
\le4\tau\sum_{\sigma\in\{-1,1\}}
\operatorname{vec}(X^{\tau-1}V)^\top
(Y_\sigma^{-1}\otimes Y_\sigma^{-1})
\operatorname{vec}(X^{\tau-1}V).
\]
\item
Assume $\kappa_{1}>8$, $\kappa_{0}(\kappa_{1}-8)>16$, and $\tau^2\rho\le\kappa_3^2/4$. For $t\in[0,1]$, put $X:=\mathcal{A}_t(\xi)$. Then, for $\sigma_1,\sigma_2\in\{-1,1\}$ and uniformly for $t\in[0,1]$,
\[
\mathcal E_{\sigma_1,\sigma_2}(X^{\tau-1}\mathcal{A}_{\mathrm{off}})\le N(c_3\tau^2\rho)^\tau,
\qquad
\mathcal E_{\sigma_1,\sigma_2}(\mathcal{A}_{\mathrm{off}})\le c_3N\rho,
\]
and
\[
\E_{\mu_t}[D^2\Phi(\mathcal{A}_t)[\mathcal{A}_{\mathrm{off}},\mathcal{A}_{\mathrm{off}}]]
\le c_4\kappa_{0} N\rho+c_4\kappa_{1}\tau N(c_3\tau^2\rho)^\tau.
\]
\end{enumerate}
\end{lemma}

\begin{proof}
\emph{Proof of (i).}
Item (i) is the tail comparison in part~\hyperlink{first-sb-part-6}{6} of the proof of Proposition~\ref{prop:first_small_ball}, whose argument uses only that $X$ and $V$ are symmetric with $\|X\|<1$.
In an eigenbasis of $X$ it applies Lemma~\ref{lem:smallball_divided_difference}(ii) through Fact~\ref{fact:spectral_hessian}, then $\max\{u,v\}^{2\tau-2}\le u^{2\tau-2}+v^{2\tau-2}$ for $u,v\ge0$.
The symmetry of $V$ makes the two resulting sums equal, and their common value is the displayed resolvent form at $X^{\tau-1}V$, so the coefficient $2\tau$ of Lemma~\ref{lem:smallball_divided_difference}(ii) becomes $4\tau$.

\emph{Proof of (ii).}
We use $c_2$ from Lemma~\ref{lem:smallball_weighted_poincare} without changing its value. Its defining denominator lies in $(0,1]$ and its numerator is $2(1+\rho)$, so $c_2\ge2$. We verify the stated choices of $c_3$ and $c_4$ in the calculations below.

\emph{Step 1: variance proxies.}
By Lemma~\ref{lem:smallball_variance_proxies}, the $\widehat{A}_i(t)$ have variance proxy $\rho I$, and every family of coefficient matrices that occurs below has variance proxy at most $4\rho I$.

\emph{Step 2: the polynomial moments.}
For all derivative tensors we use the norm $\|\nabla^jF\|_{\mathrm{HS}}$ of Section~\ref{sec:preliminaries}, which sums over all ordered index tuples.
We apply the defective Poincar\'e estimate of
Lemma~\ref{lem:smallball_weighted_poincare} to every derivative field of order $j$ and sum over all ordered indices. Since
$\nabla^{\tau+1}(X^{\tau-1}\mathcal{A}_{\mathrm{off}})=0$, induction on the degree gives
\[
\mathcal E_{\sigma_1,\sigma_2}(X^{\tau-1}\mathcal{A}_{\mathrm{off}})
\le\sum_{j=0}^\tau c_2^{j+1}
\E_{\mu_t}[\|\nabla^j(X^{\tau-1}\mathcal{A}_{\mathrm{off}})\|_{\mathrm{HS}}^2].
\]
For $0\le j\le\tau$, put $s:=\tau-j$. We claim
\begin{equation}
\E_{\mu_t}[\|\nabla^j(X^{\tau-1}\mathcal{A}_{\mathrm{off}})\|_{\mathrm{HS}}^2]\le N(4\rho)^\tau\tau^{2j}(2s)^s\le N(4\tau^2\rho)^\tau.
\label{eq:smallball_polynomial_moment}
\end{equation}
Here $0^0:=1$. The first step will follow from the word estimates below. The second uses $2s\le2\tau\le\tau^2$, since $\tau\ge2$, and $j+s=\tau$. Let $g_1,\ldots,g_j$ be independent standard Gaussian
directions, also independent of $\xi$.
Gaussian contraction of the derivative tensor and the product rule are part~\hyperlink{first-sb-part-5}{5} of the proof of Proposition~\ref{prop:first_small_ball}, read with $\mathcal{A}_{\mathrm{off}}$ for $V$.
They write $\|\nabla^j(X^{\tau-1}\mathcal{A}_{\mathrm{off}})(\xi)\|_{\mathrm{HS}}^2$ as $\E_g$ of the squared Frobenius norm of a sum of at most $(\tau)_j:=\tau(\tau-1)\cdots(\tau-j+1)\le \tau^j$ words. Each word has $\tau$ factors, exactly $j$ evaluated at
the distinct directions $g_1,\ldots,g_j$, and $s$ evaluated at the
common vector $\xi$. Every factor is a copy of either $X$ or $\mathcal{A}_{\mathrm{off}}$.

For a direction factor $G:=\sum_{i=1}^m g_iR_i$, we define the completely positive map
\[
\Phi_G(U):=\E_g[GUG]=\sum_{i=1}^mR_iUR_i.
\]
The coefficient matrices are symmetric and satisfy
$\sum_{i=1}^mR_i^2\preceq4\rho I_N$ by Step~1.
Fact~\ref{fact:cp_schatten_shift} with $\ell=\infty$ and $\|\Phi_G(I_N)\|\le4\rho$ gives
\begin{equation}
\|\Phi_G(U)\|_{S_q}\le4\rho\|U\|_{S_q},
\qquad 1\le q\le\infty.
\label{eq:smallball_cp_schatten}
\end{equation}

Fix one word $\mathcal{B}:=\widetilde{B}_1\cdots \widetilde{B}_\tau$, and let $G_1(\xi),\ldots,G_s(\xi)$ denote its common-$\xi$ factors in their order.
Part~\hyperlink{first-sb-part-5}{5} of the proof of Proposition~\ref{prop:first_small_ball} writes $\|\mathcal{B}\|_F^2=\tr[B_\tau]$ with $B_0:=I_N$ and $B_k:=\widetilde{B}_kB_{k-1}\widetilde{B}_k$, and averages this recursion conditionally on $\xi$.
Each independent-direction factor acts by $\Phi_{\widetilde{B}_k}$ and preserves the current Schatten exponent at cost $4\rho$ by Eq.~\eqref{eq:smallball_cp_schatten}.
Schatten H\"older at the $\ell$-th common-$\xi$ factor passes from $S_{s/(\ell-1)}$ to $S_{s/\ell}$ at cost $\|G_\ell(\xi)\|_{S_{2s}}^2$, with $s/0:=\infty$.
Starting in $S_\infty$ and ending in $S_1$, we obtain, conditionally on $\xi$ and for $s\ge1$,
\begin{equation}
\E_g[\|\mathcal{B}\|_F^2]
\le(4\rho)^j\prod_{\ell=1}^s\|G_\ell(\xi)\|_{S_{2s}}^2.
\label{eq:smallball_one_word}
\end{equation}
We treat the case $s=0$ separately in Step~3 below.

\emph{Step 3: Gaussian moments under the tilt.}
Suppose first that $s\ge1$. We use scalar H\"older under $\mu_t$, Harg\'e's theorem
applied separately to each even convex trace moment, and the Gaussian
trace-moment inequality for a Gaussian matrix series with variance proxy $\nu I$. Here $\E_\gamma$ is expectation under $\gamma_m$, and Fact~\ref{fact:gaussian_trace_moments} gives $\E_\gamma[\tr[|G|^{2s}]]\le N(2s-1)!!\nu^s\le N(2s\nu)^s$. Together they give
\begin{align*}
\E_{\mu_t}[\prod_{\ell=1}^s\|G_\ell(\xi)\|_{S_{2s}}^2]
&\le\prod_{\ell=1}^s
(\E_{\mu_t}[\tr[|G_\ell(\xi)|^{2s}]])^{1/s}\\
&\le\prod_{\ell=1}^s
(\E_\gamma[\tr[|G_\ell(\xi)|^{2s}]])^{1/s}\\
&\le N(8s\rho)^s.
\end{align*}
The first step is scalar H\"older, the second is Harg\'e's theorem, and the last uses the variance bound $\nu\le4\rho$ from Step~1 and $(2s-1)!!\le(2s)^s$.
Fact~\ref{fact:even_convex_tilts} applies to $\mu_t$ and to the even convex function $\xi\mapsto\tr[|G(\xi)|^{2s}]=\|G(\xi)\|_{S_{2s}}^{2s}$.

For $s\ge1$, we combine this bound with Eq.~\eqref{eq:smallball_one_word} and obtain
\[
\E_{\mu_t}[\E_g[\|\mathcal{B}\|_F^2]]\le N(4\rho)^j(8s\rho)^s=N(4\rho)^\tau(2s)^s.
\]
The first step combines the two estimates, and the second uses $j+s=\tau$.
If $s=0$, all factors are independent-direction factors. Applying the $\tau$ maps in $S_\infty$ and then using $\tr[U]\le N\|U\|$, we obtain
\[
\E_g[\|\mathcal{B}\|_F^2]\le N(4\rho)^\tau.
\]
This uses the bound $4\rho$ for each map and agrees with the preceding word bound under the convention $0^0=1$.
There are at most $(\tau)_j\le\tau^j$ words in the derivative. Cauchy--Schwarz for their sum therefore gives
\begin{align*}
\E_{\mu_t}[\|\nabla^j(X^{\tau-1}\mathcal{A}_{\mathrm{off}})\|_{\mathrm{HS}}^2]
&\le(\tau)_j^2N(4\rho)^\tau(2s)^s\\
&\le N(4\rho)^\tau\tau^{2j}(2s)^s\\
&\le N(4\tau^2\rho)^\tau.
\end{align*}
The first step combines the word bound with Cauchy--Schwarz. The second uses $(\tau)_j\le\tau^j$. The last uses $2s\le2\tau\le\tau^2$ and $j+s=\tau$. This proves Eq.~\eqref{eq:smallball_polynomial_moment}.

\emph{Step 4: the linear field and the resolvent moments.}
For the linear field $\mathcal{A}_{\mathrm{off}}$, Harg\'e's theorem in Fact~\ref{fact:even_convex_tilts} and Step~1 give
\[
\E_{\mu_t}[\|\mathcal{A}_{\mathrm{off}}\|_F^2]
\le\E_\gamma[\|\mathcal{A}_{\mathrm{off}}\|_F^2]
=\sum_{i=1}^m\tr[\widehat{A}_{i,\mathrm{off}}^2]
\le4N\rho.
\]
The first step is Harg\'e's theorem, the second is the Gaussian second-moment identity, and the last is the variance bound from Step~1.
Since $\partial_i\mathcal{A}_{\mathrm{off}}=\widehat{A}_{i,\mathrm{off}}$ is constant, Lemma~\ref{lem:smallball_weighted_poincare} gives
\[
\sum_{i=1}^m\mathcal E_{\sigma_1,\sigma_2}(\partial_i\mathcal{A}_{\mathrm{off}})
\le c_2\sum_{i=1}^m\|\widehat{A}_{i,\mathrm{off}}\|_F^2
\le4c_2N\rho.
\]
The first step uses that all derivatives of the constant fields vanish, and the second uses Step~1.
Applying Lemma~\ref{lem:smallball_weighted_poincare} once more, we obtain
\begin{align*}
\mathcal E_{\sigma_1,\sigma_2}(\mathcal{A}_{\mathrm{off}})
&\le c_2\E_{\mu_t}[\|\mathcal{A}_{\mathrm{off}}\|_F^2]
+c_2\sum_{i=1}^m\mathcal E_{\sigma_1,\sigma_2}(\partial_i\mathcal{A}_{\mathrm{off}})\\
&\le4c_2(1+c_2)N\rho=c_3N\rho.
\end{align*}
The first step is Lemma~\ref{lem:smallball_weighted_poincare}, the second uses the two preceding bounds, and the equality is the definition of $c_3$.
For the degree-$\tau$ field $X^{\tau-1}\mathcal{A}_{\mathrm{off}}$, the coefficients in the iteration from Step~2 satisfy
\[
\sum_{j=0}^\tau c_2^{j+1}\le c_2(1+c_2)^\tau\le[c_2(1+c_2)]^\tau=(c_3/4)^\tau.
\]
The first step uses the binomial expansion of $(1+c_2)^\tau$ and $\binom{\tau}{j}\ge1$ for $0\le j\le\tau$. The second uses $c_2\ge1$ and $\tau\ge2$. The last step is the definition of $c_3$.
Consequently,
\begin{align*}
\mathcal E_{\sigma_1,\sigma_2}(X^{\tau-1}\mathcal{A}_{\mathrm{off}})
&\le\sum_{j=0}^\tau c_2^{j+1}\E_{\mu_t}[\|\nabla^j(X^{\tau-1}\mathcal{A}_{\mathrm{off}})\|_{\mathrm{HS}}^2]\\
&\le N(4\tau^2\rho)^\tau\sum_{j=0}^\tau c_2^{j+1}\\
&\le N(c_3\tau^2\rho)^\tau.
\end{align*}
The first step is the iteration from Step~2, the second uses Eq.~\eqref{eq:smallball_polynomial_moment}, and the last uses the preceding bound on the sum.
Thus
\begin{equation}
\mathcal E_{\sigma_1,\sigma_2}(X^{\tau-1}\mathcal{A}_{\mathrm{off}})
\le N(c_3\tau^2\rho)^\tau,
\qquad
\mathcal E_{\sigma_1,\sigma_2}(\mathcal{A}_{\mathrm{off}})\le c_3N\rho.
\label{eq:smallball_resolvent_moments}
\end{equation}

\emph{Step 5: the interpolation Hessian.}
The logarithmic identity is
\[
D^2\tr[\phi_0(X)][\mathcal{A}_{\mathrm{off}},\mathcal{A}_{\mathrm{off}}]
=\sum_{\sigma\in\{-1,1\}}
\operatorname{vec}(\mathcal{A}_{\mathrm{off}})^\top
(Y_\sigma^{-1}\otimes Y_\sigma^{-1})\operatorname{vec}(\mathcal{A}_{\mathrm{off}}).
\]
Therefore
\begin{align}
\E_{\mu_t}[D^2\Phi(X)[\mathcal{A}_{\mathrm{off}},\mathcal{A}_{\mathrm{off}}]]
&\le\kappa_0\sum_{\sigma\in\{-1,1\}}\mathcal E_{\sigma,\sigma}(\mathcal{A}_{\mathrm{off}})
+4\tau(\kappa_1-\kappa_0)\sum_{\sigma\in\{-1,1\}}\mathcal E_{\sigma,\sigma}(X^{\tau-1}\mathcal{A}_{\mathrm{off}})\nonumber\\
&\le2c_3\kappa_0N\rho+8\kappa_1\tau N(c_3\tau^2\rho)^\tau\nonumber\\
&\le c_4\kappa_0N\rho+c_4\kappa_1\tau N(c_3\tau^2\rho)^\tau.
\label{eq:smallball_interpolation_hessian}
\end{align}
The first step uses the logarithmic identity, part~(i), and $\phi=\kappa_0\phi_0+(\kappa_1-\kappa_0)\phi_{1,\tau}$ with $\kappa_1\ge\kappa_0>0$. The second uses Eq.~\eqref{eq:smallball_resolvent_moments}, $\kappa_1-\kappa_0\le\kappa_1$, and the two values of $\sigma$, giving the factors $2$ and $4\cdot2=8$. The last uses $c_4=2c_3=8c_2(1+c_2)\ge48\ge8$, since $c_2\ge2$. Since $X=\mathcal{A}_t(\xi)$, this is the last assertion of (ii).
\end{proof}

\subsection{Explicit numerical specialization}
\label{subsec:smallball_explicit}
In this subsection we fix numbers, not new estimates. Most inequalities below are instances of the preceding lemmas whose constants we track. The exceptions are Step~1, which sharpens the cutoff, and Step~4, which applies Fact~\ref{fact:gaussian_trace_moments}. We obtain the radius $279\sqrt m\ell_{\rm asp}$ with the exponent $3711824m/10^9$.

\begin{lemma}[Explicit hereditary small ball]
\label{lem:explicit_small_ball}
Let $1\le m\le N$, and let $A_1,\ldots,A_m\in \R^{N \times N}$ be symmetric
matrices satisfying $\|A_i\|\le1$. Let $\xi_1,\ldots,\xi_m$ be independent
standard Gaussian random variables. Then
\[
\Pr[
\|\sum_{i=1}^m \xi_iA_i\|
\le279\sqrt m\max\{1,\log(N/m)\}
]
\ge\exp(-3711824m/10^9).
\]
\end{lemma}

\begin{proof}
\emph{Step 1: a sharper edge cutoff.}
The general cutoff satisfies $0<\kappa_3<1/20$. We repeat the curvature computation with the sharper cutoff below, which may exceed $1/20$, and retain convenient sufficient parameter thresholds for the numerical specialization. For any $0<b<1/2$, put
\[
\widehat{\kappa}_3:=\frac{b^2}{1+b},
\qquad
\widehat c_0:=\frac{1-2b}{2}.
\]
The proof of Lemma~\ref{lem:smallball_divided_difference}(i), part~\hyperlink{first-sb-part-2}{2} of the proof of Proposition~\ref{prop:first_small_ball}, extends to cutoff $w=\widehat{\kappa}_3$ with coefficient $\widehat c_0$ in place of $1/4$. We split at $b/\tau$, and the near case uses
$u^{2\tau-2}\ge1-2b$. In the far case, if
$\delta_{\mathrm{gap}}\le\widehat{\kappa}_3/\tau$, then
\[
x^{2\tau-1}\ge1-2\widehat{\kappa}_3,
\qquad
\frac{\delta_{\mathrm{gap}}}{b/\tau}\le\frac{\widehat{\kappa}_3}{b},
\]
and
\[
1-2\widehat{\kappa}_3-\frac{2\widehat{\kappa}_3}{b}=1-2b.
\]
Combining the two one-sided estimates ($\sigma=\pm1$) costs a factor of two. Thus the lower
divided-difference estimate is valid with
$\widehat{\kappa}_3,\widehat c_0$, while its upper constant remains $2$.
With spectral coefficient $\widehat c_0$, the curvature calculation of Lemma~\ref{lem:resolvent_poincare}, carried out in part~\hyperlink{first-sb-part-3}{3} of the proof of Proposition~\ref{prop:first_small_ball}, applies whenever $\widehat c_0\kappa_{1}>2$ and $\kappa_{0}(\widehat c_0\kappa_{1}-2)>4$. The calculation uses Young's inequality with parameter $\widehat c_0\kappa_{1}/2-1$ to bound the $00$ and mixed products in each slot by one half of the barrier Hessian quadratic form. For the numerical specialization, we retain the sufficient thresholds $\kappa_{1}\ge A_*$ and $\kappa_{1}\kappa_{0}\ge D_*$, with
\[
A_*:=8/\widehat c_0,
\qquad
D_*:=32/\widehat c_0,
\qquad
\tau^2\rho\le\widehat{\kappa}_3^2/4.
\]
Indeed, these thresholds give $\widehat c_0\kappa_{1}\ge8$ and
\[
\kappa_{0}(\widehat c_0\kappa_{1}-2)=(\widehat c_0\kappa_{1}\kappa_{0})(1-2/(\widehat c_0\kappa_{1}))\ge24>4,
\]
where the first step follows by factoring, the second step follows from $\widehat c_0\kappa_{1}\kappa_{0}\ge32$ and $\widehat c_0\kappa_{1}\ge8$, and the third step is numerical.

\emph{Step 2: the parameters and the curvature checks.}
Take
\[
b:=\frac{10021}{25000},
\quad
D:=323,
\quad
A:=4700000,
\quad
c_{\rm cut}=\frac{256}{17},
\quad
c_{\rm rad}:=279,
\quad
\tau=\lceil c_{\rm cut} \ell_{\rm asp}\rceil.
\]
Then $\widehat c_0=2479/25000$,
$\widehat{\kappa}_3=100420441/875525000$, and $\tau/\ell_{\rm asp}\le16$. Indeed, $\tau=16$
on the first cell $[1,17/16]$. On the cell of any $\tau\ge17$ the ratio $\tau/\ell_{\rm asp}$ is at most $c_{\rm cut}\tau/(\tau-1)$, whose largest value is
\[
\frac{256}{17}\frac{17}{16}=16.
\]
The numerical curvature checks are
\[
\frac{16^2}{279^2}<\frac{\widehat{\kappa}_3^2}{4},
\qquad
\rho\le\frac1{279^2}<\frac1{6400},
\qquad
A>A_*,
\qquad
D>D_*.
\]

\emph{Step 3: the explicit Poincar\'e constants.}
We now use the normalized interpolation framework of
Lemma~\ref{lem:smallball_interpolation} and track every analytic constant in
the weighted Poincar\'e and resolvent estimates of
Lemma~\ref{lem:smallball_weighted_poincare}.
The curvature lower bound is $1/2$, by the computation of Lemma~\ref{lem:resolvent_poincare} with spectral coefficient $\widehat c_0$, cutoff $\widehat{\kappa}_3$, and the sufficient thresholds $A_*$, $D_*$. The weighted Poincar\'e coefficient
is therefore at most $2$. We use this uniform upper bound throughout the numerical calculation. Table~\ref{tab:explicit_poincare_constants} instantiates the proof of Lemma~\ref{lem:smallball_weighted_poincare} at $c_1=2$ and $\rho\le1/6400$, row by row in the order of that proof.

\begin{table}[t]
\centering
\caption{The explicit constants. The first seven rows instantiate the proof of Lemma~\ref{lem:smallball_weighted_poincare} at $c_1=2$ and $\rho\le1/6400$, and the third column names the step or equation of that proof. The last three rows instantiate Eqs.~\eqref{eq:smallball_resolvent_moments} and~\eqref{eq:smallball_interpolation_hessian} with $c_\Pi=2.02$ for $c_2$ and with the moment bound $N(8\tau^2\rho)^\tau$ of Step~4 of the proof of Lemma~\ref{lem:explicit_small_ball}.}
\label{tab:explicit_poincare_constants}
\begin{tabular}{llll}
\toprule
quantity & bound in that proof & where & value here\\
\midrule
$1-L_\sigma^{-1}$ & $c_1\rho L_\sigma$ & Step~1 & $\le2\rho L_\sigma$\\
$L_\sigma=\|S_\sigma\|$ & $2/(1+\sqrt{1-4c_1\rho})$ & Eq.~\eqref{eq:smallball_one_slot_resolvent} & $<1.001$\\
$\sum_{i=1}^m\E_{\mu_t}[(\partial_i\mathcal R)\mathcal Y(\partial_i\mathcal R)]$ & $8(1+2c_1\rho)\rho I$ & Step~3 & $\preceq8.008\rho I\preceq8.01\rho I$\\
$T-S^{-1}$ & $8c_1(1+2c_1\rho)\rho I$ & Step~3 & $\preceq16.02\rho I$\\
$\|S^{-1}-I\|$ & $\eta$ & Eq.~\eqref{eq:smallball_two_slot_resolvent} & $\le17.02\rho$\\
$\|S-I\|$ & $\eta/(1-\eta)$ & Eq.~\eqref{eq:smallball_two_slot_resolvent} & $<17.07\rho$\\
$\E_{\mu_t}[(\mathcal Y-I)\mathcal Y^{-1}(\mathcal Y-I)]$ & $(\rho+\eta/(1-\eta))I$ & Step~4 & $\preceq18.07\rho I$\\
\midrule
$\mathcal E_{\sigma_1,\sigma_2}(\mathcal{A}_{\mathrm{off}})$ & $4c_2(1+c_2)N\rho$ & Eq.~\eqref{eq:smallball_resolvent_moments} & $<25N\rho$\\
$\kappa_0\sum_{\sigma}\mathcal E_{\sigma,\sigma}(\mathcal{A}_{\mathrm{off}})$ & $2c_3\kappa_0N\rho$ & Eq.~\eqref{eq:smallball_interpolation_hessian} & $<50\kappa_0N\rho$\\
$4\tau(\kappa_1-\kappa_0)\sum_{\sigma}\mathcal E_{\sigma,\sigma}(X^{\tau-1}\mathcal{A}_{\mathrm{off}})$ & $8\kappa_1\tau N(c_3\tau^2\rho)^\tau$ & Eq.~\eqref{eq:smallball_interpolation_hessian} & $\le8\kappa_1\tau N(8\tau^2\rho)^\tau$\\
\bottomrule
\end{tabular}
\end{table}

The fifth row gives $\|S^{-1}-I\|\le17.02\rho\le17.02/6400$, so $S^{-1}\preceq1.003I$. Using this bound, the seventh row, and $\|a+b\|_2^2\le2\|a\|_2^2+2\|b\|_2^2$ in the projected mean estimate, we obtain $u^\top S^{-1}u\le2.006\E_{\mu_t}[\|U\|_F^2]+36.25\rho\mathcal E_{\sigma_1,\sigma_2}(U)$. Hence
\[
(1-36.25\rho)\mathcal E_{\sigma_1,\sigma_2}(U)
\le2.006\E_{\mu_t}[\|U\|_F^2]
+2\sum_{i=1}^m\mathcal E_{\sigma_1,\sigma_2}(\partial_iU).
\]
At $\rho\le1/6400$, after we divide by $1-36.25\rho$, both coefficients are smaller than $c_\Pi:=2.02$.

\emph{Step 4: the explicit moment constants.}
A word $W$ in the $j$th derivative of $X^{\tau-1}\mathcal{A}_{\mathrm{off}}$, polarized in its $j$ independent directions by independent standard Gaussians $h$, contains exactly $j$
independent-direction factors and $s:=\tau-j$ common-$\xi$ factors. There is
only one $\mathcal{A}_{\mathrm{off}}$ factor in the word. The variance proxy of $X$ is at most
$\rho I$, and that of $\mathcal{A}_{\mathrm{off}}$ is at most $4\rho I$. The completely positive
word-by-word estimate therefore applies, with Harg\'e's theorem in Fact~\ref{fact:even_convex_tilts} passing from $\mu_t$ to $\gamma$. Together with Fact~\ref{fact:gaussian_trace_moments}, it gives
\[
\E_{\mu_t,h}[\|W\|_F^2]
\le4N\rho^j(2s\rho)^s.
\]
There are at most $\tau^j$ words, so Cauchy--Schwarz gives
\[
\E_{\mu_t}[\|\nabla^j(X^{\tau-1}\mathcal{A}_{\mathrm{off}})\|_{\mathrm{HS}}^2]
\le4N\rho^j(2s\rho)^s\tau^{2j}
\le N(2\tau^2\rho)^\tau.
\]
For the last step, after we cancel $N\rho^\tau$ it is enough to check
$4\le2^j(\tau^2/s)^s$. For $j=0$ this is $4\le \tau^\tau$. For $j=1$ it follows
from $\tau\ge2$. For $j\ge2$ it follows from $2^j\ge4$.

For $\tau\ge2$,
\[
\sum_{j=0}^\tau c_\Pi^{j+1}<4^\tau.
\]
Indeed, the inequality is direct at $\tau=2$, and the induction step uses
$c_\Pi^{\tau+2}<3\mathbin{\cdot}4^\tau$. Iterating the defective Poincar\'e
estimate, we therefore obtain
\[
\mathcal E_{\sigma_1,\sigma_2}(X^{\tau-1}\mathcal{A}_{\mathrm{off}})
\le N(8\tau^2\rho)^\tau.
\]
The remaining rows of Table~\ref{tab:explicit_poincare_constants} instantiate the proof of Lemma~\ref{lem:smallball_tail_hessian_comparison} with these proxies and $c_\Pi$, and the two checks above are the only new steps. Since $8\kappa_{1} \tau N(8\tau^2\rho)^\tau\le50\kappa_{1} \tau N(50\tau^2\rho)^\tau$, the interpolation estimate holds with the single explicit constant
\[
C_I:=50.
\]

\emph{Step 5: the endpoint reduction and the exponent bound.}
Let $r:=N/m$, $\rho:=(279^2\ell_{\rm asp}^2)^{-1}$, and choose the potential
parameters
\[
\kappa_{1,\mathrm{pot}}:=A\max\{1,r\rho\},
\qquad
\kappa_{0,\mathrm{pot}}:=D/\kappa_{1,\mathrm{pot}}.
\]
The interpolation Hessian estimate~\eqref{eq:smallball_interpolation_hessian}, with the constants we tracked above, gives uniformly for $t\in[0,1]$
\[
a''(t)\ge-50\kappa_{0,\mathrm{pot}}N\rho-50\kappa_{1,\mathrm{pot}}\tau N(50\tau^2\rho)^\tau.
\]
We next bound the diagonal endpoint. Write $\mathcal{A}_0=\operatorname{diag}(\zeta_1,\ldots,\zeta_N)$, where $\zeta$ is a centered Gaussian vector with covariance $\Sigma$. The normalization gives $\tr[\Sigma]\le N\rho$ and $\Sigma_{jj}\le\rho$ for every $j$. Since $\kappa_{1,\mathrm{pot}}\ge\kappa_{0,\mathrm{pot}}$, for $|x|\le1/2$ we have
\[
\phi(x)\le2\kappa_{0,\mathrm{pot}}x^2+\frac{2\kappa_{1,\mathrm{pot}}}{\tau}4^{-\tau}.
\]
Tilting the law of $\zeta$ by $e^{-2\kappa_{0,\mathrm{pot}}\|\zeta\|_2^2}$, we obtain a centered Gaussian vector with covariance
\[
\widetilde\Sigma=\Sigma^{1/2}(I+4\kappa_{0,\mathrm{pot}}\Sigma)^{-1}\Sigma^{1/2}\preceq\Sigma,
\]
and
\[
\E[e^{-2\kappa_{0,\mathrm{pot}}\|\zeta\|_2^2}]=\det(I+4\kappa_{0,\mathrm{pot}}\Sigma)^{-1/2}\ge e^{-2\kappa_{0,\mathrm{pot}}N\rho}.
\]
We use the Gaussian correlation inequality for the symmetric strips $\{|\zeta_j|\le1/2\}$, in the form proved by \v{S}id\'ak~\cite[Theorem~1]{sidak67}. The general inequality is Royen's~\cite{royen14}. Together with the scalar Gaussian tail bound, it gives
\[
\Pr_{\widetilde\Sigma}[|\zeta_j|\le1/2\text{ for every }j]\ge(1-2e^{-1/(8\rho)})^N\ge e^{-4N e^{-1/(8\rho)}};
\]
here $2e^{-1/(8\rho)}\le1/2$ follows from $\rho\le1/6400$. Consequently,
\[
a(0)\ge-2\kappa_{0,\mathrm{pot}}N\rho-\frac{2\kappa_{1,\mathrm{pot}}N}{\tau}4^{-\tau}-4N e^{-1/(8\rho)}.
\]
By Lemma~\ref{lem:smallball_interpolation}(iv), $a'(0)=0$. Taylor's formula therefore gives
\[
a(1)=a(0)+\int_0^1(1-t)a''(t)\,\d t.
\]
Using $r=N/m\le e^{\ell_{\rm asp}}$ and $\max\{1,r\rho\}r\le e^{\ell_{\rm asp}}+e^{2\ell_{\rm asp}}\rho$, and tracking the factor $\int_0^1(1-t)\,\d t=1/2$, we obtain $a(1)\ge-mE(\ell_{\rm asp})$, where
\begin{align}
E(\ell_{\rm asp})\le{}&
\frac{27D}{A}\nonumber\\
&+25A\tau
 (e^{\ell_{\rm asp}}+\frac{e^{2\ell_{\rm asp}}}{279^2\ell_{\rm asp}^2})
 (\frac{50\tau^2}{279^2\ell_{\rm asp}^2})^\tau\nonumber\\
&+\frac{2A}{\tau}4^{-\tau}
 (e^{\ell_{\rm asp}}+\frac{e^{2\ell_{\rm asp}}}{279^2\ell_{\rm asp}^2})
+4\exp(\ell_{\rm asp}-279^2\ell_{\rm asp}^2/8).
\label{eq:explicit_smallball_ledger}
\end{align}
The first line combines the low-order diagonal loss $2D/A$ with the
integrated interpolation loss $25D/A$.

\emph{Step 6: maximization over $\ell_{\rm asp}$.}
We write Steps~6 and~7 with the symbol $c_{\rm rad}=279$ of Step~2, since Section~\ref{subsec:smallball_regimes} reuses them with the other rows of Table~\ref{tab:six_regimes}. Put
\[
P(\ell_{\rm asp}):=e^{\ell_{\rm asp}}+\frac{e^{2\ell_{\rm asp}}}{c_{\rm rad}^2\ell_{\rm asp}^2},
\]
write $E_{\mathrm{low}}:=27D/A$ for the first line of
Eq.~\eqref{eq:explicit_smallball_ledger}, and denote its second and third terms by $I(\tau,\ell_{\rm asp})$ and $J(\tau,\ell_{\rm asp})$.
Since $(\log P)'\le2$,
\[
\partial_{\ell_{\rm asp}}\log I(\tau,\ell_{\rm asp})\le2-2\tau/\ell_{\rm asp}<0,
\]
while $J(\tau,\ell_{\rm asp})$ increases on $\ell_{\rm asp}\ge1$ because $P$ does. The last term of Eq.~\eqref{eq:explicit_smallball_ledger}, the strip term, decreases because
\[
\frac{\d}{\d \ell_{\rm asp}}(\ell_{\rm asp}-c_{\rm rad}^2\ell_{\rm asp}^2/8)=1-c_{\rm rad}^2\ell_{\rm asp}/4<0.
\]
Hence the maximum of $I$ on a
cell of $\tau$ is bounded by its value at the left endpoint, and the maximum of $J$ is at its
right endpoint.

The first ratio of consecutive cell maxima of $I$ is
\[
\frac{I(17,17/16)}{I(16,1)}<0.187.
\]
For every later transition, with $\tau\ge17$, the ratio is at most
\[
\frac{\tau+1}{\tau}e^{2/c_{\rm cut}}
\frac{50c_{\rm cut}^2}{c_{\rm rad}^2}
((1+\frac1\tau)^{\tau+1}(1-\frac1\tau)^\tau)^2
\le\frac{18}{17}e^{2/c_{\rm cut}}\frac{50c_{\rm cut}^2}{c_{\rm rad}^2}
<0.177.
\]
Here the parenthesized factor is at most one because
$\log(1-u)\le-u$ and $\log(1+u)\le u$ imply
$(1-\tau^{-2})^\tau\le(1+\tau^{-1})^{-1}$. Thus the global interpolation maximum
is $I(16,1)$. The ratio of consecutive cell maxima of $J$ is at most
\[
\frac{e^{2/c_{\rm cut}}}4<0.286,
\]
so the global diagonal maximum is the right endpoint
$J(16,17/16)$.

\emph{Step 7: rational evaluation.}
We need no numerical program for the remaining evaluation.
For rational $x\in(0,17/8]$, the exponential rule of Definition~\ref{def:rational_enclosures} with $K_{\mathrm{series}}:=30$ encloses $e^x$, since $17/8<K_{\mathrm{series}}+2$.
At $x=17/128,1,2,17/16,17/8$, this gives
\[
e^{17/128}<1.142035847,\qquad
e<2.718281829,\qquad
e^2<7.389056099,
\]
\[
e^{17/16}<2.893595945,\qquad
e^{17/8}<8.372897489.
\]
We substitute using exact rational arithmetic and obtain
\[
10^9E_{\mathrm{low}}<1855532.0,
\qquad
10^9I(16,1)<1460469.6,
\qquad
10^9J(16,17/16)<395822.3.
\]
Finally, $c_{\rm rad}^2/8-1>25$, while
$e>2.7$ and $2.7^{25}>4\mathbin{\cdot}10^{10}$, so
\[
10^9\mathbin{\cdot}4\exp(1-c_{\rm rad}^2/8)<0.1.
\]
The four bounds are strict and add up to exactly $3711824$, so
\[
E(\ell_{\rm asp})<\frac{3711824}{10^9}
\]
for every $\ell_{\rm asp}\ge1$. The Taylor estimate of Step~5 then gives $a(1)>-3711824m/10^9$, and Lemma~\ref{lem:smallball_interpolation}(v) yields
\[
\Pr[\|\mathcal{A}(\xi)\|<1]\ge a_1=e^{a(1)}>\exp(-3711824m/10^9).
\]
Undoing the normalization, we obtain the asserted radius.
\end{proof}

\begin{remark}[Reading the bound]
\label{rem:smallball_reading_the_ledger}
The bound $E(\ell_{\rm asp})<3711824/10^9$ is a maximum over the cells of $\tau$, not a sum over them. On the binding cell $[1,17/16]$ the strip term is below $10^{-10}$. The rows of Table~\ref{tab:six_regimes} behave alike, and Table~\ref{tab:six_regime_endpoint} in the next subsection records the components. In every row the interpolation term is matched against $27D/A$, to within one percent for $20\le\tau\le43$. Two margins are thin by construction rather than by accident. In every row $c_{\rm rad}$ was chosen as the least integer for which the curvature check $(\tau_0/(c_{\rm rad}\ell_0))^2<\widehat{\kappa}_3^2/4$ holds, so the margins $\Delta_{\mathrm{curv}}$ of the last column lie between $1.18\cdot10^{-7}$ and $2.1\cdot10^{-6}$. Moreover, $D$ exceeds its threshold $32/\widehat c_0$ by less than $0.2$ percent. Both checks are exact rational comparisons, so the thin margins are not near-failures.
\end{remark}

\subsection{Proof of the six-regime lemma}
\label{subsec:smallball_regimes}
One parameter set is not optimal across aspect ratios: the six rows of Table~\ref{tab:six_regimes} suffice for the constant $156000$ of Section~\ref{sec:proof}.

\begin{proof}[Proof of Lemma~\ref{lem:scale_small_ball}]
The exponent bound of Lemma~\ref{lem:explicit_small_ball} allows the parameters to change when the moment order
changes. Assign
$(c_{\rm rad},b,D,A)$ from Table~\ref{tab:six_regimes} according to $\tau(\ell_{\rm asp})=\lceil c_{\rm cut} \ell_{\rm asp}\rceil$.
The cell of $\tau$ is $\ell_{\rm asp}\in((\tau-1)/c_{\rm cut},\tau/c_{\rm cut}]$, so $c_{\rm rad}(\ell_{\rm asp})$ is the step function of the lemma.
We also record the endpoint exponent component by component. Every entry
in Table~\ref{tab:six_regime_endpoint} is an outward integer upper bound of its own quantity, so the first three columns of a row may add up to slightly more than the last (by at most $2$).

\begin{table}[t]
\centering
\caption{The first-cell endpoint calculation, scaled by $10^9$.}
\label{tab:six_regime_endpoint}
\begin{tabular}{crrrr}
\toprule
$\tau$ & $10^9(27D/A)$ & $10^9I_0$ & $10^9J_0$ & $10^9E_0$\\
\midrule
$16$       & $1855532$ & $1460470$ & $395823$ & $3711824$\\
$17$--$19$ & $1759822$ & $1639172$ & $118589$ & $3517582$\\
$20$--$26$ & $1187523$ & $1184870$ & $3879$   & $2376271$\\
$27$--$30$ & $1491968$ & $1488990$ & $1$      & $2980959$\\
$31$--$43$ & $1368735$ & $1361304$ & $1$      & $2730038$\\
$\tau\ge44$   & $26172$   & $25543$   & $1$      & $51714$\\
\bottomrule
\end{tabular}
\end{table}

\emph{Step 1: the first-cell checks.}
For one row of the table, let $\tau_0$ be the first moment order in that row and
put
\[
\ell_0=\max\{1,(\tau_0-1)/c_{\rm cut}\},
\qquad
r_0=\tau_0/c_{\rm cut}.
\]
For the row's fixed parameters define
\[
P_{c_{\rm rad}}(x)=e^x+\frac{e^{2x}}{c_{\rm rad}^2x^2},
\]
\[
I_0=25A\tau_0P_{c_{\rm rad}}(\ell_0)
(\frac{50\tau_0^2}{c_{\rm rad}^2\ell_0^2})^{\tau_0},
\qquad
J_0=\frac{2A}{\tau_0}4^{-\tau_0}P_{c_{\rm rad}}(r_0),
\]
\[
E_0=\frac{27D}{A}+I_0+J_0
+4\exp(\ell_0-c_{\rm rad}^2\ell_0^2/8).
\]
With $\widehat{\kappa}_3=b^2/(1+b)$ and $\widehat c_0=(1-2b)/2$ as in the proof of Lemma~\ref{lem:explicit_small_ball}, we verify by direct rational comparison that
\[
A>\frac8{\widehat c_0},
\qquad
D>\frac{32}{\widehat c_0},
\qquad
(\frac{\tau_0}{c_{\rm rad}\ell_0})^2<\frac{\widehat{\kappa}_3^2}{4},
\qquad
\frac1{c_{\rm rad}^2\ell_0^2}<\frac1{6400}.
\]
Here
\[
\Delta_{\mathrm{curv}}
=\frac{\widehat{\kappa}_3^2}{4}-(\frac{\tau_0}{c_{\rm rad}\ell_0})^2.
\]
Its lower bounds are the last column of Table~\ref{tab:six_regimes}. The
smallest value of
$10^6(1/6400-1/(c_{\rm rad}^2\ell_0^2))$ is greater than $143.4$.
On a fixed cell $\tau/\ell_{\rm asp}$ decreases with $\ell_{\rm asp}$. At successive left limits its
supremum is $c_{\rm cut}\tau/(\tau-1)$, which decreases with $\tau$. Likewise,
$1/(c_{\rm rad}^2\ell_{\rm asp}^2)$ decreases with $\ell_{\rm asp}$. Thus from these first-cell checks we deduce
the two inequalities throughout the row.

\emph{Step 2: the first cell controls the row.}
Step~6 of the proof of Lemma~\ref{lem:explicit_small_ball} applies with the row's $c_{\rm rad}$ in place of $279$. From one full cell to the next, the interpolation maximum is multiplied by at most $\frac{\tau+1}{\tau}e^{2/c_{\rm cut}}\frac{50c_{\rm cut}^2}{c_{\rm rad}^2}$ and the diagonal maximum by at most $e^{2/c_{\rm cut}}/4<0.286$. The first bound is below $0.387$ in every row and decreases with $\tau$. Thus we bound the normalized negative exponent throughout the row by $E_0$.

\emph{Step 3: rational evaluation.}
Using rational arithmetic, we can check every numerical entry in the last two columns of Table~\ref{tab:six_regimes}. The exponential rule of Definition~\ref{def:rational_enclosures} applies with $K_{\mathrm{series}}:=40$ on $[0,6]$, since $6<K_{\mathrm{series}}+2$. Substituting the exact table rationals, we then obtain Table~\ref{tab:six_regime_endpoint} and the displayed bounds for $10^9E_0$. For the strip term, we check by exact rational comparison that $c_{\rm rad}^2\ell_0^2/8-\ell_0>1000$ in every row. Since $e>2$ and $2^{10}>10^3$, the strip term is less than $10^{-100}$. Hence
\[
E_0<\frac{3711824}{10^9}<0.003712.
\]
We repeat the endpoint and Taylor calculation from the proof of Lemma~\ref{lem:explicit_small_ball} with the parameters of the current row and obtain $a(1)\ge-mE_0$. Hence Lemma~\ref{lem:smallball_interpolation}(v) gives $\Pr[\|\mathcal{A}(\xi)\|<1]\ge e^{-mE_0}$. The entry $10^9E$ of the row in Table~\ref{tab:six_regimes} is the outward bound of $10^9E_0$ in Table~\ref{tab:six_regime_endpoint}, so $E_0\le10^{-9}E_{\rm tab}$. Undoing the normalization, with $r_{\rm sb}=c_{\rm rad}(\ell_{\rm asp})\sqrt m\ell_{\rm asp}$, we prove the lemma.
\end{proof}

\section{Signing by repeated partial coloring}\label{sec:proof}

In this section, we give our second proof of the conjecture, after the first proof of Section~\ref{sec:first_existence}, and this proof is also algorithmic. We prove Theorem~\ref{thm:matrix_spencer_bound} from the six-regime small-ball estimate, Lemma~\ref{lem:scale_small_ball}, by repeated partial coloring. Our proof consumes two tools, the refreshed projection step of Lemma~\ref{lem:refreshed_projection} and the separated-sum estimate of Lemma~\ref{lem:bv_packing}. In Section~\ref{subsec:matrix_spencer_discrepancy} we state Theorem~\ref{thm:matrix_spencer_bound}, the formal version of Theorem~\ref{thm:matrix_spencer} with the constant $156000$, and we prove it in four steps. In one phase we project a Gaussian point onto the operator-norm body cut by the current coloring box. A successful trial turns at least a fixed fraction $\delta$ of the active coordinates into signs. We then sum the phase increments over the phases whose active set has size at most $n/e$, and over the at most $35$ phases with a larger active set. We sign the fewer than $N_0$ coordinates that remain by a cleanup rounding, using the matrix Hoeffding bound of Fact~\ref{fact:matrix_hoeffding}. We prove Lemma~\ref{lem:refreshed_projection}, the refreshed projection step, in Section~\ref{subsec:refreshed_projection}. One Gaussian projection trial succeeds with probability greater than $0.12$, and on success at least a $\delta$ fraction of the active coordinates of the updated coloring are signs. We compare the distance from the Gaussian point to the coloring box with its distances to the coordinate sections, using Fact~\ref{fact:gaussian_section_transport} and the relaxation argument of Proposition~\ref{prop:first_covering}. We state and prove our second tool, Lemma~\ref{lem:bv_packing}, in Section~\ref{subsec:bv_packing}. The phase parameters keep a minimum spacing but fall where the projection outcomes place them, so we bound their sum by a deterministic sum of bin suprema. The per-round structure of this proof is what the four algorithms of Sections~\ref{sec:algorithm}--\ref{sec:cubic_time} implement, down to $n^{3+o(1)}$ arithmetic operations (Theorem~\ref{thm:cubic_dense_runtime}). The small-constant algorithm of Section~\ref{sec:small_algorithm_constant} keeps the same structure (Theorem~\ref{thm:algorithm_constant_13}). Here we write the argument in exact arithmetic, and the implementations live in those sections.

\subsection{Matrix Spencer discrepancy}
\label{subsec:matrix_spencer_discrepancy}

\begin{theorem}[Matrix Spencer discrepancy bound, formal version of Theorem~\ref{thm:matrix_spencer}]
\label{thm:matrix_spencer_bound}
For all symmetric matrices $A_1,\ldots,A_n\in \R^{n \times n}$ with
$\|A_i\|\le1$, there is a vector $\varepsilon\in\{-1,1\}^n$ satisfying
\[
\|\sum_{i=1}^n\varepsilon_iA_i\|\le156000\sqrt n.
\]
\end{theorem}

\begin{proof}
\emph{Step 1: one phase.}
We start with the zero fractional coloring. At each phase, let $S$ be the active
set, $m:=|S|$, and $y\in(-1,1)^S$ the restriction of the current
fractional coloring to $S$. Put
\[
u:=\log(n/m),
\qquad
\ell_{\rm asp}:=\max\{1,u\},
\]
and take $c_{\rm rad}(\ell_{\rm asp})$ from Table~\ref{tab:six_regimes}. The set
\[
K:=\{x\in\R^S:\|\sum_{i\in S}x_iA_i\|
\le c_{\rm rad}(\ell_{\rm asp})\sqrt m\ell_{\rm asp}\}
\]
is closed, convex, and symmetric. Lemma~\ref{lem:scale_small_ball}, applied to the $m$ contractions $A_i$, $i\in S$, in dimension $N=n$, gives $\gamma_m(K)\ge\exp(-0.003712m)$. Since $0.003712<0.00372$, the set $K$ satisfies the hypothesis of
Lemma~\ref{lem:refreshed_projection} as long as $m\ge N_0$. We call a trial of that lemma successful when at least $\delta m$ coordinates of $y+2x/\varepsilon_{\mathrm{box}}$ become signs. On a successful trial the lemma gives $x\in K$. We then
update $y$ to $y+2x/\varepsilon_{\mathrm{box}}$ and remove from $S$ every coordinate that has become a
sign. If $m'$ is the new active-set size, then $m'\le(1-\delta)m$, and hence
the next value $u':=\log(n/m')$ satisfies
\[
u'-u\ge\lambda,
\qquad
\lambda:=-\log(1-\delta).
\]
The operator norm of the signed sum grows in this phase by at most $(2/\varepsilon_{\mathrm{box}})c_{\rm rad}(\ell_{\rm asp})\sqrt m\ell_{\rm asp}$, the bound on $\|\sum_{i\in S}(2x_i/\varepsilon_{\mathrm{box}})A_i\|$. With $\sqrt m=e^{-u/2}\sqrt n$, this bound is at most
\begin{equation}
\frac{2c_{\rm rad}(\ell_{\rm asp})}{\varepsilon_{\mathrm{box}}}e^{-u/2}\ell_{\rm asp}\sqrt n.
\label{eq:explicit_round_increment}
\end{equation}
We stop the phases when $m<N_0$.

\emph{Step 2: the phases starting at $u\ge1$.}
Let $u_0<u_1<\cdots$ be the values of $u$ at the starts of the phases. Then $u_0=0$, and Step~1 gives $u_{j+1}-u_j\ge\lambda$. Hence it remains to sum the increments~\eqref{eq:explicit_round_increment} over an arbitrary $\lambda$-separated sequence. Put
\[
\Gamma(u):=c_{\rm rad}(u)ue^{-u/2}\qquad(u\ge1),
\]
the coefficient of $2\sqrt n/\varepsilon_{\mathrm{box}}$ in~\eqref{eq:explicit_round_increment} for $u\ge1$. For $a\ge1$ and $k\ge0$, put
\[
M_k(a):=\sup_{a+k\lambda\le u<a+(k+1)\lambda}\Gamma(u).
\]
Lemma~\ref{lem:bv_packing} gives
\[
\sum_j\Gamma(u_j)\le\sum_{k=0}^{\infty}M_k(a)
\]
for every $\lambda$-separated sequence all of whose terms are at least $a$. The six values of
$c_{\rm rad}$ apply on the $\ell_{\rm asp}$ intervals listed in Table~\ref{tab:six_regimes}, and $c_{\rm rad}$ is constant on each of them.
The function $ue^{-u/2}$ increases on $[1,2]$ and decreases on $[2,\infty)$. Hence the supremum of $\Gamma$ over a bin is attained or approached at a bin endpoint, at $u=2$, or at one of
the five regime endpoints contained in that bin. After the bin containing
$731/256$, the function $\Gamma$ is decreasing. Then we evaluate the remaining sum in closed form
from $\sum_{j\ge0}\rho^j$ and $\sum_{j\ge0}j\rho^j$, where
$\rho:=e^{-\lambda/2}$. We combine the order-$40$ rational Taylor enclosures of the proof of Lemma~\ref{lem:scale_small_ball} with the logarithmic series and the exact-squaring rule for square roots of Definition~\ref{def:rational_enclosures}. Applying these tools to $\lambda=-\log(1-\delta)$ and $\rho=\sqrt{1-\delta}$, we obtain
\[
\sum_{k=0}^{\infty}M_k(1)<25020.2,
\qquad
\sum_{k=0}^{\infty}M_k(35\lambda)<25020.2.
\]
Hence the phases starting at or above either threshold contribute less than
\[
\frac{2}{\varepsilon_{\mathrm{box}}}\mathbin{\cdot}25020.2<118159.2
\]
times $\sqrt n$ to the operator norm.

\emph{Step 3: the phases starting at $u<1$.}
The same enclosures give
\[
34\lambda<1<35\lambda<1.0001.
\]
Since $u_j\ge j\lambda$, a phase with $u_j<1$ has $j\le34$, so at most $35$ phases start below $1$. On such a phase $\ell_{\rm asp}=1$ and $c_{\rm rad}(\ell_{\rm asp})=279$. Moreover, $e^{-u_j/2}\le\rho^j$ with $\rho=e^{-\lambda/2}=\sqrt{1-\delta}$, so these phases are dominated by a geometric series. By exact geometric summation we obtain
\[
\frac{2\mathbin{\cdot}279}{\varepsilon_{\mathrm{box}}}
\sum_{j=0}^{33}\rho^j<35739.4,
\qquad
\frac{2\mathbin{\cdot}279}{\varepsilon_{\mathrm{box}}}
\sum_{j=0}^{34}\rho^j<36550.1.
\]
If at most $34$ phases start below $1$, they contribute less than $35739.4$. Every later phase then starts at or above $1$, so Step~2 with $a=1$ bounds their contribution by $118159.2$. The total is less than
$153898.6$. If $35$ phases start below $1$, they contribute less than $36550.1$. The continuation then starts at $u_{35}\ge35\lambda$, so Step~2 with $a=35\lambda$ bounds it by $118159.2$. The
total is less than
\[
154709.3.
\]
In either case, all projection phases together contribute less than $154709.3\sqrt n$.

\emph{Step 4: cleanup rounding.}
When fewer than $N_0$ coordinates remain, we round independently with the
current fractional coloring as the mean. For $i\in S$, choose
$\varepsilon_i\in\{-1,1\}$ independently with $\E[\varepsilon_i]=y_i$ and
put $X_i:=(\varepsilon_i-y_i)A_i$. Since $\sum_{i\in S}A_i^2\preceq mI$, Fact~\ref{fact:matrix_hoeffding} with $\sigma^2=m$ and $N=n$ gives
$\Pr[\|\sum_{i\in S}X_i\|\ge t]\le2n\exp(-t^2/(2m))$.
We take $t:=\sqrt{2m(\log(2n)+\log2)}$, which makes the right-hand side $1/2$, so with probability at least
$1/2$ the rounding contributes at most $t=\sqrt{2m\log(4n)}\le6.5\sqrt n$. Indeed, if
$n\ge N_0$, then $m<N_0$ and $(N_0/n)\log(4n)$ decreases with $n$, so $t^2/n<2(N_0/n)\log(4n)\le2\log(4N_0)<6.5^2$. If
$n<N_0$, no projection phase runs, $m=n$, and the same endpoint bound
applies. Each projection trial succeeds with probability greater than $0.12$ and the rounding with probability at least $1/2$. A run in which every trial succeeds therefore exists, since we discard and repeat failed projection or rounding trials. Hence
\[
\|\sum_{i=1}^n\varepsilon_iA_i\|
<(154709.3+6.5)\sqrt n
<156000\sqrt n.\qedhere
\]
\end{proof}

\begin{remark}[Smaller ambient dimension]
The same conclusion holds for symmetric $N\times N$ contractions with
$N\le n$ by embedding $A_i$ as $A_i\oplus0_{n-N}$.
\end{remark}

\subsection{Refreshed projection}
\label{subsec:refreshed_projection}

Throughout this subsection, $G\sim \mathcal N(0,1)$ is a standard normal variable. Its density is $\phi(t):=(2\pi)^{-1/2}\exp(-t^2/2)$, and its upper tail is $\overline\Phi(t):=\Pr[G\ge t]$. Fact~\ref{fact:gaussian_section_transport} supplies the two Gaussian inputs of this subsection, the section inequality and the distance bound.

The projection box encodes the current fractional coloring. We compare the distance from a Gaussian point to this box with its distances to coordinate sections, and the comparison forces many box constraints to be tight.

\begin{lemma}[Refreshed projection]
\label{lem:refreshed_projection}
Let $m\ge N_0=3\mathbin{\cdot}10^8$, let $K\subseteq\R^m$ be a
symmetric closed convex set with
\[
\gamma_m(K)\ge\exp(-0.00372m),
\]
and put
\[
\delta:=0.0281672,
\qquad
\varepsilon_{\mathrm{box}}:=0.4235.
\]
Given $y\in(-1,1)^m$, put
\[
R_i:=\frac{\varepsilon_{\mathrm{box}}}2(1-y_i),
\qquad
L_i:=\frac{\varepsilon_{\mathrm{box}}}2(1+y_i)
\qquad(i\in[m]),
\qquad
Q_y:=\prod_{i=1}^m[-L_i,R_i],
\]
sample $\xi\sim \mathcal N(0,I_m)$, and let $x$ be the Euclidean projection of $\xi$ onto $K\cap Q_y$. This is one Gaussian projection trial. Then, with probability
greater than $0.12$, the point $x\in K$ satisfies
\[
y+2x/\varepsilon_{\mathrm{box}}\in[-1,1]^m
\]
and at least $\delta m$ coordinates of $y+2x/\varepsilon_{\mathrm{box}}$ are in $\{-1,1\}$.
\end{lemma}

\begin{proof}
\emph{Step 1: reduction to tight box constraints.}
Since $y_i\in(-1,1)$, the numbers $L_i$ and $R_i$ lie in $(0,\varepsilon_{\mathrm{box}})$, so $Q_y\subseteq[-\varepsilon_{\mathrm{box}},\varepsilon_{\mathrm{box}}]^m$ and
\[
d(\xi,Q_y)\ge d(\xi,[-\varepsilon_{\mathrm{box}},\varepsilon_{\mathrm{box}}]^m).
\]
The projection $x$ lies in $K\cap Q_y$, so $-L_i\le x_i\le R_i$ for every $i$, which is $y+2x/\varepsilon_{\mathrm{box}}\in[-1,1]^m$. The $i$th coordinate of $y+2x/\varepsilon_{\mathrm{box}}$ is a sign exactly when $x_i\in\{-L_i,R_i\}$. This happens exactly when one of the two box constraints of coordinate $i$ is tight at $x$. Since $L_i,R_i>0$, at most one constraint per coordinate is tight, so the tight constraints are indexed by a set of coordinates. It remains to show that, with probability greater than $0.12$, more than $\delta m$ coordinates carry a tight constraint.

\emph{Step 2: the distance to the cube.}
Put
\[
\psi(t):=\E[(|G|-t)_+^2]
=2((1+t^2)\overline\Phi(t)-t\phi(t))\qquad(t>0),
\]
and let $h(t):=-t\log t-(1-t)\log(1-t)$ be the binary entropy. The function $z\mapsto d(z,[-\varepsilon_{\mathrm{box}},\varepsilon_{\mathrm{box}}]^m)$ is
$1$-Lipschitz. Since $d(z,[-\varepsilon_{\mathrm{box}},\varepsilon_{\mathrm{box}}]^m)^2=\sum_i(|z_i|-\varepsilon_{\mathrm{box}})_+^2$,
\[
\E[d(\xi,[-\varepsilon_{\mathrm{box}},\varepsilon_{\mathrm{box}}]^m)^2]=m\psi(\varepsilon_{\mathrm{box}}).
\]
The Gaussian Poincar\'e inequality bounds the variance of a $1$-Lipschitz function of $\xi$ by $1$, so $(\E[d])^2\ge\E[d^2]-1$ and
\[
\E[d(\xi,[-\varepsilon_{\mathrm{box}},\varepsilon_{\mathrm{box}}]^m)]
\ge\sqrt{m\psi(\varepsilon_{\mathrm{box}})-1}.
\]

\emph{Step 3: the distances to the coordinate sections.}
For $I\subseteq[m]$, put $H(I):=\{z\in\R^m:z_i=0\text{ for every }i\in I\}$, the coordinate subspace in which the coordinates of $I$ are frozen. We write $\xi_{H(I)}$ for the orthogonal projection of $\xi$ onto $H(I)$, a standard Gaussian vector on $H(I)$. Fact~\ref{fact:gaussian_section_transport}(i) with $C=K$ and $H=H(I)$ gives
$\gamma_{H(I)}(K\cap H(I))\ge\gamma_m(K)\ge\exp(-0.00372m)$. Fact~\ref{fact:gaussian_section_transport}(ii) with $C=K\cap H(I)\subseteq H(I)$ then gives
\[
\E[d(\xi_{H(I)},K\cap H(I))]\le\sqrt{2(0.00372)m}.
\]
Because $K\cap H(I)\subseteq H(I)$, Pythagoras gives $d(\xi,K\cap H(I))^2=d(\xi_{H(I)},K\cap H(I))^2+\sum_{i\in I}\xi_i^2$. The first term is the square of a $1$-Lipschitz function of $\xi_{H(I)}$, so the Gaussian Poincar\'e inequality bounds its expectation by $(\E[d(\xi_{H(I)},K\cap H(I))])^2+1\le2(0.00372)m+1$. The second term has expectation $|I|$. Jensen's inequality $\E[d]\le(\E[d^2])^{1/2}$ then gives,
uniformly over $|I|\le\delta m$,
\[
\E[d(\xi,K\cap H(I))]
\le\sqrt{(\delta+2(0.00372))m+1}.
\]

\emph{Step 4: the separation constant and its enclosures.}
We divide the lower bound of Step~2 and the upper bound of Step~3 by $\sqrt m$, and reserve a further $\sqrt{2h(\delta)}$ for the union bound of Step~5. The normalized separation at $m=N_0$ is then
\[
r:=\sqrt{\psi(\varepsilon_{\mathrm{box}})-1/N_0}
-\sqrt{\delta+2(0.00372)+1/N_0}
-\sqrt{2h(\delta)}.
\]
The first square-root term increases with $m$ and the second decreases, so
the same lower separation persists for every $m\ge N_0$.
Direct rational enclosures give
\[
0.36472389700743<\phi(\varepsilon_{\mathrm{box}})<0.36472389700744,
\]
\[
0.33596524927926<\overline\Phi(\varepsilon_{\mathrm{box}})
<0.33596524927928,
\]
\[
\psi(0.4235)>0.4835216045,
\qquad
0.1283122818<h(0.0281672)<0.1283122819,
\]
\[
10^6r>77.2,
\qquad
10^6(r-6\mathbin{\cdot}10^{-5})>17.2.
\]
The displayed enclosures are self-contained. We can use
\[
\overline\Phi(x)=\frac12-\frac1{\sqrt{2\pi}}
\sum_{k=0}^{\infty}
\frac{(-1)^kx^{2k+1}}{(2k+1)2^kk!}
\]
for $0\le x\le1/2$, truncating after an even and an odd term.
Alternating upper and lower truncations with the first omitted term give the
displayed Gaussian enclosures, and we apply the same rule to $e^{-x}$ to enclose
$\phi(\varepsilon_{\mathrm{box}})$. For the logarithms and $\pi$ we follow the series rules of
Definition~\ref{def:rational_enclosures} with twenty terms. The logarithmic
series have squared ratio at most $1/9$, so the errors are below $10^{-20}$.
By exact rational squaring we enclose every square root.

\emph{Step 5: the good event.}
We choose the cube and section deviations both equal to $3\mathbin{\cdot}10^{-5}$.
Their sum is less than $r$. The cube event is $d(\xi,[-\varepsilon_{\mathrm{box}},\varepsilon_{\mathrm{box}}]^m)\le\E[d(\xi,[-\varepsilon_{\mathrm{box}},\varepsilon_{\mathrm{box}}]^m)]-3\mathbin{\cdot}10^{-5}\sqrt m$. The section event is that some $I$ with $|I|\le\delta m$ has $d(\xi,K\cap H(I))\ge\E[d(\xi,K\cap H(I))]+(\sqrt{2h(\delta)}+3\mathbin{\cdot}10^{-5})\sqrt m$. We apply Gaussian concentration for $1$-Lipschitz functions once to the cube distance and once to each of the at most $\sum_{k=0}^{\lfloor\delta m\rfloor}\binom mk\le\exp(h(\delta)m)$ section distances. The two
failure probabilities are then at most
\[
\exp(-m(3\mathbin{\cdot}10^{-5})^2/2)
\]
and
\[
\exp(-m(3\mathbin{\cdot}10^{-5}\sqrt{2h(\delta)}
+(3\mathbin{\cdot}10^{-5})^2/2)).
\]
At $m=N_0$ their sum is at most
$e^{-0.135}+e^{-4558}<0.874$, and it decreases with $m$.
Hence, with probability greater than $0.126$, neither event occurs. When neither occurs, Steps~2 and~3 and the monotonicity of Step~4 give
\[
d(\xi,Q_y)\ge d(\xi,[-\varepsilon_{\mathrm{box}},\varepsilon_{\mathrm{box}}]^m)>d(\xi,K\cap H(I))
\qquad\text{for every }I\subseteq[m]\text{ with }|I|\le\delta m,
\]
the two sides being separated by at least $(r-6\mathbin{\cdot}10^{-5})\sqrt m>0$.

\emph{Step 6: on the good event more than $\delta m$ constraints are tight.}
Suppose that the tight box constraints at $x$ have coordinate set $I$ with $|I|\le\delta m$.
The relaxation argument of part~\hyperlink{first-cov-part-4}{4} of the proof of Proposition~\ref{prop:first_covering},
with $\xi$ for $g$, $x$ for $c$, $\delta m$ for $m/100$ and the box $Q_y$ of this lemma, gives
\[
d(\xi,K\cap Q_y)\le d(\xi,K\cap H(I)),
\]
contradicting $d(\xi,K\cap Q_y)\ge d(\xi,Q_y)>d(\xi,K\cap H(I))$. Thus more than
$\delta m$ constraints are tight, and by Step~1 the affine rescaling $y+2x/\varepsilon_{\mathrm{box}}$ turns their coordinates
into signs.
\end{proof}

\subsection{Separated sums}
\label{subsec:bv_packing}

The phase parameters have a minimum spacing, but their exact locations depend on the projection outcomes. We use interval suprema to obtain a deterministic bound on the accumulated discrepancy.

\begin{lemma}[Separated sum against bin suprema]
\label{lem:bv_packing}
Let $F:[a,\infty)\to[0,\infty)$, and let
$a\le x_0<x_1<\cdots$ satisfy $x_{k+1}-x_k\ge\lambda>0$. Then
\[
\sum_{k=0}^{\infty}F(x_k)
\le\sum_{j=0}^{\infty}
\sup_{a+j\lambda\le u<a+(j+1)\lambda}F(u).
\]
\end{lemma}

\begin{proof}
Every $x_k$ lies in exactly one of the half-open bins $[a+j\lambda,a+(j+1)\lambda)$, $j\ge0$. Each bin contains at most one
point $x_k$, because consecutive points are at least $\lambda$ apart. Bounding each $F(x_k)$ by the supremum of $F$ over its bin and summing over the bins, we obtain the claim.
\end{proof}

\section{An existence constant below 7.8795}
\label{sec:existence_constant_sub10}

In this section, we combine a smooth spectral barrier, covariance-sensitive Gaussian moments, and a lossless conversion from Gaussian measure to partial signs. We certify the scalar inequalities needed by the proof on their entire domains. Every terminating decimal below denotes an exact rational number.

\begin{theorem}[Existence constant below $7.8795$]
\label{thm:existence_constant_sub10}
Let $1\le N\le n$, and let $A_1,\ldots,A_n\in\R^{N\times N}$ be symmetric matrices with $\|A_i\|\le1$. There are signs $\epsilon_1,\ldots,\epsilon_n\in\{-1,1\}$ such that
\[
\|\sum_{i=1}^n\epsilon_iA_i\|\le7.879493715947\sqrt n<7.8795\sqrt n.
\]
\end{theorem}

In Section~\ref{subsec:1070_coding} we prove Lemma~\ref{lem:1070_lossless_coding}, the lossless conversion of a Gaussian small-ball bound into a partial signing, from the peaked translated density of Lemma~\ref{lem:1070_peaked_density}. Section~\ref{subsec:sub10_smooth_barriers} introduces the two-component smooth barriers and the spectral allocation profile of Definition~\ref{def:1070_allocation}. Their curvature yields the one- and two-slot weighted Poincar\'e bounds of Lemma~\ref{lem:1070_weighted_poincare}. We estimate the resolvent means in Section~\ref{subsec:sub10_resolvent_means}: Lemma~\ref{lem:1070_resolvent_means} bounds them by the scalar roots of Definition~\ref{def:1070_roots}, Lemma~\ref{lem:sub10_mean_gap} controls the arithmetic--harmonic gap, and Lemma~\ref{lem:1070_parity} separates odd and even fields. The polynomial resolvent energies of Lemma~\ref{lem:1070_polynomial_energy} then follow from the recurrence of Definition~\ref{def:1070_energy_recurrence}. We integrate along a radial path from a reference Gaussian in Section~\ref{subsec:sub10_covariance_radial}: Lemma~\ref{lem:1070_covariance_power} bounds covariance-power moments, Lemma~\ref{lem:fixed_reference_radial} sets up the fixed-reference path, and Lemma~\ref{lem:1070_radial} retains the covariance credit. The return to the standard Gaussian is Lemma~\ref{lem:1070_transfer} in Section~\ref{subsec:sub10_gaussian_transfer}, which pays the determinant factor through a scalar cap and records the hereditary scaling of the exponent. Finite rational tests replace the spectral and profile conditions in Section~\ref{subsec:sub10_scalar_certificates}, through Lemmas~\ref{lem:scalar_profile_enclosures} and~\ref{lem:1070_spectral_certificate} under the certification rules of Definition~\ref{def:rational_certificate_rules}. Lemma~\ref{lem:1070_certificate} then certifies the seven two-component stages of Definition~\ref{def:1070:stage-parameters}. Lemma~\ref{lem:1070_infinite_tail} in Section~\ref{subsec:sub10_uniform_tail} is a uniform small-ball bound at large aspect ratios, and it closes every finite schedule of partial signings. We sharpen three shared estimates in Section~\ref{subsec:sub10_refinements}: the projection-aware variance of Lemma~\ref{lem:sub10_projection_variance}, the mixed linear moments of Lemma~\ref{lem:sub10_operator_words}, and the coupled spectral triples of Lemma~\ref{lem:sub10_triple}. Section~\ref{subsec:sub10_finite_refinement} defines the finite operator-energy refinement, Definition~\ref{def:sub10_operator_refinement}, whose validity is Lemma~\ref{lem:sub10_operator_refinement}. Its covariance-sensitive derivative means (Lemma~\ref{lem:sub10_derivative_means}) feed the trace recurrence of Definition~\ref{def:sub10_trace_recurrence}, the affine bounds of Lemma~\ref{lem:sub10_trace_tangents}, and the reflected radial remainder of Lemma~\ref{lem:sub10_reflected_remainder}. Positive scalar majorants, Lemma~\ref{lem:sub10_majorants} in Section~\ref{subsec:sub10_radial_transfer}, bound the radial integrand, and we close the Gaussian transfer with the rank-sensitive scalar condition Eq.~\eqref{eq:sub10_rank_profile}. We fix the existence-stage parameters of Definition~\ref{def:sub10_parameters} and the finite majorant construction of Definition~\ref{def:sub10:majorant-construction} in Section~\ref{subsec:sub10_shared_certificates}, and Lemma~\ref{lem:sub10_certificate} certifies every stage on its entire domain. Centering enters in Section~\ref{subsec:sub10_centered_gaps}, where we use the subtraction in the arithmetic--harmonic gap bound to center the leading quadratic field. Lemma~\ref{lem:slack_centered_updates} gives the centered one- and two-slot estimates, and Lemma~\ref{lem:slack_row_certificate} certifies the centered rows of Definition~\ref{def:slack_rows}. We add trace sensitivity in Section~\ref{subsec:sub10_trace_small_balls} through Lemma~\ref{lem:slack_trace_interface}, which scales a row's exponent with the trace parameter and turns each centered row into a hereditary small-ball bound. Repeated slack is the subject of Section~\ref{subsec:sub10_repeated_slack}. Lemma~\ref{lem:slack_refined_coding} refines the coding amplitude, Lemma~\ref{lem:envelope_joint_coding} imposes moment constraints at every active size, and Lemmas~\ref{lem:slack_transform} and~\ref{lem:slack_schedule_completion} propagate the slack and complete a finite schedule. The common-envelope moments of Section~\ref{subsec:envelope_moments}, Lemma~\ref{lem:envelope_moments}, exploit the positive congruence factor that all residual coefficients share after the first slack step. Section~\ref{subsec:iterated_envelope_powers} retains that envelope after every further congruence. Lemma~\ref{lem:iterated_envelope_update} is the hereditary update, built on the trace inequality of Lemma~\ref{lem:iterated_envelope_trace_power}, and Lemma~\ref{lem:iterated_envelope_cubic} adds the shifted cubic constraints for the initial envelope. The independent one-slot reduction of Section~\ref{subsec:iterated_envelope_rows}, Definition~\ref{def:iterated_envelope_one_slot} with Lemma~\ref{lem:iterated_envelope_one_slot}, runs the calculus with one inverse-weight slot and without a two-slot root. Section~\ref{subsec:sub81_barriers} presents the two one-slot barriers $\mathsf I$ and $\mathsf R$, their whole-domain triple certificates (Lemma~\ref{lem:sub81_triples}), and the contracted coordinates of Lemma~\ref{lem:sub10_contracted_coordinates}. We build the first small-ball row in Section~\ref{subsec:sub81_first_row} with barrier $\mathsf I$, retaining covariance credit through the tangents of Lemma~\ref{lem:sub81_covariance_tangents} and a contracted derivative metric. That subsection also fixes the shared statistic $f_3$ and its mean bound $M_3$. Section~\ref{subsec:sub81_residual} treats the residual rows with barrier $\mathsf R$, using sharper ordinary means (Lemmas~\ref{lem:sub81_low_means} and~\ref{lem:quartic_trace_means}) and the joint transfer profile of Lemma~\ref{lem:joint_profile}. Its output is Lemma~\ref{lem:joint_first_residual}, the joint first residual row, and Lemma~\ref{lem:sub81_residual_rows}, the ten hereditary residual rows. We give the proof of Theorem~\ref{thm:existence_constant_sub10} in Section~\ref{subsec:sub81_signing}. There Lemma~\ref{lem:unit_log_moment} replaces the Markov charge by a unit charge, Lemma~\ref{lem:unit_first_row} restates the first row, and eleven prefix phases, four centered rows, and the uniform tail form the exact bound. We collect the exact data in Section~\ref{subsec:sub81_data}: the fixed schedule, the moment-function linear programs, the radial polynomials and majorants, and the rounding and generation rules that replay the construction. The analytic tools and certified rows of the earlier subsections also support Section~\ref{sec:small_algorithm_constant}.

\subsection{Lossless Gaussian-to-sign coding}
\label{subsec:1070_coding}

Put $h(x):=-x\log x-(1-x)\log(1-x)$ for $0<x<1$, with $h(0):=0$ and $h(1):=0$, and fix
\[
s_\star:=0.67448,\qquad a_\star:=0.635,\qquad
b_\star:=\frac{1-2s_\star a_\star}{s_\star^2}.
\]
Let $Z$ have the even density
\[
f_Z(z):=\begin{cases}
a_\star,&|z|\le s_\star,\\
b_\star(2s_\star-|z|),&s_\star<|z|\le2s_\star,\\
0,&|z|>2s_\star.
\end{cases}
\]
The identity $2s_\star a_\star+s_\star^2b_\star=1$ normalizes this density.

By translating this density by an independent sign, we obtain the Gaussian measure comparison needed for coding.

\begin{lemma}[A peaked translated density]
\label{lem:1070_peaked_density}
If $\sigma$ is a uniform sign independent of $Z$, the law $P$ of $Z+s_\star\sigma$ has an even density nonincreasing on $[0,\infty)$, and
\[
P([-x,x])\ge\gamma_1([-x,x])\qquad(x\ge0).
\]
Consequently, for every symmetric convex Borel set $K\subseteq\R^m$,
$P^{\otimes m}(K)\ge\gamma_m(K)$.
\end{lemma}

\begin{proof}
For this proof write $s:=s_\star$, $a:=a_\star$, $b:=b_\star$, $F(x):=\gamma_1([-x,x])$ and $g(x):=F'(x)=\sqrt{2/\pi}e^{-x^2/2}$. Let $F_P(x):=P([-x,x])$ for $x\ge0$. Direct integration gives
\[
F_P(x)=\begin{cases}
(a+bs)x-bx^2/2,&0\le x\le s,\\
1/2+a(x-s),&s\le x\le2s,\\
1-b(3s-x)^2/2,&2s\le x\le3s,\\
1,&x\ge3s.
\end{cases}
\]
The following sufficient conditions apply to any $s,a$, with $b:=(1-2sa)/s^2$, and any $0<\delta<s<1$:
\begin{equation}
\begin{aligned}
&b>0,\qquad a>bs,\qquad a+bs>g(0),\qquad g(s)>a>g(s+\delta),\\
&\frac12-F(s)>\delta(g(s)-a)>0,\\
&1-F(5s/2)>bs^2/2,\qquad 1-F(3s)>bs^2/8.
\end{aligned}
\label{eq:slack_coding_criterion}
\end{equation}
We verify these conditions, for the original parameters and for a refined choice used later, by the rational checks at the end of this proof. The first two make the density even and nonincreasing, including at its jumps. On $[0,s]$, the derivative of $F_P-F$ is convex, positive at zero, and negative at $s$, so $F_P-F$ is minimized at an endpoint. Its values there are $0$ and $1/2-F(s)>0$. On $[s,2s]$, the derivative $a-g(x)$ increases and becomes positive before $s+\delta$, losing at most $\delta(g(s)-a)$ in the meantime. On $[2s,5s/2]$ and $[5s/2,3s]$, we use the respective lower bounds $1-bs^2/2$ and $1-bs^2/8$ for $F_P$, together with the last two comparisons. For $x\ge3s$, $F_P(x)=1$.

For tensorization, we express each even nonincreasing density as a mixture of uniform laws on centered intervals. With the other intervals fixed, Pr\'ekopa's theorem makes the integral of $\mathbbm{1}[K]$ over those coordinates and any remaining Gaussian coordinates even and log-concave. Its level sets are centered intervals, so the interval comparison permits replacing one Gaussian coordinate by $P$. We average over the mixtures and replace coordinates successively. Truncation and monotone convergence cover unbounded $K$. This proves the conclusion for every choice satisfying Eq.~\eqref{eq:slack_coding_criterion}.

For the numerical checks, we use the two polynomial families
\begin{align}
\mathsf e_n(x)&:=\sum_{j=0}^{n}\frac{(-1)^jx^{2j}}{2^j j!},
\\
\mathsf f_n(x)&:=\sum_{j=0}^{n}\frac{(-1)^jx^{2j+1}}{2^j j!(2j+1)}.
\end{align}
Taylor's integral remainder gives $\mathsf e_{17}(x)\le e^{-x^2/2}\le\mathsf e_{16}(x)$. Integration from $0$ to $x$ gives the corresponding bounds with $\mathsf f_{17}$ and $\mathsf f_{16}$.

The $\pi$ rule of Definition~\ref{def:rational_enclosures} at index $32$ and exact squaring give $\alpha_-<\sqrt{2/\pi}<\alpha_+$, where
$\alpha_-:=0.7978845608028653$ and $\alpha_+:=0.7978845608028654$.
Define $\underline g:=\alpha_-\mathsf e_{17}$, $\overline g:=\alpha_+\mathsf e_{16}$,
$\underline F:=\alpha_-\mathsf f_{17}$ and $\overline F:=\alpha_+\mathsf f_{16}$. Thus, for every $x\ge0$ and in particular on $0\le x\le3(0.674489750196)$,
\begin{align}
\underline g(x)&\le g(x)\le\overline g(x),
\\
\underline F(x)&\le F(x)\le\overline F(x).
\end{align}

These polynomials have degree at most $35$. The original parameter choice is $(s,a,\delta):=(0.67448,0.635,0.01)$, and the refined choice is $(s,a,\delta):=(0.674489750196,0.635553145368,10^{-12})$.
For $b:=(1-2sa)/s^2$, we consider the eight rational residuals
\[
\begin{gathered}
b,\quad a-bs,\quad a+bs-\overline g(0),\\
\underline g(s)-a,\quad a-\overline g(s+\delta),\\
1/2-\overline F(s)-\delta(\overline g(s)-a),\\
1-\overline F(5s/2)-bs^2/2,\quad 1-\overline F(3s)-bs^2/8.
\end{gathered}
\]
By exact rational substitution, we check that each residual exceeds $6\cdot10^{-7}$ for the original choice and $5\cdot10^{-14}$ for the refined choice. Their positivity verifies Eq.~\eqref{eq:slack_coding_criterion}.
Both choices also satisfy $0<\delta<s<1$. Thus the coding argument applies with either amplitude. We retain $s_\star$ in all constructions unless the refined amplitude is explicitly specified.

\end{proof}

The translated-density comparison converts a Gaussian mass bound into a partial signing through an entropy argument.

\begin{lemma}[Lossless partial signs]
\label{lem:1070_lossless_coding}
Let $K\subseteq\R^m$ be symmetric and convex, with $\gamma_m(K)\ge e^{-Em}$. If $0<q<1$ and
\[
\log2-E>h((1-q)/2),
\]
then there is $y\in\{-1,0,1\}^m$ supported on more than $(1-q)m$ coordinates such that $s_\star y\in K$. The same statement holds for any amplitude $s$ satisfying Eq.~\eqref{eq:slack_coding_criterion}. In particular, $E:=q^2/2$ is sufficient for every $m\ge1$.
\end{lemma}

\begin{proof}
For independent copies of $Z$, we average over $z:=(Z_1,\ldots,Z_m)$ and obtain
\[
\E_z[|\{\sigma\in\{-1,1\}^m:z+s_\star\sigma\in K\}|]
=2^mP^{\otimes m}(K)\ge e^{m(\log2-E)},
\]
where the first step follows by averaging over uniform signs, and the second step follows from Lemma~\ref{lem:1070_peaked_density} and the Gaussian mass bound. We choose a translate with at least this many signs. Kleitman's cube-diameter theorem~\cite[Theorem~1.1]{hkp20} bounds a family of Hamming diameter $D<m$ by
\[
\sum_{i=0}^r\binom mi\quad(D=2r),\qquad
2\sum_{i=0}^r\binom{m-1}i\quad(D=2r+1).
\]
These are at most $e^{mh(D/(2m))}$. For even $D$ this is the binomial entropy bound. For odd $D$ we use that bound in dimension $m-1$ and concavity of $h$ with $\frac{D}{2m}=\frac{m-1}{m}\frac{r}{m-1}+\frac1m\frac12$. The case $m=1$ is immediate. Thus two members $\sigma,\sigma'$ differ on more than $(1-q)m$ coordinates. Their half-difference $y:=(\sigma-\sigma')/2$ has the required support and $s_\star y\in K$, by symmetry and convexity. The translate cancels, and the argument uses no further property of $s_\star$. Finally, $\log2-h((1-q)/2)=\sum_{r=1}^\infty\frac{q^{2r}}{2r(2r-1)}>q^2/2$, where the first step follows by expanding the binary entropy, and the second step follows by retaining its positive quadratic term.
\end{proof}

\subsection{Smooth barriers and spectral curvature}\label{subsec:sub10_smooth_barriers}

We use the divided-difference convention of Definition~\ref{def:first_divided_difference}. For the two-component families below, we denote the additive barrier parameters by $\kappa_0>0$ and $\kappa_{\rm tail}>0$. Thus the coefficient $\kappa_1$ in the original barrier notation is $\kappa_0+\kappa_{\rm tail}$. For an integer $\tau\ge2$, set
\[
\phi(u):=\kappa_0\phi_0(u)+\kappa_{\rm tail}\phi_{1,\tau}(u),\qquad
\phi_0(u):=-\log(1-u^2),\qquad
\phi_{1,\tau}(u):=\sum_{j=\tau}^\infty\frac{u^{2j}}j,
\]
for $|u|<1$, and extend $\phi$ by $+\infty$ outside this interval. Write $\Phi(X):=\tr[\phi(X)]$. Let $X:=\sum_{i=1}^m \xi_iN_i$ with symmetric $N_i\in\R^{N\times N}$ and $\sum_{i=1}^mN_i^2\preceq aI$. We allow the even tilted measure
\[
\d\mu:=Z_\mu^{-1}e^{-\Phi(X)-Q(\xi)}\d\gamma_m(\xi),
\]
where $Q$ is any nonnegative quadratic form. In the applications below $\kappa_0+\kappa_{\rm tail}>2$. For $\sigma\in\{-1,1\}$ write $Y_\sigma:=I+\sigma X$.

We allocate the scalar barrier curvature across spectral values to control the resolvent weights near the boundary.

\begin{definition}[Spectral allocation profile]
\label{def:1070_allocation}
For $0<\zeta,\theta<1$, put
\begin{align*}
d_\zeta(u)&:=\min\{1,\sqrt{(1-u)/\zeta}\},\\
F_\theta(v)&:=\max\{2(1+\theta-v)/(1+\theta),
3(1-v)+(1/\theta-1)(1-v)^2\}.
\end{align*}
The all-pairs curvature condition is
\begin{equation}
\label{eq:1070_curvature_condition}
(1-u)(1-v)\phi'[u,v]\ge F_\theta(d_\zeta(u)d_\zeta(v))
\qquad(-1<u,v<1).
\end{equation}
Reflection gives the analogous condition at the other edge.
\end{definition}

The spectral allocation condition controls the curvature of the resolvent weights, which is the input to the weighted variance estimate.

\begin{lemma}[One- and two-slot weighted Poincar\'e bounds]
\label{lem:1070_weighted_poincare}
Assume Eq.~\eqref{eq:1070_curvature_condition} and $2a<\zeta^2$. Put
\[
p_1:=(1-a/\zeta^2)^{-1},\qquad p_2:=(1-2a/\zeta^2)^{-1}.
\]
For a one-slot weight $W:=Y_\sigma^{-1}$, or a two-slot weight $W:=Y_{\sigma_1}^{-1}\otimes Y_{\sigma_2}^{-1}$, respectively,
\begin{equation}
\label{eq:1070_weighted_variance}
\inf_v\E_\mu[(f-v)^\top W(f-v)]
\le p_r\sum_{i=1}^m\E_\mu[(\partial_i f)^\top W(\partial_i f)],\qquad r=1,2.
\end{equation}
Here $v$ is a constant vector of the appropriate dimension. Direct sums of these weights are also allowed.
\end{lemma}

\begin{proof}
For a positive matrix density $A$, the curvature blocks are $\mathcal K_{i,j}(A):=(\partial_iA)A^{-1}(\partial_jA)-\partial_{i,j}A$. If $A=e^{-V}W$, then $e^V\mathcal K_{i,j}(A)=W_iW^{-1}W_j-W_{i,j}+V_{i,j}W$. For $W=Y^{-1}$, congruence by $Y^{1/2}$ turns this into
$V_{i,j}I-C_jC_i$, where $C_i:=Y^{-1/2}(\partial_iY)Y^{-1/2}$. The order is $C_jC_i$. For two inverse-weight slots, mixed derivatives cancel mixed products, leaving one negative block per slot.

Consider $Y:=I-X$. Reflection treats $I+X$. In its eigenbasis let $D:=\min\{I,\sqrt{Y/\zeta}\}$, with eigenvalues $d_a$, and put $C_{i,1}:=DC_iD$. Since $DY^{-1/2}\preceq\zeta^{-1/2}I$, $\sum_{i=1}^mC_{i,1}^2\preceq\frac a{\zeta^2}I$. The tensor Gram estimate of Fact~\ref{fact:tensor_gram_inequalities} therefore bounds the regular curvature loss by $a/\zeta^2$ in each slot. To estimate the remaining loss, we fix test vectors $z_i$ and set $T_{ab,c}:=\sum_{i=1}^m(C_i)_{ab}(z_i)_c$, so $T_{ab,c}=T_{ba,c}$. Write
\[
Q_0:=\sum_{i=1}^m\sum_{j=1}^m\langle z_i,C_jC_i z_j\rangle,\qquad
Q_1:=\sum_{i=1}^m\sum_{j=1}^m\langle z_i,C_{j,1}C_{i,1} z_j\rangle.
\]
Weighted arithmetic--geometric mean, with weights $(1+\theta-d_ad_b)^{-1}$ on the cyclic products of $T$, gives
\[
2(Q_0-Q_1)\le\sum_{a=1}^N\sum_{b=1}^N\sum_{c=1}^N T_{ab,c}^2(1+\theta-v)
(\frac{1-vd_bd_c}{1+\theta-d_bd_c}
+\frac{1-vd_ad_c}{1+\theta-d_ad_c}),\quad v:=d_ad_b.
\]
Put $C:=1+\theta$. The derivative of the displayed coefficient in $d_c$ has the sign of $1-Cv$. We distinguish two cases. If $v\ge1/C$, the coefficient is nonincreasing in $d_c$, so its maximum is attained at $d_c=0$ and equals $2(C-v)/C$, the first branch of $F_\theta(v)$.

If $v\le1/C$, the coefficient is nondecreasing in $d_c$, so take $d_c=1$. Its value is $(C-v)[2v+(1-Cv)(\frac1{C-d_a}+\frac1{C-d_b})]$. For $s:=d_a+d_b$ and $d_ad_b=v$, the reciprocal sum in brackets is $(2C-s)/(C^2-Cs+v)$. Its derivative with respect to $s$ is $\frac{C^2-v}{(C^2-Cs+v)^2}>0$. Since $1-Cv\ge0$ and $(1-d_a)(1-d_b)\ge0$ gives $s\le1+v$, the maximum is attained at $\{d_a,d_b\}=\{1,v\}$. Substitution gives $3(1-v)+(1/\theta-1)(1-v)^2$, the second branch of $F_\theta(v)$. At $v=1/C$ either case applies. Thus the coefficient is at most $F_\theta(v)$ throughout the domain, including repeated indices. The spectral Hessian formula and Eq.~\eqref{eq:1070_curvature_condition} imply
\[
\sum_{i=1}^m\sum_{j=1}^m(\partial_{i,j}\Phi)\langle z_i,z_j\rangle
\ge\sum_{a=1}^N\sum_{b=1}^N\sum_{c=1}^NF_\theta(d_ad_b)T_{ab,c}^2
\ge2(Q_0-Q_1).
\]
Half the barrier Hessian absorbs each slot's nonregular loss. Adding the Gaussian Hessian $I$ and $\nabla^2Q\succeq0$ gives curvature at least $1-ra/\zeta^2$ for $r=1,2$.

The weight is $A:=e^{-V}W$ on the open convex set $\{\xi:\|X(\xi)\|<1\}$. Here $e^{-V}$ is the Lebesgue density of $\mu$ up to a constant factor, which cancels between the two sides. The curvature bound just proved is Eq.~\eqref{eq:first_1} for this weight, with $1-ra/\zeta^2=p_r^{-1}$ in the role of the constant there. Hence Eq.~\eqref{eq:1070_weighted_variance} is Eq.~\eqref{eq:first_2} of Section~\ref{sec:first_existence}, whose left side is $\inf_v\int(f-v)^\top A(f-v)$. The proof there runs through the Neumann problem for $Lu:=-A^{-1}\sum_{i=1}^m\partial_i(A\partial_iu)$ on a smooth bounded convex exhaustion, the Green identity, and the Bochner identity Eq.~\eqref{eq:first_3}, whose boundary term is nonnegative by convexity. That proof needs only that $A$ is smooth and integrable and that the fields have finite energy. At the spectral boundary, the vanishing order $\kappa_0+\kappa_{\rm tail}>2$ controls both inverse factors. Gaussian decay controls infinity, including kernel directions of $\xi\mapsto X(\xi)$. Thus $A$ is integrable, the polynomial and inverse-weighted fields have finite energies, and integration by parts is justified without differentiating $D$.
\end{proof}

\subsection{Resolvent means and parity-sensitive polynomial energies}\label{subsec:sub10_resolvent_means}

We collect the scalar roots that bound the one-slot and two-slot resolvent means.

\begin{definition}[Scalar root bounds]
\label{def:1070_roots}
With $p_1,p_2$ as above, define
\begin{align*}
L_1&:=\frac2{1+\sqrt{1-4p_1a}},&c_1&:=L_1-1,\\
A_J&:=1-c_1^2,&B_J&:=1+2c_1,\\
\Delta_J&:=A_J^2-4p_2aB_J,&J&:=\frac{2B_J}{A_J+\sqrt{\Delta_J}},\qquad c:=1-1/J.
\end{align*}
We require $1-4p_1a>0$, $A_J>0$, and $\Delta_J>0$. These requirements will be checked at the largest variance at each stage.
\end{definition}

We apply the weighted variance bounds to resolvent fields and obtain scalar quadratic inequalities for their means.

\begin{lemma}[Resolvent mean bounds]
\label{lem:1070_resolvent_means}
Under the preceding assumptions,
\[
I\preceq S_1:=\E_\mu[Y_+^{-1}]\preceq L_1I,\qquad
I\preceq S:=\E_\mu[Y_+^{-1}\otimes Y_+^{-1}]\preceq JI.
\]
The same conclusions hold with both signs negative.
\end{lemma}

\begin{proof}
We apply the one-slot estimate to $(Y_+-I)z$. Its optimized variance is $z^\top (I-S_1^{-1})z$, and its gradient energy is at most $a\|S_1\|\|z\|^2$. Thus
$I-S_1^{-1}\preceq p_1\sum_{i=1}^mN_iS_1N_i$. Jensen and evenness give $S_1\succeq I$. Put $l:=\|S_1\|$. Then $p_1al^2-l+1\ge0$. Continuity from $a=0$ selects the smaller root $L_1$.

For two slots, put $K_1:=I\otimes S_1$, $K_2:=S_1\otimes I$. We apply the two-slot estimate separately to $((Y_+-I)\otimes I)z$ and $(I\otimes(Y_+-I))z$ and add. The uncentered second moments give $U_1+U_2\succeq2I$, since pointwise
$Y_+\otimes Y_+^{-1}+Y_+^{-1}\otimes Y_+\succeq2I$. Hence, with $s:=\|S\|$, $2I-K_1S^{-1}K_1-K_2S^{-1}K_2\preceq2p_2asI$. Evenness gives $S\succeq I$: in a common eigenbasis of $X$, the geometric mean of the two entries of $Y_+^{-1}\otimes Y_+^{-1}$ and $Y_-^{-1}\otimes Y_-^{-1}$ is at least one. Put $D_i:=K_i-I$, so $0\preceq D_i\preceq c_1I$. For a unit top eigenvector $z$ of $S$, $z^\top K_iS^{-1}K_i z\le\frac{1+2c_1}s+c_1^2$. It follows that $p_2as^2-A_Js+B_J\ge0$, whose smaller root is $J$. To justify both root choices, we replace $X$ by $rX$, $0\le r\le1$, keeping $Q$ fixed. The strict discriminant conditions persist for smaller variance, and the means vary continuously from $I$. They cannot cross the gap between the two roots. The integrability we established in the preceding proof justifies this continuity.
\end{proof}

We separate weighted energy into its mean and centered parts, using the arithmetic--harmonic gap to control the mean contribution.

\begin{lemma}[Weighted means and the arithmetic--harmonic gap]
\label{lem:sub10_mean_gap}
Let $W\succeq I$ be a random positive matrix, $S:=\E[W]\preceq JI$, $T:=\E[W^{-1}]$, and $k:=\|S^{1/2}TS^{1/2}-I\|$. For a finite-energy field define $\mathcal R_W(f):=\E[f^\top Wf]$, $\mathcal P_W(f):=\|\E[Wf]\|_{S^{-1}}^2$, and $\mathcal D_W(f):=\mathcal R_W(f)-\mathcal P_W(f)$. Put $D:=I-T$ and $B_W:=\E[(W-I)W^{-1}(W-I)]$. Then
\begin{equation}
0\preceq S-T^{-1}=B_W-DT^{-1}D\preceq B_W-D^2\preceq B_W.
\label{eq:slack_arithmetic_harmonic_identity}
\end{equation}
If $\|\E[f]\|^2\le M$ and $\mathcal D_W(f)\le B$, then
\begin{equation}
\begin{aligned}
\mathcal P_W(f)&\le(\sqrt{JM}+\sqrt{kB})^2,\\
k&\le\min\{J-1,\|B_W-D^2\|\}\le\min\{J-1,\|B_W\|\}.
\end{aligned}
\end{equation}
\end{lemma}
\begin{proof}
Put $v:=S^{-1}\E[Wf]$, $g:=f-v$, and $b:=\E[f]$. Since $\E[Wg]=0$, weighted Cauchy--Schwarz gives $(b-v)(b-v)^\top\preceq\mathcal D_W(f)(T-S^{-1})$. Congruence by $S^{1/2}$ and the triangle inequality imply $\|S^{1/2}v\|\le\sqrt{JM}+\sqrt{kB}$. Its square is $\mathcal P_W(f)$.

Jensen and $W\succeq I$ give $S^{-1}\preceq T\preceq I$, so $0\le k\le J-1$. In Eq.~\eqref{eq:slack_arithmetic_harmonic_identity}, the first step follows from Jensen, the second step follows by expansion, the third step follows from $T^{-1}\succeq I$, and the fourth step follows from $D^2\succeq0$.

We have $k=\|T^{1/2}(S-T^{-1})T^{1/2}\|\le\|B_W-D^2\|\le\|B_W\|$, where the first step follows because $S^{1/2}TS^{1/2}$ and $T^{1/2}ST^{1/2}$ have the same eigenvalues, the second step follows from Eq.~\eqref{eq:slack_arithmetic_harmonic_identity} and $T\preceq I$, and the third step follows from $0\preceq B_W-D^2\preceq B_W$.
\end{proof}

For an array $f$ of vectorized matrix fields, let
\[
W_\pm:=Y_\pm^{-1}\otimes Y_\pm^{-1},\qquad
\mathcal R(f):=\E_\mu[f^\top W_+f],\qquad
\mathcal D(f):=\E_\mu[\|f\|^2],
\]
where we understand a block-diagonal copy of $W_+$ for an array.

Evenness of the tilted measure gives different controls on the projected means of odd and even fields.

\begin{lemma}[Odd and even fields]
\label{lem:1070_parity}
Under an even measure, let $X=\sum_{i=1}^m\xi_iN_i$ with $\|X\|<1$, $\sum_{i=1}^mN_i^2\preceq aI$, and $\operatorname{Cov}(\xi)\preceq I$. In one slot take $W_\pm:=(I\pm X)^{-1}$ and $(J_*,A_*):=(L,L)$. In two slots take $W_\pm:=(I\pm X)^{-1}\otimes(I\pm X)^{-1}$ and $(J_*,A_*):=(J,J(1+a))$. In either case assume $\E[W_e]\preceq J_*I$.
Here $L$ denotes any valid one-slot mean bound, for instance $L_1$ of Definition~\ref{def:1070_roots}, and $J$ denotes any valid two-slot one. Set $W_e:=(W_++W_-)/2$, $S:=\E[W_e]$, $\mathcal R(f):=\E[f^\top W_+f]$, and $\mathcal P(f):=\|\E[W_+f]\|_{S^{-1}}^2$. Let $B\ge\mathcal R(f)-\mathcal P(f)$ and let $k_*$ bound the arithmetic--harmonic gap of $W_e$. If $f$ is odd, then
\[
(\mathcal R(f),\mathcal P(f))\le(A_*B,(A_*-1)B).
\]
If $f$ is even and $\|\E[f]\|^2\le M$, put $Z:=(\sqrt{J_*M}+\sqrt{k_*B})^2$. Then $(\mathcal R(f),\mathcal P(f))\le(B+Z,Z)$. One may always take $k_*:=J_*-1$. Constants satisfy $\mathcal R(f)=\mathcal P(f)\le J_*\|f\|^2$.

In two slots, with $c:=1-1/J$, the following conservative form is also valid:
\begin{equation}
\begin{aligned}
(\mathcal R(f),\mathcal P(f))&\le(\mathcal F_c(M,B),\mathcal F_c(M,B)-B),\\
\mathcal F_c(M,B)&:=M+\frac{(\sqrt{cM}+\sqrt{cM+(1-c)B})^2}{(1-c)^2}.
\end{aligned}
\end{equation}
All bounds are componentwise, nonnegative and nondecreasing in their nonnegative inputs. The one-slot assertion uses no two-slot mean or root. Direct sums are allowed.
\end{lemma}
\begin{proof}
Put $W_o:=(W_+-W_-)/2$. In both cases $W_e\succeq I$, and
\[
W_oW_e^{-1}W_o=W_e-R,\qquad
R:=\begin{cases}I,&\text{one slot},\\(I+X\otimes X)^{-1},&\text{two slots}.\end{cases}
\]
In two slots the tensor Gram bound (Fact~\ref{fact:kronecker_cauchy_schwarz}) gives $\|\E[X\otimes X]\|\le a$, so Jensen for inversion gives $\E[R]\succeq(1+a)^{-1}I$. In one slot $\E[R]=I$. Hence in either case $\E[W_oW_e^{-1}W_o]\preceq(1-A_*^{-1})S$.

For odd $f$, parity gives $\E[W_+f]=\E[W_of]$ and $\mathcal R(f)=\E[f^\top W_ef]$. Weighted Cauchy--Schwarz therefore gives $\mathcal P(f)\le(1-A_*^{-1})\mathcal R(f)$. We absorb this term using $\mathcal R(f)-\mathcal P(f)\le B$ and obtain the odd bounds. For even $f$, parity identifies its full and projected energies with those for $W_e$. Lemma~\ref{lem:sub10_mean_gap} then proves the bound involving $Z$.

For the conservative two-slot form take $k_*:=J-1$. We have
\begin{equation}
\mathcal F_c(M,B)-B
=M+2\alpha M+\frac{cB}{1-c}
+2\sqrt{\alpha M}\sqrt{\alpha M+B/(1-c)},\qquad
\alpha:=\frac{c}{(1-c)^2}.
\label{eq:algorithm_13_projected_energy}
\end{equation}
Since $\alpha=J(J-1)$ and $J\ge1$, the right-hand side is at least $JM+(J-1)B+2\sqrt{J(J-1)MB}=Z$. Thus both even bounds follow. Eq.~\eqref{eq:algorithm_13_projected_energy} also proves their nonnegativity and monotonicity. The odd pair is manifestly monotone. Constants have zero centered energy. Each argument applies to direct sums.
\end{proof}

Successive derivatives alternate parity, so we encode their energy bounds in a backward recurrence.

\begin{definition}[Polynomial energy recurrence]
\label{def:1070_energy_recurrence}
Let $\operatorname{df}(r):=(2r-1)!!$ for $r\ge1$, and $\operatorname{df}(0):=1$. For an integer $\ell\ge1$, start with $V_\ell:=(\ell!)^2J$. For $j=\ell-1,\ldots,0$, set $r:=\ell-j$, $(\ell)_j:=\ell!/r!$, and
\[
V_j:=\begin{cases}
p_2J(1+a)V_{j+1},&r\text{ odd},\\
\mathcal F_c(M_j,p_2V_{j+1}),&r\text{ even},
\end{cases}\qquad
M_j:=(\ell)_j^2\begin{cases}1,&r=2,\\\operatorname{df}(r),&r\ne2.\end{cases}
\]
Write $C_\ell(a):=V_0$, with the stage parameter $\zeta$ held fixed.
\end{definition}

We apply the parity bounds to all derivative arrays of a matrix power to bound its energy with two resolvent factors.

\begin{lemma}[Polynomial resolvent energies]
\label{lem:1070_polynomial_energy}
For the preceding measure and variance bound,
\begin{equation}
\label{eq:1070_polynomial_energy}
\frac1N\E_\mu[\tr[X^{2\ell}Y_+^{-2}]]\le a^\ell C_\ell(a).
\end{equation}
The scalar function $C_\ell(a)$ is nondecreasing while the strict root conditions hold.
\end{lemma}

\begin{proof}
We take the full array of ordered derivatives $\nabla^j(X^\ell)$. Set $r:=\ell-j$. Then
\begin{equation}
\label{eq:1070_derivative_moment}
\frac1N\E_\mu[\|\nabla^j(X^\ell)\|_{\rm HS}^2]
\le a^\ell(\ell)_j^2\operatorname{df}(r).
\end{equation}
We polarize the $j$ derivatives by independent standard Gaussian vectors $h_1,\ldots,h_j$. Their second moment sums the ordered derivative indices. Each word has $r$ copies of $X$ and one copy of each direction matrix. The map $T\mapsto\sum_{i=1}^mN_iTN_i$ has norm at most $a$ on every Schatten class, by Fact~\ref{fact:cp_schatten_shift} at $\ell=\infty$ with $\sum_{i=1}^mN_i^2\preceq aI$. We integrate the independent directions one at a time, with Schatten H\"older between the remaining powers, and bound the conditional squared Hilbert--Schmidt norm of one word by $a^j\tr[|X|^{2r}]$. There are $(\ell)_j$ words, so Cauchy--Schwarz supplies $(\ell)_j^2$. Harg\'e's Gaussian convex-domination theorem, applied to the centered even convex tilt, and the Gaussian trace-moment bound of Fact~\ref{fact:gaussian_trace_moments} give $\E_\mu[\tr[|X|^{2r}]]\le N\operatorname{df}(r)a^r$. This proves Eq.~\eqref{eq:1070_derivative_moment}.

For the mean of a degree-two derivative array there is a sharper bound
\[
\frac1N\|\E_\mu[\nabla^{\ell-2}(X^\ell)]\|_{\rm HS}^2
\le a^\ell(\ell)_{\ell-2}^2.
\]
{\setlength{\emergencystretch}{2em}\hbadness=10000 Indeed, the two undifferentiated occurrences form $L X U X R$ in a word. Their mean is $L\Psi_\Sigma(U)R$. Here $\Sigma:=\operatorname{Cov}_\mu(\xi)\preceq I$ by Brascamp--Lieb (Fact~\ref{fact:even_convex_tilts}), as $\Phi+Q$ is convex, and
$\Psi_\Sigma(U):=\sum_{i=1}^m\sum_{j=1}^m\Sigma_{i,j}N_iUN_j$. This positive map has Schatten-2 norm at most $a$. Integrating each independent direction matrix, inside or outside $\Psi_\Sigma$, costs another factor $a$ in the squared norm. Thus one word costs at most $Na^\ell$. We sum words by Cauchy--Schwarz and obtain the assertion. Odd derivative arrays have mean zero. For other even arrays we use Jensen followed by Eq.~\eqref{eq:1070_derivative_moment}. At $j=\ell$ the array is constant and its weighted energy is at most $Na^\ell(\ell!)^2J$. We apply Lemma~\ref{lem:1070_parity} backward through the derivative arrays and divide by $Na^\ell$. This is exactly Definition~\ref{def:1070_energy_recurrence}. For $j=0$, vectorization identifies its energy with the left side of Eq.~\eqref{eq:1070_polynomial_energy}.\par}

Finally $p_1,p_2,L_1,c_1$ increase with $a$. Write $J=(B_J/A_J)\mathcal C(p_2aB_J/A_J^2)$, where $\mathcal C(x):=2/(1+\sqrt{1-4x})$ has a power series with nonnegative coefficients. Thus $J$ and $c$ increase too. Every operation in the backwards recurrence is nondecreasing in these nonnegative arguments, proving the last assertion. At $a=0$ all formulas mean their continuous limits.
\end{proof}

\subsection{Covariance-power moments and radial integration}\label{subsec:sub10_covariance_radial}

Let $Y:=\sum_{i=1}^m \xi_iB_i$ with symmetric $B_i$ and $\sum_{i=1}^mB_i^2\preceq\eta I$. Define the coefficient Gram matrix $\Gamma_{i,j}:=\tr[B_iB_j]$, and put
\[
C_\Gamma:=(I+2\kappa_0\Gamma)^{-1},\qquad
\nu:=\mathcal N(0,C_\Gamma),\qquad
\omega_j:=\frac{\tr[\Gamma C_\Gamma^j]}{N\eta}\quad(j\ge1).
\]
Then $0\le\omega_j\le1$. Write $\psi(u):=\phi(u)-\kappa_0u^2$, an even nonnegative convex function, and use the radial measures
\[
\d\mu_t:=Z_t^{-1}e^{-\tr[\psi(tY)]}\d\nu,\qquad
Z_t:=\E_\nu[e^{-\tr[\psi(tY)]}],\qquad0\le t\le1.
\]
Here $Z_0=1$. Relative to $\gamma_m$, the non-Gaussian potential is
$\Phi(tY)+\kappa_0(1-t^2)\xi^\top \Gamma \xi$, so all preceding weighted estimates apply with $a:=\eta t^2$ and a nonnegative quadratic $Q$.

By retaining powers of the preconditioned covariance, we sharpen the ordinary moments used in the radial integral.

\begin{lemma}[Covariance-power trace bound]
\label{lem:1070_covariance_power}
For every integer $j\ge1$,
\[
\frac1N\E_{\mu_t}[\tr[(tY)^{2j}]]
\le\operatorname{df}(j)(\eta t^2)^j\omega_j.
\]
\end{lemma}

\begin{proof}
We first rotate the coefficient coordinates to diagonalize $\Gamma$, with eigenvalues $\lambda_i$ and $c_i:=(1+2\kappa_0\lambda_i)^{-1}$. The rotated matrices still satisfy $\sum_{i=1}^mB_i^2\preceq\eta I$. The variance matrix under $\nu$ is $\Sigma_C:=\sum_{i=1}^m c_iB_i^2$. In an eigenbasis $u_a$ of $\Sigma_C$, set $w_{a,i}:=u_a^\top B_i^2u_a/\eta$. These weights are nonnegative and sum to at most one. We add the missing mass at zero and apply scalar Jensen to $x\mapsto x^j$. This gives $(u_a^\top \Sigma_Cu_a)^j\le\eta^j\sum_{i=1}^mw_{a,i}c_i^j$. Summing over $a=1,\ldots,N$, we obtain
\[
\tr[\Sigma_C^j]\le\eta^{j-1}\sum_{i=1}^m\lambda_i c_i^j
=\eta^{j-1}\tr[\Gamma C_\Gamma^j].
\]
We do not use operator convexity of $x^j$ here. The sharp Gaussian noncommutative Khintchine inequality gives $\E_\nu[\tr[Y^{2j}]]\le\operatorname{df}(j)\tr[\Sigma_C^j]$. We use the sharp double-factorial constant of Buchholz~\cite[Theorem~5]{b01}. See also Tropp~\cite[Eq.~(4.17)]{t12}.

Finally $e^{-\tr[\psi(tY)]}$ is log-concave and even. Harg\'e's centered Gaussian convex-domination theorem~\cite{harge04}, after whitening $\nu$, applies to the convex function $\tr[(tY)^{2j}]$. Combining the two bounds proves the lemma.
\end{proof}

Keeping the Gaussian reference fixed, we express the small-ball loss as a radial integral with explicit cell contributions.

\begin{lemma}[Fixed-reference radial integration]
\label{lem:fixed_reference_radial}
Let $Y:=\sum_{i=1}^m\xi_iB_i$ with symmetric $B_i\in\R^{N\times N}$ and Gram matrix $\Gamma_{i,j}:=\tr[B_iB_j]$. Suppose
\[
\phi(u):=\sum_{j=1}^{K-1}b_ju^{2j}/j+b_T\sum_{j=K}^{\infty}u^{2j}/j,
\qquad b_j\ge0,\quad\kappa:=b_1>0,\quad b_T>2,
\]
for an integer $K\ge2$, on $|u|<1$, extended by $+\infty$ outside. Put $\psi(u):=\phi(u)-\kappa u^2$ and $g(x):=\sum_{j=2}^{K-1}b_jx^j+b_Tx^K/(1-x)$. Keep $\nu:=\mathcal N(0,(I+2\kappa\Gamma)^{-1})$ fixed and define $Z_t:=\E_\nu[e^{-\tr[\psi(tY)]}]$ and $\d\mu_t:=Z_t^{-1}e^{-\tr[\psi(tY)]}\d\nu$. Then
\begin{equation}
-N^{-1}\log\nu[\|Y\|<1]\le-N^{-1}\log Z_1
=\int_0^1N^{-1}\E_{\mu_{\sqrt v}}[\tr[g(vY^2)]]\d v/v.
\label{eq:fixed_reference_radial}
\end{equation}
For a finite partition, if the normalized integrand on cell $[v_-,v_+]$ is bounded by $\sum_{j\ge1}c_jv^j$ with fixed nonnegative coefficients, that cell contributes at most $\sum_{j\ge1}c_j(v_+^j-v_-^j)/j$.
\end{lemma}
\begin{proof}
Relative to $\gamma_m$, the tilted potential is $\tr[\phi(tY)]+\kappa(1-t^2)\xi^\top\Gamma\xi$, an even convex potential with a nonnegative quadratic term. Thus any applicable weighted estimate allowing such a quadratic term remains valid on this path. We have $Z_0=1$ and $Z_1\le\nu[\|Y\|<1]$, since $\psi\ge0$ and $e^{-\tr[\psi(Y)]}$ vanishes outside the spectral domain. Differentiating, we obtain $-\d\log Z_t/\d t=2\E_{\mu_t}[\tr[g(t^2Y^2)]]/t$. The boundary order $b_T>2$ and Gaussian decay justify differentiation by the exhaustion argument in Lemma~\ref{lem:1070_weighted_poincare}. Near zero, $g(vY^2)=O(v^2\|Y\|^4)$ away from the spectral edge, and the boundary factor absorbs its pole. Gaussian moments therefore give integrability against $\d v/v$. Integration and $v=t^2$ give Eq.~\eqref{eq:fixed_reference_radial}, where the first step follows from the probability comparison and the second step follows from radial differentiation. The cell formula is exact power integration against $\d v/v$.
\end{proof}

In the radial integration we combine the polynomial moment bounds with a resolvent estimate for the untruncated tail.

\begin{lemma}[Radial bound with the covariance credit retained]
\label{lem:1070_radial}
Fix integers $K\ge\tau\ge2$ and $\ell\ge1$ with $L:=2K-\ell\ge1$. Assume the preceding curvature and root conditions at variance $\eta$. Put $N_{\mathrm{grid}}:=512$, $u_i:=(i/N_{\mathrm{grid}})^2$, and
\begin{align*}
p_j&:=\frac{\operatorname{df}(j)\eta^j}{j}
(\kappa_0+\kappa_{\rm tail}\mathbbm{1}[j\ge\tau]),\quad 2\le j<K,\\
T&:=\frac{(\kappa_0+\kappa_{\rm tail})\eta^K\sqrt{\operatorname{df}(L)}}K
\sum_{i=1}^{N_{\mathrm{grid}}}(u_i^K-u_{i-1}^K)\sqrt{C_\ell(\eta u_i)}.
\end{align*}
Then
\begin{equation}
\label{eq:1070_radial_bound}
\nu[\|Y\|<1]\ge Z_1\ge
\exp[-N(\sum_{j=2}^{K-1}p_j\omega_j+T\sqrt{\omega_L})].
\end{equation}
\end{lemma}

\begin{proof}
Let $X:=tY$ and $R_j(t):=N^{-1}\E_{\mu_t}[\tr[X^{2j}(I-X^2)^{-1}]]$. Lemma~\ref{lem:fixed_reference_radial} gives
\[
-\frac{\d}{\d t}\log Z_t=\frac{2N}{t}(\kappa_0R_2(t)+\kappa_{\rm tail}R_\tau(t)).
\]
We expand
\[
R_j(t)=\sum_{r=j}^{K-1}N^{-1}\E[\tr[X^{2r}]]+R_K(t).
\]
By evenness, $R_K(t)=N^{-1}\E[\tr[X^{2K}Y_+^{-1}]]$. Cauchy--Schwarz with the two matrix factors $X^{2K-\ell}$ and $X^\ell Y_+^{-1}$, followed by Lemmas~\ref{lem:1070_polynomial_energy} and~\ref{lem:1070_covariance_power}, gives
\[
R_K(t)\le(\eta t^2)^K\sqrt{\operatorname{df}(2K-\ell)C_\ell(\eta t^2)}\sqrt{\omega_{2K-\ell}}.
\]
The finite terms integrate exactly to $p_j\omega_j$. For the remainder we use monotonicity of $C_\ell$ on each cell in $u:=t^2$. The cell formula of Lemma~\ref{lem:fixed_reference_radial} gives exactly $T$ and proves Eq.~\eqref{eq:1070_radial_bound}, including the probability comparison.
\end{proof}

\subsection{Returning to the standard Gaussian and hereditary scaling}\label{subsec:sub10_gaussian_transfer}

A change of Gaussian covariance introduces a determinant factor. We bound this factor together with the radial contribution by a scalar profile.

\begin{lemma}[Gaussian transfer with a scalar cap]
\label{lem:1070_transfer}
Let $\kappa,\eta>0$, and let $\Gamma\succeq0$ be an $m\times m$ matrix with $\tr\Gamma\le N\eta$ and $\operatorname{rank}\Gamma\le wN$. Let $\nu$ be the probability measure proportional to $e^{-\kappa\xi^\top\Gamma\xi}\d\gamma_m(\xi)$, and put $\omega_j:=\tr[\Gamma(I+2\kappa\Gamma)^{-j}]/(N\eta)$. Suppose a Borel set $\mathcal K$ satisfies
\[
-N^{-1}\log\nu(\mathcal K)\le C+\sum_jQ_j\omega_j,
\qquad C,Q_j\ge0,
\]
with finitely many coefficients. For $H>1$, set
\[
L_H(x):=\frac{H\log(1+x)-\log(1+Hx)}{(H-1)x},\qquad L_H(0):=0.
\]
If $b\ge0$, $B_{\rm cap}>0$, and
\[
L_H(x)+\frac{H}{(H-1)\kappa\eta}\sum_j\frac{Q_j}{(1+x)^j}-\frac b x\le B_{\rm cap}\qquad(x>0),
\]
then
\begin{equation}
\gamma_m(\mathcal K)\ge e^{-Ne},\qquad
e:=\kappa\eta B_{\rm cap}+bw/2+HC/(H-1).
\label{eq:sub10_exponent}
\end{equation}
For $b=0$ include the continuous value at zero. The same cap gives exponent at most $\alpha e$ after replacing $\eta,w,C,Q_j$ by $\alpha\eta,\alpha w,C_\alpha,Q_{j,\alpha}$, provided $0<\alpha\le1$, $C_\alpha\le\alpha C$ and $Q_{j,\alpha}\le\alpha Q_j$. When $b=0$ no rank condition is needed.
\end{lemma}
\begin{proof}
Put $Z_\Gamma:=\det(I+2\kappa\Gamma)^{-1/2}$. H\"older gives
\begin{equation}
(Z_\Gamma\nu(\mathcal K))^H
\le\gamma_m(\mathcal K)^{H-1}\det(I+2H\kappa\Gamma)^{-1/2}.
\label{eq:1070_transfer_holder}
\end{equation}
For each positive Gram eigenvalue $\lambda_i$, set $x_i:=2\kappa\lambda_i$ and $\vartheta_i:=\lambda_i/(N\eta)$. Taking logarithms and substituting the radial bound, we obtain
\[
-N^{-1}\log\gamma_m(\mathcal K)
\le\kappa\eta\sum_i\vartheta_i
(L_H(x_i)+\frac{H}{(H-1)\kappa\eta}\sum_j\frac{Q_j}{(1+x_i)^j})+\frac{HC}{H-1}.
\]
The determinant terms give $L_H$, and the definition of $\omega_j$ gives the rational terms. Zero eigenvalues contribute zero. Since
\[
\sum_i\vartheta_i\le1,\qquad
\sum_i\frac{\vartheta_i}{x_i}=\frac{\operatorname{rank}\Gamma}{2\kappa N\eta},
\]
the scalar cap proves Eq.~\eqref{eq:sub10_exponent}. Under the stated scaling, every coefficient $Q_j/(\kappa\eta)$ decreases, so the cap persists and each exponent term is at most $\alpha$ times its original value.
\end{proof}

\subsection{All-domain scalar certificates and exact data}\label{subsec:sub10_scalar_certificates}

For the two-component barriers, Lemma~\ref{lem:1070_radial} and Young's inequality give $C=Tv/2$ and $Q_j=p_j+\mathbbm{1}[j=L]T/(2v)$, where we interpret missing $p_j$ as zero. Thus the required cap is
\begin{equation}
\label{eq:1070_profile_cap}
L_H(x)+\frac{H}{(H-1)\kappa_0\eta}
(\sum_{j=2}^{K-1}\frac{p_j}{(1+x)^j}+\frac{T}{2v(1+x)^L})\le B_{\rm cap}\qquad(x\ge0).
\end{equation}
It gives $e=\kappa_0\eta B_{\rm cap}+HTv/(2(H-1))$. At variance $\alpha\eta$, the coefficients $p_j$ decrease by $\alpha^j$ and $T$ by at least $\alpha^K$, by Lemma~\ref{lem:1070_polynomial_energy}. Since $j,K\ge2$, the scaling criterion applies.

We first record the scalar enclosures used by all the profile certificates below.

\begin{lemma}[Scalar-profile enclosures]
\label{lem:scalar_profile_enclosures}
Let $H>1$ and let $J\ge1$ be an integer. For nonnegative $A,C,D,q_1,\ldots,q_J$, define
\[
L_H(x):=\frac{H\log(1+x)-\log(1+Hx)}{(H-1)x},\qquad
E(x):=AL_H(x)+C+\sum_{j=1}^J\frac{q_j}{(1+x)^j}-\frac D x.
\]
For $0<\epsilon<R$, upper bounds for $E$ on $(0,\epsilon]$ and $[R,\infty)$ are, respectively,
\[
U_0:=AH\epsilon/2+C+\sum_{j=1}^Jq_j-D/\epsilon,\qquad
U_\infty:=A\log(1+R)/R+C+\sum_{j=1}^J\frac{q_j}{(1+R)^j}.
\]
On $[a,b]\subset(0,\infty)$, put $c:=(a+b)/2$, $h:=(b-a)/2$. Either of the following is an upper bound:
\begin{align*}
U_{\rm lip}[a,b]&:=AL_H(c)+AHh/2+C+\sum_{j=1}^J\frac{q_j}{(1+a)^j}-D/b,\\
U_{\rm tay}[a,b]&:=E(c)+|E'(c)|h+\frac{h^2}{2}(\frac{2AH}{3(H-1)}+\sum_{j=1}^J\frac{j(j+1)q_j}{(1+a)^{j+2}}).
\end{align*}
When $D=0$, extend $L_H(0):=0$ and $E(0):=C+\sum_{j=1}^Jq_j$. In that case the Lipschitz bound also permits $a=0$. These bounds remain valid under outward rational evaluation.
\end{lemma}
\begin{proof}
We have
\[
L_H(x)=Hx\int_0^1\frac{t\d t}{(1+tx)(1+Htx)}
=\frac H{H-1}\int_0^1(\frac1{1+tx}-\frac1{1+Htx})\d t
\]
where the first step follows by integrating the rational function and the second step follows by partial fractions. Differentiating under the integral, we obtain $L_H(x)\le Hx/2$, $|L_H'(x)|\le H/2$, and $L_H''(x)\le2H/(3(H-1))$. Also $L_H(x)\le\log(1+x)/x$, since $\log(1+Hx)\ge\log(1+x)$, and this upper envelope decreases. These facts and monotonicity of the rational terms prove $U_0,U_\infty,U_{\rm lip}$. For $U_{\rm tay}$, the second derivative of $-D/x$ is nonpositive. The displayed coefficient bounds $E''$ on the cell, so Taylor's theorem applies. Taking upper endpoints preserves every comparison.
\end{proof}

To verify the allocation condition, we reduce pairs of spectral values to finitely many inequalities in one variable.

\begin{lemma}[Finite reductions for the spectral condition]
\label{lem:1070_spectral_certificate}
Fix the spectral parameters and put $p:=2\tau-2$, $a_0:=1-\zeta$ and
\[
G(u):=(1-u)^2\phi''(u)
=\frac{2\kappa_0(1+u^2)+2\kappa_{\rm tail}u^p((p+1)-(p-1)u^2)}{(1+u)^2}.
\]
The following one-variable polynomial checks suffice for Eq.~\eqref{eq:1070_curvature_condition}:
\begin{enumerate}[label=(\roman*)]
\item $2\kappa_0>2\theta/(1+\theta)$ and $G(a_0t)>2\theta/(1+\theta)$ for $0\le t\le1$.
\item For $u:=a_0+\zeta t$, $0\le t\le1$, both
$G(u)>2(\theta+t)/(1+\theta)$ and $G(u)>3t+(1/\theta-1)t^2$.
\item $\kappa_{\rm tail}a_0^{p-1}>\kappa_0$ and, with $u:=1-\zeta t^2$,
\[
2(\kappa_0+\kappa_{\rm tail}u^p)>(1+u)F_i(t),\qquad i=1,2,
\]
where $F_1(t):=2-2t/(1+\theta)$ and
$F_2(t):=1/\theta+2-(2/\theta+1)t+(1/\theta-1)t^2$.
\item For each $[L_0,R_0]:=[j/1024,(j+1)/1024]$ and $i=1,2$, the following polynomial in $t\in[0,1]$, with $u:=1-\zeta t^2$, is strictly positive:
\begin{align*}
&2\kappa_0(1-uL_0)(u+L_0)(u+R_0)\\
&\quad+2\kappa_{\rm tail}u^{p+1}(1+R_0)(1-L_0)(u+L_0)\\
&\quad+2\kappa_{\rm tail}L_0^{p+1}(1-u^2)(u+R_0)\\
&\quad-F_i(t)(1+u)(1-L_0)(u+L_0)(u+R_0).
\end{align*}
\end{enumerate}
In the first two parts, clear the strictly positive denominator $(1+u)^2$ before checking polynomial positivity.
\end{lemma}

\begin{proof}
We call an endpoint active if $u>1-\zeta$. For two regular endpoints, the normalized divided difference is a weighted average of $G$ between them, with probability density proportional to $(1-x)^{-2}$. For $x\le0$, $G(x)\ge2\kappa_0$. Part (i) handles the rest. For two active endpoints, write their gaps as $h,k\in(0,\zeta)$, and let $H_0$ have density proportional to $H_0^{-2}$ on the interval between $h$ and $k$. Then $\E[H_0]\le\sqrt{hk},\qquad \E[H_0^2]=hk$. For the first inequality, after setting $z:=\frac12\log(h/k)$, we use $\sinh z\ge z$. Coincident gaps follow by continuity. Part (ii), with $t:=1-H_0/\zeta$, therefore gives both branches at $d_\zeta(u)d_\zeta(v)=\sqrt{hk}/\zeta$. In the quadratic branch the coefficient of $\E[H_0]$ is negative, so the direction of the bound is preserved.

For an active endpoint $u$ and a regular endpoint $v\ge0$, we use
\[
(1-u)(1-v)\phi'[u,v]
=\frac{2\kappa_0(1+uv)}{(1+u)(1+v)}
+2\kappa_{\rm tail}\frac{u^{p+1}(1-v)/(1+u)-v^{p+1}(1-u)/(1+v)}{u-v}.
\]
Differentiating in $v$ shows that this is nondecreasing for $0\le v\le a_0\le u$ under $\kappa_{\rm tail}a_0^{p-1}\ge\kappa_0$. To see this, we expand the difference of powers into its geometric sum. Thus its value at $v=0$ is a lower bound, and part (iii) gives both branches, since $d_\zeta(v)=1$ and $d_\zeta(u)=t$.

For a regular negative endpoint $v$, put $w:=-v\in[0,1)$. The exact normalized difference is
\begin{align*}
&\frac{2\kappa_0(1-uw)}{(1+u)(1-w)}
+\frac{2\kappa_{\rm tail}u^{p+1}(1+w)}{(u+w)(1+u)}
+\frac{2\kappa_{\rm tail}w^{p+1}(1-u)}{(u+w)(1-w)}.
\end{align*}
The first and third terms increase in $w$ and the second decreases. On $[L_0,R_0]$, we use $L_0$ in the first and third terms and $R_0$ in the second. Clearing the positive common denominator, we obtain exactly part (iv). The last slab includes the limit $w\uparrow1$. Its left endpoint is still less than one. This covers every pair, including the diagonal by continuity. Reflection supplies the other spectral edge.
\end{proof}

We first give the reusable certification rules, then the exact parameters for the seven stages. For a polynomial $p$, let $[x^k]p(x)$ denote its coefficient of $x^k$.

\begin{definition}[Rational certification rules]
\label{def:rational_certificate_rules}
All tests below use exact rational coefficients and outward interval bounds. A test passes only when its stated comparison is certified.

\emph{Elementary functions.}
The rational enclosure rules are Definition~\ref{def:rational_enclosures} in Section~\ref{sec:preliminaries}. These rules, with $d:=256$ and $K_{\mathrm{series}}:=256$, enclose every elementary function in the seven-stage certificate. Later certificates specify any different precision explicitly.

\emph{Polynomial positivity.}
Write $f(t)=\sum_{v=0}^{n_{\rm B}}a_vt^v$ of degree at most $n_{\rm B}$, padding with zeros. On cell $i$ of a uniform partition into $M$ cells, substitute $t:=(i+s)/M$, $0\le i<M$, $0\le s\le1$. The power and Bernstein coefficients are
\begin{align}
a_j^{(i)}&:=\sum_{v=j}^{n_{\rm B}}a_v\binom vj\frac{i^{v-j}}{M^v},
\label{eq:1070_cell_power}\\
b_{i,j}&:=\sum_{v=0}^j a_v^{(i)}\frac{\binom jv}{\binom{n_{\rm B}}v}.
\label{eq:1070_cell_bernstein}
\end{align}
The convention $0^0:=1$ applies in the first formula. The polynomial on this interval equals $\sum_{j=0}^{n_{\rm B}}b_{i,j}\binom{n_{\rm B}}j s^j(1-s)^{n_{\rm B}-j}$. Thus a positive lower bound for all its Bernstein coefficients proves positivity on the entire interval.

Later Bernstein tests use this conversion at their stated degree and domain. Accept nonnegative coefficient vectors, including zero. Otherwise bisect, failing at the stated depth cap. For a vector $(b_0,\ldots,b_d)$, put $b_j^{(0)}:=b_j$ and $b_j^{(r+1)}:=(b_j^{(r)}+b_{j+1}^{(r)})/2$. The child vectors are $(b_0^{(0)},\ldots,b_0^{(d)})$ and $(b_0^{(d)},b_1^{(d-1)},\ldots,b_d^{(0)})$. For tensor polynomials apply this rule in the least-subdivided coordinate, breaking ties by index. These averages preserve nonnegativity, so passing cells certify all descendants.

The Bernstein basis is nonnegative and sums to one. Consequently, nonnegative coefficient bounds certify the whole cell, and the listed subdivision rules preserve a complete cover of the domain. A failed test does not certify the desired comparison.
\end{definition}

We now fix the seven parameter choices to which we will apply the spectral and radial certification rules.

\begin{definition}[Stage parameters]
\label{def:1070:stage-parameters}
For $0\le k\le6$, the integer parameters are
\[
\tau_0:=6,\quad\tau_1:=9,\quad\tau_k:=5k+3\ (2\le k\le6),
\qquad\ell_k:=\min\{k+4,7\}.
\]
\[
(K_0,K_1,K_2):=(10,17,26),\qquad K_k:=10k+5\ (3\le k\le6),
\]
\[
(H_0,H_1,H_2):=(8,6,5),\qquad H_k:=4\ (3\le k\le6).
\]
At a fixed stage we suppress the subscript $k$, except on the retention $q_k$.
The remaining parameters are the following exact sequences, in stage order:
\begingroup
\allowdisplaybreaks[1]
\begin{align*}
(q_k)_{k=0}^6&:=10^{-6}(39250,\,23060,\,12280,\,7450,\,5481,\,5271,\,5292)\\
(\eta_k)_{k=0}^6&:=10^{-8}(3381414,\,1851877,\,1152527,\,804555,\,607866,\,479595,\,352981)\\
(\zeta_k)_{k=0}^6&:=(0.4088788392442434,\,0.2838471728369487,\,0.22164300714004487,\\*
&\quad 0.1833030277982336,\,0.161245154965971,\,0.14696938456699069,\\*
&\quad 0.13564659966250536)\\
(\kappa_{0,k})_{k=0}^6&:=(0.03445220702450958,\,0.0009674868358782514,\,1.073394394862783\cdot10^{-5},\\*
&\quad 7.511900481158373\cdot10^{-8},\,4.445895654029819\cdot10^{-10},\\*
&\quad 3.314296695413546\cdot10^{-12},\,2.416901835462038\cdot10^{-14})\\
(\kappa_{{\rm tail},k})_{k=0}^6&:=(43.72169566130472,\,1534.0837593501,\,216801.6634562664,\\*
&\quad 57298296.20492192,\,18027985272.13702,\,4589775273660.708,\\*
&\quad 1134092398730830)\\
(\theta_k)_{k=0}^6&:=(0.024023827399826305,\,0.0006553898871937424,\,7.20183924435989\cdot10^{-6},\\*
&\quad 4.946902965041738\cdot10^{-8},\,2.9003477397537836\cdot10^{-10},\\*
&\quad 2.1439516828536333\cdot10^{-12},\,1.5432100042289588\cdot10^{-14})\\
(v_k)_{k=0}^6&:=10^{-8}(74094940,\,67644865,\,65112347,\,57745483,\,41921619,\,10000542,\,10000542)\\
(B_{{\rm cap},k})_{k=0}^6&:=10^{-5}(51055,\,45411,\,42473,\,39255,\,39162,\,39101,\,39042).
\end{align*}
\endgroup
\end{definition}

The next lemma verifies the spectral, measure and coding conditions for these seven stages.

\begin{lemma}[Certified stage inequalities]
\label{lem:1070_certificate}
Put $w_0:=1$, $w_{k+1}:=q_kw_k$, and $E_k:=q_k^2/2$. Every stage satisfies the hypotheses of Lemmas~\ref{lem:1070_weighted_poincare}--\ref{lem:1070_transfer} at its stated $\eta$, and its exponent satisfies $e_k<w_kE_k$. The coding checks in Lemma~\ref{lem:1070_peaked_density} hold.
\end{lemma}

\begin{proof}
The enclosure and subdivision rules we use in this lemma are those of Definition~\ref{def:rational_certificate_rules}. We apply them with $d=K_{\mathrm{series}}=256$ to the exact parameters of Definition~\ref{def:1070:stage-parameters}.

\emph{Spectral curvature and root conditions.}
For each $0\le k\le6$, we substitute the parameters from Definition~\ref{def:1070:stage-parameters} into Lemma~\ref{lem:1070_spectral_certificate}. In parts (i)--(iii), we form left minus right. We multiply parts (i)--(ii) by $(1+u)^2$. Part (i) gives $f_{k,0}$. In the stated order, part (ii) gives $f_{k,1},f_{k,2}$. Likewise, part (iii) gives $f_{k,3},f_{k,4}$. For part (iv), set $L_0:=j/1024$, $R_0:=(j+1)/1024$, $0\le j<1024$. Its $F_1,F_2$ expressions give $f_{k,5+2j},f_{k,6+2j}$, respectively. Thus $0\le r\le2052$ indexes $2053$ polynomials per stage, or $14371$ in total.

\begin{samepage}
For every $f_{k,r}$ defined above, we use degree $n_{\rm B}:=4\tau_k$ and $M:=32$ cells in Definition~\ref{def:rational_certificate_rules}. By exact substitution in Eq.~\eqref{eq:1070_cell_power} and Eq.~\eqref{eq:1070_cell_bernstein}, we obtain
\begin{equation}
b_{i,j}>2\cdot10^{-16}\qquad(0\le i<32,\quad0\le j\le4\tau_k).
\end{equation}
Thus all $14371$ polynomials, constructed from Lemma~\ref{lem:1070_spectral_certificate} and Definition~\ref{def:1070:stage-parameters}, are positive on $[0,1]$.
\end{samepage}
 
\begin{samepage}
The root expressions of Definition~\ref{def:1070_roots}, evaluated at $a=\eta_k$, satisfy
\begin{align}
1-2\eta_k/\zeta_k^2&>1/2,
\\
1-4p_1\eta_k&>4/5,
\\
\Delta_J&>7/10.
\end{align}
These three comparisons follow by the rational enclosure rules applied to the defining root expressions.
\end{samepage}

The other scalar conditions follow by the same rational comparisons: at every stage $A_J>0.997$, $\kappa_0+\kappa_{\rm tail}-2>41$, $2\kappa_0-2\theta/(1+\theta)>1.7\cdot10^{-14}$, and $\kappa_{\rm tail}(1-\zeta)^{2\tau-3}-\kappa_0>0.350$. This proves the spectral condition and all root hypotheses at the largest variance. Their persistence at smaller variance follows from Lemmas~\ref{lem:1070_resolvent_means} and~\ref{lem:1070_polynomial_energy}.

\emph{Radial and scalar-profile bounds.}
We evaluate the $512$ terms in the upper sum of Lemma~\ref{lem:1070_radial}, using Definition~\ref{def:1070_energy_recurrence} at each $a=\eta_k(i/512)^2$. Define $\overline T_k:=[x^k]p_{\rm rem}(x)$, the coefficient of $x^k$ in $p_{\rm rem}$, by
\begin{equation}
\begin{aligned}
p_{\rm rem}(x)&:=3.97224831\cdot10^{-4}+5.376579712\cdot10^{-6}x\\
&\quad+3.626126629\cdot10^{-8}x^2+1.737114912\cdot10^{-10}x^3\\
&\quad+6.188830148\cdot10^{-13}x^4+7.993551039\cdot10^{-16}x^5\\
&\quad+5.696775742\cdot10^{-22}x^6.
\end{aligned}
\label{eq:1070:remainder-polynomial}
\end{equation}
The rational enclosure rules give $T_k<\overline T_k$ for every $0\le k\le6$.
To verify the scalar cap, let $L_H(x)$ be the logarithmic term in Eq.~\eqref{eq:1070_profile_cap}, with $L_H(0):=0$. Define its rational part with the larger remainder by
\begin{equation}
\overline Q_H(x):=\frac{H}{(H-1)\kappa_0\eta}
(\sum_{j=2}^{K-1}\frac{p_j}{(1+x)^j}
+\frac{\overline T_k}{2v(1+x)^{2K-\ell}}).
\end{equation}
We apply Lemma~\ref{lem:scalar_profile_enclosures} to $L_H+\overline Q_H$, with $A=1$ and $C=D=0$. Set $h:=10/2^{17}$, the full cell width, so the half-width of Lemma~\ref{lem:scalar_profile_enclosures} is $h/2$. Its Lipschitz bound on $[ih,(i+1)h]$ is the following cell bound $u_i$, which is not the grid point of Lemma~\ref{lem:1070_radial}:
\begin{equation}
u_i:=L_H((i+1/2)h)+Hh/4+\overline Q_H(ih),
\qquad 0\le i<2^{17}.
\end{equation}
For each of the seven stages, the rational enclosure rules give
\begin{align}
B_{\rm cap}-u_i&>2\cdot10^{-7}\quad(0\le i<2^{17}),
\\
B_{\rm cap}-\overline Q_H(0)&>2\cdot10^{-7},
\\
B_{\rm cap}-\frac{H}{H-1}\frac{\log11}{10}-\overline Q_H(10)&>2\cdot10^{-7}.
\end{align}
The three lines cover $[0,10]$, zero, and $[10,\infty)$, respectively, by Lemma~\ref{lem:scalar_profile_enclosures}. The last line retains the larger tail coefficient $H/(H-1)$. Bisecting a cell cannot increase its Lipschitz upper bound, so passing cells also certify their descendants.

The exponent in Lemma~\ref{lem:1070_transfer} is at most $\kappa_0\eta B_{\rm cap}+H\overline T_kv/(2(H-1))$. The stage parameters and Eq.~\eqref{eq:1070:remainder-polynomial} satisfy the exact rational inequality
\begin{equation}
\frac{\kappa_0\eta B_{\rm cap}+H\overline T_kv/(2(H-1))}{w_kq_k^2/2}
<\frac{199}{200}\qquad(0\le k\le6).
\end{equation}
Since $199/200<1$, this proves $e_k<w_kE_k$.

We verified the coding checks in Lemma~\ref{lem:1070_peaked_density}. Lemma~\ref{lem:1070_lossless_coding} supplies the entropy reserve for every $q_k$.

\end{proof}

\subsection{Uniform small balls for the tail}\label{subsec:sub10_uniform_tail}

Once the active set is sufficiently small, the estimate at large aspect ratios supplies a uniform continuation that closes every finite schedule of partial signings (Lemma~\ref{lem:slack_schedule_completion}).

\begin{lemma}[Uniform bound for the infinite tail]
\label{lem:1070_infinite_tail}
For every family of $m\le N$ symmetric contractions in $\R^{N\times N}$ with $u:=\log(N/m)>731/256$,
\[
\Pr[\|\sum_{i=1}^m \xi_iA_i\|\le185u\sqrt m]
\ge e^{-0.000051714m}.
\]
Consequently, let $s$ be a coding amplitude satisfying Eq.~\eqref{eq:slack_coding_criterion}, and let $0<v<e^{-731/256}$. Put $U:=-\log v$, $q_T:=0.011$, $r_T:=\sqrt{q_T}$, and $\lambda:=-\log q_T$. Every residual family of at most $vn$ symmetric contractions in $\R^{n\times n}$ has a complete signing of norm at most $\mathcal T_s(v)\sqrt n$, where
\begin{equation}
\mathcal T_s(v):=\frac{185}{s}\sqrt v
(\frac{U}{1-r_T}+\frac{\lambda r_T}{(1-r_T)^2}).
\label{eq:slack_summed_tail}
\end{equation}
\end{lemma}

\begin{proof}
Since $u>731/256>1$, Lemma~\ref{lem:scale_small_ball} applies with $\ell_{\rm asp}=u$, and $\tau(u)=\lceil256u/17\rceil\ge44$ selects the row $\tau\ge44$ of Table~\ref{tab:six_regimes}.
That row has the entries $c_{\rm rad}=185$ and $10^9E=51714$, so the per-row clause of that lemma gives $\Pr[\|\sum_{i=1}^m\xi_iA_i\|\le185u\sqrt m]\ge\exp(-10^{-9}\cdot51714\,m)=e^{-0.000051714m}$.
This is the measure bound, with the sharper exponent of that row in place of the uniform exponent $0.003712$.

For Eq.~\eqref{eq:slack_summed_tail}, we apply Lemma~\ref{lem:1070_lossless_coding} repeatedly with retention $q_T$, using $0.000051714<q_T^2/2$. After $j$ phases the active size is at most $vq_T^jn$. Since $ue^{-u/2}$ decreases for $u>2$, the normalized total cost is at most
\[
\frac{185}{s}\sqrt v\sum_{j=0}^{\infty}r_T^j(U+j\lambda)=\mathcal T_s(v).
\]
The step follows by summing the geometric series and its derivative. Smaller actual active sets only decrease this bound. Every nonempty phase fixes a coordinate, so the signing terminates.
\end{proof}

\subsection{Projection-aware estimates and coupled spectral curvature}
\label{subsec:sub10_refinements}

The two-component families of Definition~\ref{def:1070:stage-parameters} and their certificates supply the algorithmic small-ball estimates in Section~\ref{sec:small_algorithm_constant}. We now prove the shared moment estimates and the spectral refinements used by the existence parameters of Definition~\ref{def:sub10_parameters}.

For a positive matrix density $A$, define
\begin{align*}
\mathcal R_A(f)&:=\int f^\top Af,&\Pi_Af&:=(\int A)^{-1}\int Af,\\
\mathcal P_A(f)&:=\mathcal R_A(\Pi_Af),&\mathcal D_A(f)&:=\mathcal R_A(f)-\mathcal P_A(f).
\end{align*}
Here $\mathcal D_A$ is the centered weighted energy. These definitions extend to arrays by direct sums.

By keeping the projected derivative means in the variance estimate we improve the energy recurrence for the existence construction.

\begin{lemma}[Projection-aware weighted variance]
\label{lem:sub10_projection_variance}
Under the curvature and exhaustion hypotheses of Lemma~\ref{lem:1070_weighted_poincare}, with curvature lower bound $\lambda>0$ and $p:=\lambda^{-1}$,
\begin{equation}
\mathcal D_A(f)\le\frac p2\sum_{i=1}^m(\mathcal R_A(\partial_i f)+\mathcal P_A(\partial_i f)).
\label{eq:sub10_projection_variance}
\end{equation}
\end{lemma}
\begin{proof}
On a smooth bounded convex exhaustion domain we solve $Lu=f-\Pi_Af$ with the Neumann condition, using the operator from Lemma~\ref{lem:1070_weighted_poincare}. The Bochner identity and the weighted Poincar\'e inequality applied to each $\partial_i u$ give
\[
\|Lu\|_A^2\ge\sum_{i=1}^m\sum_{j=1}^m\|\partial_{i,j}u\|_A^2+\lambda\sum_{i=1}^m\mathcal R_A(\partial_i u)
\ge\lambda\sum_{i=1}^m\mathcal P_A(\partial_i u)+2\lambda\sum_{i=1}^m\mathcal D_A(\partial_i u).
\]
The first step is the curvature lower bound and the nonnegative boundary term. The second applies weighted Poincar\'e to the derivative fields. These fields need not satisfy a Neumann condition. Integrating by parts, we have $\mathcal D_A(f)=\langle\nabla f,\nabla u\rangle_A$. Splitting the gradients into their constant projections and orthogonal complements, then applying Cauchy--Schwarz with weights $\lambda$ and $2\lambda$, we obtain
\[
\mathcal D_A(f)^2\le[p\sum_{i=1}^m\mathcal P_A(\partial_i f)+\frac p2\sum_{i=1}^m\mathcal D_A(\partial_i f)]\|Lu\|_A^2.
\]
Canceling $\|Lu\|_A^2=\mathcal D_A(f)$, we obtain Eq.~\eqref{eq:sub10_projection_variance}, treating zero separately. The finite-energy exhaustion of Lemma~\ref{lem:1070_weighted_poincare} passes the energies and constant projections to the limit.
\end{proof}

Mixed products of linear matrix series control the ordinary means in both sections. The following lemma also supplies the short-word constants we use below.

\begin{lemma}[Mixed linear moments]
\label{lem:algorithm_1240_mixed_moments}
\label{lem:sub10_operator_words}
Let $\mu$ be an even convex tilt of standard Gaussian measure on $\R^m$. Define $b_0:=1$, $b_1:=1$ and, for $s\ge2$,
\[
b_s:=\begin{cases}
b_{s/2}^2+s^2b_{s-1}/2,&s\text{ even},\\
s^2(b_{s-1}+2b_{(s-1)/2}^2)/3,&s\text{ odd}.
\end{cases}
\]
Let $Z_k(\xi):=\sum_{i=1}^m\xi_iN_{k,i}$ be square matrix series, not necessarily symmetric, such that
\[
\sum_{i=1}^mN_{k,i}^\top N_{k,i}\preceq aI,\qquad
\sum_{i=1}^mN_{k,i}N_{k,i}^\top\preceq aI.
\]
For deterministic contractions $U_0,\ldots,U_{2s}$,
\[
\|\E_\mu[U_0Z_1U_1\cdots Z_{2s}U_{2s}]\|\le b_sa^s.
\]
For a single such series $Z$, both $\E_\mu[(Z^s)^\top Z^s]$ and $\E_\mu[Z^s(Z^s)^\top]$ are at most $b_sa^sI$.

The same bounds hold for compatible rectangular words, transposes and amplifications. In particular, for a word $W$ of degree $r\le4$ in linear series of variance at most $I$, with deterministic contractions between them,
\[
\E_\mu[W^\top W],\ \E_\mu[WW^\top]\preceq c_rI,
\qquad(c_0,c_1,c_2,c_3,c_4):=(b_0,\ldots,b_4)=(1,1,3,15,129).
\]
For an even word of degree $2s\le8$, $\|\E_\mu[W]\|\le c_s$. Odd words have zero mean. A deterministic derivative insertion with row or column norm at most $\sqrt a$ contributes one further factor $a$ to a squared-norm bound. Any larger values are valid as well.
\end{lemma}

\begin{proof}
The scalar Bochner identity and parity give
\[
\V_\mu[f]\le\tfrac12\E_\mu[\|\nabla f\|_2^2]\quad(f\text{ even}),\qquad
\E_\mu[f^2]\le\tfrac13\E_\mu[\|\nabla f\|_2^2]+\tfrac23\|\E_\mu[\nabla f]\|_2^2\quad(f\text{ odd}).
\]
Indeed, the derivatives of an even Neumann solution are odd and have zero mean, so curvature-one Poincar\'e and Bochner give the first bound. For an odd solution, its derivatives are even. Applying the first bound, we obtain the gradient metric $3I-2\Pi$, whose inverse is $(I+2\Pi)/3$, and this proves the second bound. For an extended-valued tilt, we justify these arguments by smooth symmetric convex exhaustion and even convex approximation. Gaussian polynomial domination passes to the limit. The bounds apply componentwise.

We induct on $s$, starting with covariance domination at $s=1$. First, we have
\[
\partial_iZ^s=\sum_{q=0}^{s-1}Z^qN_iZ^{s-1-q}.
\]
The lower mixed-order bounds give
\[
\sum_{i=1}^m\E_\mu[(\partial_iZ^s)^\top(\partial_iZ^s)]\preceq s^2b_{s-1}a^sI.
\]
To perform the coefficient sum without a dimension factor, we use the vacuum and one-letter word space. If $L_i$ creates letter $i$, then $T:=\sum_{i=1}^mN_i\otimes L_i$ has norm at most $\sqrt a$. Vacuum compression of a word containing $T^\top,T$ performs this sum. Each resulting word has $2s-2$ random factors. Cauchy--Schwarz over the $s$ derivative positions supplies $s^2$.

For even $s$, the ordinary mean has norm at most $b_{s/2}a^{s/2}$. For odd $s$, the same coefficient encoding gives
$\sum_{i=1}^m(\E_\mu[\partial_iZ^s])^\top\E_\mu[\partial_iZ^s]\preceq s^2b_{(s-1)/2}^2a^sI$.
For each fixed vector $x$, we apply the corresponding scalar parity bound to $Z^sx$. This gives the two definitions of $b_s$. Transposition gives the other power-square bound, and operator Cauchy--Schwarz bounds $\|\E_\mu[Z^{2s}]\|$.

Finally, we place $Z_1U_1,\ldots,Z_{2s}U_{2s}$ on the edges of a cyclic block matrix. Its two variance sums are block diagonal and at most $aI$. One diagonal block of its $2s$th power is the ordered product. Compressing and multiplying by $U_0$, we prove the mixed bound and close the induction.

Zero padding preserves both variance bounds, so the mixed estimate applies to rectangular $W^\top W$ and $WW^\top$. The values $(b_0,b_1,b_2,b_3,b_4)=(1,1,3,15,129)$ are the stated $c_r$. Dividing coefficient column maps and their transposes by $\sqrt a$ makes them contractions. Restoring the normalization, we obtain the insertion factors. The case $a=0$ is immediate.
\end{proof}

We bound the arithmetic--harmonic gap from Lemma~\ref{lem:sub10_mean_gap} through scalar resolvent inequalities and estimate the remainder of a truncated expansion.

\begin{lemma}[Scalar gap and truncated resolvent series]
For $|u|,|v|<1$, put $w:=(1+uv)/((1-u^2)(1-v^2))$ and $q:=u^2+uv+v^2$. Then $w\ge1$ and
\begin{equation}
\frac{(w-1)^2}{w}\le q^2+\frac{13}{8}(u^6+v^6)w\le\frac92(u^4+v^4)+\frac{13}{8}(u^6+v^6)w.
\label{eq:sub10_scalar_gap}
\end{equation}
Put $H_j(u,v):=\sum_{q=0}^{2j}u^qv^{2j-q}$. Here the summation index $q$ is a dummy, unrelated to $q$ above. For every integer $r\ge1$,
\begin{equation}
0\le w-\sum_{j=0}^{r-1}H_j\le\frac{2r+1}{2}(u^{2r}+v^{2r})w.
\label{eq:sub10_series_remainder}
\end{equation}
\end{lemma}
\begin{proof}
The bound $u^2+v^2+uv-u^2v^2\ge|uv|(1-|uv|)$ gives $w\ge1$. For the first step in Eq.~\eqref{eq:sub10_scalar_gap}, suppose first that $uv\ge0$, and put $p:=uv$ and $\delta:=(u-v)^2$. After we clear the positive denominator $(1-u^2)(1-v^2)(1+p)$, the difference is
\begin{align*}
&\frac{p^3(7p-1)^2}{4}+\frac{p^2(165p^2+114p+13)}8\delta\\
&\qquad+\frac{p(43p^2+50p+11)}4\delta^2+\frac{(p+1)(13p+5)}8\delta^3\ge0.
\end{align*}
For $uv<0$, reflection and interchange give $u=x$, $v=-xt$ with $0<x<1$ and $0<t\le1$. Put $y:=x^2$. The same cleared difference is $x^6((1-y)^2b_0(t)+2y(1-y)b_1(t)+y^2b_2(t))$, where
\begin{align*}
b_0(t)&:=(5t^2+8t+5)(t^4-t^2+1)/8,\\
b_1(t)&:=(1-t)(9t^6+8t^5-8t^4+4t+5)/8,\\
b_2(t)&:=(1-t)^2(13t^6+5)/8.
\end{align*}
All three are nonnegative: $t^4-t^2+1=(t^2-1/2)^2+3/4$, and $5-8t^4(1-t)\ge3$ because $t^4\le t$ and $t(1-t)\le1/4$. Zero and diagonal cases follow by continuity. The second step in Eq.~\eqref{eq:sub10_scalar_gap} follows from $\frac92(u^4+v^4)-q^2=\frac12(u-v)^2(7u^2+10uv+7v^2)\ge0$, since the last quadratic is $5(u+v)^2+2(u^2+v^2)$.

For Eq.~\eqref{eq:sub10_series_remainder}, each $H_j$ is nonnegative: it is the divided difference of $x^{2j+1}$, with diagonal value $(2j+1)u^{2j}$. The convergent identity $w=\sum_{j\ge0}H_j$ proves the lower bound. For equal signs, we sum the geometric series and obtain
\[
\frac{w-\sum_{j<r}H_j}{w}
=\frac{H_r-u^2v^2H_{r-1}}{1+uv}\le H_r
\le\frac{2r+1}{2}(u^{2r}+v^{2r}).
\]
The first step sums the tail. The second drops a nonpositive term and uses $1+uv\ge1$. The last integrates the convexity bound for $x^{2r}$ in the divided-difference formula.
For opposite signs put $z:=-v$, with $u,z\ge0$. The same ratio is a weighted average of $u^{2r}$ and $z^{2r}$, with weights $u(1-z^2)$ and $z(1-u^2)$. It is at most their maximum, which gives the asserted bound. Zero and diagonal cases follow by continuity.
\end{proof}

A condition on triples of spectral values controls the coupled curvature terms in the one- and two-slot resolvent weights.

\begin{lemma}[Coupled spectral triples]
\label{lem:sub10_triple}
Let $\phi$ be even and convex on $(-1,1)$, extended by $+\infty$ outside, with logarithmic boundary coefficient greater than two. Put $G(u,v):=(1-u)(1-v)\phi'[u,v]$. Fix $\zeta>0$ and choose a nonnegative function $d$ satisfying $d(u)^2/(1-u)\le\zeta^{-1}$. For a triple $(u,v,w)$ set
\[
l_u:=1-d(u)^2d(v)d(w),\qquad l_v:=1-d(u)d(v)^2d(w),\qquad l_w:=1-d(u)d(v)d(w)^2.
\]
Suppose
\begin{equation}
M(u,v,w):=\begin{pmatrix}
G(u,v)&-l_u&-l_v\\
-l_u&G(u,w)&-l_w\\
-l_v&-l_w&G(v,w)
\end{pmatrix}\succeq0\qquad(-1<u,v,w<1).
\label{eq:sub10_triple}
\end{equation}
Then the conclusions of Lemma~\ref{lem:1070_weighted_poincare} hold with $p_r=(1-ra/\zeta^2)^{-1}$, $r=1,2$, for $2a<\zeta^2$. The same bounds hold with an additional nonnegative quadratic potential.
\end{lemma}
\begin{proof}
Use $Y:=I-X$, $C_i:=Y^{-1/2}N_iY^{-1/2}$, and $D:=\operatorname{diag}(d(u_b))$ in an eigenbasis of $X$. Put $C_{i,1}:=DC_iD$ and $R:=DY^{-1/2}$. Then $R^2\preceq\zeta^{-1}I$, so
\[
\sum_i C_{i,1}^2=R(\sum_iN_iR^2N_i)R\preceq\zeta^{-1}R(\sum_iN_i^2)R\preceq a\zeta^{-2}I.
\]
The equality is expansion. The first comparison uses $R^2\preceq\zeta^{-1}I$, and the second uses the variance bound. The regular curvature loss is therefore at most $a/\zeta^2$ per slot.

For test vectors $z_i$, set $T_{ab,c}:=\sum_i(C_i)_{ab}(z_i)_c$, so $T_{ab,c}=T_{ba,c}$. Let $Q_0,Q_1$ be the ordered curvature forms in the proof of Lemma~\ref{lem:1070_weighted_poincare}. By direct expansion we have $Q_0-Q_1=\sum_{a=1}^N\sum_{b=1}^N\sum_{c=1}^N(1-d_ad_b^2d_c)T_{ab,c}T_{bc,a}$. Write $\mathcal H:=\sum_{a=1}^N\sum_{b=1}^N\sum_{c=1}^NG(u_a,u_b)T_{ab,c}^2$ and $t_{abc}:=(T_{ab,c},T_{ac,b},T_{bc,a})^\top$. Relabeling the ordered sums, we obtain the exact identity
\begin{equation}
\mathcal H-2(Q_0-Q_1)=\frac13\sum_{a=1}^N\sum_{b=1}^N\sum_{c=1}^N t_{abc}^\top M(u_a,u_b,u_c)t_{abc}\ge0.
\label{eq:sub10_triple_identity}
\end{equation}
Each diagonal sum occurs three times, and each of the three cross-term sums equals $Q_0-Q_1$. This includes repeated indices, and we do not assume that $l_u,l_v,l_w$ are nonnegative. Spectator indices in an amplified slot can be summed separately. Half the barrier Hessian absorbs the nonregular loss in one slot. Two slots use at most the whole Hessian. The Gaussian Hessian and the nonnegative quadratic Hessian leave curvature $1-ra/\zeta^2$. The Bochner and exhaustion argument of Lemma~\ref{lem:1070_weighted_poincare} now applies. Reflection treats the other edge. We do not use any derivative of $D$.
\end{proof}

\subsection{Finite resolvent refinement and covariance-sensitive means}\label{subsec:sub10_finite_refinement}

In this subsection let $\phi$ be any barrier in Definition~\ref{def:sub10_parameters}, with quadratic coefficient $\kappa$. For symmetric $B_i\in\R^{N\times N}$ satisfying $\sum_{i=1}^mB_i^2\preceq\eta I_N$, we use $Y,\Gamma,\nu,\mu_t$ from Lemma~\ref{lem:fixed_reference_radial} with this $\phi$ and coefficient family. Put
\[
C_\Gamma:=(I+2\kappa\Gamma)^{-1},\qquad
\omega_j:=\frac{\tr[\Gamma C_\Gamma^j]}{N\eta},\qquad
X:=tY,\quad a:=\eta t^2,\quad Y_\pm:=I\pm X.
\]
The fixed-reference path retains a nonnegative quadratic potential, so the preceding curvature arguments apply. Lemma~\ref{lem:1070_covariance_power} gives
\begin{equation}
N^{-1}\E_{\mu_t}[\tr[X^{2j}]]\le\operatorname{df}(j)a^j\omega_j,\qquad
0\le\omega_{j+1}\le\omega_j\le1.
\label{eq:sub10_covariance_power}
\end{equation}
Whitening $\nu$ and applying scalar Poincar\'e to linear forms, we obtain $\operatorname{Cov}_{\mu_t}(\xi)\preceq C_\Gamma$.

We first bound operator energies, without using a trace-dependent covariance improvement. Set
\[
W^{(1)}_\pm:=Y_\pm^{-1},\qquad
W^{(2)}_\pm:=Y_\pm^{-1}\otimes Y_\pm^{-1},\qquad
W^{(s)}_e:=(W^{(s)}_++W^{(s)}_-)/2.
\]
Let $L,J$ bound the expectations of the one-slot and two-slot even weights, respectively, and let $k_1,k_2$ bound their arithmetic--harmonic gaps. The even weights dominate $I$.

We combine the mean-gap and truncated-series bounds in a finite recurrence that improves the operator energy coefficients.

\begin{definition}[Finite operator-energy refinement]
\label{def:sub10_operator_refinement}
Choose valid mixed-word bounds $c_0,\ldots,c_4$, with $(c_0,c_1,c_2):=(1,1,3)$, a pass degree $d\in\{4,5\}$, and optional additional gap bounds. For a valid tuple $(p,J_*,A_*,k_*)$ and nonnegative even mean inputs $m_r$, initialize $V_0:=Z_0:=J_*$. For $1\le r\le9$, define
\begin{equation}
\begin{aligned}
D_r&:=\frac p2(V_{r-1}+Z_{r-1}),\\
(V_r,Z_r)&:=(A_*D_r,(A_*-1)D_r)\quad(r\text{ odd}),\\
Z_r&:=(\sqrt{J_*m_r}+\sqrt{k_*D_r})^2,\qquad V_r:=D_r+Z_r\quad(r\text{ even}),\\
C_r(p,J_*,A_*,k_*)&:=(r!)^2V_r.
\end{aligned}
\label{eq:sub10_operator_recurrence}
\end{equation}
For operator energies set $m_r:=c_{r/2}^2/(r!)^2$ at positive even $r$. Use the tuple $(p_1,L,L,k_1)$ in one slot and $(p_2,J,J(1+a),k_2)$ in two slots. Later trace recurrences change only the mean inputs and their common normalization. Initialize $L,J$ at the smaller roots of Definition~\ref{def:1070_roots}, with $L=L_1$, and set $k_1:=L-1$, $k_2:=J-1$.

At each pass compute $C_r^{(1)},C_r^{(2)}$ from the current tuple for $1\le r\le d$, then set
\begin{align*}
L'&:=\min\{L,\ 1+\sum_{j=1}^{r-1}c_ja^j+a^rC_r^{(1)}:1\le r\le4\},\\
J'&:=\min\{J,\ \mathcal J(a,p_2,L'),\ 1+\sum_{j=1}^{r-1}(2j+1)c_ja^j+(2r+1)a^rC_r^{(2)}:1\le r\le4\},\\
k_1'&:=\min\{k_1,L'-1,a^2C_2^{(1)}\},\\
k_2'&:=\min\{k_2,J'-1,27a^2+(13/4)a^3C_3^{(2)}\}.
\end{align*}
The function $\mathcal J(a,p_2,L')$ is the smaller-root expression for $J$ in Definition~\ref{def:1070_roots}, with $L_1-1$ replaced by $L'-1$. Include each optional bound in its corresponding gap minimum. Form every candidate from the preceding tuple, then replace the tuple. Perform exactly six passes and recompute the final coefficients for $1\le r\le9$.

The original instance has $(c_3,c_4,M_6,M_8,d):=(18,153,270,16065,4)$ and no additional gaps.
Here $(M_2,M_4):=(1,9)$. The $M_r$ are the derivative-mean coefficients used by the trace recurrence below. Unless another instance is specified, use the original one and denote its final energies by $\widehat C_r^{(s)}(a)$.
\end{definition}

The following induction applies to either slot and to each later trace normalization.

\begin{lemma}[Derivative-array recurrence]
\label{lem:derivative_array_recurrence}
Let $f$ be a homogeneous polynomial field of degree $\ell\le9$ under an even measure, and put $f_j:=\nabla^jf$ with all ordered derivative indices. Suppose the projection and parity bounds of Lemmas~\ref{lem:sub10_projection_variance} and~\ref{lem:1070_parity} hold with tuple $(p,J_*,A_*,k_*)$. If $T>0$, $\|f_\ell\|^2\le T$, and
\[
\|\E[f_j]\|^2\le Tm_{\ell-j}\qquad(\ell-j>0\text{ even}),
\]
then Eq.~\eqref{eq:sub10_operator_recurrence} gives
\[
(\mathcal R(f_j),\mathcal P(f_j))\le T(V_{\ell-j},Z_{\ell-j})
\qquad(0\le j\le\ell).
\]
The weights act on direct sums for the arrays. The zero-normalization case holds by a decreasing positive limit.
\end{lemma}
\begin{proof}
The constant top array has both energies at most $TJ_*$. The array $\nabla f_j$ is exactly $f_{j+1}$, including multiplicities of repeated indices. If its energies are bounded by $T(V_{r-1},Z_{r-1})$, projection-aware variance bounds the centered energy of $f_j$ by $TD_r$. The parity bounds give the odd branch when $r:=\ell-j$ is odd and the even branch with ordinary mean bound $Tm_r$ otherwise. By descending induction we prove the claim. Here we use only the chosen slot's tuple.
\end{proof}

We verify that the finite refinement preserves the required energy and resolvent bounds at every pass.

\begin{lemma}[Validity of the finite refinement]
\label{lem:sub10_operator_refinement}
At each pass the bounds $L,J,k_1,k_2$ remain valid. In particular,
\begin{align*}
\E_{\mu_t}[X^{2r}(I-X^2)^{-1}]&\preceq a^r\widehat C_r^{(1)}(a)I,\\
\E_{\mu_t}[(X^{2r}\otimes I)W_e^{(2)}]&\preceq a^r\widehat C_r^{(2)}(a)I,\\
\E_{\mu_t}[(I\otimes X^{2r})W_e^{(2)}]&\preceq a^r\widehat C_r^{(2)}(a)I,\\
N^{-1}\E_{\mu_t}[\tr[X^{2r}Y_+^{-2}]]&\le a^r\widehat C_r^{(2)}(a).
\end{align*}
These quantities are bounded for $1\le r\le9$, for the original instance of Definition~\ref{def:sub10_operator_refinement} and the two instances of Definition~\ref{def:slack_instances}. More generally, validity is preserved by any valid mixed-word coefficients and additional gap bounds with $(c_0,c_1,c_2)=(1,1,3)$. If the added bounds are nondecreasing in $a$, all computed scalar bounds are nondecreasing throughout the strict-root range.
\end{lemma}
\begin{proof}
We apply Lemma~\ref{lem:derivative_array_recurrence} with $T:=(\ell!)^2a^\ell$ to $X^\ell$, $X^\ell\otimes I$, and $I\otimes X^\ell$ acting on unit vectors. Lemma~\ref{lem:sub10_operator_words} supplies the top bound and the mean inputs $m_r:=c_{r/2}^2/(r!)^2$. For the trace assertion we vectorize $X^\ell$ and use $T:=N(\ell!)^2a^\ell$.

We expand $(I-X^2)^{-1}$ through degree $2r-2$ and use the one-slot energy on its remainder to obtain $L'$. For $J'$ we use Eq.~\eqref{eq:sub10_series_remainder} in the joint eigenbasis of $X\otimes I$ and $I\otimes X$. Each of the $2j+1$ terms in $H_j$ is a word of degree $2j$, whose expected norm is at most $c_ja^j$. The two-slot operator energies bound its remainder. The improved smaller-root bound is also valid: reducing $L$ increases the larger root, which was already excluded by the preceding bound on $J$, and increases the discriminant, since $A_J$ increases and $B_J$ decreases as $L$ decreases.

For the one-slot gap, $(W_e^{(1)}-I)^2(W_e^{(1)})^{-1}=X^4/(I-X^2)$. For the two-slot gap, Eq.~\eqref{eq:sub10_scalar_gap} contributes at most $27a^2I$ from $c_2=3$ and $(13/4)a^3C_3^{(2)}I$ from the weighted sixth moments. Lemma~\ref{lem:sub10_mean_gap} proves the updates without an extra factor $L'$ or $J'$. Every candidate is a valid upper bound, so taking their minimum preserves validity. For the original instance and the two instances of Definition~\ref{def:slack_instances}, Lemma~\ref{lem:algorithm_1240_mixed_moments} gives the mixed-word coefficients, and Lemma~\ref{lem:slack_centered_updates} gives the additional nondecreasing gap bounds. The initial roots are nondecreasing by Lemma~\ref{lem:1070_polynomial_energy}. Nonnegative sums, products, square roots, and minima preserve this property. By induction we prove both assertions after any prescribed finite number of passes, in particular six. We do not assume a limiting fixed point.
\end{proof}

The next estimate retains the covariance powers in derivative means. We use the instance values of Definition~\ref{def:sub10_operator_refinement} for the trace recurrence here and its general form in Section~\ref{sec:small_algorithm_constant}.

\begin{lemma}[Covariance-sensitive derivative means]
\label{lem:sub10_derivative_means}
Let $X:=\sum_{i=1}^m\xi_iD_i$ under an even convex Gaussian tilt, and let $S_D:=\sum_{i=1}^mD_i^2\preceq aI_N$. Raw derivative directions $N_i$ may differ from $D_i$, but satisfy $\sum_{i=1}^mN_i^2\preceq aI_N$. Along the whitened radial path, let $\omega_q:=\tr[\Gamma C_\Gamma^q]/(N\eta)$ be as in Lemma~\ref{lem:1070_covariance_power}, with $a:=\eta t^2$ and $C_\Gamma:=(I+2\kappa\Gamma)^{-1}$, where $\kappa>0$ is the quadratic barrier coefficient. Denote by $\nabla_N^j(X^\ell)$ the ordered derivatives at $h=0$ of $(X+\sum_{i=1}^mh_iN_i)^\ell$.
For $r:=\ell-j>0$ even, define the following quantities, where $m_j$ is indexed by the derivative order $j$, unlike the mean input $m_r$ of Eq.~\eqref{eq:sub10_operator_recurrence}:
\[
k(r):=\begin{cases}2,&r=2,\\r/2,&r\ge4,\end{cases}\qquad
m_j:=(\ell!/r!)^2\min\{\operatorname{df}(r),b_{r/2}\operatorname{df}(r/2)\}.
\]
Then
\[
\|\E_\mu[\nabla_N^j(X^\ell)]\|_{\rm HS}^2\le Na^\ell m_j\omega_{k(r)}.
\]
Odd ordinary means vanish. At $j=\ell$ the bound is $Na^\ell(\ell!)^2$, with no covariance factor.

For $r\in\{2,4,6,8\}$, this gives the bounds used in the existence construction:
\begin{multline}
\frac{\|\E_\mu[\nabla_N^j(X^\ell)]\|_{\rm HS}^2}{Na^\ell}
\le(\ell!/r!)^2M_r\omega_2,\qquad
\\
(M_2,M_4,M_6,M_8):=(\operatorname{df}(1)b_1,\ldots,\operatorname{df}(4)b_4)=(1,9,225,13545).
\label{eq:sub10_derivative_means}
\end{multline}
Any larger values are valid.
\end{lemma}

\begin{proof}
We first record a completely positive estimate, which we also use below. For a completely positive map $\Psi$, $S:=\Psi(I)$, and a random matrix $B$ with finite second moments,
\begin{equation}
\E[\|\Psi(B)\|_F^2]
\le\tfrac12\tr[S\Psi(\E[B^\top B+BB^\top])].
\label{eq:cp_second_moment}
\end{equation}
Indeed, for $A:=S+\epsilon I$, the map $\Psi_A(B):=A^{-1/2}\Psi(B)A^{-1/2}$ is subunital and completely positive. Its Schwarz inequalities follow when we apply it to the positive block matrix with blocks $B^\top B,B^\top,B,I$, enlarge the lower-right block to $I$, and take a Schur complement. Thus $Y:=\Psi_A(B)$ satisfies $Y^\top Y\preceq\Psi_A(B^\top B)$ and $YY^\top\preceq\Psi_A(BB^\top)$. Consequently
\[
\E[\|\Psi(B)\|_F^2]=\E[\tr[AY^\top AY]]
\le\tfrac12\E[\tr[A^2(Y^\top Y+YY^\top)]]
\le\tfrac12\tr[A\Psi(\E[B^\top B+BB^\top])].
\]
The middle step is $\|AY-YA\|_F^2\ge0$. The last step uses Schwarz and cyclicity. Let $\epsilon\downarrow0$. In particular, $\E[B^\top B],\E[BB^\top]\preceq bI$ imply the bound $b\tr[S^2]$.

Write $\Sigma:=\E_\mu[\xi\xi^\top]\preceq I$ and $\Psi(B):=\sum_{i=1}^m\sum_{j=1}^m\Sigma_{i,j}D_iBD_j$. Here $j$ is a summation dummy, not the derivative order. This map has symmetric Kraus matrices and $\Psi(I)\preceq S_D$. In a word $AXBXC$ with two common factors $X$, let every other factor be a distinct independent raw Gaussian direction. Integrating the outer factors and applying the preceding bound to the middle product, we obtain $\E[\|\E_\mu[AXBXC]\|_F^2]\le a^{\ell-2}\tr[S_D^2]\le Na^\ell\omega_2$. The first step also uses $\tr[\Psi(I)^2]\le\tr[S_D^2]$, which follows by taking the trace of $(S_D-\Psi(I))(S_D+\Psi(I))$. The second is the covariance-power trace bound of Lemma~\ref{lem:1070_covariance_power}.

For $r=2s\ge4$, we encode the raw derivative labels in the orthonormal word space of grades $0,\ldots,j$, with $T:=\sum_{i=1}^mN_i\otimes L_i$ and $\|T\|\le\sqrt a$. Let $V$ embed the original space into the vacuum grade. We split a coefficient word as $LR$ with $s$ common factors on each side, and $j_L+j_R=j$ raw factors. The expected block Gram matrix gives
\[
\|\E_\mu[LRV]\|_{\rm HS}^2
\le\|\E_\mu[LL^\top]\|\tr[V^\top\E_\mu[R^\top R]V]
\le Nb_s\operatorname{df}(s)a^\ell\omega_s.
\]
The first step is its Schur-complement bound. For the second, Lemma~\ref{lem:algorithm_1240_mixed_moments} bounds the first factor by $b_sa^{s+j_L}$. Vacuum compression of the second is exactly contraction against distinct independent raw directions. The completely positive word argument in Lemma~\ref{lem:1070_polynomial_energy} bounds it by $a^{j_R}\E_\mu[\tr[X^{2s}]]$, and Lemma~\ref{lem:1070_covariance_power} bounds this trace by $N\operatorname{df}(s)a^s\omega_s$.

There are $\binom\ell j$ placements. The full ordered derivative array is $j!$ times their orthogonal symmetrization, including repeated derivative indices. The triangle inequality therefore contributes $(j!\binom\ell j)^2=(\ell!/r!)^2$. The ordinary second-moment bound also gives coefficient $(\ell!/r!)^2\operatorname{df}(r)\omega_r$, and $\omega_r\le\omega_{r/2}$. Taking the smaller bound, we prove the assertion. At the terminal derivative no common random factor remains, so no $\omega_q$ is inserted.
For $r=2,4,6,8$, the coefficients before the factorial factor are at most $1,9,225,13545$, respectively. Since $\omega_{k(r)}\le\omega_2$, these give Eq.~\eqref{eq:sub10_derivative_means}. Any larger values remain valid.
\end{proof}

We insert the covariance-sensitive derivative means into the trace recurrence while retaining the established operator bounds.

\begin{definition}[Covariance-sensitive trace recurrence]
\label{def:sub10_trace_recurrence}
Fix the final operator tuple and the derivative-mean coefficients of an instance of Definition~\ref{def:sub10_operator_refinement}, and let $\rho\ge0$. Use Eq.~\eqref{eq:sub10_operator_recurrence} with
\[
(p,J_*,A_*,k_*):=(p_2,J,J(1+a),k_2),\qquad
m_r:=M_r\rho/(r!)^2\quad(r\text{ even}).
\]
Define $F_\ell(a,\rho):=(\ell!)^2V_\ell$ for $1\le\ell\le9$.
\end{definition}

Concavity in the covariance parameter provides affine upper bounds that we can integrate and certify with nonnegative coefficients.

\begin{lemma}[Concave trace energies and affine bounds]
\label{lem:sub10_trace_tangents}
For $1\le\ell\le9$,
\begin{equation}
N^{-1}\E_{\mu_t}[\tr[X^{2\ell}Y_+^{-2}]]\le a^\ell F_\ell(a,\omega_2).
\label{eq:sub10_trace_energy}
\end{equation}
For fixed $a$, $F_\ell(a,\rho)$ is nonnegative, increasing, and concave in $\rho\ge0$. For fixed $\rho$ it is nondecreasing in $a$. At any $\rho_0>0$, put
\begin{equation}
\mathcal A_\ell:=F_\ell(a,\rho_0)-\rho_0\partial_\rho F_\ell(a,\rho_0),\qquad
\mathcal B_\ell:=\partial_\rho F_\ell(a,\rho_0).
\end{equation}
Both coefficients are nonnegative and $F_\ell(a,\rho)\le\mathcal A_\ell+\mathcal B_\ell\rho$ for all $\rho\ge0$. More generally, the same scalar concavity and tangent conclusions hold for Eq.~\eqref{eq:sub10_operator_recurrence} with any fixed nonnegative tuple $(p,J_*,A_*,k_*)$ satisfying $A_*\ge1$ and even mean inputs $m_r=M_r\rho/(r!)^2$, where $M_r\ge0$.
\end{lemma}
\begin{proof}
Lemma~\ref{lem:derivative_array_recurrence}, with $T:=N(\ell!)^2a^\ell$ and the means from Lemma~\ref{lem:sub10_derivative_means}, gives Definition~\ref{def:sub10_trace_recurrence} and Eq.~\eqref{eq:sub10_trace_energy}.

For the general scalar assertion, we expand the even branch as $Z_r=J_*m_r+k_*D_r+2\sqrt{J_*k_*m_rD_r}$. The geometric mean is concave and nondecreasing in its nonnegative arguments. By induction we prove nonnegativity, monotonicity, and concavity in $\rho$. For the preceding two-slot instance, the same recurrence is nondecreasing in the valid operator tuple, proving monotonicity in $a$. A differentiable concave function lies below its tangent. Its increasing property gives $\mathcal B_\ell\ge0$, and evaluation at zero gives $\mathcal A_\ell\ge F_\ell(a,0)\ge0$. At a degenerate zero variance we use continuous limits.
\end{proof}

On a radial cell we first increase $a$ to its right endpoint and then take the tangent there. We do not need separate monotonicity of $\mathcal A_\ell$ and $\mathcal B_\ell$. Replacing either coefficient by an outward upper bound preserves the affine estimate.

Combining the two spectral edges gives a sharper bound for the untruncated radial remainder.

\begin{lemma}[Reflected radial remainder]
\label{lem:sub10_reflected_remainder}
Let $d:=2K-\ell\ge2$ and $q:=d-1$. If $D_\ell(a)$ is either $\widehat C_\ell^{(2)}(a)$ or $F_\ell(a,\omega_2)$, then
\begin{equation}
R_K(t):=N^{-1}\E_{\mu_t}[\tr[X^{2K}(I-X^2)^{-1}]]
\le a^K\sqrt{\frac{\operatorname{df}(d)D_\ell(a)}{1+(2d-1)a}}\sqrt{\omega_q}.
\label{eq:sub10_reflected_remainder}
\end{equation}
\end{lemma}
\begin{proof}
Let $\varrho$ be the expected normalized absolute-eigenvalue distribution of $X$, and put $\mu_j:=\int r^{2j}\d\varrho(r)$. Evenness and the trace energy bound give $\int\frac{r^{2\ell}(1+r^2)}{(1-r^2)^2}\d\varrho(r)\le a^\ell D_\ell(a)$. By Cauchy--Schwarz we bound $R_K(t)^2$ by this quantity times $\int r^{2d}/(1+r^2)\d\varrho$. Concavity of $z/(1+z)$ under the probability measure $r^{2d-2}\d\varrho/\mu_{d-1}$ gives
\[
\int\frac{r^{2d}}{1+r^2}\d\varrho\le\frac{\mu_{d-1}\mu_d}{\mu_{d-1}+\mu_d}
\le\frac{\operatorname{df}(d)a^d}{1+(2d-1)a}\omega_{d-1}.
\]
The second step uses monotonicity of $xy/(x+y)$, Eq.~\eqref{eq:sub10_covariance_power}, and $\omega_d\le\omega_{d-1}$. The zero-moment case is immediate. Multiplying the bounds and using $d+\ell=2K$, we prove Eq.~\eqref{eq:sub10_reflected_remainder}.
\end{proof}

\subsection{Positive radial majorants and a rank-sensitive Gaussian transfer}
\label{subsec:sub10_radial_transfer}

Here we use the barrier, radial function $g$, and fixed Gaussian path of Lemma~\ref{lem:fixed_reference_radial}. Positive radial majorants bound the integrand in Eq.~\eqref{eq:fixed_reference_radial}.

Positive scalar majorants decompose the radial integrand into terms controlled by ordinary moments and resolvent energies.

\begin{lemma}[Positive scalar majorants]
\label{lem:sub10_majorants}
Suppose nonnegative rational coefficients $\alpha_j,\beta_\ell,\gamma_\ell$ satisfy
\begin{equation}
g(x)\le\sum_j\alpha_jx^j+\sum_\ell\beta_\ell\frac{x^\ell(1+x)}{(1-x)^2}
+\sum_\ell\gamma_\ell\frac{x^\ell}{1-x}\qquad(0\le x<1),
\label{eq:sub10_majorant}
\end{equation}
where ordinary degrees satisfy $j\ge2$ and weighted degrees satisfy $1\le\ell\le9$. On a radial interval $[v_-,v_+]$, evaluate the tangents of Lemma~\ref{lem:sub10_trace_tangents} at $a_+:=\eta v_+$ and any fixed positive tangent point. The contributions to an upper bound on the radial integral over this interval are
\begin{align*}
&\alpha_j\operatorname{df}(j)\eta^j(v_+^j-v_-^j)\omega_j/j,\\
&\beta_\ell\eta^\ell(v_+^\ell-v_-^\ell)(\mathcal A_\ell+\mathcal B_\ell\omega_2)/\ell,\\
&\gamma_\ell\eta^\ell(v_+^\ell-v_-^\ell)\widehat C_\ell^{(1)}(a_+)/\ell.
\end{align*}
\end{lemma}
\begin{proof}
To the ordinary terms we apply Eq.~\eqref{eq:sub10_covariance_power}. Evenness gives
\[
N^{-1}\E[\tr[X^{2\ell}(I+X^2)(I-X^2)^{-2}]]
=N^{-1}\E[\tr[X^{2\ell}Y_+^{-2}]].
\]
For the two-slot terms we use Lemma~\ref{lem:sub10_trace_tangents}, first increasing the variance to $a_+$ and then taking its tangent. For one-slot terms we use Lemma~\ref{lem:sub10_operator_refinement}. Integrating the remaining powers exactly against $\d v/v$ gives the three expressions. All coefficients are nonnegative.

To verify Eq.~\eqref{eq:sub10_majorant}, it suffices to check the following polynomial on the entire closed interval $[0,1]$:
\begin{align*}
&(1-x)^2\sum_j\alpha_jx^j+(1+x)\sum_\ell\beta_\ell x^\ell+(1-x)\sum_\ell\gamma_\ell x^\ell\\
&\qquad-(1-x)^2\sum_{j=2}^{K-1}b_jx^j-b_Tx^K(1-x)\ge0.
\end{align*}
This is multiplication by the positive denominator $(1-x)^2$ on $[0,1)$. Removing a common nonnegative power of $x$ and applying the Bernstein criterion from Definition~\ref{def:rational_certificate_rules} proves the inequality without omitting the spectral edge.
\end{proof}

For the first three stages, we apply Lemma~\ref{lem:sub10_majorants} on $[1/4,1]$. On $[0,1/4]$ we retain the finite moments and use Lemma~\ref{lem:sub10_reflected_remainder} with $\widehat C_\ell^{(2)}$. For $v_i:=(i/256)^2/4$, the early remainder is at most $T_{\rm early}\sqrt{\omega_q}$, where $d:=2K-\ell$, $q:=d-1$, and
\[
T_{\rm early}:=\frac{b_T\eta^K\sqrt{\operatorname{df}(d)}}K
\sum_{i=1}^{256}(v_i^K-v_{i-1}^K)
\sqrt{\frac{\widehat C_\ell^{(2)}(\eta v_i)}{1+(2d-1)\eta v_{i-1}}}.
\]
Its finite moment coefficients are $b_j\operatorname{df}(j)\eta^j4^{-j}/j$. We bound the numerator at the right endpoint and the reciprocal denominator at the left. The power integral is exact. Young's inequality gives $T_{\rm early}\sqrt{\omega_q}\le T_{\rm early}\omega_q/(2z)+T_{\rm early}z/2$ for $z>0$.

For the last four stages, we use $2048$ quadratic cells $v_i:=(i/2048)^2$ on $[0,1]$. We take tangents at $\rho_0=0.65$. Let $A_i,B_i$ be nonnegative upper bounds for the degree-$\ell$ tangent coefficients at $\eta v_i$, and put
\begin{align*}
h_i&:=\frac{b_T\eta^K\sqrt{\operatorname{df}(d)}}K
\frac{v_i^K-v_{i-1}^K}{\sqrt{1+(2d-1)\eta v_{i-1}}},\qquad T_i:=h_i\sqrt{A_i+B_i},\\
T&:=\sum_iT_i,\qquad T_A:=\sum_i\frac{T_iA_i}{A_i+B_i},\qquad
T_B:=\sum_i\frac{T_iB_i}{A_i+B_i}.
\end{align*}
A zero denominator contributes zero, and we omit it. Lemma~\ref{lem:sub10_reflected_remainder} bounds the cell remainder by $T_i\sqrt{\omega_q(A_i+B_i\omega_2)/(A_i+B_i)}$. Young's inequality on each cell and summation therefore give
\begin{equation}
\text{remainder}\le\frac{T}{2z}\omega_q+\frac z2(T_A+T_B\omega_2).
\label{eq:sub10_covariance_young}
\end{equation}
The finite terms have coefficients $b_j\operatorname{df}(j)\eta^j/j$. Combining either construction gives
\begin{equation}
-N^{-1}\log\nu[\|Y\|<1]\le C+\sum_{j\ge2}P_j\omega_j+\frac{T}{2z}\omega_q,
\qquad C,P_j,T\ge0.
\end{equation}
For a majorant stage, $C$ includes $T_{\rm early}z/2$ and $T=T_{\rm early}$. For a reflected stage, $C=zT_A/2$ and we add $zT_B/2$ to $P_2$. There is then no further $Tz/2$ term. The probability comparison is part of Lemma~\ref{lem:fixed_reference_radial}.

For either radial construction, we apply Lemma~\ref{lem:1070_transfer} with $Q_j=P_j+\mathbbm{1}[j=q]T/(2z)$ and $\operatorname{rank}\Gamma\le m\le wN$. Define
\[
\mathcal P(x):=L_H(x)+\frac{H}{(H-1)\kappa\eta}
(\sum_{j\ge2}\frac{P_j}{(1+x)^j}+\frac{T}{2z(1+x)^q}).
\]
The scalar condition is
\begin{equation}
\mathcal P(x)-b/x\le B_{\rm cap}\qquad(x>0).
\label{eq:sub10_rank_profile}
\end{equation}
The constant $C$ remains outside the spectral average. The exponent is Eq.~\eqref{eq:sub10_exponent}.

\subsection{Shared spectral and radial certificates}\label{subsec:sub10_shared_certificates}

The spectral parameters and radial constructions below supply the centered rows that we use in Section~\ref{sec:small_algorithm_constant}. We specify the first three majorant families by finite rational linear systems after the stage parameters.

\begin{definition}[Existence-stage parameters]
\label{def:sub10_parameters}
In the remainder of this section, the stage parameters refer to the following separate family. For $k=0,1,2$, define
\[
\phi^{[k]}(u):=\sum_{j=1}^{K_k-1}b_{k,j}u^{2j}/j+b_{k,T}\sum_{j=K_k}^\infty u^{2j}/j.
\]
The nonzero finite coefficients and tail coefficients are
\begin{align*}
k=0:&\quad b_{0,1}=0.07085499,\quad b_{0,3}=0.58222347,\\
&\quad b_{0,6}=22.10545334,\quad b_{0,7}=14.50511282,\quad b_{0,T}=22.83787582;\\
k=1:&\quad b_{1,1}=0.001322,\quad b_{1,3}=0.02099,\quad b_{1,14}=61900,\quad b_{1,T}=1057;\\
k=2:&\quad b_{2,1}=0.00001267,\quad b_{2,2}=0.00003561,\quad b_{2,3}=0.00004397,\\
&\quad b_{2,22}=58162000,\quad b_{2,23}=54687000,\quad b_{2,T}=91750.
\end{align*}
For stage zero use $\zeta=0.47982971$, $\gamma=1.02221683$, and
\[
d(u):=\min\{\gamma,\sqrt{(1-u)/\zeta}\},\qquad a_0:=1-\zeta\gamma^2=0.498612861844063305535181.
\]
For stages one and two use Definition~\ref{def:1070_allocation} with
\[
(\zeta_1,\theta_1):=(0.32288046,0.0009507852),\qquad
(\zeta_2,\theta_2):=(0.27359432,0.00001093325).
\]
For $k=3,4,5,6$, use exactly the two-component spectral barrier parameters $(\tau_k,\zeta_k,\kappa_{0,k},\kappa_{{\rm tail},k},\theta_k)$ of Definition~\ref{def:1070:stage-parameters}. Only these spectral parameters are inherited. At such a stage, set $b_{k,j}:=\kappa_{0,k}+\kappa_{{\rm tail},k}\mathbbm{1}[j\ge\tau_k]$ and $b_{k,T}:=\kappa_{0,k}+\kappa_{{\rm tail},k}$. At every stage $\kappa=b_{k,1}$.

The radial degrees are
\[
(K_k,\ell_k)_{k=0}^6:=((8,4),(17,7),(25,9),(34,9),(44,9),(56,9),(63,5)).
\]
The variance, retention, cost, and transfer parameters are the following exact sequences:
\begingroup
\allowdisplaybreaks[1]
\begin{align*}
(\eta_k)_{k=0}^6&:=(0.04419261,\,0.0217136,\,0.01367132,\,0.00869674,\,0.00646514,\,0.00493509,\\*
&\quad 0.0005)\\
(q_k)_{k=0}^6&:=(0.0578850350,\,0.0223838957,\,0.0119572705,\,0.0067284670,\,0.0046741066,\\*
&\quad 0.0040355636,\,0.0016523300)\\
(c_k)_{k=0}^6&:=(7.052711159929,\,2.420740060029,\,0.456432632054,\,0.062577714042,\\*
&\quad 0.005953427019,\,0.000465862518,\,0.000092976331)\\
(H_k)_{k=0}^6&:=(5.45444597,\,5.53920133,\,5.54909934,\,5.33109076,\,2.61968753,\,1.15232869,\\*
&\quad 1.02752192)\\
(z_k)_{k=0}^6&:=(0.73257320,\,0.73298851,\,0.69131064,\,0.72583701,\,0.56550013,\,0.18305585,\\*
&\quad 0.03589337)\\
(B_{{\rm cap},k})_{k=0}^6&:=(0.461415,\,0.446524,\,0.464189,\,0.439229,\,0.338323,\,0.238652,\,0.222120).
\end{align*}
\endgroup
Define the rank-correction parameters
\begin{align*}
b_0^{\rm rk}&:=8.257020779696646\cdot10^{-6},\\
b_1^{\rm rk}&:=1.260355005129572\cdot10^{-6},\\
b_2^{\rm rk}&:=1.598055954762951\cdot10^{-7},
\end{align*}
and $b_3^{\rm rk}=b_4^{\rm rk}=b_5^{\rm rk}=b_6^{\rm rk}:=0$.

For stages zero, one, and two, use respectively $192,64,64$ uniform majorant intervals on $[1/4,1]$, subdividing each into $8,16,16$ equal radial subintervals for the energy upper sums. Stages three through six use Eq.~\eqref{eq:sub10_covariance_young} with $2048$ quadratic cells and tangent point $0.65$. Every operator refinement uses exactly six passes.
\end{definition}

In the next definition we fix the tangent points and specify the radial coefficients by three rational linear systems. Their constraints combine polynomial domination with the scalar budgets needed by the Gaussian transfer.

\begin{definition}[A finite construction of the radial majorants]
\label{def:sub10:majorant-construction}
Set $(m_0,m_1,m_2):=(192,64,64)$, $(d_0,d_1,d_2):=(24,40,56)$, and $(s_0,s_1,s_2):=(8,16,16)$. The tangent points are
\[
(\rho_0,\rho_1,\rho_2):=(0.6709927841045108,0.625,0.625).
\]
Here $m_k$ is the number of radial intervals, $d_k$ bounds the ordinary degree, and $s_k$ is the number of energy subintervals. At a fixed stage suppress $k$ and put $v_i:=1/4+3i/(4m)$ for $0\le i\le m$. The unknowns on interval $i$ are
\[
0\le\alpha_{i,j},\beta_{i,\ell},\gamma_{i,\ell}\le10^{20}
\quad(2\le j\le d,\quad1\le\ell\le9,\quad0\le i<m).
\]
Every unlisted coefficient is zero. Define the cleared polynomial
\begin{equation}
\begin{aligned}
p_i(x):={}&(1-x)^2\sum_{j=2}^{d}\alpha_{i,j}x^{j-1}
 +(1+x)\sum_{\ell=1}^9\beta_{i,\ell}x^{\ell-1}
 +(1-x)\sum_{\ell=1}^9\gamma_{i,\ell}x^{\ell-1}\\
&-(1-x)^2\sum_{j=2}^{K-1}b_jx^{j-1}-b_Tx^{K-1}(1-x).
\end{aligned}
\end{equation}
Require every degree-$(d+1)$ Bernstein coefficient of $p_i$ on every interval
$[a2^{-48},(a+1)2^{-48}]$, $0\le a<2^{48}$, to be nonnegative. These coefficients are the rational linear expressions obtained from Definition~\ref{def:rational_certificate_rules} with cell count $M:=2^{48}$ and degree $n_{\rm B}:=d+1$.

We next specify the rational coefficients of the radial constraints. For a real number $y$, put $\mathcal U(y):=10^{-60}\lceil10^{60}y\rceil$. The enclosure rules of Definition~\ref{def:rational_certificate_rules}, refined as needed, determine these upward endpoints. Let $\mathcal A_\ell(a;\rho)$ and $\mathcal B_\ell(a;\rho)$ be the exact tangent coefficients of Lemma~\ref{lem:sub10_trace_tangents}, and put
\[
t_{i,r}:=v_i+(v_{i+1}-v_i)r/s\qquad(0\le r\le s).
\]
Define the following nonnegative rational weights:
\begin{align}
o_{i,j}&:=\mathcal U(\operatorname{df}(j)\eta^j(v_{i+1}^j-v_i^j)/j),
\\
a_{i,\ell}&:=\mathcal U(\frac{\eta^\ell}{\ell}\sum_{r=1}^{s}
 (t_{i,r}^\ell-t_{i,r-1}^\ell)\mathcal A_\ell(\eta t_{i,r};\rho)),
\\
b_{i,\ell}^{\rm lin}&:=\mathcal U(\frac{\eta^\ell}{\ell}\sum_{r=1}^{s}
 (t_{i,r}^\ell-t_{i,r-1}^\ell)\mathcal B_\ell(\eta t_{i,r};\rho)),
\\
g_{i,\ell}&:=\mathcal U(\frac{\eta^\ell}{\ell}\sum_{r=1}^{s}
 (t_{i,r}^\ell-t_{i,r-1}^\ell)\widehat C_\ell^{(1)}(\eta t_{i,r})).
\end{align}
With the early remainder from Section~\ref{subsec:sub10_radial_transfer}, set $T^+:=\mathcal U(T_{\rm early})$, $L:=2K-\ell_k-1$, and
\begin{align}
P_j&:=\mathbbm{1}[j<K]b_j\operatorname{df}(j)\eta^j4^{-j}/j
 +\sum_{i=0}^{m-1}o_{i,j}\alpha_{i,j}
 +\mathbbm{1}[j=2]\sum_{i=0}^{m-1}\sum_{\ell=1}^9b_{i,\ell}^{\rm lin}\beta_{i,\ell},
\\
C&:=T^+z/2+\sum_{i=0}^{m-1}\sum_{\ell=1}^9
 (a_{i,\ell}\beta_{i,\ell}+g_{i,\ell}\gamma_{i,\ell}).
\end{align}
Here $2\le j\le d$, and $b_j:=0$ for $K\le j\le d$ in the first term. Define
\[
Q(x):=\frac{H}{(H-1)\kappa\eta}
 (\sum_{j=2}^{d}\frac{P_j}{(1+x)^j}+\frac{T^+}{2z(1+x)^L}),
\qquad \varepsilon:=10^{-8},\qquad \delta:=10^{-12}.
\]
For $x_a:=\varepsilon+(100-\varepsilon)a2^{-28}$, $0\le a\le2^{28}$, impose
\begin{align}
H\varepsilon/2+Q(0)-b_k^{\rm rk}/\varepsilon&\le B_{\rm cap}-\delta,
\\
\mathcal U(L_H((x_a+x_{a+1})/2))+H(x_{a+1}-x_a)/4
 +Q(x_a)-b_k^{\rm rk}/x_{a+1}&\le B_{\rm cap}-\delta
 \quad(0\le a<2^{28}),
\\
\mathcal U(\log(101)/100)+Q(100)&\le B_{\rm cap}-\delta.
\end{align}
The final linear constraint is
\begin{equation}
\kappa\eta B_{\rm cap}+b_k^{\rm rk}w_k/2+HC/(H-1)\le\overline e_k,
\label{eq:sub10:linear-exponent-budget}
\end{equation}
where $w_0:=1$, $w_{k+1}:=q_kw_k$, and
\[
(\overline e_0,\overline e_1,\overline e_2)
:=(0.00167627421655443,\ 0.0000145025346222408,\ 9.26288824926953\cdot10^{-8}).
\]

These constraints define the bounded rational polytope $\mathfrak P_k$. Order coordinates by increasing $i$, then by $\alpha,\beta,\gamma$, then by degree. Select the lexicographically least point by successive coordinate minimization. Equivalently, solve every full-rank active-constraint system and select the least feasible vertex. This finite rule gives a unique rational coefficient vector.
\end{definition}

In the next lemma we prove each polytope nonempty. The certificate checks curvature on complete spectral domains and verifies the rational systems that define the radial bounds. The resulting exponents fit the fixed coding budgets.

\begin{lemma}[All-domain certificates for the existence stages]
\label{lem:sub10_certificate}
Put $w_0:=1$, $w_{k+1}:=q_kw_k$, and $E_k:=q_k^2/2+q_k^4/12$, with the parameters of Definition~\ref{def:sub10_parameters}. The polytopes of Definition~\ref{def:sub10:majorant-construction} are nonempty. Every stage satisfies the curvature and strict-root hypotheses, the majorant or reflected radial bounds, and Eq.~\eqref{eq:sub10_rank_profile} on its entire domain. Its exponent in Eq.~\eqref{eq:sub10_exponent}, using $b=b_k^{\rm rk}$ and $w=w_k$, satisfies $e_k<w_kE_k$. The stage costs satisfy $c_k>\sqrt{w_k/\eta_k}/s_\star$.
\end{lemma}
\begin{proof}
We give bounded recursions for the curvature checks and rational feasibility conditions for the radial construction. All arithmetic comparisons use the enclosure rules of Definition~\ref{def:rational_certificate_rules}.

\emph{Step 1: curvature on all spectral triples.}
For stage zero we use Lemma~\ref{lem:sub10_triple}. The multiplier obeys $d(u)^2/(1-u)\le\zeta^{-1}$ even though $\gamma>1$. If all coordinates are regular, meaning at most $a_0$, then $l_u=l_v=l_w=1-\gamma^4$. Set $t_0:=\gamma^4-1$. The bound $G(u,v)\ge t_0$ on regular pairs implies
\[
M=t_0\mathbf1\mathbf1^\top+\operatorname{diag}(G(u,v)-t_0,G(u,w)-t_0,G(v,w)-t_0)\succeq0.
\]
As in Lemma~\ref{lem:1070_spectral_certificate}, $G$ is a weighted average of $(1-u)^2\phi''(u)$. For $u\le0$ this quantity is at least $2b_{0,1}>t_0$. For $0\le u\le a_0$, clearing $(1+u)^2$ gives the polynomial
\begin{align*}
&\sum_{j<K}2b_j(2j-1)u^{2j-2}(1-u^2)^2\\
&\quad+2b_Tu^{2K-2}((2K-1)-(2K-3)u^2)-t_0(1+u)^2.
\end{align*}
After $u=a_0s$, its Bernstein coefficients on the two half-intervals are positive.

For triples with an active coordinate, we use the parameterizations
\[
\mathrm N:\ u=-s,\qquad \mathrm P:\ u=a_0s,\qquad
\mathrm A:\ u=1-\zeta\gamma^2s^2,\qquad 0\le s\le1.
\]
The corresponding multipliers are $\gamma,\gamma,\gamma s$. Permutations reduce the domain to the six cubes $\mathrm{NNA},\mathrm{NPA},\mathrm{PPA},\mathrm{NAA},\mathrm{PAA},\mathrm{AAA}$. We give explicit diagonal lower bounds on each box. Write $p:=2K-2$ and $Q_r(x,y):=\sum_{i=0}^r x^iy^{r-i}$. If the nonnegative moduli lie in $[x_0,x_1]$ and $[y_0,y_1]$, set
\[
q_*:=\min\{x_0y_0(1-x_0y_0),x_1y_1(1-x_1y_1)\},\qquad
B_*:=x_0^p+y_0^p+q_*Q_{p-2}(x_0,y_0).
\]
For two nonnegative eigenvalues a lower bound for $G$ is
\[
2(1-x_1)(1-y_1)\sum_{j<K}b_jQ_{2j-2}(x_0,y_0)+\frac{2b_TB_*}{(1+x_1)(1+y_1)}.
\]
For two nonpositive eigenvalues a lower bound is
\[
2(1+x_0)(1+y_0)\sum_{j<K}b_jQ_{2j-2}(x_0,y_0)+\frac{2b_TB_*}{(1-x_0)(1-y_0)}.
\]
These follow from $Q_p-x^2y^2Q_{p-2}=x^p+y^p+xy(1-xy)Q_{p-2}$, positivity of the coefficients, and concavity of $z(1-z)$ on the product interval. For a nonnegative eigenvalue $x$ and a nonpositive eigenvalue $-y$, we use
\begin{align*}
&2(1-x_1)(1+y_0)(b_1+\sum_{2\le j<K}b_j\frac{x_0^{2j-1}+y_0^{2j-1}}{x_1+y_1})\\
&\quad+2b_T(\frac{x_0^{p+1}(1+y_1)}{(x_0+y_1)(1+x_0)}
+\frac{y_0^{p+1}(1-x_1)}{(x_1+y_0)(1-y_0)}).
\end{align*}
The two tail terms are respectively increasing in $x$, decreasing in $y$, and increasing in $y$, decreasing in $x$. A zero endpoint numerator contributes the lower bound zero. All denominators evaluated on accepted nondegenerate cells are positive.

Let $D_1,D_2,D_3>0$ be the resulting lower bounds for $G(u,v),G(u,w),G(v,w)$. We enclose $l_u,l_v,l_w$ in intervals $I_1,I_2,I_3$, and put $s_i:=\max_{x\in I_i}x^2$. Let $P_*$ be the largest product of three interval endpoints, one from each $I_i$. A leaf is accepted only if
\begin{equation}
\begin{gathered}
D_1D_2>s_1,\qquad D_1D_3>s_2,\qquad D_2D_3>s_3,\\
D_1D_2D_3>D_1s_3+D_2s_2+D_3s_1+2P_*.
\end{gathered}
\label{eq:sub10_triple_leaf}
\end{equation}
These comparisons prove positivity of all principal minors with diagonal $D_i$ and the actual off-diagonal entries. Raising the diagonal to the actual $G$ preserves positive definiteness. The product $P_*$ retains its sign.

Set $\chi(B):=1$ if all denominators and $D_i$ are positive and Eq.~\eqref{eq:sub10_triple_leaf} holds on $B$. Otherwise set $\chi(B):=0$. If $B_{j,-}$ and $B_{j,+}$ are its two halves in coordinate $j$, define
\begin{equation}
\begin{aligned}
\mathsf v(B,0)&:=\chi(B),\\
\mathsf v(B,d)&:=\max\{\chi(B),\ \max_{j\in\{1,2,3\}}
 \mathsf v(B_{j,-},d-1)\mathsf v(B_{j,+},d-1)\}\qquad(d\ge1).
\end{aligned}
\label{eq:sub10:bounded-triple-recursion}
\end{equation}
By induction, $\mathsf v(B,d)=1$ exactly when $B$ has a verified cover of depth at most $d$. We retain the region parameterization, accept passing boxes immediately, and otherwise split along the first successful coordinate in the order $1,2,3$.

Exact rational box comparisons give
\begin{equation}
\mathsf v([0,1]^3,d_{\mathcal R})=1,\qquad
(d_{\mathrm{NNA}},d_{\mathrm{NPA}},d_{\mathrm{PPA}},d_{\mathrm{NAA}},d_{\mathrm{PAA}},d_{\mathrm{AAA}})
:=(15,20,28,34,23,17).
\end{equation}
We start from each entire cube. Failed comparisons or zero denominators force subdivision, and return zero at depth zero.

\emph{Step 2: the remaining spectral barriers.}
At stages one through six it suffices to verify Eq.~\eqref{eq:1070_curvature_condition}. For the custom barriers, multiplying its left side by $(1+u)(1+v)$ gives
\[
2(1-u^2)(1-v^2)\sum_{j<K}b_jQ_{2j-2}(u,v)+2b_T(Q_{2K-2}(u,v)-u^2v^2Q_{2K-4}(u,v)).
\]
For a two-component barrier it gives $2\kappa_0(1+uv)+2\kappa_{\rm tail}(Q_{2\tau-2}-u^2v^2Q_{2\tau-4})$. Set $u_{\rm reg}(s):=-1+(2-\zeta)s$ and $u_{\rm act}(s):=1-\zeta s^2$. We subtract both cleared branches of $F_\theta$ on the three squares
\[
 (u,v)\in\{(u_{\rm reg}(s),u_{\rm reg}(t)),\ (u_{\rm act}(s),u_{\rm reg}(t)),\ (u_{\rm act}(s),u_{\rm act}(t))\},\quad 0\le s,t\le1.
\]
The corresponding products $d_\zeta(u)d_\zeta(v)$ are $1,s,st$, respectively.
The common tensor Bernstein test of Definition~\ref{def:rational_certificate_rules} certifies the $36$ polynomials within total depth $40$. Boundary zeros are allowed. Symmetry and reflection cover the other orderings and edge.

\emph{Step 3: root and radial certificates.}
The initial root quantities at the largest variance of every stage satisfy
\[
1-2\eta/\zeta^2>0.4823,\qquad 1-4p_1\eta>0.7812,\qquad
A_J>0.9962,\qquad \Delta_J>0.6701.
\]
Thus, by the monotonicity in $a$ (Definition~\ref{def:1070_roots}), all roots and six refinement passes are valid at every smaller radial variance. For the first three stages, the systems in Definition~\ref{def:sub10:majorant-construction} have respectively $7872$, $3648$, and $4672$ variables. Exact rational feasibility gives $\mathfrak P_0\ne\varnothing$, $\mathfrak P_1\ne\varnothing$, and $\mathfrak P_2\ne\varnothing$. To construct the coefficients, we apply the finite vertex-selection rule of Definition~\ref{def:sub10:majorant-construction}.

The common Bernstein test verifies the depth-$48$ constraints. Multiplication by $x\ge0$ gives the criterion of Lemma~\ref{lem:sub10_majorants}.

For the radial weights, we differentiate the recurrence in $\rho$ with the operator tuple fixed. Product and square-root interval rules determine the upward endpoints $\mathcal U$. Lemma~\ref{lem:sub10_trace_tangents} proves their nonnegativity. Lemma~\ref{lem:sub10_majorants} gives $C,P_j$ on $[1/4,1]$, and the early reflected sums cover $[0,1/4]$. For the last four stages we use the $2048$ cells of Section~\ref{subsec:sub10_radial_transfer}.

\emph{Step 4: the complete Gaussian-transfer half-line.}
We apply Lemma~\ref{lem:scalar_profile_enclosures} to $E(x):=\mathcal P(x)-b/x$, with $A=1$, $C=0$, $D=b$, and the coefficients of $Q:=\mathcal P-L_H$. Its origin, Lipschitz, and tail bounds are precisely the constraints of Definition~\ref{def:sub10:majorant-construction}, with $\varepsilon:=10^{-8}$, $R:=100$, and reserve $10^{-12}$. By bisection from $[\varepsilon,100]$ we evaluate the $2^{28}$ cell constraints. A passing cell certifies its descendants by the Lipschitz bound and monotonicity of the rational terms, allowing $10^{-60}$ for their upward-rounded logarithmic terms. For $b=0$, we start at zero and include $Q(0)$.

For the last four stages, midpoint bisection of unresolved cells in $[0,100]$ terminates by depth $28$ under outward evaluation. The endpoint bounds cover zero and $x\ge100$. We refine the rational enclosures until all strict comparisons and prescribed upward roundings are determined.

\emph{Step 5: entropy and costs.}
We use $\overline e_0,\overline e_1,\overline e_2$ from Definition~\ref{def:sub10:majorant-construction} and the outward rational bounds
\begin{align*}
\overline e_3&:=3.50703055350984\cdot10^{-10},&
\overline e_4&:=1.13872520877497\cdot10^{-12},\\
\overline e_5&:=3.96760383059514\cdot10^{-15},&
\overline e_6&:=2.68421117846414\cdot10^{-18}.
\end{align*}
Eq.~\eqref{eq:sub10:linear-exponent-budget} for $k\le2$ and outward evaluation for $k\ge3$ give $e_k\le\overline e_k$. Exact comparison gives $\overline e_k<w_k(q_k^2/2+q_k^4/12)$ and $c_k^2s_\star^2\eta_k>w_k$ at every stage. The entropy series of Lemma~\ref{lem:1070_lossless_coding} gives $\log2-h((1-q)/2)>q^2/2+q^4/12$ for $0<q<1$, completing the coding and cost checks.
\end{proof}

\subsection{Centered resolvent gaps}\label{subsec:sub10_centered_gaps}

Lemma~\ref{lem:sub10_mean_gap} bounds the arithmetic--harmonic gap by $\|B_W-D^2\|$. We use this subtraction to center the leading quadratic field in the finite operator refinement.

Under the even radial tilt we retain $X=\sum_i\xi_iN_i$ and $\sum_iN_i^2\preceq aI$. Write
\[
Q:=X^2\otimes I+X\otimes X+I\otimes X^2,\qquad
h(a):=\frac{72a^3(1+a)}{(1+3a)^2}.
\]

\begin{lemma}[Centered one- and two-slot estimates]
\label{lem:slack_centered_updates}
For $0\le a<1/4$, let $C_r^{(1)},C_r^{(2)}$ be valid operator-energy coefficients from a preceding tuple in Definition~\ref{def:sub10_operator_refinement}. Let $c_3,c_4\ge0$ satisfy $\E[X^6]\preceq c_3a^3I$ and $\E[X^8]\preceq c_4a^4I$. The preceding rows use $(c_3,c_4)=(18,153)$. The one-slot gap is at most each of
\[
2a^2+a^3C_3^{(1)},\qquad
2a^2+c_3a^3+a^4C_4^{(1)},\qquad
2a^2+c_3a^3+c_4a^4+a^5C_5^{(1)}.
\]
The two-slot gap is at most $18a^2+h(a)+\min\{U_0,U_1,U_2,U_3\}$, where
\begin{align*}
U_0&:=\tfrac{13}{4}a^3C_3^{(2)},&
U_1&:=2c_3a^3+4c_4a^4+2a^3C_3^{(2)},\\
U_2&:=3c_3a^3+14c_4a^4+2a^4C_4^{(2)},&
U_3&:=3c_3a^3+14c_4a^4+6a^5C_5^{(2)}.
\end{align*}
\end{lemma}
\begin{proof}
We apply the even scalar Poincar\'e inequality in Lemma~\ref{lem:sub10_operator_words} to $X^2z$ and $Qz$ for fixed vectors $z$ and obtain
\[
\E[X^4]-(\E[X^2])^2\preceq2a^2I,\qquad
\E[Q^2]-(\E[Q])^2\preceq18a^2I,\qquad \|\E[Q]\|\le3a.
\]
The derivatives of $X^2$ and $Q$ have respectively two and six terms, each with stacked second moment at most $a^2I$. Cauchy--Schwarz and the even Poincar\'e factor $1/2$ give $2^2a^2/2$ and $6^2a^2/2$. For the last bound we use $\|\E[X\otimes X]\|\le a$, from tensor Gram and covariance domination (Fact~\ref{fact:kronecker_cauchy_schwarz}).

For one slot, $W=(I-X^2)^{-1}$ gives $D=\E[X^2]$ and $B-D^2=\E[X^4]-(\E[X^2])^2+\E[X^6(I-X^2)^{-1}]$. We expand the last resolvent through zero, one, or two further terms and apply the $c_3,c_4$ bounds and Lemma~\ref{lem:sub10_mean_gap}.

For two slots let $W=W_e^{(2)}$, $R:=Q-(I-W^{-1})$, $A:=\E[Q]$, and $R_0:=\E[R]$. Its joint spectral entry is
\[
r(u,v)=\frac{uv(u+v)^2}{1+uv},\qquad
|r(u,v)|\le2(\frac{u^4}{1+u^2}+\frac{v^4}{1+v^2}).
\]
To check the inequality when $uv\ge0$, put $p:=uv$, $s:=u^2+v^2=2p+\delta$. Clearing the positive denominator, we obtain
\[
2(s^2-2p^2+p^2s)(1+p)-p(s+2p)(1+s+p^2)
=\delta((p+2)\delta+p(p^2+4p+7))\ge0.
\]
For opposite signs put $p:=|uv|$. The comparison $p(s-2p)/(1-p)\le p(s+2p)/(1+p)$ follows from $s\le2$, so the same bound holds. For a fixed vector, concavity of $y/(1+y)$ under its $X^2$-weighted spectral measure gives $\E[X^4(I+X^2)^{-1}]\preceq\frac{3a^2}{1+3a}I$. Here we used $\E[X^2]\preceq aI$, $\E[X^4]\preceq3a^2I$, and monotonicity of $xy/(x+y)$. Thus $\|R_0\|\le r_0:=12a^2/(1+3a)$. Since $r_0\le3a$, $AR_0+R_0A-R_0^2\preceq(6ar_0-r_0^2)I=h(a)I$. For a unit vector we obtain this by bounding the quadratic form by $6a\|R_0z\|-\|R_0z\|^2$. This expression is increasing for $0\le\|R_0z\|\le r_0\le3a$. We need no sign assumption on $R_0$.

Put $w:=(1+uv)/((1-u^2)(1-v^2))$, $q:=u^2+uv+v^2$, $S_j:=u^{2j}+v^{2j}$, and $g(w):=(w-1)^2/w$. The four scalar inequalities
\[
g(w)\le q^2+\sum_j\alpha_jS_j+\sum_j\beta_jS_jw
\]
hold for models zero through three with the respective tuples
\[
 (\alpha_3,\alpha_4;\beta_3,\beta_4,\beta_5)\in
 \{(0,0;13/8,0,0),\ (1,2;1,0,0),\ (3/2,7;0,1,0),\ (3/2,7;0,0,3)\},
\]
and all other coefficients zero.
Model zero is the first step in Eq.~\eqref{eq:sub10_scalar_gap}. For models one through three, write $u=x$, $v=\sigma xt$ after a simultaneous sign change and interchange, with $0\le x,t\le1$ and $\sigma\in\{-1,1\}$. Set $y:=x^2$, $p:=\sigma yt$, $q_0:=1+\sigma t+t^2$, and $B_0:=(1-y)(1-yt^2)$. The cleared polynomial divided by $x^4$ is
\[
(q_0^2+\sum_j\alpha_jy^{j-2}(1+t^{2j}))B_0(1+p)
+\sum_j\beta_jy^{j-2}(1+t^{2j})(1+p)^2-(q_0-yt^2)^2.
\]
We remove its common power of $y$ and apply the tensor Bernstein test of Definition~\ref{def:rational_certificate_rules}. For models one through three with signs $+,-$, respectively, the covers have $2,1,3,1,1,1$ leaves and depth at most two. We include boundary zeros, and $x=0$ follows by continuity.

Finally, $D=A-R_0$ gives $B-D^2=(\E[Q^2]-A^2)+\E[g(W)-Q^2]+AR_0+R_0A-R_0^2$. We have just bounded the first and last terms. In the middle term, we use $\E[S_j]\preceq2c_ja^jI$ for $j=3,4$ and $\E[S_jW]\preceq2a^jC_j^{(2)}I$. The four scalar models give the displayed $U_r$, and Lemma~\ref{lem:sub10_mean_gap} completes the proof. All added candidates are nondecreasing in $a$, since the preceding energies are and $h'(a)=72a^2(3+7a+6a^2)/(1+3a)^3\ge0$.
\end{proof}

We incorporate the centered gap estimates into the refinement and specify the resulting small-ball rows.

\begin{definition}[Centered and sharpened refinement instances]
\label{def:slack_instances}
The centered instance of Definition~\ref{def:sub10_operator_refinement} changes $d$ to $5$ and adds the gaps of Lemma~\ref{lem:slack_centered_updates}. The sharpened instance further replaces $(c_3,c_4,M_6,M_8)$ by $(15,129,225,13545)$. The additional gaps are all three one-slot candidates and the two-slot candidate of Lemma~\ref{lem:slack_centered_updates}, with the selected $c_3,c_4$.
\end{definition}

\begin{definition}[Centered refinement and analytic rows]
\label{def:slack_rows}
Retain the spectral barriers, allocations, and degrees $(K_j,\ell_j)$ of Definition~\ref{def:sub10_parameters}. Use the centered instance of Definition~\ref{def:slack_instances}, denote its energies by $C_r^{(s),c}$, and use its tuple and coefficients in the trace recurrence and its tangents.

The separate variance and transfer parameters are
\begingroup
\allowdisplaybreaks[1]
\begin{align*}
(\eta_j)_{j=0}^6&:=(0.0442852429,\,0.0217270503,\,0.0136732551,\,0.0087116729,\,0.0065347088,\\*
&\quad 0.0051558179,\,0.0017485685)\\
(H_j)_{j=0}^6&:=(5.5503660858,\,5.6060022395,\,5.5568912562,\,5.6997395822,\,3.8742111245,\\*
&\quad 1.8377834232,\,1.0536662495)\\
(z_j)_{j=0}^6&:=(0.7371923736,\,0.8246603556,\,0.7,\,0.7348520358,\,0.6520287961,\,0.4456621531,\\*
&\quad 0.1046782494)\\
(B_{{\rm cap},j})_{j=0}^6&:=(0.463265,\,0.447682,\,0.463844,\,0.448853,\,0.393242,\,0.291352,\,0.227610).
\end{align*}
\endgroup
Set $(\beta_0,\beta_1,\beta_2):=(8.033111630413584\cdot10^{-6},1.305387270607384\cdot10^{-6},1.671417693169337\cdot10^{-7})$ and $\beta_j:=0$ for $3\le j\le6$. The rank-free exponent caps are
\[
\begin{aligned}
(\overline e_0^{\circ},\ldots,\overline e_6^{\circ}):={}&(0.00166795657847907232,\ 0.0000144489294827552684,\\
&9.22009562386846628\cdot10^{-8},\ 3.61505180575163530\cdot10^{-10},\\
&1.41033178138237065\cdot10^{-12},\ 5.60018055966625035\cdot10^{-15},\\
&9.61906812911829762\cdot10^{-18}).
\end{aligned}
\]
For the first three rows define the majorants by the finite rational construction of Definition~\ref{def:sub10:majorant-construction}, with these parameters, the centered energies, and $(s_0,s_1,s_2):=(16,32,32)$. Keep its degrees, tangent points, and coefficient bounds. Replace the profile depth $28$ by $45$, $b_j^{\rm rk}$ by $\beta_j$, and its final budget by
\[
\kappa\eta_jB_{{\rm cap},j}+\frac{H_j}{H_j-1}C\le\overline e_j^{\circ}.
\]
Use the same lexicographic vertex-selection rule. The early reflected interval retains $256$ quadratic cells. Rows three through six use the centered covariance-reflected bound with $2048$ quadratic cells and tangent point $0.65$.
\end{definition}

We need the centered rows to satisfy the same curvature and transfer conditions as the original construction.

\begin{lemma}[Centered row certificate]
\label{lem:slack_row_certificate}
The finite constructions in Definition~\ref{def:slack_rows} are feasible. They satisfy the curvature, strict-root, radial, and full-half-line profile requirements, and their rank-free exponents are at most $\overline e_j^{\circ}$.
\end{lemma}
\begin{proof}
The spectral parameters retain the polynomial and triple certificates of Lemma~\ref{lem:sub10_certificate}. Uniformly over the seven new variance caps, the regular-curvature reserve $1-2\eta/\zeta^2$ exceeds $0.48$, the one-slot discriminant $1-4p_1\eta$ exceeds $0.78$, $A_J=1-(L_1-1)^2>0.99$, and the two-slot discriminant $\Delta_J$ exceeds $0.66$. All caps are below $1/4$. Lemmas~\ref{lem:slack_centered_updates} and~\ref{lem:sub10_operator_refinement} therefore give validity and monotonicity through the six passes. Lemma~\ref{lem:sub10_trace_tangents} supplies the trace tangents.

\setlength{\emergencystretch}{2em}The three coefficient systems in Definition~\ref{def:slack_rows} have $7872$, $3648$, and $4672$ variables and nonempty feasible vertex sets. We apply the lexicographic construction of Definition~\ref{def:sub10:majorant-construction}. The polynomial, radial and half-line tests are those of Lemma~\ref{lem:sub10_certificate}, with Bernstein depth $48$, radial rounding to $10^{-60}$, profile depth $45$, and the centered parameters. For the last four rows we use the specified $2048$ reflected cells and $\beta_j=0$. The endpoint tests cover zero and infinity. These rational checks give all seven rank-free caps in Definition~\ref{def:slack_rows}.
\end{proof}

\subsection{Trace-sensitive hereditary small balls}\label{subsec:sub10_trace_small_balls}

The trace improvement enters the trace energies and Gaussian transfer. The centered operator bounds remain valid.

\begin{lemma}[Trace-sensitive row interface]
\label{lem:slack_trace_interface}
Fix a row obtained from the positive or reflected radial bounds of Section~\ref{subsec:sub10_radial_transfer}, using a refinement from Definition~\ref{def:sub10_operator_refinement} valid throughout its radial path. Suppose its moment coefficients satisfy $M_{2s}\le c_s^2$, and write its normalized radial bound, for a finite integer $J\ge2$, as
\[
-n^{-1}\log\nu[\|Y\|<1]\le C+\sum_{j=2}^{J}P_j\omega_j,\qquad C,P_j\ge0.
\]
Here a reflected term $T\omega_L/(2z)$ is included in $P_L$. Let $\eta,\kappa>0$, $H>1$, $\beta\ge0$, and $e^{\circ}>HC/(H-1)$ be the row parameters, and suppose
\[
\kappa\eta L_H(x)+\frac H{H-1}(C+\sum_{j=2}^J\frac{P_j}{(1+x)^j})-\frac{\kappa\eta\beta}{x}\le e^{\circ}\qquad(x>0).
\]
Then for $0<\rho\le1$, symmetric contractions $B_i\in\R^{n\times n}$ with $\tr[B_i^2]\le\rho n$, every $w>0$, and every nonempty subfamily $S$ of size $m\le wn$,
\[
\gamma_m[\|\sum_{i\in S}\xi_iB_i\|\le\sqrt{wn/\eta}]
\ge\exp[-m(\rho e^{\circ}/w+\beta/2)].
\]
In particular, Lemma~\ref{lem:slack_row_certificate} verifies these hypotheses for every centered row $j$ of Definition~\ref{def:slack_rows}, with $(\eta,e^{\circ},\beta):=(\eta_j,\overline e_j^{\circ},\beta_j)$. A retention $q\in(0,1)$ is valid for lossless coding whenever
\begin{equation}
\rho\overline e_j^{\circ}+\beta_jw/2<w(q^2/2+q^4/12).
\label{eq:slack_coding_reserve}
\end{equation}
The partial-signing increment is at most $s_\star^{-1}\sqrt{w/\eta_j}\sqrt n$.
\end{lemma}
\begin{proof}
We first normalize a linear series $Y=\sum_i\xi_iD_i$ by $\sum_iD_i^2\preceq\eta I_n,\qquad \tr[\sum_iD_i^2]\le\rho n\eta$. For $X=tY$, $a=\eta t^2$, and $N_i=tD_i$, the completely positive map $\mathcal T(U):=\sum_iN_iUN_i$ gives
\[
\sum_{i_1=1}^m\cdots\sum_{i_\ell=1}^m\|N_{i_1}\cdots N_{i_\ell}\|_F^2
=\tr[\mathcal T^\ell(I)]\le a^{\ell-1}\tr[\mathcal T(I)]\le\rho na^\ell.
\]
The first step follows by expanding the iterates of $\mathcal T$, the second step follows from positivity and $\mathcal T(I)\preceq aI$, and the third step follows from the trace bound. Summing the $\ell!$ permutations in each top derivative and then all ordered indices, we obtain $\|\nabla^\ell(X^\ell)\|_{\rm HS}^2\le(\ell!)^2\rho na^\ell$. Repeated indices retain their multiplicities.

Put $\widehat\omega_r:=\tr[\Gamma C_\Gamma^r]/(\rho n\eta)$, so $\omega_r=\rho\widehat\omega_r$ and $0\le\widehat\omega_{r+1}\le\widehat\omega_r\le1$. Lemma~\ref{lem:sub10_derivative_means} supplies the same factor $\rho$ in every nonzero ordinary derivative mean. The trace recurrence is homogeneous of degree one in its initial energies and squared-mean inputs: its even branch is $(\sqrt{Jm}+\sqrt{kD})^2$ and its odd branch is linear. We therefore divide by $\rho$ and obtain the selected instance's trace function at $\widehat\omega_2$, so its trace energy is bounded by $\rho a^\ell F_\ell(a,\widehat\omega_2)$. The corresponding one-slot and non-covariance trace bounds also gain $\rho$, since $M_{2s}\le c_s^2$. The operator tuple itself does not gain this factor.

Ordinary moments and positive radial majorants carry $\rho$, as does the reflected Cauchy--Schwarz bound because both trace factors carry $\rho$. The unchanged radial calculation consequently gives
\[
-n^{-1}\log\nu[\|Y\|<1]\le\rho(C+\sum_{j=2}^JP_j\widehat\omega_j).
\]
We apply Lemma~\ref{lem:1070_transfer} with $N:=n$, trace parameter $\rho\eta$, radial coefficients $\rho C,\rho P_j$, rank parameter $m/n$, $b:=\beta$, and
\[
B_{\rm cap}:=\frac{e^{\circ}-HC/(H-1)}{\kappa\eta}>0.
\]
Its covariance weights are $\widehat\omega_j$, and its scalar hypothesis is the stated profile. Its exponent gives $-\log\gamma_m[\|Y\|<1]\le n\rho e^{\circ}+\beta m/2$. The rank term is not multiplied by $\rho$.

For the stated subfamily take $D_i:=B_i/\sqrt{wn/\eta}$ and $\alpha:=m/(wn)\le1$. The variance is at most $\alpha\eta I$ and its trace is at most $\rho n\alpha\eta$. We freeze the majorants and affine bounds at their nominal radial endpoints. Monotonicity keeps them valid. Ordinary contributions gain $\alpha^j$, $j\ge2$, and weighted contributions gain $\alpha^\ell$, $\ell\ge1$, even when assigned to $\widehat\omega_2$. For reflected remainders, $1+c\alpha\eta v\ge\alpha(1+c\eta v)$ gives a gain of $\alpha^{K-1/2}$ separately for $T,T_A,T_B$. Since $K\ge2$, every radial coefficient decreases by at least $\alpha$. The scaling clause of Lemma~\ref{lem:1070_transfer}, starting from rank parameter $w$ and replacing it by $\alpha w=m/n$, gives
\[
-\log\gamma_m[\|Y\|<1]\le n\rho\alpha e^{\circ}+\beta m/2=m(\rho e^{\circ}/w+\beta/2),
\]
where the first step follows from the scaled profile and the second step follows from $n\alpha=m/w$. Passing to the closed ball, we prove the measure bound. The zero Gram case is immediate. The centered-row assertion follows from Lemma~\ref{lem:slack_row_certificate}, and Eq.~\eqref{eq:slack_coding_reserve} and Lemma~\ref{lem:1070_lossless_coding} give its coding consequence.
\end{proof}

\subsection{Repeated spectral slack}\label{subsec:sub10_repeated_slack}

We can retain the trace improvement through several preconditioning steps. At each step we impose one joint moment constraint, chosen for the next spectral majorant.

In this subsection, $j$ denotes an integer row of Definition~\ref{def:slack_rows}. The original coding amplitude $s_\star$ remains unchanged in all preceding constructions. We use the following larger amplitude only for the final signing.

\begin{lemma}[Refined coding amplitude]
\label{lem:slack_refined_coding}
Put $s_+:=0.674489750196$. The conclusions of Lemmas~\ref{lem:1070_peaked_density} and~\ref{lem:1070_lossless_coding} hold with $s_+$ in place of $s_\star$, using the same density family with $a:=0.635553145368$ and $b:=(1-2s_+a)/s_+^2$.
\end{lemma}
\begin{proof}
The refined row in the proof of Lemma~\ref{lem:1070_peaked_density} verifies Eq.~\eqref{eq:slack_coding_criterion} for these parameters. Lemma~\ref{lem:1070_lossless_coding} then gives the partial-signing conclusion.
\end{proof}

For the constructions below put
\[
\delta_{\rm mom}:=10^{-8},\qquad
d_8(q):=q^2/2+q^4/12+q^6/30+q^8/56.
\]
The entropy series in Lemma~\ref{lem:1070_lossless_coding} gives $\log2-h((1-q)/2)>d_8(q)$ for $0<q<1$.

The entropy reserve pays for simultaneous moment constraints while allowing us to skip phases with small active sets.

\begin{lemma}[Moment constraints at every active size]
\label{lem:envelope_joint_coding}
Let $0<q<1$, $w,\eta,L>0$, $a,\beta,c,\theta\ge0$, $m\le wn$, and let $b$ be a nonnegative integer. For symmetric matrices $B_i$, put $G(x):=\sum_{i=1}^m x_iB_i$. Let $K_0$ be a symmetric convex Borel subset of the norm body $\{x:\|G(x)\|\le\sqrt{wn/\eta}\}$, for instance the norm body itself, and suppose
\[
\gamma_m[K_0]\ge\exp[-m(a/w+\beta/2)-\theta].
\]
Let $K_1,\ldots,K_b\subseteq\R^m$ be symmetric convex Borel sets of Gaussian mass at least $e^{-c}$. If
\[
L^2s_+^2\eta\ge w,\qquad n(w(d_8(q)-\beta/2)-a)>cb+\theta,
\]
there is $y\in\{-1,0,1\}^m$ with $\|G(y)\|\le L\sqrt n$, $s_+y\in\bigcap_{r=0}^bK_r$, and at most $qwn$ zero coordinates. If $m\le qwn$, one may take $y=0$.
\end{lemma}
\begin{proof}
The skipped case follows because each $K_r$ contains zero. Otherwise put $\alpha:=m/(wn)>q$. By Gaussian correlation we bound the mass of $K_0$ intersected with the $K_r$ below by $\exp[-m(a/w+\beta/2)-\theta-cb]$. We have
\[
m(d_8(q/\alpha)-a/w-\beta/2)
\ge n(wd_8(q)-a-\beta w/2)>cb+\theta,
\]
where the first step follows from $\alpha\le1$ and the positive even-power coefficients of $d_8$, and the second step follows from the assumed reserve. Lemma~\ref{lem:slack_refined_coding}, with retention $q/\alpha$, gives $s_+y$ in the intersection and fewer than $(q/\alpha)m=qwn$ zero coordinates. Since $K_0$ lies in the norm body, the radius condition gives the stated norm bound.
\end{proof}

In the applications with $b=0$ we take $K_0$ to be the norm body and $\theta:=0$. In every application below, a moment body has the form $\{x:F(x)\le(1+\delta_{\rm mom})\mu\}$, where $F$ is a nonnegative even convex trace polynomial and $\E[F(\xi)]\le\mu$. Markov gives mass at least $\delta_{\rm mom}/(1+\delta_{\rm mom})>e^{-19}$. When $\mu=0$, the mass is one. The strict comparison follows from $e>27/10$ and $(27/10)^{19}>100000001$.

A scalar majorant of the available spectral slack controls the trace bounds after we precondition the residual family.

\begin{lemma}[Hereditary spectral slack]
\label{lem:slack_transform}
Let $B_i\in\R^{n\times n}$ be symmetric contractions with $\tr[B_i^2]\le\rho n$, and let $M$ be a partial signed sum with $\|M\|\le L\sqrt n$, where $0<L<C$. Put $g:=C-L$ and $U:=|M|/(L\sqrt n)$. Suppose $p(u):=\sum_{j=0}^d a_ju^{2j}$ has nonnegative coefficients and
\begin{equation}
\frac{(C-L)^2}{(C-Lu)^2}\le p(u)\quad(0\le u\le1),\qquad
\tr[p(U)-a_0I]\le n\mu.
\nonumber
\end{equation}
For every remaining coefficient put
\[
H:=C\sqrt n I-|M|,\qquad
D_i:=g\sqrt n H^{-1/2}B_iH^{-1/2}.
\]
Then $D_i$ is a symmetric contraction and $\tr[D_i^2]\le\rho'n$, where $\rho':=a_0\rho+\mu$.
Moreover, if the transformed residual family has a complete signing of norm at most $F\sqrt n$ with $0\le F\le g$, then the current untransformed family has a complete signing of norm at most $(L+F)\sqrt n$.
\end{lemma}
\begin{proof}
Since $H\succeq g\sqrt n I$, congruence of $-I\preceq B_i\preceq I$ proves $\|D_i\|\le1$. We have
\begin{align*}
\tr[D_i^2]&\le g^2n\tr[B_i^2H^{-2}]\\
&\le a_0\tr[B_i^2]+\tr[p(U)-a_0I]\le\rho'n,
\end{align*}
where the first step follows from the identity $\tr[B_iRB_iR]=\tr[B_i^2R^2]-\|B_iR-RB_i\|_F^2/2$ with $R:=H^{-1}$, the second step follows from the majorant by functional calculus and $0\preceq B_i^2\preceq I$, and the third step follows from the trace and moment bounds. This includes $M=0$ and holds for every remaining coefficient.

For the last assertion, let $\Delta$ be the residual signed sum before transformation and put $r:=F/g\le1$. The transformed norm bound gives $-rH\preceq\Delta\preceq rH$. We have
\[
M+\Delta\preceq rC\sqrt n I+M-r|M|
\preceq(rC+(1-r)L)\sqrt n I=(L+F)\sqrt n I,
\]
where the first step follows from the congruence bound, the second step follows from functional calculus for $M$ and $r\le1$, and the third step follows from $r(C-L)=F$. The lower bound follows in the same way from $M+r|M|\succeq-(1-r)L\sqrt n I$. We use no commutativity between a coefficient and $M$.
\end{proof}

We combine successive slack transformations with the uniform tail bound to complete a finite signing schedule.

\begin{lemma}[Completing a finite schedule]
\label{lem:slack_schedule_completion}
Start from symmetric contractions. Suppose $r$ partial signings $M_i$ satisfy $\|M_i\|\le L_i\sqrt n$, where $L_i>0$ and $C_i:=C_0-\sum_{h=1}^iL_h>0$. After step $i$, replace every remaining coefficient matrix $B$ by the matrix displayed next. The letter $B$ also denotes the scalar bound below. The replacement is
\[
C_i\sqrt n H_i^{-1/2}BH_i^{-1/2},\qquad
H_i:=C_{i-1}\sqrt n I-|M_i|.
\]
Suppose the final active fraction is at most $v_0$. Suppose $s$ subsequent phases satisfy Lemma~\ref{lem:envelope_joint_coding} with $(w,L,q,b):=(v_{k-1},c_k,q_k,0)$, and put $v_k:=q_kv_{k-1}$. If $0<v_s<e^{-731/256}$ and
\[
B:=\sum_{i=1}^rL_i+\sum_{k=1}^sc_k+\mathcal T_{s_+}(v_s)<C_0,
\]
the original family has a complete signing of norm at most $B\sqrt n$.
\end{lemma}
\begin{proof}
Congruence and $H_i\succeq C_i\sqrt n I$ preserve symmetry and contraction. Lemma~\ref{lem:envelope_joint_coding} with no moment bodies performs the suffix phases, including skipped phases. Lemma~\ref{lem:1070_infinite_tail} then completes the residual signing at cost $\mathcal T_{s_+}(v_s)\sqrt n$. Its geometric continuation terminates for finite $n$, since each nonempty call fixes a positive number of coordinates.

At level $i$ the sum of later costs is less than $C_i$. The last paragraph of the proof of Lemma~\ref{lem:slack_transform} uses only this reserve, the displayed congruence, and the norm bound on $M_i$, not the trace hypothesis. We apply that argument backwards and obtain the stated sum.
\end{proof}

\subsection{Common-envelope moments}
\label{subsec:envelope_moments}

After the first spectral-slack step, all residual coefficients share one positive congruence factor. By retaining its higher trace powers we improve the next small-ball estimate.

\begin{lemma}[Common-envelope powers and derivative arrays]
\label{lem:envelope_moments}
Suppose $B_i=T^{1/2}A_iT^{1/2}$, where $0\preceq T\preceq I$ and the $A_i\in\R^{n\times n}$ are symmetric contractions. For every subfamily of size $m$ and every integer $\ell\ge1$,
\[
\tr[(\sum_{i=1}^mB_i^2)^\ell]\le m^\ell\tr[T^{2\ell}].
\]
Each $B_i$ is a symmetric contraction and $\tr[B_i^2]\le\tr[T^2]$.

More generally, let $X:=\sum_{i=1}^m\xi_iN_i$ with symmetric $N_i$, and put $S:=\sum_{i=1}^mN_i^2$. Under an even convex tilt $\mu$ of the standard Gaussian, for $0\le j\le\ell$ and $r:=\ell-j$,
\begin{equation}
\E_\mu[\|\nabla^j(X^\ell)\|_{\rm HS}^2]
\le(\ell!/r!)^2\operatorname{df}(\ell)^{r/\ell}\tr[S^\ell].
\label{eq:envelope_derivatives}
\end{equation}
The same bound holds for $\|\E_\mu[\nabla^j(X^\ell)]\|_{\rm HS}^2$, and this mean vanishes when $r$ is odd.
\end{lemma}
\begin{proof}
The contraction assertion follows by congruence of $-I\preceq A_i\preceq I$. Put $V:=\sum_{i=1}^mB_i^2$. We have
\[
\|V\|_\ell
\le\|T^{1/2}\|_{4\ell}^2\|\sum_{i=1}^mA_iTA_i\|_{2\ell}
\le m\|T\|_{2\ell}^2,
\]
where the first step follows from $V=T^{1/2}(\sum_{i=1}^mA_iTA_i)T^{1/2}$ and Schatten H\"older, and the second step follows from the triangle inequality and $\|A_iTA_i\|_{2\ell}\le\|T\|_{2\ell}$. Raising to the $\ell$th power, we prove the trace bound. The case $m=\ell=1$ gives the individual bound.

For the derivative estimate define $\mathcal T(U):=\sum_{i=1}^mN_iUN_i$. This $\mathcal T$ is a map on matrices, not the tail cost $\mathcal T_s$.
Fact~\ref{fact:cp_schatten_shift} with $S=\mathcal T(I)$ gives Eq.~\eqref{eq:cp_schatten_shift} for $\ell>1$.

We polarize the $j$ derivatives by standard Gaussian vectors $h_1,\ldots,h_j$, independent of each other and of $\xi$. Their second moment sums the ordered indices exactly, including repeated indices. The product rule has $\ell!/r!$ words, each containing $r$ copies of $X$ and one copy of each direction matrix. For one word $\mathcal B$, we condition on $X$ and evaluate $\mathcal B^\top\mathcal B$ from the inside out. Integrating a direction applies $\mathcal T$, and a common factor applies $U\mapsto XUX$. Along the exponents $\infty,\ell,\ell/2,\ldots,1$, Eq.~\eqref{eq:cp_schatten_shift} and H\"older therefore give
\[
\E_h[\|\mathcal B\|_F^2]\le\|S\|_\ell^j\|X\|_{2\ell}^{2r}.
\]
For $r>0$, we have
\[
\E_\mu[\|X\|_{2\ell}^{2r}]
\le(\E_\mu[\tr[X^{2\ell}]])^{r/\ell}
\le(\operatorname{df}(\ell)\tr[S^\ell])^{r/\ell},
\]
where the first step follows from concavity of $x^{r/\ell}$, and the second step follows from Harg\'e's convex domination~\cite{harge04} and the Gaussian trace bound~\cite[Eq.~(4.17)]{t12}. Since $j+r=\ell$, each word contributes at most $\operatorname{df}(\ell)^{r/\ell}\tr[S^\ell]$. For $r=0$, we apply Eq.~\eqref{eq:cp_schatten_shift} successively and obtain $\tr[\mathcal T^\ell(I)]\le\tr[S^\ell]$. The case $\ell=1$ follows directly from covariance domination. Cauchy--Schwarz over the words proves Eq.~\eqref{eq:envelope_derivatives}. Jensen gives the ordinary-mean bound and parity gives its vanishing assertion.
\end{proof}

\subsection{Retaining the envelope after further congruences}
\label{subsec:iterated_envelope_powers}

We can retain the common envelope after each preconditioning, not only after the first. First we prove the trace inequality needed for this update. All Schatten norms below are unnormalized.

\begin{lemma}[Trace powers under a positive congruence]
\label{lem:iterated_envelope_trace_power}
For positive semidefinite matrices $P,Q\in\R^{n\times n}$ and every real $r\ge1$,
\[
\tr[(Q^{1/2}PQ^{1/2})^r]\le\tr[P^rQ^r].
\]
\end{lemma}
\begin{proof}
This is the Araki--Lieb--Thirring inequality, Bhatia~\cite[Theorem~IX.2.10]{bhatia97} in the trace norm with $(A,B,p):=(P,Q^{1/2},r)$.
We cover the semidefinite case by replacing $P,Q$ by $P+\epsilon I,Q+\epsilon I$ and letting $\epsilon\downarrow0$.
\end{proof}

The trace-power inequality tracks a common envelope and its moment bounds through each further congruence. By retaining the envelope inside the quadratic term we improve the moment budget for each congruence update.

\begin{lemma}[A hereditary iterated envelope]
\label{lem:iterated_envelope_update}
Suppose $B_i=T^{1/2}A_iT^{1/2}$, where $0\preceq T\preceq I$, each $A_i$ is a real symmetric contraction, and $\tr[T^{2j}]\le n\tau_j$ for every integer $j\ge1$. Let a partial sum $M$ satisfy $\|M\|\le L\sqrt n$, choose $0<L<C$, and put
\[
g:=C-L,\quad H:=C\sqrt n I-|M|,\quad K:=g\sqrt n H^{-1},\quad
T':=K^{1/2}TK^{1/2}.
\]
For every remaining index, $D_i:=K^{1/2}B_iK^{1/2}$ has the common representation
\[
D_i=(T')^{1/2}A_i'(T')^{1/2},\qquad 0\preceq T'\preceq I,
\]
where all $A_i'$ are symmetric contractions.

For an integer $k\ge1$, let $p_k(u):=\sum_{j=0}^{d_k}a_{k,j}u^{2j}$ have nonnegative coefficients and satisfy
\begin{equation}
(C-Lu)^{2k}p_k(u)\ge(C-L)^{2k}\qquad(0\le u\le1).
\label{eq:iterated_envelope_polynomial}
\end{equation}
If the partial sum also obeys
\[
\tr[p_k(|M|/(L\sqrt n))-a_{k,0}I]\le(1+\delta_{\rm mom})n\mu_k,
\]
then
\begin{equation}
\tr[(T')^{2k}]\le n(a_{k,0}\tau_k+(1+\delta_{\rm mom})\mu_k).
\label{eq:iterated_envelope_update}
\end{equation}
When $d_k\ge1$, the same conclusion holds if instead the weighted statistic
\[
F_k(X):=a_{k,1}\tr[T^{2k}X^2]+\sum_{j=2}^{d_k}a_{k,j}\tr[X^{2j}]
\]
at $X:=|M|/(L\sqrt n)$ is at most $(1+\delta_{\rm mom})n\mu_k$. For a Gaussian series $G:=\sum_{i=1}^m\xi_iB_i$ with $m\le wn$ and $L^2s_+^2\eta\ge w$, put $X:=G/(s_+L\sqrt n)$. The unweighted statistic $\tr[p_k(X)-a_{k,0}I]$ and the weighted statistic $F_k(X)$ are nonnegative, even and convex functions of $\xi$. Their expectations are at most $n\mu_k$ for the respective choices
\[
\mu_k:=\sum_{j=1}^{d_k}a_{k,j}\operatorname{df}(j)\eta^j\tau_j
\quad\text{and}\quad
\mu_k:=a_{k,1}\eta\tau_{k+1}+\sum_{j=2}^{d_k}a_{k,j}\operatorname{df}(j)\eta^j\tau_j.
\]
All these assertions hold for a skipped phase with $M=0$.
\end{lemma}
\begin{proof}
We have $H\succeq g\sqrt n I$, so $0\prec K\preceq I$ and $T'\preceq K\preceq I$. The real polar decomposition, extended orthogonally across a kernel if necessary, gives
\[
K^{1/2}T^{1/2}=(T')^{1/2}O,\qquad O^\top O=I.
\]
Consequently $A_i':=OA_iO^\top$ gives the common representation and preserves symmetry and contraction. The same $O$ works for every index.

Write $U:=|M|/(L\sqrt n)$. Then
\begin{equation}
\tr[(T')^{2k}]\le\tr[T^{2k}K^{2k}]
\le a_{k,0}\tr[T^{2k}]+\sum_{j=1}^{d_k}a_{k,j}\tr[T^{2k}U^{2j}].
\label{eq:envelope_common_trace_bound}
\end{equation}
The first step is Lemma~\ref{lem:iterated_envelope_trace_power}. The second step is Eq.~\eqref{eq:iterated_envelope_polynomial} by functional calculus, followed by tracing against $T^{2k}\succeq0$. Since $T^{2k}\preceq I$, we discard this weight in every nonconstant term and apply the moment constraint, which proves Eq.~\eqref{eq:iterated_envelope_update}. We assume no commutation. The common representation persists under subfamily selection. For $M=0$, every nonconstant term vanishes and the scalar inequality at zero gives the same conclusion.

For the weighted statistic, we retain $T^{2k}$ in the quadratic term of Eq.~\eqref{eq:envelope_common_trace_bound} and discard it only for $j\ge2$. The assumed weighted constraint then gives Eq.~\eqref{eq:iterated_envelope_update}. The common representation and Eq.~\eqref{eq:envelope_common_trace_bound} use no moment constraint, and the same argument covers kernels, later subfamilies and $M=0$, without any commutation assumption.

For a single coefficient, Schatten H\"older with exponents $(2k+2)/(2k+1)$ and $2k+2$ gives
\[
 \begin{aligned}
 \tr[T^{2k}B_i^2]
 &=\tr[T^{2k+1}A_iTA_i]\\
 &\le\|T^{2k+1}\|_{(2k+2)/(2k+1)}\|A_iTA_i\|_{2k+2}\\
 &\le\|T^{2k+1}\|_{(2k+2)/(2k+1)}\|T\|_{2k+2}
 =\tr[T^{2k+2}].
 \end{aligned}
\]
The first step is cyclicity, the second is H\"older, the third uses $\|A_i\|\le1$, and the last evaluates powers of the same positive matrix. Independence gives $\E[G^2]=\sum_{i=1}^mB_i^2$, so the radius condition bounds the weighted quadratic expectation by $n\eta\tau_{k+1}$. We obtain every unweighted term from the ordinary Gaussian trace-moment bound and the common-envelope variance powers of Lemma~\ref{lem:envelope_moments}.

For symmetric $X$, $\tr[T^{2k}X^2]=\|XT^k\|_F^2$, a squared norm of a linear map. Every other statistic is an unweighted convex even spectral trace. Thus their nonnegative linear combination is convex and even. We assert no convexity of an arbitrary weighted fourth or higher trace power.
\end{proof}

Markov gives each moment body at threshold $(1+\delta_{\rm mom})n\mu_k$ Gaussian mass at least $\delta_{\rm mom}/(1+\delta_{\rm mom})>e^{-19}$, and mass one when the expectation vanishes.

To construct the initial envelope, we use shifted cubic trace statistics as additional moment constraints.

\begin{lemma}[Shifted cubic constraints for the initial envelope]
\label{lem:iterated_envelope_cubic}
For $a>0$, $t\ge0$, and a standard Gaussian $\xi$, define
\[
G_3(a,t):=\E[(\sqrt a|\xi|-t)_+^3].
\]
Let $N_1,\ldots,N_m\in\R^{n\times n}$ be symmetric, and put $Z:=\sum_{i=1}^m\xi_iN_i$, where the $\xi_i$ are independent standard Gaussians. Suppose $\sum_{i=1}^mN_i^2\preceq aI$. Then
\[
\E[\tr[(|Z|-tI)_+^3]]\le nG_3(a,t).
\]
If $b:=t/\sqrt a$, $\varphi(b):=(2\pi)^{-1/2}e^{-b^2/2}$ and $\overline\Phi(b):=\Pr[\xi\ge b]$, then
\begin{equation}
G_3(a,t)=2((t^2+2a)\sqrt a\varphi(b)-(t^3+3at)\overline\Phi(b)).
\label{eq:iterated_envelope_cubic_value}
\end{equation}
\end{lemma}
\begin{proof}
It suffices to prove the comparison for $f(x):=(|x|-t)_+^3$ and its Gaussian smoothings. Each has an even, convex, nonnegative second derivative and at most cubic growth. Put $f_s(x):=\E[f(x+\sqrt{(1-s)a}\xi)]$ and $Q(s):=\E[\tr[f_s(\sqrt sZ)]]$. The heat equation and Gaussian integration by parts give
\[
Q'(s)=\frac12\sum_{i=1}^m\E[D^2\tr[f_s(\sqrt sZ)][N_i,N_i]]
-\frac a2\E[\tr[f_s''(\sqrt sZ)]].
\]
For spectral values $u,v$, we have
\[
f_s'[u,v]=\int_0^1 f_s''((1-h)u+hv)\d h
\le\frac{f_s''(u)+f_s''(v)}2.
\]
The first step is the divided-difference integral formula of Fact~\ref{fact:spectral_hessian}, and the second step is the convex chord bound integrated in $h$. In an eigenbasis of $\sqrt sZ$, the spectral Hessian formula of the same Fact gives
\[
\sum_{i=1}^m D^2\tr[f_s(\sqrt sZ)][N_i,N_i]
\le\tr[f_s''(\sqrt sZ)\sum_{i=1}^mN_i^2]
\le a\tr[f_s''(\sqrt sZ)],
\]
where the first step follows from the divided-difference bound and symmetry of each $N_i$, and the second step follows from $f_s''\ge0$ and the variance bound. Hence $Q'(s)\le0$. Since $Q(0)=n\E[f(\sqrt a\xi)]$ and $Q(1)=\E[\tr[f(Z)]]$, we obtain the comparison. For $f(x)=(|x|-t)_+^3$, its second derivative is $6(|x|-t)_+$. We justify the calculation and endpoint limits for this function by Gaussian smoothing and polynomial domination. Expanding the shifted cube in the scalar Gaussian integral and integrating its four monomials by parts, we obtain Eq.~\eqref{eq:iterated_envelope_cubic_value}.
\end{proof}

In the initial phase we use
\begin{equation}
p_k(u):=a_{k,0}+a_{k,1}u^2+\sum_{t\in\mathcal H_k}b_{k,t}(u-t)_+^3,
\qquad a_{k,0},a_{k,1},b_{k,t}\ge0,
\nonumber
\end{equation}
where $\mathcal H_k\subseteq\{j/40:0\le j<40\}$. The initial envelope is $I$. If Eq.~\eqref{eq:iterated_envelope_polynomial} holds piecewise for this $p_k$, functional calculus for $|M|/(L\sqrt n)$ gives $\tr[(T')^{2k}]\le n\tau_k'$ for any cap satisfying
\[
\tau_k'\ge a_{k,0}+(1+\delta_{\rm mom})\mu_k,\qquad
\mu_k:=a_{k,1}\eta+\sum_{t\in\mathcal H_k}b_{k,t}\mathcal U_{45}(G_3(\eta,t)).
\]
Put $\mathcal U_{45}(x):=10^{-45}\lceil10^{45}x\rceil$ for $x\ge0$. We determine these ceilings from Eq.~\eqref{eq:iterated_envelope_cubic_value} by the rational enclosure rules in Section~\ref{subsec:sub81_data}. Thus $\mu_k$ is a rational upper bound on the expectation cost. The cap is valid once we impose the moment hypothesis $\tr[p_k(|M|/(L\sqrt n))-a_{k,0}I]\le(1+\delta_{\rm mom})n\mu_k$ of Lemma~\ref{lem:iterated_envelope_update}, which the following convexity remark and Gaussian correlation permit. The moment body is convex: for an even convex scalar $f$, we choose an eigenbasis of a convex combination of two symmetric matrices, apply scalar Jensen to each diagonal quadratic form, and sum to obtain convexity of $\tr[f(X)]$. Gaussian correlation therefore applies simultaneously to the moment bodies of a phase (Section~\ref{subsec:sub81_signing}). There are six such bodies in each of the first ten phases and one in phase eleven.

\subsection{Independent one-slot calculus}
\label{subsec:iterated_envelope_rows}

The $7.8795$ construction uses one inverse-weight slot for both its first and residual rows. We state the common calculus for an arbitrary certified allocation parameter. In Section~\ref{subsec:sub81_barriers} we verify its hypotheses for the two barriers used below.

\begin{definition}[Independent one-slot refinement]
\label{def:iterated_envelope_one_slot}
Fix $\zeta>0$ and retain the mixed-word constants $(c_0,c_1,c_2,c_3,c_4):=(1,1,3,15,129)$. For
\[
0<a<\frac{\zeta^2}{1+4\zeta^2},\qquad
p:=(1-a/\zeta^2)^{-1},
\]
initialize $L:=2/(1+\sqrt{1-4ap})$ and $k:=L-1$. In Eq.~\eqref{eq:sub10_operator_recurrence}, use $(p,J_*,A_*,k_*):=(p,L,L,k)$ and, for positive even $r$,
\begin{equation}
 m_r:=\frac1{(r!)^2}\begin{cases}c_{r/2}^2,&\text{operator energy},\\
 b_{\ell,r},&\text{trace energy of degree }\ell,\end{cases}
 \qquad b_{\ell,r}:=\lceil\operatorname{df}(\ell)^{r/\ell}\rceil.
\label{eq:iterated_envelope_one_slot_recurrence}
\end{equation}
For operator energies write $C_r:=(r!)^2V_r$. At each pass compute $C_1,\ldots,C_5$ from the preceding tuple and put
\begin{align*}
L'&:=\min\{L,\ 1+\sum_{j=1}^{r-1}c_ja^j+a^rC_r:1\le r\le4\},\\
k'&:=\min\{k,L'-1,a^2C_2,2a^2+a^3C_3,\\
&\hspace{25mm}2a^2+15a^3+a^4C_4,2a^2+15a^3+129a^4+a^5C_5\}.
\end{align*}
Replace the tuple only after forming all candidates, and perform exactly six passes. No two-slot quantity is initialized or used.

For the trace energy of degree $1\le\ell\le9$, use the trace inputs of Eq.~\eqref{eq:iterated_envelope_one_slot_recurrence} with the final one-slot tuple, and denote $(\ell!)^2V_\ell$ by $\mathcal C_\ell(a)$.
\end{definition}

Using the spectral allocation and derivative recurrence, we justify the one-slot refinement and preserve its envelope-dependent trace bounds.

\begin{lemma}[Validity without a two-slot root]
\label{lem:iterated_envelope_one_slot}
Let $\varphi$ be an even convex barrier satisfying the spectral-triple condition in Eq.~\eqref{eq:sub10_triple} with allocation parameter $\zeta$ and a nonnegative function $d$ satisfying $d(u)^2/(1-u)\le\zeta^{-1}$. Put $\phi:=\theta\varphi$, where $\theta\ge1/2$, and assume that the logarithmic boundary coefficients of $\varphi$ and $\phi$ both exceed two. Let $X:=\sum_{i=1}^m\xi_iN_i$, where $N_i\in\R^{n\times n}$ are symmetric and $\sum_{i=1}^mN_i^2\preceq aI$. Let $\mu$ have density proportional to $e^{-\tr[\phi(X)]-\xi^\top Q\xi}$ relative to the standard Gaussian, where $Q\succeq0$. All expectations in this lemma are under $\mu$. Under the conditions in Definition~\ref{def:iterated_envelope_one_slot}, the six-pass tuple bounds the mean of $(I-X^2)^{-1}$ by $LI$ and its arithmetic--harmonic gap by $k$. For $1\le\ell\le9$, if also $\tr[(\sum_{i=1}^mN_i^2)^\ell]\le na^\ell\tau_\ell$, then
\begin{equation}
n^{-1}\E[\tr[X^{2\ell}(I-X^2)^{-1}]]\le a^\ell\tau_\ell\mathcal C_\ell(a).
\label{eq:iterated_envelope_one_slot_trace}
\end{equation}
The scalar bounds are nonnegative and nondecreasing in $a$. In particular, for $N_i:=\sqrt{\eta v/(wn)}B_i$ and $m=\alpha wn$, with $B_i=T^{1/2}A_iT^{1/2}$ as in Lemma~\ref{lem:envelope_moments}, the right side gains $\alpha^\ell$ at nominal variance $a:=\eta v$. These conclusions hold with the additional nonnegative quadratic potential along the fixed Gaussian radial path.
\end{lemma}
\begin{proof}
In the notation of Lemma~\ref{lem:sub10_triple}, convexity and Eq.~\eqref{eq:sub10_triple_identity} give
\[
\mathcal H_\phi\ge\tfrac12\mathcal H_{\varphi}\ge Q_0-Q_1.
\]
The first step uses $\theta\ge1/2$ and convexity of $\varphi$, and the second step uses the assumed spectral-triple certificate. A single inverse weight loses $Q_0$, while the regular Gram bound gives $Q_1\le a\zeta^{-2}\sum_{i=1}^m\|z_i\|_2^2$. Thus the normalized one-slot curvature is at least $1-a/\zeta^2$. We treat the other edge by reflection, and $Q\succeq0$ only increases the curvature. The assumed boundary coefficient of $\phi$ permits the exhaustion and integration-by-parts arguments of Lemma~\ref{lem:1070_weighted_poincare}.

The same curvature bound holds at every smaller variance. The one-slot part of Lemma~\ref{lem:1070_resolvent_means} therefore gives $1-L^{-1}\le apL$, and by continuous radial scaling from variance zero we select the smaller root while $1-4ap>0$. The condition $1-4ap>0$ is equivalent to the displayed range, which also implies $a<\min\{1/4,\zeta^2\}$. Thus its initial $L$ is valid without evaluating a two-slot root.

Lemma~\ref{lem:derivative_array_recurrence} with tuple $(p,L,L,k)$, normalization $(\ell!)^2a^\ell$, and the means from Lemma~\ref{lem:sub10_operator_words} gives the operator recurrence, without any two-slot quantity.

Expanding $(I-X^2)^{-1}$ with its weighted remainder, we obtain every candidate for $L'$. The mean-gap bound gives $k\le L'-1$ and the candidate $a^2C_2$. The one-slot part of Lemma~\ref{lem:slack_centered_updates}, with $c_3=15$ and $c_4=129$, gives the other three candidates. Its centered variance estimate is $\E[X^4]-(\E[X^2])^2\preceq2a^2I$ and does not use a two-slot mean. Hence each finite pass preserves validity. The initial smaller root has a nonnegative power series. The recurrence and the updates use nonnegative sums, products, square roots and minima. By induction we prove monotonicity. We require no limiting fixed point.

For the trace bound, we apply Lemma~\ref{lem:derivative_array_recurrence} with the scalar $T:=(\ell!)^2na^\ell\tau_\ell$ of that lemma, which is not the envelope $T$ of the statement, and with $m_r:=b_{\ell,r}/(r!)^2$, supplied by Lemma~\ref{lem:envelope_moments}. We prove Eq.~\eqref{eq:iterated_envelope_one_slot_trace} by summing column energies. At actual size $\alpha wn$ the normalization gains $\alpha^\ell$. The operator tuple never receives a factor $\tau_\ell$.
\end{proof}

\subsection{Two one-slot barriers and their complete curvature certificates}
\label{subsec:sub81_barriers}

Both the first small-ball row and the residual rows use one resolvent slot. The first row retains covariance credit in its ordinary and derivative moments. For the residual rows we use common-envelope trace powers. We supply these rows with the following two barriers and their complete curvature certificates.

For either row family, with the coefficients below, define
\[
 \phi_b(u):=\sum_{j=1}^7 b_ju^{2j}/j+b_8\sum_{j=8}^{\infty}u^{2j}/j,
 \qquad \widehat\phi_b:=\phi_b/2\quad(|u|<1).
\]
We extend these functions by $+\infty$ outside $(-1,1)$. Put
\[
 d(u):=\min\{\gamma,\sqrt{(1-u)/\zeta}\},\quad
 a_0:=1-\zeta\gamma^2,\quad \kappa:=b_1/2.
\]
The two families have $(\zeta,\gamma)=(0.65,1.025)$ for $\mathsf I$ and
$(\zeta,\gamma)=(0.9,1)$ for $\mathsf R$. Their coefficients are
$b_j:=[z^j]\mathcal B_{\mathsf F}(z)$ for $1\le j\le8$ and
$\mathsf F\in\{\mathsf I,\mathsf R\}$, where
\begin{align*}
\mathcal B_{\mathsf I}(z)&:=0.319z+2.36z^3+16.37z^5+5.885z^8,\\
\mathcal B_{\mathsf R}(z)&:=0.19809566z+17.64890782z^3+6.93662749z^4\\
&\quad+8.0322205z^5+13.00332072z^7+6.99335308z^8.
\end{align*}
In both rows we set $b_2:=0$ and $b_6:=0$. The actual half-barrier boundary coefficients are $2.9425$ and $3.49667654$, respectively, both strictly greater than two. The first row has only four nonzero full-barrier coefficients.

We verify the spectral-triple condition for both barriers before applying the one-slot calculus.

\begin{lemma}[Whole-domain triple certificates]
\label{lem:sub81_triples}
For each of the two displayed barriers, let
\[
 G(u,v):=(1-u)(1-v)\phi_b'[u,v],\qquad
 l_u:=1-d(u)^2d(v)d(w),
\]
with $l_v,l_w$ defined by permutation. Then
\[
 \begin{pmatrix}G(u,v)&-l_u&-l_v\\-l_u&G(u,w)&-l_w\\-l_v&-l_w&G(v,w)\end{pmatrix}\succeq0
 \qquad(-1<u,v,w<1).
\]
\end{lemma}
\begin{proof}
We use the pairwise box bounds and bounded recursion in Step 1 of the proof of Lemma~\ref{lem:sub10_certificate}, specifically Eqs.~\eqref{eq:sub10_triple_leaf}--\eqref{eq:sub10:bounded-triple-recursion}, with $K:=8$, $b_T:=b_8$ and the displayed coefficients. In the all-regular region $u,v,w\le a_0$, put $t:=\gamma^4-1$. For $\gamma=1$, convexity gives $G(u,v)\ge2b_1\zeta^2>0$. For $\gamma=1.025$, the regular-region polynomial in that step, with $t_0:=t$, is positive on $[0,a_0]$ by the Bernstein test of Definition~\ref{def:rational_certificate_rules}, and $2b_1>t$ handles negative arguments. The weighted-average argument there gives $G(u,v)\ge t$, so the matrix is positive semidefinite.

For the other triples we use the same endpoint parametrizations, allocation values and six closed cubes, in the order $\mathrm{NNA},\mathrm{NPA},\mathrm{PPA},\mathrm{NAA},\mathrm{PAA},\mathrm{AAA}$. On each dyadic box we form the three diagonal lower bounds $D_i$ from that step. We round their products down on the $10^{-160}$ grid and positive denominators up. Zero endpoint numerators contribute zero. A type-zero leaf is one passing the first complete-box test. That test requires positive denominators and $D_i$, together with Eq.~\eqref{eq:sub10_triple_leaf}, using outward enclosures of $l_u,l_v,l_w$. The principal-minor argument there proves positivity on the entire box and retains the sign of their product.

The second complete-box test retains the positive diagonal and two-by-two tests of the first, but bounds the exact determinant by Taylor's theorem. Put $Q_r(u,v):=\sum_{j=0}^r u^jv^{r-j}$. The exact rational identity is
\[
 G(u,v)=\frac{2(1-u^2)(1-v^2)\sum_{j=1}^7b_jQ_{2j-2}(u,v)
 +2b_8\{u^{14}+v^{14}+uv(1-uv)Q_{12}(u,v)\}}{(1+u)(1+v)}.
\]
We obtain it by clearing the divided-difference denominator. Substitution of the above parametrizations gives the exact determinant $f(s_1,s_2,s_3)$. A type-four leaf is one passing the second complete-box test. For a leaf centered at $c$, with half-widths $h_i$, that test requires
\[
 \underline f(c)-\sum_{i=1}^3\overline{|\partial_i f(c)|}h_i
 -\tfrac12\sum_{i=1}^3\sum_{j=1}^3\overline{|\partial_{ij}f|}_{\rm box}h_ih_j>0.
\]
We compute value, gradient and Hessian enclosures by the product and inverse rules, outwards on the $10^{-50}$ grid. A box on which a denominator enclosure contains zero is not accepted by this test. The inverse rules are
\[
 \nabla(q^{-1})=-q^{-2}\nabla q,\qquad
 \nabla^2(q^{-1})=2q^{-3}(\nabla q)(\nabla q)^\top-q^{-2}\nabla^2q.
\]
We prove the interval inclusions by induction through the rational expression. Taylor's theorem on the segment from $c$ to every point of the box proves positivity on the entire box.

A zero denominator or an unresolved comparison is a failed leaf test. Set $\chi_b(B):=1$ when either complete-box test passes, and set $\chi_b(B):=0$ otherwise. We use the bounded recursion in Eq.~\eqref{eq:sub10:bounded-triple-recursion} with $\chi_b$ in place of $\chi$. In the displayed order of regions, we take depth budgets
\[
 \begin{aligned}
 d^{\mathsf I}&:=(10,10,13,20,17,21),\\
 d^{\mathsf R}&:=(11,8,12,20,18,25).
 \end{aligned}
\]
For each region and each barrier, these rational tests give $\mathsf v([0,1]^3,d)=1$. Successful covers have respectively $77,146,696,8465,9550,6937$ leaves for $\mathsf I$ and $88,54,88,2384,1858,4028$ leaves for $\mathsf R$. These counts describe successful covers and are not additional acceptance conditions. The formulas and depth budgets specify a finite certificate. Induction in the bounded recursion proves coverage of each entire cube, including its faces. The all-regular argument and permutation symmetry complete the proof.
\end{proof}

\paragraph{Application of the one-slot calculus.}
For either barrier, we apply Lemma~\ref{lem:iterated_envelope_one_slot} with $\varphi:=\phi_b$ and $\theta:=1/2$. The nonnegative coefficients give evenness and convexity. The displayed allocation satisfies $d(u)^2/(1-u)\le\zeta^{-1}$, and Lemma~\ref{lem:sub81_triples} gives the triple condition. The boundary coefficients of both $\phi_b$ and $\widehat\phi_b$ exceed two. Under the coefficient and quadratic-potential hypotheses of Lemma~\ref{lem:iterated_envelope_one_slot}, its conclusions therefore hold for $\phi:=\widehat\phi_b$ throughout $0<a<\zeta^2/(1+4\zeta^2)$. The operator tuple and its six refinement passes are those of Definition~\ref{def:iterated_envelope_one_slot}, with no trace factor or two-slot quantity.

For both row families we change coordinates by the same contraction of the coefficient metric, which we record once in the next lemma.

\begin{lemma}[Contracted coordinates for the one-slot radial tilt]
\label{lem:sub10_contracted_coordinates}
Let $E_1,\ldots,E_m\in\R^{n\times n}$ be symmetric with $S:=\sum_{i=1}^mE_i^2\preceq\eta I$ and $\Gamma_{ij}:=\tr[E_iE_j]$. For either displayed row family, use the fixed-reference setup of Lemma~\ref{lem:fixed_reference_radial} with $N:=n$, $B_i:=E_i$, $\phi:=\widehat\phi_b$ and $\kappa:=b_1/2$. Its reference Gaussian is $\nu=\mathcal N(0,C_\Gamma)$ with $C_\Gamma:=(I+2\kappa\Gamma)^{-1}$, and $\mu_v$ denotes its measure at $t=\sqrt v$, with $X:=\sqrt vY$. Assume $\eta\le\zeta^2$ and fix a right endpoint $b\in(0,1]$ with $\eta b<\zeta^2/(1+4\zeta^2)$. Define
\[
 \lambda:=1-\eta b/\zeta^2,\qquad
 \beta:=\frac{1-b}{\lambda},\qquad
 D:=(I+2\kappa\beta\Gamma)^{-1},\qquad
 S(D):=\sum_{i=1}^m\sum_{j=1}^mD_{ij}E_iE_j.
\]
Assume the spectral certificate of Lemma~\ref{lem:sub81_triples} for the chosen barrier.
Then $0\le\beta\le1$, and for $0\le v\le b$ the constant coordinates $\xi=D^{1/2}z$ have three properties.
\begin{enumerate}[label=(\roman*)]
\item The law of $z$ under $\mu_v$ is an even convex tilt of standard Gaussian measure, and $X=\sum_{i=1}^mz_iN_i'$ with symmetric $N_i':=\sqrt v\sum_{j=1}^mE_j(D^{1/2})_{ji}$ and $\sum_{i=1}^m(N_i')^2=vS(D)\preceq vS\preceq\eta bI$.
\item The one-slot curvature in the $z$ coordinates is bounded below by $\lambda I$. The weighted Bochner, Poincar\'e and projection-aware variance arguments of Lemmas~\ref{lem:1070_weighted_poincare} and \ref{lem:sub10_projection_variance} therefore apply with $p:=\lambda^{-1}$ and one inverse-weight slot.
\item The law of $X$ is unchanged. Any valid one-slot mean and gap bounds $L,k$ at variance $\eta b$ therefore bound the mean of $(I-X^2)^{-1}$ and its arithmetic--harmonic gap under $\mu_v$, and they receive no covariance or envelope factor.
\end{enumerate}
The matrix $D$ is held fixed on $[0,b]$ and is never differentiated in radial integration.
\end{lemma}
\begin{proof}
Since $\eta\le\zeta^2$, $1-b\le\lambda$ and $0\le\beta\le1$. Relative to standard Gaussian measure in $\xi$, the radial potential is $\tr[\widehat\phi_b(X)]+\kappa(1-v)\xi^\top\Gamma\xi$. By Lemma~\ref{lem:iterated_envelope_one_slot} under the certificate of Lemma~\ref{lem:sub81_triples}, with the nonnegative quadratic Hessian retained, its one-slot curvature in coefficient indices is bounded below by
\[
 (1-\eta v/\zeta^2)I+2\kappa(1-v)\Gamma\succeq\lambda I+2\kappa(1-b)\Gamma,
 \qquad
 D^{1/2}\{\lambda I+2\kappa(1-b)\Gamma\}D^{1/2}=\lambda I.
\]
The equality uses $\lambda\beta=1-b$. The first bound is the half-Hessian absorption and regular tensor-Gram estimate in the proof of that lemma, and its second term is the Hessian of the quadratic potential. The invertible constant map $\xi=D^{1/2}z$ preserves convexity of the exhaustion domains and the finite-energy assertions, the boundary coefficient of $\widehat\phi_b$ exceeds two, and reflection gives the other spectral edge. This proves (ii).

For (i), simultaneous functional calculus for $\Gamma$ and $\beta\le1$ give
\[
 D^{1/2}C_\Gamma^{-1}D^{1/2}=I+2\kappa(1-\beta)\Gamma D\succeq I,
\]
so $\nu$ is an even convex tilt of standard Gaussian measure in the $z$ coordinates. The remaining potential $\tr[\widehat\phi_b(X)-\kappa X^2]$ is even and convex, since removing the quadratic term leaves the nonnegative higher even powers of the barrier. Substituting $\xi=D^{1/2}z$ in $X=\sqrt v\sum_{j=1}^m\xi_jE_j$, we obtain $X=\sum_{i=1}^mz_iN_i'$, and $\sum_{i=1}^m(N_i')^2=vS(D)$ because $D^{1/2}D^{1/2}=D$. In a coefficient basis diagonalizing $\Gamma$, the matrix $D$ is diagonal with entries in $(0,1]$ and $S$ is unchanged, so $S(D)\preceq S\preceq\eta I$.

For (iii), the mean of $(I-X^2)^{-1}$ and its arithmetic--harmonic gap depend only on the law of $X$, which the coordinate change does not alter. The contraction enters the derivative recurrences only through $p$ and the $N_i'$. Since $D$ depends on $b$ alone, no $v$ derivative of $D$ appears in the radial integration.
\end{proof}

\subsection{A covariance-sensitive one-slot first row}
\label{subsec:sub81_first_row}

For the first row we use barrier $\mathsf I$ and
\[
 \eta_0:=0.065,\quad\zeta_0:=0.65,\quad\kappa_0:=0.1595,\quad H_0:=4.338,
 \quad e_0:=0.007955,\quad\beta_0:=0.
\]
The additional quadratic Hessian supplies covariance credit to the top derivative array. We retain the fixed reference Gaussian throughout the radial path.

Let $E_1,\ldots,E_m\in\R^{n\times n}$ be symmetric with $\sum_{i=1}^mE_i^2\preceq\eta I$. We use the fixed-reference setup of Lemma~\ref{lem:fixed_reference_radial} with the lemma's $N:=n$ and its coefficient matrices $B_i:=E_i$, with $\phi:=\widehat\phi_b$, and quadratic coefficient $\kappa:=b_1/2$. Denote its measure at $t=\sqrt v$ by $\mu_v$, and put
\[
 C_\Gamma:=(I+2\kappa\Gamma)^{-1},\quad
 \omega_j:=\frac{\tr[\Gamma C_\Gamma^j]}{n\eta},\quad
 X:=\sqrt vY,\quad a:=\eta v,\qquad0\le v\le1.
\]
Write $Z_v:=\E_\nu[e^{-\tr[\widehat\phi_b(X)-\kappa X^2]}]$, so $\d\mu_v=Z_v^{-1}e^{-\tr[\widehat\phi_b(X)-\kappa X^2]}\d\nu$ and $Z_0=1$. All expectations in the next lemma are under $\mu_v$. The additional quadratic potential is nonnegative, and we do not reset the reference $\nu$ at radial endpoints.

Covariance-sensitive derivative means yield affine trace-energy bounds for the one-slot radial integral.

\begin{lemma}[Covariance tangents with one inverse slot]
\label{lem:sub81_covariance_tangents}
Let $(p,L,k)$ be any valid one-slot tuple at variance $a$. Use Eq.~\eqref{eq:sub10_operator_recurrence} for $1\le r\le9$ with
\[
(p,J_*,A_*,k_*):=(p,L,L,k),\qquad
m_r:=M_rz/(r!)^2\quad(r\text{ even}),
\]
where $(M_2,M_4,M_6,M_8):=(1,9,225,13545)$. For $1\le\ell\le9$, put $F_\ell(a,z):=(\ell!)^2V_\ell$. Then
\[
 n^{-1}\E[\tr[X^{2\ell}(I-X^2)^{-1}]]\le a^\ell F_\ell(a,\omega_2).
\]
For fixed tuple, $F_\ell$ is nonnegative, increasing and concave in $z\ge0$. At $z_*:=0.67$, the coefficients
\[
 A_\ell:=F_\ell(a,z_*)-z_*\partial_zF_\ell(a,z_*),\qquad
 B_\ell:=\partial_zF_\ell(a,z_*)
\]
are nonnegative and $F_\ell(a,z)\le A_\ell+B_\ell z$ for every $z\ge0$.
\end{lemma}
\begin{proof}
The raw derivative directions are $N_i:=\sqrt vE_i$. To estimate their ordinary means, we whiten the fixed Gaussian $\nu$, representing the same random $X$ by coefficient matrices $\sqrt v\sum_{j=1}^mE_j(C_\Gamma^{1/2})_{ji}$. The tilt in those coordinates is by $\tr[\widehat\phi_b(X)-\kappa X^2]$, an even convex function. Lemma~\ref{lem:sub10_derivative_means} explicitly permits raw directions different from the coefficient family of the random field. It therefore bounds every even ordinary derivative mean, at residual degree $r>0$, by
\[
 \|\E[\nabla_N^j(X^\ell)]\|_{\rm HS}^2
 \le na^\ell(\ell!/r!)^2 M_r\omega_2,\qquad r:=\ell-j.
\]
These four $M_r$ are the values of Eq.~\eqref{eq:sub10_derivative_means}. With no covariance credit, we bound the top constant array by $na^\ell(\ell!)^2$. Odd ordinary means vanish.

Lemma~\ref{lem:derivative_array_recurrence}, with $T:=na^\ell(\ell!)^2$ and tuple $(p,L,L,k)$, now gives the trace bound by summing column energies. Here we do not multiply the operator tuple by $\omega_2$.

The scalar assertion of Lemma~\ref{lem:sub10_trace_tangents} gives concavity and the nonnegative global tangent at $z_*=0.67$. Increasing a valid tuple before taking this tangent is permissible. We do not use separate monotonicity of tangent intercept and slope in $a$.
\end{proof}

For $j\ge1$ and $\beta\ge0$, define
\[
 \omega_j^{(\beta)}:=\frac{\tr[\Gamma(I+2\kappa\beta\Gamma)^{-j}]}{n\eta}.
\]
Thus $\omega_j^{(1)}=\omega_j$ and $\omega_j^{(0)}=\tr[\Gamma]/(n\eta)\le1$.

The quadratic part of the radial tilt contracts the derivative metric, retaining an additional covariance factor in the energy bound.

\begin{lemma}[A contracted derivative metric]
Suppose $0<\eta<\zeta^2$, $0<v\le1$, and $a:=\eta v<\zeta^2/(1+4\zeta^2)$. Assume the spectral certificate of Lemma~\ref{lem:sub81_triples} for the chosen barrier. Define
\[
 \lambda_v:=1-a/\zeta^2,\qquad p:=\lambda_v^{-1},\qquad
 \beta_v:=\frac{1-v}{\lambda_v},\qquad
 D_v:=(I+2\kappa\beta_v\Gamma)^{-1}.
\]
Then $0\le\beta_v\le1$. Let $L,k$ be any valid one-slot mean and
gap bounds at variance $a$. Let $A_\ell,B_\ell$ be the nonnegative global
covariance tangent coefficients from
Lemma~\ref{lem:sub81_covariance_tangents}, computed with this tuple.
For $1\le\ell\le9$,
\begin{equation}
 \frac1n\E[\tr[X^{2\ell}(I-X^2)^{-1}]]
 \le a^\ell\{A_\ell\omega_\ell^{(\beta_v)}+B_\ell\omega_2\}.
 \label{eq:metric_energy}
\end{equation}
The operator bounds $L,k$ receive no covariance or envelope factor.
The matrix $D_v$ is only a change of coordinates in the energy proof.
\end{lemma}
\begin{proof}
First we apply Lemma~\ref{lem:sub10_contracted_coordinates} to barrier $\mathsf I$ with right endpoint $b:=v$, so that $\eta b=a$, $\lambda=\lambda_v$, $\beta=\beta_v$ and $D=D_v$. By part (ii), the weighted Bochner, Poincar\'e and projection-aware variance arguments of Lemmas~\ref{lem:1070_weighted_poincare} and \ref{lem:sub10_projection_variance} apply in the coordinates $\xi=D_v^{1/2}z$ with constant $p$. By part (i), the derivative directions in these coordinates are
\[
 N_i':=\sqrt v\sum_{j=1}^m E_j(D_v^{1/2})_{ji}.
\]
They are symmetric and $\sum_{i=1}^m(N_i')^2\preceq aI$. To bound their higher
variance traces, we rotate the coefficient basis to diagonalize $\Gamma$,
with eigenvalues $\gamma_i$ and $d_i:=(1+2\kappa\beta_v\gamma_i)^{-1}$.
Then $\sum_{i=1}^m(N_i')^2=v\sum_{i=1}^m d_iE_i^2$. In an eigenbasis of this last
matrix, scalar Jensen with weights $u^\top E_i^2u/\eta$ and missing
weight at zero gives
\[
 \tr[(\sum_{i=1}^m(N_i')^2)^\ell]
 \le v^\ell\eta^{\ell-1}\sum_{i=1}^m d_i^\ell\tr[E_i^2]
 =na^\ell\omega_\ell^{(\beta_v)}.
\]
The first step is scalar Jensen as in
Lemma~\ref{lem:1070_covariance_power}. The second uses
$\tr[E_i^2]=\gamma_i$ in the rotated coefficient basis. Here we use no operator
convexity of a high power. Therefore we bound the constant top derivative array
in squared Hilbert--Schmidt norm by
$n(\ell!)^2a^\ell\omega_\ell^{(\beta_v)}$, using the top-array estimate
in Lemma~\ref{lem:envelope_moments}. Its weighted energy is at most $L$
times this quantity.

For the lower ordinary means, we whiten the original, fixed Gaussian $\nu$
by $\xi=C_\Gamma^{1/2}g$. The law of $g$ is an even convex tilt by
$\tr[\widehat\phi_b(X)-\kappa X^2]$. Its random coefficient family is
$\sqrt v\sum_{j=1}^mE_j(C_\Gamma^{1/2})_{ji}$. In contrast, the raw derivative family
is $N_i'$. Lemma~\ref{lem:sub10_derivative_means} explicitly allows these
families to differ and requires only the raw variance bound just proved.
Consequently, for $r:=\ell-j>0$ even,
\[
 \|\E[\nabla_{N'}^j(X^\ell)]\|_{\rm HS}^2
 \le na^\ell(\ell!/r!)^2M_r\omega_2,
 \quad (M_2,M_4,M_6,M_8):=(1,9,225,13545).
\]
Odd means vanish. These are statements about the same random field and
the specified raw perturbations, so the two coordinate descriptions do
not require a second change of probability measure.

Next we normalize by $na^\ell(\ell!)^2$. The trace recurrence now starts with
$V_0=Z_0=tL$, where $t:=\omega_\ell^{(\beta_v)}$. Its even mean inputs are
$M_r z/(r!)^2$. Here the scalar $z:=\omega_2$ is not the coordinate vector $z$ above. The projection-aware variance
inequality in the contracted coordinates and the unchanged odd/even
projection bounds prove this recurrence. If $F_\ell(a,z)$ denotes the
recurrence starting at $L$, its two-input version equals
$tF_\ell(a,z/t)$ for $t>0$, by homogeneity of every branch. The global
tangent inequality gives
\[
 tF_\ell(a,z/t)\le A_\ell t+B_\ell z.
\]
This remains meaningful even when $z/t>1$, since the scalar recurrence
and its concavity statement hold for every nonnegative argument. If
$t=0$, then $\Gamma=0$ and the random field is zero. By summing column
energies we prove Eq.~\eqref{eq:metric_energy}.

Finally $\beta_v=(1-v)/(1-\eta v/\zeta^2)$ decreases with $v$ because
$\eta/\zeta^2<1$. Therefore we can use a right endpoint to freeze
$\beta_v$ from below. We can increase the original valid operator tuple
to its right-endpoint value and round it upward before forming its
tangents. The resulting tangent remains a global upper bound. We
assert no separate monotonicity of its intercept or slope, and we use no
$v$ derivative of $D_v$.
\end{proof}

We define the first-row coefficients by the rational program in Section~\ref{subsec:sub81_data}. Its Bernstein constraints give
\[
 g_0(x):=(2.36x^3+16.37x^5+5.885x^8/(1-x))/2
 \le\sum_{j=2}^{24}\alpha_{i,j}x^j+
        \sum_{\ell=1}^9\gamma_{i,\ell}x^\ell/(1-x)
 \quad(0\le x<1)
\]
for each of $16$ blocks, by clearing the denominator with $(1-x)/x$. 
\paragraph{Radial integration and a shared constraint.}
For $i=0,\ldots,15$ and
$h=1,\ldots,32$, put $v_{i,h}:=(32i+h)/512$ and
$v_{i,0}:=i/16$. At $a=\eta_0v_{i,h}$ we freeze the six-pass tuple at its uniquely isolated
$45$-place upward ceilings, then form and upward-round
its tangent coefficients $A_{i,h,\ell},B_{i,h,\ell}$. Define
\[
 \beta_{i,h}:=\frac{1-v_{i,h}}{1-\eta_0v_{i,h}/\zeta_0^2},\qquad
 Q_{i,h,\ell}:=\mathcal U_{45}
 (\gamma_{i,\ell}\frac{\eta_0^\ell}{\ell}
       (v_{i,h}^\ell-v_{i,h-1}^\ell)A_{i,h,\ell}).
\]
We retain only nonzero terms. The termwise ceiling retains each covariance contribution. More explicitly, put
\[
 P_{i,j}:=\mathcal U_{45}(\alpha_{i,j}\operatorname{df}(j)\eta_0^j(v_{i,32}^j-v_{i,0}^j)/j)
 \qquad(2\le j\le24),
\]
and define
\[
 \begin{aligned}
 P_j&:=\sum_{i=0}^{15}P_{i,j}&& (3\le j\le24),\\
 P_2&:=\sum_{i=0}^{15}\mathcal U_{45}(P_{i,2}
 +\sum_{h=1}^{32}\sum_{\ell=1}^9\gamma_{i,\ell}\eta_0^\ell
 (v_{i,h}^\ell-v_{i,h-1}^\ell)B_{i,h,\ell}/\ell).
 \end{aligned}
\]
These are the first-row accumulation rules of Section~\ref{subsec:sub81_data}.
There are at most $4608$ nonzero $Q$ terms.

Let $L_H$ be the logarithmic profile of Lemma~\ref{lem:1070_transfer}. For $x\ge0$ define the scalar profile $\mathcal R(x)$, which is distinct from the energy functional $\mathcal R(f)$, by
\[
 \mathcal R(x):=\sum_{j=2}^{24}\frac{P_j}{(1+x)^j}
  +\sum_{i=0}^{15}\sum_{h=1}^{32}\sum_{\ell=1}^9\frac{Q_{i,h,\ell}}
                          {(1+\beta_{i,h}x)^\ell}.
\]
The final endpoint has $\beta=0$, and we include it. Let $\gamma_1,\ldots,\gamma_m$ be the eigenvalues of $\Gamma$. With
$x_s:=2\kappa_0\gamma_s$ and $\theta_s:=\gamma_s/(n\eta_0)$, fixed-reference
radial integration and Eq.~\eqref{eq:metric_energy} give
\[
 -n^{-1}\log Z_1\le\sum_{s=1}^m\theta_s\mathcal R(x_s),\qquad
 \sum_{s=1}^m\theta_s\le1.
\]
The first step integrates every remaining positive power exactly after
right-endpoint freezing. The second is the Gram trace bound. Zero Gram
eigenvalues have zero weight. The displayed scalar majorants hold on the complete spectral interval.

The whole-half-line certificate is
\begin{equation}
 \kappa_0\eta_0L_{H_0}(x)+\frac{H_0}{H_0-1}\mathcal R(x)<e_0
 \qquad(x\ge0).
 \label{eq:metric_profile}
\end{equation}
The endpoint constraints and interpolation bound in Section~\ref{subsec:sub81_data}
verify it, with certified reserve at least $1.6\cdot10^{-7}$.
Our construction includes the final endpoint $\beta=0$ and bounds all
accumulation-rounding errors.

The first row imposes one shared statistic. Put
\[
 f_3(u):=\frac{u^6}{1-u^2}\quad(|u|<1),\qquad
 f_3(u):=+\infty\quad(|u|\ge1).
\]
At the terminal point, let $(p,L,k)$ be the same frozen six-pass one-slot
tuple that we used above, and define
\[
 d:=p^2L(2L-1)/2,\qquad z:=(\sqrt L/2+\sqrt{kd})^2.
\]
Define
\[
 M_3:=\mathcal U_{45}(18\eta_0^3pL(d+2z)).
\]
The degree-three instance of Lemma~\ref{lem:derivative_array_recurrence},
with the ordinary second-derivative mean bound from
Lemma~\ref{lem:sub81_low_means}, gives
$\E_{\mu_1}[\tr[f_3(Y)]]\le nM_3$.
Indeed, its top normalized energies are $(L,L)$, and the first odd step gives
$(pL^2,pL(L-1))$. The next variance is $d$, and the even projection is
at most $z$. The last odd step, multiplied by $(3!)^2\eta_0^3$, is
$18\eta_0^3pL(d+2z)$. Here we use no covariance improvement in the operator
mean or gap. The exact terminal ceiling is
\[
 M_3=0.011088253617755063971661323336071274275573242.
\]
The expansion $f_3(u)=\sum_{j=0}^{\infty}u^{6+2j}$, interpreted as an
increasing extended-valued limit, proves that $f_3$ and its spectral trace
are even and convex.
Its finite sublevel sets are closed and exclude the spectral boundary.

\begin{remark}[The Markov-charge route not taken]
\label{rem:sub81_markov_route}
A Markov-charge route to the first row exists and the final signing does not take it. It adds the hinge $h_*(u):=(|u|-t_*)_+^3$, with $t_*:=37/40$ and $r_*:=0.00000776$, for which $h_*(u)\le r_*u^8/(1-u^2)$ on $|u|<1$, and pays the charge $H_0\chi/(H_0-1)$ with the scalar $\chi:=\log((1+\delta_{\rm mom})/\delta_{\rm mom})<19$ in place of the unit charge $H_0/(H_0-1)$.
\end{remark}

\subsection{Sharper residual moments and quadratic-retaining envelopes}
\label{subsec:sub81_residual}

In this subsection we use barrier $\mathsf R$ from Section~\ref{subsec:sub81_barriers}. The first residual row uses variance $0.14$, the second $0.178$, and the remaining eight $0.19$. All lie below $81/424$, the one-slot bound for $\zeta=0.9$. At the largest cap, the curvature reserve is $62/81$ and the one-slot discriminant is $11/1550$.

The residual trace recurrence uses sharper mean estimates at degrees two, three and four.

\begin{lemma}[Two low-degree ordinary mean bounds]
\label{lem:sub81_low_means}
Let $X:=\sum_{i=1}^m\xi_iN_i$, where $N_i=N_i^\top$ and the centered law satisfies $\E[\xi\xi^\top]\preceq I$. Set $S:=\sum_{i=1}^mN_i^2$. Then
\[
 \|\E[X^2]\|_F^2\le\tr[S^2],\qquad
 \sum_{i=1}^m\|\E[\partial_i(X^3)]\|_F^2\le9\tr[S^3].
\]
\end{lemma}
\begin{proof}
Put $V:=\E[X^2]\preceq S$ and $\Psi(U):=\E[XUX]$. This map is completely positive, trace self-adjoint, and $\Psi(I)=V$. The first step follows from the Schatten $2$-norm monotonicity for $0\preceq V\preceq S$.

Next we apply Eq.~\eqref{eq:cp_second_moment} to each deterministic symmetric $N_i$ and sum:
\[
 \sum_{i=1}^m\tr[\Psi(N_i)^2]\le\tr[V\Psi(S)].
\]
The Schatten shift of Fact~\ref{fact:cp_schatten_shift}, applied to $\Psi$ with exponent three, gives
\[
 \tr[V\Psi(S)]=\tr[S\Psi(V)]
 \le\|S\|_3\|\Psi(V)\|_{3/2}
 \le\|S\|_3\|V\|_3^2\le\tr[S^3].
\]
The first step uses trace self-adjointness, the second H\"older, the third the shift inequality, and the fourth $V\preceq S$. Also
\[
 \sum_{i=1}^m\|N_iV\|_F^2=\sum_{i=1}^m\|VN_i\|_F^2
 =\tr[SV^2]\le\|S\|_3\|V\|_3^2\le\tr[S^3].
\]
Finally $\E[\partial_i(X^3)]=N_iV+\Psi(N_i)+VN_i$. Cauchy--Schwarz for this sum of three terms, followed by summation in $i$, gives the factor nine. We use no commutativity assumption.
\end{proof}

The degree-four trace recurrence also requires bounds for the ordinary mean and the mean of its second derivative.

\begin{lemma}[Fourth-degree ordinary means]
\label{lem:quartic_trace_means}
Let $N_1,\ldots,N_m\in\R^{n\times n}$ be symmetric, and put
$S:=\sum_{i=1}^mN_i^2$ and $X:=\sum_{i=1}^m\xi_iN_i$. Suppose the
centered law of $\xi$ has covariance at most $I_m$. Then
\begin{equation}
 \sum_{i=1}^m\sum_{j=1}^m
 \|\E[\partial_i\partial_j(X^4)]\|_F^2
 \le144\tr[S^4].
 \label{eq:quartic_second_derivative}
\end{equation}
If the law is an even convex tilt of standard Gaussian measure, then also
\begin{equation}
 \|\E[X^4]\|_F^2\le9\tr[S^4].
 \label{eq:quartic_zeroth_derivative}
\end{equation}
Both constants are sharp for these two inequalities. No commutation between
the coefficient matrices is assumed.
\end{lemma}
\begin{proof}
Put $V:=\E[X^2]$, $\mathcal T(U):=\sum_{i=1}^mN_iUN_i$, and
$\Psi(U):=\E[XUX]$. Covariance domination gives $0\preceq V\preceq S$.
The maps $\mathcal T$ and $\Psi$ are completely positive and self-adjoint
for the trace pairing, with symmetric Kraus matrices. They satisfy
$\mathcal T(I)=S$ and $\Psi(I)=V$. We use the self-adjoint completely
positive Schatten shift of Fact~\ref{fact:cp_schatten_shift}, here
with exponent four. It gives
\begin{equation}
 \|\mathcal T(U)\|_2\le\|S\|_4\|U\|_4,
 \qquad \|\Psi(U)\|_2\le\|V\|_4\|U\|_4
                         \le\|S\|_4\|U\|_4.
 \label{eq:quartic_cp_shift}
\end{equation}
The proof of that shift uses only complete positivity and trace
self-adjointness, so it applies to both maps. In particular,
$\|\Psi(S)\|_2,\|\mathcal T(S)\|_2\le\|S\|_4^2$.

Every $N_i$ annihilates $\ker S$: for $z\in\ker S$,
$\sum_{i=1}^m\|N_i z\|^2=z^\top Sz=0$. Now we restrict to the support of $S$.
The discarded block is zero in all the expressions under consideration.
We may therefore suppose $S$ positive definite, with the zero case immediate.
The map $\Psi_S(U):=S^{-1/2}\Psi(U)S^{-1/2}$ is subunital and completely
positive. Its Schwarz inequalities are
\[
 \Psi_S(U)^\top\Psi_S(U)\preceq\Psi_S(U^\top U),\qquad
 \Psi_S(U)\Psi_S(U)^\top\preceq\Psi_S(UU^\top).
\]
We obtain them by applying the map to the positive block matrix with diagonal
blocks $U^\top U,I$, enlarging the lower-right block to $I$, and taking its
Schur complement. Transposition gives the second inequality.

Set $Y_i:=\Psi_S(N_i)$. These matrices are symmetric, and
$\sum_{i=1}^mY_i^2\preceq\Psi_S(S)$. In an eigenbasis of $S$, the scalar
inequality $s^2t+st^2\le s^3+t^3$ proves
$\tr[S^2YSY]\le\tr[S^3Y^2]$ for every symmetric $Y$. Consequently,
\begin{equation}
 \begin{aligned}
 \sum_{i=1}^m\tr[S\Psi(N_i)^2]
 &=\sum_{i=1}^m\tr[S^2Y_iSY_i]\\
 &\le\tr[S^3\sum_{i=1}^mY_i^2]
 \le\tr[S^2\Psi(S)]\\
 &\le\|S^2\|_2\|\Psi(S)\|_2\le\tr[S^4].
 \end{aligned}
 \label{eq:quartic_middle_word}
\end{equation}
The first step expands the normalization. The second is the preceding
scalar comparison. The third is Schwarz and cyclicity. The fourth is
trace Cauchy--Schwarz. The fifth uses Eq.~\eqref{eq:quartic_cp_shift}.

Let $G:=\sum_{i=1}^m \xi'_iN_i$ and $H:=\sum_{i=1}^m \xi''_iN_i$ for two
independent standard Gaussian vectors, independent also of $\xi$.
For $U:=GH$ we have $\E_{\xi',\xi''}[U^\top U]=\E_{\xi',\xi''}[UU^\top]=\mathcal T(S)$.
If $Y:=\Psi_S(U)$, the identity
\[
 \tfrac12\tr[S^2(Y^\top Y+YY^\top)]-\tr[SY^\top SY]
 =\tfrac12\|SY-YS\|_F^2\ge0
\]
and the two Schwarz inequalities imply
\begin{equation}
 \begin{aligned}
 \E_{\xi',\xi''}[\|\Psi(GH)\|_F^2]
 &\le\tr[S\Psi(\mathcal T(S))]\\
 &=\tr[\Psi(S)\mathcal T(S)]
 \le\|\Psi(S)\|_2\|\mathcal T(S)\|_2
 \le\tr[S^4].
 \end{aligned}
\end{equation}
The first step applies the identity and Schwarz before taking the
Gaussian expectation. The second uses trace self-adjointness. The third
is Cauchy--Schwarz. The fourth uses Eq.~\eqref{eq:quartic_cp_shift}.
The same arguments give
\[
 \begin{aligned}
 \E_{\xi',\xi''}[\|VGH\|_F^2]
 &=\tr[V^2\mathcal T(S)]
 \le\|V\|_4^2\|\mathcal T(S)\|_2\le\tr[S^4],\\
 \E_{\xi',\xi''}[\|GVH\|_F^2]
 &=\tr[VSVS]\le\tr[V^2S^2]
 \le\|V\|_4^2\|S\|_4^2\le\tr[S^4].
 \end{aligned}
\]
In the first line the first step integrates the two independent directions,
the second is H\"older, and the third uses $V\preceq S$ and the shift.
In the second line the second step is the nonnegative commutator-square
identity for $V,S$. The remaining steps are H\"older and norm monotonicity.
Transposition also treats $GHV$. Equation~\eqref{eq:quartic_middle_word}
bounds both $\E_{\xi',\xi''}[\|\Psi(G)H\|_F^2]$ and
$\E_{\xi',\xi''}[\|G\Psi(H)\|_F^2]$ by $\tr[S^4]$.

The full second directional derivative $D^2(X^4)[G,H]$ has twelve words.
After taking expectation in $\xi$, the six placements in which $G$
precedes $H$ are
\[
 VGH,\quad GVH,\quad GHV,\quad
 \Psi(G)H,\quad G\Psi(H),\quad\Psi(GH).
\]
The other six exchange $G,H$. Every word has squared $L^2(\xi',\xi'')$
Hilbert--Schmidt norm at most $\tr[S^4]$, as just proved. Thus
Cauchy--Schwarz over the twelve words gives
\[
 \E_{\xi',\xi''}[\|\E_\xi[D^2(X^4)[G,H]]\|_F^2]
 \le12^2\tr[S^4].
\]
Gaussian polarization identifies the left side exactly with the full
ordered sum in Eq.~\eqref{eq:quartic_second_derivative}, including repeated
indices. This proves the first assertion.

For the second assertion, we use the even Poincar\'e bound
$\V_\mu[f]\le\tfrac12\E_\mu[\|\nabla f\|^2]$ from
the proof of Lemma~\ref{lem:algorithm_1240_mixed_moments}. Then we apply it componentwise to
$X^2z$ for every fixed vector $z$. Since
$\partial_i(X^2)=N_iX+XN_i$, we obtain
\[
 \begin{aligned}
 \E[X^4]-V^2
 &\preceq\tfrac12\sum_{i=1}^m
       \E[(N_iX+XN_i)^\top(N_iX+XN_i)]\\
 &\preceq\Psi(S)+\mathcal T(V)\preceq2\mathcal T(S).
 \end{aligned}
\]
The first step is even Poincar\'e, the second uses
$(A+B)^\top(A+B)\preceq2A^\top A+2B^\top B$, and the third uses
$V\preceq S$ and $\Psi(S)\preceq\mathcal T(S)$. The second of these comparisons
follows directly from covariance domination, applied to the vectors
$S^{1/2}N_i z$ for every $z$.
Therefore
\[
 \|\E[X^4]\|_2
 \le\|V^2+2\mathcal T(S)\|_2
 \le\|V\|_4^2+2\|\mathcal T(S)\|_2
 \le3\|S\|_4^2.
\]
The first step is Schatten norm monotonicity on positive matrices, the
second is the triangle inequality, and the third is the shift and
$V\preceq S$. Squaring proves Eq.~\eqref{eq:quartic_zeroth_derivative}.
The even Poincar\'e assertion includes extended-valued convex tilts by
the same symmetric convex exhaustion as in its source proof.
Finally, in dimension one with $m=1$, $N_1=1$ and standard Gaussian
$\xi_1$, the two left sides are $12^2$ and $3^2$, respectively. Thus
both constants are attained.
\end{proof}

For degree $1\le\ell\le9$ and positive even $r\le\ell$, define
\[
 \widehat b_{\ell,r}:=
 \begin{cases}
 1,&r=2,\ \ell\in\{2,3,4\},\\
 9,&(\ell,r)=(4,4),\\
 10^{-30}\lceil10^{30}\operatorname{df}(\ell)^{r/\ell}\rceil,&\text{otherwise}.
 \end{cases}
\]
In the one-slot trace recurrence we use $m_r:=\widehat b_{\ell,r}/(r!)^2$ and leave the operator recurrence unchanged. The top constant energy is $L$. The four special cases follow from Lemmas~\ref{lem:sub81_low_means} and~\ref{lem:quartic_trace_means}. All other cases follow from Lemma~\ref{lem:envelope_moments}. Odd means vanish. The resulting trace energy, denoted $\widehat{\mathcal C}_\ell(a)$, therefore satisfies
\[
 n^{-1}\E[\tr[X^{2\ell}(I-X^2)^{-1}]]
 \le a^\ell\tau_\ell\widehat{\mathcal C}_\ell(a)
 \quad\text{if }\tr[(\sum_{i=1}^mN_i^2)^\ell]\le na^\ell\tau_\ell.
\]
This is Lemma~\ref{lem:derivative_array_recurrence} with $T:=na^\ell\tau_\ell(\ell!)^2$. At actual active fraction $\alpha$ the normalization gains $\alpha^\ell$. We multiply neither $L$ nor $k$ by $\tau_\ell$. Nonnegative arithmetic and finite minima preserve monotonicity in the nominal variance.

For the input caps $\tau_1,\ldots,\tau_6$ that we construct below, we set $\tau_j:=\tau_6$ at larger integer indices. This is valid because the common envelope is at most $I$. The ten residual rows use the nonnegative arrays defined in
Section~\ref{subsec:sub81_data}, with
\[
 g_{\rm R}(x):=\tfrac12(b_3x^3+b_4x^4+b_5x^5+b_7x^7+b_8x^8/(1-x))
 \le\sum_{j=1}^{24}\alpha_{i,j}x^j+
            \sum_{\ell=1}^9\gamma_{i,\ell}x^\ell/(1-x).
\]
On $0\le x<1$ we certify each of the $16$ blocks by the numerator on the closed interval $[0,1]$. Ordinary powers have costs $\operatorname{df}(j)\eta^j\tau_j$, and weighted powers have costs $\eta^\ell\tau_\ell\widehat{\mathcal C}_\ell(\eta v)$. Unlike the first-row majorant, the ordinary sum here starts at $j=1$, and such a term is admissible at cost $\operatorname{df}(1)\eta\tau_1$. The one-slot radial procedure of Section~\ref{subsec:sub81_first_row}, with the cell data of Section~\ref{subsec:sub81_data}, with residual-row accumulation, defines their upper sum $R$.

From the residual radial bounds we obtain small-ball estimates that retain the envelope caps for every active subfamily.

\paragraph{The joint first residual row.}
We retain the contraction from the quadratic tilt throughout the one-slot trace recurrence.

\begin{lemma}[Contraction of the complete one-slot trace energy]
\label{lem:joint_metric_trace}
Let $E_1,\ldots,E_m\in\R^{n\times n}$ be symmetric. Set
$S:=\sum_{i=1}^m E_i^2\preceq\eta I$, $Y:=\sum_{i=1}^m\xi_iE_i$, and
$\Gamma_{ij}:=\tr[E_iE_j]$. Use the half barrier $\widehat\phi$ of
$\mathsf R$ in Section~\ref{subsec:sub81_barriers}, with quadratic
coefficient $\kappa:=0.09904783$, allocation $\zeta:=0.9$, and boundary
coefficient $3.49667654>2$. Keep
\[
 C_\Gamma:=(I+2\kappa\Gamma)^{-1},\qquad
 \nu:=\mathcal N(0,C_\Gamma)
\]
fixed throughout the radial path. At parameter $v\in[0,1]$, put $X:=\sqrt vY$
and let $\mu_v$ have density proportional to
$\exp[-\tr[\widehat\phi(X)-\kappa X^2]]$ relative to $\nu$.
Assume $\eta\le\zeta^2$. Fix a right endpoint $b\in(0,1]$ with
$\eta b<\zeta^2/(1+4\zeta^2)$, and define
\[
 \lambda:=1-\eta b/\zeta^2,\qquad
 \beta:=\frac{1-b}{\lambda},\qquad
 D_\beta:=(I+2\kappa\beta\Gamma)^{-1}.
\]
Then $0\le\beta\le1$. For $0\le v\le b$ and every degree $1\le\ell\le9$,
\begin{equation}
 \E_{\mu_v}[\tr[X^{2\ell}(I-X^2)^{-1}]]
 \le v^\ell\widehat{\mathcal C}_\ell(\eta b)\tr[S(D_\beta)^\ell],
 \qquad
 S(D):=\sum_{i=1}^m\sum_{j=1}^mD_{ij}E_iE_j.
 \label{eq:joint_contracted_trace}
\end{equation}
Here $\widehat{\mathcal C}_\ell(\eta b)$ is the same residual trace recurrence:
compute the exact six-pass one-slot tuple at $\eta b$, round its entries
upward to $45$ places, evaluate the quartic-refined degree-$\ell$ recurrence,
and round its trace energy upward to $45$ places.
\end{lemma}
\begin{proof}
We apply Lemma~\ref{lem:sub10_contracted_coordinates} to barrier $\mathsf R$ with the right endpoint $b$, so that $D=D_\beta$ and $0\le\beta\le1$. By part (ii), the weighted Bochner and projection-aware variance arguments apply in the coordinates $\xi=D_\beta^{1/2}z$ with $p=\lambda^{-1}$. By part (i), the law of $z$ is an even convex tilt of standard Gaussian measure and
\[
 X=\sum_{i=1}^m z_iN_i',\qquad
 N_i':=\sqrt v\sum_{j=1}^mE_j(D_\beta^{1/2})_{ji}.
\]
Thus all ordinary-mean estimates, including the even Poincar\'e
quartic estimate, apply to the actual transformed series.
In particular, the raw derivative directions here are the coefficients
$N_i'$ of this same series. We do not use covariance domination alone
to assert a stronger Poincar\'e inequality.

By part (iii), the old operator mean and
arithmetic--harmonic gap are still bounded by the nominal tuple
$L,k$. Their upward-frozen values, and the upward-frozen $p$, are valid.
Moreover $\sum_{i=1}^m(N_i')^2=vS(D_\beta)\preceq vS\preceq\eta bI$ by part (i).
We normalize the full ordered derivative arrays by
$(\ell!)^2\tr[(vS(D_\beta))^\ell]$. A zero denominator is the zero-series
case. The top derivative, the ordinary means and the one-slot
odd/even projection bounds give exactly the residual recurrence.
This proves Eq.~\eqref{eq:joint_contracted_trace}. We insert no trace factor
in $L$ or $k$. The matrix $D_\beta$ stays
fixed on the cell, and we never differentiate it in radial integration.
The reference Gaussian remains $\nu$.
\end{proof}

A common set of scalar weights links the contracted variance bounds to the determinant contribution.

\begin{lemma}[A joint spectral weighting]
\label{lem:joint_gram_variance_weights}
Let $S,\Gamma,E_i$ be as above. Rotate the coefficient index orthogonally so
that $\Gamma$ is diagonal, with eigenvalues $\gamma_i\ge0$, and put
$x_i:=2\kappa\gamma_i$. Let $u_a$, $1\le a\le n$, be an orthonormal eigenbasis
of $S$, with eigenvalues $\eta y_a$, $0\le y_a\le1$. There are numbers
$\pi_{ia}\ge0$, with $\sum_{i=1}^m\pi_{ia}=1$, such that, for every integer
$j\ge1$ and $\beta\ge0$,
\begin{align}
 \tr[S(D_\beta)^j]
 &\le\eta^j\sum_{a=1}^n\sum_{i=1}^m
       \pi_{ia}\frac{y_a^j}{(1+\beta x_i)^j},
 \label{eq:joint_variance_kernel}\\
 \sum_{i=1}^m\gamma_iL_H(x_i)
 &=\eta\sum_{a=1}^n\sum_{i=1}^m\pi_{ia}y_aL_H(x_i).
 \label{eq:joint_determinant_weights}
\end{align}
The same weights work for every $j,\beta$ and radial parameter. No
commutation of the coefficient matrices with $S$ is assumed.
\end{lemma}
\begin{proof}
For $y_a>0$, define
\[
 \pi_{ia}:=\frac{\|E_i u_a\|^2}{\eta y_a}.
\]
Their sum in $i$ is one because $\sum_{i=1}^mE_i^2=S$. If $y_a=0$, every $E_i u_a$
is zero. We choose any probability vector for that $a$. All terms multiplied
by $y_a$ then vanish. Orthogonal rotation preserves $S$. We do not assume
that individual rotated coefficients retain the original contraction
representation.

Write $\mathcal B:=[E_1\ \cdots\ E_m]$. For a positive coefficient matrix
$D$, Lemma~\ref{lem:iterated_envelope_trace_power} gives
\[
 \begin{aligned}
 \tr[S(D)^j]
 &=\tr[(\mathcal B(D\otimes I)\mathcal B^\top)^j]\\
 &\le\tr[(\mathcal B^\top\mathcal B)^j(D^j\otimes I)]\\
 &=\tr[S^{j-1}\mathcal B(D^j\otimes I)\mathcal B^\top].
 \end{aligned}
\]
The first step expands $S(D)$. For the second step, the nonzero eigenvalues
are those of $(D^{1/2}\otimes I)\mathcal B^\top\mathcal B(D^{1/2}\otimes I)$.
We apply the trace-power inequality in that larger space. The third step uses
$(\mathcal B^\top\mathcal B)^j=\mathcal B^\top S^{j-1}\mathcal B$ and cyclicity.
It includes $j=1$ and singular matrices.

For $D=D_\beta$, the last expression is
\[
 \sum_{a=1}^n\sum_{i=1}^m
 (\eta y_a)^{j-1}\frac{\|E_i u_a\|^2}{(1+\beta x_i)^j},
\]
which proves Eq.~\eqref{eq:joint_variance_kernel}, including zero eigenvalues.
Also $\gamma_i=\tr[E_i^2]=\sum_{a=1}^n\|E_i u_a\|^2$, proving
Eq.~\eqref{eq:joint_determinant_weights}. The physical variance moments
satisfy $n^{-1}\sum_{a=1}^n y_a^j=\tr[S^j]/(n\eta^j)$.
\end{proof}

\paragraph{Exact positive radial kernels.}
For the first residual row we use $\eta:=7/50$, $\zeta:=9/10$,
$\kappa:=0.09904783$, and the $16$ scalar majorants assigned to
residual row one in Section~\ref{subsec:sub81_data}. Their coefficients obey
\[
 g_{\rm R}(t)\le\sum_{j=1}^{24}\alpha_{i,j}t^j
             +\sum_{\ell=1}^9\gamma_{i,\ell}\frac{t^\ell}{1-t}
 \qquad(0\le t<1).
\]
The physical scalar majorants and the transfer profile use different
variables. We check the physical scalar majorants on their full closed cleared domain.
Put $v_h:=h/512$, $\beta_h:=(1-v_h)/(1-\eta v_h/\zeta^2)$, and let $E_{h,j}$
be the $45$-place energy from Lemma~\ref{lem:joint_metric_trace} at $v_h$.
Define the nonnegative exact coefficients
\begin{align*}
 A_j&:=\mathcal U_{45}(
  \sum_{i=0}^{15}\frac{\alpha_{i,j}\operatorname{df}(j)\eta^j}{j}
       \{((i+1)/16)^j-(i/16)^j\}),\\
 Q_{h,j}&:=\mathcal U_{45}(
  \frac{\gamma_{\lfloor(h-1)/32\rfloor,j}\eta^j E_{h,j}}{j}
       (v_h^j-v_{h-1}^j)).
\end{align*}
We omit exact zero coefficients. The upper-sum convention is that we
aggregate ordinary coefficients by degree before rounding, and round each
weighted cell coefficient separately. Tuple-before-energy
rounding is unchanged. By outward $90$-place arithmetic we isolate every
named ceiling. Define
\[
 r_j(x):=\frac{A_j}{(1+x)^j}
       +\mathbbm{1}[j\le9]\sum_{h=1}^{512}
                 \frac{Q_{h,j}}{(1+\beta_h x)^j}
 \qquad(1\le j\le24).
\]
The finite list has $520$ nonzero terms. We retain terms with $\beta_h=0$, in
particular at $h=512$.

The joint weighting reduces Gaussian transfer to a scalar inequality for these kernels.

\begin{lemma}[The complete joint transfer profile]
\label{lem:joint_profile}
Set $H:=17/5$ and
\[
 p(y):=0.0003494+0.1605y^3+0.4636y^4.
\]
The exact kernels just defined satisfy
\begin{equation}
 \kappa\eta yL_H(x)
 +\frac{H}{H-1}\sum_{j=1}^{24}y^jr_j(x)
 \le p(y)
 \qquad(x\ge0,\ 0\le y\le1),
 \label{eq:joint_scalar_profile}
\end{equation}
where $L_H$ is the profile of Lemma~\ref{lem:1070_transfer}.
\end{lemma}
\begin{proof}
For $j\ge4$ we replace $y^j$ by $y^4$. All kernel coefficients are nonnegative,
so this is an upper bound. The expression to be bounded is now a polynomial
of degree four in $y$, with rational/logarithmic functions of $x$ as
coefficients. We split its $x$-domain into
$[0,10^{-8}]$, $[10^{-8},100]$, and $[100,\infty)$.

At the origin we use $L_H(x)\le H10^{-8}/2$ and $r_j(x)\le r_j(0)$.
At infinity we use $L_H(x)\le\log101/100$ and $r_j(x)\le r_j(100)$.
We subtract each resulting degree-four upper polynomial from $p$ and verify
nonnegativity on the complete interval $0\le y\le1$.

For a middle cell $[a,b]$, put $c:=(a+b)/2$ and $h:=(b-a)/2$.
Write the folded upper function as $f(x,y)$. Its second $x$-derivative is
bounded above by the nonnegative polynomial
\[
 M_a(y):=\frac{2\kappa\eta H}{3(H-1)}y
 +\frac{H}{H-1}\sum_{(g,j,\beta,q)}
       \frac{j(j+1)\beta^2q}{(1+\beta a)^{j+2}}y^g,
\]
where $(g,j,\beta,q)$ enumerates the $520$ positive kernel terms and
$g:=\min\{j,4\}$. The first term is the scalar-profile bound on
$L_H''$. The other terms follow by exact differentiation of the positive
rational kernels. Taylor's theorem gives
\[
 f(x,y)\le f(c,y)+h|\partial_x f(c,y)|+h^2M_a(y)/2.
\]
It therefore suffices to certify both degree-four polynomials
\[
 p(y)-f(c,y)-\sigma h\partial_xf(c,y)-h^2M_a(y)/2
 \qquad(\sigma\in\{-1,1\})
\]
on $[0,1]$. In the upper polynomials we apply outward rounding
coefficientwise. This is valid since $y^g\ge0$.

The exact Bernstein test, with midpoint subdivision in $y$, proves both
polynomials on every accepted $x$-cell. We bisect unresolved $x$-cells.
Starting from the complete middle interval, the tests accept every
terminal cell within $x$-depth $12$ and $y$-depth $8$, including the two
endpoint ranges. The generated endpoints form a partition. The
logarithms use the rational atanh-series enclosures. Kernel and
Taylor arithmetic is outward. In Section~\ref{subsec:sub81_data} we specify the outward arithmetic. These complete-domain checks
prove Eq.~\eqref{eq:joint_scalar_profile}.
\end{proof}

Combining the joint transfer profile with the envelope trace bounds, we obtain a small-ball estimate for every active subfamily.

\begin{lemma}[Hereditary joint first residual row]
\label{lem:joint_first_residual}
Suppose $B_i=T^{1/2}A_iT^{1/2}$, $0\preceq T\preceq I$, and all $A_i$ are
real symmetric contractions. Suppose $\tr[T^{2j}]\le n\tau_j$ for the
available envelope powers, including $j=3,4$. Put
\[
 e_{\rm joint}:=0.0003494+0.1605\tau_3+0.4636\tau_4.
\]
For every $w>0$ and every nonempty subfamily of actual size $m\le wn$,
\begin{equation}
 \gamma_m[\|\sum_{i=1}^m\xi_iB_i\|
                  \le\sqrt{wn/0.14}]
 \ge\exp[-me_{\rm joint}/w].
 \label{eq:joint_hereditary_row}
\end{equation}
There is no rank term or additional moment-body charge in this row.
\end{lemma}
\begin{proof}
We first take a normalized family with $S\preceq\eta I$ and
$\tr[S^j]\le n\eta^j\tau_j$. Ordinary moments along the fixed-reference
path are at most $\operatorname{df}(j)v^j\tr[S(C_\Gamma)^j]$, by the retained
Gaussian convex comparison and Gaussian trace moments. The weighted
moments are bounded by Lemma~\ref{lem:joint_metric_trace} on each cell, freezing
at its right endpoint. Lemma~\ref{lem:joint_gram_variance_weights} bounds
both sorts of moments using the same numbers $(x_i,y_a,\pi_{ia})$.
Fixed-reference radial integration consequently gives
\[
 -\log\nu[\|Y\|<1]
 \le\sum_{a=1}^n\sum_{i=1}^m\pi_{ia}
                     \sum_{j=1}^{24}y_a^jr_j(x_i).
\]
The first step is the nonnegative partition-function comparison. The
second integrates every remaining power against $\d v/v$, with only
upper endpoint bounds and upward rounding. All degrees are positive.
The boundary coefficient ensures the same finite-energy justification
as in the radial formula.

The H\"older step Eq.~\eqref{eq:1070_transfer_holder} in the proof of Lemma~\ref{lem:1070_transfer}, with $Z_\Gamma:=\det(I+2\kappa\Gamma)^{-1/2}$
and $\mathcal K:=\{\|Y\|<1\}$, and Eq.~\eqref{eq:joint_determinant_weights} yield
\[
 \begin{aligned}
 -\log\gamma_m[\|Y\|<1]
 &\le\sum_{a=1}^n\sum_{i=1}^m\pi_{ia}
  \{\kappa\eta y_aL_H(x_i)
       +\frac{H}{H-1}\sum_{j=1}^{24}y_a^jr_j(x_i)\}\\
 &\le\sum_{a=1}^n p(y_a)
 \le n(0.0003494+0.1605\tau_3+0.4636\tau_4).
 \end{aligned}
\]
The first step keeps determinant and radial contributions together,
rather than maximizing them separately. The second applies the joint
scalar certificate of Lemma~\ref{lem:joint_profile} and $\sum_{i=1}^m\pi_{ia}=1$. The third uses the physical
variance trace caps. Zero Gram eigenvalues and zero physical eigenvalues
are included. We charge the constant $0.0003494$ against the full physical
dimension. It is not a rank penalty and is not multiplied by $\tau_1$.

For the stated hereditary assertion put
$E_i:=\sqrt{\eta/(wn)}B_i$ and $\alpha:=m/(wn)\le1$.
The common-envelope estimate of Lemma~\ref{lem:envelope_moments} gives
\[
 S\preceq\alpha\eta I,\qquad
 \tr[S^j]\le n(\alpha\eta)^j\tau_j.
\]

We keep every nominal metric $D_{\beta_h}$, tuple, energy, and scalar
majorant fixed. The original weighted curvature is now at least
$(1-\alpha\eta v/\zeta^2)I+2\kappa(1-v)\Gamma$, which still dominates
the same nominal cell lower bound used in Lemma~\ref{lem:joint_metric_trace}.
The transformed law is again explicitly an even convex Gaussian tilt.
We normalize the physical eigenvalues by $S u_a=\alpha\eta y_a u_a$.
Each ordinary or weighted degree-$j$ contribution gains $\alpha^j$, and
the determinant term gains $\alpha$. Since every $j\ge1$, the whole
joint expression is at most $\alpha$ times its nominal upper profile.
Thus even its constant cap is multiplied by $\alpha$ after this comparison:
\[
 -\log\gamma_m[\|Y\|<1]
 \le n\alpha e_{\rm joint}=me_{\rm joint}/w.
\]
We apply this scaling to the complete integrand before its scalar
majorization, rather than inferring it by scaling a trace cap. Passing to the closed
norm body proves Eq.~\eqref{eq:joint_hereditary_row}. The zero-series case
has probability one.
\end{proof}

\paragraph{Actual-family convention.}
For each actual subfamily, we form $\Gamma$ from that actual family. The
scalars $\beta_h$ and the operator/trace endpoint tuples remain nominal. The
matrix $D_{\beta_h}$ uses this same actual $\Gamma$ and is held fixed on its
radial cell. Thus in the hereditary proof we do not compare different families'
Gram matrices or reset the reference Gaussian.

\begin{lemma}[The ten hereditary residual rows]
\label{lem:sub81_residual_rows}
Fix a printed row, with its variance $\eta$ and exponent $e$, and fix $w>0$. Suppose $B_i=T^{1/2}A_iT^{1/2}$, $0\preceq T\preceq I$, $A_i=A_i^\top$, $\|A_i\|\le1$, and $\tr[T^{2j}]\le n\tau_j$ for its generated input caps. For every nonempty subfamily of size $m\le wn$,
\[
 \gamma_m[\|\sum_{i=1}^m\xi_iB_i\|\le\sqrt{wn/\eta}]
 \ge\exp[-me/w].
\]
The row exponent already includes the envelope caps and has no rank term.
\end{lemma}
\begin{proof}
The first row is Lemma~\ref{lem:joint_first_residual}, with $e:=\mathcal U_{45}(e_{\rm joint})$. For rows two through ten, we proceed as follows.
Lemma~\ref{lem:iterated_envelope_one_slot}, applied to barrier $\mathsf R$ as in Section~\ref{subsec:sub81_barriers}, supplies the valid tuple. Lemmas~\ref{lem:sub81_low_means}, \ref{lem:envelope_moments}, and~\ref{lem:derivative_array_recurrence} give the refined trace bounds. With the exact arrays and rounding rules of Section~\ref{subsec:sub81_data}, the radial sum obeys $R\le R_*$, and the complete determinant profile of row $r$ satisfies $L_{H_r}(x)<B_r$ for $x>0$. The values $H_r$, $B_r$, $R_{*,r}$, $\eta_{r+1}$, $\tau_{r,1}$, $e_{r+1}$ of that section stand in place of $H$, $B$, $R_*$, $\eta$, $\tau_1$, $e$ here and below, with
\[
 \kappa\eta\tau_1 B+H R_* /(H-1)<e,
 \qquad\kappa=0.09904783.
\]
We use $D_i:=\sqrt{\eta/(wn)}B_i$, $Y:=\sum_{i=1}^m\xi_iD_i$, $\Gamma_{ij}:=\tr[D_iD_j]$ and the fixed Gaussian $\nu:=\mathcal N(0,(I+2\kappa\Gamma)^{-1})$. Its extra radial quadratic potential is nonnegative. Lemma~\ref{lem:fixed_reference_radial}, with the common-envelope moment bounds and the refined one-slot trace estimates, gives $-n^{-1}\log\nu[\|Y\|<1]\le R_*$. Its cell formula integrates the complete-domain majorants with nominal right-endpoint bounds. If $\Gamma=0$, all coefficient matrices vanish and the conclusion is immediate. Otherwise we apply Lemma~\ref{lem:1070_transfer} with $N:=n$, trace parameter $\eta\tau_1$, $C:=R_*$, $Q_j:=0$, $b:=0$ and $B_{\rm cap}:=B$. Since $\tr[\Gamma]\le n\eta\tau_1$, this gives
\[
 -\log\gamma_m[\|Y\|<1]
 \le n\{\kappa\eta\tau_1B+H R_* /(H-1)\}<ne.
\]
The first step follows from Gaussian transfer and the second from the exact row comparison. At actual size $m=\alpha wn$, ordinary degree-$j$ and weighted degree-$\ell$ terms gain $\alpha^j$ and $\alpha^\ell$, while the Gram trace gains $\alpha$. We keep nominal endpoint bounds, and the scaling clause of Lemma~\ref{lem:1070_transfer} gives exponent $n\alpha e=me/w$. Passing to the closed ball proves the result.
\end{proof}

\subsection{The complete signing below 7.8795}
\label{subsec:sub81_signing}

\paragraph{Mean-level moment bodies.}
The following consequence of log-concavity replaces the Markov charge in
this construction only. All preceding moment-body statements remain valid.

\begin{lemma}[A unit logarithmic charge at the mean]
\label{lem:unit_log_moment}
Let $\mu$ have a log-concave probability density on a finite-dimensional
Euclidean space, and let $F$ be a nonnegative lower-semicontinuous convex
function with finite mean. If $a\ge\E_\mu[F]$, then
\[
 \mu[F\le a]\ge e^{-1}.
\]
The function may take the value $+\infty$ on a $\mu$-null set. If $a=0$,
the displayed probability is one.
\end{lemma}
\begin{proof}
Write $Z:=F(X)$ for $X$ of law $\mu$ and, within this proof, $G(t):=\Pr[Z\le t]$.
The nonnegative function
\[
 (x,t)\longmapsto \rho(x)e^{-t}\mathbbm{1}[F(x)\le t]
\]
is log-concave, where $\rho$ is the density of $\mu$: the epigraph of $F$
is convex. It is integrable because $F\ge0$. Pr\'ekopa's marginal theorem~\cite[Theorem~6]{prekopa73}
therefore makes $e^{-t}G(t)$, and hence $G(t)$, log-concave.
For a uniform $U$ on $(0,1)$, its generalized inverse $G^{-1}(U)$ has the
law of $Z$ and $G(G^{-1}(U))\ge U$. Thus
\[
 \E[\log G(Z)]\ge\int_0^1\log u\d u=-1.
\]
This argument includes atoms. We do not assert that $G(Z)$ is uniform.
Concavity of $\log G$ and Jensen's inequality give
\[
 \log G(\E[Z])\ge\E[\log G(Z)]\ge-1.
\]
The logarithm is integrable by the preceding lower bound and its upper
bound zero. The constant case is immediate. Otherwise the mean is in the
interior of the interval where $G$ is positive, so Jensen applies there.
Monotonicity of $G$ proves the assertion for $a\ge\E[Z]$.
When $a=0$, nonnegativity and zero mean give $F=0$ almost surely.
\end{proof}

\begin{lemma}[The first row with a unit-charge shared constraint]
\label{lem:unit_first_row}
Let $B_i$ be symmetric contractions, $w,L>0$, and $0<m\le wn$.
Set $G:=\sum_{i=1}^m\xi_iB_i$ and $r_0:=\sqrt{wn/\eta_0}$, and suppose
$L^2s_+^2\eta_0\ge w$. With the unchanged $f_3,M_3,\eta_0,H_0,e_0$
of Section~\ref{subsec:sub81_first_row}, the symmetric convex set
\[
 \mathcal K_3:=\{\xi:\|G\|\le r_0,\quad
 \tr[f_3(G/(s_+L\sqrt n))]\le(1+\delta_{\rm mom})nM_3\}
\]
has standard Gaussian measure at least
\[
 \gamma_m[\mathcal K_3]\ge
 \exp\{-me_0/w-H_0/(H_0-1)\}.
\]
\end{lemma}
\begin{proof}
We first use the nominal variance normalization $E_i:=B_i/r_0$,
$Y:=G/r_0$ and the fixed Gaussian $\nu$ and terminal tilted law $\mu_1$
of Section~\ref{subsec:sub81_first_row}. The nominal variance bound
$\sum_{i=1}^mE_i^2\preceq\eta_0I$ holds since $m\le wn$. The terminal recurrence there proves
$\E_{\mu_1}[\tr[f_3(Y)]]\le nM_3$. Put
$c:=r_0/(s_+L\sqrt n)\le1$. Monotonicity of $f_3$ in the absolute value
gives $\tr[f_3(cY)]\le\tr[f_3(Y)]$. Its spectral trace is nonnegative,
even, lower-semicontinuous and convex, with the extended-value convention
already proved for $f_3$.

The law $\mu_1$ has a log-concave density: its negative logarithm is the
Gaussian quadratic form plus the convex spectral barrier. We apply
Lemma~\ref{lem:unit_log_moment} under this one law to the one shared
statistic. Its body has $\mu_1$-mass at least $e^{-1}$. Since
$\d\mu_1=Z_1^{-1}e^{-V}\d\nu$, with
$V=\tr[\widehat\phi_b(Y)-\kappa_0Y^2]\ge0$, and $\mu_1$ is supported
on $\|Y\|<1$, it follows that $\nu[\mathcal K_3]\ge Z_1e^{-1}$.
Under the tilted law we make no correlation assertion.

We then apply the H\"older step Eq.~\eqref{eq:1070_transfer_holder} in the proof of
Lemma~\ref{lem:1070_transfer} with $\kappa:=\kappa_0$, $H:=H_0$ and
$\mathcal K:=\mathcal K_3$. Taking logarithms, using the partition bound, and averaging
Eq.~\eqref{eq:metric_profile} against the weights $\theta_s$, we obtain
\[
 -\log\gamma_m[\mathcal K_3]\le ne_0+\frac{H_0}{H_0-1}
\]
at nominal size.
The profile is nonnegative and the weights sum to at most one, so missing
weight causes no extra charge. At actual size $m=\alpha wn$, we retain all nominal endpoint
tuples, scalar majorants and auxiliary metric parameters, using the Gram
matrix of that actual family. The actual curvature
is at least the same nominal lower bound with the actual Gram matrix,
since $1-\alpha\eta_0v/\zeta_0^2\ge\lambda_v$.
The contracted raw variance has trace powers at most
$n(\alpha\eta_0v)^\ell\omega_\ell^{(\beta_v)}$, now with normalizers
$n\alpha\eta_0$. Ordinary degree-$j$ terms gain $\alpha^j$ and weighted
degree-$\ell$ terms gain $\alpha^\ell$. All degrees are positive. After
division by $\alpha$, each coefficient is at most the nominal one,
and the determinant charge also scales by $\alpha$. The terminal
expectation does not increase. Hence the first term becomes
$n\alpha e_0=me_0/w$, while the positive conditional charge
$H_0/(H_0-1)$ remains fixed. This proves the assertion, including a zero
series. We will use Gaussian correlation only after this transfer to
standard Gaussian measure.
\end{proof}

\paragraph{Coding with the new mass bound.}
Each ordinary nonnegative even convex moment body, at its certified mean upper bound or at $(1+\delta_{\rm mom})$ times that bound, has standard Gaussian mass at least $e^{-1}$ by Lemma~\ref{lem:unit_log_moment}. We therefore apply Lemma~\ref{lem:envelope_joint_coding} with the charge $c:=1$.
For a first-row body from Lemma~\ref{lem:unit_first_row}, taken as $K_0$, we add the fixed charge $\theta:=H_0/(H_0-1)$.
Zero-mean statistics have mass one.

\begin{proof}[Proof of Theorem~\ref{thm:existence_constant_sub10}]
We first assume $N=n$. Set $C_0:=7.8795$, $w_0:=1$, $T_0:=I$ and
$n_*:=2^{43}$. We retain $s_+$ of Lemma~\ref{lem:slack_refined_coding} and the $\delta_{\rm mom}$, $d_8$ defined after it.
Here we use the eleven prefix phases generated in Section~\ref{subsec:sub81_data},
with $C_i:=C_{i-1}-L_i$ and $w_i:=q_iw_{i-1}$. The first ten phases
impose six ordinary moment bodies each, and the last imposes one. All eleven
norm rows have rank correction zero. The first phase also imposes the
single shared $f_3$ constraint of Lemma~\ref{lem:unit_first_row}.

At phase one we use the six functions defined in Section~\ref{subsec:sub81_data}
\[
 p_k(u):=a_{k,0}+a_{k,1}u^2+
 \sum_{t\in\mathcal H_k}b_{k,t}(u-t)_+^3+c_kf_3(u).
\]
Their defining Bernstein constraints imply Eq.~\eqref{eq:iterated_envelope_polynomial},
with $C:=C_0$ and $L:=L_1$, on every closed cell after multiplication
by $1-u^2$. The proof of Lemma~\ref{lem:iterated_envelope_update} uses
only $p_k\ge a_{k,0}$, the scalar majorant and the moment constraint, so
it applies to these functions. Lemma~\ref{lem:iterated_envelope_cubic}
gives six ordinary Gaussian mean bounds
\[
 \mu_k:=a_{k,1}\eta_0+
 \sum_{t\in\mathcal H_k}b_{k,t}\mathcal U_{45}(G_3(\eta_0,t)).
\]
Their bodies have mass at least $e^{-1}$ by
Lemma~\ref{lem:unit_log_moment}. Lemma~\ref{lem:unit_first_row} gives the
one joint norm/$f_3$ body. All six coefficients $c_k$ use this same event.
We apply Gaussian correlation only after return to standard Gaussian
measure. Functional calculus gives direct caps
\[
 \widehat\tau_{1,k}:=\mathcal U_{45}
 (a_{k,0}+(1+\delta_{\rm mom})(\mu_k+c_kM_3)).
\]
Put $\tau_{1,k}:=\min_{1\le h\le k}\widehat\tau_{1,h}$.

We retain these actual certified caps throughout the later phases, with no
reset to larger reference caps. For phases two through eleven set
\[
 \begin{aligned}
 \mu_{i,k}&:=a^{[i]}_{k,1}\eta_i\tau_{i-1,k+1}
 +\sum_{j=2}^{d_{i,k}}a^{[i]}_{k,j}\operatorname{df}(j)\eta_i^j\tau_{i-1,j},\\
 \widehat\tau_{i,k}&:=\mathcal U_{45}
 (a^{[i]}_{k,0}\tau_{i-1,k}+(1+\delta_{\rm mom})\mu_{i,k}),\qquad
 \tau_{i,k}:=\min_{1\le h\le k}\widehat\tau_{i,h}.
 \end{aligned}
\]
Here $d_{i,k}$ is the degree of the generated polynomial $P^{[i]}_k$.
Indices above six use the sixth cap, and phase eleven needs only the first
cap. The defining constraints of each of the $55$ later polynomials use
its fixed pair $(C_{i-1},L_i)$. The weighted case of Lemma~\ref{lem:iterated_envelope_update} proves the displayed bounds
by forward induction. Its quadratic statistic is a squared norm of a
linear map. Its higher-power statistics are unweighted convex spectral
traces. Lemma~\ref{lem:unit_log_moment} therefore applies to the actual
ordinary Gaussian statistics we use here. We apply monotone closure only after
imposing all bodies: $T_i\preceq I$ gives
$\tr[T_i^{2k}]\le\tr[T_i^{2h}]$ for $h\le k$. This deterministic closure
adds no event or charge. We regenerate every residual radial sum from
its incoming caps with the stated outward-rounding order.

The exact radius and entropy tests are
\[
 L_i^2s_+^2\eta_i\ge w_{i-1},\qquad
 n_*\{w_{i-1}d_8(q_i)-e_i\}>b_i,\qquad
 b_i:=\begin{cases}8,&i=1,\\7,&2\le i\le10,\\2,&i=11.\end{cases}
\]
Here $e_1:=e_0$, and $e_i$ is the certified exponent budget of residual row
$i-1$ that we specify in Section~\ref{subsec:sub81_data}, for $2\le i\le11$. We bound the actual logarithmic charges by
\[
 6+H_0/(H_0-1)<8,\qquad 6<7,\qquad1<2,
\]
respectively. In particular, we retain the positive first-row conditional transfer
cost.

For $n>n_*$ and actual count $m=\alpha wn>qwn$, we use retention $q/\alpha$.
The positive even coefficients of $d_8$ and $\alpha\le1$ give
\[
 m\{d_8(q/\alpha)-e/w\}
 =n\{\alpha w d_8(q/\alpha)-\alpha e\}
 \ge n\{wd_8(q)-e\}.
\]
We combine this inequality with Lemma~\ref{lem:unit_first_row} in the first
phase and Lemma~\ref{lem:sub81_residual_rows} thereafter. The coding proof
with the unit-charge mass bounds, as given immediately above, supplies
every partial signing at its actual integer count. In the norm-row proof we impose
no extra moment body. We do not multiply the first conditional charge
by $m$. If $m\le qwn$, we take $M=0$ and still perform the
prescribed common congruence. All nonconstant statistics, including $f_3$,
vanish at zero. Empty families terminate.

After phase eleven we retain $\rho:=\tau_{11,1}$ for every coefficient and
use the four reflected rows $\mathsf A_3,\ldots,\mathsf A_6$ (rows three through six) of
Definition~\ref{def:slack_rows}, with parameters $\widehat\eta_j$ and
$\overline e_j^\circ$, $1\le j\le4$. We replay their spectral, reflected radial and
profile certificates with the generated suffix schedule.
At each capacity $v$, which is not the radial parameter of the preceding proofs, the exact tests are
\[
 c_j^2s_+^2\widehat\eta_j\ge v,\qquad
 vd_8(\widehat q_j)>\rho\overline e_j^\circ.
\]
All four rank parameters are zero. Lemma~\ref{lem:slack_trace_interface}
and Lemma~\ref{lem:envelope_joint_coding}, here with $b:=0$, apply to
every actual subfamily. The final exact capacity satisfies
$-\log v>731/256$, and the entire uniform-tail bound is
\[
 \mathcal T_{s_+}(v)\le 0.000000090723.
\]
The exact total is
\[
 \begin{aligned}
 B&:=\sum_{i=1}^{11}L_i+\sum_{j=1}^4c_j+0.000000090723\\
  &=7.879227923700+0.000265701524+0.000000090723\\
  &=7.879493715947<C_0.
 \end{aligned}
\]
The common nested reserve is $C_0-B=0.000006284053>0$.
Lemma~\ref{lem:slack_schedule_completion} now gives norm at most
$B\sqrt n$, including finite termination and backward lifting at every
level. The number $C_0$ defines positive slack matrices. We do not assume it
as a signing theorem.

For $n\le2^{43}$, the unchanged random-sign branch gives a signing below
$7.82\sqrt n$, because
\[
 7.82^2/2-44\log2>0.0777240553>0.
\]
Fact~\ref{fact:matrix_hoeffding} with uniform signs ($y_i=0$), $\sigma^2:=n$ and $t:=7.82\sqrt n$ bounds its failure probability
by $2ne^{-7.82^2/2}<1$. We enclose the logarithm outward. Since
$7.82<B$, both branches together cover every finite $n$. Zero-block
padding covers $1\le N\le n$ without adding variables.
\end{proof}

\subsection{Exact data and replay rules}
\label{subsec:sub81_data}

Every displayed decimal is exact. In the displayed polynomials, unlisted coefficients are zero. The
following formulas and finite linear programs determine the full construction.
For a polynomial $P$, write $[z^j]P$ for its degree-$j$ coefficient, and $\mathcal U_p(x):=10^{-p}\lceil10^px\rceil$ for the $p$-place upward ceiling.

\paragraph{Parameters and fixed schedule.}
We retain $C_0:=7.8795$, $n_*:=2^{43}$, $s_+:=0.674489750196$,
$\delta_{\rm mom}:=10^{-8}$ and the small-dimension coefficient $7.82$.
The first-row parameters are
\[
 (\eta_0,H_0,e_0,z_*,t_*,r_*):=(0.065,4.338,0.007955,0.67,0.925,0.00000776).
\]
The terminal tuple and $M_3$ use the formulas preceding
Remark~\ref{rem:sub81_markov_route}. Set $\eta_1:=\eta_0$, $\eta_2:=0.14$,
$\eta_3:=0.178$, and $\eta_i:=0.19$ for $4\le i\le11$.
Set $e_1:=e_0$ and
\[
 e_2:=\mathcal U_{45}(0.0003494+0.1605\tau_{1,3}+0.4636\tau_{1,4}).
\]
For residual row $r=2,\ldots,10$, we use the exact values below. Its
input caps are the propagated $\tau_{r,j}$. The reconstructed radial sum obeys
$R_r\le R_{*,r}$, and the whole-half-line determinant profile obeys
$L_{H_r}(x)<B_r$ for $x>0$. We define the residual exponent below from the quantity
\[
 0.09904783\eta_{r+1}\tau_{r,1}B_r
          +H_rR_{*,r}/(H_r-1)<e_{r+1}.
\]
{
\begin{align*}
(H_2,\ldots,H_{10})&:=(2.402939,2.394716,1.992972,1.758147,1.587473,\\
&\quad 1.438139,1.32436,1.247245,1.201003),\\
(B_2,\ldots,B_{10})&:=(0.32089952,0.32045803,0.29711174,0.28155265,0.2691329,\\
&\quad 0.25735114,0.24770819,0.24080055,0.23650001).
\end{align*}
}
Use the displayed rational upper bounds $R_{*,r}$ below. After the
prescribed $16$ blockwise $45$-place ceilings, we verify $R_r\le R_{*,r}$ and put
\[
 e_{r+1}:=\mathcal U_{30}
 (0.09904783\eta_{r+1}\tau_{r,1}B_r+
                         \frac{H_rR_{*,r}}{H_r-1})+10^{-30}.
\]
This recurrence defines the exact rational exponents $e_3,\ldots,e_{11}$.
The first-row and joint-row accumulation rules are unchanged.

{
\begin{align*}
(R_{*,2},\ldots,R_{*,10})&:=(0.000021700182992514369305829458,0.000003115558059222399674886095,\\
&\quad 0.000000242987494089016347385876,0.000000022525511824903699911079,\\
&\quad 0.00000000241867796694625338079,0.000000000280728903744786009913,\\
&\quad 0.000000000039674458198944364857,0.00000000000646248110466016591,\\
&\quad 0.000000000001122400908996784123).
\end{align*}
}

For $1\le k\le6$, define the initial caps from the generated initial functions by
$\widehat\tau_{1,k}:=\mathcal U_{45}(a_{k,0}+(1+\delta_{\rm mom})(\mu_k+c_kM_3))$
and $\tau_{1,k}:=\min_{1\le h\le k}\widehat\tau_{1,h}$. The nonnegative recursion
in the proof generates all subsequent caps without a reference-cap reset.

Fix the prefix and suffix retentions by
\[
\begin{aligned}
10^{10}(q_1,\ldots,q_{11})&:=(1259673254,1206067791,1199787480,1283079925,\\
&\quad1187122124,1160707142,1183201784,1254839036,\\
&\quad1433634986,1758229716,2270194732),\\
10^{10}(\widehat q_1,\ldots,\widehat q_4)&:=(841040,5728112,15081553,16094908).
\end{aligned}
\]
Starting at $w_0:=1$, set
\[
 L_i:=\mathcal U_{12}(\sqrt{w_{i-1}/(s_+^2\eta_i)}),\qquad
 w_i:=q_iw_{i-1},\qquad C_i:=C_0-\sum_{h=1}^iL_h.
\]
The moment construction below determines the caps and exponents for these
fixed retentions. The exact entropy comparisons are
$n_*\{w_{i-1}d_8(q_i)-e_i\}>b_i$, with $b_1:=8$, $b_i:=7$ for
$2\le i\le10$, and $b_{11}:=2$.
The first ten phases have six moment bodies each, and phase eleven has one.
Each ordinary body has charge at most one. The single shared $f_3$
constraint in phase one has charge $H_0/(H_0-1)$.
All rank parameters are zero.

For the four suffix rows $\mathsf A_3,\ldots,\mathsf A_6$ of
Definition~\ref{def:slack_rows}, put $v_0:=w_{11}$,
$\rho:=\tau_{11,1}$ and
\[
 c_j:=\mathcal U_{12}(\sqrt{v_{j-1}/(s_+^2\widehat\eta_j)}),\qquad
 v_j:=\widehat q_jv_{j-1}.
\]
The fixed retentions satisfy $v_{j-1}d_8(\widehat q_j)>\rho\overline e_j^\circ$.
The recurrences give $\sum_{i=1}^{11}L_i=7.879227923700$
and $\sum_{j=1}^4c_j=0.000265701524$. At the final capacity, we use $q_T:=0.011$ in
Eq.~\eqref{eq:slack_summed_tail}. Its full infinite cost is at most
$0.000000090723$.

\begingroup
\paragraph{Moment functions from finite linear programs.}
The six initial functions and the $55$ later polynomials use one selection
rule. Put $\psi_j(u):=(u-j/40)_+^3$, and use the families
\begin{align*}
 p^{[1]}_k(u)&:=a_{k,0}+a_{k,1}u^2+
                   \sum_{j=1}^{31}b_{k,j/40}\psi_j(u)+c_kf_3(u),\\
 p^{[i]}_k(u)&:=P^{[i]}_k(u^2),\qquad
 P^{[i]}_k(z):=\sum_{j=0}^{d_i}a^{[i]}_{k,j}z^j\quad(i\ge2),
\end{align*}
where $1\le k\le6$ for $i\le10$, $k=1$ for $i=11$, and
$d_2:=5$, $d_3:=4$, $d_i:=3$ for $i\ge4$. Set $p_k:=p^{[1]}_k$ and
$\mathcal H_k:=\{j/40:1\le j\le31\}$. Every coefficient lies in $[0,64]$.
The coordinate order is
$(a_{k,0},a_{k,1},b_{k,1/40},\ldots,b_{k,31/40},c_k)$ initially and
$(a^{[i]}_{k,0},\ldots,a^{[i]}_{k,d_i})$ thereafter.

Put $t_i:=L_i/C_{i-1}$ and define the cleared majorant
\[
 Q_{i,k}(u):=\omega_i(u)
       ((1-t_i u)^{2k}p^{[i]}_k(u)-(1-t_i)^{2k}),\qquad
 \omega_1(u):=1-u^2,\quad \omega_i(u):=1\ (i\ge2).
\]
Use degrees $n_{1,k}:=2k+6$ and $n_{i,k}:=2k+2d_i$ for $i\ge2$.
Set $m_1:=40\cdot2^{13}$ and $m_i:=2^{18}$ for $i\ge2$.
On each cell $[h/m_i,(h+1)/m_i]$, $0\le h<m_i$, we require every degree-$n_{i,k}$
Bernstein coefficient of $Q_{i,k}$ to be nonnegative. The initial mesh
contains every hinge endpoint, so each cleared expression is polynomial on its cell.
These are rational linear constraints in the coefficients, using the
conversion of Definition~\ref{def:rational_certificate_rules}.
The feasible polytope is nonempty: the constant function $p^{[i]}_k=1$
satisfies all constraints. Indeed, the Bernstein coefficients of
$(1-t_i u)^{2k}-(1-t_i)^{2k}$ on $[0,1]$ are nonnegative, and multiplication
by $1-u^2$ preserves this property. The fixed schedule has $0<t_i<1$.

At each stage, we minimize the exact moment cost
\begin{align*}
 J_{1,k}&:=a_{k,0}+(1+\delta_{\rm mom})
   (a_{k,1}\eta_0+\sum_{j=1}^{31}b_{k,j/40}
             \mathcal U_{45}(G_3(\eta_0,j/40))+c_kM_3),\\
 J_{i,k}&:=a^{[i]}_{k,0}\tau_{i-1,k}+(1+\delta_{\rm mom})
   (a^{[i]}_{k,1}\eta_i\tau_{i-1,k+1}
    +\sum_{j=2}^{d_i}a^{[i]}_{k,j}\operatorname{df}(j)\eta_i^j\tau_{i-1,j})
    \quad(i\ge2).
\end{align*}
We choose the lexicographically least minimizer in the coordinate order above,
and set
$\widehat\tau_{i,k}:=\mathcal U_{45}(J_{i,k})$ and
$\tau_{i,k}:=\min_{1\le h\le k}\widehat\tau_{i,h}$.
Indices above six use the sixth cap. This specifies the functions in stage
order by rational linear programs with $34$ variables initially and at most
six thereafter. Equivalently, enumerate the feasible vertices by independent
active constraints and minimize the cost followed by the coordinates.

Bernstein nonnegativity proves the cleared inequalities on every closed cell
and hence the required scalar majorants, with endpoint values interpreted
by limits. To certify an upper bound
on a minimum, it suffices to evaluate a feasible rational point. Since the
costs are nondecreasing in the incoming caps, these upper bounds propagate
through all eleven stages. The resulting rational comparisons give the
prefix and suffix entropy tests in the proof and the residual bounds below.
\endgroup

\begingroup\setlength{\jot}{1.5pt}
\paragraph{First-row radial polynomials.}
One rational linear program specifies all $16$ pairs. For each block
$0\le i\le15$, take
$(\alpha^{[0]}_{i,2},\ldots,\alpha^{[0]}_{i,24},
\gamma^{[0]}_{i,1},\ldots,\gamma^{[0]}_{i,9})\in[0,2^{14}]^{32}$ and put
\[
 A_i(z):=\sum_{j=2}^{24}\alpha^{[0]}_{i,j}z^j,\qquad
 G_i(z):=\sum_{\ell=1}^9\gamma^{[0]}_{i,\ell}z^\ell,\qquad
 \Pi_i(z):=\frac{(1-z)(A_i(z)-g_0(z))+G_i(z)}{z}.
\]
We require every degree-$24$ Bernstein coefficient of $\Pi_i$ on
$[j/128,(j+1)/128]$, $0\le j<128$, to be nonnegative.
Here $g_0$ is the first-row barrier majorant above, so $\Pi_i$ extends
to a polynomial. Use Definition~\ref{def:rational_certificate_rules}
for the Bernstein conversion.

Let $\mathcal R^\circ$ be the radial sum defined above with only the
ceilings in $P_{i,j},P_2,Q_{i,h,\ell}$ omitted. The frozen tuple and tangent
coefficients retain their prescribed ceilings. Thus $\mathcal R^\circ(x)$
is linear in the $512$ variables. In the display below, $\rho$ denotes the profile reserve rather than the retained cap $\tau_{11,1}$. Set
\[
 \rho:=1.6\cdot10^{-7},\qquad
 \varepsilon:=5000\frac{H_0}{H_0-1}10^{-45},\qquad
 c:=\frac{2\kappa_0\eta_0H_0^3}{3(H_0-1)}.
\]
Let $\overline L(x)$ and $\overline\ell$ be the upper rational enclosures
of $L_{H_0}(x)$ and $\log101$ from the logarithm rule below, with
$\overline L(0):=0$. We partition $[0,1]$ into cells of length $2^{-10}$
and $[1,100]$ into cells of length $2^{-4}$. For each cell $[a,b]$,
we impose at both endpoints $x\in\{a,b\}$, and then impose at the tail,
\begin{align*}
 \kappa_0\eta_0\overline L(x)+\frac{H_0}{H_0-1}\mathcal R^\circ(x)
 &\le e_0-\rho-\varepsilon-c(b-a)^2/8,\\
 \kappa_0\eta_0\overline\ell/100+
 \frac{H_0}{H_0-1}\mathcal R^\circ(100)
 &\le e_0-\rho-\varepsilon.
\end{align*}
These are rational linear constraints. Exact rational feasibility checks
give a nonempty bounded polytope. We choose its lexicographically least point,
ordering blocks by $i$ and coordinates as displayed. Equivalently,
enumerate intersections of $512$ independent active constraints, retain
the feasible vertices, and take the lexicographically least one.
This fixes all coefficients by finite rational operations.

The Bernstein constraints prove $g_0(z)\le A_i(z)+G_i(z)/(1-z)$.
For the profile, the integral formula in
Lemma~\ref{lem:scalar_profile_enclosures} gives
$L_{H_0}''\ge-2H_0^3/(3(H_0-1))$, and $\mathcal R^\circ$ is convex.
Hence the profile lies at most $c(b-a)^2/8$ above its endpoint chord.
The tail uses the decreasing envelope $\log(1+x)/x$ and decreasing
radial terms. Restoring the at most $4992$ accumulation ceilings increases
the profile by less than $\varepsilon$. Thus
Eq.~\eqref{eq:metric_profile} holds with reserve at least $\rho$.

\paragraph{Residual radial majorants.}
The joint first residual row uses the following fixed majorants. For blocks
$i=0,1$, put
\begin{align*}
 A^{[1]}_0(z)&:=0.0000000088247z+8.8242z^3+3.511z^4+3.0231z^5+7.1505z^6+4.5456z^{11},\\
 A^{[1]}_1(z)&:=0.0000000088247z+8.8244z^3+3.4852z^4+3.6708z^5+2.6456z^6+21.903z^9,\\
 G^{[1]}_0(z)&:=G^{[1]}_1(z):=3.4967z^9.
\end{align*}
For $2\le i\le15$, set $A^{[1]}_i(z):=10^{-6}a_i^\circ z^2+b_i^\circ z^3$ and
$G^{[1]}_i(z):=3.4967z^4$. Here $(a_i^\circ,b_i^\circ):=(3.5198,8.824)$ for $2\le i\le8$, and
\begin{align*}
 (a_i^\circ)_{i=9}^{15}&:=(12.599,29.912,51.71,96.24,163.26,231.72,352.1),\\
 (b_i^\circ)_{i=9}^{15}&:=(8.8238,8.8236,8.8233,8.8228,8.8221,8.8214,8.8203).
\end{align*}
These formulas specify $\alpha_{i,j}:=[z^j]A^{[1]}_i$ and
$\gamma_{i,\ell}:=[z^\ell]G^{[1]}_i$ in Lemma~\ref{lem:joint_profile}.

We generate the remaining $144$ block majorants by the same rational linear
program, with cost weights indexed by the row and block. For
$\theta:=(\alpha_1,\ldots,\alpha_{12},\gamma_2,\gamma_3,\gamma_4,\gamma_9)
\in[0,64]^{16}$, define
\[
 A_\theta(z):=\sum_{j=1}^{12}\alpha_jz^j,\qquad
 G_\theta(z):=\sum_{\ell\in\{2,3,4,9\}}\gamma_\ell z^\ell,\qquad
 \Pi_\theta(z):=\frac{(1-z)(A_\theta(z)-g_{\rm R}(z))+G_\theta(z)}{z}.
\]
The last expression extends to a polynomial of degree at most $12$.
Let $\mathcal P$ consist of those $\theta$ for which every degree-$12$
Bernstein coefficient of $\Pi_\theta$ on every interval
$[h/16,(h+1)/16]$, $0\le h<16$, is nonnegative, using
Definition~\ref{def:rational_certificate_rules}. Thus $\mathcal P$ is a fixed
rational polytope. It is nonempty: take
$A_\theta(z):=(b_3z^3+b_4z^4+b_5z^5+b_7z^7)/2$ and
$G_\theta(z):=b_8z^4/2$. Then
$\Pi_\theta(z)=(b_8/2)z^3(1-z)(1+z+z^2+z^3)$ has nonnegative Bernstein
coefficients already on $[0,1]$.

For row $r$, put $\eta:=\eta_{r+1}$, $v_i:=i/16$ and
$v_{i,h}:=(32i+h)/512$. Form the exact rational weights
\begin{align*}
 c_{r,i,j}&:=\frac{\operatorname{df}(j)\eta^j\tau_{r,j}}{j}
                         (v_{i+1}^j-v_i^j),\\
 d_{r,i,\ell}&:=\frac{\eta^\ell\tau_{r,\ell}}{\ell}
   \sum_{h=1}^{32}(v_{i,h}^\ell-v_{i,h-1}^\ell)
                \widehat{\mathcal C}_\ell(\eta v_{i,h}),\\
 J_{r,i}(\theta)&:=\sum_{j=1}^{12}c_{r,i,j}\alpha_j+
                 \sum_{\ell\in\{2,3,4,9\}}d_{r,i,\ell}\gamma_\ell.
\end{align*}
First round the one-slot tuple, then the endpoint energies, as specified below.
We choose $\theta_{r,i}$ to be the lexicographically least minimizer of
$J_{r,i}$ on $\mathcal P$, in the displayed coordinate order, and set
$(A^{[r]}_i,G^{[r]}_i):=(A_{\theta_{r,i}},G_{\theta_{r,i}})$.
This is a finite rational construction: enumerate intersections of $16$
independent active constraints, retain the feasible vertices, and minimize
$(J_{r,i}(\theta),\theta)$ lexicographically. Nonemptiness and boundedness
ensure that the minimum exists. These formulas determine every coefficient
without a catalog or address list.

Bernstein nonnegativity gives
$g_{\rm R}(z)\le A^{[r]}_i(z)+G^{[r]}_i(z)/(1-z)$ on $[0,1)$.
Exact rational upper bounds for the $144$ programs give
\[
 R_r:=\sum_{i=0}^{15}\mathcal U_{45}(J_{r,i}(\theta_{r,i}))
       <(1-2\cdot10^{-5})R_{*,r}\qquad(2\le r\le10).
\]
The printed budgets $R_{*,r}$ therefore remain valid. The exponents, moment
caps, schedule and final signing bound use those same budgets.
\endgroup

\paragraph{Rounding and generation rules.}
Write $\mathcal U_p(x):=10^{-p}\lceil10^p x\rceil$. We evaluate linear programs and their rational objective values exactly. For the other scalar computations, we enclose elementary algebraic operations outward on the $10^{-90}$ rational grid. A finite minimum uses the minimum of the lower endpoints and the minimum of the upper endpoints. Square roots use exact integer squaring. We accept a named $45$-place ceiling only when both enclosing endpoints have the same ceiling. Before enclosure, we form the rational value $p=(1-a/\zeta^2)^{-1}$ by exact division.

For the one-slot radial procedure in Sections~\ref{subsec:sub81_first_row} and~\ref{subsec:sub81_residual}, put $v_i:=i/16$ for $0\le i\le16$. Here $v_i$ denotes a grid point, not a suffix capacity. Subdivide each block $[v_i,v_{i+1}]$, $0\le i\le15$, into $32$ equal cells with right endpoints $t_{i,h}:=(32i+h)/512=v_{i,h}$, $1\le h\le32$. At each $a:=\eta t_{i,h}$ compute the exact six-pass one-slot tuple and freeze its three entries at their $45$-place upward ceilings. Only then evaluate the first-row covariance tangents or residual trace energies and take their $45$-place upward ceilings. With these endpoint bounds fixed, integrate the remaining powers exactly as in Lemma~\ref{lem:fixed_reference_radial}. We evaluate all $512$ endpoints. This order applies only to these one-slot rows and leaves the preceding two-slot rules unchanged.

For the first row, retain every weighted intercept term separately as $Q_{i,h,\ell}$, with the termwise $45$-place ceiling specified above. We generate the nonzero coefficients and their rational $\beta_{i,h}$ values by that formula, including the endpoint with $\beta=0$. For each ordinary $P_j$, first round its block contribution upward to $45$ places. Add the weighted $B_\ell$ contributions to $P_2$ and round that block value upward again. Finally sum the blocks. For residual rows two through ten, sum all ordinary and weighted contributions in a block, round the complete sum upward to $45$ places, and sum the $16$ rounded blocks. This defines $R_r$. Compare it with the printed $R_{*,r}$ and then check the strict exponent inequality. For the joint first residual row, use instead the degreewise ordinary ceiling and the termwise weighted ceilings in Lemma~\ref{lem:joint_profile}. Do not combine its metric terms into block constants.

For $\widehat b_{\ell,r}$ outside the four explicit low-degree cases, let $v$ be the least integer with $v^\ell\ge\operatorname{df}(\ell)^r10^{30\ell}$. Exact comparisons also check $(v-1)^\ell<\operatorname{df}(\ell)^r10^{30\ell}$. Then $\widehat b_{\ell,r}=v10^{-30}$. The special low-degree values are proved analytically, not certified by these integer comparisons.

Evaluate the Gaussian cubic moments by Eq.~\eqref{eq:iterated_envelope_cubic_value}. For $\pi$, use the Machin series of Definition~\ref{def:rational_certificate_rules}, retaining indices $0$ through $80$ and the next alternating term. With $y:=t^2/a<36$, the consecutive Taylor sums of degrees $256,257$ enclose $e^{-y/2}$ and, after termwise integration, $\int_0^1e^{-yz^2/2}\d z$. Outward $90$-place arithmetic then isolates each $45$-place ceiling. No floating-point tail value is substituted.

For logarithms use Eqs.~\eqref{eq:1070_log_partial}--\eqref{eq:1070_log_remainder} with $K_{\mathrm{series}}:=109$, including the same enclosure of $\log2$ when reversing range reduction. For $L_H$ alone use $U_0,U_\infty,U_{\rm tay}$ from Lemma~\ref{lem:scalar_profile_enclosures}, with $A:=1$, $C:=D:=0$, all $q_j:=0$, $\epsilon:=10^{-8}$ and $R:=100$. For Eq.~\eqref{eq:metric_profile}, use the first-row program and its chord and tail bounds.

The moment and radial programs impose their own Bernstein constraints. For the other univariate polynomials we use the Bernstein conversion and midpoint subdivision of Definition~\ref{def:rational_certificate_rules}, at the actual degree and with depth cap $40$. We accept an identically zero numerator directly. For a nonzero numerator, we remove only exact endpoint factors with nonnegative orientation. A separate Sturm calculation reconstructs the explicitly printed first-residual-row radial polynomials, the all-regular-region polynomial used in the proof of Lemma~\ref{lem:sub81_triples}, and the degree-eight hinge polynomial independently. It proves each remaining polynomial positive at both endpoints and root-free on its interval. For the joint profile, we use the two quartic Taylor polynomials in the proof of Lemma~\ref{lem:joint_profile}, with midpoint subdivision in $y$ to depth $8$ before bisecting an unresolved $x$-cell. The complete middle interval is $[10^{-8},100]$ and its depth cap is $12$. The origin and tail quartics are included. We apply outward coefficientwise arithmetic before the exact Bernstein tests.

The spectral certificate is the bounded recursion of Lemma~\ref{lem:sub81_triples}. The preceding formulas and printed rational data also specify all radial, moment, entropy, suffix and tail tests. The matrix-weighted, covariance-mean, coding and uniform-tail analytic implications are the cited lemmas. The finite checks supply their stated numerical hypotheses.

\section{The \texorpdfstring{$n^{2+\omega/2}$}{n\textasciicircum(2+omega/2)} algorithm}\label{sec:algorithm}

Remark~\ref{rem:bit_section6} records why the diagonalization step of
Algorithm~\ref{alg:dense_matrix_spencer} has no bit-complexity bound.

\subsection{Statement and algorithm}\label{subsec:dense_statement}

\begin{theorem}[Dense arithmetic implementation, formal version of the
$n^{2+\omega/2+o(1)}$ variant in Remark~\ref{rem:dense_runtime_informal}]
\label{thm:dense_runtime}
Let $A_1,\ldots,A_n\in\R^{n\times n}$ be symmetric with $\|A_i\|\le1$, and let $p\in(0,1/2)$. Algorithm~\ref{alg:dense_matrix_spencer}
returns, with failure probability at most $p$, a signing satisfying the
conclusion of Theorem~\ref{thm:matrix_spencer_bound}, using
\[
n^{2+\omega/2+o(1)}\operatorname{polylog}(1/p)
\]
arithmetic operations in the real-arithmetic model of
Section~\ref{sec:preliminaries}. With the $n^{o(1)}$ carried by $\omega$
absorbed into the soft-$O$, this is $\widetilde O(n^{2+\omega/2})$. The
working accuracy is inverse polynomial, and its contribution to the running
time is polylogarithmic.
\end{theorem}

Algorithm~\ref{alg:dense_matrix_spencer} is the implementation, and in the rest
of the section we prove Theorem~\ref{thm:dense_runtime}. Both sit in the present
Section~\ref{subsec:dense_statement}, and in
Remark~\ref{rem:dense_reading_runtime} we read the exponent $2+\omega/2$ off
the two cost terms of one block query. We fix the smoothed body and its
constants in Section~\ref{subsec:dense_body}: the constraint function $h$, a
smooth proxy for $\|B(x)\|/R$, the body $D=\{h\le b\}$ inside $K$, and the
trial projection Eq.~\eqref{eq:smooth_projection}. There we also record the
Gaussian measure of $D$, the success probability of one attempt, and the
gradient and Hessian bounds on $h$ that the solver uses. In
Section~\ref{subsec:dense_solver} we solve one projection by an inexact
block-coordinate method inside a robust bisection on the multiplier, and we
make the computed point exactly feasible. The block size $\ell$ is free
there, and the cost is
$\widetilde O((m/\sqrt\ell)(n^{\omega+o(1)}+\ell n^2)+mn^2)$ arithmetic
operations given the oracle. We implement in
Section~\ref{subsec:dense_oracle} the oracle that the solver queries. One
diagonalization forms a signed Gibbs matrix, and the $\ell$ block derivatives
are Frobenius products with it, so a query costs $n^{\omega+o(1)}+O(\ell n^2)$.
We snap the computed point back into the original body $K$ in
Section~\ref{subsec:dense_rounds}, fix the block size
$\ell=\min\{m,\max\{1,\lceil n^{\omega-2}\rceil\}\}$, and sum the round costs
to $n^{2+\omega/2+o(1)}\operatorname{polylog}(1/p)$. We handle the cleanup in
Section~\ref{subsec:dense_cleanup} by independent biased rounding whose
error one diagonalization certifies (Algorithm~\ref{alg:certified_cleanup}),
and we assemble the proof of Theorem~\ref{thm:dense_runtime} there.

\begin{algorithm}[!ht]
\caption{$n^{2+\omega/2}$ time algorithm for Matrix Spencer signing}
\label{alg:dense_matrix_spencer}
\begin{algorithmic}[1]
\Procedure{MainDense}{$A_1,\ldots,A_n,n,p$} \Comment{Theorem~\ref{thm:dense_runtime}}
\State $y\gets0$, $S\gets[n]$, and $N_0\gets3\mathbin{\cdot}10^8$
\State Fix the per-phase repetition cap $O(\log\log n+\log(1/p))$ and the cleanup cap $O(\log(1/p))$ of Sections~\ref{subsec:dense_rounds} and~\ref{subsec:dense_cleanup}
\While{$|S|\ge N_0$}
    \State $m\gets|S|$, $u\gets\log(n/m)$, $\ell_{\rm asp}\gets\max\{1,u\}$, and $R\gets c_{\rm rad}(\ell_{\rm asp})\sqrt m\ell_{\rm asp}$
    \State $\ell\gets\min\{m,\max\{1,\lceil n^{\omega-2}\rceil\}\}$
    \Repeat
        \State Sample $\xi\sim\mathcal N(0,I_m)$, rejecting draws with $\|\xi\|_2^2>C_gm$ ($C_g$ is the universal constant of Section~\ref{subsec:dense_solver}), and solve Eq.~\eqref{eq:smooth_projection} to the prescribed inverse-polynomial accuracy using block size $\ell$
        \State Rescale the computed point toward $0$ to restore exact smooth feasibility, and snap every coordinate within $2\rho$ of a box facet; call the result $\overline x$
    \Until{at least $\delta m$ coordinates of $\overline x$ are snapped, or the repetition cap is reached}
    \If{the repetition cap is reached without success}
        \State \Return \textsc{Fail}
    \EndIf
    \State $y_i\gets y_i+2\overline x_i/\varepsilon_{\mathrm{box}}$ for every $i\in S$
    \State $S\gets\{i\in[n]:|y_i|<1\}$
\EndWhile
\State \Return \Call{CertifiedCleanup}{$A_1,\ldots,A_n,y,S,p/2$} \Comment{Algorithm~\ref{alg:certified_cleanup}, cleanup cap $\lceil\log_2(2/p)\rceil$}
\EndProcedure
\end{algorithmic}
\end{algorithm}

\begin{remark}[Reading the running time]
\label{rem:dense_reading_runtime}
One projection at active dimension $m$ costs $\widetilde O(m/\sqrt\ell)$
block queries. Each query is a diagonalization at $n^{\omega+o(1)}$ plus $\ell$
Frobenius products at $O(\ell n^2)$. The block size $\ell=n^{\omega-2}$
balances the two terms at $m\,n^{1+\omega/2+o(1)}$, and the active dimensions
sum to $O(n)$ over the rounds. The exponent $2+\omega/2$ is the cost of one
diagonalization per query, which the algorithms of
Sections~\ref{sec:faster_algorithm}--\ref{sec:cubic_time} avoid.
\end{remark}

\subsection{The smoothed body and its constants}
\label{subsec:dense_body}

\emph{Step 1: the smoothed body.}
We first fix the constants. At an active set $S$ of size $m$, put
\[
u:=\log(n/m),
\qquad
\ell_{\rm asp}:=\max\{1,u\},
\qquad
R:=c_{\rm rad}(\ell_{\rm asp})\sqrt m\ell_{\rm asp},
\]
with the coefficient $c_{\rm rad}(\ell_{\rm asp})$ of Lemma~\ref{lem:scale_small_ball}
(Table~\ref{tab:six_regimes}). We use the same projection parameters at
every phase:
\[
(\delta,\varepsilon_{\mathrm{box}}):=(0.0281672,0.4235).
\]
Put $B(x):=\sum_{i\in S}x_iA_i$ and
\[
\chi:=s:=5\mathbin{\cdot}10^{-7},
\qquad
a:=2\mathbin{\cdot}10^6\log(2n),
\qquad
b:=1-s,
\]
\[
h(x):=\frac1a\log(\tr[\exp(aB(x)/R)]+\tr[\exp(-aB(x)/R)]),
\qquad
D:=\{x\in\R^S:h(x)\le b\}.
\]
The function $h$ is the constraint function of the whole section. Since
$\tr[\exp(aZ)]+\tr[\exp(-aZ)]$ lies between $e^{a\|Z\|}$ and $2ne^{a\|Z\|}$
for symmetric $Z$, and $\log(2n)/a=\chi$,
\[
\frac{\|B(x)\|}{R}\le h(x)\le\frac{\|B(x)\|}{R}+\chi;
\]
in particular $h(0)=\chi$, and with $K:=\{x:\|B(x)\|\le R\}$,
\[
(1-s-\chi)K\subseteq D\subseteq(1-s)K.
\]
Radial monotonicity of the Gaussian density and
Lemma~\ref{lem:scale_small_ball} give
\[
\gamma_m(D)
\ge(1-10^{-6})^m\gamma_m(K)
\ge\exp(-0.00372m),
\qquad
\text{because }0.003712-\log(1-10^{-6})<0.00372.
\]

At the current fractional coloring $y$, define
\[
R_i:=\frac{\varepsilon_{\mathrm{box}}}2(1-y_i),
\qquad
L_i:=\frac{\varepsilon_{\mathrm{box}}}2(1+y_i),
\qquad
Q:=\prod_{i\in S}[-L_i,R_i].
\]
One trial is the projection
\begin{equation}
\min\{f_\xi(x):x\in Q,\ h(x)\le b\},
\qquad
f_\xi(x):=\frac12\|x-\xi\|_2^2.
\label{eq:smooth_projection}
\end{equation}
Lemma~\ref{lem:refreshed_projection}, applied to the symmetric closed convex
body $D$, gives exact-projection success probability greater than $0.12$.
We choose the Gaussian-norm rejection and the numerical failure budgets below
with total below $0.02$, which leaves success probability greater
than $0.1$ per attempt.

\emph{Step 2: smoothness of $h$.}
For a symmetric matrix $Z$, let
\[
\Phi(Z):=\frac1a\log\tr[\exp(aZ)],
\qquad
\rho_Z:=\frac{\exp(aZ)}{\tr[\exp(aZ)]}.
\]
Differentiating the trace exponential (Duhamel's formula), we obtain the Fr\'echet derivatives
\[
D\Phi(Z)[H]=\tr[\rho_ZH],
\qquad
D^2\Phi(Z)[H,H]
=a\int_0^1\tr[\rho_Z^{1-\theta}H\rho_Z^\theta H]\,\d\theta
-a(\tr[\rho_ZH])^2.
\]
Fact~\ref{fact:trace_exp_hessian} applied to $aZ$ bounds the integral by
$\|H\|^2$, and
\[
0\le D^2\Phi(Z)[H,H]\le a\|H\|^2.
\]
We apply this to the two-block representation of $h$ and put $G:=\sqrt m/R$.
Since $\|A_i\|\le1$,
\[
|\partial_ih(x)|\le\frac1{R},
\qquad
\|\nabla h(x)\|_2\le G,
\qquad
\partial_{ii}^2h(x)\le\frac{a}{R^2}.
\]

\subsection{The block solver and robust bisection}
\label{subsec:dense_solver}

In this subsection we solve Eq.~\eqref{eq:smooth_projection}, for every block
size $1\le\ell\le m$, to inverse-polynomial Euclidean accuracy in
\[
\widetilde O(\frac{m}{\sqrt\ell}(n^{\omega+o(1)}+\ell n^2)+mn^2)
\]
arithmetic operations, given a block-gradient oracle whose error and cost
we supply in Section~\ref{subsec:dense_oracle}.

\emph{Step 3: the multiplier bracket and the path bounds.}
Let $x_*$ be the minimizer in Eq.~\eqref{eq:smooth_projection}. If the box
projection of $\xi$ is feasible, it equals $x_*$ and there is nothing to
prove. Otherwise $0$ is a strict Slater point, because $h(0)=\chi<b$. Hence
there is a multiplier $\lambda_*>0$ for which $x_*$ is the unique minimizer
over $Q$ of
\[
F_\lambda(x):=f_\xi(x)+\lambda h(x)
\]
at $\lambda=\lambda_*$, and $h(x_*)=b$. Write $x_\lambda$ for the unique
minimizer of $F_\lambda$ over $Q$. Comparing with $0$, we obtain
$\lambda_*\le f_\xi(0)/(b-\chi)$. Define
\[
\Lambda:=\frac{2f_\xi(0)}{b-\chi}.
\]
From optimality of $x_\Lambda$ and comparison with $0$ we obtain
\[
f_\xi(x_\Lambda)+\Lambda(h(x_\Lambda)-b)
\le f_\xi(0)+\Lambda(\chi-b)=-f_\xi(0),
\]
so $h(x_\Lambda)\le(b+\chi)/2<b$, and $[0,\Lambda]$ brackets $\lambda_*$.
We reject and resample draws with $\|\xi\|_2^2>C_gm$, an event of
probability $e^{-\Omega(m)}$ for a suitable universal $C_g$. On the accepted
event, $\Lambda=O(m)$.

The map $\lambda\mapsto h(x_\lambda)$ is nonincreasing. Moreover, for
$\lambda,\mu\ge0$,
\begin{equation}
\|x_\lambda-x_\mu\|_2\le G|\lambda-\mu|.
\label{eq:multiplier_path}
\end{equation}
Indeed, let $\lambda\ge\mu$ and $\Delta:=x_\lambda-x_\mu$. Subtracting the two
variational inequalities and taking the inner product with $\Delta$, we obtain, by
monotonicity of $\nabla h$ and of the normal cone of $Q$,
\[
\|\Delta\|_2^2
\le-(\lambda-\mu)\langle\nabla h(x_\mu),\Delta\rangle
\le G(\lambda-\mu)\|\Delta\|_2.
\]
Strong convexity also gives the residual estimate
\begin{equation}
\|x_\lambda-x_*\|_2^2
\le|\lambda-\lambda_*||h(x_\lambda)-b|:
\label{eq:multiplier_residual}
\end{equation}
we apply $1$-strong convexity once to $F_\lambda$ and once to $F_{\lambda_*}$,
at their respective minimizers, and add the two inequalities.

\emph{Step 4: the fixed-multiplier block solver.}
It remains to minimize a fixed $F_\lambda$. Fix an integer $1\le\ell\le m$,
and partition the active coordinates into $q:=\lceil m/\ell\rceil$ blocks
$I_1,\ldots,I_q$ of size at most $\ell$. We use the Euclidean norm on each
block. The Hessian estimate of Step~2 implies that, for every vector $v$
supported on one block,
\[
D^2h(x)[v,v]
\le\frac{a}{R^2}\|\sum_{i\in S}v_iA_i\|^2
\le\frac{a\ell}{R^2}\|v\|_2^2.
\]
Consequently every block gradient of $F_\lambda$ has the common Lipschitz
constant
\[
L_\ell:=1+\frac{\Lambda a\ell}{R^2}
=\widetilde O(\ell),
\qquad
\text{by }\Lambda=O(m),\ R^2=\Omega(m),\ a=O(\log n).
\]
The function $F_\lambda$ is $1$-strongly convex in the Euclidean norm, hence
$1/L_\ell$-strongly convex in the weighted norm
$\|v\|_E^2:=L_\ell\|v\|_2^2$. The feasible set $Q$ is a product over the
blocks, and its Euclidean block proximal map is coordinatewise clipping.

Our solver is the restarted accelerated random block-coordinate method with
inexact block derivatives of Dvurechensky, Gasnikov, Tyurin, and
Zholobov~\cite[Corollary~3 and Theorem~2]{dvurechenskyetal23}. Under the
preceding normalization its block-count parameter is $q$, its
strong-convexity parameter is $1/L_\ell$, and one restart epoch has
\[
K_\ell=O(q\sqrt{L_\ell})
\]
block queries. Suppose that every returned block gradient has Euclidean
error at most $\Delta_{\mathrm{or}}$. After $J$ epochs the cited theorem
gives
\[
\E[F_\lambda(y^{(J)})-F_\lambda(x_\lambda)]
\le\varepsilon_{\mathrm{init}}2^{-J}
+8(K_\ell+2q)^2\frac{\Delta_{\mathrm{or}}^2}{L_\ell},
\]
where $\varepsilon_{\mathrm{init}}$ is an upper bound on the initial objective gap. On the
accepted Gaussian event we can take $\varepsilon_{\mathrm{init}}=n^{O(1)}$. For a target
objective error $\varepsilon_{\mathrm{obj}}$ and a failure probability
$p_0$, we choose
\[
J:=\lceil\log_2\frac{2\varepsilon_{\mathrm{init}}}{p_0\varepsilon_{\mathrm{obj}}}\rceil,
\qquad
\Delta_{\mathrm{or}}
\le\frac{\sqrt{L_\ell p_0\varepsilon_{\mathrm{obj}}}}{4(K_\ell+2q)}.
\]
The expected objective gap is then at most $p_0\varepsilon_{\mathrm{obj}}$,
and Markov's inequality shows that the gap exceeds
$\varepsilon_{\mathrm{obj}}$ with probability at most $p_0$. All target
errors below are inverse polynomial, so the required block-gradient
accuracy is inverse polynomial as well. The number of block queries is
\[
\widetilde O(q\sqrt{L_\ell})
=\widetilde O(m/\sqrt\ell).
\]

The accelerated recurrence forms affine combinations of feasible vectors
and changes only the selected block in each proximal update. We maintain the
corresponding matrix images, such as $B(x)$ and the auxiliary image
$B(w)$ of the auxiliary iterate $w$, together with the coordinate vectors.
An affine combination of matrix images costs $O(n^2)$ operations, and a
selected-block change costs $O(\ell n^2)$ operations. Hence acceleration
introduces no hidden $O(mn^2)$ matrix reconstruction per query.

\emph{Step 5: robust bisection and exact feasibility.}
We make the multiplier search robust to the inexact solves as follows. Let
$\rho>0$ be the desired Euclidean accuracy, put $d:=\rho/4$, and set
\[
\tau:=\min\{\frac{d^2}{1+\Lambda},\frac{d}{16\sqrt m}\},
\qquad
\sigma:=\min\{\frac d8,\frac{\tau}{16G}\}.
\]
At each bisection point, we run the fixed-multiplier solver with target
objective gap $\sigma^2/2$, which it attains except with the failure
probability allocated above. Strong convexity then gives
$\|\widehat x_\lambda-x_\lambda\|_2\le\sigma$. We evaluate $h(\widehat x_\lambda)$
with absolute numerical error at most $\tau/16$. Because $h$ is
$G$-Lipschitz, these two bounds give a certified interval
$\mathcal I_\lambda$ of length at most $\tau/4$ containing $h(x_\lambda)$.

If $\mathcal I_\lambda$ lies strictly above or strictly below $b$,
monotonicity determines the half interval containing $\lambda_*$. If
$b\in\mathcal I_\lambda$, then $|h(x_\lambda)-b|\le\tau/4$, and
Eq.~\eqref{eq:multiplier_residual} gives $\|x_\lambda-x_*\|_2\le d/2$.
Otherwise we continue until the multiplier bracket has length at most
$d/(4G)$ and solve at its feasible endpoint. Then Eq.~\eqref{eq:multiplier_path}
gives $\|x_\lambda-x_*\|_2\le d/4$. The number of fixed-multiplier
solves is $O(\log(\Lambda G/d))=O(\log(n/\rho))$.

We can make the computed point $\widehat x$ exactly feasible for the smooth
constraint. Compute a certified upper bound $\overline h\ge h(\widehat x)$
with numerical error at most $\tau/16$. If $\overline h\le b$, set
$\alpha:=1$. Otherwise set
\[
\alpha:=\frac{b-h(0)}{\overline h-h(0)},
\]
and in either case replace $\widehat x$ by $\alpha\widehat x$. Since
$0,\widehat x\in Q$, the new point remains in $Q$, and convexity gives
$h(\alpha\widehat x)\le b$. In both stopping cases, after the preceding
constant choices, $\overline h-b\le\tau/2$, hence
\[
(1-\alpha)\|\widehat x\|_2
\le\tau\|\widehat x\|_2
\le2\tau\sqrt m
\le\frac d8.
\]
Combining the multiplier, optimization, and rescaling errors, we conclude that the
final feasible point $z$ satisfies $\|z-x_*\|_2\le\rho$.

\subsection{The diagonalization oracle}
\label{subsec:dense_oracle}

The solver of Section~\ref{subsec:dense_solver} needs the value $h(x)$ and
the $\ell$ derivatives $\partial_ih(x)$, $i\in I$, of a selected block, each
to inverse-polynomial accuracy. Here we supply them at cost
$n^{\omega+o(1)}+O(\ell n^2)$ per query.

\emph{Step 6: the value and derivative oracle.}
At the current query point, we diagonalize
\[
B(x)=U\operatorname{diag}(\beta_1,\ldots,\beta_n)U^\top
\]
to spectral residual $\zeta:=\|B(x)-\widetilde B\|$ at most an
inverse-polynomial tolerance fixed below, where $\widetilde B$ is the matrix
represented by the computed factorization. We apply the diagonalization
after rescaling $B(x)$ by an upper bound on its norm. The
near-orthogonality defect $\|U^\top U-I\|$ of the computed factor is inverse
polynomial, and we charge it to the matrix-arithmetic budget below. Put
\[
t_j:=\frac{a\beta_j}{R},
\qquad
M:=\max\{|t_1|,\ldots,|t_n|\},
\qquad
w_j^+:=e^{t_j-M},
\qquad
w_j^-:=e^{-t_j-M},
\]
so that
\[
h(x)=\frac1a(M+\log\sum_{j=1}^n(w_j^++w_j^-)).
\]
Define the common signed Gibbs matrix
\[
W_x:=
U\operatorname{diag}(
\frac{w_j^+-w_j^-}{\sum_{k=1}^n(w_k^++w_k^-)}
)_{j=1}^nU^\top,
\qquad
\text{so that}
\qquad
\partial_ih(x)=\frac1{R}\tr[W_xA_i]
\]
for every active coordinate. Thus we form $W_x$ with one diagonalization and one matrix
multiplication. After that, all derivatives in a selected block
$I$ are the $\ell$ Frobenius products $\tr[W_xA_i]$, $i\in I$, at
$O(\ell n^2)$ operations. This batching is what the exponent depends on:
forming $U^\top A_iU$ separately for every selected coordinate would lose
the improvement.

Sobczyk's Hermitian diagonalization algorithm~\cite[Corollary~2.4]{sobczyk25}
computes the required approximate orthogonal diagonalization in
\[
O(n^\omega\log n+n^2\operatorname{polylog}(n/\zeta))
=n^{\omega+o(1)}
\]
arithmetic operations, and forming $W_x$ has the same asymptotic cost. The
quadratic part of the block gradient costs $O(\ell)$ operations, and
maintaining a matrix image after a block update costs $O(\ell n^2)$.
Consequently one block query costs
\[
n^{\omega+o(1)}+O(\ell n^2).
\]

The numerical diagonalization supplies the inexact oracle that Step~4
requires. Put
\[
C_B:=\operatorname{diag}(B/R,-B/R),
\qquad
C_{\widetilde B}:=\operatorname{diag}(\widetilde B/R,-\widetilde B/R).
\]
The symmetrized relative-entropy identity and the quantum Pinsker inequality
for the corresponding Gibbs states give
\[
\|\rho_{C_B}-\rho_{C_{\widetilde B}}\|_1
\le2a\|C_B-C_{\widetilde B}\|
\le\frac{2a}{R}\|B-\widetilde B\|.
\]
Therefore the spectral residual $\zeta=\|B-\widetilde B\|$ causes error at
most
\[
\frac{2\Lambda a\sqrt\ell}{R^2}\zeta
\]
in the selected block gradient of $F_\lambda$, and error at most $\zeta/R$
in $h$. It suffices to request
\[
\zeta\le
\frac{\Delta_{\mathrm{or}}R^2}{4\Lambda a\sqrt\ell}
\]
and to allocate the remaining half of $\Delta_{\mathrm{or}}$ to scalar and
matrix-arithmetic errors. All requested tolerances are inverse polynomial,
so the spectral cost remains $n^{\omega+o(1)}$. No eigengap is required.

The shift by $M$ prevents overflow. Let $\varepsilon_{\mathrm{sc}}$ be the
requested scalar accuracy, which is inverse polynomial. Terms below
$e^{-T}$, with $T=O(\log(n/\varepsilon_{\mathrm{sc}}))$, can be discarded.
Range reduction followed by Taylor approximation evaluates the remaining
scalar exponentials and the logarithm in
$\operatorname{polylog}(n/\varepsilon_{\mathrm{sc}})$ arithmetic operations.
Including the one-time $O(mn^2)$ construction of a matrix image, a
fixed-multiplier solve therefore costs
\[
\widetilde O(
\frac{m}{\sqrt\ell}(n^{\omega+o(1)}+\ell n^2)+mn^2
),
\]
and multiplier bisection changes this bound only by a polylogarithmic
factor.

\subsection{Snapping and round costs}
\label{subsec:dense_rounds}

\emph{Step 7: snapping.}
Let $x_*$ be the exact minimizer and let $z$ be a feasible point with
$\|z-x_*\|_2\le\rho$, where
\[
\rho:=\min\{\frac18,\frac{sR}{4(2m+\sqrt m)}\}.
\]
We snap each coordinate within $2\rho$ of a box facet. Every truly tight
coordinate is snapped, and the snapped point $\overline x$ satisfies
\[
\|B(\overline x)\|
\le(1-s)R+2m\rho<R.
\]
Thus every accepted numerical increment belongs to the original body $K$
and preserves the discrepancy certificate. The Euclidean accuracy we request
is inverse polynomial and does not change the asymptotic solver bound. The
repetition cap for one phase is
\[
O(\log\log n+\log(1/p)),
\]
and there are $O(\log n)$ phases.

\emph{Step 8: round costs.}
One successful round at active dimension $m$ uses a constant number of
projections, up to the repetition factor we just recorded. Set
\[
\ell_0:=\max\{1,\lceil n^{\omega-2}\rceil\},
\qquad
\ell(m):=\min\{m,\ell_0\}.
\]
If $m\ge\ell_0$, the projection bound gives
\[
\widetilde O(
\frac{m}{\sqrt{\ell_0}}(n^\omega+\ell_0n^2)+mn^2
)
=m\,n^{1+\omega/2+o(1)}
\]
operations for the round. Each round freezes at least a fixed fraction of
the active coordinates, so the active dimensions satisfy
$m_{k+1}\le(1-\delta)m_k$ and $\sum_{k=0}^{\infty}m_k=O(n)$. The rounds with
$m_k\ge\ell_0$ therefore cost
\[
n^{2+\omega/2+o(1)}.
\]
For the remaining rounds, we take one block of size $\ell=m$. We bound their
total cost by
\[
\widetilde O(
n^\omega\sqrt{\ell_0}+n^2\ell_0^{3/2}+n^2\ell_0
)
=n^{3\omega/2-1+o(1)},
\]
which is lower order than $n^{2+\omega/2+o(1)}$ because $\omega<3$. The
one-time matrix-image constructions sum to at most $\widetilde O(n^3)$ and
are of no larger order. Hence all partial-coloring rounds together use
\[
n^{2+\omega/2+o(1)}\operatorname{polylog}(1/p)
\]
arithmetic operations.

\subsection{Cleanup and the proof of Theorem~\ref{thm:dense_runtime}}
\label{subsec:dense_cleanup}

\emph{Step 9: cleanup.}
When the active dimension first falls below $N_0:=3\mathbin{\cdot}10^8$, we use
independent biased rounding, and certify the cleanup error
$E:=\sum_{i\in S}(\varepsilon_i-y_i)A_i$ as follows. The matrix $E/(2m)$
is symmetric with $\|E/(2m)\|\le1$, because $|\varepsilon_i-y_i|\le2$ and
$\|A_i\|\le1$. We diagonalize it by Sobczyk's
algorithm~\cite[Corollary~2.4]{sobczyk25} to spectral residual
$\zeta\le\min\{10^{-2},\sqrt n/(1000m)\}$ and orthogonality defect at most
$\zeta$. The output is $\widetilde U$ and $\beta_1,\ldots,\beta_n$ with
$\|E/(2m)-\widetilde U\operatorname{diag}(\beta)\widetilde U^\top\|\le\zeta$
and $\|\widetilde U^\top\widetilde U-I\|\le\zeta$, at cost
$n^{\omega+o(1)}$. The scalar $U$ below is the spectral certificate, not the orthogonal factor of Section~\ref{subsec:dense_oracle}. Put
\[
U:=2m\max_j|\beta_j|+0.01\sqrt n.
\]
Then
\[
\|E\|\le U\le\|E\|+0.03\sqrt n.
\]
Indeed, write $\widetilde U=VP$ with $V$ orthogonal and
$P:=(\widetilde U^\top\widetilde U)^{1/2}$, so that $\|P-I\|\le\zeta$ and
$\|P\operatorname{diag}(\beta)P-\operatorname{diag}(\beta)\|
\le(2+\zeta)\zeta\max_j|\beta_j|\le3\zeta\max_j|\beta_j|$. Hence
$\|E/(2m)-V\operatorname{diag}(\beta)V^\top\|\le\zeta(1+3\max_j|\beta_j|)$,
and Weyl's inequality for the orthogonally diagonalizable
$V\operatorname{diag}(\beta)V^\top$ gives
$|\,\|E/(2m)\|-\max_j|\beta_j|\,|\le\zeta(1+3\max_j|\beta_j|)$. Since
$\|E/(2m)\|\le1$ and $\zeta\le10^{-2}$, this forces $\max_j|\beta_j|\le1.05$
and then $|\,\|E\|-2m\max_j|\beta_j|\,|\le4.2\cdot2m\zeta\le0.0084\sqrt n$,
which gives both inequalities. We accept when $U\le6.5\sqrt n$.

The cleanup signs are independent with $\E\varepsilon_i=y_i$, so
Fact~\ref{fact:matrix_hoeffding} with $\sigma^2=\|\sum_{i\in S}A_i^2\|\le m<N_0$ gives
\[
\Pr[\|E\|\ge t]\le2n\exp(-\frac{t^2}{2\|\sum_{i\in S}A_i^2\|}),
\qquad
\|\sum_{i\in S}A_i^2\|\le m<N_0.
\]
Setting the right side equal to $1/2$, and using that
$(N_0/n)(\log(2n)+\log2)$ decreases for $n\ge N_0$, we obtain, together with
the $0.03\sqrt n$ certification allowance,
\[
\sqrt{2(\log(2N_0)+\log2)}+0.03<6.5.
\]
For $n<N_0$ the algorithm rounds immediately with $m=n$ and
$\log(4n)\le\log(4N_0)$, so the bound is uniform in $n$, and one cleanup
trial succeeds with probability at least $1/2$. Repetition contributes only
$O(\log(1/p))$.

\begin{algorithm}[H]
\caption{Certified bounded-dimension cleanup}
\label{alg:certified_cleanup}
\begin{algorithmic}[1]
\Procedure{CertifiedCleanup}{$A_1,\ldots,A_n,y,S,p_{\mathrm{clean}}$}
\State $m\gets|S|$
\Repeat
    \State Set $\varepsilon_i\gets y_i$ for $i\notin S$, and independently sample $\varepsilon_i\in\{-1,1\}$ with $\Pr[\varepsilon_i=1]=(1+y_i)/2$ for $i\in S$
    \State Compute the deterministic spectral certificate $U$ of the cleanup error $E=\sum_{i\in S}(\varepsilon_i-y_i)A_i$ as in Step~9 of Section~\ref{subsec:dense_cleanup} (diagonalize $E/(2m)$ to residual $\zeta\le\min\{10^{-2},\sqrt n/(1000m)\}$ and orthogonality defect at most $\zeta$, and put $U:=2m\max_j|\beta_j|+0.01\sqrt n$, so that $\|E\|\le U\le\|E\|+0.03\sqrt n$)
\Until{$U\le6.5\sqrt n$, or the cleanup cap prescribed by $p_{\mathrm{clean}}$ is reached}
\If{$U>6.5\sqrt n$}
    \State \Return \textsc{Fail}
\EndIf
\State \Return $\varepsilon$
\EndProcedure
\end{algorithmic}
\end{algorithm}

\begin{proof}[Proof of Theorem~\ref{thm:dense_runtime}]
Every accepted numerical increment lies in the body $K$ (Step~7) and
freezes at least $\delta m$ coordinates, exactly as the exact projection of
Lemma~\ref{lem:refreshed_projection} does. Hence the increment bound
Eq.~\eqref{eq:explicit_round_increment} and the constant calculation in the
proof of Theorem~\ref{thm:matrix_spencer_bound} apply unchanged. The
cleanup of Step~9 adds at most $6.5\sqrt n$. The returned signing has
discrepancy at most $156000\sqrt n$. There are $N_{\mathrm{ph}}=O(\log n)$
phases, as we recorded at the end of Step~7, by the geometric decrease of
Step~8. One attempt succeeds with probability at least $0.1$ (Step~1), so
the repetition cap $\lceil\log(2N_{\mathrm{ph}}/p)/\log(10/9)\rceil=O(\log\log n+\log(1/p))$
of Step~7 makes a phase fail with probability at most $p/(2N_{\mathrm{ph}})$. One cleanup
trial succeeds with probability at least $1/2$ (Step~9), so the cleanup cap
$\lceil\log_2(2/p)\rceil=O(\log(1/p))$ makes the cleanup fail with
probability at most $p/2$. The numerical failure budgets of Steps~4 and~5
are part of the $0.02$ allowance of Step~1. We conclude that the total failure
probability is at most $N_{\mathrm{ph}}\cdot p/(2N_{\mathrm{ph}})+p/2=p$. The arithmetic count is the sum of the round
costs of Step~8 and the $O(\log(1/p))$ cleanup trials of Step~9, each of
cost $n^{\omega+o(1)}$, which is
$n^{2+\omega/2+o(1)}\operatorname{polylog}(1/p)$.
\end{proof}

\section{The selected-coordinate algorithm}
\label{sec:faster_algorithm}

In this section we change only the solver of each smooth projection
Eq.~\eqref{eq:smooth_projection} from Section~\ref{sec:algorithm}.  We retain
the outer partial-coloring trajectory, the smoothed spectral body, the
projection accuracy, the snapping rule, and the cleanup.

We state Theorem~\ref{thm:faster_dense_runtime} in
Section~\ref{subsec:selected_statement}, give the implementation as
Algorithm~\ref{alg:faster_dense_matrix_spencer}, and explain in a remark what
one selected-coordinate query costs.  We set up the fixed-multiplier problem
$F_\nu$ in Section~\ref{sec:selected_setup}, together with its
selected-coordinate oracle, a dense snapshot combined with a range-finder
sketch of the increment.  We bound the oracle error by the objective gaps in
the same subsection, and prove the one-step contraction that a loopless
coordinate method obtains from this oracle.  We then analyze one guarded
candidate phase in two steps, with exact and then with finite matrix actions.
In Section~\ref{sec:selected_guard} we localize the spectrum, guard every
proposal by a trace estimate that keeps accepted iterates below the level
$H$, and bound the crossing probability.  Passing to finite actions in
Section~\ref{sec:selected_finite}, we add the bias and variance of the
surrogates and conclude that one phase halves the gap, up to an additive
term, with probability above $0.9$.  We give the selection rule
\textsc{CompareValueExact} and the multiplier wrapper in
Section~\ref{sec:selected_wrapper}.  There we prove
Lemma~\ref{lem:selected_accepted_draw}, the lower bound on the accepted draw,
and run a certified bisection over the multiplier.  We then correct the
output radially, so that it is exactly feasible and within $\rho$ of the
exact projection.  We transfer correctness, allocate the failure budget, and
count operations in Section~\ref{sec:selected_proof}, which completes the
proof of Theorem~\ref{thm:faster_dense_runtime}.

\subsection{Statement and pseudocode}\label{subsec:selected_statement}

\begin{theorem}[Selected-coordinate dense implementation, formal version of the $n^{1+\omega(1,1/2,1)+o(1)}$ variant in Remark~\ref{rem:dense_runtime_informal}]
\label{thm:faster_dense_runtime}
For symmetric $A_1,\ldots,A_n\in\R^{n\times n}$ with $\|A_i\|\le1$ and $p\in(0,1/2)$, a signing satisfying the conclusion of
Theorem~\ref{thm:matrix_spencer_bound} can be found with failure probability at
most $p$ using
\[
n(\Tmat(n,n,\lceil\sqrt n\rceil)+\Tmat(n,\lceil\sqrt n\rceil,n))
n^{o(1)}\operatorname{polylog}(1/p)
\]
arithmetic operations in the real-arithmetic model. In particular, the running
time is at most
\[
n^{1+\omega(1,1/2,1)+o(1)}\operatorname{polylog}(1/p).
\]
Using the bound $\omega(1,1/2,1)<2.042776$ from
\cite[Table~1]{almanetal25asymmetry} gives the numerical exponent
$3.042776+o(1)$.
\end{theorem}

Algorithm~\ref{alg:faster_dense_matrix_spencer} is the implementation.  Its
outer loop is that of Algorithm~\ref{alg:dense_matrix_spencer}.  We solve each
projection by the certified multiplier wrapper of
Section~\ref{sec:selected_wrapper} around the fixed-multiplier procedure
\textsc{CoordinateMultiplier}.  The procedure \textsc{CompareValueExact}
compares the candidates, and for the final rounding we call the shared
procedure \textsc{CertifiedCleanup} of Algorithm~\ref{alg:certified_cleanup}
with its own share $p_{\mathrm{clean}}$ of the failure budget.  In the rest of
the section we prove Theorem~\ref{thm:faster_dense_runtime}.
In Section~\ref{sec:selected_setup} we set up the fixed-multiplier problem and
its selected-coordinate oracle.  We then analyze one guarded candidate phase in
Sections~\ref{sec:selected_guard} and~\ref{sec:selected_finite}, with exact
and then with finite matrix actions.  In Section~\ref{sec:selected_wrapper} we
give the selection rule and the wrapper.  We transfer correctness, allocate the
failure budget, and count operations in Section~\ref{sec:selected_proof}.

\begin{algorithm}[H]
\caption{$n^{1+\omega(1,1/2,1)+o(1)}\operatorname{polylog}(1/p)$ time algorithm for Matrix Spencer signing}
\label{alg:faster_dense_matrix_spencer}
\begin{algorithmic}[1]
\Procedure{MainSelected}{$A_1,\ldots,A_n,n,p$} \Comment{Theorem~\ref{thm:faster_dense_runtime}}
\State $y\gets0$, $S\gets[n]$, and $N_0\gets3\mathbin{\cdot}10^8$
\State Allocate from $p$ the per-solve budget $\delta_\nu$, the trial caps, and the cleanup budget $p_{\mathrm{clean}}$ as in the proof of Theorem~\ref{thm:faster_dense_runtime}
\While{$|S|\ge N_0$}
    \State $m\gets|S|$, $u\gets\log(n/m)$, $\ell_{\rm asp}\gets\max\{1,u\}$, and $R\gets c_{\rm rad}(\ell_{\rm asp})\sqrt m\ell_{\rm asp}$
    \Repeat
        \State Sample an accepted $\xi\sim\mathcal N(0,I_m)$ and keep it fixed through all multiplier tests
        \State If the certified interval for $\Phi(\xi_Q)$ lies below $1$, retain the box projection $\xi_Q$; otherwise solve Eq.~\eqref{eq:smooth_projection} by certified bisection of the multiplier over $[\underline\nu,\Lambda]$
        \State At each tested $\nu$: $\widehat x_\nu\gets$\Call{CoordinateMultiplier}{$F_\nu,Q,m,a,R,\varepsilon_{\mathrm{opt}},\delta_\nu$}; update the bracket only from the outward-rounded interval for $\Phi(\widehat x_\nu)$, or retain $\widehat x_\nu$ if that interval meets $1$
        \State Apply the downward-rounded radial correction to the retained point, snap within $2\rho$, and call the result $\overline x$
    \Until{at least $\delta m$ coordinates of $\overline x$ are snapped, or the prescribed trial cap is reached}
    \If{the trial cap is reached without success}
        \State \Return \textsc{Fail}
    \EndIf
    \State $y_i\gets y_i+2\overline x_i/\varepsilon_{\mathrm{box}}$ for every $i\in S$
    \State $S\gets\{i\in[n]:|y_i|<1\}$
\EndWhile
\State \Return \Call{CertifiedCleanup}{$A_1,\ldots,A_n,y,S,p_{\mathrm{clean}}$} \Comment{Algorithm~\ref{alg:certified_cleanup}}
\EndProcedure
\Procedure{CoordinateMultiplier}{$F_\nu,Q,m,a,R,\varepsilon_{\mathrm{opt}},\delta_\nu$} \Comment{$\delta_\nu$ is the failure budget of one fixed-multiplier solve, written $p_0$ in Section~\ref{sec:selected_setup}}
\State $F_0\gets F_\nu(0)$, $H\gets16F_0$, $J\gets\max\{1,\lceil\log_2(4F_0/\varepsilon_{\mathrm{opt}})\rceil\}$, $R_{\mathrm{rep}}\gets\lceil C_{\mathrm{rep}}\log(J/\delta_\nu)\rceil$, and $z\gets0$
\For{$j=1,\ldots,J$}
    \State Run $R_{\mathrm{rep}}$ independent guarded candidate phases from $z$, each of $T$ selected-coordinate queries with $k,s=\widetilde\Theta(\sqrt m)$, finite positive semidefinite exponential actions, and fresh trace guards; let $\mathcal C$ be $z$ together with the returned points
    \State $z\gets$\Call{CompareValueExact}{$\mathcal C,\varepsilon_{\mathrm{opt}}/(128J)$}
\EndFor
\State \Return $z$, or \textsc{Fail} if a prescribed cap is reached
\EndProcedure
\Procedure{CompareValueExact}{$\mathcal C,w$} \Comment{$\mathcal C$ contains the old point $z$; $w$ is the interval width}
\State Enclose $F_\nu(x)$ for every $x\in\mathcal C$ in an outward-rounded interval of width at most $w$, by one dense matrix-function evaluation per point
\State \Return the $x\in\mathcal C\setminus\{z\}$ with the least upper endpoint among those whose upper endpoint lies below the lower endpoint of $z$; if there is none, \Return $z$
\EndProcedure
\end{algorithmic}
\end{algorithm}

\begin{remark}[What a selected-coordinate query costs]
Remark~\ref{rem:dense_reading_runtime} derives the exponent $2+\omega/2$ from
one diagonalization per block query.
Theorem~\ref{thm:faster_dense_runtime} instead makes $\widetilde O(m)$
single-coordinate queries per projection and never forms a Gibbs matrix
inside a query.  A query applies polynomial functions of two dense matrices to
$\widetilde O(\sqrt m)$ vectors, at cost
$\Tmat(n,n,\lceil\sqrt m\rceil)+\Tmat(n,\lceil\sqrt m\rceil,n)+n^2$ up to
subpolynomial factors.  This cost is $n^{\omega(1,1/2,1)+o(1)}$ at $m=n$.
Remark~\ref{rem:accelerated_exponent} shows by convexity of
$b\mapsto\omega(1,b,1)$ that the new exponent never exceeds $2+\omega/2$.
\end{remark}

\subsection{Setup and the fixed-multiplier problem}
\label{sec:selected_setup}

From the proof of Theorem~\ref{thm:dense_runtime} we retain, at an active set
$S$ of size $m$ with fractional coloring $y$, the box $Q$ of
Eq.~\eqref{eq:smooth_projection}, the matrix $B(x):=\sum_{i\in S}x_iA_i$, and
the scale $R=c_{\rm rad}(\ell_{\rm asp})\sqrt m\ell_{\rm asp}$.  We also retain
the smoothing constants $a$, $b$, and $\chi=5\mathbin{\cdot}10^{-7}$ together
with the bound $h(x)\le\|B(x)\|/R+\chi$.  Likewise we retain the norm rejection
that discards a Gaussian draw $\xi$ unless $\|\xi\|_2^2\le C_gm$, the
projection accuracy $\rho$ of the snapping step, and the snapping rule.  We
retain the approximate Hermitian diagonalization primitive and the cleanup.
Every object called \emph{retained} below is one of these.

Fix one instance of Eq.~\eqref{eq:smooth_projection}.  Define
\[
C(x):=\operatorname{diag}(B(x)/R,-B(x)/R),
\qquad
D_x:=aC(x)-abI_{2n},
\]
\[
P_x:=\exp(D_x),
\qquad
\Phi(x):=\tr[P_x],
\qquad
q(x):=\frac12\|x-\xi\|_2^2.
\]
In this notation, $h(x)=b+a^{-1}\log\Phi(x)$.  The trace $\Phi(x)$ here is not
the log-partition $\Phi(Z)$ of Section~\ref{sec:algorithm}.  The smooth
constraint is therefore equivalent to
\[
h(x)\le b
\quad\Longleftrightarrow\quad
\Phi(x)\le1,
\]
and at the origin
\[
\phi_0:=\Phi(0)=2n\exp(-ab)<\frac14.
\]
The eigenvalues of $C(x)$ come in pairs $\pm\lambda$ and
$e^{a\lambda}+e^{-a\lambda}\ge2$, so $\Phi(x)\ge\phi_0$ for every $x$.

\paragraph{The fixed-multiplier problem.}
For a multiplier $\nu>0$ consider the strongly convex box problem
\[
F_\nu(x):=q(x)+\nu\Phi(x),
\qquad
x_\nu:=\operatorname*{argmin}_{x\in Q}F_\nu(x),
\]
and write
\[
\Delta_x:=F_\nu(x)-F_\nu(x_\nu)
\]
for the objective gap.  The wrapper of Section~\ref{sec:selected_wrapper}
reduces one projection to $O(\log(n/\rho))$ such problems.  In this subsection
and the next two we solve one of them.

\paragraph{Constants.}
We first fix the constants.  Put
\[
\eta:=10^{-3},
\qquad
d_0:=2n,
\qquad
d:=\frac\rho4,
\]
where $\eta$ is the relative accuracy of the coordinate oracle.  We use only
$\eta\le1/4$ and the value $10^{-3}$ in the crossing bounds of
Sections~\ref{sec:selected_guard} and~\ref{sec:selected_finite}.
Here $d_0$ is the order of $D_x$, and $d$ is the radius the wrapper works to.  For a fixed multiplier $\nu$ put
\[
F_0:=F_\nu(0),
\qquad
H:=16F_0;
\]
$H$ is the level below which we keep every guarded iterate.  For a requested
objective accuracy $\varepsilon_{\mathrm{opt}}>0$ and a failure budget
$p_0\in(0,1/2)$ put
\[
J:=\max\{1,\lceil\log_2(4F_0/\varepsilon_{\mathrm{opt}})\rceil\},
\qquad
R_{\mathrm{rep}}:=\lceil C_{\mathrm{rep}}\log(J/p_0)\rceil
\]
for a sufficiently large universal constant $C_{\mathrm{rep}}$.  Here $J$ is
the number of halving phases, and $R_{\mathrm{rep}}$ is the number of
independent candidates per phase.  Unsubscripted constants $C$ are universal and may
change from line to line.

\paragraph{The failure budget and the accepted draw.}
Let $p_0\in(0,1/2)$ denote the failure budget we allocate to one
fixed-multiplier solve.  The $O(\log(n/\rho))$ solves of one projection call
share the numerical allowance of Section~\ref{sec:algorithm}, and we do the
accounting in Section~\ref{sec:selected_proof}.  Before the multiplier
search, we draw $\xi$ once and apply the retained norm rejection.  The acceptance
probability is at least a universal constant, so we cap the sampler after
$O(\log(1/p_0))$ draws.  The cap fails with probability at most $p_0$ and
costs a logarithmic factor.  We fix the accepted $\xi$ through all multiplier
tests.  Every statement below is conditional on it.  On the accepted draw
$q(0)\le C_gm/2$, and for every multiplier the wrapper tests
$F_0\le q(0)+\Lambda\phi_0=O(m)$ with the bracket endpoint $\Lambda=O(m)$ of
Section~\ref{sec:selected_wrapper}.

\paragraph{The selected-coordinate Gibbs oracle.}
Put
\[
E_i:=\partial_iC(x)=\operatorname{diag}(A_i/R,-A_i/R),
\]
so that $\|E_i\|\le1/R$ by the normalization $\|A_i\|\le1$, and
\[
\partial_iF_\nu(x)=x_i-\xi_i+\nu a\tr[P_xE_i].
\]
The oracle estimates this derivative for one uniformly random coordinate $i$.
The estimate draws on a dense snapshot at a point $w\in Q$, whose full
gradient $\nu a\tr[P_wE_i]$, $i\in S$, we have formed, and on a sketch of the
increment
\[
E:=P_x-P_w.
\]
Choose integers $k,s\ge2$.  This $s$ is the number of Gaussian probes,
unrelated to the constant $s=5\mathbin{\cdot}10^{-7}$ of
Section~\ref{sec:algorithm}.  Draw a standard Gaussian matrix
$\Omega\in\R^{2n\times2k}$, let $Y$ have orthonormal columns spanning the
range of $E\Omega$, and put $\Pi:=YY^\top$.  Define
\[
E_{\mathrm{low}}:=\Pi E+E\Pi-\Pi E\Pi,
\qquad
E_{\mathrm{res}}:=(I-\Pi)E(I-\Pi),
\]
so that $E=E_{\mathrm{low}}+E_{\mathrm{res}}$ exactly.  For fresh independent
standard Gaussian vectors $z_1,\ldots,z_s\in\R^{2n}$, define
\[
\widehat t_i:=\tr[E_iE_{\mathrm{low}}]
+\frac1{2s}\sum_{j=1}^s
z_j^\top(E_iE_{\mathrm{res}}+E_{\mathrm{res}}E_i)z_j
\]
and
\[
\widehat g_i:=x_i-\xi_i+\nu a(\tr[P_wE_i]+\widehat t_i).
\]
Conditional on $x,w,i$ and $\Omega$, Gaussian isotropy $\E[zz^\top]=I$ gives
\[
\E[\widehat t_i]=\tr[E_iE_{\mathrm{low}}]+\tr[E_iE_{\mathrm{res}}]=\tr[E_iE],
\]
so the estimator is conditionally unbiased when the exponential actions are
exact.

For every matrix $M$ and a standard Gaussian vector $z$,
\begin{equation}
\E[z^\top Mz]=\tr[M],
\qquad
\V[z^\top Mz]=2\|\tfrac12(M+M^\top)\|_F^2,
\label{eq:gaussian_quadratic_form}
\end{equation}
and $\|M\|_F\le\tr[M]$ when $M$ is positive semidefinite.
The matrix $E_iE_{\mathrm{res}}+E_{\mathrm{res}}E_i$ is symmetric with
Frobenius norm at most $2\|E_i\|\|E_{\mathrm{res}}\|_F$, so the $s$
independent probes give
\[
\V[\widehat g_i\mid x,w,i,\Omega]
\le\frac{2\nu^2a^2}{sR^2}\|E_{\mathrm{res}}\|_F^2.
\]
The residual is controlled by the Gaussian range-finder estimate underlying
Hutch++ of Meyer, Musco, Musco, and Woodruff~\cite{meyeretal21hutchpp}:
\[
\E_\Omega[\|E_{\mathrm{res}}\|_F^2]
\le\frac3k\|E\|_1^2.
\]
Since we quote the constant $3$ in Section~\ref{sec:bit_complexity}, we give
its derivation from the deterministic projector bound of Halko, Martinsson,
and Tropp~\cite[Theorem~9.1]{hmt11}.  Let $E=W\Sigma V^\top$ be a singular
value decomposition with nonincreasing singular values.  We partition
$\Sigma=\operatorname{diag}(\Sigma_1,\Sigma_2)$ with $\Sigma_1$ of size
$k\times k$, and partition $V=[V_1\ V_2]$ conformally.  Put
$\Omega_1:=V_1^\top\Omega\in\R^{k\times2k}$ and $\Omega_2:=V_2^\top\Omega$.
Since $\|(I-\Pi)E(I-\Pi)\|_F\le\|(I-\Pi)E\|_F$, the projector bound gives
\[
\|E_{\mathrm{res}}\|_F^2\le\|\Sigma_2\|_F^2+
\|\Sigma_2\Omega_2\Omega_1^\dagger\|_F^2.
\]
Here $\Omega_1$ is a $k\times2k$ standard Gaussian matrix independent of
$\Omega_2$.  Conditioning on $\Omega_1$, the identity
$\E[\Omega_2^\top\Sigma_2^2\Omega_2]=\|\Sigma_2\|_F^2I$
\cite[Proposition~A.1]{hmt11} bounds the expectation of the second term by
$\|\Sigma_2\|_F^2\,\E[\|\Omega_1^\dagger\|_F^2]$.  The Gaussian
pseudoinverse identity $\E[\|\Omega_1^\dagger\|_F^2]=k/(k-1)$
\cite[Appendix~A]{hmt11} then makes the whole expectation at most
$(1+k/(k-1))\|\Sigma_2\|_F^2\le3\|\Sigma_2\|_F^2$ for $k\ge2$.  Finally,
\[
\|\Sigma_2\|_F^2
\le\sigma_{k+1}(E)\|E\|_1
\le\frac1k\|E\|_1^2,
\]
because the singular values are nonincreasing and $\sigma_{k+1}(E)$ is at
most the average of the $k$ larger ones, which proves the stated estimate.

\paragraph{From trace norm to objective gaps.}
For positive definite $P,P'$ of possibly different traces, the relative entropy
$\mathcal D(P\|P'):=\tr[P(\log P-\log P')-P+P']$ satisfies the generalized Pinsker
inequality $\|P-P'\|_1^2\le4\max\{\tr[P],\tr[P']\}\mathcal D(P\|P')$ of Fact~\ref{fact:generalized_pinsker}.
Let $P_\nu:=P_{x_\nu}$.  Since $\log P_x=D_x$ is affine in $x$, the
divergence $\mathcal D(P_\nu\|P_x)$ equals
$\Phi(x)-\Phi(x_\nu)-\langle\nabla\Phi(x_\nu),x-x_\nu\rangle$, the Bregman
divergence of $\Phi$.  Adding the nonnegative Bregman divergence of $q$,
we obtain
\[
\nu\mathcal D(P_\nu\|P_x)
\le D_{F_\nu}(x,x_\nu),
\]
where the right side is the Bregman divergence of $F_\nu$.  The
box-constrained optimality condition
\[
\langle\nabla F_\nu(x_\nu),x-x_\nu\rangle\ge0
\qquad(x\in Q)
\]
shows $D_{F_\nu}(x,x_\nu)=\Delta_x-\langle\nabla F_\nu(x_\nu),x-x_\nu\rangle\le\Delta_x$.

Suppose $F_\nu(x),F_\nu(w),F_\nu(x_\nu)\le H$ for the level $H$ fixed above,
then $\nu\tr[P_x],\nu\tr[P_w],\nu\tr[P_\nu]\le H$, and the Pinsker inequality
for the pairs $(P_\nu,P_x)$ and $(P_\nu,P_w)$, the trace-norm triangle
inequality, and $(u+v)^2\le2u^2+2v^2$ give
\[
\nu^2\|P_x-P_w\|_1^2
\le8H(\Delta_x+\Delta_w).
\]
Write $\mathfrak{e}_i:=\widehat g_i-\partial_iF_\nu(x)$ for the oracle error.
Combining the variance bound, the range-finder estimate, and this display,
we obtain
\[
\E[|\mathfrak{e}_i|^2\mid x,w]
\le\frac\eta m(\Delta_x+\Delta_w)
\]
whenever
\[
ks\ge C\frac{a^2mH}{R^2\eta}.
\]

\paragraph{A loopless coordinate method.}
Set
\[
\overline L:=\max\{1+ea^2H/R^2,2a/R\},
\qquad
\alpha:=\frac1{m\overline L}.
\]
The first term in $\overline L$ is the coordinate smoothness constant on the
level set, which we prove below.  The second term is a trust-step correction that
makes the comparison displacement admissible.  Given the selected coordinate $i$ and
the estimate $\widehat g_i$, we minimize
\[
\widehat g_i\gamma+\frac{\overline L}{2}\gamma^2
\]
over all displacements $\gamma$ with $x_i+\gamma\in Q_i$ and
$|\gamma|\le R/a$, and put $y:=x+\gamma e_i$.  If $F_\nu(x)\le H$, then for
every $\vartheta$ between $0$ and $\gamma$,
\[
\|a(C(x+\vartheta e_i)-C(x))\|=a|\vartheta|\|E_i\|\le1,
\]
so $D_{x+\vartheta e_i}\preceq D_x+I$ and $\Phi(x+\vartheta e_i)\le e\Phi(x)$
by monotonicity of $\tr\exp$.  By Fact~\ref{fact:trace_exp_hessian} with
$Z=D_x$ and $H=aE_i$, the second derivative of
$\vartheta\mapsto\Phi(x+\vartheta e_i)$ is at most
$a^2\|E_i\|^2\Phi(x+\vartheta e_i)\le a^2\Phi(x+\vartheta e_i)/R^2$, and with
$\nu\Phi(x)\le F_\nu(x)\le H$,
\[
\partial_{ii}^2F_\nu(x+\vartheta e_i)
\le1+\frac{ea^2H}{R^2}
\le\overline L.
\]
This proves coordinate smoothness along the proposed segment before the
proposed point has passed any level test.

Take any comparison displacement $\gamma'$ that is admissible, meaning
$x_i+\gamma'\in Q_i$ and $|\gamma'|\le R/a$.  The one-dimensional proximal
three-point inequality for the minimized quadratic, the smoothness bound
along the segment, and Young's inequality on the error term
$\mathfrak{e}_i(\gamma'-\gamma)$ give
\[
F_\nu(y)
\le F_\nu(x)+\partial_iF_\nu(x)\gamma'
+\frac{\overline L}{2}(\gamma')^2
+\frac{\mathfrak{e}_i^2}{2\overline L}.
\]
The choice
\[
\gamma':=\frac{x_{\nu,i}-x_i}{\overline L}
\]
is admissible.  It is a convex-combination displacement inside $Q_i$, as
$\overline L\ge1$.  Also $|\gamma'|\le2/\overline L\le R/a$, as the box has
width below $2$ and $\overline L\ge2a/R$.  Averaging over the uniform
coordinate $i$ and using the $1$-strong convexity of $F_\nu$ in the form
$\sum_i(\partial_iF_\nu(x)(x_{\nu,i}-x_i)+\frac12(x_{\nu,i}-x_i)^2)\le-\Delta_x$,
we obtain
\[
\E[\Delta_y\mid x,w]
\le(1-\alpha)\Delta_x+
\frac1{2\overline L}\E[|\mathfrak{e}_i|^2\mid x,w].
\]
With the oracle bound and $\frac1{2\overline L}\cdot\frac\eta m=\frac{\eta\alpha}2$
this reads
\[
\E[\Delta_y\mid x,w]
\le A_0\Delta_x+B_0\Delta_w,
\qquad
A_0:=1-(1-\eta/2)\alpha,
\qquad
B_0:=\eta\alpha/2.
\]

\paragraph{The unguarded contraction.}
After forming $y$, set $x^+:=y$ and, independently, set
\[
w^+:=
\begin{cases}
y,&\text{with probability }\alpha,\\
w,&\text{with probability }1-\alpha.
\end{cases}
\]
When a refresh occurs, we form the dense snapshot and all its coordinate
derivatives.  For
\[
V(x,w):=\Delta_x+\frac12\Delta_w
\]
direct substitution gives
\[
\E[V(x^+,w^+)\mid x,w]
\le(1-\alpha/4)V(x,w)
\qquad(\eta\le1/4).
\]
Indeed, $\E[\Delta_{w^+}\mid x,w]\le\alpha\E[\Delta_y\mid x,w]+(1-\alpha)\Delta_w$,
so the coefficients of $\Delta_x$ and $\Delta_w$ on the right are
$(1+\alpha/2)A_0$ and $(1+\alpha/2)B_0+(1-\alpha)/2$.  For $\eta\le1/4$
and $\alpha\in(0,1]$,
\[
(1+\alpha/2)A_0\le1-\alpha/4,
\qquad
(1+\alpha/2)B_0+\frac{1-\alpha}2
\le\frac{1-\alpha/4}2.
\]
Consequently, $T:=\lceil16m\overline L\rceil$ coordinate queries reduce the
potential by the factor $(1-\alpha/4)^T\le e^{-4}$, and the expected number
of dense snapshot refreshes is at most $T\alpha\le17$.  The candidate phase
of Section~\ref{sec:selected_guard} uses this update only when its level
guard accepts the proposal.  If the guard rejects, the phase terminates and
returns its starting point.

\begin{remark}[What the sketch buys]
The oracle's mean-square error is $(\eta/m)(\Delta_x+\Delta_w)$, a $1/m$
fraction of the objective gaps, at sketch size $ks=\widetilde O(m)$.  Here
$k,s=\widetilde\Theta(\sqrt m)$.  The factor $1/m$ is what a coordinate
method needs.  Its step $\alpha=1/(m\overline L)$ contracts the gap by
$1-\Theta(\alpha)$ per query, so an error of order $\alpha\Delta$ per query
is absorbed.  Thus $\widetilde O(m)$ queries reduce the gap by a constant
factor.  The same argument without the range finder, with Hutchinson probes
on $P_x-P_w$ alone, needs $s=\widetilde\Theta(m)$ probes for the same
variance.  A query then costs $\Tmat(n,n,m)=n^{\omega+o(1)}$, the dense
cost of Section~\ref{sec:algorithm}.  The range finder of rank
$k=\widetilde\Theta(\sqrt m)$ removes the $k$ leading singular directions of
$P_x-P_w$ and leaves a residual of squared Frobenius norm at most $3/k$ times
$\|P_x-P_w\|_1^2$.  Hence $s=\widetilde\Theta(\sqrt m)$ probes suffice.  The
speedup comes from this cheaper certified oracle, not from a new
acceleration theorem.
\end{remark}

\subsection{The level guard, the crossing bound, and localization}
\label{sec:selected_guard}

A fixed-multiplier solve starts at $z_0:=0$, so $F_0=F_\nu(z_0)$ and
$H=16F_0$ are the level constants we fixed above.  Each candidate phase
starts from $x=w=z$ with $F_\nu(z)\le H/16$ and proposes $T$ coordinate
steps, every one of which we guard.

\paragraph{Localization of the spectrum.}
On the level $F_\nu(x)\le H$, the bound
$\nu e^{\lambda_{\max}(D_x)}\le\nu\Phi(x)\le H$ and the $\pm$ symmetry of
the spectrum of $C(x)$ give
\[
\lambda_{\max}(D_x)\le\log(H/\nu),
\qquad
\lambda_{\min}(D_x)\ge-2ab-\log(H/\nu).
\]
A trust step $|\gamma|\le R/a$ moves $D_x$ by at most $\|a\gamma E_i\|\le1$
in operator norm, so at a proposed point the spectrum lies in the interval
enlarged by one at each end.  We choose a polynomial $q_r$ uniformly
approximating $\lambda\mapsto e^{\lambda/2}$ on this enlarged interval
$[\lambda_-,\lambda_+]$ and put
\[
\widetilde P_x:=q_r(D_x)^2.
\]
The square makes $\widetilde P_x$ positive semidefinite.  An explicit degree
bound follows.  Put $c:=(\lambda_-+\lambda_+)/2$ and truncate the Taylor series,
\[
q_r(\lambda):=e^{c/2}\sum_{j=0}^r\frac{((\lambda-c)/2)^j}{j!}.
\]
The Taylor remainder and $(r+1)!\ge((r+1)/e)^{r+1}$ show that its uniform
error on the interval is at most
\[
e^{\lambda_+/2}\Bigl(\frac{e(\lambda_+-\lambda_-)}{4(r+1)}\Bigr)^{r+1}.
\]
We choose $r$ so that this error is at most
$\varepsilon_{\mathrm{op}}/(3(1+e^{\lambda_+/2}))$.  Then
$|q_r(\lambda)^2-e^\lambda|\le\varepsilon_{\mathrm{op}}$ throughout the
interval, since
$|q_r^2-e^\lambda|\le|q_r-e^{\lambda/2}|(2e^{\lambda/2}+|q_r-e^{\lambda/2}|)$.
As $\lambda_+-\lambda_-=O(ab+|\log(H/\nu)|)$, it is enough to take
\[
r=O(ab+|\log(H/\nu)|+\log(1/\varepsilon_{\mathrm{op}})),
\]
and the spectral theorem gives $\|\widetilde P_x-P_x\|\le\varepsilon_{\mathrm{op}}$.

\paragraph{The trace guard.}
By Eq.~\eqref{eq:gaussian_quadratic_form}, $\E[z^\top Mz]=\tr[M]$ and
$\V[z^\top Mz]\le2(\tr[M])^2$ for the positive semidefinite $M:=\widetilde P_y$.
Set
\[
s_{\mathrm{tr}}:=\lceil C\log(TJR_{\mathrm{rep}}/p_0)\rceil.
\]
We split $s_{\mathrm{tr}}$ fresh probes into batches of a fixed size, apply
Chebyshev's inequality to each batch mean, and take the median of the batch
means.  The resulting estimate $\widehat\tau_y$ obeys
\[
|\widehat\tau_y-\tr[\widetilde P_y]|\le\tr[\widetilde P_y]/4
\]
except with probability $p_0/(100TJR_{\mathrm{rep}})$.  We choose
$\varepsilon_{\mathrm{op}}$ with $d_0\varepsilon_{\mathrm{op}}\le\phi_0/8$.  The
choice we make in Section~\ref{sec:selected_finite} is stronger.  Define
\[
U(y):=\frac43\widehat\tau_y+d_0\varepsilon_{\mathrm{op}}.
\]
Since $\Phi(y)\ge\phi_0$,
$|\tr[\widetilde P_y]-\Phi(y)|\le d_0\varepsilon_{\mathrm{op}}\le\Phi(y)/8$, and
the successful certificate satisfies
\[
\Phi(y)\le U(y)\le2\Phi(y):
\]
the lower bound is
$\frac43\cdot\frac34\tr[\widetilde P_y]+d_0\varepsilon_{\mathrm{op}}\ge\Phi(y)$,
and the upper bound is
$\frac43\cdot\frac54\cdot\frac98\Phi(y)+\frac18\Phi(y)=2\Phi(y)$.  We accept
the proposal only when
\[
q(y)+\nu U(y)\le H.
\]
Every accepted point is in the $H$-level set, while we accept every proposal
with $F_\nu(y)\le H/3$.  Because
$\Phi(y)\ge\phi_0=2n\exp(-ab)=n^{-O(1)}$, the required additive trace
accuracy is inverse polynomial.  If the test rejects, we terminate the
candidate phase and return $z$.  A conditional union bound is valid for the
adaptive queries because every guard uses fresh probes.  The $T$ guards of one
candidate phase are all correct with probability at least $0.99$, since
$T\cdot p_0/(100TJR_{\mathrm{rep}})\le0.01$.

\paragraph{The crossing bound.}
A crossing means that the first proposed point $y$ with $F_\nu(y)>H/3$ has
been formed.  Taking $\gamma'=0$ in the pathwise coordinate inequality, we obtain
\[
F_\nu(y)\le F_\nu(x)+\frac{\mathfrak{e}_i^2}{2\overline L}.
\]
Before the first crossing of $H/3$, both $x$ and $w$ lie below $H/3$, so
$\Delta_x+\Delta_w\le2H/3$ and the oracle bound gives
\[
\E[|\mathfrak{e}_i|^2]\le\frac{2\eta H}{3m}.
\]
Summing the pathwise inequality up to the crossing, we see that a crossing from the
initial level $H/16$ within $T$ steps requires
\[
\sum_{t<T}\mathfrak{e}_{i_t}^2\ge2\overline L\Bigl(\frac H3-\frac H{16}\Bigr)
=\frac{13}{24}\overline LH.
\]
Since $T\le17m\overline L$, the expected stopped sum is at most
$17m\overline L\cdot2\eta H/(3m)=34\eta\overline LH/3$, and Markov's
inequality gives
\[
\Pr[\text{a crossing before time }T]
\le\frac{34/3}{13/24}\,\eta=\frac{272}{13}\eta<0.021
\qquad(\eta=10^{-3}).
\]
On the event $\mathrm{safe}$ of no crossing and correct guards, every proposal is accepted.
The guarded process then coincides with the unguarded one, with $x$ and $w$
below $H/3\le H$ at every step.  Therefore the killed-process version of the
unguarded contraction, started from $V(z,z)=\frac32\Delta_z$, gives
\[
\E[\mathbbm{1}[\mathrm{safe}]V_T]
\le\frac32e^{-4}\Delta_z.
\]
This is the exact-oracle contribution to the phase estimate.  In the next
subsection we add the error of finite exponential actions.

\subsection{Finite exponential actions}
\label{sec:selected_finite}

We now run the gradient oracle on the surrogates.  Apply the deflated
construction of Section~\ref{sec:selected_setup} to
\[
\widetilde E:=\widetilde P_x-\widetilde P_w
\]
in place of $E$, with the snapshot baseline $\tr[\widetilde P_wE_i]$, so
that the estimator has conditional expectation
$x_i-\xi_i+\nu a\tr[\widetilde P_xE_i]$.  With $d_0=2n$ the order of the
matrices,
\[
\|\widetilde E-(P_x-P_w)\|_1\le2d_0\varepsilon_{\mathrm{op}}
\qquad\text{and}\qquad
\|\widetilde P_x-P_x\|_1\le d_0\varepsilon_{\mathrm{op}}.
\]
The bias of the estimator is therefore
$\nu a|\tr[(\widetilde P_x-P_x)E_i]|\le\nu ad_0\varepsilon_{\mathrm{op}}/R$,
and the range-finder estimate applied to $\widetilde E$ gives
\[
\E[|\mathfrak{e}_i|^2\mid x,w]
\le\frac{C\nu^2a^2}{ksR^2}\|\widetilde E\|_1^2
+\Bigl(\frac{\nu ad_0}{R}\varepsilon_{\mathrm{op}}\Bigr)^2.
\]
Moreover, by the triangle inequality and $(u+v)^2\le2u^2+2v^2$,
\[
\|\widetilde E\|_1^2
\le2\|P_x-P_w\|_1^2+8d_0^2\varepsilon_{\mathrm{op}}^2.
\]
After we increase $ks$ by a universal constant, the $\|P_x-P_w\|_1^2$ term
is again at most $\frac\eta m(\Delta_x+\Delta_w)$.  Put
\[
\delta_g^2:=\min\{\eta H/(100m),
\varepsilon_{\mathrm{opt}}/(C_{\mathrm g}Jm)\},
\]
where $C_{\mathrm g}$ is a sufficiently large universal constant, and choose
\[
\varepsilon_{\mathrm{op}}
\le\min\{\phi_0/(16d_0),R\delta_g/(4\nu ad_0),
R\sqrt{ks}\delta_g/(C\nu ad_0)\}.
\]
The first constraint is the guard's requirement with headroom.  The second
bounds the squared bias by $\delta_g^2/16$.  The third bounds the
$\varepsilon_{\mathrm{op}}^2$ term of the variance by $\delta_g^2/2$ for a
suitable universal $C$.  The resulting total mean-square error satisfies
\[
\E[|\mathfrak{e}_i|^2\mid x,w]
\le\frac\eta m(\Delta_x+\Delta_w)+\delta_g^2.
\]
The one-step Lyapunov calculation then acquires only the additive term
$C\delta_g^2/\overline L$, and over $T\le17m\overline L$ iterations its
total contribution is $O(m\delta_g^2)$.  We choose $C_{\mathrm g}$ so that the noisy
killed-process estimate is
\[
\E[\mathbbm{1}[\mathrm{safe}]V_T]
\le\frac32e^{-4}\Delta_z+\frac{\varepsilon_{\mathrm{opt}}}{1024J}.
\]
The first term in the definition of $\delta_g$ changes the crossing bound by
less than $0.005$: the expected stopped sum gains at most
$T\delta_g^2\le17\eta\overline LH/100$, which raises the Markov bound by
$\frac{17/100}{13/24}\eta<0.005$.  Applying Markov's inequality to the
noisy estimate, with $\Delta_{x_T}\le V_T$ and
$\max\{3e^{-4},32/1024\}<0.055$, we now obtain
\[
\Pr[\mathbbm{1}[\mathrm{safe}]=1\text{ and }
\Delta_{x_T}>\Delta_z/2+\varepsilon_{\mathrm{opt}}/(32J)]<0.055.
\]
Together with the crossing probability, which is below $0.021+0.005$, and the guard
failure $0.01$, one candidate phase returns a point with gap at most
$\Delta_z/2+\varepsilon_{\mathrm{opt}}/(32J)$ with probability greater than
$1-0.026-0.01-0.055>0.9$.  Since all parameters multiplying
$\varepsilon_{\mathrm{op}}$ are polynomially bounded, the displayed degree $r$
is polylogarithmic.  A constant-relative trace guard uses only
$O(\log(TJR_{\mathrm{rep}}/p_0))$ fresh Gaussian probes, so its work is
absorbed by the polylogarithmic factors.

\subsection{Candidate selection and the certified wrapper}
\label{sec:selected_wrapper}

\paragraph{Candidate selection.}
{\setlength{\emergencystretch}{2em}We repeat each candidate phase independently $R_{\mathrm{rep}}$ times and pass
the returned points, together with the old point $z$, to
\textsc{CompareValueExact} with width $\varepsilon_{\mathrm{opt}}/(128J)$.  The
procedure encloses every objective in an outward-rounded interval of width
at most $\varepsilon_{\mathrm{opt}}/(128J)$, using one dense,
inverse-polynomial-accuracy matrix-function evaluation per point.  Here we evaluate $q(x)$
exactly in the real-arithmetic model, and we enclose $\Phi(x)$ to
absolute error at most $\varepsilon_{\mathrm{opt}}/(256\nu J)$.  Multiplying
that interval by $\nu$, we obtain the stated width.  These
evaluations are deterministic given their accuracy.  If we use a randomized trace
certificate instead, we allocate failure probability $p_0/(100J)$ among
them.  Among the candidates whose certified upper endpoint lies below the old
point's certified lower endpoint, the procedure retains the one with the
least upper endpoint.  If there is none, it retains the old point.  Thus the
accepted phase-start objectives never increase.  If
$\Delta_z>\varepsilon_{\mathrm{opt}}$, a successful candidate, with gap at
most $\Delta_z/2+\varepsilon_{\mathrm{opt}}/(32J)$, is separated from the
old point by more than
$\Delta_z/2-\varepsilon_{\mathrm{opt}}/(32J)>15\varepsilon_{\mathrm{opt}}/32$,
which exceeds the two certificate widths.  The retained point pays at
most one further width.  The exceptional event has probability at most $p_0/J$: all
$R_{\mathrm{rep}}$ candidates fail with probability
$0.1^{R_{\mathrm{rep}}}\le p_0/(2J)$ for $C_{\mathrm{rep}}$ large, and the
internal budgets add at most $p_0/(2J)$.  Hence, outside this event, either the old point already has
gap at most $\varepsilon_{\mathrm{opt}}$ or the selected point satisfies
\[
\Delta_{j+1}\le\Delta_j/2+\varepsilon_{\mathrm{opt}}/(16J),
\]
as $\frac1{32}+\frac1{128}<\frac1{16}$.  Unrolling the recurrence over the
$J$ phases, with $2^{-J}F_0\le\varepsilon_{\mathrm{opt}}/4$, we see that the
returned point has objective gap at most $\varepsilon_{\mathrm{opt}}$, except
with probability $J\cdot p_0/J=p_0$.  We cap snapshot refreshes at a
logarithmic multiple of their mean.  A binomial tail bound absorbs the cap
failure into $p_0$.\par}

\paragraph{The multiplier wrapper.}
Let $\xi_Q$ be the coordinatewise box projection of $\xi$ onto $Q$.  We certify an
interval of length at most $\tau/4$ for $\Phi(\xi_Q)$.  Here $\tau$ is the tolerance we fix below, not the moment order $\tau(\ell_{\rm asp})$ of Table~\ref{tab:six_regimes}.
If it lies below $1$, the box projection $\xi_Q$ is feasible and is the
required exact projection.  Otherwise $\Phi(\xi_Q)\ge1-\tau/4$, and Slater's
condition at the origin gives a multiplier $\nu_*\ge0$ such that $x_{\nu_*}$
is the exact projection $x_*$.  Here $\Phi(x_*)\le1$, and $\nu_*>0$ only if
$\Phi(x_*)=1$.  In the estimates below we use only $\nu_*\ge0$ and $\Phi(x_*)\le1$.
As the analogue for the selected-coordinate solver of the bracket endpoint $\Lambda$ of Section~\ref{sec:algorithm}, define
\[
\Lambda:=\frac{2q(0)}{1-\phi_0}.
\]

\begin{lemma}[Lower bound on the accepted draw]
\label{lem:selected_accepted_draw}
On the multiplier branch, $\|\xi\|_2\ge\|\xi_Q\|_2\ge184$.  Consequently
$\Lambda=\|\xi\|_2^2/(1-\phi_0)=\Omega(1)$, and $\Lambda=O(m)$ on the
accepted draw.
\end{lemma}

\begin{proof}
On this branch $\Phi(\xi_Q)\ge1-\tau/4$, so
$h(\xi_Q)=b+a^{-1}\log\Phi(\xi_Q)\ge b-\tau/(2a)$, using $\log(1-u)\ge-2u$ for
$u\le1/2$.  The retained bound $h(x)\le\|B(x)\|/R+\chi$ with
$\chi=5\mathbin{\cdot}10^{-7}$, the estimate
$\|B(x)\|\le\sum_{i\in S}|x_i|=:\|x\|_1\le\sqrt m\|x\|_2$, and
$|(\xi_Q)_i|\le|\xi_i|$ (because $0\in Q$) give
\[
\|\xi\|_2\ge\|\xi_Q\|_2\ge\frac{(b-\chi-\tau/(2a))R}{\sqrt m}
=(b-\chi-\tau/(2a))c_{\rm rad}(\ell_{\rm asp})\ell_{\rm asp}\ge184,
\]
where in the last step we use $c_{\rm rad}(\ell_{\rm asp})\ell_{\rm asp}\ge279$ for every $\ell_{\rm asp}\ge1$ by
Table~\ref{tab:six_regimes}, $b-\chi=1-10^{-6}$, and $\tau/(2a)\le10^{-9}$.
The bound on the tolerance term holds as $\tau\le d^2\le1/1024$ and
$a\ge2\mathbin{\cdot}10^6$.  Hence
$\Lambda=\|\xi\|_2^2/(1-\phi_0)\ge184^2$, and on the accepted draw
$\Lambda\le\frac43\|\xi\|_2^2\le\frac43C_gm$, as $\phi_0<1/4$.
\end{proof}

By Lemma~\ref{lem:selected_accepted_draw}, every tolerance below that
involves $\Lambda$ or $1/\underline\nu$ (which we define below) is inverse polynomial.  Comparing
$x_*$ and $x_\Lambda$ with the origin in $F_{\nu_*}$ and $F_\Lambda$
respectively, using $q\ge0$ and $\Phi(x_*)=1$ when $\nu_*>0$, we obtain
\[
\nu_*\le\frac{q(0)}{1-\phi_0}=\Lambda/2,
\qquad
\Phi(x_\Lambda)\le\frac{1+\phi_0}{2}<1.
\]
For $\nu>\mu>0$, optimality of $x_\nu$ and $x_\mu$ gives
\[
F_\nu(x_\nu)\le F_\nu(x_\mu),
\qquad
F_\mu(x_\mu)\le F_\mu(x_\nu),
\]
and adding the two inequalities yields
$(\nu-\mu)(\Phi(x_\nu)-\Phi(x_\mu))\le0$.  Hence $\nu\mapsto\Phi(x_\nu)$ is
nonincreasing.  Adding the strong-convexity inequalities
$F_\nu(x_*)\ge F_\nu(x_\nu)+\frac12\|x_\nu-x_*\|_2^2$ and
$F_{\nu_*}(x_\nu)\ge F_{\nu_*}(x_*)+\frac12\|x_\nu-x_*\|_2^2$, we obtain
$\|x_\nu-x_*\|_2^2\le(\nu-\nu_*)(\Phi(x_*)-\Phi(x_\nu))$.  The right
side is at most $|\nu-\nu_*||\Phi(x_\nu)-1|$ in both cases: for
$\nu\ge\nu_*$ by monotonicity and $\Phi(x_*)\le1$, for $\nu<\nu_*$ because
then $\Phi(x_*)=1$.  Thus
\[
\|x_\nu-x_*\|_2^2
\le|\nu-\nu_*||\Phi(x_\nu)-1|.
\]
For $\nu\ge\nu_*$, convexity of $\Phi$ and $\Phi(x_*)\le1$ further give,
through $\Phi(x_*)-\Phi(x_\nu)\le\|\nabla\Phi(x_*)\|_2\|x_*-x_\nu\|_2$ and
$|\partial_i\Phi(x)|=a|\tr[P_xE_i]|\le a\Phi(x)/R$ for every $i$,
\[
\|x_\nu-x_*\|_2
\le(\nu-\nu_*)G_*,
\qquad
G_*:=a\sqrt m/R.
\]

Put
\[
\nu_{\min}:=\frac\rho{8G_*},
\qquad
\underline\nu:=\min\{\nu_{\min},\Lambda/2\}.
\]
We first solve at $\underline\nu$ and certify an interval of length at most
$\tau/4$ for $\Phi(x_{\underline\nu})$ as below.  If the interval lies below
$1$, then $\Phi(x_{\underline\nu})<1$, which forces
$\underline\nu\ge\nu_*$.  In the contrary case $\nu_*>0$, $\Phi(x_*)=1$, and
monotonicity would give $\Phi(x_{\underline\nu})\ge1$.  The path
estimate then makes $x_{\underline\nu}$ $\rho/8$-close to $x_*$.  If the
interval meets $1$, the residual estimate
Eq.~\eqref{eq:phi_multiplier_residual} below gives distance at most $d/2$
directly.  Otherwise $\Phi(x_{\underline\nu})>1\ge\Phi(x_*)$, so
$\underline\nu<\nu_*\le\Lambda/2$ and $[\underline\nu,\Lambda]$ brackets
$\nu_*$.

\paragraph{Certified bisection.}
Put
\[
H_{\max}:=16(q(0)+\Lambda\phi_0+1),
\qquad
M_{\mathrm{lev}}:=H_{\max}/\underline\nu,
\qquad
G_{\mathrm{lev}}:=G_*M_{\mathrm{lev}},
\]
\[
\tau:=\min\{d^2/(1+\Lambda),d(1-\phi_0)/(16\sqrt m)\}.
\]
At a tested multiplier $\nu\in[\underline\nu,\Lambda]$, we solve to Euclidean
error at most
\[
\sigma:=\min\{d/8,\tau/(16G_{\mathrm{lev}})\}
\]
by invoking the fixed-multiplier solver with
\[
\varepsilon_{\mathrm{opt}}\le\sigma^2/2;
\]
the $1$-strong convexity of $F_\nu$ then gives
$\|\widehat x-x_\nu\|_2\le\sigma$ for the returned point $\widehat x$.
We evaluate $\Phi(\widehat x)$ with an outward-rounded absolute error at most
$\tau/16$.  The wrapper requires of the fixed-multiplier solver the interface
hypothesis
\[
F_\nu(\widehat x)\le H_{\max}.
\]
The solver of this section meets it: $\widehat x$ is an accepted guarded
iterate or the old point, so
$F_\nu(\widehat x)\le H=16F_\nu(0)=16(q(0)+\nu\phi_0)\le H_{\max}$ as
$\nu\le\Lambda$.  In Sections~\ref{sec:accelerated_block_algorithm}
and~\ref{sec:cubic_time} we verify it for the solvers of those sections.  Since also
$F_\nu(x_\nu)\le F_\nu(0)\le H_{\max}$, both endpoints of the segment from
$\widehat x$ to $x_\nu$ lie in the $H_{\max}$-level set of $F_\nu$, and
convexity puts the entire segment there.  Every point $v$ of the segment
therefore has $\nu\Phi(v)\le H_{\max}$, so $\Phi(v)\le M_{\mathrm{lev}}$ and
\[
\|\nabla\Phi(v)\|_2
\le\frac{a\sqrt m}{R}\Phi(v)
\le G_*M_{\mathrm{lev}}=G_{\mathrm{lev}},
\]
whence
\[
|\Phi(\widehat x)-\Phi(x_\nu)|
\le G_{\mathrm{lev}}\sigma\le\tau/16.
\]
The evaluation error $\tau/16$ on each side and the transfer error
$\tau/16$ on each side produce a certified interval for $\Phi(x_\nu)$ of
length at most $\tau/4$.  If the interval lies above or below $1$,
monotonicity chooses the correct half of the bracket.  If it meets $1$,
then $|\Phi(x_\nu)-1|\le\tau/4$ and $|\nu-\nu_*|\le\Lambda$, so the
residual estimate gives
\begin{equation}
\|x_\nu-x_*\|_2^2
\le|\nu-\nu_*||\Phi(x_\nu)-1|
\le\Lambda\tau/4\le d^2/4,
\label{eq:phi_multiplier_residual}
\end{equation}
distance at most $d/2$, and we retain $\widehat x$.  Otherwise we stop when
the multiplier bracket has length at most $d/(4G_*)$ and use its feasible
upper endpoint.  The path estimate gives distance at most $d/4$ from $x_*$,
and the fixed-multiplier optimization error adds at most $\sigma\le d/8$.
The number of tested multipliers is $O(\log(\Lambda G_*/d))=O(\log(n/\rho))$.

\paragraph{Exact feasibility.}
Let $\widehat x$ be the approximate output of the wrapper and let
$\overline\Phi\ge\Phi(\widehat x)$ be its certified upper bound.  Define
\[
\beta_0:=
\begin{cases}
1,&\overline\Phi\le1,\\
(1-\phi_0)/(\overline\Phi-\phi_0),&\overline\Phi>1,
\end{cases}
\]
and compute a downward-rounded number $\beta\le\beta_0$ with
$0\le\beta_0-\beta\le d/(32\sqrt m)$.  Since $0,\widehat x\in Q$ and $\Phi$
is convex with $\Phi(0)=\phi_0$,
$\Phi(\beta\widehat x)\le(1-\beta)\phi_0+\beta\overline\Phi\le1$ for
$\beta\le\beta_0$.  Moreover,
$1-\beta_0=(\overline\Phi-1)/(\overline\Phi-\phi_0)\le(\overline\Phi-1)/(1-\phi_0)$
when $\overline\Phi>1$, and $\|\widehat x\|_2\le\sqrt m$ because every
coordinate of $Q$ has modulus below $1$, so
\[
(1-\beta_0)\|\widehat x\|_2
\le\frac{\overline\Phi-1}{1-\phi_0}\sqrt m.
\]
Our accuracy choices above ensure $\overline\Phi-1\le\tau/2$.  In the
retained case $\Phi(x_\nu)\le1+\tau/4$, and the evaluation and transfer errors add
$\tau/8$.  In every other case $\overline\Phi\le1$.  The second term in
the definition of $\tau$ then makes the preceding display at most $d/32$.
Downward rounding moves the point by at most
$(\beta_0-\beta)\|\widehat x\|_2\le d/32$.  The point $\beta\widehat x$ is
therefore exactly feasible for $\Phi\le1$, equivalently for $h\le b$.  Its
distance to the exact projection is at most
\[
\frac d2+\frac d8+\frac d{32}+\frac d{32}=\frac{11d}{16}<d=\frac\rho4\le\rho,
\]
the four terms being the residual or path bound, the fixed-multiplier
error, the radial correction, and the rounding.

\subsection{Correctness transfer, failure budget, and running time}
\label{sec:selected_proof}

\begin{proof}[Proof of Theorem~\ref{thm:faster_dense_runtime}]
\emph{Step 1: correctness transfer.}
Every projection accepted by the wrapper is exactly feasible for $\Phi\le1$,
equivalently for $h\le b$, and lies within the retained accuracy $\rho$ of
the exact projection.  Therefore the snapping proof, the success probability
of each refreshed projection, the discrepancy constant $156000$, and the
cleanup of the proof of Theorem~\ref{thm:dense_runtime} are unchanged.

\emph{Step 2: the failure budget.}
Every randomized ingredient is capped and returns \textsc{Fail} when its cap
is reached, so an uncertified signing is never output, and the theorem's $p$
is the sum of the cap failures.  We allocate it as in the proof of
Theorem~\ref{thm:dense_runtime}.  One fixed-multiplier solve fails with
probability at most $p_0$ (Section~\ref{sec:selected_wrapper}).  One
projection call performs $O(\log(n/\rho))$ solves and one capped rejection
sampler.  We choose the parameter $p_0$ so that these budgets together stay within the
$0.02$ numerical allowance per attempt, so that an attempt succeeds with
probability greater than $0.1$.  The per-phase trial caps
$O(\log\log n+\log(1/p))$ over the $O(\log n)$ outer phases and the cleanup
cap $O(\log(1/p))$, which is prescribed by the budget $p_{\mathrm{clean}}$ passed to \textsc{CertifiedCleanup} in Algorithm~\ref{alg:faster_dense_matrix_spencer}, then bound the total failure probability by $p$.  Fresh
randomness at every call makes the conditional union bound valid under
adaptivity.  The quantity $p_0$ enters the running time only through
$\log(1/p_0)$, so the caps contribute the factor
$\operatorname{polylog}(1/p)$.

\emph{Step 3: cost of one selected-coordinate query.}
On every fixed-multiplier level, $H=O(m)$, $R^2=\widetilde\Omega(m)$, and
$a=O(\log n)$, so $\overline L=\operatorname{polylog}(n)$ and we meet the sketch
requirement $ks\ge Ca^2mH/(R^2\eta)$ with $ks=\widetilde O(m)$.  We take
integers
\[
k,s=\widetilde\Theta(\sqrt m),
\]
both at least $2$ as the range-finder estimate requires, which is automatic
since $m\ge N_0$ throughout the outer loop.  One query applies polynomial
functions of the two dense $n\times n$ matrices $B(x)$ and $B(w)$, through
which the surrogates $\widetilde P_x$ and $\widetilde P_w$ act blockwise, to
$\widetilde O(\sqrt m)$ vectors.  These vectors are the columns of $\Omega$ and $Y$, the probes
$z_j$, and the guard probes.  This step costs
$\widetilde O(\Tmat(n,n,\lceil\sqrt m\rceil))$.  The query then materializes the two
dense matrices $\Pi\widetilde E+\widetilde E\Pi-\Pi\widetilde E\Pi$ and
$\sum_{j=1}^s(\widetilde E_{\mathrm{res}}z_j)z_j^\top$, where $\widetilde E_{\mathrm{res}}:=(I-\Pi)\widetilde E(I-\Pi)$.  We form them from their
rank-$O(k+s)$ factors, at cost $\widetilde O(\Tmat(n,\lceil\sqrt m\rceil,n))$.
The query reads the requested coordinate derivative as a Frobenius contraction
of $E_i$ with these two matrices, at cost $O(n^2)$.  The cost of one query is
\[
\widetilde O(
\Tmat(n,n,\lceil\sqrt m\rceil)
+\Tmat(n,\lceil\sqrt m\rceil,n)
+n^2).
\]
QR factorizations, Gaussian generation, projections onto the range of $Y$,
and the maintenance of $B(x)$ after a coordinate update are covered by the
$n^2$ term.  A snapshot refresh forms $\widetilde P_w$ densely by the
retained approximate Hermitian diagonalization primitive and then all $m$
coordinate derivatives, at cost
\[
n^{\omega+o(1)}+O(mn^2).
\]
Refreshes, dense objective evaluations for \textsc{CompareValueExact}, multiplier
tests, and candidate repetitions are polylogarithmically many.  The dense
inverse-polynomial-accuracy evaluations of $\Phi$ use the same primitive at
cost $n^{\omega+o(1)}$ each.  A fixed-multiplier solve makes
$TJR_{\mathrm{rep}}=\widetilde O(m)$ queries and a projection call makes
$O(\log(n/\rho))$ solves, so a projection at active size $m$ costs
\[
\widetilde O(
m\Tmat(n,n,\lceil\sqrt m\rceil)
+m\Tmat(n,\lceil\sqrt m\rceil,n)
+mn^2+n^\omega).
\]

\emph{Step 4: the sum over rounds.}
The active dimensions in the outer partial-coloring algorithm decrease
geometrically, so $\sum_tm_t=O(n)$.  Monotonicity of $\Tmat$ in each
argument gives
\[
\sum_tm_t\Tmat(n,n,\lceil\sqrt{m_t}\rceil)
\le O(n\Tmat(n,n,\lceil\sqrt n\rceil)),
\]
and the transposed orientation satisfies the analogous bound.  Also,
\[
\sum_tm_tn^2=O(n^3),
\]
while all dense snapshots and certified value evaluations together cost
$n^{\omega+o(1)}$.  Input construction, snapping, the final signed sum, and
the retained bounded-dimension cleanup cost at most $n^{3+o(1)}$.

Producing the dense $n\times n$ output gives
$\Tmat(n,\lceil\sqrt n\rceil,n)=\Omega(n^2)$, so
$n\Tmat(n,\lceil\sqrt n\rceil,n)$ absorbs the $n^3$ term.  Since
$\omega<3$, it also absorbs $n^\omega$.  Thus the global cost is
\[
n(\Tmat(n,n,\lceil\sqrt n\rceil)+\Tmat(n,\lceil\sqrt n\rceil,n))
n^{o(1)}\operatorname{polylog}(1/p).
\]
By the definition of the rectangular exponent and the permutation invariance
of $\omega(\cdot,\cdot,\cdot)$, which we recorded in Section~\ref{sec:preliminaries},
both $\Tmat(n,n,\lceil\sqrt n\rceil)$ and $\Tmat(n,\lceil\sqrt n\rceil,n)$
are $n^{\omega(1,1/2,1)+o(1)}$, which is the second bound of the theorem.
The numerical exponent follows from $\omega(1,1/2,1)<2.042776$.
\end{proof}

\section{The accelerated block algorithm}
\label{sec:accelerated_block_algorithm}

Theorem~\ref{thm:faster_dense_runtime} spends one sketch of width
$\widetilde\Theta(\sqrt m)$ on every coordinate query. In this section we let
$\ell$ coordinates share one sketch of width $\widetilde\Theta(\sqrt{m/\ell})$,
pay for the sharing with a block curvature of order $\ell$, and recover that
cost by Catalyst acceleration, which needs only $\widetilde O(\sqrt\ell)$
regularized subproblems. We obtain the speedup from a cheaper certified Gibbs
oracle and from a safe-level invariant for the extrapolated centers, not from a
new acceleration theorem. The Catalyst centers can leave the Gibbs level set on
which we analyze the deflated oracle, and an energy argument
(Lemma~\ref{lem:accelerated_catalyst}) keeps every regularized inner solve on
that level.

\begin{theorem}[Accelerated block-coordinate dense implementation]
\label{thm:accelerated_dense_runtime}
For symmetric $A_1,\ldots,A_n\in\R^{n\times n}$ with $\|A_i\|\le1$ and every $p\in(0,1/2)$, a signing satisfying the
conclusion of Theorem~\ref{thm:matrix_spencer_bound} can be found with failure
probability at most $p$ using
\[
n^{3+\gamma_\star+o(1)}\operatorname{polylog}(1/p)
\]
arithmetic operations in the real-arithmetic model, where
\[
\gamma_\star:=
\min_{0\le b\le1/2}
\max\{\omega(1,b,1)+b-5/2,1/2-b\}.
\]
\end{theorem}

In Algorithms~\ref{alg:accelerated_dense_matrix_spencer} and
\ref{alg:accelerated_multiplier} we record the implementation. Their subroutines
\textsc{GuardedBlockTrial} and \textsc{CompareValue} are
Algorithms~\ref{alg:accelerated_guarded_trial} and
\ref{alg:accelerated_compare_value}, listed in
Section~\ref{subsec:accelerated_trial}. We prove
Theorem~\ref{thm:accelerated_dense_runtime} in the rest of the section. In
Section~\ref{sec:accelerated_constants} we fix one projection subproblem at a
fixed multiplier, its block partition, the regularized objective $G$, and the
level $H$. There we also set the parameters of the solver: block curvature,
step and phase length, bias floor, sketch width, proximal tolerance, Catalyst
schedule, and failure budgets. We prove Lemma~\ref{lem:accelerated_variance} in
Section~\ref{subsec:accelerated_variance}. It bounds the mean-square error of
one block query whose $\ell$ coordinates share a sketch of width
$\widetilde\Theta(\sqrt{m/\ell})$. In Section~\ref{subsec:accelerated_step} we
prove Lemma~\ref{lem:accelerated_step}. The block curvature is at most
$\widehat L$ on the trust ball, and one proximal step contracts the expected
gap by $1-\gamma$ up to oracle error and residual. The third part of the lemma
is the guard, a probe certificate of the objective at the new point. We couple
one guarded trial to an ideal trial in Section~\ref{subsec:accelerated_trial}
and prove Lemma~\ref{lem:accelerated_trial}. A phase of repeated trials halves
the gap up to an additive term, and $J$ phases solve one regularized subproblem
with $\widetilde O(q\widehat L/\mu)$ block queries and $\widetilde O(1)$ dense
snapshot constructions. We run Catalyst on these subproblems in
Section~\ref{subsec:accelerated_catalyst} and prove
Lemma~\ref{lem:accelerated_catalyst}. Every subproblem starts below $H/16$,
which is the safe-level invariant, and $K=\widetilde O(\sqrt\ell)$ subproblems
deliver the accuracy $\varepsilon_{\mathrm{opt}}$ and the level that the
multiplier wrapper requires. We count arithmetic operations in
Lemma~\ref{lem:accelerated_cost} of Section~\ref{subsec:accelerated_cost}, from
one block query and one snapshot construction up to all rounds. The lemma
expresses the total projection cost through the rectangular exponent
$\omega(1,b,1)$. We assemble the proof of
Theorem~\ref{thm:accelerated_dense_runtime} in
Section~\ref{subsec:accelerated_proof}: correctness, the running time, and the
numerical value of $\gamma_\star$.

\begin{remark}[Reading the exponent]
\label{rem:accelerated_exponent}
The dual matrix-multiplication exponent $\alpha_{\mathrm{dual}}$ is defined in
Section~\ref{sec:model}. If
$\alpha_{\mathrm{dual}}\ge1/2$, then $b=1/2$ gives $\gamma_\star=0$ and the
running time $n^{3+o(1)}$. This holds in particular if $\omega=2$. The
rectangular matrix-multiplication bounds of Alman, Duan, Vassilevska Williams,
Xu, Xu and Zhou~\cite[Table~1]{almanetal25asymmetry} give
$\gamma_\star\le0.0183197999\ldots$. The exponent $3.018320+o(1)$ therefore
improves on the exponent $3.042776+o(1)$ of
Theorem~\ref{thm:faster_dense_runtime} and on the exponent
$2+\omega/2+o(1)<3.185670$ of Theorem~\ref{thm:dense_runtime}. That
exponent uses $\omega<2.371339$ from the same source. The comparison with
Theorem~\ref{thm:dense_runtime} holds for every value of $\omega$. Convexity
of $b\mapsto\omega(1,b,1)$, proved in the proof of
Theorem~\ref{thm:accelerated_dense_runtime} below, gives
$\omega(1,1/2,1)\le(\omega(1,0,1)+\omega(1,1,1))/2=1+\omega/2$. Hence
$\gamma_\star\le\omega(1,1/2,1)-2\le(\omega-2)/2$.
\end{remark}

\begin{algorithm}[H]
\caption{$n^{3+\gamma_\star+o(1)}\operatorname{polylog}(1/p)$ time algorithm for Matrix Spencer signing}
\label{alg:accelerated_dense_matrix_spencer}
\begin{algorithmic}[1]
\Procedure{MainBlock}{$A_1,\ldots,A_n,n,p$} \Comment{Theorem~\ref{thm:accelerated_dense_runtime}}
\State Choose $b\in[0,1/2]$ attaining the exponent in Theorem~\ref{thm:accelerated_dense_runtime} up to $o(1)$, and set $\ell\gets\max\{2,\lceil n^{1-2b}\rceil\}$
\State Run Algorithm~\ref{alg:faster_dense_matrix_spencer} with the same fixed failure-budget allocation, multiplier certificates, snapping rule, and cleanup
\State Replace every fixed-multiplier solve by \textsc{AcceleratedMultiplier}; return its signing or \textsc{Fail}
\EndProcedure
\end{algorithmic}
\end{algorithm}

\begin{algorithm}[H]
\caption{Accelerated fixed-multiplier solver}
\label{alg:accelerated_multiplier}
\begin{algorithmic}[1]
\Procedure{AcceleratedMultiplier}{$F_\nu,Q,m,\ell,a,R,\varepsilon_{\mathrm{opt}},\delta_\nu$} \Comment{Lemmas~\ref{lem:accelerated_trial} and \ref{lem:accelerated_catalyst}}
\If{$m<Ca^2\ell^2$}
    \State \Return \Call{CoordinateMultiplier}{$F_\nu,Q,m,a,R,\varepsilon_{\mathrm{opt}},\delta_\nu$} \Comment{Algorithm~\ref{alg:faster_dense_matrix_spencer}}
\EndIf
\State $U\gets F_\nu(0)$, $H\gets64U$, $\kappa\gets\ell$, $\mu\gets1+\kappa$, $\alpha\gets1/\sqrt{1+\kappa}$, and $\vartheta\gets(1-\alpha)/(1+\alpha)$
\State $K\gets\lceil C\alpha^{-1}\log(U/(\alpha^2\varepsilon_{\mathrm{opt}}))\rceil$, $x_0\gets0$, $y_0\gets0$, and $h_0\gets0$
\State Partition $[m]$ into $q\gets\lceil m/\ell\rceil$ fixed blocks of size at most $\ell$
\State $\widehat L\gets\max\{1+\kappa+ea^2H\ell/R^2,2a(1+\kappa)\ell/R\}$, $\gamma\gets\mu/(q\widehat L)$, and $T\gets\lceil16/\gamma\rceil$
\State Choose $k,s\ge2$ with $ks\ge Ca^2H\ell q/(\eta\mu R^2)$ and $k=s=\widetilde\Theta(\sqrt{m/\ell})$
\For{$j=1,\ldots,K$}
    \State $G_j(x)\gets F_\nu(x)+\kappa\|x-y_{j-1}\|_2^2/2$ and $\varepsilon_{\mathrm{ph},j}\gets c_0\alpha^3U(1-\alpha/2)^j$
    \State $J_j\gets\lceil\log_2(4H/\varepsilon_{\mathrm{ph},j})\rceil$, $\delta_{\mathrm{ph}}\gets\delta_\nu/(KJ_j)$, $R_{\mathrm{rep}}\gets\lceil C_{\mathrm{rep}}\log(1/\delta_{\mathrm{ph}})\rceil$, and $\delta_{\mathrm{tr}}\gets\delta_{\mathrm{ph}}/(3R_{\mathrm{rep}})$
    \State $\delta_{g,j}^2\gets\min\{\eta\mu H/(Cm),\mu\varepsilon_{\mathrm{ph},j}/(CJ_jm)\}$, and choose $\varepsilon_{\mathrm{op}}$ by the three-term bound of Section~\ref{sec:accelerated_constants} with $\delta_g=\delta_{g,j}$
    \State $\mathcal P_j\gets(H,a,R,\ell,q,\widehat L,\gamma,T,k,s,\delta_{g,j},\varepsilon_{\mathrm{op}},c_{\mathrm{prox}}\varepsilon_{\mathrm{ph},j}/J_j,\delta_{\mathrm{tr}})$ and $z\gets x_{j-1}$
    \For{$r=1,\ldots,J_j$}
        \State $\mathcal C\gets\{z\}$
        \For{$R_{\mathrm{rep}}$ independent repetitions}
            \State $x\gets\textsc{GuardedBlockTrial}(G_j,z,\mathcal P_j)$; if the trial did not abort, set $\mathcal C\gets\mathcal C\cup\{x\}$
        \EndFor
        \State $z\gets\textsc{CompareValue}(\mathcal C,\varepsilon_{\mathrm{ph},j}/(128J_j),\delta_{\mathrm{ph}}/3)$
    \EndFor
    \State $x_j\gets z$, $h_j\gets x_j-x_{j-1}$, and $y_j\gets x_j+\vartheta h_j$
\EndFor
\State \Return $x_K$
\EndProcedure
\end{algorithmic}
\end{algorithm}

\subsection{Setup and constants}
\label{sec:accelerated_constants}

Fix an active set of size $m$, a multiplier $\nu>0$, and one projection
subproblem. We retain the notation of the proof of
Theorem~\ref{thm:faster_dense_runtime}. The box is $Q$, whose coordinate
intervals have length $\varepsilon_{\mathrm{box}}$, and the matrices are $B(x)$, $C(x)$, $D_x$ and
$E_i$. The parameters are $R$, $a$ and $b$. This $b$ is the smoothing constant of that proof, unrelated to the exponent parameter $b$ of $\gamma_\star$. The Gibbs trace is $\Phi$, its value
is $\phi_0=\Phi(0)$, the dimension is $d_0=2n$, and the accepted Gaussian draw is $\xi$.
Put
\[
F(x):=F_\nu(x)=\frac12\|x-\xi\|_2^2+\nu\Phi(x),
\qquad
\Phi(x):=\tr[\exp(D_x)],
\qquad x\in Q.
\]
The function $F$ is $1$-strongly convex. Write $x_*$ for its unique minimizer
over $Q$ and $F_*:=F(x_*)$. We start the solver at $x_0:=0$. Put
\[
U:=F(x_0)=O(m),
\qquad
H:=64U,
\]
where the bound $U=O(m)$ holds uniformly over every multiplier tested by the
certified search of Section~\ref{sec:faster_algorithm}. The level $H$ is four
times the level $16F_\nu(0)=16U$ of Section~\ref{sec:faster_algorithm}. Catalyst
subproblems start at objective values up to $3U$
(Lemma~\ref{lem:accelerated_catalyst}), and every phase must start below
$H/16$. We use the unsubscripted letter $C$ to denote a sufficiently large universal
constant whose value may change from line to line.

\paragraph{Blocks and the regularized objective.}
Fix a block size $\ell\ge2$. We run the accelerated solver only at active sizes
$m\ge Ca^2\ell^2$. At smaller active sizes we use the selected-coordinate solver of
Theorem~\ref{thm:faster_dense_runtime}. In Section~\ref{sec:algorithm} we fix
$R=c_{\rm rad}(\ell_{\rm asp})\sqrt m\ell_{\rm asp}$ with $\ell_{\rm asp}\ge1$ and $c_{\rm rad}(\ell_{\rm asp})$ bounded below by a
universal constant, so $R^2\ge c_Rm$ for a universal constant $c_R>0$. Then
$m\ge Ca^2\ell^2$ gives
\[
\ell\le\min\{m,R/(4a)\}
\]
once $C\ge16/c_R$. Partition $[m]$ into $q:=\lceil m/\ell\rceil$ fixed
blocks of size at most $\ell$. Then $q\ell\le m+\ell\le2m$. Put
\[
\kappa:=\ell,
\qquad
\mu:=1+\kappa,
\qquad
\alpha:=\frac1{\sqrt{1+\kappa}},
\qquad
\vartheta:=\frac{1-\alpha}{1+\alpha}.
\]
For a center $y\in\R^m$ define on $Q$ the regularized objective
\[
G(x):=F(x)+\frac\kappa2\|x-y\|_2^2 .
\]
It is $\mu$-strongly convex. Write $x_G$ for its minimizer over $Q$ and
$\Delta_v:=G(v)-G(x_G)$ for the gap at $v\in Q$. The center enters only the
quadratic part of $G$, which we evaluate exactly. Every matrix-function
calculation below is one for $F$.

\paragraph{Block curvature, step, and phase length.}
Put
\[
\widehat L:=\max\{1+\kappa+ea^2H\ell/R^2,
2a(1+\kappa)\ell/R\},
\qquad
\gamma:=\frac\mu{q\widehat L},
\qquad
T:=\lceil16/\gamma\rceil.
\]
Since $\widehat L\ge\mu$ and $q\ge1$, $\gamma\le1$ and $T\gamma\le17$.
Because $\ell\le R/(4a)$, the second term of $\widehat L$ is at most $\mu/2$.
Because $H=O(m)$, $R^2\ge c_Rm$ and $a=O(\log n)$, the first term is
$\widetilde O(1+\kappa)$. Hence $\widehat L/(1+\kappa)=\widetilde O(1)$.

\paragraph{Accuracy, bias floor, sketch width, and proximal tolerance.}
We solve a regularized subproblem to a target accuracy
$\varepsilon_{\mathrm{ph}}\in(0,U]$ in
\[
J:=\lceil\log_2(4H/\varepsilon_{\mathrm{ph}})\rceil
\]
phases. Let $\eta\le1/4$ be a sufficiently small universal constant. The
oracle carries a bias floor $\delta_g>0$ with
\[
\delta_g^2
\le\min\{\eta\mu H/(Cm),\mu\varepsilon_{\mathrm{ph}}/(CJm)\},
\]
a range finder of width $k$ and $s$ Hutchinson probes, which is the sketch of
Section~\ref{sec:faster_algorithm}, $k,s\ge2$, with
\[
ks\ge C\frac{a^2H\ell q}{\eta\mu R^2},
\qquad
k=s=\widetilde\Theta(\sqrt{m/\ell}),
\]
and the polynomial precision
\[
\varepsilon_{\mathrm{op}}
\le\min\{\phi_0/(16d_0),R\delta_g/(4\nu ad_0),
R\sqrt{ks}\delta_g/(C\nu ad_0)\}
\]
of Section~\ref{sec:faster_algorithm}. The first term of this precision bound also forces
$\nu d_0\varepsilon_{\mathrm{op}}\le\nu\phi_0/16\le U/16=H/1024$. We stop the scalar
searches inside the proximal steps at residuals $\rho_t$ with
\[
\sum_{t<T}\rho_t
\le c_{\mathrm{prox}}\min\{H,\varepsilon_{\mathrm{ph}}/J\}
\]
in every phase, where $c_{\mathrm{prox}}\le1/8$ is a sufficiently small
universal constant.

\paragraph{Catalyst schedule and failure budgets.}
We run $K$ regularized Catalyst subproblems with centers $y_{j-1}$ and target
accuracies
\[
\varepsilon_{\mathrm{ph},j}:=c_0\alpha^3U(1-\alpha/2)^j,
\qquad
K:=\lceil C\alpha^{-1}\log(U/(\alpha^2\varepsilon_{\mathrm{opt}}))\rceil,
\]
where $c_0\le1/4$ is a sufficiently small universal constant and
$\varepsilon_{\mathrm{opt}}$ is the accuracy requested by the multiplier
wrapper of Section~\ref{sec:faster_algorithm}. For subproblem $j$ we use
$\varepsilon_{\mathrm{ph}}=\varepsilon_{\mathrm{ph},j}$, $J=J_j$ and $\delta_g=\delta_{g,j}$ in the
displays above. The wrapper gives the fixed-multiplier solve a failure budget
$\delta_\nu$. To each of the $J_j$ phases of subproblem $j$ we give the budget
\[
\delta_{\mathrm{ph}}:=\frac{\delta_\nu}{KJ_j},
\]
so that the phase budgets sum to $\delta_\nu$. In each phase we run
$R_{\mathrm{rep}}:=\lceil C_{\mathrm{rep}}\log(1/\delta_{\mathrm{ph}})\rceil$
independent trials, give each trial the budget
\[
\delta_{\mathrm{tr}}:=\frac{\delta_{\mathrm{ph}}}{3R_{\mathrm{rep}}},
\]
spend $\delta_{\mathrm{ph}}/3$ on the objective comparison of the phase, and keep
$\delta_{\mathrm{ph}}/3$ for the event that no trial improves. A trial spends
$\delta_{\mathrm{tr}}/2$ on its snapshot cap and $\delta_{\mathrm{tr}}/(2T)$
on each of its $T$ guard certificates.

\subsection{Block oracle variance}\label{subsec:accelerated_variance}

Conditional on the history, one block query selects a block uniformly, draws
one fresh range finder and one fresh family of Hutchinson probes, and evaluates
the finite-action oracle simultaneously for all coordinates of the block. The
finite-action oracle is the gradient oracle of
Section~\ref{sec:faster_algorithm} implemented with finite matrix actions,
Section~\ref{sec:selected_finite}. We share the sketch and the probes only
among the coordinates of that query.

\begin{lemma}[Block oracle variance]
\label{lem:accelerated_variance}
Let $G$ be a regularized objective as in
Section~\ref{sec:accelerated_constants}, let $\mathcal F_t$ contain the
history before query $t$, and let the iterate $x$ and the snapshot $w$ be
$\mathcal F_t$-measurable points of $Q$ with $G(x)\le H$ and $G(w)\le H$.
Conditional on $\mathcal F_t$, select the block $I_t$ uniformly from the fixed
partition, and draw a fresh range finder of width $k$ and $s$ fresh Hutchinson
probes. Let $\mathfrak{e}_{I_t}$ be the error vector of the finite-action oracle
with snapshot $w$ and polynomial precision $\varepsilon_{\mathrm{op}}$ on the
coordinates of $I_t$. Then
\begin{equation}
\E[\|\mathfrak{e}_{I_t}\|_2^2\mid\mathcal F_t]
\le\frac{\eta\mu}q(\Delta_x+\Delta_w)+\ell\delta_g^2.
\label{eq:accelerated_block_variance}
\end{equation}
\end{lemma}

\begin{proof}
Put $P_v:=\exp(D_v)$ for $v\in Q$. In the proof of
Theorem~\ref{thm:faster_dense_runtime} we bound the mean-square error of one
coordinate before converting the trace norm into optimizer gaps. Under the
snapshot convention, we apply that raw estimate to the deflated difference
$\widetilde P_x-\widetilde P_w$ of the polynomial operators. Together with its
finite-action perturbation bound, it gives, uniformly for every fixed coordinate
$i$,
\[
\E[|\mathfrak{e}_i|^2\mid\mathcal F_t]
\le\frac{C\nu^2a^2}{ksR^2}
\|P_x-P_w\|_1^2+\delta_g^2.
\]
The bound does not depend on which other coordinates share the sketch. Since
the squared Euclidean norm is the sum of the coordinate squares, correlations
between the coordinates of a block do not enter. Summing over all $m$
coordinates, we obtain
\[
\sum_{i=1}^m\E[|\mathfrak{e}_i|^2\mid\mathcal F_t]
\le\frac{C\nu^2a^2m}{ksR^2}\|P_x-P_w\|_1^2+m\delta_g^2.
\]

\emph{Step 1: the trace norm is controlled by the gaps.} Let $\mathcal D(P\|P')$ be
the matrix relative entropy we defined in Section~\ref{sec:selected_setup}. Fact~\ref{fact:generalized_pinsker} bounds
$\|P-P'\|_1^2$ by $4\max\{\tr[P],\tr[P']\}\mathcal D(P\|P')$. Since $x\mapsto\log P_x=D_x$ is
affine, $\mathcal D(P_{x_G}\|P_v)$ is the Bregman divergence of $\Phi$
between $v$ and its base point $x_G$. Write $D_F(v,u):=F(v)-F(u)-\langle\nabla F(u),v-u\rangle$ and $D_G$ likewise for the Bregman divergences of $F$ and $G$. The quadratic parts of $F$ and $G$ add
nonnegative Bregman terms, and the constrained optimality condition
$\langle\nabla G(x_G),v-x_G\rangle\ge0$ removes the linear term of $D_G$.
Hence, for every $v\in Q$,
\[
\nu\mathcal D(P_{x_G}\|P_v)
\le D_F(v,x_G)
\le D_G(v,x_G)
\le\Delta_v .
\]
The level hypothesis gives $\nu\tr[P_v]=\nu\Phi(v)\le G(v)\le H$ for $v=x$
and $v=w$, and also for the reference point $v=x_G$, because
$G(x_G)\le G(w)\le H$. Applying the trace-norm triangle inequality through $P_{x_G}$,
the bound $(\beta_1+\beta_2)^2\le2\beta_1^2+2\beta_2^2$, and Pinsker at
the trace level $H/\nu$, we therefore obtain
\[
\nu^2\|P_x-P_w\|_1^2
\le8H(\Delta_x+\Delta_w).
\]

\emph{Step 2: averaging over the block.} Every coordinate lies in exactly one
block, and we select each block with probability $1/q$. Hence
$\E[\|\mathfrak{e}_{I_t}\|_2^2\mid\mathcal F_t]$ is $1/q$ times the coordinate sum.
Combining the coordinate sum with Step~1 and $m/q\le\ell$, we have
\[
\E[\|\mathfrak{e}_{I_t}\|_2^2\mid\mathcal F_t]
\le\frac{Ca^2\ell H}{ksR^2}(\Delta_x+\Delta_w)
+\ell\delta_g^2.
\]
The sketch condition $ks\ge Ca^2H\ell q/(\eta\mu R^2)$ makes the first
coefficient at most $\eta\mu/q$, which is
Eq.~\eqref{eq:accelerated_block_variance}. The finite-action term
$\delta_g^2$ per coordinate comes from the fixed polynomial operator
$\widetilde P_v=q_r(D_v)^2$ of Section~\ref{sec:faster_algorithm} under the
three-term bound on $\varepsilon_{\mathrm{op}}$. The Catalyst center changes only
the exactly evaluated quadratic part of $G$ and does not enter this
calculation.
\end{proof}

\subsection{The guarded block step}\label{subsec:accelerated_step}

For a block $I$ and a point $x\in Q$, the trust ball is the set of directions
$d$ with $\operatorname{supp}(d)\subseteq I$ and $\|d\|_2\le R/(a\sqrt\ell)$.
We use $\mathcal D_I(x)$ to denote the set of such $d$ with $x+d\in Q$. The
subscript distinguishes this set from the relative entropy
$\mathcal D(\cdot\|\cdot)$ above. Given the oracle
output $\widehat\nabla_IG(x)$, put
\[
\phi_I(u):=
\langle\widehat\nabla_I G(x),u\rangle
+\frac{\widehat L}{2}\|u\|_2^2,
\qquad u\in\mathcal D_I(x),
\]
and write $\mathfrak{e}_I:=\widehat\nabla_IG(x)-\nabla_IG(x)$ for the oracle error. We
say that a displacement $d\in\mathcal D_I(x)$ has three-point residual $\rho\ge0$ if
\[
\phi_I(u)\ge
\phi_I(d)+\frac{\widehat L}{2}\|u-d\|_2^2-\rho
\qquad\text{for every }u\in\mathcal D_I(x);
\]
the exact minimizer of $\phi_I$ over the convex set $\mathcal D_I(x)$ has
residual $0$, which is the strong three-point inequality.

\begin{lemma}[Guarded block step]
\label{lem:accelerated_step}
Let $G$, $\widehat L$ and $\gamma$ be as in
Section~\ref{sec:accelerated_constants}, let $x\in Q$ with $G(x)\le H$, and
let $I$ be a block.
\begin{enumerate}[label=(\roman*)]
\item For every $d$ in the trust ball of $I$ at $x$ with $x+d\in Q$ and every
$t\in[0,1]$, $\Phi(x+td)\le e\Phi(x)$ and
\[
\frac{\d^2}{\d t^2}G(x+td)
\le(1+\kappa+ea^2H\ell/R^2)\|d\|_2^2
\le\widehat L\|d\|_2^2.
\]
\item Let $d\in\mathcal D_I(x)$ have three-point residual $\rho_t$ and put
$x^+:=x+d$. For every $d'\in\mathcal D_I(x)$,
\[
G(x^+)\le G(x)+\langle\nabla_I G(x),d'\rangle
+\frac{\widehat L}{2}\|d'\|_2^2
+\frac1{2\widehat L}\|\mathfrak{e}_I\|_2^2+\rho_t ,
\]
and, when $I=I_t$ is selected uniformly conditional on a history
$\mathcal F_t$ containing $x$,
\begin{equation}
\E[\Delta_{x^+}\mid\mathcal F_t]
\le(1-\gamma)\Delta_x
+\frac1{2\widehat L}\E[\|\mathfrak{e}_{I_t}\|_2^2\mid\mathcal F_t]
\label{eq:accelerated_block_step}
\end{equation}
holds with the additive term $\rho_t$ on the right. Such a $d$ is computed
by a scalar bisection stopped at the tolerance $\rho_t$ in
$O(\ell\log(1/\rho_t))$ arithmetic operations, all scales being polynomial
in $n$.
\item Suppose $G(x)\le H/3$ and let $x^+=x+d$ with $d$ in the trust ball of
$I$ at $x$. Given $\delta\in(0,1)$, a certificate $\widehat G(x^+)$ computed
from $\lceil8\log(1/\delta)\rceil$ fresh batches of a universal constant
number of Gaussian probes satisfies, except with probability $\delta$,
\[
G(x^+)\le\widehat G(x^+)\le G(x^+)+\frac{9H}{512},
\]
and $9H/512<H/48$. Its cost is $O(\log(1/\delta))$ polynomial actions of
degree $r$ on vectors, which is $\widetilde O(n^2\log(1/\delta))$
arithmetic operations.
\end{enumerate}
\end{lemma}

\begin{proof}
\emph{Step 1: block smoothness on the trust ball.} For $d$ in the trust ball,
$\|E_i\|\le1/R$ and $\sum_{i\in I}|d_i|\le\sqrt\ell\|d\|_2$ give
\[
\|D_{x+d}-D_x\|
=a\|\sum_{i\in I}d_iE_i\|
\le a\sqrt\ell\|d\|_2/R\le1 .
\]
Hence along the segment the matrix exponent moves by at most one in operator
norm, $D_{x+td}\preceq D_x+I_{2n}$, and monotonicity of the trace exponential
gives $\Phi(x+td)\le e\Phi(x)$. In particular
$\nu\Phi(x+td)\le eG(x)\le eH$. Write $E_d:=\sum_{i\in I}d_iE_i$, so that
$\|E_d\|\le\sqrt\ell\|d\|_2/R$. Applying Fact~\ref{fact:trace_exp_hessian} to
$M_t:=D_{x+td}=D_x+taE_d$, we get
\[
\frac{\d^2}{\d t^2}\Phi(x+td)
\le a^2\|E_d\|^2\Phi(x+td)
\le\frac{a^2\ell\|d\|_2^2}{R^2}e\Phi(x).
\]
Multiplying by $\nu$, using $\nu\Phi(x)\le H$, and adding the second
derivative $(1+\kappa)\|d\|_2^2$ of the two quadratic terms of $G$, we obtain the
first inequality of (i). The second inequality is the first term in the
definition of $\widehat L$.

\emph{Step 2: the proximal step and the comparison displacement.} A
displacement with the required residual is inexpensive: for a scalar
multiplier $\lambda\ge0$, we clip each coordinate of
$-\widehat\nabla_IG(x)/(\widehat L(1+\lambda))$ to the corresponding box
interval and bisect on $\lambda$ until the ball constraint is met. We
certify the residual $\rho_t$ by the same bisection at logarithmic additional
cost, and every scale involved is polynomial in $n$. The cost is therefore
$O(\ell\log(1/\rho_t))$ at tolerance $\rho_t$. Let first $d$ be the exact
minimizer of $\phi_I$ over $\mathcal D_I(x)$ and $d'\in\mathcal D_I(x)$. By
(i), $G(x+d)-G(x)\le\langle\nabla_IG(x),d\rangle+\frac{\widehat L}2\|d\|_2^2
=\phi_I(d)-\langle\mathfrak{e}_I,d\rangle$. The strong three-point inequality at
$u=d'$ and Young's inequality
$\langle\mathfrak{e}_I,d'-d\rangle\le\frac{\widehat L}2\|d'-d\|_2^2
+\frac1{2\widehat L}\|\mathfrak{e}_I\|_2^2$ then give
\[
\begin{split}
G(x+d)-G(x)
&\le\phi_I(d)-\langle\mathfrak{e}_I,d\rangle\\
&\le\langle\nabla_I G(x),d'\rangle
+\frac{\widehat L}{2}\|d'\|_2^2
+\langle\mathfrak{e}_I,d'-d\rangle
-\frac{\widehat L}{2}\|d'-d\|_2^2\\
&\le\langle\nabla_I G(x),d'\rangle
+\frac{\widehat L}{2}\|d'\|_2^2
+\frac1{2\widehat L}\|\mathfrak{e}_I\|_2^2.
\end{split}
\]
If $d$ has residual $\rho_t$ instead, the residual inequality replaces the
strong three-point inequality, and the last two lines acquire the additive
term $\rho_t$. This is the pathwise inequality of (ii). The calculation uses
only the second moment of $\mathfrak{e}_I$. We do not assume that the finite-action
error is unbiased.

The comparison displacement
\[
d':=\frac\mu{\widehat L}(x_G-x)_I
\]
is admissible. It stays in the box because $\mu/\widehat L\le1$, so each
coordinate of $x+d'$ is a convex combination of the corresponding coordinates
of $x$ and $x_G$. Every coordinate interval of $Q$ has length $\varepsilon_{\mathrm{box}}<2$,
so
\[
\|d'\|_2
\le\frac{2\mu\sqrt\ell}{\widehat L}
\le\frac R{a\sqrt\ell}
\]
by the second term in the definition of $\widehat L$. We select each block
with probability $1/q$, and the blockwise pieces of $d'$ assemble to
$\frac\mu{\widehat L}(x_G-x)$. The conditional expectation of the two
comparison terms is therefore
\[
\frac\mu{q\widehat L}
(\langle\nabla G(x),x_G-x\rangle
+\frac\mu2\|x_G-x\|_2^2)
\le-\gamma\Delta_x,
\]
where the inequality is $\mu$-strong convexity of $G$ between $x$ and $x_G$.
Taking conditional expectations in the pathwise inequality and subtracting
$G(x_G)$, we obtain Eq.~\eqref{eq:accelerated_block_step}, with the additive
term $\rho_t$ when we stop the search early.

\emph{Step 3: the additive guard.} Suppose $G(x)\le H/3$ and let
$x^+=x+d$ with $d$ in the trust ball. We compute the quadratic part
$\frac12\|x^+-\xi\|_2^2+\frac\kappa2\|x^+-y\|_2^2$ of $G(x^+)$
exactly, so only $\nu\Phi(x^+)$ needs a certificate. By (i),
$\nu\Phi(x^+)\le e\nu\Phi(x)\le eG(x)\le eH/3$. Let
$\widetilde P_{x^+}=q_r(D_{x^+})^2$ be the polynomial operator of
Section~\ref{sec:faster_algorithm} on the interval enlarged by one. Then
$\|\widetilde P_{x^+}-P_{x^+}\|\le\varepsilon_{\mathrm{op}}$, so
$|\tr[\widetilde P_{x^+}]-\Phi(x^+)|\le d_0\varepsilon_{\mathrm{op}}$.
Section~\ref{sec:accelerated_constants} gives
$\nu d_0\varepsilon_{\mathrm{op}}\le H/1024$. Hence
\[
\nu\tr[\widetilde P_{x^+}]\le\frac e3H+\frac H{1024}<0.91H .
\]
By Eq.~\eqref{eq:gaussian_quadratic_form}, for a standard Gaussian vector $z$ and
the positive semidefinite matrix $M:=\widetilde P_{x^+}$, $\E[z^\top Mz]=\tr[M]$ and
$\V[z^\top Mz]=2\|M\|_F^2\le2(\tr[M])^2$. The quantity $z^\top Mz=\|q_r(D_{x^+})z\|_2^2$
is one polynomial action. We average $z^\top Mz$ over a batch of $2^{17}$ fresh
probes. By Chebyshev's inequality the batch mean lies within $\tr[M]/128$ of
$\tr[M]$ with probability at least $3/4$, because $2\cdot128^2/2^{17}=1/4$.
Let $\widehat\tau$ be the median of $\lceil8\log(1/\delta)\rceil$ independent
batch means. Hoeffding's inequality for the number of batches outside this
window gives
\[
|\widehat\tau-\tr[\widetilde P_{x^+}]|\le\tr[\widetilde P_{x^+}]/128
\]
except with probability $\delta$. Define
\[
\widehat G(x^+):=\frac12\|x^+-\xi\|_2^2+\frac\kappa2\|x^+-y\|_2^2
+\nu\widehat\tau+\frac{9H}{1024}.
\]
On the good event the statistical error is
$\nu|\widehat\tau-\tr[\widetilde P_{x^+}]|\le\nu\tr[\widetilde P_{x^+}]/128
\le H/128$ and the bias is $\nu|\tr[\widetilde P_{x^+}]-\Phi(x^+)|\le H/1024$.
Their sum $9H/1024$ is the shift we built into $\widehat G$. Hence
$\nu\Phi(x^+)\le\nu\widehat\tau+9H/1024$, which is
$G(x^+)\le\widehat G(x^+)$. Also, $\nu\widehat\tau\le\nu\Phi(x^+)+9H/1024$,
which is $\widehat G(x^+)\le G(x^+)+18H/1024=G(x^+)+9H/512$. Finally
$9/512<1/48$ because $9\cdot48=432<512$. The certificate uses
$2^{17}\lceil8\log(1/\delta)\rceil$ polynomial actions, each $O(rn^2)$
arithmetic operations with $r$ polylogarithmic, since
$D_{x^+}=aC(x^+)-abI_{2n}$ is available from the $B(x^+)$ we maintain.
\end{proof}

\subsection{One guarded trial and the regularized subproblem}\label{subsec:accelerated_trial}

A trial is Algorithm~\ref{alg:accelerated_guarded_trial}, started at a point
$z$ with $G(z)\le H/16$. It makes $T$ guarded block queries with the oracle of
Lemma~\ref{lem:accelerated_variance} and the step and guard of
Lemma~\ref{lem:accelerated_step}. The ideal trial that we couple to it uses
the same block selections, oracle randomness and refresh coins. It never
aborts, and we kill it at the first crossing of the level $5H/16$.

\begin{lemma}[One guarded trial and the regularized subproblem]
\label{lem:accelerated_trial}
Let $G$, $\varepsilon_{\mathrm{ph}}\in(0,U]$, $J$, $\delta_g$, $k$, $s$,
$\varepsilon_{\mathrm{op}}$ and the residuals $\rho_t$ be as in
Section~\ref{sec:accelerated_constants}.
\begin{enumerate}[label=(\roman*)]
\item Let $z\in Q$ with $G(z)\le H/16$ and $\delta_{\mathrm{tr}}\in(0,1)$.
\textnormal{\textsc{GuardedBlockTrial}}$(G,z,\mathcal P)$ makes $T$ block queries and at
most $C_{\mathrm{snap}}\log(2/\delta_{\mathrm{tr}})+2$ dense snapshot
constructions. Every point it accepts satisfies $G\le H/3$. Except on an
event of probability at most $\delta_{\mathrm{tr}}$, it agrees with the
coupled ideal trial. This event is a wrong guard certificate or a reached
snapshot cap. With probability at least a universal constant $c_\star>0$,
the ideal trial is not killed and returns a point $x_T$ with
\[
\Delta_{x_T}\le\frac{\Delta_z}2+\frac{\varepsilon_{\mathrm{ph}}}{32J}.
\]
\item Let $z$ be as in (i) and $\delta_{\mathrm{ph}}\in(0,1)$. Run
$R_{\mathrm{rep}}=\lceil C_{\mathrm{rep}}\log(1/\delta_{\mathrm{ph}})\rceil$
independent trials from $z$ with budgets
$\delta_{\mathrm{tr}}=\delta_{\mathrm{ph}}/(3R_{\mathrm{rep}})$. Let
$z^+$ be the output of \textnormal{\textsc{CompareValue}} on $z$ and the returned points,
with width $\varepsilon_{\mathrm{ph}}/(128J)$ and budget $\delta_{\mathrm{ph}}/3$. Except
with probability $\delta_{\mathrm{ph}}$,
\[
\Delta_{z^+}\le\frac{\Delta_z}2+\frac{\varepsilon_{\mathrm{ph}}}{16J},
\qquad
G(z^+)\le G(z)+\frac{\varepsilon_{\mathrm{ph}}}{128J}.
\]
\item Let $z_0\in Q$ with $G(z_0)\le3U$. Starting from $z_0$, $J$ phases as
in (ii) return an $\varepsilon_{\mathrm{ph}}$-minimizer of $G$ over $Q$ except with
probability $J\delta_{\mathrm{ph}}$. They use $\widetilde O(q\widehat L/\mu)$
block queries and $\widetilde O(1)$ dense snapshot constructions, and both
counts are capped before the phases start. The hidden factors are logarithmic in
$1/\varepsilon_{\mathrm{ph}}$ and $1/\delta_{\mathrm{ph}}$.
\end{enumerate}
\end{lemma}

\begin{proof}
\emph{Step 1: the Lyapunov potential and the snapshot cap.} Put
\[
V(x,w):=\Delta_x+\frac12\Delta_w .
\]
Suppose the level hypothesis of Lemma~\ref{lem:accelerated_variance} holds for
$x$ and $w$. From Eqs.~\eqref{eq:accelerated_block_variance} and
\eqref{eq:accelerated_block_step} we obtain
$\E[\Delta_{x^+}\mid\mathcal F_t]\le A_0\Delta_x+B_0\Delta_w
+\ell\delta_g^2/(2\widehat L)+\rho_t$ with
\[
A_0:=1-(1-\eta/2)\gamma,
\qquad
B_0:=\eta\gamma/2,
\]
because $\eta\mu/(2q\widehat L)=\eta\gamma/2$. The refresh rule gives
$\E[\Delta_{w^+}\mid\mathcal F_t,x^+]=\gamma\Delta_{x^+}+(1-\gamma)\Delta_w$.
Hence the coefficient of $\Delta_x$ in $\E[V(x^+,w^+)\mid\mathcal F_t]$ is
$(1+\gamma/2)A_0\le1-\gamma/4$, and that of $\Delta_w$ is
$(1+\gamma/2)B_0+(1-\gamma)/2\le(1-\gamma/4)/2$. In both bounds we use $\eta\le1/4$
and $\gamma\le1$. We therefore obtain
\[
\E[V(x^+,w^+)\mid\mathcal F_t]
\le(1-\gamma/4)V(x,w)
+C\ell\delta_g^2/\widehat L+C\rho_t.
\]
Thus $T=\lceil16/\gamma\rceil$ block queries reduce the potential by a
constant factor, up to $O(m\delta_g^2/\mu)+O(\sum_{t<T}\rho_t)$ additive
error, since $q\ell\le2m$. The expected number of snapshot refreshes in these
$T$ queries is at most $T\gamma\le17$. The refresh count is stochastically
dominated by a binomial random variable with mean at most $17$. By a Chernoff
bound we then bound the probability of exceeding
$C_{\mathrm{snap}}\log(2/\delta_{\mathrm{tr}})$ refreshes by
$\delta_{\mathrm{tr}}/2$, after increasing $C_{\mathrm{snap}}$. Each guard
certificate uses the budget $\delta_{\mathrm{tr}}/(2T)$ in
Lemma~\ref{lem:accelerated_step}(iii), so all $T$ certificates are correct
except with probability $\delta_{\mathrm{tr}}/2$. Together the two failure sources
have probability at most $\delta_{\mathrm{tr}}$, as we claimed in (i).

\emph{Step 2: the coupling and the crossing estimate.} Index the queries by
$t=0,\ldots,T-1$ and define
\[
\tau:=\min\{t+1:G(x_{t+1})>5H/16\},
\]
with $\tau:=T+1$ if there is no crossing. We kill the ideal trial at $\tau$.
Before $\tau$, both the iterate and the snapshot, which is a past iterate, are
below $5H/16$. Hence the level hypotheses of
Lemmas~\ref{lem:accelerated_variance} and \ref{lem:accelerated_step} hold.
Lemma~\ref{lem:accelerated_step}(i) then keeps every intervening
matrix-exponential path below the $H$ trace level because $5e/16<1$. On the
event that no crossing occurs and all certificates are correct, every proposal
has true value at most $5H/16$. Its certificate is then at most
$5H/16+9H/512=169H/512<H/3$. Thus the implemented trial never aborts on a guard,
and the two trials agree. Conversely, an accepted proposal satisfies
$G(x^+)\le\widehat G(x^+)\le H/3$, which is the level claim of (i). In the
Lyapunov calculation we do not condition on the global guard-success event.
Afterwards we add the certificate failures by the union bound.

Taking $d'=0$ in the pathwise inequality of
Lemma~\ref{lem:accelerated_step}(ii), we obtain
\[
G(x^+)\le G(x)+\frac1{2\widehat L}\|\mathfrak{e}_I\|_2^2+\rho_t.
\]
Since the allocated proximal errors sum to at most
$c_{\mathrm{prox}}H\le H/8$, a crossing from $H/16$ to $5H/16$ requires
\[
\sum_{t<T}\mathbbm{1}[\tau>t]\|\mathfrak{e}_{I_t}\|_2^2
\ge\frac14\widehat LH.
\]
The event $\{\tau>t\}$ is $\mathcal F_t$-measurable, and on it
$\Delta_{x_t},\Delta_{w_t}\le5H/16$. By Tonelli's theorem and
Eq.~\eqref{eq:accelerated_block_variance}, with $T\le17q\widehat L/\mu$ and
$q\ell\le2m$, we bound the expectation of the stopped nonnegative sum by
\[
C\eta\widehat LH+C m\widehat L\delta_g^2/\mu.
\]
By Markov's inequality and the first term in the bound on $\delta_g^2$, we bound the
crossing probability by $C\eta$, a small universal constant once $\eta$
is small enough.

\emph{Step 3: the killed potential.} Put $A_t:=\{\tau>t\}$ for
$0\le t\le T$ and define
\[
W_t:=\mathbbm{1}[A_t]V(x_t,w_t),
\qquad
\beta:=C\ell\delta_g^2/\widehat L.
\]
The event $A_t$ is $\mathcal F_t$-measurable, $A_{t+1}\subseteq A_t$, and
$V\ge0$. Hence $W_{t+1}\le\mathbbm{1}[A_t]V(x_{t+1},w_{t+1})$, and the one-step
inequality of Step~1 applies on $A_t$:
\[
\E[W_{t+1}\mid\mathcal F_t]
\le \mathbbm{1}[A_t]((1-\gamma/4)V(x_t,w_t)+\beta+C\rho_t).
\]
Iterating, with $(1-\gamma/4)^T\le e^{-4}$ because $T\gamma\ge16$ and a
geometric sum for the bias terms, we obtain
\[
\E[W_T]\le e^{-4}V(z,z)+4\beta/\gamma+C\sum_{t<T}\rho_t.
\]
Since $q\ell\le2m$, the bias floor and the proximal tolerance make
$4\beta/\gamma+C\sum_{t<T}\rho_t\le\varepsilon_{\mathrm{ph}}/(512J)$. Here we
increase the universal constant in the upper bound for $\delta_g$ and
decrease $c_{\mathrm{prox}}$. By Markov's inequality we therefore bound the probability of
the event
\[
A_T\cap\{V(x_T,w_T)>V(z,z)/3+\varepsilon_{\mathrm{ph}}/(32J)\}
\]
by $\max\{3e^{-4},1/16\}=1/16$. The stopped noise-energy estimate of Step~2
separately bounds $\Pr[A_T^c]$. Hence the ideal trial has no crossing and
satisfies $V(x_T,w_T)\le V(z,z)/3+\varepsilon_{\mathrm{ph}}/(32J)$ with probability at
least $c_\star:=1-1/16-C\eta>0$. Since $V(z,z)=3\Delta_z/2$ and
$\Delta_{x_T}\le V(x_T,w_T)$, this is the improvement in (i). The additive
term $\varepsilon_{\mathrm{ph}}/(32J)$ covers the near-target case.

{\setlength{\emergencystretch}{2em}\emph{Step 4: one phase.} The $R_{\mathrm{rep}}$ trials are independent.
\textsc{CompareValue} encloses $G$ at each of the at most $R_{\mathrm{rep}}+1$
candidates in a fresh outward-rounded interval of width at most
$\varepsilon_{\mathrm{ph}}/(128J)$, which we compute as in Section~\ref{sec:faster_algorithm}. We
compute the quadratic part exactly, and we enclose $\Phi$ by one dense
inverse-polynomial-accuracy matrix-function evaluation with its certificate.
Each interval is correct except
with probability $\delta_{\mathrm{ph}}/(3(R_{\mathrm{rep}}+1))$. The procedure returns
a candidate $z^+$ with the smallest upper endpoint. Three events cover a
failure of the phase. One event is that some trial has a wrong certificate or
reaches its cap, which has probability at most $R_{\mathrm{rep}}\delta_{\mathrm{tr}}
=\delta_{\mathrm{ph}}/3$ by (i). Another event is that no ideal trial succeeds,
which has probability at
most $(1-c_\star)^{R_{\mathrm{rep}}}\le\delta_{\mathrm{ph}}/3$ once
$C_{\mathrm{rep}}$ is large. The remaining event is that some comparison
interval is wrong, which has probability
at most $\delta_{\mathrm{ph}}/3$. Off these events the implemented trials
agree with their ideal versions, some returned point $x^{\mathrm{best}}$ has
$\Delta_{x^{\mathrm{best}}}\le\Delta_z/2+\varepsilon_{\mathrm{ph}}/(32J)$, and every
interval is correct. Hence $G(z^+)$ is at most the upper endpoint of
$x^{\mathrm{best}}$, which is at most
$G(x^{\mathrm{best}})+\varepsilon_{\mathrm{ph}}/(128J)$. We thus obtain
$\Delta_{z^+}\le\Delta_z/2+\varepsilon_{\mathrm{ph}}/(32J)+\varepsilon_{\mathrm{ph}}/(128J)
\le\Delta_z/2+\varepsilon_{\mathrm{ph}}/(16J)$. Because $z$ is among the candidates,
$G(z^+)$ is also at most the upper endpoint of $z$, hence at most
$G(z)+\varepsilon_{\mathrm{ph}}/(128J)$.\par}

\emph{Step 5: the phases.} Let $z_0,\ldots,z_J$ be the phase starts and put
$\Delta_r:=\Delta_{z_r}$. By the second bound of (ii),
$G(z_r)\le G(z_0)+r\varepsilon_{\mathrm{ph}}/(128J)\le3U+\varepsilon_{\mathrm{ph}}/128\le3U+U/128<4U=H/16$
for $r\le J$. Hence every phase starts below $H/16$, and (ii) applies to it. The
phase failures add to $J\delta_{\mathrm{ph}}$. On the complementary event,
$\Delta_{r+1}\le\Delta_r/2+\varepsilon_{\mathrm{ph}}/(16J)$ for every $r<J$. Unrolling
the recurrence, we obtain
\[
\Delta_J\le2^{-J}\Delta_0+\frac{\varepsilon_{\mathrm{ph}}}{8J}
\le\frac{\varepsilon_{\mathrm{ph}}}4+\frac{\varepsilon_{\mathrm{ph}}}8<\varepsilon_{\mathrm{ph}},
\]
because $\Delta_0\le G(z_0)\le3U\le H$ and $2^{-J}\le\varepsilon_{\mathrm{ph}}/(4H)$. The
block queries number $JR_{\mathrm{rep}}T=\widetilde O(q\widehat L/\mu)$. The
dense snapshot constructions number
$JR_{\mathrm{rep}}(C_{\mathrm{snap}}\log(2/\delta_{\mathrm{tr}})+2)
+J(R_{\mathrm{rep}}+1)=\widetilde O(1)$, with logarithms in $1/\varepsilon_{\mathrm{ph}}$
and $1/\delta_{\mathrm{ph}}$. We fix both counts before the phases start. All
polynomial actions use one fixed positive semidefinite operator per oracle
call, and every guard and every comparison uses fresh probes. Hence the
conditional calculations above remain valid under adaptivity.
\end{proof}

\begin{algorithm}[H]
\caption{One guarded block trial}
\label{alg:accelerated_guarded_trial}
\begin{algorithmic}[1]
\Procedure{GuardedBlockTrial}{$G,z,\mathcal P$} \Comment{Lemma~\ref{lem:accelerated_trial}(i); $\mathcal P=(H,a,R,\ell,q,\widehat L,\gamma,T,k,s,\delta_g,\varepsilon_{\mathrm{op}},\rho_{\mathrm{tot}},\delta_{\mathrm{tr}})$}
\State $x\gets z$, $w\gets z$, build the dense snapshot data at $w$, and set the refresh counter to $0$
\State Fix residual tolerances $\rho_0,\ldots,\rho_{T-1}\ge0$ with $\sum_{t<T}\rho_t\le\rho_{\mathrm{tot}}$
\For{$t=0,\ldots,T-1$}
    \State Select a block $I_t$ uniformly from the fixed partition, independently of the past
    \State Draw a fresh range finder of width $k$ and $s$ fresh Hutchinson probes, and compute the finite-action oracle $\widehat\nabla_{I_t}G(x)$ with snapshot $w$ at polynomial precision $\varepsilon_{\mathrm{op}}$
    \State Compute $d\in\mathcal D_{I_t}(x)$ with three-point residual $\rho_t$ by scalar bisection, and put $x^+\gets x+d$
    \State Compute the additive guard certificate $\widehat G(x^+)$ with fresh probes and budget $\delta_{\mathrm{tr}}/(2T)$
    \If{$\widehat G(x^+)>H/3$}
        \State \Return \textsc{Abort}
    \EndIf
    \State $x\gets x^+$; with probability $\gamma$, independently of everything else, set $w\gets x$, rebuild the snapshot data, and increment the refresh counter
    \If{the refresh counter exceeds $C_{\mathrm{snap}}\log(2/\delta_{\mathrm{tr}})$}
        \State \Return \textsc{Abort}
    \EndIf
\EndFor
\State \Return $x$
\EndProcedure
\end{algorithmic}
\end{algorithm}

\begin{algorithm}[H]
\caption{Certified objective comparison}
\label{alg:accelerated_compare_value}
\begin{algorithmic}[1]
\Procedure{CompareValue}{$\mathcal C,\theta,\delta$} \Comment{Lemma~\ref{lem:accelerated_trial}(ii); $\mathcal C$ contains the phase start $z$, $\theta$ is the interval width, $\delta$ the failure budget}
\For{each $u\in\mathcal C$}
    \State Evaluate the quadratic part of $G$ at $u$ exactly, and enclose $\Phi(u)$ by one dense inverse-polynomial-accuracy matrix-function evaluation with certified absolute error at most $\theta/(2\nu)$, failing with probability at most $\delta/|\mathcal C|$
    \State Record the outward-rounded interval $[\underline G(u),\overline G(u)]$ of width at most $\theta$
\EndFor
\State \Return a $u\in\mathcal C$ with the smallest upper endpoint $\overline G(u)$
\EndProcedure
\end{algorithmic}
\end{algorithm}

\subsection{Catalyst acceleration and the safe-level invariant}\label{subsec:accelerated_catalyst}

Start with $y_0:=x_0=0$ and $h_0:=0$. For $j\ge1$, let $x_j$ be an
$\varepsilon_{\mathrm{ph},j}$-minimizer over $Q$ of
\[
G_j(x):=F(x)+\frac\kappa2\|x-y_{j-1}\|_2^2,
\]
which we compute by the phases of Lemma~\ref{lem:accelerated_trial} started at
$x_{j-1}$, and set
\[
h_j:=x_j-x_{j-1},
\qquad
y_j:=x_j+\vartheta h_j.
\]
{\setlength{\emergencystretch}{2em}These are the Catalyst parameters for the strongly convex case of Lin, Mairal and
Harchaoui~\cite{linetal18catalyst}. The usual quantity
$q_{\mathrm{cat}}:=1/(1+\kappa)$ satisfies $\alpha=\sqrt{q_{\mathrm{cat}}}$,
and the momentum coefficient is $\vartheta$.\par}

\begin{lemma}[Catalyst acceleration and the safe-level invariant]
\label{lem:accelerated_catalyst}
Let $\kappa$, $\alpha$, $\vartheta$, $\varepsilon_{\mathrm{ph},j}$ and $K$ be as in
Section~\ref{sec:accelerated_constants}, and suppose that
$G_j(x_j)-\min_{x\in Q}G_j(x)\le\varepsilon_{\mathrm{ph},j}$ for $j=1,\ldots,K$.
\begin{enumerate}[label=(\roman*)]
\item For every $j\ge0$, $G_{j+1}(x_j)\le3U<4U=H/16$: every regularized
subproblem starts below $H/16$, and every accepted inner iterate satisfies
$\nu\Phi\le H$.
\item With $\varrho:=\alpha/2$,
\[
F(x_K)-F_*
\le\frac8{(\alpha-\varrho)^2}(1-\varrho)^{K+1}U ;
\]
hence the choice of $K$ in Section~\ref{sec:accelerated_constants}, which is
$O(\alpha^{-1}\log(U/(\alpha^2\varepsilon_{\mathrm{opt}})))
=\widetilde O(\sqrt{1+\kappa})$, gives $F(x_K)-F_*\le\varepsilon_{\mathrm{opt}}$.
\item If $\nu\le\Lambda$ is a multiplier tested by the wrapper of
Section~\ref{sec:faster_algorithm} and
$\varepsilon_{\mathrm{opt}}\le\sigma^2/2$ is the accuracy it requests, then
\[
F(x_K)\le F_*+\varepsilon_{\mathrm{opt}}\le U+\varepsilon_{\mathrm{opt}}<H_{\max},
\]
where $H_{\max}=16(q(0)+\Lambda\phi_0+1)$ is the level of that wrapper. Here $q(x)=\frac12\|x-\xi\|_2^2$ is its quadratic part, not the block count $q$.
\end{enumerate}
\end{lemma}

\begin{proof}
\emph{Step 1: the rate.} We apply Catalyst with the box constraint $Q$ as
the composite term. Theorem~3 and the proof of Proposition~5 of Lin, Mairal
and Harchaoui~\cite{linetal18catalyst} are
homogeneous in an upper bound for the initial gap and in the prescribed
absolute-error sequence. Rerunning that proof with $U\ge F(x_0)-F_*$ in place
of the unknown initial gap, and using $c_0\alpha^3U\le2U/9$ for sufficiently
small $c_0$, we obtain the bound in (ii). Since
$8/(\alpha-\varrho)^2=32\alpha^{-2}$ and
$(1-\varrho)^{K+1}\le\exp(-(K+1)\alpha/2)$, an objective error at most
$\varepsilon_{\mathrm{opt}}$ follows once
$K\ge2\alpha^{-1}\log(32U/(\alpha^2\varepsilon_{\mathrm{opt}}))$. The
choice $K=\lceil C\alpha^{-1}\log(U/(\alpha^2\varepsilon_{\mathrm{opt}}))\rceil$
of Section~\ref{sec:accelerated_constants} satisfies this for a large enough $C$,
because $U\ge\frac12\|\xi\|_2^2\ge1$ on the multiplier branch of Section~\ref{sec:faster_algorithm} and $\varepsilon_{\mathrm{opt}}<1$ on every solve the wrapper
requests. We invoke Catalyst only after we have solved each regularized subproblem
to its prescribed absolute accuracy. Its outer theorem therefore needs the
pathwise inexactness event of Lemma~\ref{lem:accelerated_trial}, not an
unconditional linear-rate recurrence for the killed inner process.

\emph{Step 2: the energy.} The generic theorem does not supply the
localization. Define
\[
\mathcal E_j:=F(x_j)-F_*+
\frac{\kappa\vartheta}{2}\|h_j\|_2^2.
\]
Let $x_j^{*}$ be the exact minimizer of $G_j$ over $Q$ and put $e_j:=x_j-x_j^{*}$.
Since $G_j$ is $(1+\kappa)$-strongly convex and $x_j^{*}$ is its constrained
minimizer, $G_j(x_j)\ge G_j(x_j^{*})+\frac{1+\kappa}2\|e_j\|_2^2$. Combining
this with $G_j(x_j)-G_j(x_j^{*})\le\varepsilon_{\mathrm{ph},j}$, we obtain
\[
\|e_j\|_2^2\le\frac{2\varepsilon_{\mathrm{ph},j}}{1+\kappa}.
\]
Using the same strong convexity between $x_j^{*}$ and $x_{j-1}$, followed by
$G_j(x_j)\le G_j(x_j^{*})+\varepsilon_{\mathrm{ph},j}$, we have
\[
G_j(x_j)+\frac{1+\kappa}{2}
\|x_j^{*}-x_{j-1}\|_2^2
\le G_j(x_{j-1})+\varepsilon_{\mathrm{ph},j}.
\]
We substitute $y_{j-1}=x_{j-1}+\vartheta h_{j-1}$, so that
$G_j(x_j)=F(x_j)+\frac\kappa2\|h_j-\vartheta h_{j-1}\|_2^2$ and
$G_j(x_{j-1})=F(x_{j-1})+\frac{\kappa\vartheta^2}2\|h_{j-1}\|_2^2$. Then we write
$x_j^{*}-x_{j-1}=h_j-e_j$, expand the squared norms, and bound the cross term by
\[
\kappa\vartheta\langle h_j,h_{j-1}\rangle
\le\frac{\kappa\vartheta}{2}
(\|h_j\|_2^2+\|h_{j-1}\|_2^2).
\]
Subtracting $F_*$ from both sides and writing $\psi:=2\kappa(1-\vartheta)+1$, we obtain
\[
\mathcal E_j+\frac \psi2\|h_j\|_2^2
-(1+\kappa)\langle h_j,e_j\rangle
+\frac{1+\kappa}{2}\|e_j\|_2^2
\le\mathcal E_{j-1}+\varepsilon_{\mathrm{ph},j}.
\]
Completing the square in $h_j$, we have
\[
\frac \psi2\|h_j\|_2^2-(1+\kappa)\langle h_j,e_j\rangle
\ge-\frac{(1+\kappa)^2}{2\psi}\|e_j\|_2^2 .
\]
Dropping the nonnegative term $\frac{1+\kappa}2\|e_j\|_2^2$ and inserting the
bound on $\|e_j\|_2^2$, we prove
\begin{equation}
\mathcal E_j
\le\mathcal E_{j-1}
+(1+\frac{1+\kappa}{2\kappa(1-\vartheta)+1})
\varepsilon_{\mathrm{ph},j}.
\label{eq:catalyst_safe_energy}
\end{equation}
Here $1+\kappa=\alpha^{-2}$. Since $1-\vartheta=2\alpha/(1+\alpha)$ and
$\kappa=(1-\alpha)(1+\alpha)/\alpha^2$, we have $\psi=4/\alpha-3$. The coefficient of
$\varepsilon_{\mathrm{ph},j}$ in Eq.~\eqref{eq:catalyst_safe_energy} is therefore
$1+1/(4\alpha-3\alpha^2)\le2/\alpha$, because
$3\alpha^2-10\alpha+7=(3\alpha-7)(\alpha-1)\ge0$ for $\alpha\le1$. Since
$\mathcal E_0=F(x_0)-F_*\le U$ and
$\sum_{r\ge1}\varepsilon_{\mathrm{ph},r}\le c_0\alpha^3U\cdot2/\alpha=2c_0\alpha^2U$,
\[
\mathcal E_j
\le\mathcal E_0+\frac2\alpha\sum_{r=1}^j\varepsilon_{\mathrm{ph},r}
\le2U
\]
when $c_0\le1/4$. Finally, $\vartheta^2\le\vartheta$ gives
\[
G_{j+1}(x_j)-F_*
=F(x_j)-F_*+
\frac{\kappa\vartheta^2}{2}\|h_j\|_2^2
\le\mathcal E_j\le2U,
\]
and $F_*\le F(x_0)=U$ gives $G_{j+1}(x_j)\le3U<4U=H/16$, the invariant of
(i). Because $G_j\ge F\ge\nu\Phi$ and every accepted inner iterate has
$G_j\le H/3$ (Lemma~\ref{lem:accelerated_trial}(i)), every accepted inner
iterate satisfies $\nu\Phi\le H$. The sketch width and the matrix-polynomial
degree therefore remain those of the $O(m)$ level, rather than growing with
$\kappa$.

\emph{Step 3: the hand-off.} By (ii),
$F(x_K)\le F_*+\varepsilon_{\mathrm{opt}}\le F(x_0)+\varepsilon_{\mathrm{opt}}
=U+\varepsilon_{\mathrm{opt}}$. The wrapper of
Section~\ref{sec:faster_algorithm} tests multipliers $\nu\le\Lambda$ only,
since its first solve is at $\underline\nu\le\Lambda/2$ and its bisection
stays inside $[\underline\nu,\Lambda]$. Hence
$U=q(0)+\nu\phi_0\le q(0)+\Lambda\phi_0$. Its accuracy satisfies
$\varepsilon_{\mathrm{opt}}\le\sigma^2/2\le2^{-17}<1$, because $\sigma\le d/8$
and $d\le1/32$ in the notation of Section~\ref{sec:faster_algorithm}. Therefore
$U+\varepsilon_{\mathrm{opt}}<q(0)+\Lambda\phi_0+1=H_{\max}/16\le H_{\max}$,
which is (iii).
\end{proof}

\subsection{Arithmetic cost}\label{subsec:accelerated_cost}

Sharing a sketch reduces the work per block query, while Catalyst increases the number of dense snapshot constructions. We count both contributions before summing over the geometrically decreasing active dimensions.

\begin{lemma}[Arithmetic cost]
\label{lem:accelerated_cost}
Let $\ell\ge2$ and let $C$ be the universal constant of
Section~\ref{sec:accelerated_constants}.
\begin{enumerate}[label=(\roman*)]
\item At an active size $m\ge Ca^2\ell^2$, one selected-block query costs
\[
\widetilde O(
\Tmat(n,n,\lceil\sqrt{m/\ell}\rceil)
+\Tmat(n,\lceil\sqrt{m/\ell}\rceil,n)
+\ell n^2)
\]
arithmetic operations. One dense snapshot construction costs
$n^{\omega+o(1)}+O(mn^2)$. One fixed-multiplier solve uses
$\widetilde O(m/\sqrt\ell)$ selected-block queries and
$\widetilde O(\sqrt\ell)$ dense snapshot constructions. With
\[
M_{m,\ell}:=\Tmat(n,n,\lceil\sqrt{m/\ell}\rceil)
+\Tmat(n,\lceil\sqrt{m/\ell}\rceil,n),
\]
one projection at active size $m$ costs
\begin{equation}
\widetilde O(
\frac m{\sqrt\ell}(M_{m,\ell}+\ell n^2)
+\sqrt\ell(n^\omega+mn^2)).
\label{eq:accelerated_projection_cost}
\end{equation}
\item Using the accelerated solver at active sizes $m\ge Ca^2\ell^2$ and the
selected-coordinate solver of Theorem~\ref{thm:faster_dense_runtime} below
that size, the accelerated rounds of the outer algorithm cost
\begin{equation}
\widetilde O(
\frac n{\sqrt\ell}
(\Tmat(n,n,\lceil\sqrt{n/\ell}\rceil)
+\Tmat(n,\lceil\sqrt{n/\ell}\rceil,n))
+n^3\sqrt\ell+n^\omega\sqrt\ell)
\label{eq:accelerated_global_cost}
\end{equation}
and the late rounds cost $\widetilde O(n^2\ell^3+n^\omega)$.
\item For $b\in[1/3,1/2]$ and $\ell:=\max\{2,\lceil n^{1-2b}\rceil\}$, the
projections of all rounds together cost
\[
n^{3+\max\{\omega(1,b,1)+b-5/2,1/2-b\}+o(1)}
\]
arithmetic operations.
\end{enumerate}
\end{lemma}

\begin{proof}
\emph{Step 1: one block query and one snapshot.} By
Section~\ref{sec:accelerated_constants}, $H=O(m)$, $R^2\ge c_Rm$ with
$R^2=\widetilde\Theta(m)$, and $\ell q/\mu=\Theta(m/\ell)$, so we meet the sketch
condition by
\[
ks=\widetilde\Theta(m/\ell),
\qquad
k=s=\widetilde\Theta(\sqrt{m/\ell}),
\qquad k,s\ge2,
\]
{\setlength{\emergencystretch}{2em}with integer ceilings understood. A block query is the selected-coordinate query
of Step~3 in the proof of Theorem~\ref{thm:faster_dense_runtime}, with batch width
$r_{\mathrm{query}}:=\widetilde O(\sqrt{m/\ell})$. Its matrix work is
$\widetilde O(\Tmat(n,n,r_{\mathrm{query}}))$ for the polynomial actions and
$\widetilde O(\Tmat(n,r_{\mathrm{query}},n))$ for the two materializations. Thereafter
every requested coordinate is one Frobenius contraction with the two materialized matrices,
costing $O(n^2)$ per coordinate, so sharing the sketch across the block
introduces neither an extra probe factor nor an extra block factor. The
$\ell n^2$ term covers the block contractions, the update of $B(x)$, and the
proximal step of Lemma~\ref{lem:accelerated_step}(ii). It also covers the guard of
Lemma~\ref{lem:accelerated_step}(iii). Although the Catalyst center $y_j$ is
generally dense, it occurs only in the quadratic term of $G_j$, whose gradient
we evaluate coordinatewise. Hence we need no matrix image of $y_j$, and we incur no
$O(mn^2)$ matrix reconstruction at a block query. This is the
query cost in (i). A dense snapshot construction costs
$n^{\omega+o(1)}+O(mn^2)$, as the snapshot refresh of that Step~3 does.\par}

\emph{Step 2: one projection.} By Lemma~\ref{lem:accelerated_trial}(iii) and
$\widehat L/(1+\kappa)=\widetilde O(1)$, each regularized subproblem uses
$\widetilde O(q)=\widetilde O(m/\ell)$ block queries and $\widetilde O(1)$
dense snapshot constructions. By Lemma~\ref{lem:accelerated_catalyst}(ii),
Catalyst uses $\widetilde O(\sqrt\ell)$ subproblems. Hence one
fixed-multiplier solve uses $\widetilde O(m/\sqrt\ell)$ selected-block
queries and $\widetilde O(\sqrt\ell)$ dense snapshot constructions, and its
cost is $\frac m{\sqrt\ell}(M_{m,\ell}+\ell n^2)+\sqrt\ell(n^\omega+mn^2)$ up
to polylogarithmic factors. The certified multiplier search, the polynomial
precision, the guard amplification, the objective comparisons, and the
allocation of the failure budget over all inner solves multiply this by
polylogarithmic factors only, so we obtain
Eq.~\eqref{eq:accelerated_projection_cost}.

\emph{Step 3: all rounds.} For a fixed block size $\ell$, we use the accelerated
solver while $m\ge Ca^2\ell^2$, where $\ell\le R/(4a)$ by
Section~\ref{sec:accelerated_constants}. We use the selected-coordinate solver of
Theorem~\ref{thm:faster_dense_runtime} once $m<Ca^2\ell^2$. The active
dimensions decrease geometrically and have sum $O(n)$. By monotonicity of
$\Tmat$ and Eq.~\eqref{eq:accelerated_projection_cost}, we obtain
Eq.~\eqref{eq:accelerated_global_cost} for the accelerated rounds: the
query terms sum to $\widetilde O(\frac n{\sqrt\ell}(M_{n,\ell}+\ell n^2))$,
and the $O(\log n)$ rounds contribute
$\widetilde O(\sqrt\ell(n^\omega+n^3))$ through their snapshots. The last
term of Eq.~\eqref{eq:accelerated_global_cost} is dominated by
$n^3\sqrt\ell$ because $\omega<3$. In the late rounds the largest active
dimension is $\widetilde O(\ell^2)$ and the width in the rectangular products
is $\widetilde O(\sqrt m)=\widetilde O(\ell)$. Using classical multiplication
for these thin products, we bound the geometric sum of their cost by
\[
\widetilde O(n^2\ell^3+n^\omega).
\]

{\setlength{\emergencystretch}{2em}\hbadness=10000 \emph{Step 4: the exponent.} Put $\ell:=\max\{2,\lceil n^{1-2b}\rceil\}$ for
$b\in[1/3,1/2]$. Then $n/\sqrt\ell=O(n^{1/2+b})$ and
$\lceil\sqrt{n/\ell}\rceil=O(n^b)$. Permutation invariance of
$\omega(\cdot,\cdot,\cdot)$, which we recorded in Section~\ref{sec:preliminaries},
makes both rectangular products in Eq.~\eqref{eq:accelerated_global_cost}
cost $n^{\omega(1,b,1)+o(1)}$, so the accelerated rounds cost
$n^{3+\max\{\omega(1,b,1)+b-5/2,1/2-b\}+o(1)}$. The late rounds cost
$n^{5-6b+o(1)}+n^{\omega+o(1)}\le n^{3+o(1)}$ because $b\ge1/3$ and
$\omega<3$. Since $3\le3+(1/2-b)$, they do not add a term to the maximum.\par}
\end{proof}

\subsection{Proof of Theorem~\ref{thm:accelerated_dense_runtime}}\label{subsec:accelerated_proof}

\begin{proof}
\emph{Correctness.} Algorithm~\ref{alg:accelerated_dense_matrix_spencer} runs
the body, projection accuracy, certified multiplier search, snapping rule, and
outer partial-coloring construction of Theorems~\ref{thm:dense_runtime} and
\ref{thm:faster_dense_runtime}. It replaces every fixed-multiplier solve by
Algorithm~\ref{alg:accelerated_multiplier}, which at active sizes
$m<Ca^2\ell^2$ is the solver of Theorem~\ref{thm:faster_dense_runtime}.
Consider one accelerated solve at a tested multiplier $\nu\le\Lambda$, with
the accuracy $\varepsilon_{\mathrm{opt}}\le\sigma^2/2$ and the budget
$\delta_\nu$ of the wrapper. By Lemma~\ref{lem:accelerated_catalyst}(i),
every regularized subproblem starts at $G_j(x_{j-1})\le3U$, and
$\varepsilon_{\mathrm{ph},j}\le c_0U\le U$. Hence Lemma~\ref{lem:accelerated_trial}(iii)
applies to subproblem $j$ with $\varepsilon_{\mathrm{ph}}=\varepsilon_{\mathrm{ph},j}$ and the phase
budget $\delta_{\mathrm{ph}}=\delta_\nu/(KJ_j)$. The subproblem fails with
probability at most $J_j\delta_{\mathrm{ph}}=\delta_\nu/K$, and the $K$
subproblems fail with probability at most $\delta_\nu$ in total. On the
complementary event every inexactness condition
$G_j(x_j)-\min_QG_j\le\varepsilon_{\mathrm{ph},j}$ holds pathwise, and
Lemma~\ref{lem:accelerated_catalyst}(ii) and (iii) give
\[
F(x_K)\le F_*+\varepsilon_{\mathrm{opt}}\le U+\varepsilon_{\mathrm{opt}}<H_{\max}.
\]
This is the interface hypothesis stated by the multiplier wrapper of
Section~\ref{sec:faster_algorithm}: an $\varepsilon_{\mathrm{opt}}$-accurate
fixed-multiplier solution in the $H_{\max}$-level set. Hence the wrapper of
Section~\ref{sec:selected_wrapper} applies as written, and the transfer to the
outer algorithm is Step~1 of the proof of Theorem~\ref{thm:faster_dense_runtime}.
We distribute the failure budget over all active rounds, multiplier tests,
fixed-multiplier solves, and cleanup trials, with fresh randomness at every
call, and this contributes the factor $\operatorname{polylog}(1/p)$.

{\setlength{\emergencystretch}{2em}\emph{Running time.} By Lemma~\ref{lem:accelerated_cost}(iii), for every
$b\in[1/3,1/2]$ the block size $\ell:=\max\{2,\lceil n^{1-2b}\rceil\}$ makes
all projections cost $n^{3+\max\{\omega(1,b,1)+b-5/2,1/2-b\}+o(1)}$. Input
construction, the bounded-dimension cleanup, and all remaining work cost at
most $n^{3+o(1)}$, and the failure allocation contributes
$\operatorname{polylog}(1/p)$. We first choose $b=1/2$ and use the known bound
$\omega(1,1/2,1)<2.042776$ to show that $\gamma_\star\le0.042776$. Since
the second term in the defining maximum is $1/2-b$, we may therefore take a
minimizer with $b\ge0.457224$, which lies in $[1/3,1/2]$. Minimizing over such
$b$, we obtain the exponent $3+\gamma_\star+o(1)$ of the theorem.\par}

\emph{The numerical value.} The function $b\mapsto\omega(1,b,1)$ is convex on
$[0,1]$. It is nondecreasing, because $\Tmat(n,\lceil n^b\rceil,n)$ is
nondecreasing in $b$. It is midpoint convex, because algorithms for
$\langle n,n^{b_1},n\rangle$ and $\langle n,n^{b_2},n\rangle$ compose by
tensor product into an algorithm for $\langle n^2,n^{b_1+b_2},n^2\rangle$, so
that $2\omega(1,(b_1+b_2)/2,1)\le\omega(1,b_1,1)+\omega(1,b_2,1)$. A
nondecreasing midpoint-convex function is convex. With the bounds
\[
\omega(1,0.4,1)<2.009280,
\qquad
\omega(1,0.5,1)<2.042776
\]
of Alman, Duan, Vassilevska Williams, Xu, Xu and
Zhou~\cite[Table~1]{almanetal25asymmetry}, we obtain by convex interpolation
\[
\omega(1,b,1)
\le2.009280+0.33496(b-0.4)
\]
for $0.4\le b\le0.5$. Thus the first term in the defining maximum is at most
\[
1.33496b-0.624704,
\]
and the second term is $1/2-b$. Balancing these terms, we get
\[
b_\star:=\frac{1.124704}{2.33496}
=0.4816802000\ldots
\]
and therefore
\[
\gamma_\star\le\frac12-b_\star=0.0183197999\ldots,
\]
the value quoted in Remark~\ref{rem:accelerated_exponent}.
\end{proof}

\section{The cubic-time algorithm}
\label{sec:cubic_time}

In this section, we remove the exponent loss in
Theorem~\ref{thm:accelerated_dense_runtime} by tracking the dense Gibbs
matrix recursively.  We keep the tracking error from accumulating by a random
refresh.  The variance of every nonrefresh increment is proportional
to the squared length of the coordinate step, which the negative step-energy
term of coordinate descent pays for.  We change the fixed-multiplier solver, not the partial-coloring
procedure around it, which is that of
Section~\ref{sec:faster_algorithm}.

\begin{theorem}[Recursive Gibbs-tracking implementation]
\label{thm:cubic_dense_runtime}
There is an algorithm, given by procedure
\textnormal{\textsc{MainCubic}} in Algorithm~\ref{alg:cubic_matrix_spencer},
with the following guarantee.  For every $p\in(0,1/2)$, it finds a signing
satisfying the conclusion of Theorem~\ref{thm:matrix_spencer_bound} with
failure probability at most $p$ using
\[
n^{3+o(1)}\operatorname{polylog}(1/p)
\]
arithmetic operations in the real-arithmetic model.
\end{theorem}

For dense input, whose size is $n^3$ (Section~\ref{sec:model}), the running time is within an $n^{o(1)}\operatorname{polylog}(1/p)$ factor of the time needed to read the input.

\begin{proof}
Lemma~\ref{lem:cubic_tracked_phase} and independent repetition give
Lemma~\ref{lem:cubic_recursive_multiplier}, including its accuracy and
certificate claims.  Lemma~\ref{lem:cubic_outer_reduction} then gives the
required signing and failure probability, and
Lemma~\ref{lem:cubic_global_runtime} gives the claimed running time.
\end{proof}

In the rest of the section we state the three procedures (Algorithms~\ref{alg:cubic_matrix_spencer}--\ref{alg:cubic_tracked_phase}) and prove the four lemmas the proof composes. We record the outer procedure \textsc{MainCubic}, Algorithm~\ref{alg:cubic_matrix_spencer}, in Section~\ref{subsec:cubic_signing} and prove Lemma~\ref{lem:cubic_outer_reduction}, the outer-wrapper reduction. That lemma shows the partial-coloring procedure of Section~\ref{sec:faster_algorithm} keeps its discrepancy constant and failure probability once the inner solver returns accurate, certified solutions. In Section~\ref{subsec:cubic_multiplier} we give the recursive solver \textsc{RecursiveMultiplier}, Algorithm~\ref{alg:cubic_recursive_multiplier}, which runs certified halving phases with fresh randomness. In Lemma~\ref{lem:cubic_recursive_multiplier}, the fixed-multiplier reduction, we derive its accuracy and certificates from the guarantee of a single phase. We fix the notation of one phase with exact matrix functions in Section~\ref{subsec:cubic_phase}, where we state Algorithm~\ref{alg:cubic_tracked_phase} and prove Lemma~\ref{lem:cubic_tracked_phase}, the tracked-phase guarantee. Its proof composes the three subsections that follow. We build the randomized estimator of one Gibbs increment $P_y-P_x$ in Section~\ref{subsec:cubic_increment}, where Lemma~\ref{lem:cubic_local_increment} bounds its variance by the squared length of the coordinate step. We then couple the accumulated tracking error to the progress of randomized coordinate descent in Section~\ref{subsec:cubic_coupling}. Lemma~\ref{lem:cubic_lyapunov}, the coupled contraction, shows a joint potential contracts while the step energy pays for the increment variance. Its localization step bounds the probability that the phase leaves the guarded level. We replace every exact exponential by a common polynomial surrogate in Section~\ref{subsec:cubic_finite}. Lemma~\ref{lem:cubic_finite_stability}, the finite-surrogate stability, fits the resulting bias into the additive error of the phase and certifies the guards, the objective intervals, and the $\Phi$-interval. Finally, in Section~\ref{subsec:cubic_runtime} we prove Lemma~\ref{lem:cubic_global_runtime}, the $n^{3+o(1)}\operatorname{polylog}(1/p)$ bound: the refresh probability trades sketch width against dense reconstructions, and the fixed-multiplier cost sums over the outer rounds.

\subsection{The signing algorithm}\label{subsec:cubic_signing}

We record the outer signing procedure in Algorithm~\ref{alg:cubic_matrix_spencer}.

\begin{algorithm}[H]
\caption{$n^{3+o(1)}\operatorname{polylog}(1/p)$ time algorithm for Matrix Spencer signing}
\label{alg:cubic_matrix_spencer}
\begin{algorithmic}[1]
\Procedure{MainCubic}{$A_1,\ldots,A_n,n,p$} \Comment{Theorem~\ref{thm:cubic_dense_runtime}}
\State Run Algorithm~\ref{alg:faster_dense_matrix_spencer} with the same failure-budget allocation, multiplier certificates, radial correction, snapping rule, and cleanup
\State Replace every fixed-multiplier solve by \Call{RecursiveMultiplier}{$F_\nu,Q,m,n,a,R,\varepsilon_{\mathrm{opt}},\delta_\nu$}, Algorithm~\ref{alg:cubic_recursive_multiplier}, and return its signing or \textsc{Fail}
\EndProcedure
\end{algorithmic}
\end{algorithm}

The outer algorithm uses each fixed-multiplier solution to certify a projection before snapping coordinates. Its discrepancy and progress estimates therefore depend on the accuracy and level certificates supplied by the inner solver.

\begin{lemma}[Outer-wrapper reduction]
\label{lem:cubic_outer_reduction}
Suppose Algorithm~\ref{alg:cubic_recursive_multiplier} returns each
fixed-multiplier solution with the accuracy, certificate, and failure
probability required by the certified multiplier wrapper of
Section~\ref{sec:faster_algorithm}.  Then
Algorithm~\ref{alg:cubic_matrix_spencer} returns a signing satisfying
Theorem~\ref{thm:matrix_spencer_bound}, with discrepancy constant $156000$
and failure probability at most $p$.
\end{lemma}

\begin{proof}
The smoothed body, certified multiplier search, feasibility correction,
snapping rule, and outer partial-coloring construction are exactly those of
Sections~\ref{sec:algorithm}, \ref{sec:faster_algorithm}, and
\ref{sec:accelerated_block_algorithm}.  The hypothesis supplies the only
interface we use there.  That interface is an accurately solved
fixed-multiplier problem, returned with a valid certificate inside the level
set on which the wrapper works.  Hence the snapping argument and the progress
in every outer round are unchanged.  The same terminal cleanup contributes at
most $6.5\sqrt n$, and the same union-bound allocation over the prescribed
caps gives total failure probability at most $p$.  The discrepancy constant
therefore remains $156000$.
\end{proof}

\subsection{The recursive fixed-multiplier solver}
\label{subsec:cubic_multiplier}

In Algorithm~\ref{alg:cubic_recursive_multiplier} we use certified
halving phases.  Each candidate phase starts from the currently certified point, but its
random probes, refresh coins, and guard certificates are fresh.

\begin{algorithm}[H]
\caption{Recursive fixed-multiplier solver}
\label{alg:cubic_recursive_multiplier}
\begin{algorithmic}[1]
\Procedure{RecursiveMultiplier}{$F_\nu,Q,m,n,a,R,\varepsilon_{\mathrm{opt}},\delta_\nu$} \Comment{Lemma~\ref{lem:cubic_recursive_multiplier}}
\State $U\gets F_\nu(0)$, $H\gets64U$, $\varepsilon_{\mathrm{opt}}\gets\min\{\varepsilon_{\mathrm{opt}},H\}$, and $z\gets0$
\State $J\gets\lceil\log_2(\max\{2,4H/\varepsilon_{\mathrm{opt}}\})\rceil$
\State $L\gets4\max\{1+ea^2H/R^2,2a/R,1\}$ and $\alpha\gets1/(mL)$
\State $\pi\gets\min\{1,\max\{n^{-1/2},4\alpha\}\}$
\State $k,s\gets\lceil C\pi^{-1/2}\rceil$ and $T\gets\lceil16/\alpha\rceil$
\For{$j=1,\ldots,J$}
    \State $\varepsilon_{\mathrm{ph},j}\gets\max\{\varepsilon_{\mathrm{opt}},2^{-j}H\}$ and $\delta_{g,j}^2\gets c\min\{H/m,\varepsilon_{\mathrm{ph},j}/(Jm)\}$
    \State Choose the finite-action and proximal tolerances from Lemma~\ref{lem:cubic_finite_stability}
    \State $\mathcal C\gets\{z\}$ and $R_{\mathrm{rep}}\gets\lceil C\log(J/\delta_\nu)\rceil$
    \State $\delta_{\mathrm{ph}}\gets\delta_\nu/(8JR_{\mathrm{rep}})$
    \For{$h=1,\ldots,R_{\mathrm{rep}}$}
        \State $x\gets\textsc{TrackedPhase}(F_\nu,z,Q,H,L,\pi,k,s,T,\delta_{g,j},\delta_{\mathrm{ph}},\varepsilon_{\mathrm{ph},j},J)$ \Comment{Algorithm~\ref{alg:cubic_tracked_phase}}
        \State If $x\ne\textsc{Fail}$, set $\mathcal C\gets\mathcal C\cup\{x\}$
    \EndFor
    \State $z\gets\textsc{CompareValue}(\mathcal C,\varepsilon_{\mathrm{ph},j}/(128J),\delta_\nu/(8J))$ using fresh outward intervals and the smallest upper endpoint \Comment{Algorithm~\ref{alg:accelerated_compare_value}}
\EndFor
\State \Return $z$ with fresh outward-rounded intervals for $F_\nu(z)$ and $\Phi(z)$, using failure budget $\delta_\nu/8$
\EndProcedure
\end{algorithmic}
\end{algorithm}

Certified comparison lets us combine independent phase trials while controlling the starting level for the next phase. This converts a constant-probability phase improvement into the accuracy and interval certificates required by the multiplier search.

\begin{lemma}[Fixed-multiplier reduction]
\label{lem:cubic_recursive_multiplier}
Assume that every independent call to
Algorithm~\ref{alg:cubic_tracked_phase} satisfies
Lemma~\ref{lem:cubic_tracked_phase}.  Then
Algorithm~\ref{alg:cubic_recursive_multiplier} returns an
$\varepsilon_{\mathrm{opt}}$-accurate fixed-multiplier solution $z$ with
$F_\nu(z)<H/16$, with probability at least $1-\delta_\nu$.  It also returns
outward-rounded intervals for its objective value and for $\Phi$ at the
accuracy requested by the certified multiplier wrapper.  Its number of
phases and candidate repetitions is polylogarithmic in
$n/\varepsilon_{\mathrm{opt}}$ and $1/\delta_\nu$.
\end{lemma}

\begin{proof}
Put $\Delta_j:=F_\nu(z_j)-F_\nu(x_\nu)$, where $z_j$ is the point retained
after phase $j$ and $x_\nu$ is the exact minimizer.
Lemma~\ref{lem:cubic_tracked_phase} and the choice of $\delta_{g,j}$ show
that each candidate satisfies
\[
\Delta(x)\le\frac12\Delta_{j-1}+\frac{\varepsilon_{\mathrm{ph},j}}{64J}
\]
with a universal positive probability.  Hence $R_{\mathrm{rep}}$ independent
candidates make the probability that no such candidate is returned at most
$\delta_\nu/(4J)$ once we increase the universal constant.  The internal
failure probabilities sum to at most $\delta_\nu/(8J)$.

The old point belongs to $\mathcal C$, and \textsc{CompareValue} chooses the
smallest certified upper endpoint.  From the interval widths we therefore obtain, on
the certificate-success event,
\begin{equation}
\Delta_j\le\frac12\Delta_{j-1}+\frac{\varepsilon_{\mathrm{ph},j}}{16J}.
\label{eq:cubic_halving_recurrence}
\end{equation}
Since $\Delta_0\le U=H/64$ (as $F_\nu\ge0$) and
$H2^{-J}\le\varepsilon_{\mathrm{opt}}/4$ by our choice of $J$, we unroll
Eq.~\eqref{eq:cubic_halving_recurrence} and obtain
\[
\Delta_J
\le2^{-J}\Delta_0+
\frac1{16J}\sum_{j=1}^J2^{-(J-j)}
\max\{\varepsilon_{\mathrm{opt}},H2^{-j}\}
\le\varepsilon_{\mathrm{opt}}.
\]

By the same comparison rule we prove the guard invariant needed by the
next phase and the level bound in the statement.  Indeed, the retained objective
can exceed the preceding objective by at most $\varepsilon_{\mathrm{ph},j}/(64J)$.  In
all uses by the multiplier wrapper, $\varepsilon_{\mathrm{opt}}\le H$, and
hence
\[
F_\nu(z_j)
\le U+\frac1{64J}\sum_{r=1}^j\varepsilon_{\mathrm{ph},r}
\le\frac H{64}+\frac H{32}<\frac H{16}.
\]
Thus Lemma~\ref{lem:cubic_tracked_phase} applies at every phase, and the
returned point $z=z_J$ satisfies $F_\nu(z)<H/16$.  We give
\textsc{CompareValue} failure budget $\delta_\nu/(8J)$ in phase $j$, and give
the final objective and $\Phi$ evaluations total budget $\delta_\nu/8$.
Lemma~\ref{lem:cubic_finite_stability} supplies their requested widths.
Together with the candidate and internal phase budgets above, a final union
bound gives failure probability at most $\delta_\nu$.  Our displayed choices
of $J$ and $R_{\mathrm{rep}}$ give the claimed polylogarithmic counts.
\end{proof}

\subsection{The recursive Gibbs-tracking phase}\label{subsec:cubic_phase}

We first analyze the phase called by
Algorithm~\ref{alg:cubic_recursive_multiplier} with exact matrix functions.
In Section~\ref{subsec:cubic_finite} we supply their finite implementations.

Fix an active set of size $m$, a multiplier $\nu>0$, and one projection
subproblem.  Retain from Section~\ref{sec:faster_algorithm} the box $Q$
of Eq.~\eqref{eq:smooth_projection}, the accepted draw $\xi$, the parameters
$a$, $b$, and $R$, the matrices $B(x)$, $C(x)$, $D_x$, $P_x=\exp(D_x)$, and
$E_i=\partial_iC(x)=\operatorname{diag}(A_i/R,-A_i/R)$, and the Gibbs trace
$\Phi(x)=\tr[P_x]$.  We write the fixed-multiplier objective without its
subscript,
\[
F(x):=\frac12\|x-\xi\|_2^2+\nu\Phi(x),
\qquad x\in Q;
\]
$x_*$ is its minimizer on $Q$, and
\[
\Delta(x):=F(x)-F(x_*)
\]
is the objective gap.  The normalization $\|A_i\|\le1$ of the input matrices
gives
\begin{equation}
\|E_i\|\le\frac1R,
\qquad
\sum_{i=1}^m E_i^2\preceq\frac m{R^2}I_{2n},
\label{eq:cubic_E_bounds}
\end{equation}
and differentiating $F$ in the $i$th coordinate gives
$\partial_iF(x)=x_i-\xi_i+\nu a\tr[P_xE_i]$, as in Section~\ref{sec:selected_setup}.

Fix the level $H:=64U$, $U:=F(0)$, as in the accelerated solver of
Section~\ref{sec:accelerated_block_algorithm}.
Lemma~\ref{lem:cubic_recursive_multiplier} needs the factor $64$, since its
recurrence keeps every retained point below $3H/64<H/16$, the level at which
Lemma~\ref{lem:cubic_tracked_phase} starts a phase, and the level $16F(0)$
of Section~\ref{sec:faster_algorithm} would not.  Define the coordinate
curvature parameter
\begin{equation}
L:=4\max\{1+ea^2H/R^2,2a/R,1\}.
\label{eq:cubic_L}
\end{equation}
On every fixed-multiplier level used by the wrapper,
$H=O(m)$, $R^2=\widetilde\Omega(m)$, and $a=O(\log n)$, so
$L=n^{o(1)}$.

\begin{algorithm}[H]
\caption{One guarded recursive Gibbs-tracking phase}
\label{alg:cubic_tracked_phase}
\begin{algorithmic}[1]
\Procedure{TrackedPhase}{$F_\nu,z,Q,H,L,\pi,k,s,T,\delta_g,\delta_{\mathrm{ph}},\varepsilon_{\mathrm{ph}},J$} \Comment{Lemma~\ref{lem:cubic_tracked_phase}}
\State Choose the polynomial and proximal accuracies from Lemma~\ref{lem:cubic_finite_stability}
\State $M_{\mathrm{ref}}\gets\lceil C(T\pi+\log(1/\delta_{\mathrm{ph}}))\rceil$
\State $x_0\gets z$, $\widehat P_0\gets\widetilde P_z$, and $N_{\mathrm{ref}}\gets0$
\For{$t=0,\ldots,T-1$}
    \State Draw $i_t$ uniformly from $[m]$ and set $\widehat g_{t,i_t}\gets x_{t,i_t}-\xi_{i_t}+\nu a\tr[\widehat P_tE_{i_t}]$
    \State Compute the certified trust-capped proximal step $d_t$ and set $x'\gets x_t+d_te_{i_t}$
    \State Compute a fresh certificate $F_\nu(x')\le\overline F_\nu(x')\le F_\nu(x')+H/48$
    \If{$\overline F_\nu(x')>H/3$}
        \State \Return \textsc{Fail}
    \EndIf
    \If{an independent $\pi$-refresh coin succeeds}
        \State $\widehat P_{t+1}\gets\widetilde P_{x'}$ and $N_{\mathrm{ref}}\gets N_{\mathrm{ref}}+1$
    \Else
        \State Construct $\widehat\Delta_t$ from Eq.~\eqref{eq:cubic_increment_estimator} for $\widetilde P_{x'}-\widetilde P_{x_t}$
        \State $\widehat P_{t+1}\gets\widehat P_t+\widehat\Delta_t$
    \EndIf
    \State $x_{t+1}\gets x'$
    \If{$N_{\mathrm{ref}}>M_{\mathrm{ref}}$}
        \State \Return \textsc{Fail}
    \EndIf
\EndFor
\State \Return $x_T$ together with a fresh outward-rounded objective interval
\EndProcedure
\end{algorithmic}
\end{algorithm}

In the phase analysis we couple descent with control of the tracking error, while the guard preserves the spectral level needed for each matrix-increment estimate. Our additive error allowance accommodates finite matrix actions and inexact proximal steps.

\begin{lemma}[Tracked-phase guarantee]
\label{lem:cubic_tracked_phase}
Suppose $F(z)\le H/16$, $\pi\ge4/(mL)$, and $ks\ge K_0/\pi$, where
$K_0$ is a sufficiently large universal constant.  With the accuracies in
Section~\ref{subsec:cubic_finite}, one call to
Algorithm~\ref{alg:cubic_tracked_phase} returns a candidate $x$ satisfying
\begin{equation}
\Delta(x)\le\frac12\Delta(z)+C_3m\delta_g^2
\label{eq:cubic_phase_additive}
\end{equation}
with a universal positive probability, where $C_3$ is a universal constant
fixed in Section~\ref{subsec:cubic_finite}.  Except on the assigned event of
probability at most $\delta_{\mathrm{ph}}$, every returned candidate satisfies the
true objective guard and has a valid outward-rounded objective interval.
\end{lemma}

\begin{proof}
We prove the step-sensitive variance bound in
Lemma~\ref{lem:cubic_local_increment}.  The coupled estimate in
Lemma~\ref{lem:cubic_lyapunov} then gives constant-factor contraction and
bounds the probability of leaving the guarded level.  By
Lemma~\ref{lem:cubic_finite_stability}, polynomial actions and inexact scalar
minimization add at most $C_3m\delta_g^2$ to that contraction.  The guard has
an $H/48$ margin, and the refresh cap has failure probability at most
$\delta_{\mathrm{ph}}/4$ by a binomial tail bound.  We assign the other parts of $\delta_{\mathrm{ph}}$ to the remaining
certificate failures.  By Markov's inequality and a union bound we therefore
obtain Eq.~\eqref{eq:cubic_phase_additive} with a universal positive
probability, together with the asserted conditional guard and certificate
guarantee.  We choose the universal constant $c$ in the definition of
$\delta_{g,j}$ so that
$C_3m\delta_{g,j}^2\le\varepsilon_{\mathrm{ph},j}/(64J)$.
\end{proof}

\subsection{The local matrix-increment estimator}
\label{subsec:cubic_increment}

Let $x,y\in Q$ with $y=x+de_i$, where $e_i$ is the $i$th standard basis vector,
$F(x)\le H$, and $|d|\le R/a$.  Put
$E:=P_y-P_x$.  For integers $k,s\ge2$, draw a standard Gaussian matrix
$\Omega\in\R^{2n\times2k}$.  Let $Y$ have orthonormal columns spanning
$E\Omega$, set $\Pi:=YY^\top$, and define
\[
E_{\mathrm{low}}:=\Pi E+E\Pi-\Pi E\Pi,
\qquad
E_{\mathrm{res}}:=(I_{2n}-\Pi)E(I_{2n}-\Pi).
\]
Expanding the second product, we obtain $E=E_{\mathrm{low}}+E_{\mathrm{res}}$.
For independent standard Gaussian vectors $z_1,\ldots,z_s\in\R^{2n}$, set
\begin{equation}
\widehat E:=E_{\mathrm{low}}+
\frac1{2s}\sum_{j=1}^s
\{(E_{\mathrm{res}}z_j)z_j^\top+z_j(E_{\mathrm{res}}z_j)^\top\}.
\label{eq:cubic_increment_estimator}
\end{equation}
Conditional on $\Pi$, Gaussian isotropy $\E[z_jz_j^\top]=I_{2n}$ gives
$\E[\widehat E\mid\Pi]=E$.

For a symmetric matrix $Z$, define the seminorm
\begin{equation}
\|Z\|_{\mathcal G}^2:=
\frac{\nu^2a^2}{m}\sum_{r=1}^m\tr[E_rZ]^2.
\label{eq:cubic_gradient_seminorm}
\end{equation}
It is the mean squared coordinate-gradient perturbation induced by $Z$.  By
the coordinate derivative of Section~\ref{sec:selected_setup}, replacing $P_x$ by $P_x+Z$ changes
$\partial_rF(x)$ by $\nu a\tr[E_rZ]$, and the average of the squares over a
uniform $r\in[m]$ is the right-hand side.

\begin{lemma}[Local increment variance]
\label{lem:cubic_local_increment}
Let $x,y\in Q$ with $y=x+de_i$, $F(x)\le H$, and $|d|\le R/a$.  Put
$E:=P_y-P_x$, and let $\widehat E$ be the estimator of
Eq.~\eqref{eq:cubic_increment_estimator} with integers $k,s\ge2$.  Then
\[
\E[\|\widehat E-E\|_{\mathcal G}^2]
\le\frac{C L^2}{ks}d^2
\]
for a universal constant $C$.
\end{lemma}

\begin{proof}
Conditional on $\Pi$, the Gaussian quadratic-form variance identity and
Eq.~\eqref{eq:cubic_E_bounds} give
\begin{align}
\E[\|\widehat E-E\|_{\mathcal G}^2\mid\Pi]
&\le\frac{2\nu^2a^2}{ms}
\sum_{r=1}^m\|E_{\mathrm{res}}E_r\|_F^2 \notag\\
&=\frac{2\nu^2a^2}{ms}
\tr[E_{\mathrm{res}}^2\sum_{r=1}^mE_r^2] \notag\\
&\le\frac{2\nu^2a^2}{sR^2}\|E_{\mathrm{res}}\|_F^2.
\label{eq:cubic_conditional_increment_variance}
\end{align}
Here the first inequality applies the variance identity to the symmetric
part of $E_rE_{\mathrm{res}}$, whose Frobenius norm is at most
$\|E_{\mathrm{res}}E_r\|_F$.  The equality is cyclicity of the trace.  The
last inequality is the second bound of Eq.~\eqref{eq:cubic_E_bounds}
against $E_{\mathrm{res}}^2\succeq0$.  The Gaussian range-finder estimate
we proved in Section~\ref{sec:faster_algorithm} yields
\[
\E_\Omega[\|E_{\mathrm{res}}\|_F^2]
\le\frac Ck\|E\|_1^2.
\]

Let $\mathcal D(P\|P'):=\tr[P(\log P-\log P')-P+P']$ be the matrix relative entropy of Fact~\ref{fact:generalized_pinsker}.
Since $\log P_x=D_x$ is affine in $x$, this is the Bregman divergence of the
convex function $\Phi$ at the base point $x$:
\[
\mathcal D(P_x\|P_y)
=\Phi(y)-\Phi(x)-\langle\nabla\Phi(x),y-x\rangle.
\]
The trust condition $|d|\le R/a$ gives $\|D_y-D_x\|=a|d|\,\|E_i\|\le1$, so
monotonicity of $\tr\exp$ and $\nu\Phi(x)\le F(x)\le H$ give
$\Phi(y)\le e\Phi(x)\le eH/\nu$.  Fact~\ref{fact:generalized_pinsker} therefore gives
\begin{equation}
\nu^2\|P_y-P_x\|_1^2
\le4eH\nu\mathcal D(P_x\|P_y).
\label{eq:cubic_local_pinsker}
\end{equation}
Along the segment from $x$ to $y$, the coordinate second derivative of
$F$ is at most $1+ea^2H/R^2\le L/4$ by Eq.~\eqref{eq:cubic_L}, and the
quadratic part of $F$ contributes $d^2/2\ge0$ to the Bregman remainder.
Consequently,
\[
\nu\mathcal D(P_x\|P_y)
\le F(y)-F(x)-\partial_iF(x)d
\le\frac L8d^2.
\]
\setlength{\emergencystretch}{2em}Combining this estimate with
Eq.~\eqref{eq:cubic_conditional_increment_variance}, the range-finder
bound, and Eq.~\eqref{eq:cubic_local_pinsker}, and using $a^2H/R^2\le L$
from Eq.~\eqref{eq:cubic_L} in the last step, we obtain
\[
\E[\|\widehat E-E\|_{\mathcal G}^2]
\le\frac{C}{ks}\frac{a^2H}{R^2}L d^2
\le\frac{CL^2}{ks}d^2.
\qedhere
\]
\end{proof}

\subsection{Coupled contraction and localization}\label{subsec:cubic_coupling}

Here we couple the recursive tracking error to the progress of
randomized coordinate descent.

At time $t$, we maintain a symmetric dense matrix $\widehat P_t$ and an
iterate $x_t$.  We select $i_t$ uniformly from $[m]$, independently of the
past, and use
\begin{equation}
\widehat g_{t,i_t}:=
x_{t,i_t}-\xi_{i_t}+\nu a\tr[\widehat P_tE_{i_t}]
\label{eq:cubic_tracked_gradient}
\end{equation}
as the selected derivative.  After the coordinate step produces
$x_{t+1}$, we independently refresh with probability $\pi$ by setting
$\widehat P_{t+1}:=P_{x_{t+1}}$.  Otherwise, we construct an independent copy
$\widehat\Delta_t$ of the estimator in
Eq.~\eqref{eq:cubic_increment_estimator} for
$P_{x_{t+1}}-P_{x_t}$ and set
\[
\widehat P_{t+1}:=\widehat P_t+\widehat\Delta_t.
\]
The matrix $\widehat P_t$ need not be positive semidefinite.  We use it only
in Eq.~\eqref{eq:cubic_tracked_gradient}.  The objective guards continue to
use fresh positive-semidefinite surrogates.

Put
\[
Z_t:=\widehat P_t-P_{x_t},
\qquad
\mathcal E_t:=\|Z_t\|_{\mathcal G}^2.
\]
Let $\mathcal F_t$ be the history before step $t$ and write
$\E_t[\cdot]:=\E[\cdot\mid\mathcal F_t]$.  At a nonrefresh step, the increment error has conditional matrix expectation
zero, so its cross term with $Z_t$ vanishes in the seminorm.
Lemma~\ref{lem:cubic_local_increment}, applied at the base point $x_t$ when
$F(x_t)\le H$, gives
\begin{equation}
\E_t[\mathcal E_{t+1}]
\le(1-\pi)\{\mathcal E_t+
CL^2\E_t[d_t^2]/(ks)\}.
\label{eq:cubic_tracker_recurrence}
\end{equation}

\paragraph{The coordinate step and its energy.}
For the selected coordinate $i:=i_t$, define
\[
\mathfrak{e}_{t,i}:=\nu a\tr[E_iZ_t].
\]
By Eq.~\eqref{eq:cubic_gradient_seminorm} and the independent uniform choice
of $i_t$,
\begin{equation}
\E_{i_t}[\mathfrak{e}_{t,i_t}^2\mid\mathcal F_t]=\mathcal E_t.
\label{eq:cubic_coordinate_error}
\end{equation}
Choose $d_t$ to minimize
\[
\widehat g_{t,i}d+\frac L2d^2
\]
over $x_{t,i}+d\in Q_i:=[-L_i,R_i]$, the $i$th factor of $Q$, and $|d|\le R/a$, and put
$x_{t+1}:=x_t+d_te_i$.  The comparison displacement
\[
d_i':=(x_{*,i}-x_{t,i})/L
\]
is feasible.  Indeed, every coordinate interval of $Q$ has length
$\varepsilon_{\mathrm{box}}<1$, while Eq.~\eqref{eq:cubic_L} gives $L\ge1$ and
$L\ge2a/R$.

The strong three-point inequality for the one-dimensional proximal step,
the coordinate Hessian bound $L/4$, and Young's inequality give
\begin{equation}
F(x_{t+1})-F(x_t)
\le\partial_iF(x_t)d_i'+\frac L2(d_i')^2
+\frac{\mathfrak{e}_{t,i}^2}{2L}-\frac L4d_t^2.
\label{eq:cubic_three_point}
\end{equation}
The negative term arises because the model has curvature $L$, while the
true coordinate curvature is at most $L/4$.  Young's inequality uses at most
half of the remaining model curvature.

Averaging the comparison terms over the uniform coordinate (write $x:=x_t$),
and using the
$1$-strong convexity of $F$ in the last step, we obtain
\begin{align*}
\E_i[\partial_iF(x)d_i'+L(d_i')^2/2]
&=\frac1{mL}
\{\langle\nabla F(x),x_*-x\rangle
+\|x-x_*\|_2^2/2\}\\
&\le-\frac{\Delta(x)}{mL}.
\end{align*}
With
\begin{equation}
\alpha:=\frac1{mL},
\label{eq:cubic_alpha}
\end{equation}
Eqs.~\eqref{eq:cubic_coordinate_error} and
\eqref{eq:cubic_three_point} imply
\begin{equation}
\E_t[\Delta(x_{t+1})]
\le(1-\alpha)\Delta(x_t)
+\frac{\mathcal E_t}{2L}
-\frac L4\E_t[d_t^2].
\label{eq:cubic_gap_recurrence}
\end{equation}

We weight the tracking error by $1/(\pi L)$, which allows the decrease from refreshes to absorb its contribution to the objective gap. The negative step-energy term then absorbs the variance of the matrix increments and controls the probability of leaving the analytical level.

\begin{lemma}[Coupled contraction]
\label{lem:cubic_lyapunov}
Suppose
\[
\pi\ge4\alpha,
\qquad
ks\ge K_0/\pi,
\]
where $K_0$ is a sufficiently large universal constant.  Then, at every step
$t$ with $F(x_t)\le H$, the potential
\[
V_t:=\Delta(x_t)+\frac{\mathcal E_t}{\pi L}
\]
satisfies
\begin{equation}
\E_t[V_{t+1}]
\le(1-\alpha)V_t-\frac L8\E_t[d_t^2].
\label{eq:cubic_lyapunov}
\end{equation}
Moreover, if $x_0:=z$, $\widehat P_0:=P_z$, $F(z)\le H/16$, and
$T:=\lceil16/\alpha\rceil$, then the stopped ideal phase crosses the
$5H/16$ analytical level with probability at most $1/20$.  A candidate phase therefore
halves the objective gap with a universal positive probability.
\end{lemma}

\begin{proof}
\emph{Step 1: the two coefficients.}
Combining Eqs.~\eqref{eq:cubic_tracker_recurrence} and
\eqref{eq:cubic_gap_recurrence}, the coefficient of $\mathcal E_t$ is
\[
\frac1{2L}+\frac{1-\pi}{\pi L}
=\frac{1-\pi/2}{\pi L}
\le\frac{1-\alpha}{\pi L}.
\]
The coefficient of $\E_t[d_t^2]$ is at most
\[
-\frac L4+\frac{CL(1-\pi)}{\pi ks}
\le-\frac L8
\]
after increasing $K_0$.  These are exactly the two coefficients in
Eq.~\eqref{eq:cubic_lyapunov}.

\emph{Step 2: contraction and step energy over a phase.}
We initialize a reduction phase at $z$ by setting $x_0:=z$ and
$\widehat P_0:=P_z$.  Thus $\mathcal E_0=0$.  For
\[
T:=\lceil16/\alpha\rceil=O(mL),
\]
iterating Eq.~\eqref{eq:cubic_lyapunov} and using $(1-\alpha)^T\le e^{-16}$,
we obtain $\E[V_T]\le e^{-16}V_0$ as long as the process stays on the guarded
level.  Summing Eq.~\eqref{eq:cubic_lyapunov} over $t<T$ and dropping
$V_T\ge0$ also gives the step-energy estimate
\begin{equation}
L\sum_{t=0}^{T-1}\E[d_t^2]\le8V_0.
\label{eq:cubic_step_energy}
\end{equation}

\emph{Step 3: localization.}
In this step we justify the qualification in Step~2.  Start a phase at $z$ with
$F(z)\le H/16$, and define
\[
\tau:=\min\{t+1:F(x_t+d_te_{i_t})>5H/16\}.
\]
We stop the ideal process after the crossing proposal.  Before this time, every
current iterate is below $5H/16$.  The trust
condition changes the matrix exponential by at most a factor $e$, so the
entire proposed segment remains in the $H$-trace level because $5e/16<1$.
Thus all preceding local estimates apply through the crossing proposal.

Taking the comparison displacement to be zero in the proximal inequality
gives, before stopping,
\[
F(x_{t+1})-F(x_t)\le\mathfrak{e}_{t,i_t}^2/L.
\]
Therefore a crossing requires
\[
\sum_{t=0}^{\tau-1}\mathfrak{e}_{t,i_t}^2/L\ge cH
\]
for a universal constant $c>0$.  Summing the stopped version of
Eq.~\eqref{eq:cubic_tracker_recurrence}, using $\mathcal E_0=0$, gives
\[
\pi\sum_{t=0}^{T-1}
\E[\mathbbm{1}[t<\tau]\mathcal E_t]
\le\frac{CL^2}{ks}
\sum_{t=0}^{T-1}\E[\mathbbm{1}[t<\tau]d_t^2].
\]
The stopped version of Eq.~\eqref{eq:cubic_step_energy}, together with
$V_0\le H/16$, now yields
\begin{align*}
\E[\sum_{t=0}^{T-1}\mathbbm{1}[t<\tau]\mathfrak{e}_{t,i_t}^2/L]
&=\sum_{t=0}^{T-1}
\E[\mathbbm{1}[t<\tau]\mathcal E_t/L]\\
&\le\frac{CL}{\pi ks}
\sum_{t=0}^{T-1}\E[\mathbbm{1}[t<\tau]d_t^2]\\
&\le\frac{CH}{K_0}.
\end{align*}
Markov's inequality shows that $\Pr[\tau\le T]\le C/K_0$.  We choose $K_0$
so that this probability is at most $1/20$.

\emph{Step 4: agreement with the implemented process.}
On the event of no analytical crossing, every proposal has true objective at
most $5H/16$.  A correct $H/48$-accurate upper certificate is therefore at
most $H/3$, so every proposal is retained and the implemented process agrees
with the ideal process.  We then apply the killed-process argument of
Section~\ref{sec:faster_algorithm} to
Eq.~\eqref{eq:cubic_lyapunov} and conclude that one candidate phase halves
the gap with a universal positive probability.  Independent candidate
repetitions and outward-rounded objective comparisons amplify this to the
failure budget we assign to the phase.
\end{proof}

\begin{remark}[Positivity of the tracker is not used]
Positivity of $\widehat P_t$ plays no role in the proof of
Lemma~\ref{lem:cubic_lyapunov}.  The tracker enters only through the selected
derivative of Eq.~\eqref{eq:cubic_tracked_gradient}.  The one-sided guard
uses a fresh positive-semidefinite approximation of $P_{x_{t+1}}$ at the
proposed point, not the tracker.
\end{remark}

\subsection{Finite matrix-function implementation}
\label{subsec:cubic_finite}

For finite matrix actions, we center the recursive tracker on a common polynomial surrogate, so its stochastic increment error remains conditionally unbiased. The difference between that surrogate and the exact exponential contributes a deterministic bias that must fit within the phase's additive error allowance.

\begin{lemma}[Finite-surrogate stability]
\label{lem:cubic_finite_stability}
There are inverse-polynomial action accuracies and certified scalar
minimization tolerances, computable from the phase parameters, with the
following properties.
\begin{enumerate}[label=(\roman*)]
\item The finite implementation of Algorithm~\ref{alg:cubic_tracked_phase}
satisfies Eq.~\eqref{eq:cubic_phase_additive} with a universal positive
probability, the killed-process bound of Lemma~\ref{lem:cubic_lyapunov}.
\item Except on an event of probability at most $\delta_{\mathrm{ph}}$, the
fresh guard of Algorithm~\ref{alg:cubic_tracked_phase} obeys
$F_\nu(x')\le\overline F_\nu(x')\le F_\nu(x')+H/48$ and its phase-end
objective interval has its requested outward-rounded width.
\item Outside the event of (ii), the fixed-multiplier $\Phi$-interval
returned by Algorithm~\ref{alg:cubic_recursive_multiplier} has its requested
outward-rounded width.
\end{enumerate}
The required polynomial degrees and scalar iteration counts are
polylogarithmic in the inverse target accuracy.
\end{lemma}

\begin{proof}
\emph{Step 1: the polynomial surrogate and its bias.}
We replace every exact exponential by the same finite polynomial
surrogate used in Section~\ref{sec:faster_algorithm}.  On the guarded
spectral interval, let $q_r$ approximate $t\mapsto\exp(t/2)$ and define
\begin{equation}
\widetilde P_x:=q_r(D_x)^2.
\label{eq:cubic_polynomial_surrogate}
\end{equation}

We now define the recursive tracker relative to $\widetilde P_x$.  A refresh
sets $\widehat P_{t+1}:=\widetilde P_{x_{t+1}}$, and a nonrefresh step adds
an unbiased copy of the estimator in
Eq.~\eqref{eq:cubic_increment_estimator} with $E$ replaced by
$\widetilde E:=\widetilde P_{x_{t+1}}-\widetilde P_{x_t}$.  This convention
preserves exact conditional unbiasedness for the finite
algorithm.  We treat the difference between $\widetilde P_x$ and $P_x$
separately as deterministic gradient bias.

Let $d_0:=2n$, and let $\varepsilon_{\mathrm{op}}$ bound $\|\widetilde P_x-P_x\|$ uniformly on the guarded interval.
As in Section~\ref{sec:selected_finite}, the trace norm of a $d_0\times d_0$ matrix is at most $d_0$ times its operator norm and $\|E_i\|\le1/R$, so $\|\widetilde P_x-P_x\|_1\le d_0\varepsilon_{\mathrm{op}}$, $|\nu a\tr[E_i(\widetilde P_x-P_x)]|\le\nu ad_0\varepsilon_{\mathrm{op}}/R$,
and by the triangle inequality $\|\widetilde P_y-\widetilde P_x-(P_y-P_x)\|_1\le2d_0\varepsilon_{\mathrm{op}}$.
The phase accuracy $\varepsilon_{\mathrm{ph}}$ is the argument of
Algorithm~\ref{alg:cubic_tracked_phase}, and it equals $\varepsilon_{\mathrm{ph},j}$ in phase
$j$ of Algorithm~\ref{alg:cubic_recursive_multiplier}.  For this accuracy and the number $J$
of halving phases, we choose an inverse-polynomial $\delta_g$ with
\begin{equation}
\delta_g^2\le c\min\{H/m,\varepsilon_{\mathrm{ph}}/(Jm)\},
\label{eq:cubic_delta_g}
\end{equation}
and take
\begin{equation}
\varepsilon_{\mathrm{op}}\le
\min\{\phi_0/(16d_0),
H/(192\nu d_0),
R\delta_g/(C\nu ad_0),
R\sqrt{ks}\delta_g/(C\nu ad_0)\}.
\label{eq:cubic_epsilon_op}
\end{equation}
The certified wrapper in Section~\ref{sec:faster_algorithm} has
$\nu\ge\underline\nu=n^{-O(1)}$, and all other quantities in
Eq.~\eqref{eq:cubic_epsilon_op} are polynomially bounded.  The guarded
spectral interval and the required polynomial degree are therefore
polylogarithmic in the inverse target accuracy.

\emph{Step 2: the finite recurrences.}
The three bias bounds of Step~1, the range-finder estimate, and Young's
inequality change the two exact recurrences only to
\begin{align}
\E_t[\Delta(x_{t+1})]
&\le(1-\alpha)\Delta(x_t)
+\frac{C_1\mathcal E_t}{L}
-\frac L4\E_t[d_t^2]
+\frac{C_1\delta_g^2}{L},
\label{eq:cubic_finite_gap}\\
\E_t[\mathcal E_{t+1}]
&\le(1-\pi)\{\mathcal E_t
+CL^2\E_t[d_t^2]/(ks)
+C\delta_g^2/(ks)\}.
\label{eq:cubic_finite_tracker}
\end{align}
Here $\mathcal E_t$ measures the stochastic tracking error relative to
$\widetilde P_{x_t}$, and $C_1$ is universal.  To see the only new term in
Eq.~\eqref{eq:cubic_finite_tracker}, we use the bound
$\|\widetilde P_y-\widetilde P_x\|_1^2\le2\|P_y-P_x\|_1^2+8d_0^2\varepsilon_{\mathrm{op}}^2$ of Section~\ref{sec:selected_finite}
in the proof of Lemma~\ref{lem:cubic_local_increment}.  The bias in
Eq.~\eqref{eq:cubic_finite_gap} follows directly from
the gradient bias bound of Step~1.

\emph{Step 3: the finite potential.}
We choose a sufficiently large universal $c_1$ and use the finite potential
\[
V_t^{\mathrm{fin}}:=\Delta(x_t)+\frac{c_1\mathcal E_t}{\pi L}.
\]
Because $\pi\ge4\alpha$, increasing $c_1$ absorbs the
$C_1\mathcal E_t/L$ term.  Increasing $K_0$ in $ks\ge K_0/\pi$ absorbs the
step-variance term.  Thus we obtain
\begin{equation}
\E_t[V_{t+1}^{\mathrm{fin}}]
\le(1-\alpha)V_t^{\mathrm{fin}}
-\frac L8\E_t[d_t^2]
+\frac{C_2\delta_g^2}{L}.
\label{eq:cubic_finite_lyapunov}
\end{equation}
Over $T=O(mL)$ queries, the total additive contribution is
$O(m\delta_g^2)$.  Equation~\eqref{eq:cubic_delta_g} makes this smaller
than the allocated phase error and a sufficiently small fraction of the
crossing budget.  The exact killed-process argument therefore gives
Eq.~\eqref{eq:cubic_phase_additive}, rather than a pure multiplicative
halving statement.

\emph{Step 4: the scalar minimization.}
We can perform the scalar coordinate minimization to a certified
variational residual $\rho_t$.  As in
Section~\ref{sec:accelerated_block_algorithm}, we choose the bisection
tolerances so that
\[
\sum_{t=0}^{T-1}\rho_t
\le c\min\{H,\varepsilon_{\mathrm{ph}}/J\}.
\]
This contributes another admissible additive term to
Eq.~\eqref{eq:cubic_finite_lyapunov}.

\emph{Step 5: the certificates.}
It remains to record the certificates.  For every proposal, we use fresh probes
of the positive-semidefinite surrogate and a median-of-means trace estimate,
with enough samples that its contribution to the objective is at most
$H/192$.  The polynomial bias is at most $H/192$ by
Eq.~\eqref{eq:cubic_epsilon_op}, so the two-sided width is $H/48$ and
\[
F_\nu(x')\le\overline F_\nu(x')
\le F_\nu(x')+H/48.
\]
To each guard we assign conditional failure probability at most
$\delta_{\mathrm{ph}}/(4T)$.  Fresh randomness makes a union bound
valid under adaptivity.  The cap
$M_{\mathrm{ref}}=\lceil C(T\pi+\log(1/\delta_{\mathrm{ph}}))\rceil$ fails with
probability at most $\delta_{\mathrm{ph}}/4$ by a binomial tail bound.

We compute the much narrower phase-end objective intervals and the multiplier
$\Phi$-interval of Algorithm~\ref{alg:cubic_recursive_multiplier}
separately, exactly as in
Section~\ref{sec:faster_algorithm}.  For each candidate we use one fresh dense polynomial
evaluation.  We choose its operator error so that the trace error is at most
$\varepsilon_{\mathrm{ph}}/(256\nu J)$ or the smaller accuracy requested
by the multiplier wrapper, and round the resulting interval outward.  This
requires only polylogarithmic polynomial degree.  We allocate the remaining
parts of $\delta_{\mathrm{ph}}$ to these evaluations and to the finite proximal
calls.  The finite phase therefore satisfies
Eq.~\eqref{eq:cubic_phase_additive} with the probability of
Lemma~\ref{lem:cubic_lyapunov}, and the certificate claims (ii) and
(iii) hold as stated.  We do not assert any additional pure-halving step.
\end{proof}

\subsection{Running time}\label{subsec:cubic_runtime}

The refresh probability trades sketch width against the frequency of dense matrix reconstructions. At the scale we choose, thin rectangular products control the cost of each coordinate query, and dense refreshes contribute a lower-order term after summing over the active dimensions.

\begin{lemma}[Global running time]
\label{lem:cubic_global_runtime}
With the parameters $\pi$, $k$, $s$, and $T$ set by
Algorithm~\ref{alg:cubic_recursive_multiplier} at every active dimension,
Algorithm~\ref{alg:cubic_matrix_spencer} uses
\[
n^{3+o(1)}\operatorname{polylog}(1/p)
\]
arithmetic operations in the real-arithmetic model.
\end{lemma}

\begin{proof}
\emph{Step 1: the parameters.}
Section~\ref{sec:model} quotes $\alpha_{\mathrm{dual}}>0.321334$ and
$\omega<2.371339$, and here we use only $\alpha_{\mathrm{dual}}>1/4$ and
$\omega<5/2$.  At active dimension
$m$, Algorithm~\ref{alg:cubic_recursive_multiplier} sets
\begin{equation}
\pi_m:=\min\{1,\max\{n^{-1/2},4/(mL)\}\},
\qquad
k,s:=\lceil C\pi_m^{-1/2}\rceil,
\end{equation}
where the universal constant is large enough that $ks\ge K_0/\pi_m$.
Since $\pi_m\ge n^{-1/2}$, $k+s\le n^{1/4+o(1)}$.

\emph{Step 2: the cost of one query.}
A nonrefresh query applies $\widetilde P_x$ and $\widetilde P_y$ to batches
of $O(k+s)$ vectors, together with the polylogarithmically many fresh guard
probes.  The same query constructs $E\Omega$, $EY$, and the residual actions, forms the low-rank products in
Eq.~\eqref{eq:cubic_increment_estimator}, contracts
$\tr[\widehat P_tE_{i_t}]$, and updates $B(x_t)$.  Polynomial actions cost
only a polylogarithmic number of thin matrix products.  Since
$1/4<\alpha_{\mathrm{dual}}$, both orientations of an
$n\times n$ by $n\times n^{1/4+o(1)}$ product, and the corresponding
$n\times n^{1/4+o(1)}$ by $n^{1/4+o(1)}\times n$ product, cost
$n^{2+o(1)}$.  The QR factorization costs
$O(n(k+s)^2)=n^{3/2+o(1)}$, while the contraction and the update of $B(x_t)$
cost $O(n^2)$.  Therefore a nonrefresh query costs
\begin{equation}
n^{2+o(1)}.
\end{equation}

A refresh constructs the full dense matrix
$\widetilde P_{x_{t+1}}$ by a polylogarithmic number of dense matrix
products, and hence costs
\begin{equation}
n^{\omega+o(1)}.
\end{equation}
In contrast with the snapshot implementation in
Section~\ref{sec:faster_algorithm}, a refresh does not compute all $m$
coordinate derivatives.  The algorithm stores the dense matrix and
computes only the subsequently selected contraction.

\emph{Step 3: one fixed-multiplier solve.}
A constant-reduction phase uses $T=O(mL)=mn^{o(1)}$ coordinate queries.
Its nonrefresh work is therefore
\[
mn^{2+o(1)}.
\]
The expected number of refreshes is $O(mL\pi_m+1)$, and a standard binomial
tail bound caps it at a logarithmic multiple of this quantity with the
failure probability we assign.  If $\pi_m=n^{-1/2}$, the refresh work is
\[
m n^{\omega-1/2+o(1)}.
\]
If the term $4/(mL)$ determines $\pi_m$, there are only
$\operatorname{polylog}(n)$ refreshes, whose total cost is
$n^{\omega+o(1)}$.  Consequently, including the polylogarithmically many
halving phases and accuracy repetitions, one fixed-multiplier solve costs
\begin{equation}
mn^{2+o(1)}+m n^{\omega-1/2+o(1)}+n^{\omega+o(1)}.
\label{eq:cubic_fixed_multiplier_cost}
\end{equation}
The phase-end objective comparisons and the certified $\Phi$-intervals use
only a polylogarithmic number of fresh dense polynomial evaluations.  Their
cost is absorbed by the $n^{\omega+o(1)}$ term in
Eq.~\eqref{eq:cubic_fixed_multiplier_cost}.  Their probe counts and outward
rounding contribute only the displayed polylogarithmic factors.
Certified multiplier bisection contributes only another polylogarithmic
factor.  We leave its finite objective intervals, exact-feasibility radial
correction, and snapping step unchanged.

\emph{Step 4: all outer rounds.}
Let $N_{\mathrm{out}}$ be the number of outer rounds.  Denote their active
dimensions by $m_0,m_1,\ldots,m_{N_{\mathrm{out}}-1}$.  These dimensions decrease
geometrically, so $N_{\mathrm{out}}=O(\log n)$ and
\[
\sum_{t=0}^{N_{\mathrm{out}}-1}m_t=O(n).
\]
Summing Eq.~\eqref{eq:cubic_fixed_multiplier_cost} over the outer rounds
gives
\begin{align*}
\sum_{t=0}^{N_{\mathrm{out}}-1}m_tn^{2+o(1)}
&=n^{3+o(1)},\\
\sum_{t=0}^{N_{\mathrm{out}}-1}m_tn^{\omega-1/2+o(1)}
&=n^{\omega+1/2+o(1)},\\
\sum_{t=0}^{N_{\mathrm{out}}-1}n^{\omega+o(1)}
&=n^{\omega+o(1)}.
\end{align*}
The input construction, final signed sum, snapping, and bounded-dimension
cleanup cost at most $n^{3+o(1)}$.  Since $\omega<5/2$,
$\omega+1/2<3$.  The total arithmetic cost is therefore
\[
(n^3+n^{\omega+1/2}+n^\omega)n^{o(1)}
\operatorname{polylog}(1/p)
=n^{3+o(1)}\operatorname{polylog}(1/p).
\]

\emph{Step 5: the hand-off to the wrapper.}
Lemma~\ref{lem:cubic_recursive_multiplier} supplies the objective
accuracy and outward-rounded $\Phi$-interval required by the certified
multiplier wrapper, and its returned point satisfies
$F_\nu(z)<H/16=4U\le H_{\max}$, because $U=F_\nu(0)=\frac12\|\xi\|_2^2+\nu\phi_0$
with $\nu\le\Lambda$ while $H_{\max}=16(\frac12\|\xi\|_2^2+\Lambda\phi_0+1)$.
This is the interface hypothesis of the wrapper of
Section~\ref{sec:faster_algorithm}, so its segment estimate applies to the
returned point.  The wrapper therefore chooses the correct bracket half
or stops with the prescribed residual, and its downward-rounded radial
correction makes the selected projection exactly feasible.
We allocate $p$ over all outer rounds,
multiplier tests, candidate phases, guards, refresh-count caps, value
certificates, and cleanup trials, which contributes only the displayed
polylogarithmic factor.
\end{proof}

\section{A smaller constant for the cubic-time algorithm}
\label{sec:small_algorithm_constant}

In this section, we combine small-ball estimates with uniform-target projections and deterministic binary repair. We construct the auxiliary helpers in Sections~\ref{subsec:algorithm_centered_box}--\ref{subsec:cubic_refined_certificate}, prove the final small-ball row in Section~\ref{subsec:cubic_localized_means}, and certify the complete algorithm in Section~\ref{subsec:cubic_global_profiles}. We begin in Section~\ref{subsec:algorithm_centered_box} with Lemma~\ref{lem:algorithm_centered_box}, the ball-target projection. We project a uniform ball sample onto a symmetric convex body of Gaussian measure at least $e^{-Em}$, cut by a shifted box. With probability above $0.999$, the projection fixes more than $\delta m$ sign coordinates. Section~\ref{subsec:algorithm_small_implementation} turns this step into the procedure \textsc{SuffixSigning} of Algorithm~\ref{alg:algorithm_suffix_signing}. Lemma~\ref{lem:appendix_fixed_target_projection} certifies its projection subroutine, and Lemma~\ref{lem:algorithm_certified_implementation} proves its failure budget and its $n^{3+o(1)}\operatorname{polylog}(1/p)$ cost. Section~\ref{subsec:algorithm_1240_moments} introduces the polynomial barriers with logarithmic tail of Definition~\ref{def:algorithm_1240_barrier}. There we prove Lemma~\ref{lem:algorithm_1240_curvature}, which reduces the all-pairs curvature condition to finitely many one-variable polynomial tests. We prove Lemma~\ref{lem:algorithm_21_rounding}, the unbiased dyadic repair, in Section~\ref{subsec:cubic_binary_repair}: it rounds the remaining fractional coordinates one binary digit at a time and tracks support sizes and accumulated discrepancy. In Section~\ref{subsec:cubic_refined_uniform} we fix the coding amplitude $s_{\rm alg}$ and pass to uniform targets. Lemma~\ref{lem:cubic_refined_projection} gives a feasible projection with a prefix test, and Lemma~\ref{lem:cubic_refined_potential} bounds the cost of deterministic binary repair with a common orientation. Section~\ref{subsec:cubic_refined_functional} supplies the spectral estimates we need for the radial argument: the odd functional inequalities of Lemma~\ref{lem:cubic_refined_functional}, their centered refinement in Lemma~\ref{lem:cubic_centered_functional}, and the shifted cubic moments of Lemma~\ref{lem:cubic_refined_cubic}. Section~\ref{subsec:cubic_localized_radial} follows a Gaussian radial path and proves Lemma~\ref{lem:cubic_refined_radial}, a local radial majorant on each variance interval from a piecewise-polynomial witness. Lemma~\ref{lem:cubic_refined_transfer} then converts the integrated radial bound into a small-ball row $(r,e)$. We collect the concrete data in Section~\ref{subsec:cubic_refined_data}: the rows and their witnesses, the $17$-row registry of Definition~\ref{def:cubic_refined_trace_rows}, and the capacity-matched suffix schedules of Definition~\ref{def:cubic_refined_suffix}. Its Lemma~\ref{lem:cubic_refined_suffix}, the suffix interface, bounds the suffix discrepancy at capacity $w_i$ by $C_i^{\rm suf}\sqrt n$. In Section~\ref{subsec:cubic_actual_size_profiles} we assign each helper state a profile over the actual input sizes in its capacity band, Definition~\ref{def:cubic_refined_profiles}. Lemma~\ref{lem:cubic_refined_accuracy} shows that its projection actions and fallbacks are implementable at every dimension within the allocated failure budget. Section~\ref{subsec:cubic_refined_certificate} makes the auxiliary policy an explicit finite object, Definition~\ref{def:cubic_refined_policy_construction}. We verify it in Lemma~\ref{lem:cubic_refined_certificate} and implement the suffixes and four repair levels in Lemma~\ref{lem:cubic_refined_complete}. In Section~\ref{subsec:cubic_localized_means} we localize the derivative means and weighted means away from zero, Lemma~\ref{lem:cubic_localized_means}, sharpening the functional coefficients. With the capped spectral tests that follow, Lemma~\ref{lem:cubic_localized_certificate} certifies the localized root row of Definition~\ref{def:cubic_localized_row}, the final small-ball row. Finally, in Section~\ref{subsec:cubic_global_profiles} we dilate two existing rows, add nine helper layers and a root, and select the global policy of Definition~\ref{def:cubic_global_policy}. Lemma~\ref{lem:cubic_global_certificate} certifies the constant $11.503394001765$, and Lemma~\ref{lem:cubic_global_complete} proves Theorem~\ref{thm:algorithm_constant_13}. For all projections we use the multiplier wrapper and recursive solver of Section~\ref{sec:cubic_time}. The arithmetic count in Theorem~\ref{thm:algorithm_constant_13} does not assert new fixed-point word bounds for Section~\ref{sec:bit_complexity}.

\begin{theorem}[A cubic-time signing with constant below $11.504$]
\label{thm:algorithm_constant_13}
There is a randomized algorithm, procedure \textnormal{\textsc{MainCubicSmallConstant}} in Algorithm~\ref{alg:main_cubic_small_constant}, with the following guarantee. For symmetric contractions $A_1,\ldots,A_n\in\R^{n\times n}$ and $p\in(0,1/2)$, it returns signs $\varepsilon_1,\ldots,\varepsilon_n\in\{-1,1\}$ satisfying
\[
\|\sum_{i=1}^n\varepsilon_iA_i\|\le11.503394001765\sqrt n<11.504\sqrt n
\]
with failure probability at most $p$. It uses $n^{3+o(1)}\operatorname{polylog}(1/p)$ arithmetic operations in the real-arithmetic model. The same statement holds for $N\times N$ input matrices with $N\le n$.
\end{theorem}

\begin{proof}
Lemma~\ref{lem:cubic_global_complete} proves the discrepancy, failure, and arithmetic bounds for the policy certified in Lemma~\ref{lem:cubic_global_certificate}. For $N<n$ we use zero padding.
\end{proof}

Throughout this section, terminating decimals are exact rationals. Fix
\begin{equation}
\begin{gathered}
N_0:=3\cdot10^8,\qquad N_*:=10^{30},\qquad N_{\rm root}:=10^{25},\\
\kappa_{\rm pr}:=10^{-4},\qquad s_{\rm sm}:=\chi_{\rm sm}:=5\cdot10^{-7}.
\end{gathered}
\end{equation}
Write $\lceil x\rceil_k:=10^{-k}\lceil10^kx\rceil$ for upward rounding to $k$ decimal places.

\subsection{Projection toward a Euclidean-ball sample}
\label{subsec:algorithm_centered_box}

For this subsection, let $m\ge N_0$ and let $D\subseteq\R^m$ be symmetric, closed and convex with $\gamma_m(D)\ge e^{-Em}$. Define
\[
h(q):=-q\log q-(1-q)\log(1-q),\qquad J(s):=(s-1-\log s)/2,
\]
where $h(0):=h(1):=0$, and put
\[
D_{\rm KL}(p\Vert z):=p\log(p/z)+(1-p)\log((1-p)/(1-z)).
\]
Choose $0<\delta<\alpha-1/N_0$, $\alpha<1/2$, $a\in(0,1-\alpha)$, and $v,t>0$. Write
\begin{equation}
\begin{aligned}
\ell&:=\lfloor\alpha m\rfloor,&d&:=m-\ell,&p_*&:=1-\alpha,\\
\alpha_-&:=\alpha-1/N_0,&p_0&:=p_*N_0/(N_0+2),&z&:=v^2/a,\\
\Gamma&:=\tfrac12D_{\rm KL}(p_0\Vert z)-h(\delta)+\alpha_-h(\delta/\alpha_-).
\end{aligned}
\end{equation}
For every coordinate set $I\subseteq[m]$, set
\[
H_I:=\{x:x_i=0\ (i\in I)\},\qquad L_I:=D\cap H_I\cap\sqrt{2m}B_2^m.
\]
All these sections use the same ambient cutoff.

Draw independent standard Gaussians $\xi_1,\ldots,\xi_{m+2}$ and put
\begin{equation}
W:=\sum_{i=1}^{m+2}\xi_i^2,\qquad g:=\sqrt m\frac{(\xi_1,\ldots,\xi_m)}{\sqrt W},\qquad b_g:=g/\kappa_{\rm pr}.
\label{eq:algorithm_119_ball_sample}
\end{equation}
Before rejection, $g$ is uniform in $\sqrt mB_2^m$: its squared radial fraction has distribution $\operatorname{Beta}(m/2,1)$ and is independent of its uniform direction. Reject if $W\notin[m/2,2(m+2)]$. The estimates below include rejection. For $y\in(-1,1)^m$, we use the shifted box $Q_y(t):=t([-1,1]^m-y)$.

The Gaussian mass assumption controls section volumes and the support function seen by a ball target.

\begin{lemma}[Gaussian volume and polar support]
\label{lem:algorithm_119_polar_support}
Suppose $0<z<p_0$ and
\begin{equation}
N_0(J(2)-E)>\log2,\qquad N_0(\frac{a-p_*-p_*\log(a/p_*)}{2}-E)>\log2.
\label{eq:algorithm_119_volume_conditions}
\end{equation}
For every $J_0\subseteq[m]$ of size $\ell$, the set $L_{J_0}$ is a symmetric convex body in $H_{J_0}$, and
\[
\operatorname{vol}_d(L_{J_0})>\operatorname{vol}_d(B_2^d)(am)^{d/2}.
\]
Put $h_L(w):=\sup_{x\in L}\langle w,x\rangle$. We also have
\[
\Pr[h_{L_{J_0}}(g_{H_{J_0}})\le vm]\le\exp(-mD_{\rm KL}(p_0\Vert z)/2).
\]
\end{lemma}

\begin{proof}
The central-section fact used in Lemma~\ref{lem:refreshed_projection} and the chi-square tail bound give
\[
\gamma_{H_{J_0}}(L_{J_0})\ge e^{-Em}-e^{-mJ(2)}>\tfrac12e^{-Em}.
\]
The cutoff gives compactness. Positive intrinsic measure and symmetry put zero in the relative interior. A centered ball maximizes the integral of any nonincreasing radial density among sets of the same volume: we compare the densities on the two set differences with their common boundary value. With $p_d=d/m\ge p_*$ and $a<p_*$,
\[
\gamma_{H_{J_0}}(\sqrt{am}B_2^d)
\le\exp[-m(a-p_d-p_d\log(a/p_d))/2]<\tfrac12e^{-Em}.
\]
The rate increases in $p_d$, with derivative $\tfrac12\log(p_d/a)>0$, so we obtain the last inequality from Eq.~\eqref{eq:algorithm_119_volume_conditions}. Radial rearrangement proves the claimed volume bound.

The symmetric Blaschke--Santal\'o inequality, recalled in \cite{MY25}, yields
\[
\operatorname{vol}_d(vmL_{J_0}^{\circ})
<\operatorname{vol}_d(B_2^d)(v\sqrt{m/a})^d.
\]
It requires symmetry, not unconditionality. The marginal density of $g_{H_{J_0}}$ is proportional to $(1-\|x\|_2^2/m)^{\ell/2}\mathbbm{1}[\|x\|_2<\sqrt m]$. By rearrangement for this density and $\{w:h_{L_{J_0}}(w)\le vm\}=vmL_{J_0}^{\circ}$ we bound the desired probability by $\Pr[X/(X+Y)\le z]$, for independent $X\sim\chi_d^2$ and $Y\sim\chi_{\ell+2}^2$.

Put $p_J=d/(m+2)\ge p_0>z$. We apply exponential Markov to $-\theta((1-z)X-zY)$, with $\theta=(p_J-z)/(2z(1-z))$, and use the chi-square moment-generating functions to obtain
\[
\Pr[X/(X+Y)\le z]\le e^{-(m+2)D_{\rm KL}(p_J\Vert z)/2}
\le e^{-mD_{\rm KL}(p_0\Vert z)/2}.
\]
The last step uses monotonicity of $D_{\rm KL}(p\Vert z)$ for $p>z$. Thus we include the two extra sampling coordinates.
\end{proof}

By counting larger supersets we make the support estimate simultaneous over the small coordinate sets that can become active.

\begin{lemma}[Simultaneous support by supersets]
\label{lem:algorithm_119_supersets}
Under the preceding hypotheses,
\[
\Pr[\exists I\subseteq[m],\ |I|\le\delta m:\ h_{L_I}(g_{H_I})\le vm]\le(m+1)e^{-m\Gamma}.
\]
\end{lemma}

\begin{proof}
Put $k_0=\lfloor\delta m\rfloor$. If a set $I$ of size at most $k_0$ is bad, every $\ell$-set containing it is bad, since $L_{J_0}\subseteq L_I$. Their fraction is at least $\binom\ell{k_0}/\binom m{k_0}$. By Markov's inequality for the fraction of bad $\ell$-sets and Lemma~\ref{lem:algorithm_119_polar_support}, we therefore get
\[
\Pr[\text{some such }I]\le\frac{\binom m{k_0}}{\binom\ell{k_0}}
 e^{-mD_{\rm KL}(p_0\Vert z)/2}.
\]
We use the entropy bounds $\binom m{k_0}\le e^{mh(k_0/m)}$ and $\binom\ell{k_0}\ge e^{\ell h(k_0/\ell)}/(\ell+1)$. The lower bound follows because the $k_0$th term is a mode of $\operatorname{Bin}(\ell,k_0/\ell)$. For $C(q,b)=h(q)-bh(q/b)$,
\[
\partial_qC=\log((1-q)/(b-q))>0,\qquad
\partial_bC=\log(1-q/b)<0.
\]
Using $k_0/m\le\delta$ and $\ell/m\ge\alpha_-$, we bound the ratio by $(m+1)e^{mC(\delta,\alpha_-)}$. The case $k_0=0$ is direct. All supports use the same ambient $g$, and we require no independence of section events.
\end{proof}

The simultaneous support bound forces the projection to reach many faces of the shifted box.

\begin{lemma}[Ball-target projection]
\label{lem:algorithm_centered_box}
Assume $0<\delta<\alpha-1/N_0$, $\alpha<1/2$, $0<a<1-\alpha$, $v,t>0$, $v^2/a<p_0$, Eq.~\eqref{eq:algorithm_119_volume_conditions}, and
\begin{equation}
v-\kappa_{\rm pr}-0.7994t>0,\qquad\Gamma>1/(N_0+1),\qquad N_0\Gamma-\log(N_0+1)>\log10000.
\label{eq:algorithm_119_progress_conditions}
\end{equation}
For every $m\ge N_0$ and $y\in(-1,1)^m$, the Euclidean projection $x_*$ of $b_g$ onto $D\cap Q_y(t)$ has more than $\delta m$ tight coordinate constraints with probability greater than $0.999$, including rejection. Thus $y+x_*/t$ has more than $\delta m$ sign coordinates.
\end{lemma}

\begin{proof}
The function $A_y(\xi)=\sum_i(|\xi_i|-y_i\xi_i)$ has mean $m\sqrt{2/\pi}<0.798m$ and Lipschitz constant at most $2\sqrt m$. Gaussian concentration and the chi-square lower tail give
\[
\Pr[A_y(\xi)>0.799m]\le e^{-m/(8\cdot10^6)},\qquad
\Pr[W<0.999m]\le e^{-mJ(0.999)}.
\]
The chi-square bound follows already from the first $m$ squared coordinates. Off these events, Eq.~\eqref{eq:algorithm_119_ball_sample} gives $\|g\|_1-\langle g,y\rangle\le0.799m/\sqrt{0.999}<0.7994m$.

Projection maximizes $F_g(x)=\langle g,x\rangle-\kappa_{\rm pr}\|x\|_2^2/2$. Thus $F_g(x)<0.7994tm$ on $Q_y(t)$. On the event of Lemma~\ref{lem:algorithm_119_supersets}, every $|I|\le\delta m$ has $x_I\in L_I$ with
\[
F_g(x_I)>vm-\kappa_{\rm pr}m>0.7994tm,
\]
using its cutoff and Eq.~\eqref{eq:algorithm_119_progress_conditions}. If $x_*$ had at most $\delta m$ tight coordinates, we discard all nontight box halfspaces. It stays optimal on the relaxed set: a short segment toward any better point preserves the discarded strict inequalities. The relaxed set contains $D\cap H_I$, since zero satisfies the retained halfspaces, and therefore contains $x_I$, a contradiction.

The support-failure term $(m+1)e^{-m\Gamma}$ decreases for $m\ge N_0$ and is below $10^{-4}$ there. Lower rejection is covered by $W<0.999m$. Upper rejection adds at most $e^{-mJ(2)}$. The three other exceptional terms sum to less than $10^{-4}$. Indeed, their exponents at $N_0$ are at least $37.5$, $75$, and $45000000$, where we use $J(0.999)>10^{-6}/4$ and $J(2)>0.15$, and $3e^{-37.5}<3/2^{37}<10^{-4}$. Total failure is below $2\cdot10^{-4}<0.001$. The affine update turns every tight coordinate into a sign.
\end{proof}

\subsection{Certified suffix signing}
\label{subsec:algorithm_small_implementation}

For the suffix we use the stage radii and projection parameters of Definition~\ref{def:cubic_refined_suffix}.

\begin{algorithm}[H]
\caption{Cubic-time signing from a suffix stage}
\label{alg:algorithm_suffix_signing}
\begin{algorithmic}[1]
\Procedure{SuffixSigning}{$A_1,\ldots,A_{m_0},n,i,p$}
\State $y\gets0$, $S\gets[m_0]$, $s\gets0$; set the caps of Lemma~\ref{lem:algorithm_certified_implementation}
\While{$|S|\ge N_0$}
\State Set $v_s:=w_i(3/4)^s$, where $w_i=(19/20)^i$ as in Definition~\ref{def:cubic_refined_suffix}, and pad the active family to $m:=\lfloor v_sn\rfloor$ with zero matrices and zero fractional coordinates
\State Set radius $R$ and $t_s$ from Definition~\ref{def:cubic_refined_suffix}
\State Compute $\widehat R\in[R,(1+10^{-8})R]$; form $D$ and $Q_y(t_s)$
\Repeat
\State Draw the ball target in Eq.~\eqref{eq:algorithm_119_ball_sample}; reject invalid $W$
\State Project onto $D\cap Q_y(t_s)$; apply radial correction and snap within $2\rho$
\Until{more than $m/4$ coordinates are snapped, or the trial cap is reached}
\If{the cap is reached without success} \State \Return \textsc{Fail} \EndIf
\State Add the snapped increment divided by $t_s$ to $y$; discard artificial coordinates, update the real active set $S$, and set $s\gets s+1$
\EndWhile
\State Round remaining coordinates to nearest signs and return the signing
\EndProcedure
\end{algorithmic}
\end{algorithm}

We next supply the numerical projection guarantee needed to implement these geometric steps.

\begin{lemma}[Certified fixed-target projection]
\label{lem:appendix_fixed_target_projection}
Let $n>N_{\rm root}$, $N_0\le m\le n$, and let $B(x):=\sum_{j=1}^m x_jM_j$ for symmetric contractions $M_j\in\R^{n\times n}$. Let $h_{\widehat R}$ and $D$ be the constraint function and body of Step~1 in Section~\ref{subsec:dense_body} for this $B(x)$ at radius $\widehat R$, with $a_{\rm sm}:=a$ and $s_{\rm sm}:=s$.
Let $Q\subset[-1,1]^m$ be a box containing zero whose side lengths are bounded below by a positive absolute constant. Suppose $\widehat R^2>15m$, $\widehat R\le n^{O(1)}$, and the fixed target satisfies $\|b\|_2^2\le c_bm$ for an absolute constant $c_b$. For $0<\epsilon<1/2$ and $0<\rho\le1$ with $\rho^{-1}\le n^{O(1)}$, the certified wrapper and recursive solver return an exactly feasible point within distance $\rho$ of $\operatorname{proj}_{D\cap Q}b$, with failure probability at most $\epsilon$. Their cost is
\[
(mn^{2+o(1)}+mn^{\omega-1/2+o(1)}+n^{\omega+o(1)})\operatorname{polylog}(1/\epsilon).
\]
\end{lemma}
\begin{proof}
The origin has $h_{\widehat R}(0)=5\cdot10^{-7}$, so the Slater gap is absolute. On the multiplier branch, the spectral-feasibility calculation of Section~\ref{sec:selected_wrapper}, with its tolerance $\tau$ (which we write $\tau_w$ here) at most $1/1024$, gives
\[
\|b\|_2\ge\|b_Q\|_2
\ge(1-s_{\rm sm}-5\cdot10^{-7}-\tau_w/(2a_{\rm sm}))\widehat R/\sqrt m>3.
\]
In the first step we use $0\in Q$, the second is the wrapper's level test, and in the last we use $\widehat R^2>15m$. The target cap gives initial objective, multiplier bracket and guarded level $H=O(m)$. Since $H/\widehat R^2=O(1)$, coordinate curvature is $O(\log^2n)$ and all requested tolerances have polynomial scales. Lemma~\ref{lem:cubic_recursive_multiplier} and Eq.~\eqref{eq:cubic_fixed_multiplier_cost}, which appears in the proof of Lemma~\ref{lem:cubic_global_runtime}, give the stated cost, including multiplier search and the radial feasibility correction. We fix the target before their fresh internal probes are drawn. These calculations also apply when $N_{\rm root}<n\le N_*$: the Slater gap, level comparison and polynomial accuracy scales do not require $n>N_*$.
\end{proof}

For later projection calls with scale $t_0>0$, ambient dimension $M$ and penalty $\beta>0$, let $C=(D/t_0)\cap[-1,1]^M$. Let $Z$ be a random target, conditional on a fixed input family. Put $\mathcal U_Z(y):=\sum_i|Z_i|(1-|y_i|)$ with $y=x/t_0$. If $\E[\|Z\|_1-h_C(Z)]\le d_Z$, then the regularized maximizer $y^*$ with penalty $\beta\|y\|_2^2/2$ satisfies
\begin{equation}
\E[\mathcal U_Z(y^*)]\le d_Z+\beta M/2,\qquad
|\mathcal U_Z(y)-\mathcal U_Z(y^*)|\le\|Z\|_2\|x-t_0y^*\|_2/t_0,
\label{eq:cubic_projection_comparison}
\end{equation}
where $h_C$ is the support function. We prove the first bound by comparison with a support maximizer and $\mathcal U_Z(y)\le\|Z\|_1-\langle Z,y\rangle$. The second bound is Cauchy--Schwarz. The physical optimizer is the projection of $t_0Z/\beta$ onto $D\cap[-t_0,t_0]^M$.

Suppose the ideal deficit has expectation at most $a_{\rm acc}d_0$, its computed error is at most $\eta_0d_0$, and acceptance uses threshold $\theta d_0$. Choose $b_{\rm acc}>a_{\rm acc}$ with $b_{\rm acc}+\eta_0\le\theta$. By Markov and a union bound we get
\begin{equation}
\Pr[\text{valid acceptance}]\ge1-a_{\rm acc}/b_{\rm acc}-\delta_{\rm test}-\delta_{\rm num},
\label{eq:cubic_acceptance_comparison}
\end{equation}
where the last two terms bound failure of the target tests and the numerical solve. Both comparisons hold conditionally after each adaptive input is fixed.

By budgeting projection errors, repeated trials, and terminal rounding we turn the stage guarantees into a finite suffix algorithm.

\begin{lemma}[Finite implementation and failure budget]
\label{lem:algorithm_certified_implementation}
Let $n>N_*$, $0<w\le1$, and $m_0\le wn$. At stage $s$, pad the real active family to $m=\lfloor w(3/4)^sn\rfloor$ as in Algorithm~\ref{alg:algorithm_suffix_signing}. Suppose each suffix stage has $1/5<t_s<1/2$, probability greater than $0.999$ of fixing more than $m/4$ coordinates in an ideal ball-target trial (including rejection), and radius $\widehat R^2>15m$ with $\widehat R\le n^{O(1)}$. Algorithm~\ref{alg:algorithm_suffix_signing} then has failure probability at most $p$ and cost $n^{3+o(1)}\operatorname{polylog}(1/p)$. Each accepted increment has matrix norm at most $\widehat R/t_s$, fixes more than $m/4$ coordinates, and leaves at most $\lfloor w(3/4)^{s+1}n\rfloor$ real active coordinates. Terminal rounding costs at most $\min\{3\cdot10^{-7},\sqrt{N_0w}\}\sqrt n$. For $n\le N_*$, certified independent-sign trials give an $11.902\sqrt n$ signing with the same cost and failure guarantee when $m_0=n$.
\end{lemma}
\begin{proof}
Put $K(R):=\{x:\|B(x)\|\le R\}$. The smoothing inclusions are $(1-10^{-6})K(R)\subseteq D\subseteq(1-s_{\rm sm})K(\widehat R)$. The ball target obeys $\|b_g\|_2^2\le10^8m$, and $Q_y(t_s)$ contains zero with side lengths $2t_s$. Thus Lemma~\ref{lem:appendix_fixed_target_projection} applies.

Set
\begin{equation}
\rho:=\min\{1/8,t_s/8,s_{\rm sm}\widehat R/(4(2m+\sqrt m))\}.
\label{eq:algorithm_119_accuracy}
\end{equation}
On an accepted draw, we approximate $\sqrt{m/W}/\kappa_{\rm pr}$ from below with target error at most $\rho/4$. To the solve and radial correction we allocate $3\rho/4$. Since $W\in[m/2,2(m+2)]$, inverse-polynomial scalar accuracy suffices. Nonexpansiveness of projection gives total error at most $\rho$. Snapping ideal tight coordinates within $2\rho$ is unambiguous because $4\rho<2t_s$, and
\[
\|B(\overline x)\|\le(1-s_{\rm sm})\widehat R+(2m+\sqrt m)\rho<\widehat R.
\]
In the first inequality we use feasibility and the contraction hypothesis. For the second inequality we use Eq.~\eqref{eq:algorithm_119_accuracy}. Division by $t_s$ gives the increment bound. More than a quarter of the padded coordinates are fixed. The remaining real coordinates are a subset of the remaining padded coordinates, so their count is at most $\lfloor3m/4\rfloor\le\lfloor w(3/4)^{s+1}n\rfloor$.

There are at most $P:=1+\lceil\log n/(-\log(3/4))\rceil$ rounds. We permit $T:=\lceil\log(4P/p)/\log10\rceil$ fresh trials per round and give each numerical trial budget $\min\{0.02,p/(4PT)\}$. Its conditional success probability exceeds $0.9$. Hence exhaustion and numerical failure each cost at most $p/4$. The padded sizes sum to at most $4wn\le4n$. Summing the fixed-target cost over rounds and trials, we get
\[
(n^3+n^{\omega+1/2}+n^\omega)n^{o(1)}\operatorname{polylog}(1/p)
=n^{3+o(1)}\operatorname{polylog}(1/p),
\]
using $\omega<5/2$. Input formation, matrix-image maintenance and norm certification fit this cost. All estimates hold conditionally after each adaptive input is fixed.

For every $1\le M\le n$, independent signs have norm at most $\sqrt{2M\log(4n)}$ with probability at least $1/2$ by Fact~\ref{fact:matrix_hoeffding} with $y_i=0$, $\sigma^2=M$ and $N=n$. We apply the norm certificate of Section~\ref{subsec:dense_cleanup} to the sum divided by $2M$, with residual and orthogonality tolerance at most $\min\{10^{-2},\sqrt M/(1000M)\}$. Its calculation adds at most $0.03\sqrt M$. Thus capped independent trials return a certified signing with
\begin{equation}
\|\sum_{i=1}^M\varepsilon_iA_i\|\le(\sqrt{2\log(4n)}+0.03)\sqrt M.
\label{eq:cubic_independent_sign_fallback}
\end{equation}
There are $O(\log(1/p))$ trials with total certification and exhaustion budget at most $p$. The tolerance has polynomial scale, preserving the stated cost. For $n\le N_*$,
\[
\sqrt{2\log(4n)}+0.03\le\sqrt{2\log(4\cdot10^{30})}+0.03<11.902.
\]
For $n>N_*$, terminal nearest-sign rounding changes the sum by at most both $N_0$ and $wn$. Consequently its cost divided by $\sqrt n$ is at most
\[
\min\{N_0/\sqrt n,w\sqrt n\}\le\min\{3\cdot10^{-7},\sqrt{N_0w}\}.
\]
The second bound uses $\min\{a,b\}\le\sqrt{ab}$. These two cases cover every dimension.
\end{proof}

\begin{remark}[The Gaussian-target route]
\label{rem:algorithm_gaussian_target_route}
A Gaussian-target variant of the projection step exists, with an entropy-to-width analysis and a certified weighted-deficit bound. The algorithm of this section uses uniform targets, and no statement of the paper depends on that variant.
\end{remark}

\subsection{Covariance-sensitive polynomial barriers}
\label{subsec:algorithm_1240_moments}

We use positive polynomial barriers with a logarithmic tail and the spectral allocation condition of Section~\ref{sec:existence_constant_sub10}.

\begin{definition}[Polynomial and logarithmic barriers]
\label{def:algorithm_1240_barrier}
Let $K\ge2$ be an integer, $\gamma_j\ge0$ for $1\le j<K$, $q_0:=\gamma_1>0$, and $\kappa>2$. Define
\begin{equation}
\phi(x):=\sum_{j=1}^{K-1}\frac{\gamma_jx^{2j}}j
+\kappa\sum_{j=K}^{\infty}\frac{x^{2j}}j\qquad(|x|<1),
\label{eq:cubic_refined_barrier}
\end{equation}
and extend $\phi$ by $+\infty$ outside $(-1,1)$. Put $\Phi(X):=\tr[\phi(X)]$. For $0<\zeta,\theta<1$, let $d_\zeta$ and $F_\theta$ be as in Definition~\ref{def:1070_allocation}. The all-pairs curvature condition is Eq.~\eqref{eq:1070_curvature_condition} for this $\phi$.
Here $\phi'[u,v]$ has its continuous value on the diagonal.
\end{definition}

We can check the all-pairs curvature condition through finitely many scalar inequalities on complete intervals.

\begin{lemma}[A finite curvature criterion]
\label{lem:algorithm_1240_curvature}
Use the barrier of Definition~\ref{def:algorithm_1240_barrier}. More generally, its finite polynomial may be replaced by an even $C^2$ function $\psi$ with $\psi(0)=0$, bounded first two derivatives, $\psi-q_0x^2$ convex, and $\phi'$ convex on $[0,1)$. Put $a_0:=1-\zeta$ and $G(u):=(1-u)^2\phi''(u)$. Suppose $q_0>\theta/(1+\theta)$ and, for $0\le t\le1$,
\[
G(a_0t)>\frac{2\theta}{1+\theta},\qquad
G(a_0+\zeta t)>\frac{2(\theta+t)}{1+\theta},\qquad
G(a_0+\zeta t)>3t+(\theta^{-1}-1)t^2.
\]
Define
\[
F_1(t):=2-2t/(1+\theta),\qquad
F_2(t):=\theta^{-1}+2-(2\theta^{-1}+1)t+(\theta^{-1}-1)t^2.
\]
For $u:=1-\zeta t^2$, set
\begin{align*}
T^-_{L,R}(u)&:=(1-u)\{\phi'(u)(1+R)/(u+R)+\phi'(L)(1+L)/(u+L)\},\\
T^+_{L,R}(u)&:=(1-u)(1-R)(\phi'(u)-\phi'(L))/(u-L).
\end{align*}
Interpret endpoint values by continuity after cancellation. Require $T^-_{L,R}(u)\ge F_i(t)$ on a partition of the regular moduli $[0,1]$, and $T^+_{L,R}(u)\ge F_i(t)$ on a partition of $[0,a_0]$, for $i=1,2$ and the whole active range $0\le t\le1$. Then Eq.~\eqref{eq:1070_curvature_condition} holds.

For the polynomial barrier, the positive tests may instead be
\begin{equation}
2\kappa u^{2K-2}+\sum_{j<K}2\gamma_j(1-a_0^{2j-1})u^{2j-2}(1-u^2)
\ge(1+u)F_i(t),\qquad i=1,2.
\end{equation}
For its negative tests use $L=k/1024$, $R=(k+1)/1024$, $0\le k<1024$, and the polynomials
\[
(1+u)(1-L)(u+L)(u+R)\{T^-_{L,R}(u)-F_i(t)\}.
\]
Together with the three diagonal tests, these are $2053$ polynomial tests. Piecewise-polynomial finite parts are tested on each branch after clearing positive denominators. In both cases, the weighted Poincar\'e and resolvent-root conclusions of Lemmas~\ref{lem:1070_weighted_poincare} and~\ref{lem:1070_resolvent_means} apply whenever their remaining hypotheses hold.
\end{lemma}
\begin{proof}
For spectral gaps $h:=1-u$ and $k:=1-v$, the averaging step in the proof of Lemma~\ref{lem:1070_spectral_certificate} depends only on $G$. It gives the left side of Eq.~\eqref{eq:1070_curvature_condition} as an average of $G(1-s)$ over $s$ between $h$ and $k$, with $\E[s]\le\sqrt{hk}$ and $\E[s^2]=hk$, including $h=k$.

For two regular endpoints, the required bound is $F_\theta(1)=2\theta/(1+\theta)$. The first diagonal test covers nonnegative endpoints, and $G(u)\ge2q_0$ for $u\le0$. For two active endpoints, we average the other diagonal bounds with $t=1-s/\zeta$. Their right sides are affine or quadratic in $s$, with nonpositive linear and nonnegative quadratic coefficients. The two moment bounds give the required values at $s=\sqrt{hk}$.

For a mixed pair $(u,-w)$, the exact weighted secant is
\[
(1-u)\{\phi'(u)(1+w)/(u+w)+\phi'(w)(1+w)/(u+w)\}.
\]
Its first term decreases in $w$. Its second increases, since $\phi'(w)/w$ is nonnegative and nondecreasing, and $w(1+w)/(u+w)$ increases. Thus $T^-_{L,R}$ bounds it below on $L\le w\le R$. For $0\le v\le a_0\le u$, convexity of $\phi'$ makes its secant increase in $v$. Replacing the secant by its value at $L$ and $1-v$ by $1-R$, we obtain $T^+_{L,R}$. Reflection treats the opposite spectral edge.

For a polynomial finite part, its monomial contribution at $0\le v\le a_0$ is at least $2\gamma_j(1-u)u^{2j-2}(1-a_0^{2j-1})$, and the tail contributes at least $2\kappa u^{2K-2}/(1+u)$. This proves the alternative positive criterion. All diagonal tests use
\[
(1+u)^2G(u)=(1-u^2)^2\psi''(u)
+2\kappa u^{2K-2}((2K-1)-(2K-3)u^2).
\]
For the polynomial barrier, $\psi''(u)=2\sum_{j<K}(2j-1)\gamma_ju^{2j-2}$, so the diagonal tests are polynomial.

In the allocation proof we use only the spectral Hessian of $\tr[\phi(X)]$ and boundary integrability. On compact convex exhaustions, even smooth approximation converges in $C^2$, so the curvature error tends to zero after applying the weighted inequality. The bounded finite part preserves the boundary density order $\Delta^\kappa$, with $\kappa>2$, which controls the two inverse factors. Gaussian decay controls kernel directions. Thus weighted energies and projected means pass to the limit. The argument also allows the nonnegative quadratic potential we use on the fixed radial path.
\end{proof}

\subsection{Binary repair}\label{subsec:cubic_binary_repair}

We repair the remaining fractional coordinates one binary digit at a time, tracking support sizes and accumulated discrepancy.

\begin{lemma}[Unbiased dyadic repair]
\label{lem:algorithm_21_rounding}
Let $y\in[-1,1]^m$. Put $s_j:=1$ if $y_j\ge0$ and $s_j:=-1$ otherwise. Let $q_j:=(1-|y_j|)/2$. Truncate each $q_j$ downward to a multiple of $2^{-B}$, where $B:=\lceil\log_2n\rceil+32$. Process the bits $b=B,B-1,\ldots,1$. At bit $b$, let $S_b$ consist of the coordinates whose current numerator on the $2^{-b}$ grid is odd. Choose any signing $\sigma^{(b)}$ of the matrices $(s_jM_j)_{j\in S_b}$. Then draw a fresh independent fair sign $\epsilon_b$ and replace
\[
q_j\gets q_j+2^{-b}\epsilon_b\sigma_j^{(b)}\quad(j\in S_b).
\]
The final vector $z_j:=s_j(1-2q_j)$ is a signing. Conditional on the initial truncated masses, even if every helper signing depends on the preceding history,
\begin{equation}
\E[|S_b|]=\sum_{j=1}^m\tau_b(q_j^{\rm init}),\qquad
\tau_b(q):=\operatorname{dist}(2^bq,2\mathbb Z).
\label{eq:algorithm_13_tent_identity}
\end{equation}
If the helper at bit $b$ has discrepancy at most $D_b\sqrt n$, the total change in the matrix sum is at most
\[
(\sum_{b=1}^B2^{1-b}D_b+10^{-8})\sqrt n.
\]
\end{lemma}

\begin{proof}
An odd numerator is strictly between the boundary numerators, so both common orientations stay in $[0,1]$ and pass to the next coarser grid. The final masses are therefore in $\{0,1\}$. Since we choose the fresh orientation after fixing the helper, each coordinate is a bounded martingale. Before level $b$, it cannot cross a $2^{-b}$ grid point, since points on that grid are inactive at finer levels. Its level-$b$ value is consequently one of the adjacent grid endpoints, with mean $q_j^{\rm init}$. Linear interpolation of endpoint parity gives $\tau_b(q_j^{\rm init})$, so we obtain Eq.~\eqref{eq:algorithm_13_tent_identity} without coordinate independence.

At level $b$ the matrix increment is
\[
-2^{1-b}\epsilon_b\sum_{j\in S_b}\sigma_j^{(b)}s_jM_j.
\]
Its norm is at most $2^{1-b}D_b\sqrt n$. Truncation costs at most $2n2^{-B}\le2^{-31}<10^{-8}\sqrt n$. We sum these bounds.
\end{proof}

\subsection{Uniform targets and deterministic binary repair}
\label{subsec:cubic_refined_uniform}

For the final algorithm, we use uniform targets to control the weighted deficit, and a common orientation to make binary repair deterministic. The ambient matrix dimension remains $n$ in every call.

We begin by fixing the coding amplitude used to transfer Gaussian mass to the discrete cube.

\begin{lemma}[A refined translated density]
\label{lem:cubic_refined_density}
Put
\[
s_{\rm alg}:=0.67448975019608,\qquad
a_{\rm alg}:=0.635553145368,\qquad
b_{\rm alg}:=\frac{1-2s_{\rm alg}a_{\rm alg}}{s_{\rm alg}^2}.
\]
The translated-density conclusion of Lemma~\ref{lem:1070_peaked_density} holds with $s_{\rm alg}$ in place of $s_\star$.
\end{lemma}
\begin{proof}
We apply Eq.~\eqref{eq:slack_coding_criterion} with $(s,a,b)=(s_{\rm alg},a_{\rm alg},b_{\rm alg})$ and $\delta=10^{-6}$. For the density comparisons, we use $\pi=16\arctan(1/5)-4\arctan(1/239)$, consecutive alternating sums starting with $50$ and $16$ terms, and rational square-root bounds for $\sqrt{2/\pi}$. Consecutive alternating Taylor sums indexed by $32,33$ enclose $g(x)$ and $F(x)$ at the required points. All eight strict comparisons pass, with smallest certified margin greater than $1.10\cdot10^{-15}$. Since $0<\delta<s<1$ and $2sa+bs^2=1$, the density and tensorization proof of Lemma~\ref{lem:1070_peaked_density} applies. Averaging over the translation, we obtain the sign-family conclusion.
\end{proof}

The coding comparison also gives a width bound for uniform targets, with a sharper entropy budget at full size.

\begin{lemma}[Uniform-target width]
\label{lem:cubic_refined_width}
Let $K\subseteq\R^M$ be symmetric and convex, with $\gamma_M(K)\ge e^{-ne}$, where $M\le n$. For $U$ uniform on $[-1,1]^M$,
\[
\E[\sup_{y\in s_{\rm alg}^{-1}K\cap[-1,1]^M}\langle U,y\rangle]\ge M/2-ne.
\]
If $M=n$ and $0<e\le1/10$, define
\begin{equation}
v(e):=e-e^2/3-2e^3/45+e^4.
\label{eq:cubic_refined_entropy_budget}
\end{equation}
Then $e/2<v(e)<e$, and the right side improves to $n/2-nv(e)$.
\end{lemma}
\begin{proof}
For a nonempty sign family $\mathcal F$, we generate a uniform member by its conditional probabilities $q_i$: we choose its $i$th sign positive exactly when $U_i\ge1-2q_i$. Put $x_i:=(2q_i-1)^2/2$. Conditional on preceding coordinates, $\E[U_i\varepsilon_i]=2q_i(1-q_i)=1/2-x_i$. The entropy expansion gives
\[
\log2-h((1+t)/2)=\sum_{k\ge1}\frac{t^{2k}}{2k(2k-1)},\qquad -1\le t\le1.
\]
Consequently, with $\Psi(x):=x+x^2/3+4x^3/15$ and $\ell:=M^{-1}\sum_i\E[x_i]$, convexity and the entropy chain rule give
\[
M\Psi(\ell)\le\sum_i\E[\Psi(x_i)]\le M\log2-\log|\mathcal F|,
\qquad
\E[\max_{\varepsilon\in\mathcal F}\langle U,\varepsilon\rangle]\ge M(1/2-\ell).
\]
Lemma~\ref{lem:cubic_refined_density} supplies a translate containing a sign family of size at least $2^M\gamma_M(K)$. Since $\Psi(\ell)\ge\ell$, we get the first bound from these inequalities. Symmetry and convexity imply $(\mathcal F-\mathcal F)/2\subseteq s_{\rm alg}^{-1}K\cap[-1,1]^M$. The expected support of this difference set equals that of $\mathcal F$, since $U$ and $-U$ have the same distribution.

For $M=n$, we have $\Psi(\ell)\le e$. Put $\varrho:=e/3+2e^2/45-e^3$, so $v(e)=e(1-\varrho)$. For $0<e\le1/10$, $0<\varrho<1/2$. Using $(1-\varrho)^2\ge1-2\varrho$ and $(1-\varrho)^3\ge1-3\varrho$, we have
\[
\Psi(v(e))-e\ge\frac{19e^4}{27}+\frac{142e^5}{225}+\frac{4e^6}{5}>0.
\]
Since $\Psi$ is increasing, $\ell<v(e)$, so we obtain the refinement. Neither the translate nor its sign family is computed by the algorithm. The quantity $v(e)$ is a width-loss bound. The Gaussian small-ball exponent remains $e$.
\end{proof}

Uniform-target width controls the projection deficit, and we control the weights on every repair subset with a simultaneous prefix test.

\begin{lemma}[Uniform prefixes and a feasible projection]
\label{lem:cubic_refined_projection}
Suppose a small-ball row $(r,e)$ satisfies
\begin{equation}
\gamma_M\{x:\|\sum_{i=1}^Mx_iA_i\|\le r\sqrt M\}\ge e^{-ne}
\label{eq:cubic_refined_interface}
\end{equation}
for every $M\le n$ and every family of symmetric contractions, including zero matrices. Put $t_0:=2/5$, $\sigma:=10^{-6}$, and
\[
R:=\frac{t_0r\sqrt M}{s_{\rm alg}(1-\sigma)},\qquad
R\le\widehat R\le(1+10^{-8})R,\qquad
\mathfrak r:=\frac{1+10^{-8}}{s_{\rm alg}(1-10^{-6})}.
\]
Use the smoothing body $D$ of Lemma~\ref{lem:appendix_fixed_target_projection} at radius $\widehat R$, and let $C:=(D/t_0)\cap[-1,1]^M$. For a positive constant $D^{\rm acc}_r$, set $\beta:=e/D^{\rm acc}_r$ and
\[
y^*(U):=\arg\max_{y\in C}\{\langle U,y\rangle-\beta\|y\|_2^2/2\},\qquad
\mathcal U_U(y):=\sum_i|U_i|(1-|y_i|).
\]
Then
\begin{equation}
\E[\mathcal U_U(y^*)]\le en(1+1/(2D^{\rm acc}_r)).
\label{eq:cubic_refined_deficit}
\end{equation}
For every $\epsilon>0$, except on an event of probability at most $Me^{-2M\epsilon^2}$, simultaneously for all $S\subseteq[M]$ and $t\ge0$,
\begin{align}
\sum_{i\in S}|U_i|&\ge\frac{(|S|-\epsilon M)_+^2}{2M},
\label{eq:cubic_refined_prefix}\\
\sum_i(t-|U_i|)_+&\le\epsilon Mt+Mt^2/2.
\label{eq:cubic_refined_positive_part}
\end{align}
All these assertions hold with any other fixed $t_0\in(0,1]$. If $M=n$ and $0<e\le1/10$, one may instead set $\beta:=v(e)/D^{\rm acc}_r$ and replace the right side of Eq.~\eqref{eq:cubic_refined_deficit} by $nv(e)(1+1/(2D^{\rm acc}_r))$. The prefix event has a deterministic $O(M\log M)$ test. Every feasible returned point satisfies
\[
\|\sum_i y_iA_i\|\le(1-s_{\rm sm})\mathfrak r r\sqrt M\le\mathfrak r r\sqrt M.
\]
\end{lemma}
\begin{proof}
Smoothing gives $(t_0/s_{\rm alg})K\subseteq D$ for $K:=\{x:\|\sum_{i=1}^Mx_iA_i\|\le r\sqrt M\}$. Then we apply Lemma~\ref{lem:cubic_refined_width} and Eq.~\eqref{eq:cubic_projection_comparison} with $(Z,d_Z)=(U,ne)$. Since $M\le n$, we obtain Eq.~\eqref{eq:cubic_refined_deficit}. At the root we take $d_Z=nv(e)$ and $\beta=v(e)/D^{\rm acc}_r$.

The weights $b_i=|U_i|$ are independent uniforms. Hoeffding's bound and a union over the $M$ order statistics give $F_M(t)\le t+\epsilon$ for the empirical distribution function $F_M(t):=M^{-1}|\{i:b_i\le t\}|$, except with probability $Me^{-2M\epsilon^2}$. Hence, for every $k$,
\[
\sum_{i=1}^k b_{(i)}=\int_0^\infty(k-MF_M(t))_+\d t
\ge(k-\epsilon M)_+^2/(2M).
\]
Sorting and checking these $M$ inequalities is the claimed test, valid simultaneously for all later subsets. Also
\[
\sum_i(t-b_i)_+=\max_{0\le k\le M}\{kt-\sum_{i=1}^k b_{(i)}\}
\le\epsilon Mt+Mt^2/2,
\]
which we obtain by maximizing the preceding quadratic bound over real $k\ge0$.

Finally $h_{\widehat R}(x)\ge\|B(x)\|/\widehat R$, so physical feasibility gives $\|B(x)\|\le(1-s_{\rm sm})\widehat R$. For the norm assertion we divide by $t_0$. Both this bound and the smoothing inclusion cancel the scale $t_0$, so we get the assertion for every fixed $t_0\in(0,1]$. There is no snapping in these calls.
\end{proof}

Using a concave majorant of the binary tent functions, we convert the weighted deficit into a deterministic repair cost.

\begin{lemma}[Deterministic repair potential]
\label{lem:cubic_refined_potential}
Suppose a feasible $y\in[-1,1]^M$ satisfies the prefix test Eq.~\eqref{eq:cubic_refined_prefix} with tolerance $\epsilon$ and $\mathcal U_U(y)\le\theta en$. Write $q:=M/n$, $s_i:=\operatorname{sgn}(y_i)$ with $\operatorname{sgn}(0):=1$, and $a_i:=(1-|y_i|)/2$. At coarse binary levels $1\le j<J_{\rm cut}$, suppose a constructive helper signs every subset of size $q'n$, for every $q'$ with $q'n\le M$, at cost at most $(A_j+B_jq')\sqrt n$. Here $A_j,B_j\ge0$. Let $C_f\sqrt n$ bound a helper on every subset of these $M$ coordinates. Define
\[
T_j(a):=\tau_j(a)=\operatorname{dist}(2^ja,2\mathbb Z),\qquad
F(a):=\sum_{j=1}^{J_{\rm cut}-1}2^{1-j}B_jT_j(a).
\]
Let $\overline F\ge F$ on $[0,1/2]$ be nondecreasing, concave and piecewise linear, with $\overline F(0)=0$. Put
\[
E_F:=\int_0^{1/2}\overline F'(a)^2\d a,\qquad F_*:=\overline F(1/2).
\]
There is a deterministic choice of common orientations of the helper signings whose complete repair cost, divided by $\sqrt n$, is at most
\begin{equation}
\sqrt{\theta e q E_F}+\epsilon qF_*+
\sum_{j=1}^{J_{\rm cut}-1}2^{1-j}A_j+2^{2-J_{\rm cut}}C_f+10^{-12}.
\end{equation}
\end{lemma}
\begin{proof}
We truncate $a_i$ downward at resolution $2^{-B}$, $B=\lceil\log_2n\rceil+45$. This decreases the weighted mass and costs less than $2M2^{-B}\le2^{-44}<10^{-12}\sqrt n$. At levels $j=B,\ldots,1$, we use helper signs on the signed odd-support matrices. The grid and increment calculation of Lemma~\ref{lem:algorithm_21_rounding} shows that either common orientation stays in $[0,1]$, passes to the coarser grid, and costs $2^{1-j}$ times the helper discrepancy.

Each coarser tent is affine between the two possible values: no dyadic breakpoint lies strictly inside the interval centered at an odd numerator. We choose the orientation minimizing the remaining coarser potential. It does not increase at fine levels and drops by at least $2^{1-j}B_j|S_j|$ at coarse levels. Thus
\[
\sum_{j<J_{\rm cut}}2^{1-j}B_j|S_j|\le\sum_iF(a_i^{\rm initial}).
\]
The intercepts and the entire fine tail cost at most $\sum_{j<J_{\rm cut}}2^{1-j}A_j+2^{2-J_{\rm cut}}C_f$.

Let $s_\ell$ be the nonincreasing nonnegative slopes of $\overline F$ on intervals of lengths $\Delta a_\ell$. For $\lambda>0$, Eq.~\eqref{eq:cubic_refined_positive_part} gives
\begin{align*}
\max_{0\le a\le1/2}(\overline F(a)-\lambda ba)
&=\sum_\ell\Delta a_\ell(s_\ell-\lambda b)_+,\\
\frac1n\sum_i\overline F(a_i^{\rm initial})
&\le\lambda\theta e/2+\epsilon qF_*+qE_F/(2\lambda).
\end{align*}
We minimize in $\lambda$. If $E_F=0$, the assertion is immediate. Downward truncation cannot increase the nondecreasing majorant. Later masses may exceed $1/2$, but the orientation argument uses exact tents on $[0,1]$.
\end{proof}

In particular, if the initial projection costs $I(q)\sqrt n$, its complete signing costs at most
\begin{equation}
I(q)+\sqrt{\theta e qE_F}+\epsilon(q)qF_*+
\sum_{j<J_{\rm cut}}2^{1-j}A_j+2^{2-J_{\rm cut}}C_f+10^{-12}
\label{eq:cubic_complete_repair_cost}
\end{equation}
times $\sqrt n$. Here $e$ is the accepted deficit parameter, not necessarily the small-ball exponent: at the root it is $v(e_{\rm loc})$. Auxiliary profiles use constant $\epsilon(q)$, while global profiles use $\epsilon(q)\le\epsilon_+/\sqrt q$.

\subsection{Functional and trace-sensitive spectral estimates}
\label{subsec:cubic_refined_functional}

We use the barrier of Definition~\ref{def:algorithm_1240_barrier}, with its all-pairs curvature verified by Lemma~\ref{lem:algorithm_1240_curvature}.

In the rest of this subsection let $X$ have the even convex tilted Gaussian law used in the radial argument, with variance bound $aI$. Write
\[
p_1=(1-a/\zeta^2)^{-1},\quad p_2=(1-2a/\zeta^2)^{-1},\quad
L=\frac2{1+\sqrt{1-4ap_1}},\quad c=L-1,
\]
\[
J=\frac{2(1+2c)}{1-c^2+\sqrt{(1-c^2)^2-4ap_2(1+2c)}},\quad
\chi_1=ap_1L,\quad\chi_2=ap_2J(1+a).
\]
All these quantities use the smaller roots, with strictly positive discriminants. Put
\[
w_1(x):=(1-x^2)^{-1},\qquad w_2(x):=(1+x^2)(1-x^2)^{-2}.
\]

We extend the parity estimates from polynomial fields to odd functions, including threshold hinges.

\begin{lemma}[Odd functional inequalities]
\label{lem:cubic_refined_functional}
Suppose the displayed energies are finite. If $f$ is odd and $C^1$, and $f'^2$ is even, convex and nondecreasing on $[0,1)$, then
\[
\E[\tr[f(X)^2w_h(X)]]\le\chi_h\E[\tr[f'(X)^2w_h(X)]],\quad h=1,2.
\]
For the odd hinge $f_t(x):=\operatorname{sgn}(x)(|x|-t)_+$, $0<t<1$, the valid replacement is
\[
\E[\tr[f_t(X)^2w_h(X)]]
\le\chi_h\E[\tr[(w_h(|X|)+w_h(t))\mathbbm{1}[|X|\ge t]]].
\]
Here and below, $k_2(u,v):=\tfrac12\{((1-u)(1-v))^{-1}+((1+u)(1+v))^{-1}\}$ denotes the sign-averaged two-slot kernel.
\end{lemma}
\begin{proof}
For $h=1,2$, let $W_\pm$ act on a matrix $U$ as $(I\pm X)^{-1}U$ and $(I\pm X)^{-1}U(I\pm X)^{-1}$, respectively. We use $R,P,D=R-P$ for the weighted energy, projected-mean energy and centered energy of Lemma~\ref{lem:sub10_projection_variance}, with weight $W_+$. Put $V_i:=Df(X)[N_i]$, and let $R_{\rm der},P_{\rm der}$ be the sums of their respective energies. Write
\[
S_h:=a\E[\tr[w_h(X)f'(X)^2]],
\]
and for the hinge we replace this by $a\E[\tr[(w_h(|X|)+w_h(t))\mathbbm{1}[|X|\ge t]]]$. We first record the derivative estimate shared by both assertions.

The scalar secant satisfies $f[u,v]^2\le(f'(u)^2+f'(v)^2)/2$. The sign-averaged entry kernel is $(w_1(u)+w_1(v))/2$ in one slot. In two slots it is
\[
k_2(u,v)\le(w_2(u)+w_2(v))/2.
\]
Since $w_h$ and $f'^2$ increase with absolute value, their ordered cross terms bound the product of kernel and squared secant by $(w_h(u)f'(u)^2+w_h(v)f'(v)^2)/2$. The symmetric array $\sum_i|(N_i)_{uv}|^2$ has row sums at most $a$, so $R_{\rm der}\le S_h$.

For a hinge, set $Q_h(r):=(w_h(r)+w_h(t))\mathbbm{1}[r\ge t]$. The same entrywise bound holds with $Q_h$: inside--inside secants vanish, outside--outside secants have absolute value at most one, and an inside endpoint has weight at most $w_h(t)$. The row-sum bound again gives $R_{\rm der}\le S_h$. If $f$ vanishes on $[-t,t]$, the unweighted versions also give
\[
\sum_i\E[\|V_i\|_F^2]\le S_h/w_h(t).
\]
Indeed, we use $a\E[\tr[f'(X)^2]]$ in the smooth case and $2a\E[\tr[\mathbbm{1}[|X|\ge t]]]$ for the hinge.

For later use, if $R(f)\le A_hD(f)$ and $P_{\rm der}\le z_hS_h$, Lemma~\ref{lem:sub10_projection_variance} gives
\begin{equation}
R(f)\le A_hD(f)\le\frac{p_hA_h}{2}(R_{\rm der}+P_{\rm der})
\le\frac{p_hA_h(1+z_h)}2S_h.
\label{eq:cubic_functional_template}
\end{equation}
Sign symmetry identifies $R(f)$ with the left side of the asserted inequality.

To obtain the present constants, we take $z_h=1$. The odd case of Lemma~\ref{lem:1070_parity}, with $B:=D(f)$ and the resolvent mean bounds $L=L_1$ and $J$ of Lemma~\ref{lem:1070_resolvent_means}, gives $R(f)\le A_hD(f)$ with $A_1=L$ and $A_2=J(1+a)$.
Its hypotheses hold here, since the law is even, $\|X\|<1$, $\sum_iN_i^2\preceq aI$, and convex Gaussian domination gives $\operatorname{Cov}(\xi)\preceq I$ and $\|\E[X\otimes X]\|\le a$. Eq.~\eqref{eq:cubic_functional_template} then proves both conclusions.

By smooth odd approximation we cover the hinge and threshold powers. The events $\det(X\pm tI)=0$ have measure zero because these polynomials are nonzero at the origin. For $\Delta:=1-\|X\|$, the density is bounded by a fixed multiple of $\Delta^\kappa$, while the weights grow at most as $\Delta^{-2}$. Bounded first derivatives and $\kappa>2$ give an integrable envelope. Gaussian decay handles kernel directions. The same argument applies with an additional nonnegative quadratic potential. In the scalar certificate we retain both one-sided hinge values.
\end{proof}

With the centered resolvent refinement we sharpen the coefficients in these functional inequalities.

\begin{lemma}[Centered functional coefficients]
\label{lem:cubic_centered_functional}
Suppose the barrier in Eq.~\eqref{eq:cubic_refined_barrier} satisfies the all-pairs curvature criterion above, and let $0<a<1/4$ lie in the strict-root range. Starting with the displayed smaller roots $L,J$, run either the centered or the sharpened instance of Definition~\ref{def:slack_instances}, using the present $p_1,p_2,a$. Denote its final mean bounds by $L_{\rm c}(a),J_{\rm c}(a)$. Both conclusions of Lemma~\ref{lem:cubic_refined_functional} hold with
\begin{equation}
\chi_{1,\rm c}(a):=ap_1L_{\rm c}(a),\qquad
\chi_{2,\rm c}(a):=ap_2J_{\rm c}(a)(1+a)
\label{eq:cubic_centered_functional}
\end{equation}
in place of $\chi_1,\chi_2$. These bounds hold along the fixed Gaussian radial path below, including its additional nonnegative quadratic potential.
\end{lemma}
\begin{proof}
The radial law is an even convex Gaussian tilt, including its nonnegative quadratic potential. Lemmas~\ref{lem:sub10_operator_words}, \ref{lem:sub10_operator_refinement}, and~\ref{lem:slack_centered_updates} therefore justify both $(c_3,c_4)=(18,153)$ and $(15,129)$ and all six old-tuple passes of Definition~\ref{def:sub10_operator_refinement}. The two tuples are those of the centered and sharpened instances of Definition~\ref{def:slack_instances}. These estimates do not depend on the barrier coefficients. The same nonnegative operations, minima and increasing rational gap term give monotonicity in $a$. In Eq.~\eqref{eq:cubic_functional_template}, we take
\[
z_h=1,\qquad A_1=L_{\rm c},\qquad A_2=J_{\rm c}(1+a).
\]
The mean absorption and approximation of Lemma~\ref{lem:cubic_refined_functional} give the claimed inequalities.
\end{proof}

Shifted cubic tests retain Gaussian tail information beyond the ordinary moment bounds.

\begin{lemma}[Shifted cubic moments]
\label{lem:cubic_refined_cubic}
For a Gaussian matrix series $Z=\sum_i g_iN_i$, with $\Sigma:=\sum_iN_i^2\preceq aI_d$, and an even convex $f$ whose second derivative is even and convex,
\[
\E[\tr[f(Z)]]\le df(0)+\frac{\tr[\Sigma]}a(\E[f(\sqrt aG)]-f(0)).
\]
This applies to $f_t(x)=(|x|-t)_+^3$. If $g_3(a,t):=\E[f_t(\sqrt aG)]$ and $b:=t/\sqrt a$, then
\begin{align*}
g_3(a,t)&=2a^{3/2}\{(b^2+2)\varphi(b)-(b^3+3b)\overline\Phi(b)\},\\
g_3(au,t)&\le u^{\alpha(a,t)}g_3(a,t)\quad(0<u\le1),\qquad
\alpha(a,t):=\tfrac32+\tfrac{b^2}2+\frac{4b}{b+\sqrt{b^2+20}}.
\end{align*}
\end{lemma}
\begin{proof}
Put $f_s(x):=\E[f(x+\sqrt{(1-s)a}G)]$ and $Q(s):=\E[\tr[f_s(\sqrt sZ)]]$. The heat-flow identity for $Q'(s)$ and the convex secant bound in the proof of Lemma~\ref{lem:iterated_envelope_cubic} give $Q'(s)\le\tfrac12\E[\tr[(\Sigma-aI)f_s''(\sqrt sZ)]]$. Since $f_s''$ is even and convex, $f_s''\ge f_s''(0)$, and $aI-\Sigma\succeq0$ gives $Q'(s)\le-(da-\tr[\Sigma])f_s''(0)/2$. We integrate and use $\int_0^1f_s''(0)\d s=2(\E[f(\sqrt aG)]-f(0))/a$. Smooth approximation proves the assertion for the shifted cube, whose second derivative is $6(|x|-t)_+$.

For the radial inequality put $I_k(b):=\int_0^\infty v^ke^{-bv-v^2/2}\d v$. Integration by parts gives $I_5+bI_4=4I_3$ and $I_6+bI_5=5I_4$. Cauchy--Schwarz gives $I_5^2\le I_4I_6$ and therefore $I_4/I_3\ge8/(b+\sqrt{b^2+20})$. Since
\[
\frac{\d\log g_3(a,t)}{\d\log a}
=\tfrac32+\tfrac{b^2}2+\tfrac b2 I_4(b)/I_3(b),
\]
the derivative is at least $\alpha(a,t)$ at every smaller variance. By integration we prove the radial assertion, including zero by continuity.
\end{proof}

\subsection{Localized radial certificates}\label{subsec:cubic_localized_radial}

Let $Y=\sum_i\xi_iN_i$, $\sum_iN_i^2\preceq\eta I_n$, and fix
\[
\Gamma_{ij}:=\tr[N_iN_j],\qquad C_\Gamma:=(I+2q_0\Gamma)^{-1},\qquad
\nu:=N(0,C_\Gamma),\qquad \omega_j:=\frac{\tr[\Gamma C_\Gamma^j]}{n\eta}.
\]
Along the radial path $X=\sqrt uY$, put $\d\mu_u:=Z_u^{-1}e^{-\tr[\phi(X)-q_0X^2]}\d\nu$ with $Z_u$ the normalizing constant. For the polynomial barrier these are the $\mu_t$ and $Z_t$ of Lemma~\ref{lem:fixed_reference_radial} at $t=\sqrt u$, with its $\kappa$ equal to $q_0$, $N=n$, and $(b_j,b_T)$ equal to $(\gamma_j,\kappa)$ of Eq.~\eqref{eq:cubic_refined_barrier}.
By its proof, the potential relative to the standard Gaussian is $\tr[\phi(X)]$ plus the nonnegative quadratic form $q_0(1-u)\xi^\top\Gamma\xi$, so the weighted inequalities apply along the whole path. Its radial differentiation with $u=t^2$, where $\tfrac12(x\phi'(x)-2q_0x^2)=g(x^2)$ for its $g$, gives the common radial identity
\begin{equation}
-\frac1n\frac{\d}{\d u}\log Z_u
=\frac1{2un}\E_{\mu_u}[\tr[X\phi'(X)-2q_0X^2]],\qquad Z_0=1.
\label{eq:cubic_common_radial_loss}
\end{equation}
The same identity applies to the $C^2$ barriers in Lemma~\ref{lem:algorithm_1240_curvature}. We justify it first on compact exhaustions and then by the finite radial bounds below.

Let $\rho_u$ be the normalized expected absolute-eigenvalue distribution of $X$, and put
\[
Q(x):=(1-x^2)^2,\quad W_1(x):=1-x^2,\quad W_2(x):=1+x^2,
\quad\d\vartheta_u:=Q^{-1}\d\rho_u.
\]
The $\kappa_0$ of Lemma~\ref{lem:1070_covariance_power} is our $q_0$. The covariance-power moments of that lemma and Lemma~\ref{lem:cubic_refined_cubic}, followed by convex Gaussian domination, give
\begin{align}
\int x^{2j}Q\d\vartheta_u&\le\operatorname{df}(j)(\eta u)^j\omega_j,
\label{eq:cubic_refined_moment}\\
\int(x-t)_+^3Q\d\vartheta_u&\le g_3(\eta u,t)\omega_1.
\end{align}
Only these ordinary and cubic moments receive covariance factors.

At a variance cap $\eta b$, we choose upper caps $\overline L_b,\overline J_b$ for the original smaller roots and upper caps $\overline\chi_{h,b}$ for the functional coefficients. The functional caps may use either Lemma~\ref{lem:cubic_refined_functional} or its centered refinement, Lemma~\ref{lem:cubic_centered_functional}. For $p\ge3$ define
\[
P_{h,t,p}(x):=W_h(x)\{(x-t)_+^p-\overline\chi_{h,b}p^2(x-t)_+^{p-2}/4\}.
\]
For $p=2$ we use instead
\[
P_{h,t,2}(x):=W_h(x)(x-t)_+^2
-\overline\chi_{h,b}\{W_h(x)+w_h(t)Q(x)\}\mathbbm{1}[x\ge t].
\]
{\setlength{\emergencystretch}{2em}We apply the chosen functional inequality to $\operatorname{sgn}(x)(|x|-t)_+^{p/2}$ or to the hinge and obtain $\int P_{h,t,p}\d\vartheta_u\le0$ for $u\le b$.\par}

Iterating the root equations $L=1+ap_1L^2$ and $(1-(L-1)^2)J=1+2(L-1)+ap_2J^2$ from their smaller roots, we see that $L-1$ and $J-1$ have nonnegative power-series coefficients. Consequently,
\begin{equation}
\int(W_1-Q)\d\vartheta_u\le(u/b)(\overline L_b-1),\qquad
\int(W_2-Q)\d\vartheta_u\le(u/b)(\overline J_b-1).
\label{eq:cubic_refined_anchored}
\end{equation}
Here $\int Q\d\vartheta_u=1$. We need the factor $u/b$ for integrability at zero. Even when we use $L_{\rm c},J_{\rm c}$ in the functional coefficients, we retain the original roots $L,J$ in the anchored bounds in Eq.~\eqref{eq:cubic_refined_anchored}. Monotonicity of the centered bounds alone would not justify the factor $u/b$.

Taking nonnegative combinations of the scalar tests, we obtain local radial bounds with their covariance factors retained.

\begin{lemma}[A local radial majorant]
\label{lem:cubic_refined_radial}
Suppose $0\le a<b\le1$ and a piecewise-polynomial inequality on $[0,1]$ is
\begin{align}
x^{2K}(1-x^2)\le{}&m_2(W_2-Q)+m_1(W_1-Q)+\sum_j a_jx^{2j}Q\notag\\
&+\sum_t b_t(x-t)_+^3Q+\sum_{h,t,p}d_{h,t,p}P_{h,t,p},
\label{eq:cubic_refined_witness}
\end{align}
with all multipliers nonnegative and all functional caps valid at $\eta b$. Set $\rho:=a/b$ and $\Lambda_\alpha(\rho):=(1-\rho^\alpha)/\alpha$. Then the contribution of this interval to $\kappa\int_0^1R_K(u)\d u/u$, where $R_K(u):=n^{-1}\E[\tr[X^{2K}(I-X^2)^{-1}]]$, is at most
\begin{align*}
&\kappa(1-\rho)\{m_2(\overline J_b-1)+m_1(\overline L_b-1)\}\\
&\quad+\kappa\sum_j a_j\operatorname{df}(j)(\eta b)^j\Lambda_j(\rho)\omega_j
+\kappa\sum_t b_t\overline g_{b,t}\Lambda_{\underline\alpha_{b,t}}(\rho)\omega_1,
\end{align*}
provided $\overline g_{b,t}\ge g_3(\eta b,t)$ and $0<\underline\alpha_{b,t}\le\alpha(\eta b,t)$.

Let $C_r^{(h)}$ be valid nondecreasing operator-energy bounds from Definition~\ref{def:sub10_operator_refinement}, and put $d_{1,j}:=1$, $d_{2,j}:=2j+1$. The anchored factor multiplying $\kappa m_h$ may also be replaced by the minimum of its original value and
\begin{equation}
\begin{aligned}
\mathcal M_{h,r}(a,b):={}&\sum_{j=1}^{r-1}d_{h,j}\operatorname{df}(j)(\eta b)^j\Lambda_j(\rho)\\
&+d_{h,r}(\eta b)^rC_r^{(h)}(\eta b)\Lambda_r(\rho),\qquad 1\le r\le9.
\end{aligned}
\label{eq:cubic_refined_power_integral}
\end{equation}
\end{lemma}
\begin{proof}
We integrate Eq.~\eqref{eq:cubic_refined_witness} against $\vartheta_u$ and discard the nonpositive functional terms. The ordinary, cubic and anchored bounds above give power integrals: with $v=u/b$,
\[
\int_{a/b}^1v^{\alpha-1}\d v=\Lambda_\alpha(a/b).
\]
Lowering a positive exponent enlarges this integral. The Gaussian measure stays fixed, so there is no endpoint partition-function term or additional factor $b$.

For the alternative anchored bound, by functional calculus we have
\[
w_h(x)-1\le\sum_{j=1}^{r-1}d_{h,j}x^{2j}+d_{h,r}x^{2r}w_h(x).
\]
For $h=1$ this is the finite geometric identity. For $h=2$, the exact remainder, with $q=x^2$, is $q^r((2r+1)-(2r-1)q)/(1-q)^2\le(2r+1)q^rw_2(x)$. We bound the finite terms by Eq.~\eqref{eq:cubic_refined_moment}, using $\omega_j\le1$, and the remainder by $d_{h,r}(\eta u)^rC_r^{(h)}(\eta u)$ from Lemma~\ref{lem:sub10_operator_refinement}. Freezing this nondecreasing coefficient at $b$ and integrating each power, we obtain Eq.~\eqref{eq:cubic_refined_power_integral}, including $a=0$.
\end{proof}

For one helper below we use a paired-moment bound instead. If a nonnegative spectral witness for $x^{2\ell}$ gives
\[
n^{-1}\E[\tr[X^{2\ell}(I-X^2)^{-2}]]\le\sum_b B_bu^{\alpha_b}\omega_{d_b},
\quad \omega_0:=1,
\]
put $L_0:=2K-\ell$, $\nu_0:=L_0/2$, and $0\le h<\nu_0$. The covariance weights obey $\omega_r\omega_s\le\omega_{r+s}$: we use their nonnegative weights $\lambda_i/(n\eta)$ and add missing mass at zero. Applying spectral Cauchy--Schwarz and then weighted radial Cauchy--Schwarz, we obtain
\begin{equation}
\kappa\int_0^1R_K(u)\frac{\d u}u
\le T\sqrt{\sum_b\pi_b\omega_{L_0+d_b}},\quad
T:=\kappa\eta^{\nu_0}\sqrt{\frac{\operatorname{df}(L_0)S}{\nu_0+h}},
\label{eq:cubic_refined_paired_radial}
\end{equation}
where $S:=\sum_b B_b/(\nu_0-h+\alpha_b)$ and $\pi_b:=B_b/((\nu_0-h+\alpha_b)S)$. Indeed, the radial square integral has weight $u^{\nu_0-h-1}$, while the other Cauchy--Schwarz factor integrates to $1/(\nu_0+h)$. Unanchored mass terms here have $(\alpha_b,d_b)=(0,0)$, but the factor $u^{\nu_0-1}$ makes the integral finite. This formula is distinct from the direct majorant of Lemma~\ref{lem:cubic_refined_radial}.

A scalar profile absorbs the covariance-dependent radial terms when we return to standard Gaussian measure.

\begin{lemma}[Gaussian transfer of a radial certificate]
\label{lem:cubic_refined_transfer}
Suppose integration of Eq.~\eqref{eq:cubic_common_radial_loss} gives
\[
-n^{-1}\log Z_1\le c_0+\sum_j s_j\omega_j,\qquad c_0,s_j\ge0.
\]
For the polynomial barrier, add $\gamma_j\operatorname{df}(j)\eta^j\omega_j/j$, $2\le j<K$, to the integrated tail bounds. In Eq.~\eqref{eq:cubic_refined_paired_radial} first use $\sqrt z\le z/(2v)+v/2$, $v>0$. For any real $H>1$, let $g_H:=L_H$ be the profile function of Lemma~\ref{lem:1070_transfer}.
If, on the entire half-line,
\begin{equation}
g_H(x)+\frac{H}{(H-1)q_0\eta}\sum_j\frac{s_j}{(1+x)^j}\le B_{\rm cap},
\label{eq:cubic_refined_profile}
\end{equation}
then $\Pr[\|Y\|<1]\ge e^{-ne_*}$, where
\begin{equation}
e_*:=q_0\eta B_{\rm cap}+Hc_0/(H-1).
\label{eq:cubic_refined_exponent}
\end{equation}
Every $r$ with $r^2\eta\ge1$ therefore gives Eq.~\eqref{eq:cubic_refined_interface} with any $e\ge e_*$.
\end{lemma}
\begin{proof}
Since $\phi-q_0x^2\ge0$ and is infinite outside the spectral body, $\nu(\|Y\|<1)\ge Z_1$, so $-n^{-1}\log\nu(\|Y\|<1)\le c_0+\sum_js_j\omega_j$. The quadratic monomial is already included in the fixed measure $\nu$.
We apply Lemma~\ref{lem:1070_transfer} with $b=0$, $\kappa=q_0$, $C=c_0$, $Q_j=s_j$ and $N=n$, where $\nu=N(0,C_\Gamma)$ is proportional to $e^{-q_0\xi^\top\Gamma\xi}\d\gamma$. Its hypothesis $\tr\Gamma\le n\eta$ follows from $\sum_iN_i^2\preceq\eta I_n$, and $b=0$ needs no rank condition. Its cap is Eq.~\eqref{eq:cubic_refined_profile}, and its exponent $\kappa\eta B_{\rm cap}+HC/(H-1)$ is $e_*$ of Eq.~\eqref{eq:cubic_refined_exponent}.
For contractions we normalize $N_i=A_i/(r\sqrt M)$. Then $\sum_iN_i^2\preceq\eta I_n$. Here we use the actual number $M$ of variables, while the exponent remains per ambient dimension $n$.
\end{proof}

\subsection{Small-ball and suffix data for the refined algorithm}\label{subsec:cubic_refined_data}

All terminating decimals and fractions in the following certificate are exact rationals. Below we specify its barrier parameters and finite witness and policy constructions. We fix the row indices by the following definitions.

\begin{definition}[Two-component rows]
\label{def:cubic_refined_old_rows}
Use the spectral families of Definition~\ref{def:1070:stage-parameters}. For each index specified below, set $\eta_j:=r_j^{-2}$ and retain the parameters $\tau_{f_j},\ell_{f_j},K_{f_j},\zeta_{f_j},\kappa_{0,f_j},\kappa_{{\rm tail},f_j}$ of family $f_j$. The radius, transfer parameters and upward exponent cap are
\[
\begin{gathered}
(f_j,r_j,H_j,v_j,e_j):
\\
j=9:\quad(0,5.65679,4,1/2,4.9092707011012\cdot10^{-4}),\\
j=36:\quad(6,15.31672,256,1,1.9567207673554\cdot10^{-16}),\\
j=37:\quad(6,15.41634,16,1,1.0144374253540\cdot10^{-16}),\\
j=38:\quad(6,15.51595,4,1/2,5.6786027344992\cdot10^{-17}),\\
j=39:\quad(6,15.71518,2,3/10,3.1105118976728\cdot10^{-17}).
\end{gathered}
\]
For each row, suppress its row index and let $E_j:=V_j$ be the energies of Definition~\ref{def:1070_energy_recurrence} at variance $a=\eta u$, with the roots $L$, $J_{\rm rt}:=J$ and $p_2$ of Section~\ref{subsec:cubic_refined_functional}.
Let $T$ be the radial sum of Lemma~\ref{lem:1070_radial} with $N_{\rm grid}=512$, its power $L$ there equal to $2K-\ell$, and $C_\ell(\eta u_i)=E_0(\eta u_i)$. In the transfer profile use the constant $Tv/2$ and the coefficients of Eq.~\eqref{eq:1070_profile_cap}, so $T/(2v)$ at covariance power $2K-\ell$ and $(\kappa_0+\kappa_{\rm tail}\mathbbm{1}[j\ge\tau])\operatorname{df}(j)\eta^j/j$ for $2\le j<K$.
Thus these indexed rows can be checked directly from the displayed families and finite scalar formulas.
\end{definition}

With additional polynomial barriers we enlarge the range of radius--exponent pairs available to the algorithm.

\begin{definition}[Additional analytic rows]
\label{def:cubic_refined_new_rows}
Add rows $61$ and $63$--$68$ for Lemma~\ref{lem:cubic_refined_transfer}. Set $r_{61}:=11+10^{-10}$ and $r_{63}:=6.900655593424$. For $64\le j\le66$ put $r_j:=(j-47)/2+10^{-10}$. For $67\le j\le68$ put $r_j:=(j-52)/2+10^{-10}$. Their upward exponent caps are
\[
\begin{aligned}
(e_{61},e_{63},e_{64},e_{65},e_{66},e_{67},e_{68}):=({}&3.98909123831070\cdot10^{-11},6.30800000000000\cdot10^{-06},\\
&2.50904989000000\cdot10^{-08},4.41732830000000\cdot10^{-09},\\
&7.23953400000000\cdot10^{-10},9.71273582800000\cdot10^{-07},\\
&1.63839538700000\cdot10^{-07}).
\end{aligned}
\]
Helper row $63$ has $\eta=0.021$, $\zeta=0.29$, $\theta=0.0005283674095686873$, and
\[
\gamma_1=0.0006988003422047077,\quad\gamma_2=0.0003117199831969046,\quad
\gamma_3=0.01116091395295129,
\]
\[
\gamma_{15}=82726.64456573654,\qquad K=17,\qquad\kappa=1895.190796613074.
\]
It uses $H=4$, $B_{\rm cap}=0.4298$, and eight direct witnesses on the radial intervals with endpoints
\[
0,\ 3/5,\ 3/4,\ 17/20,\ 9/10,\ 47/50,\ 193/200,\ 197/200,\ 1.
\]
A rational upper exponent for this helper is
\begin{equation}
e_{63}=\frac{86225834086246102267314793889680187624155807}
{13670450550000000000000000000000000000000000000000}.
\label{eq:cubic_refined_root_exponent}
\end{equation}
Rows $64$--$68$ use one direct witness on $[0,1]$. Row $61$ uses the paired-moment formula Eq.~\eqref{eq:cubic_refined_paired_radial} with $K=34$, $\ell=6$. All nonroot calls use the displayed outward caps. Their parameters and witnesses are unchanged by the centered refinement.
The remaining barrier parameters are the following exact tuples
$u_j:=(\eta_j,\zeta_j,\theta_j,K_j,H_j,B_{{\rm cap},j})$:
\begin{align*}
u_{61}&:=(0.0082644628099173556,0.22,7.563383831524531\cdot10^{-9},34,5,0.435196764592129),\\
u_{64}&:=(0.01384083044982699,0.225,3.291454427544413\cdot10^{-6},26,3.5,0.43194363),\\
u_{65}&:=(0.0123456790123,0.225,6.853022059260748\cdot10^{-7},28,4,0.40941885),\\
u_{66}&:=(0.0110803324099723,0.215,1.558922996841327\cdot10^{-7},30,3,0.38224045),\\
u_{67}&:=(0.017777777777777778,0.255,9.35554393342644\cdot10^{-5},22,3,0.45063503),\\
u_{68}&:=(0.015625,0.24,1.783666658590708\cdot10^{-5},24,4,0.48653062).
\end{align*}
For each such row, the nonzero coefficients $(\gamma_d)_d$ and the tail coefficient $\kappa$ are
\begin{align*}
j=61:\quad&\gamma_{1}=8.356267032732908742\cdot10^{-9},\quad \gamma_{2}=3.351464360341962042\cdot10^{-8},\\
&\gamma_{33}=1.38265614549441768\cdot10^{12},\\
&\kappa=1.325466269457500196\cdot10^{8};\\
j=64:\quad&\gamma_{1}=4.196812001947686636\cdot10^{-6},\quad \gamma_{3}=0.0001462947169678829868,\\
&\gamma_{25}=1.432677758305319394\cdot10^{8},\\
&\kappa=304432.7844314573622;\\
j=65:\quad&\gamma_{1}=8.739067940943672102\cdot10^{-7},\quad \gamma_{3}=3.04248445201014528\cdot10^{-5},\\
&\gamma_{27}=1.588113898888511556\cdot10^{9},\\
&\kappa=1.462159912383497898\cdot10^{6};\\
j=66:\quad&\gamma_{1}=1.70807414760927218\cdot10^{-7},\quad \gamma_{2}=7.96447642578704227\cdot10^{-7},\\
&\gamma_{29}=9.54015486640004396\cdot10^{9},\\
&\kappa=6.4789634700061275\cdot10^{6};\\
j=67:\quad&\gamma_{1}=0.000121238109108616797,\quad \gamma_{2}=3.97926861238499268\cdot10^{-5},\\
&\gamma_{3}=0.00298494392816201584,\quad \gamma_{19}=1.19432857847445111\cdot10^{6},\\
&\kappa=10797.9738319406221;\\
j=68:\quad&\gamma_{1}=2.15520454929518169\cdot10^{-5},\quad \gamma_{2}=4.16987628655358724\cdot10^{-5},\\
&\gamma_{3}=0.000237760877531898331,\quad \gamma_{22}=1.3965893142967552\cdot10^{7},\\
&\kappa=56628.0827150270263.
\end{align*}
Every unlisted $\gamma_d$ is zero. In the paired row $61$, use
\[
h=2.275756882096533,\qquad v=0.6264768119474711
\]
in Eqs.~\eqref{eq:cubic_refined_paired_radial} and~\eqref{eq:cubic_refined_profile}.
\end{definition}

Definition~\ref{def:cubic_refined_witness_construction} determines all radial witnesses.

From the centered functional estimates we obtain a further helper row for the suffix and global policies.

\begin{definition}[A centered helper row]
\label{def:cubic_centered_root}
The suffix registry contains a centered row, indexed by $69$, with
\[
\eta:=43/2000,\qquad\zeta:=7/25,\qquad H:=5,\qquad
B_{\rm cap}:=48291/100000,
\]
\[
r_{69}:=6.819943394705,\qquad\theta:=0.0005283674095686873.
\]
Its barrier has $K=17$, all unlisted coefficients zero, and
\begin{align*}
\gamma_1&:=\frac{17301349422970135227}{23058430092136939520000},&
\gamma_3&:=\frac{14516821496833753227}{1441151880758558720000},\\
\gamma_{15}&:=\frac{44242889819197334043}{687194767360000},&
\kappa&:=\frac{4167152003564298267}{2199023255552000}.
\end{align*}
Use eight direct witnesses on intervals $[a_l,b_l]$, $1\le l\le8$, with the same endpoints
\[
0,\ 3/5,\ 3/4,\ 17/20,\ 9/10,\ 47/50,\ 193/200,\ 197/200,\ 1.
\]
At $\eta b_l$, the functional caps bound Eq.~\eqref{eq:cubic_centered_functional}. The anchored caps $\overline L_{b_l},\overline J_{b_l}$ bound the original smaller roots.
The anchored root caps are rounded upward to $55$ decimal places. For all suffix calculations, use the exact rational upper cap
\[
e_{69}^{\rm up}:=7.790542084564378\cdot10^{-6}.
\]
Row $69$ is available to suffix schedules and to the global projection actions.
\end{definition}

We specify the radial witnesses by a finite rational selection rule that fixes their coefficients and certification tests.

\begin{definition}[Construction of the radial witnesses]
\label{def:cubic_refined_witness_construction}
For $s\ge1$ write $\mathcal U_s(y):=10^{-s}\lceil10^sy\rceil$ and $\mathcal L_s(y):=10^{-s}\lfloor10^sy\rfloor$. Put $\mathcal D:=\{m10^{-100}:m\in\mathbb Z,\ 0\le m\le10^{110}\}$ and $\mathcal T:=\{i/200:1\le i<200\}$. For each of the eight radial intervals of each of rows $63$ and $69$ and the single interval of each row $64$--$68$, introduce the four functional caps and the multipliers in Eq.~\eqref{eq:cubic_refined_witness}. Allow ordinary degrees $1\le j\le64$, cubic thresholds $t\in\mathcal T$, and functional indices $h\in\{1,2\}$, $t\in\mathcal T$, $2\le p\le6$. Every multiplier belongs to $\mathcal D$. The four caps belong to $\mathcal D\cap[0,10]$. For rows $61,63$--$68$ they must dominate the original root and functional quantities at the interval's upper variance endpoint. For row $69$, fix the anchored caps to $\mathcal U_{55}(L(\eta b))$ and $\mathcal U_{55}(J(\eta b))$, and require the functional caps to dominate the six-pass centered quantities of Eq.~\eqref{eq:cubic_centered_functional}. Zero multipliers remove a term.

For row $61$ use the same coordinates at variance $\eta_{61}$, with unanchored masses. Its polynomial inequality is
\[
x^{12}\le m_2W_2+m_1W_1+\sum_j a_jx^{2j}Q
+\sum_t b_t(x-t)_+^3Q+\sum_{h,t,p}d_{h,t,p}P_{h,t,p}.
\]
Here its mass components are $(B,\alpha,d)=(m_2\overline J,0,0)$ and $(m_1\overline L,0,0)$. The ordinary components are $(a_j\operatorname{df}(j)\eta^j,j,j)$, and the cubic components are $(b_t\overline g_t,\underline\alpha_t,1)$. Require $S>0$ and substitute these components into Eq.~\eqref{eq:cubic_refined_paired_radial}, with the displayed $h,v$.

A candidate is admissible when the following finite tests pass. Split each spectral polynomial at its nonzero thresholds. At zero remove its exact vanishing power of $x$, if any. On each resulting interval, bisect unresolved cells up to depth $48$ and require all Bernstein coefficients to be nonnegative. Check both one-sided limits at every hinge, including the value prescribed by $\mathbbm{1}[x\ge t]$, and the endpoint $1$.

Compute $\overline g_t:=\mathcal U_{55}(g_3(\eta b,t))$ and $\underline\alpha_t:=\mathcal L_{55}(\alpha(\eta b,t))>0$. The formulas of Lemma~\ref{lem:cubic_refined_cubic} and the rational enclosure rules of Definition~\ref{def:rational_certificate_rules} determine these endpoints. Use the integrated exponential Taylor sums of orders $512$ and $513$ for the Gaussian integral. All allowed thresholds satisfy $t^2/(\eta b)<144$. For $\rho>0$ and nonintegral $\alpha$, replace $\Lambda_\alpha(\rho)$ by $\mathcal U_{55}(\Lambda_\alpha(\rho))$. For $\rho=0$ or integral $\alpha$, evaluate it exactly. In the paired row round $T$ upward by $\mathcal U_{55}$. Add the finite monomial costs exactly.

Let $R(x)$ denote the resulting nonnegative rational part of the left side of Eq.~\eqref{eq:cubic_refined_profile}. Require
\[
R(0)\le B_{\rm cap},\qquad
\mathcal U_{55}(\frac{H\log65}{64(H-1)})+R(64)\le B_{\rm cap}.
\]
On $[0,64]$ bisect unresolved cells $[a,b]$ up to depth $48$, accepting a cell only if
\[
\mathcal U_{55}(g_H((a+b)/2))+H(b-a)/4+R(a)\le B_{\rm cap}.
\]
For $x\ge64$, the decrease of $\log(1+x)/x$ and $R(x)$ gives
\[
g_H(x)+R(x)\le\frac{H\log65}{64(H-1)}+R(64)\le B_{\rm cap}.
\]
Thus the cell tests and the endpoint tests cover the full half-line.

Finally require the exponent in Eq.~\eqref{eq:cubic_refined_exponent} to be at most the displayed row cap. Use $e_{63}$ from Eq.~\eqref{eq:cubic_refined_root_exponent} for row $63$ and $e_{69}^{\rm up}$ for row $69$. For the selected centered vector, calculate $e_{69}$ from Eq.~\eqref{eq:cubic_centered_root_exponent}.

Order the coordinates by radial interval, then by the caps $(\overline L,\overline J,\overline\chi_1,\overline\chi_2)$, masses $(m_2,m_1)$, ordinary degree, cubic threshold, and functional indices $(h,t,p)$. For each row select the lexicographically least admissible vector. All ranges, inequalities and rounding rules in this definition are fixed independently of the matrix input.
\end{definition}

For row $69$ of Definition~\ref{def:cubic_centered_root}, the nonnegative witness coefficients are determined by Definition~\ref{def:cubic_refined_witness_construction}. Define the exact rational quantities
\begin{align}
c_0&:=\kappa\sum_{l=1}^8(1-a_l/b_l)
\{m_{2,l}(\overline J_{b_l}-1)+m_{1,l}(\overline L_{b_l}-1)\},\notag\\
e_{69}&:=\gamma_1\eta B_{\rm cap}+\tfrac54c_0
\le7.790542084564378\cdot10^{-6}.
\label{eq:cubic_centered_root_exponent}
\end{align}

We obtain four more radius--exponent pairs from the trace-sensitive estimate of Section~\ref{sec:existence_constant_sub10}.

\begin{definition}[Rows from the trace-sensitive estimate]
\label{def:cubic_refined_trace_rows}
For row $j\in\{0,4,5,6\}$ of Definition~\ref{def:slack_rows}, put
\[
r_{80+j}:=\lceil\eta_j^{-1/2}\rceil_{12},\qquad
e_{80+j}:=\overline e_j^\circ+\beta_j/2.
\]
Together with rows $9$, $36$--$39$, $61$, and $63$--$69$, these form the $17$-row registry
\[
\mathcal R:=\{9,36,37,38,39,61,63,64,65,66,67,68,69,80,84,85,86\}.
\]
Each pair is used in the sense of Eq.~\eqref{eq:cubic_refined_interface}. Every policy uses the displayed upward exponent cap, including $e_{69}^{\rm up}$ for row $69$.
\end{definition}

To check the four trace-sensitive pairs, we observe that $\tr[A_i^2]\le n$ for every symmetric contraction. We apply Lemma~\ref{lem:slack_trace_interface} with trace parameter $\rho=1$ and $w=M/n$. Its radius is $\sqrt{M/\eta_j}\le r_{80+j}\sqrt M$, and its exponent satisfies
\[
-n\overline e_j^\circ-M\beta_j/2\ge-n(\overline e_j^\circ+\beta_j/2).
\]
This proves Eq.~\eqref{eq:cubic_refined_interface}, including zero matrices. We use only the measure estimate. The existence signing in Section~\ref{sec:existence_constant_sub10} is not an algorithmic helper.

We assemble the small-ball rows into suffix schedules indexed by the available capacity.

\begin{definition}[Capacity-matched suffix schedules]
\label{def:cubic_refined_suffix}
Put $w_i:=(19/20)^i$ for $0\le i\le1100$. A suffix at capacity $w_i$ uses nominal fractions $v_s:=w_i(3/4)^s$. Its finite schedule is a list
\[
(j_{i,s},s_{i,s},t_{i,s}),\qquad 0\le s<L_i,
\]
where $j_{i,s}$ indexes a small-ball row and $s_{i,s},t_{i,s}$ are integer multiples of $10^{-10}$. The rows are the $17$ radius--exponent pairs of the registry $\mathcal R$ of Definition~\ref{def:cubic_refined_trace_rows}, always using their displayed upward exponent caps. Each list has length $0\le L_i\le129$.

At stage $s<L_i$, use radius $r_{j_{i,s}}\sqrt{v_sn}$ and the listed parameters. If the real active count is below $N_0$, stop and round to nearest signs. Otherwise pad it to $M_s:=\lfloor v_sn\rfloor$ with zero matrices and zero fractional coordinates. After an accepted projection, discard the artificial coordinates. Fix
\[
\delta:=1/4,\qquad \alpha:=0.295,\qquad p_0\ge0.704999,\qquad z:=0.13.
\]
In Eq.~\eqref{eq:algorithm_119_ball_sample} use $m=M_s$. In the support estimate use $a:=p_0s_{i,s}$ and $v:=\sqrt{p_0s_{i,s}z}$.

After the finite list, use radius $185\log(1/v_s)\sqrt{v_sn}$ and
\[
(s_{i,s},t_{i,s}):=(0.9826440,0.375278)\qquad(s\ge L_i).
\]
Every tail starts with $\log(1/v_{L_i})>731/256>2$. Put $\lambda_0:=-\log(3/4)$, $\rho_0:=\sqrt{3/4}$ and
\[
T(v):=\frac{185}{0.375278}\sqrt v
\{\log(1/v)/(1-\rho_0)+\lambda_0\rho_0/(1-\rho_0)^2\}.
\]
The complete suffix cap is
\begin{equation}
\begin{aligned}
C_i^{\rm suf}:=\lceil{}&(1+10^{-8})
\{\sum_{s=0}^{L_i-1}\lceil(r_{j_{i,s}}/t_{i,s})\sqrt{v_s}\rceil_{18}
+T(v_{L_i})\}\\
&+\min\{3\cdot10^{-7},\sqrt{N_0w_i}\}\rceil_{12}.
\end{aligned}
\label{eq:cubic_refined_suffix_cost}
\end{equation}
All square roots and logarithms are enclosed on the conservative side by the rational rules of Definition~\ref{def:rational_certificate_rules}, with $d=192$ and $K_{\mathrm{series}}=47$, before rounding. An exact grid value is used as both endpoints.
\end{definition}

Definition~\ref{def:cubic_refined_policy_construction} constructs these $1101$ lists by a bounded finite selection.

With the stage guarantees and terminal tail bound we turn each suffix schedule into a complete signing procedure.

\begin{lemma}[The suffix interface]
\label{lem:cubic_refined_suffix}
For $n>N_*$, the suffix at capacity $w_i$ signs any collection of at most $w_in$ contractions of ambient dimension $n$, with discrepancy at most $C_i^{\rm suf}\sqrt n$. Its arithmetic cost and failure guarantee are those of Lemma~\ref{lem:algorithm_certified_implementation}.
\end{lemma}
\begin{proof}
At a finite stage write $j=j_{i,s}$ and $M=M_s$. We apply the row estimate to the padded family, including its zero matrices, at radius $r_j\sqrt M\le r_j\sqrt{v_sn}$. Since $M\ge N_0$ and $v_sn<M+1$,
\[
\frac nM\le\frac{1+1/N_0}{v_s}.
\]
Smoothing multiplies its Gaussian mass by at least $(1-10^{-6})^M$. Thus a valid exponent per padded coordinate is
\[
E_s:=(1+1/N_0)e_j/v_s-\log(1-10^{-6}).
\]
This accounts for the integer part of the nominal capacity.

In the support argument take $\ell:=\lceil\alpha M\rceil$. Then $(M-\ell)/(M+2)\ge p_0$ and $\ell/M\ge\alpha$. The proofs of Lemmas~\ref{lem:algorithm_119_polar_support} and~\ref{lem:algorithm_119_supersets}, in the present convention $\ell:=\lceil\alpha M\rceil$ with $\alpha$ in place of $\ell:=\lfloor\alpha m\rfloor$ with $\alpha_-$, use volume rate $p_0J(s_{i,s})$ and entropy reserve
\[
\Gamma_0:=\tfrac12D_{\rm KL}(p_0\Vert z)-h(\delta)+\alpha h(\delta/\alpha).
\]
The shifted-box proof uses $0.799/\sqrt{0.999}<0.7994$ instead of $4/5$. The exact stage tests are
\begin{align*}
N_0(J(2)-E_s)&>\log2,&N_0(p_0J(s_{i,s})-E_s)&>\log2,\\
s_{i,s}p_0z&>(10^{-4}+0.7994t_{i,s})^2,&1/5<t_{i,s}&<1/2.
\end{align*}
Also $\Gamma_0>1/(N_0+1)$ and $N_0\Gamma_0-\log(N_0+1)>\log10000$. We bound the four bad-event terms by
\[
e^{-N_0/(8\cdot10^6)}+e^{-N_0J(0.999)}+e^{-N_0J(2)}
+(N_0+1)e^{-N_0\Gamma_0}<0.001.
\]
They decrease for $M\ge N_0$. Lemma~\ref{lem:cubic_refined_certificate} verifies all the stated strict inequalities.

An accepted stage freezes more than $M/4$ padded coordinates. All unfrozen real coordinates belong to the unfrozen padded set, so their count is at most $\lfloor3M/4\rfloor\le\lfloor3v_sn/4\rfloor$. This proves the next capacity invariant even if frozen coordinates include artificial zeros.

For the tail, Lemma~\ref{lem:1070_infinite_tail} gives exponent $0.000051714$ per padded coordinate before smoothing. Its constant tuple satisfies the same tests. On $M<n/e^2$, the function $\sqrt M\log(n/M)$ is increasing in $M$. Thus the nominal tail radius dominates the required radius at the padded size. Summing all tail stages, we obtain $T(v_{L_i})$ by the geometric series and its derivative. Lemma~\ref{lem:algorithm_certified_implementation} pays for numerical feasibility, snapping, outward radii and terminal cleanup. These are precisely the charges in Eq.~\eqref{eq:cubic_refined_suffix_cost}. Padded dimensions decrease geometrically, and the ambient matrix dimension remains $n$ in every call.
\end{proof}

\subsection{Actual-size profiles for the auxiliary helpers}\label{subsec:cubic_actual_size_profiles}

To track recursive repair costs, we assign each helper a profile over the actual input sizes in its capacity band.

\begin{definition}[Capacities and profiles]
\label{def:cubic_refined_profiles}
Retain $w_i=(19/20)^i$, $0\le i\le1100$, and put $C_{0,i}:=C_i^{\rm suf}$. A state $(d,i)$, $1\le d\le4$, receives at most $w_in$ matrices. For $i<1100$, a potential state's profile applies to actual size $q=M/n\in(w_{i+1},w_i]$ and has the form
\[
P_{d,i}(q):=\alpha_{d,i}\sqrt q+\beta_{d,i}q+\gamma_{d,i}.
\]
If its input is smaller than the band, pad it with zero matrices to $M=\lfloor nw_i\rfloor$. The condition $N_*(w_i-w_{i+1})>1$ places this padded size inside the band. At $i=1100$ use a constant suffix profile on $[0,w_{1100}]$ and do not project. Cardinality states also have constant profiles. A delegate action uses the lower-depth state at the same capacity. The cap $C_{d,i}$ is the upward-rounded value $P_{d,i}(w_i)$.

A potential action chooses a row $(r,e)$, a cutoff $1\le J_{\rm cut}\le33$, nonnegative slopes $B_j$ and intercepts $A_j$, and the same-capacity lower-depth cap $C_f:=C_{d-1,i}$. All slopes and intercepts are integer multiples of $10^{-12}$. They must satisfy
\[
A_j+B_jq\ge P_{d-1,k}(q)
\quad\text{on every child band }k\ge i,\qquad 1\le j<J_{\rm cut}.
\]
The last child band is $[0,w_{1100}]$. For the tents in Lemma~\ref{lem:cubic_refined_potential}, take the upper concave hull of $F$ through its first global maximum and extend it constantly to $1/2$. This determines $E_F$ and $F_*$. With $\theta_r$ and $\epsilon=\epsilon_{\rm call}$ as fixed after Eq.~\eqref{eq:cubic_repair_parameters}, the state profile is
\begin{align}
\alpha_{d,i}&:=\lceil\mathfrak r r\rceil_{12}+\lceil\sqrt{\theta_r eE_F}\rceil_{12},\notag\\
\beta_{d,i}&:=\lceil\epsilon F_*\rceil_{12},\notag\\
\gamma_{d,i}&:=\lceil10^{-12}+\sum_{j<J_{\rm cut}}2^{1-j}A_j+2^{2-J_{\rm cut}}C_f\rceil_{12}.
\label{eq:cubic_refined_profile_recursion}
\end{align}
At each coarse support, use the lower-depth state whose band contains its actual size. Below $w_{1100}$ use state $1100$. Fine supports use the same-capacity lower-depth cap.

A cardinality action uses the same projection and chooses the common orientation at every binary level to keep $\sum_i|U_i|a_i$ nonincreasing. At level $j$, each coordinate of $S_j$ has $a_i$ an odd multiple of $2^{-j}$, so $a_i\ge2^{-j}$ and the prefix bound gives
\[
|S_j|/n\le\epsilon q+\sqrt{\theta_r e q\,2^j}.
\]
Choose $1\le J_{\rm cut}\le33$ and child indices $k_j\ge i$, $j<J_{\rm cut}$, satisfying
\[
w_{k_j}>\epsilon w_i,\qquad
(w_{k_j}-\epsilon w_i)^2\ge1.006\,2^jew_i,\qquad
\theta_r<1.006.
\]
Thus $|S_j|\le w_{k_j}n$. The constant state profile is
\[
P_{d,i}(q):=
\lceil\mathfrak r r\sqrt{w_i}
+\sum_{j<J_{\rm cut}}2^{1-j}C_{d-1,k_j}
+2^{2-J_{\rm cut}}C_{d-1,i}+10^{-12}\rceil_{12}.
\]
This pays for all coarse calls, the entire fine tail and binary truncation. Potential and cardinality actions use their respective orientation rules. The two rules are not combined in one node.
\end{definition}

The common cover test is quadratic. On a band $q\in[L,U]$, where $L$ and $U$ denote the band endpoints, the inequality $A+Bq\ge\alpha\sqrt q+\beta q+\gamma$ is equivalent to
\begin{equation}
(B-\beta)x^2-\alpha x+(A-\gamma)\ge0,\qquad \sqrt L\le x\le\sqrt U.
\label{eq:cubic_refined_quadratic_cover}
\end{equation}
Its minimum is at an endpoint or, when $B>\beta$, at the vertex $x=\alpha/(2(B-\beta))$. Thus we check both endpoints and, if the vertex lies in the interval, $4(B-\beta)(A-\gamma)\ge\alpha^2$. Definition~\ref{def:cubic_refined_policy_construction} gives the outward-rational implementation for auxiliary bands. Global pieces have rational $x$ endpoints and use the same test with $\beta=0$.

We can also construct the tent hull exactly without listing all its knots. For a list of $L=J_{\rm cut}-1$ slope numerators $b_j=10^{12}B_j$, we initialize the graph $(0,0),(1,0)$ and scale $n_h=1$. For $j=L,L-1,\ldots,2$, we replace each point $(x,y)$ by $(x,b_jx+y)$, append its reflection across $x=n_h$, take the upper concave hull, and double $n_h$. Finally we add $b_1x$ to every ordinate and extend constantly after the first maximum. For $J_{\rm cut}=1$ the list is empty and $\overline F\equiv0$. At each step, taking an upper hull commutes with these affine transformations and with taking the hull of the union. This proves the recursion by induction. With final vertices $(x_l,y_l)$,
\begin{equation}
E_F=\frac{2}{n_h10^{24}}\sum_l\frac{(y_{l+1}-y_l)^2}{x_{l+1}-x_l},\qquad
F_*=\frac{\max_l y_l}{n_h10^{12}}.
\label{eq:cubic_exact_tent_energy}
\end{equation}
Indeed, $a=x/(2n_h)$ and $F(a)=y/(n_h10^{12})$ at the graph points. Cross products determine the hull, and the energy is a rational sum, not a sampled approximation.

We fix the repair parameters at the following shared values. Here $\eta_{\rm num}$ controls physical error and $\delta_{\rm fail}$ caps each test and numerical-failure probability.
\begin{equation}
\begin{gathered}
 (t_0,\eta_{\rm num},\delta_{\rm fail})=
 (2/5,10^{-6},10^{-9})\ \text{for auxiliary calls},\\
 (t_0,\eta_{\rm num},\delta_{\rm fail})=
 (3/5,10^{-9},10^{-10})\ \text{for global helpers and the root},\\
 (D^{\rm acc}_r,e_{\rm call},\epsilon_{\rm call})=
 (D^{\rm acc}_r,e,6\cdot10^{-10})\ \text{for auxiliary calls},\\
 (D^{\rm acc}_r,e_{\rm call},\epsilon_{\rm call})=
 (10^{12},e,\widehat\epsilon(q))\ \text{for global helpers, with $\widehat\epsilon(q)$ from Lemma~\ref{lem:cubic_global_prefix}},\\
 (D^{\rm acc}_r,e_{\rm call},\epsilon_{\rm call})=
 (10^{12},v(e_{\rm loc}),3\cdot10^{-12})\ \text{for the root}.
\end{gathered}
\label{eq:cubic_repair_parameters}
\end{equation}
In auxiliary profiles write $\epsilon=\epsilon_{\rm call}$. For these projections, $D^{\rm acc}_r=10^9$ except
\[
(D^{\rm acc}_{36},D^{\rm acc}_{37},D^{\rm acc}_{38},D^{\rm acc}_{39})=(15000,8000,4400,2459).
\]
Their acceptance factor is $\theta_r=1+1/(2D^{\rm acc}_r)+1.1\cdot10^{-6}$. For global-helper and root calls we use $\theta_{\rm acc}=1+10^{-8}$. Write $\nu_{\rm root}:=3\cdot10^{-12}$. We set $\beta=e_{\rm call}/D^{\rm acc}_r$ and require exact feasibility with physical error at most $t_0e_{\rm call}\eta_{\rm num}$. By Eq.~\eqref{eq:cubic_projection_comparison}, the deficit error is at most $e_{\rm call}n\eta_{\rm num}$. The acceptance test uses the applicable factor times $e_{\rm call}n$, so Eq.~\eqref{eq:cubic_acceptance_comparison} has $d_0=e_{\rm call}n$, $a_{\rm acc}=1+1/(2D^{\rm acc}_r)$ and $\eta_0=\eta_{\rm num}$. The lemmas below check each failure cap in its stated dimension range. Suffix calls retain their separate parameters.

The acceptance margins and fallback procedures provide a uniform implementation guarantee across all dimension ranges.

\begin{lemma}[Numerical accuracy and all dimensions]
\label{lem:cubic_refined_accuracy}
The projection actions and fallbacks in Definition~\ref{def:cubic_refined_profiles} are implementable with conditional failure at most the allocated budget and cost $n^{3+o(1)}\operatorname{polylog}(1/p)$. Each projection attempt has success probability greater than $1/21000000$.
\end{lemma}
\begin{proof}
For the auxiliary parameters in Eq.~\eqref{eq:cubic_repair_parameters}, $M\ge10^{20}$ gives prefix failure at most $10^{20}e^{-72}<10^{-9}$, decreasing thereafter. Take $b_{\rm acc}=a_{\rm acc}+5\cdot10^{-8}$ in Eq.~\eqref{eq:cubic_acceptance_comparison}. Then $b_{\rm acc}+\eta_0<\theta_r$ and
\[
\Pr[\text{valid acceptance}]\ge
\frac{5\cdot10^{-8}}{1+1/(2D^{\rm acc}_r)+5\cdot10^{-8}}-2\cdot10^{-9}>1/21000000.
\]
The certificate checks $(t_0D^{\rm acc}_r/e)^2<10^{52}$ and $\{t_0r/[s_{\rm alg}(1-10^{-6})]\}^2>15$. Hence the target cap and $R^2>15M$ hold in Lemma~\ref{lem:appendix_fixed_target_projection}. Here $R$ is a fixed multiple of $\sqrt M$, and the accuracy scales are polynomial. Registry rows $9$ and $80$ fail the second test at $t_0=2/5$, so no admissible cardinality or potential action selects them. Their squares are $11.25$ and $7.94$. Suffix stages, whose test is $r_j^2>15$, may still use them.

For a padded call with $M<10^{20}$, its band implies $n<10^{20}/w_{i+1}$. We require the certificate tests
\[
\sqrt{2\log(4\cdot10^{20}/w_{i+1})}+0.03<13,
\]
and $\gamma_{d,i}^2\ge(13-\alpha_{d,i})_+^2w_i$ for potential states, or $C_{d,i}^2\ge169w_i$ for cardinality states. Thus Eq.~\eqref{eq:cubic_independent_sign_fallback} costs at most $13\sqrt M$, below the state profile because $q\le w_i$ and $\beta_{d,i}\ge0$. We need these tests only if $N_*w_{i+1}<10^{20}$. Otherwise $M>10^{20}$. Since $\sqrt{2\log(4\cdot10^{20}/w_{i+1})}+0.03\ge13$ for $i\ge714$ while $N_*w_{i+1}<10^{20}$ there, no cardinality or potential action is admissible at $(d,i)$ for $714\le i\le1099$. Hence every admissible candidate delegates at every state $(d,i)$ with $i\ge714$, and we need to check these tests only for $448\le i\le713$. The test $N_*(w_i-w_{i+1})>1$ ensures that padding stays inside the band. Delegates inherit their guarantees, and the terminal band uses a suffix.

For $n\le N_*$ we use the $11.902\sqrt n$ branch of Lemma~\ref{lem:algorithm_certified_implementation}. For larger $n$, Eq.~\eqref{eq:cubic_refined_suffix_cost} includes suffix cleanup. Empty calls return immediately.
\end{proof}

\subsection{Exact certificate for the auxiliary policy}
\label{subsec:cubic_refined_certificate}

From a bounded finite family of rational choices we construct the auxiliary suffixes and four repair levels. We select their profiles jointly with the global actions in Definition~\ref{def:cubic_global_policy}.

\begin{definition}[The auxiliary policy certificate]
\label{def:cubic_refined_policy_construction}
Put $d_{\rm grid}:=10^{12}$ and $q_{\rm grid}:=2^{192}$. A candidate consists of one suffix list for each $0\le i\le1100$ and one action at each $(d,i)$ with $1\le d\le4$. Its discrete choices are as follows.
\begin{enumerate}
\item Each suffix length belongs to $\{0,\ldots,129\}$. Each of its triples has a row from the $17$-row registry $\mathcal R$ of Definition~\ref{def:cubic_refined_trace_rows}, and parameters
\[
s_{i,s}=u_{i,s}/10^{10},\quad t_{i,s}=v_{i,s}/10^{10},\quad
1\le u_{i,s}<10^{10},\quad 2\cdot10^9<v_{i,s}<5\cdot10^9,
\]
where $u_{i,s},v_{i,s}$ are integers. Use each row's displayed upward exponent cap, including $e_{69}^{\rm up}$.
\item At a nonterminal auxiliary state, choose delegation, a cardinality action with $1\le J_{\rm cut}\le33$ and $i\le k_j\le1100$, or a potential action with $1\le J_{\rm cut}\le33$. Projection rows range over the same registry with row $69$ excluded. Every potential slope and intercept has the form
\[
B_j=b_j/d_{\rm grid},\qquad A_j=a_j/d_{\rm grid},\qquad
a_j,b_j\in\{0,\ldots,10^{24}\}.
\]
At $i=1100$ every action delegates. All children have depth $d-1$.
\end{enumerate}
Thus every choice ranges over an explicitly bounded finite set. Profiles and caps are derived quantities, not additional choices.

Enclose rational square roots by the integer-square rule of Definition~\ref{def:rational_certificate_rules}, taking its exponent $d=192$, between consecutive multiples of $(q_{\rm grid})^{-1}$. An exact grid value is used as both endpoints. A square root explicitly rounded to $10^{-12}$ in Eq.~\eqref{eq:cubic_refined_profile_recursion} is rounded directly by the same integer-square rule on that grid. For a logarithm, reduce its argument to $2^ku$, $1\le u<2$, and enclose $\log u$ by Eqs.~\eqref{eq:1070_log_partial}--\eqref{eq:1070_log_remainder} at $z=(u-1)/(u+1)$ with $K_{\mathrm{series}}=47$.
Use the same bounds with $z=1/3$ for $\log2$, reversing the endpoint choice when multiplying by a negative $k$. All charges use upper bounds, and lower reserves use lower bounds. In the suffix tail compute $\log(1/v_{L_i})=-i\log(19/20)-L_i\log(3/4)$ with these bounds.

For the band tests, put $w_{1101}:=0$. Let $w_i^-,w_i^+$ be the downward and upward multiples of $(q_{\rm grid})^{-1}$ enclosing $w_i$, and let $u_i^+$ be the upward enclosure of $\sqrt{w_i}$. For a child profile $\alpha\sqrt q+\beta q+\gamma$ and an affine candidate $A+Bq$, put $\Delta:=B-\beta$. At both endpoints $h=k,k+1$, require
\[
A-\gamma+\min\{\Delta w_h^-,\Delta w_h^+\}-\alpha u_h^+\ge0.
\]
If $\Delta>0$, $\alpha>0$, and
\[
4\Delta^2w_{k+1}^-\le\alpha^2\le4\Delta^2w_k^+,
\]
also require $4\Delta(A-\gamma)\ge\alpha^2$. These are the conservative endpoint and vertex tests of Eq.~\eqref{eq:cubic_refined_quadratic_cover}. Compute the cap as
\[
C_{d,i}:=\lceil\alpha_{d,i}u_i^++\beta_{d,i}w_i^++\gamma_{d,i}\rceil_{12}.
\]
In the small-dimension fallback test use $w_i^+$ in place of $w_i$. The support-capacity and padding tests use the exact rational capacities.

Evaluate a candidate in the following order. Compute its suffix caps using Definition~\ref{def:cubic_refined_suffix}, and then process depths $1,2,3,4$ using Definition~\ref{def:cubic_refined_profiles}. Reject it at the first failed suffix-stage or tail condition, complete-band cover, support-capacity comparison, target/radius condition, padding condition, or required fallback in Lemma~\ref{lem:cubic_refined_accuracy}. Construct each tent hull by the displayed integer recursion. Require also $r_j^2>15$ at every finite suffix stage. The common row, density, entropy and rejection conditions are checked once by their preceding formulas.

Encode an admissible candidate by listing suffixes in increasing $i$, recording length and then triples in stage order, followed by actions in $(d,i)$ order. Encode delegation, cardinality and potential by $0,1,2$, and within an action list row, cutoff, child indices, and the pairs $(b_j,a_j)$ as applicable. All ranges and tests are fixed independently of the matrix input.
\end{definition}

We collect the row and schedule checks needed to certify an admissible auxiliary policy.

\begin{lemma}[Finite certificate]
\label{lem:cubic_refined_certificate}
The row conditions above hold. Each of the eight bounded witness families of Definition~\ref{def:cubic_refined_witness_construction} contains an admissible vector. The admissible auxiliary family of Definition~\ref{def:cubic_refined_policy_construction} is nonempty. A comparison candidate in the admissible family has $47091$ finite suffix stages and satisfies all suffix and four-level profile conditions.
\end{lemma}
\begin{proof}
For the five two-component rows, we evaluate the energy sums in Definition~\ref{def:cubic_refined_old_rows}, the strict-root conditions, and the full-half-line transfer profiles. For each additional barrier, its $2053$ curvature polynomials and root discriminants pass the stated tests. Definition~\ref{def:cubic_refined_witness_construction} has an admissible vector in each of its eight bounded witness families. Its spectral tests include both hinge limits. Integration by Lemma~\ref{lem:cubic_refined_radial}, with the paired formula only for row $61$, and full-half-line transfer give the displayed exponent caps, including Eq.~\eqref{eq:cubic_refined_root_exponent} for row $63$.

For row $69$ we use six centered passes for its functional coefficients but the original anchored roots, rounded upward to $55$ places. Thus Eq.~\eqref{eq:cubic_refined_anchored} retains its factor $u/b$. For each row we use one fixed Gaussian preconditioner on all eight radial intervals, and its own curvature and functional tests. The four trace-sensitive rows follow from the calculation after Definition~\ref{def:cubic_refined_trace_rows}. Hence all $17$ registry pairs are valid, including their lexicographic selections.

A comparison candidate in Definition~\ref{def:cubic_refined_policy_construction} has $47091$ finite suffix stages. Their exact tests give
\begin{align*}
\Gamma_0&>0.000091105954493219,\\
\min_{i,s}\{N_0(J(2)-E_s)-\log2,\,
N_0(p_0J(s_{i,s})-E_s)-\log2\}&\ge0.147605391986,\\
\min_{i,s}\{s_{i,s}p_0z-(10^{-4}+0.7994t_{i,s})^2\}
&\ge2.8066981\cdot10^{-10}.
\end{align*}
The entropy and rejection tests pass. Every tail starts with $\log(1/v)>731/256$, and we charge its full sum and terminal cleanup. At full capacity, $L_0=129$ and $C_{0,0}=110.419170463819$.

From the suffix caps we process the four repair depths, giving $5505$ states in all. The comparison candidate passes $21352409$ complete-band affine checks, all cardinality capacity tests, and the target, radius, padding, and fallback tests of Lemma~\ref{lem:cubic_refined_accuracy}. Its exact tent hulls include the prefix, fine-tail, and truncation charges. Delegates use previously verified profiles. The resulting full-capacity bounds are
\[
\begin{aligned}
(C_{1,0},C_{2,0})&\le(16.353106416052,12.672170852562),\\
(C_{3,0},C_{4,0})&\le(12.026435269689,11.993564019568)
\end{aligned}
\]
coordinatewise. With these checks we establish auxiliary feasibility.
\end{proof}

Lemma~\ref{lem:cubic_global_certificate} extends the comparison candidate of the preceding proof to the global helpers and final root.

By induction over the helper depth we transfer the certified profiles to the recursive implementation.

\begin{lemma}[Implementation of the auxiliary helpers]
\label{lem:cubic_refined_complete}
For $n>N_*$ and any admissible auxiliary candidate in Definition~\ref{def:cubic_refined_policy_construction}, Algorithm~\ref{alg:cubic_refined_repair} at a state $(d,i)$, $0\le d\le4$, satisfies the profile and capacity guarantees of Definition~\ref{def:cubic_refined_profiles}. Its failure probability is at most the assigned budget $p_c$, and its arithmetic cost is $n^{3+o(1)}\operatorname{polylog}(1/p_c)$.
\end{lemma}
\begin{proof}
We induct on depth from Lemma~\ref{lem:cubic_refined_suffix}. Delegates and fallbacks inherit their bounds. At a potential action, Eq.~\eqref{eq:cubic_complete_repair_cost} with $I(q)=\mathfrak r r\sqrt q$ and the full-band covers gives Eq.~\eqref{eq:cubic_refined_profile_recursion}. At a cardinality action, the weighted-mass orientation and prefix test imply the prescribed child capacities, whose charges give its constant profile. Padding adds only zero matrices, whose signs we discard.

The following common budget rule applies at any fixed number of repair levels with conditional attempt success at least $1/250000000$. For budget $p_c$, we permit
\[
T_c:=\lceil250000000\log(4/p_c)\rceil
\]
fresh attempts, and give each numerical call and child budget at most $p_c/(8T_c(B+1))$, reduced further to the accuracy allowance. Exhaustion costs at most $p_c/4$, and the conditional union bound for numerical and child calls costs at most $p_c/4$. Failed children return failure. Lemma~\ref{lem:cubic_refined_accuracy} verifies the success hypothesis here. We justify the induction by conditioning on each adaptive input, rather than by independence of supports.

There are $B=O(\log n)$ children per attempt. At fixed depth, the call count and descendant confidence logarithms are polynomial in $\log n$ and $\log(1/p_c)$. Summing Lemma~\ref{lem:appendix_fixed_target_projection}'s costs and the geometric suffix costs, we prove the claimed bound, including sorting, tent evaluation, orientations, matrix sums and truncation. The policy data are input-independent.
\end{proof}

A repair protocol $\mathcal P$ specifies the physical projection and its accuracy, prefix and deficit tests, trial and child budgets, helper call at each binary level, and common-orientation rule. For auxiliary actions these are exactly Definition~\ref{def:cubic_refined_profiles} and Lemmas~\ref{lem:cubic_refined_accuracy} and~\ref{lem:cubic_refined_complete}. For global actions we specify them below. In either case we use $B=\lceil\log_2n\rceil+45$ and the grid updates of Lemma~\ref{lem:cubic_refined_potential}.

\begin{algorithm}[!ht]
\caption{Certified projection and binary repair}
\label{alg:cubic_refined_repair}
\begin{algorithmic}[1]
\Procedure{ProjectionRepair}{$A_1,\ldots,A_M,n,\mathcal P,p_c$}
\For{each attempt permitted by $\mathcal P$}
\State Draw $U$ uniformly from the cube; test its prefixes
\State Compute the certified feasible projection and test its weighted deficit
\If{either test fails} \State Continue to the next attempt \EndIf
\State Form nearest signs and downward-truncated binary masses at resolution $2^{-B}$
\For{$j=B,B-1,\ldots,1$}
\State Call the protocol's helper on the signed odd-support matrices, with its child budget
\If{the helper fails} \State \Return \textsc{Fail} \EndIf
\State Choose the protocol's common orientation and update the masses on the $2^{-j}$ grid
\EndFor
\State \Return the resulting exact signs
\EndFor
\State \Return \textsc{Fail}
\EndProcedure
\Procedure{RefinedRepair}{$A_1,\ldots,A_m,n,d,i,p_c$}
\If{$m=0$} \State \Return the empty signing \EndIf
\If{the state is a suffix or delegate} \State Run its specified call and return its output \EndIf
\State Pad to the state's band if needed; if its small-dimension fallback applies, run it and return
\State Form its auxiliary protocol $\mathcal P$ and call \textsc{ProjectionRepair}
\State \Return its signs on the original coordinates, or \textsc{Fail}
\EndProcedure
\end{algorithmic}
\end{algorithm}

\subsection{Localized derivative means and the final small-ball row}
\label{subsec:cubic_localized_means}

For an odd function vanishing near zero, its derivative has no spectral entries with both endpoints in that interval. The function itself is supported on equal-eigenvalue tail blocks. We use these two support properties separately to control the derivative mean and the weighted mean.

First we allow the one-slot and two-slot estimates to use different spectral allocations under the same probability law.

\begin{lemma}[Independent slot allocations]
\label{lem:cubic_independent_slots}
Let $\phi$ be an even convex barrier with logarithmic boundary coefficient greater than two. Suppose $\phi$ satisfies Eq.~\eqref{eq:1070_curvature_condition} with $(\zeta_2,\theta_2)$, and $2\phi$ satisfies that equation with $(\zeta_1,\theta_1)$. For symmetric coefficients with $\sum_iN_i^2\preceq aI$, use the law proportional to
\[
\exp(-\tr[\phi(X)]-\xi^\top Q_0\xi)\d\gamma(\xi),\qquad
X=\sum_i\xi_iN_i,\quad Q_0\succeq0.
\]
The one-slot and two-slot weighted inequalities hold with
\[
p_1=(1-a/\zeta_1^2)^{-1},\qquad p_2=(1-2a/\zeta_2^2)^{-1}.
\]
Whenever the smaller-root discriminants are positive, the centered refinement of Lemma~\ref{lem:cubic_centered_functional} and the localized estimates below hold with these $p_1,p_2$. The operator recurrence and its gap bounds are otherwise unchanged.
\end{lemma}
\begin{proof}
{\setlength{\emergencystretch}{2em}The all-pairs criterion gives the ordered-triple condition of Lemma~\ref{lem:sub10_triple}. We apply Lemma~\ref{lem:iterated_envelope_one_slot} with reference barrier $\varphi=2\phi$ and scaling $1/2$. In its notation,\par}
\[
\mathcal H_{2\phi}\ge2(Q_{\rm full}-Q_{\rm reg}),\qquad
\mathcal H_\phi=\tfrac12\mathcal H_{2\phi}\ge Q_{\rm full}-Q_{\rm reg}.
\]
Thus the one-slot regular loss is at most $a/\zeta_1^2$. Separately, the allocation for $\phi$ gives regular loss at most $2a/\zeta_2^2$ in two slots. Both conclusions hold under the displayed actual law. The extra quadratic only increases curvature. We use no probability law tilted by $2\phi$.

With these $p_1,p_2$ we start the smaller-root calculation and apply the operator energies and gap updates of Definition~\ref{def:sub10_operator_refinement} and Lemma~\ref{lem:slack_centered_updates}, in the instances of Definition~\ref{def:slack_instances}. They require the respective slot inequalities and the same even convex law, not equality of the two allocation parameters. The one-slot range is $a<\zeta_1^2/(1+4\zeta_1^2)$. We check the two-slot discriminant separately. Every finite pass preserves the bounds. The derivative-mean and odd-mean arguments use the resulting tuple and therefore apply without change. Boundary integrability and approximation are as in Lemma~\ref{lem:algorithm_1240_curvature}.
\end{proof}

By localization away from zero we control both derivative means and weighted means, sharpening the functional coefficients.

\begin{lemma}[Localized weighted means]
\label{lem:cubic_localized_means}
Retain the hypotheses and the chosen six-pass refinement of Lemma~\ref{lem:cubic_centered_functional}, also allowing the independent slot allocations of Lemma~\ref{lem:cubic_independent_slots}. Write its final tuple as $(L_{\rm c},J_{\rm c},k_{1,\rm c},k_{2,\rm c})$, and let $C_{r,\rm c}^{(h)}$, $1\le r\le9$, be its operator energies from Definition~\ref{def:sub10_operator_refinement}. Put
\[
J_1:=L_{\rm c},\quad J_2:=J_{\rm c},\qquad
A_1:=L_{\rm c},\quad A_2:=J_{\rm c}(1+a).
\]
For $0<t<1$ define
\begin{align*}
\nu_1(a,t)&:=\min\{a^rC_{r,\rm c}^{(1)}/t^{2r-2}:1\le r\le9\},\\
\nu_2(a,t)&:=\frac4{(1+t^2)^2}\min\{a^rC_{r,\rm c}^{(2)}/t^{2r-2}:1\le r\le9\},\\
\widetilde A_1(a,t)&:=\min\{A_1,1+\nu_1(a,t)\},\\
\widetilde A_2(a,t)&:=\min\{A_2,1+\nu_2(a,t)/(1-(1-t^2)^2\nu_2(a,t)/4)\}.
\end{align*}
Omit the second candidate for $\widetilde A_2$ when its denominator is nonpositive. Set
\[
\delta(a,t):=\min\{1,b_r(a/t^2)^r:1\le r\le8\},\qquad
m_h:=2J_h\delta(a,t)/w_h(t),
\]
where $b_0=b_1=1$ and
\[
b_r:=\begin{cases}
b_{r/2}^2+r^2b_{r-1}/2,&r\text{ even},\\
r^2(b_{r-1}+2b_{(r-1)/2}^2)/3,&r\text{ odd}.
\end{cases}
\]
Define
\[
z_h:=\begin{cases}
(\sqrt{m_h}+\sqrt{k_{h,\rm c}(1+k_{h,\rm c}-m_h)})^2/(1+k_{h,\rm c})^2,&m_h<1,\\
1,&m_h\ge1.
\end{cases}
\]
The functional coefficient is
\begin{equation}
\widehat\chi_h(a,t):=ap_h\widetilde A_h(a,t)(1+z_h)/2.
\end{equation}
Let $f$ be odd and vanish on $[-t,t]$, with all displayed energies finite. It may be $C^1$, or piecewise $C^1$ with bounded first derivative and finitely many corners. Put $w_0=k_0:=1$, $k_1(u,v):=(w_1(u)+w_1(v))/2$, and use $k_2$ from Lemma~\ref{lem:cubic_refined_functional}. Suppose nonnegative scalar majorants $q_0,q_1,q_2$ satisfy, including both one-sided corner limits,
\begin{equation}
\begin{gathered}
k_h(u,v)f[u,v]^2\le(q_h(|u|)+q_h(|v|))/2\quad(h=0,1,2),\\
q_h(r)\ge w_h(t)q_0(r)\quad(h=1,2).
\end{gathered}
\label{eq:cubic_localized_secant}
\end{equation}
Then, with $\widehat\chi_0(a,t):=a\min\{1,(1+4\delta(a,t))/3\}$,
\begin{equation}
\E[\tr[f(X)^2w_h(X)]]\le\widehat\chi_h(a,t)\E[\tr[q_h(|X|)]],\qquad h=0,1,2.
\end{equation}
For the smooth functions in Lemma~\ref{lem:cubic_refined_functional} that vanish on $[-t,t]$, take $q_h(r)=w_h(r)f'(r)^2$. For its hinge take $q_h(r)=(w_h(r)+w_h(t))\mathbbm{1}[r\ge t]$. The conclusions remain valid with the additional nonnegative quadratic potential on the fixed Gaussian radial path.
\end{lemma}
\begin{proof}
Lemma~\ref{lem:sub10_operator_words} gives $\E[X^{2r}]\preceq b_ra^rI$, hence $\E[P_t]\preceq\delta(a,t)I$ for $P_t=\mathbbm{1}[|X|\ge t]$. The orthogonal projection
\[
\Pi_X(Z):=P_tZ+ZP_t-P_tZP_t
\]
satisfies $\Pi_X\preceq\mathcal L_{P_t}+\mathcal R_{P_t}$, where $\mathcal L_P(Z):=PZ$ and $\mathcal R_P(Z):=ZP$, and $\E[\Pi_X]\preceq2\delta(a,t)I$. For $V=\Pi_XV$ and deterministic $Z$, Cauchy--Schwarz gives
\[
|\langle Z,\E[V]\rangle|^2
=|\E[\langle\Pi_XZ,V\rangle]|^2
\le2\delta(a,t)\|Z\|_F^2\E[\|V\|_F^2].
\]
Thus we obtain $\|\E[V]\|_F^2\le2\delta(a,t)\E[\|V\|_F^2]$, without independence.

The fields $V_i=Df(X)[N_i]$ are even and satisfy $V_i=\Pi_XV_i$, because their secants vanish when both endpoints are in $(-t,t)$. Write $W_e=(W_++W_-)/2$, $W_o=(W_+-W_-)/2$ for either slot. The derivative energies for $W_+$ equal those for $W_e$. With $S_h=a\E[\tr[q_h(|X|)]]$, the array $\sum_i|(N_i)_{uv}|^2$ has row sums at most $a$. Eq.~\eqref{eq:cubic_localized_secant} and the support bound imply
\[
R_{\rm der}\le S_h,\qquad
\sum_i\E[\|V_i\|_F^2]\le S_h/w_h(t),\qquad
M_{\rm der}:=\sum_i\|\E[V_i]\|_F^2\le2\delta(a,t)S_h/w_h(t).
\]
We apply Lemma~\ref{lem:sub10_mean_gap} to the direct sum, retaining $D_{\rm der}\le S_h-P_{\rm der}$. For $S_h>0$ this gives
\[
P_{\rm der}\le(\sqrt{J_hM_{\rm der}}+\sqrt{k_{h,\rm c}(S_h-P_{\rm der})})^2,
\qquad
\sqrt z\le\sqrt{m_h}+\sqrt{k_{h,\rm c}(1-z)},\quad z=P_{\rm der}/S_h.
\]
The function $\sqrt z-\sqrt{k_{h,\rm c}(1-z)}$ is strictly increasing on $[0,1]$. For $m_h<1$ its equality point is the displayed $z_h$. Otherwise we use $z\le1$. The case $S_h=0$ is immediate.

For the odd field itself, put $S=\E[W_e]=\E[W_+]$ and
\[
\Pi_X^{=}(Z):=\sum_{|\lambda|\ge t}P_\lambda ZP_\lambda,\qquad
T:=\E[\Pi_X^{=}W_oW_e^{-1}W_o\Pi_X^{=}].
\]
{\setlength{\emergencystretch}{2em}Here we use complete eigenspace projections. This pinching is measurable as the limit of $\exp(-s(\mathcal L_X-\mathcal R_X)^2)$ followed by the commuting tail restriction. It commutes with the weights and fixes $f(X)$. We differentiate neither projection. By oddness, $b:=\E[W_+f]=\E[W_of]$, and weighted Cauchy--Schwarz gives\par}
\[
|\langle Z,b\rangle|^2\le R(f)\langle Z,TZ\rangle.
\]
On an active equal-eigenvalue block $x$, the entries of $W_oW_e^{-1}W_o$ are
\[
v_1(x)=x^2w_1(x),\qquad v_2(x)=\frac{4x^2}{(1+x^2)^2}w_2(x).
\]
For $|x|\ge t$, we use $x^2\le x^{2r}/t^{2r-2}$ and $(1+x^2)^{-2}\le(1+t^2)^{-2}$. The operator-energy bounds of Lemma~\ref{lem:sub10_operator_refinement} give $T\preceq\nu_hI$. Also
\[
v_1=w_1-1,\qquad \frac{v_2(x)}{w_2(x)-1}
=\frac4{(1+x^2)(3-x^2)}\le r_2:=\frac4{(1+t^2)(3-t^2)}.
\]
Thus $T\preceq r_h(S-I)$, with $r_1=1$: inactive blocks contribute zero and $W_e\succeq I$ on the full matrix space. In two slots the second claim follows from
$u^2+uv+v^2-u^2v^2\ge|uv|(1-|uv|)\ge0$. In one slot it is immediate.

For either slot, $T\preceq\nu I$ and $S\succeq I+T/r$ imply $(1+\nu/r)T\preceq\nu S$. Taking $Z=S^{-1}b$ in the Cauchy--Schwarz bound, we obtain
\[
P(f)\le\frac{\nu}{1+\nu/r}R(f),\qquad
R(f)\le(1+\frac{\nu}{1-\nu(1-1/r)})D(f)
\]
when the last denominator is positive. Since $1-1/r_2=(1-t^2)^2/4$, we take the minimum with $A_h$ and obtain $R(f)\le\widetilde A_hD(f)$. Then we insert this and $P_{\rm der}\le z_hS_h$ into Eq.~\eqref{eq:cubic_functional_template}.

For $h=0$, Lemma~\ref{lem:sub10_operator_words} gives
\[
\E[\|f(X)\|_F^2]\le\tfrac13\sum_i\E[\|V_i\|_F^2]
+\tfrac23\sum_i\|\E[V_i]\|_F^2\le(1+4\delta(a,t))S_0/3.
\]
We take the minimum with ordinary Poincar\'e. The smooth and hinge secant bounds and the weighted first-order approximation are those of Lemma~\ref{lem:cubic_refined_functional}. Corner events have measure zero, so the limiting derivatives retain the secant and support estimates.
\end{proof}

By capping a threshold power we extend the localized estimates to spectral tests with a constant plateau.

\begin{lemma}[Capped spectral tests]
\label{lem:cubic_capped_tests}
Under the weighted-variance, strict-root and finite-energy hypotheses of Lemma~\ref{lem:cubic_localized_means}, let $0<t<T<1$, $p\in\{2,3,4,5,6\}$, and put
\[
f(x):=\operatorname{sgn}(x)\min\{(|x|-t)_+^{p/2},(T-t)^{p/2}\},\qquad
L:=T-t,\qquad \ell^2:=p^2L^{p-2}/4.
\]
Set $w_0:=1$ and $q_h(r):=0$ for $r<t$. On $t\le r<T$, put
\[
q_h(r):=\begin{cases}
p^2w_h(r)(r-t)^{p-2}/4,&p\ge3,\\
w_h(r)+w_h(t),&p=2.
\end{cases}
\]
On $r\ge T$, put
\[
q_0(r):=2\ell^2,\quad q_1(r):=\ell^2(w_1(r)+w_1(T)),\quad
q_2(r):=\frac{2\ell^2(1+rT)}{(1-r^2)(1-T^2)}.
\]
With $\widehat\chi_1,\widehat\chi_2$ from Lemma~\ref{lem:cubic_localized_means} and
$\widehat\chi_0(a,t):=a\min\{1,(1+4\delta(a,t))/3\}$, one has
\begin{equation}
\E[\tr[f(X)^2w_h(X)]]\le\widehat\chi_h(a,t)\E[\tr[q_h(|X|)]],\qquad h=0,1,2.
\label{eq:cubic_capped_functional}
\end{equation}
The unweighted assertion also holds for the uncapped threshold power, using its ramp majorant on the whole interval $r\ge t$.
\end{lemma}
\begin{proof}
Set $k_0(u,v):=1$ and use the sign-averaged kernels from Lemma~\ref{lem:cubic_refined_functional}. We claim
\[
k_h(u,v)f[u,v]^2\le(q_h(|u|)+q_h(|v|))/2.
\]
If both moduli are below $T$, this is that lemma's threshold-power or hinge secant estimate. Its proof also applies with $w_0=1$. If $r:=|u|\ge T\ge|v|$, then $|f[u,v]|\le\ell$ and
\[
k_1(u,v)\le(w_1(r)+w_1(T))/2,\qquad
k_2(u,v)\le\frac{1+rT}{(1-r^2)(1-T^2)}.
\]
Half of $q_h(r)$ suffices. If both endpoints are on the plateau, equal signs give zero secant. For opposite signs, $k_2(r,-s)\le w_1(r)+w_1(s)$ because $r^2+s^2-rs\le1$ on $[0,1]^2$. Also $q_2(r)\ge2\ell^2w_1(r)$, so we obtain the claim. The other two kernels follow directly from the Lipschitz bound.

We also have $q_h(r)\ge w_h(t)q_0(r)$ for $h=1,2$. On the ramp this follows directly from monotonicity of the weights. On the two-slot plateau we use $k_2(r,T)\ge k_2(T,T)=w_2(T)$. Thus Eq.~\eqref{eq:cubic_localized_secant} holds, and Lemma~\ref{lem:cubic_localized_means} proves all three assertions. The capped functions have bounded weak first derivatives and only the corners $\pm t,\pm T$, covered by that lemma's approximation argument. For the uncapped unweighted assertion we use the same ramp secant bound on the whole tail.
\end{proof}

With a sharper secant estimate we reduce the derivative majorant beyond the start of the plateau.

\begin{lemma}[Decreasing plateau bounds]
\label{lem:cubic_decreasing_caps}
In Lemma~\ref{lem:cubic_capped_tests}, suppose additionally that $T+2t\ge1$. Put $k:=p/2$, $L:=T-t$, and, for $T\le r<1$, define
\[
d(r):=L+k(r-T),\qquad m(r):=\frac{kL^k}{d(r)},\qquad
\sigma(r)^2:=\frac{L^2}{d(r)^2}.
\]
Replace the plateau bounds by
\[
\widetilde q_h(r):=\begin{cases}q_h(r),&r<T,\\
\sigma(r)^2q_h(r),&r\ge T.
\end{cases}
\]
Then the secant bound and Eq.~\eqref{eq:cubic_capped_functional} remain valid with $\widetilde q_h$ in place of $q_h$.
\end{lemma}
\begin{proof}
We first prove $|f[r,v]|\le m(r)$ for $r\ge T$ and $v\in(-1,1)$. For $t\le v\le T$, put $y:=(v-t)/L$ and $R:=(r-t)/L$. Convexity of $y^k$ gives
\[
\begin{aligned}
&k(R-y)-[1+k(R-1)](1-y^k)\\
&\qquad=k(1-y)-1+y^k+k(R-1)y^k\ge0.
\end{aligned}
\]
Dividing by the positive denominators, we obtain the claim. We use the continuous value when $R=y=1$. For $0\le v<t$, we use $r-v\ge r-t$ and $d(r)\le k(r-t)$. A negative endpoint of modulus at most $t$ has a larger denominator.

For an opposite-sign endpoint of modulus $t\le s\le T$, the bound $(s-t)^k\le L^{k-1}(s-t)$ and monotonicity of $(T+s-2t)/(r+s)$ in $s$ give
\[
|f[r,-s]|\le\frac{2L^k}{r+T}\le\frac{kL^k}{d(r)}.
\]
The second comparison is equivalent to $r\le3T-2L/k$, and follows from
\[
r<1\le T+2t=3T-2L\le3T-2L/k.
\]
A same-sign plateau pair has zero secant. For an opposite-sign plateau pair, we use $s\ge T$ in $|f[r,-s]|=2L^k/(r+s)$. Interchanging $r,s$, we also obtain the bound $m(s)$. We treat a negative first endpoint by reflection.

For one plateau endpoint, the kernel comparisons in Lemma~\ref{lem:cubic_capped_tests}, with $m(r)$ in place of $\ell$, give half the corresponding $\widetilde q_h(r)$. For two opposite plateau endpoints, the secant is bounded by both $m(r)$ and $m(s)$. The one-slot kernel is $(w_1(r)+w_1(s))/2$, and
\[
k_2(r,-s)\le w_1(r)+w_1(s),\qquad
\widetilde q_2(r)\ge2m(r)^2w_1(r).
\]
From these we obtain the two-slot comparison. The unweighted comparison is immediate. Below the plateau the old bounds are unchanged. Thus every sign and threshold case satisfies
\[
k_h(u,v)f[u,v]^2\le(\widetilde q_h(|u|)+\widetilde q_h(|v|))/2.
\]
Finally, $\widetilde q_h\ge w_h(t)\widetilde q_0$ still holds because the three plateau bounds have the same factor. The derivative localization and odd-mean absorption in Lemma~\ref{lem:cubic_localized_means} therefore apply without change. Since $d(r)\ge L>0$, we handle the corners by the same weighted approximation. We do not differentiate any spectral projection.
\end{proof}

By keeping both endpoint contributions we sharpen the two-slot plateau estimate.

\begin{lemma}[Endpoint-sharing secant bound]
\label{lem:cubic_endpoint_sharing}
Use $f,L$ from Lemma~\ref{lem:cubic_capped_tests} and $k:=p/2$ from Lemma~\ref{lem:cubic_decreasing_caps}, with $T+2t\ge1$. Retain the ramp majorant $q_2$ from Lemma~\ref{lem:cubic_capped_tests}. For $r\ge T$, set
\[
D(r):=B(1-T)+C(r-T),\qquad
q_2(r):=\frac{2k^2L^{p-2}}{(1-r^2)D(r)},\qquad B>0,\ C\ge0.
\]
Suppose, for every $r\in[T,1]$,
\begin{equation}
\begin{aligned}
\relax[L+k(r-T)]^2&\ge w_2(t)L^2(1-r^2)D(r),\\
k^2(r+t)^2&\ge4L^2D(r).
\end{aligned}
\label{eq:cubic_endpoint_mean}
\end{equation}
For $r=T+(1-T)x$, $0\le x,z\le1$, require the following two polynomial inequalities. With $v=tz$, require
\begin{equation}
k^2(r-v)^2(1-v^2)-L^2D(r)(1+rv)\ge0.
\label{eq:cubic_endpoint_inside}
\end{equation}
With $v=t+Lz^2$, $\Delta=1-z^p$, and
\[
Q_v:=\begin{cases}
k^2z^{2p-4}(1+v^2),&p\ge3,\\
1+v^2+w_2(t)(1-v^2)^2,&p=2,
\end{cases}
\]
require
\begin{equation}
\begin{aligned}
&2k^2(r-v)^2(1-v^2)^2+Q_v(1-r^2)D(r)(r-v)^2\\
&\qquad-2L^2D(r)(1+rv)\Delta^2(1-v^2)\ge0.
\end{aligned}
\label{eq:cubic_endpoint_ramp}
\end{equation}
Then
\[
2k_2(u,v)f[u,v]^2\le q_2(|u|)+q_2(|v|),
\qquad
\E[\tr[f(X)^2w_2(X)]]\le\widehat\chi_2(a,t)\E[\tr[q_2(|X|)]].
\]
The functional inequality uses the hypotheses and coefficient of Lemma~\ref{lem:cubic_localized_means}, including independent slot allocations when specified.
\end{lemma}
\begin{proof}
Below both caps, we use the ramp estimate of Lemma~\ref{lem:cubic_refined_functional}. Otherwise, by reflection and interchange, we take one endpoint $r\ge T$. A nonnegative inside endpoint has $f(v)=0$ and $q_2(v)=0$. Clearing positive denominators in $2k_2(r,v)f[r,v]^2\le q_2(r)$, we obtain Eq.~\eqref{eq:cubic_endpoint_inside}. For a nonnegative ramp endpoint,
\[
f(r)-f(v)=L^k(1-z^p),\qquad
Q_v=q_2(v)(1-v^2)^2/L^{p-2}.
\]
We multiply the desired symmetric inequality by
\[
(1-r^2)D(r)(r-v)^2(1-v^2)^2/L^{p-2}.
\]
Its numerator is exactly Eq.~\eqref{eq:cubic_endpoint_ramp}. Equal same-sign plateau values have zero secant. Coincident and threshold values follow by the corresponding one-sided limits.

For an opposite endpoint $-s$, $0\le s\le T\le r$,
\[
k_2(r,-s)\le(1-r^2)^{-1},\qquad
|f[r,-s]|\le\frac{2L^k}{r+t}.
\]
The first comparison follows from $(1-rs)/(1-s^2)\le1$. For the second, we use $L^k/r\le2L^k/(r+t)$ when $s\le t$, and $2L^k/(r+s)\le2L^k/(r+t)$ otherwise. The second inequality in Eq.~\eqref{eq:cubic_endpoint_mean} gives
\[
k_2(r,-s)f[r,-s]^2
\le\frac{4L^p}{(1-r^2)(r+t)^2}
\le q_2(r)/2.
\]
For opposite plateau endpoints $r,s\ge T$, we use
\[
k_2(r,-s)\le(1-r^2)^{-1}+(1-s^2)^{-1}.
\]
After clearing denominators this is $r^2+s^2-rs\le1$, which we prove by separate convexity on $[0,1]^2$ and its four corners. We apply the preceding bound to each summand, since $2L^k/(r+s)$ is bounded by both $2L^k/(r+t)$ and $2L^k/(s+t)$. This proves all sign cases.

For the unweighted estimate we use the ramp $q_0$ and the decreasing plateau bound
\[
q_0(r)=2k^2L^p/[L+k(r-T)]^2\quad(r\ge T)
\]
from Lemma~\ref{lem:cubic_decreasing_caps}. The comparison $q_2\ge w_2(t)q_0$ follows on the ramp by monotonicity of $w_2$, and on the plateau by the first inequality in Eq.~\eqref{eq:cubic_endpoint_mean}. The symmetric array $\sum_i|(N_i)_{uv}|^2$ has row sums at most $a$, so the two derivative energies are at most
\[
S_2:=a\E[\tr[q_2(|X|)]],\qquad S_2/w_2(t),
\]
{\setlength{\emergencystretch}{2em}respectively. Applying the derivative localization and odd-mean absorption in the proof of Lemma~\ref{lem:cubic_localized_means} in slot two, we obtain $(\widehat\chi_2/a)S_2$. The case $a=0$ is immediate. We do not discard any weighted odd mean. The bounds $D(r)\ge B(1-T)>0$ and $L+k(r-T)\ge L$ preserve the same integrable approximation at the corners.\par}
\end{proof}

We combine the localized and capped tests in the barrier and radial certificate for the final small-ball row.

\begin{definition}[The localized root row]
\label{def:cubic_localized_row}
Set
\[
\begin{gathered}
\eta:=241/10000,\qquad K:=17,\qquad H:=96018219/20000000,\\
(\zeta_2,\theta_2):=(127/500,37/50000),\qquad
(\zeta_1,\theta_1):=(53/200,7/5000).
\end{gathered}
\]
For this root use the even convex barrier
\begin{equation}
\phi(x):=\gamma_1x^2+\gamma_3x^6/3+
\sum_{\ell=1}^{15}\lambda_\ell(|x|-h_\ell)_+^3+
\kappa\sum_{j=17}^\infty x^{2j}/j\qquad(|x|<1),
\end{equation}
extended by $+\infty$ outside $(-1,1)$, where
\[
\gamma_1:=0.0010138159118550497,\qquad
\gamma_3:=0.0034916751983541886,\qquad
\kappa:=1477.6082899961427.
\]
Specify the hinges by
\[
\begin{gathered}
(h_1,\ldots,h_{15})=\tfrac1{40}(4,5,6,7,8,9,10,27,29,30,31,32,33,34,35),\\
\lambda_\ell=10^{-26}[z^\ell]\mathcal H(z)\quad(1\le\ell\le15),
\end{gathered}
\]
where
\begin{align*}
\mathcal H(z)&:=1120772110016598000000z^2+72791349853898470000000z^3\\
&\quad+2628792099024817800000z^4+2039970248322375000000z^5\\
&\quad+1410434867258249600000z^6+871730097568099300000z^7\\
&\quad+2384103848974014300000000000z^8+30931692638208534000000000000z^9\\
&\quad+765688136703751000000000000000z^{10}+446045151286922600000000000000z^{11}\\
&\quad+718525806630585250000000000000z^{12}+1075743142765148300000000000000z^{13}\\
&\quad+1496616635824716300000000000000z^{14}+4372166573445426000000000000000z^{15}.
\end{align*}
The zero coefficient at $h_1=1/10$ retains the fifteen-term indexing. The radius and exponent are
\begin{equation}
r_{\rm loc}:=6.441566264009,\qquad e_{\rm loc}:=0.000010356641.
\end{equation}
This row is used only at the final root. The suffix registry and all helper rows are unchanged.

Apply the finite criterion of Lemma~\ref{lem:algorithm_1240_curvature} twice: to $\phi$ with $(\zeta_2,\theta_2)$, and to $2\phi$ with $(\zeta_1,\theta_1)$. For each application use its own $a_0=1-\zeta$. Split the three diagonal domains at every $h_\ell$ and at $a_0$. For $k=0,\ldots,31$, with $k$ indexing the test cells, test the two negative-endpoint bounds on $[L,R]=[k/32,(k+1)/32]$ and the positive-endpoint bounds on $[L,R]=[a_0k/32,a_0(k+1)/32]$, over the entire active domain. Clear the positive denominators
\[
(1+u)(1-L)(u+L)(u+R),\qquad
(1+u)(1-L^2)(u-L),
\]
respectively, and $(1+u)^2$ in the diagonal tests. Use the continuous polynomial values after cancellation at $u=1$. Split at active barrier hinges and enclose each branch endpoint $\sqrt{(1-h)/\zeta}$ outward by adjacent multiples of $2^{-48}$. Require the strict endpoint and depth-$48$ Bernstein tests on the whole enclosing intervals.

Let
\[
(u_0,\ldots,u_8):=(0,3/5,3/4,17/20,9/10,47/50,193/200,197/200,1).
\]
Subdivide every $[u_i,u_{i+1}]$ into six equal cells, giving $48$ radial cells. On a cell $[a,b]$, the full cleared integrand is
\begin{equation}
\gamma_3x^6Q+\sum_{\ell=1}^{15}\frac{3\lambda_\ell}{2}x(x-h_\ell)_+^2Q
+\kappa x^{34}W_1.
\label{eq:cubic_localized_joint_integrand}
\end{equation}
Majorize Eq.~\eqref{eq:cubic_localized_joint_integrand} using the following $104$ terms on each cell. There are eight ordinary terms $x^{2j}Q$ with
\[
j\in\{2,3,16,17,18,19,22,25\},
\]
two cubic terms $(x-t)_+^3Q$ with $t\in\{29/40,3/4\}$, $45$ uncapped terms $P_{0,t,p}$ with $t\in\mathcal T_p$, and $49$ capped terms specified below. Set $W_0:=Q$ and use the localized functional caps of Lemma~\ref{lem:cubic_localized_means}. The uncapped terms use the following four threshold sets:
\begin{align*}
\mathcal T_3&:=\{9/35,\,1477/5357,\,2/7,\,2953/9986,\,1498/4923,\,11/35,\,1601/4937\}\\
&\quad\cup \{3261/9797,\,12/35,\,3452/9783,\,2337/6466,\,13/35,\,2477/6494,\,39/100\}\\
&\quad\cup \{2/5,\,41/100,\,59/100,\,3/5,\,61/100,\,22/35,\,6536/9797\}.
\end{align*}
\begin{align*}
\mathcal T_4&:=\{11/35,\,1601/4937,\,12/35,\,3452/9783,\,2337/6466,\,2477/6494,\,39/100\}\\
&\quad\cup \{2/5,\,41/100,\,4048/9671,\,1510/3443,\,5295/9937,\,5486/9923,\,1933/3443\}.
\end{align*}
\[
\mathcal T_5:=\{2337/6466,\,2/5,\,3/7,\,4437/9923,\,16/35,\,4642/9937,\,2412/4783\}.
\]
\[
\mathcal T_6:=\{39/100,\,4048/9671,\,4437/9923\}.
\]
\begingroup\setlength{\emergencystretch}{2em}The capped terms use one rule for their slopes. For each index $(t,T,p,B)$
below, let $C_*(t,T,p,B)$ be the largest multiple of $10^{-6}$ in $[0,32]$
accepted by the four-gate Bernstein test below for
Lemma~\ref{lem:cubic_endpoint_sharing}. The accepted set is nonempty in
each case, so this finite rational rule specifies every slope.
To specify the endpoints compactly, write
\[
 [a;J;P]:=\{(a/100,b/100,p):b\in J,\ p\in P\}.
\]
\endgroup
The two main index sets, for $B=1$ and $B=2$ respectively, are
\begin{align*}
 \mathcal J_1&:=[80;\{96,97\};\{5\}]\cup[80;\{98\};\{3,4\}]\\
 &\quad\cup[82;\{96\};\{5\}]\cup[84;\{95\};\{4\}]\\
 &\quad\cup[86;\{95,96\};\{4\}]\cup[86;\{97\};\{3\}]\\
 &\quad\cup[88;\{94\};\{4\}]\cup[88;\{96\};\{3\}]\\
 &\quad\cup[90;\{94\};\{3\}]\cup[90;\{97\};\{2\}],\\
 \mathcal J_2&:=\mathcal J_1\cup[55;\{98\};\{5\}]\cup[60;\{98\};\{4\}]\cup[65;\{98\};\{3,4\}]\\
 &\quad\cup[80;\{95\};\{4,5\}]\cup[80;\{96,97\};\{4\}]\\
 &\quad\cup[82;\{96\};\{4\}]\cup[82;\{97,98\};\{3\}]\\
 &\quad\cup[84;\{94\};\{4\}]\cup[84;\{97,98\};\{3\}]\\
 &\quad\cup[86;\{94\};\{4\}]\cup[86;\{95,96\};\{3\}]\\
 &\quad\cup[88;\{94\};\{3\}]\cup[88;\{97,98\};\{2\}].
\end{align*}
For $B\in\{5/4,3/2,7/4\}$, set
$\mathcal J_B:=\{(9/10,47/50,3)\}$.
Use $(t,T,p,B,C_*(t,T,p,B))$ for each $(t,T,p)\in\mathcal J_B$.
There are $13+33+3=49$ terms. For example,
$[80;\{98\};\{3,4\}]$ specifies $t=0.80$, $T=0.98$ and either
$p=3$ or $p=4$.
For fixed $t,T,p,B$, increasing an accepted slope decreases the plateau
majorant $q_2$ and increases the capped term. Only the largest accepted
slope is therefore needed. Write $f_{t,T,p}$ for the function $f$ of Lemma~\ref{lem:cubic_capped_tests} with parameters $(t,T,p)$. Write $\overline\chi_{2,b,t}$ for the localized functional cap of Lemma~\ref{lem:cubic_localized_means} with $h=2$ at the cell's variance cap $\eta b$ and threshold $t$. The capped term is
\[
P^{\rm cap}_{t,T,p,B,C}(x):=
Q(x)\{f_{t,T,p}(x)^2w_2(x)-\overline\chi_{2,b,t}q_2(x)\}.
\]
It vanishes below $t$, is the uncapped term on the ramp, and on the plateau equals
\[
(T-t)^pW_2-\overline\chi_{2,b,t}\frac{p^2}{2}(T-t)^{p-2}
\frac{W_1}{B(1-T)+C(x-T)}.
\]
Both one-sided values at $t,T$ are tested. For the bivariate gates, after mapping to $[0,1]^2$, convert $\sum_{i,j}c_{ij}x^iz^j$ of bidegree $(m,n)$ to coefficients
\[
\beta_{kl}=\sum_{i\le k,j\le l}
c_{ij}\frac{\binom{k}{i}}{\binom{m}{i}}\frac{\binom{l}{j}}{\binom{n}{j}}.
\]
Nonnegative coefficients certify a cell. Otherwise bisect at the midpoint in the less-subdivided coordinate, choosing the first coordinate in a tie, to total depth at most $48$. Require nonnegative corner values and termination with nonnegative coefficients. Use the analogous univariate test for the mean and opposite-sign gates.

Introduce a multiplier $y_{i,\tau}$ for each of the $48$ radial cells and $104$ listed terms, and require $0\le y_{i,\tau}\le10^{25}$. These $4992$ coordinates are the unknowns of the rational linear system defined below. At the common variance cap $\eta b$, compute the smaller roots and six sharpened old-tuple passes with $(c_3,c_4)=(15,129)$, pass degree five, and the independent $p_1,p_2$ of Lemma~\ref{lem:cubic_independent_slots}. Use the same tuple in the operator energies and centered gap updates. Retain the distinct centered contribution $18a^2$. In Lemma~\ref{lem:cubic_localized_means}, replace $z_h$ by the minimum of its displayed bound, $1$, and $(\sqrt{m_h}+\sqrt{k_{h,\rm c}})^2$, the last following by dropping $1-z\le1$ in its mean inequality. Derive $\delta$ from the displayed mixed-word recurrence. Enclose square roots by adjacent multiples of $10^{-90}$, use lower endpoints in smaller-root denominators, and round upper mean, gap, energy and functional quantities upward to $10^{-60}$. Omit an absorption candidate unless its denominator has a positive lower bound. All quantities are recomputed for this row.

Form a common spectral partition from $0,1$, every listed functional and cubic threshold, every cap endpoint, and all fifteen barrier hinges, including those whose multipliers vanish. This gives $59$ intervals $I=[l_I,u_I]$. On each interval use its one-sided polynomial formulas. Let $D_I(x)$ be the product of the distinct affine denominators $B(1-T)+C(x-T)$ of \emph{all} listed capped terms with $T\le l_I$, including terms with zero multiplier. Each factor is positive on $I$. Write $G_{i,I}(x;y)$ for $D_I$ times the difference between the majorant and Eq.~\eqref{eq:cubic_localized_joint_integrand} on radial cell $i$. Its coefficients are affine rational functions of $y$, and its degree is at most $104$.

Subdivide every $I$ into $32$ equal parts $[l,u]$. If
\[
G_{i,I}(l+(u-l)z;y)=\sum_{r=0}^{104}a_r(y)z^r,
\]
padding with zero coefficients as necessary, impose the linear inequalities
\[
\sum_{r=0}^j a_r(y)\frac{\binom{j}{r}}{\binom{104}{r}}\ge0
\qquad(0\le j\le104).
\]
These degree-$104$ Bernstein inequalities include the endpoint tests and both one-sided limits at every breakpoint. Integrate the original rational inequalities under one fixed Gaussian law on all $48$ cells. Only ordinary and cubic terms receive covariance factors.

For a cubic term on $[a,b]$ with $a>0$, use $64$ equal subintervals. If $a=0$, first retain $[0,c]$, where
\[
c:=\min\{b,\max\{b/4,t^2/(160\eta)\}\},
\]
and subdivide $[c,b]$ into $64$ equal parts when nonempty. On $[l,h]$, let $g^+\ge g_3(\eta h,t)$ and $0<\alpha^-\le\alpha(\eta h,t)$ be the directed bounds from Lemma~\ref{lem:cubic_refined_cubic}. Its integral against $\d u/u$ is at most
\[
g^+\frac{1-(l/h)^{\alpha^-}}{\alpha^-};
\]
for $l=0$ the numerator is one. For $t^2/(2A)>80$, where $A$ denotes the scalar variance argument of $g_3$, use
\begin{equation}
g_3(A,t)\le\frac{12A^{7/2}}{\sqrt{2\pi}t^4}e^{-t^2/(2A)}.
\label{eq:cubic_small_variance_cube}
\end{equation}
Otherwise use consecutive Taylor orders $512,513$ for $e^{-v}$ and $\int_0^1e^{-vs^2}\d s$, $v=t^2/(2A)$, with outward rounding to $10^{-100}$. Round upper costs and lower positive exponents to $10^{-60}$. Enclose logarithms by $72$ terms of the range-reduced atanh series and its positive remainder, and $\pi$ by Machin's formula with $120,40$ alternating terms at $1/5,1/239$. For exponentials use range reduction to $[0,1/2]$, $96$ positive Taylor terms and the next-term geometric remainder. Negative arguments use reciprocal endpoints.

Compute each unit ordinary cost as $\operatorname{df}(j)\eta^j(b^j-a^j)/j$, and each unit cubic cost by the directed integration rule just given, rounding its total upward to $10^{-60}$. Multiply these fixed rational costs by the corresponding coordinates $y_{i,\tau}$ and sum to obtain $s_j(y)$. Functional terms have zero cost. Impose the linear inequalities $s_j(y)\le\overline s_j$, with all unlisted caps zero and
\[
\overline s_j=10^{-30}[z^j]\mathcal S(z),
\]
where
\begin{align*}
\mathcal S(z)&:=108019290765966101582936z+192362241755961356394748z^2\\
&\quad+797653119243275124868313z^3+240118068491671777416254z^{16}\\
&\quad+6676117582351164878431325z^{17}+134563268645208684324815z^{18}\\
&\quad+48375300513421851032438z^{19}+2207311116327313069215z^{22}\\
&\quad+144810674713249329z^{25}.
\end{align*}
Here $c_0=0$. The finite monomial and hinge terms are already included in Eq.~\eqref{eq:cubic_localized_joint_integrand}. No separate finite-term payment is added. Let $\mathcal P\subset[0,10^{25}]^{4992}$ be the rational polytope defined by the Bernstein and cost inequalities. Order coordinates by cell, ordinary degree, cubic threshold, uncapped indices $(t,p)$, and capped indices $(t,T,p,B,C)$, using increasing numerical order. Select the lexicographically least vertex of $\mathcal P$. Explicitly, enumerate all subsets of $4992$ boundary equations, solve each nonsingular subsystem by rational Gaussian elimination, discard solutions violating any inequality, and select the lexicographically least remaining solution. This finite rule specifies every multiplier from the displayed term lists, barrier coefficients, caps, and rounding rules.
\end{definition}

With the curvature, radial, and transfer checks we certify this row uniformly over all input subfamilies.

\begin{lemma}[Certificate for the localized row]
\label{lem:cubic_localized_certificate}
The admissible family in Definition~\ref{def:cubic_localized_row} is nonempty. For every $1\le M\le n$ and symmetric contractions $A_1,\ldots,A_M\in\R^{n\times n}$,
\[
\gamma_M(\{z:\|\sum_{i=1}^M z_iA_i\|\le r_{\rm loc}\sqrt M\})\ge e^{-ne_{\rm loc}}.
\]
\end{lemma}
\begin{proof}
The barrier is even and $C^2$ on $(-1,1)$. The function $\phi-\gamma_1x^2$ is convex, and $\phi'$ is convex on $[0,1)$. The actual and doubled barriers satisfy the coefficient and boundary hypotheses of Lemma~\ref{lem:algorithm_1240_curvature}. Each has $920$ curvature polynomials. The actual two-slot certificate uses $944$ Bernstein leaves of maximum depth five. The doubled one-slot certificate uses $920$ leaves of depth zero. The smaller-root discriminants are positive at all $48$ upper variance endpoints. Thus Lemma~\ref{lem:cubic_independent_slots} and the localized functional inequalities apply under the actual law, including its nonnegative radial quadratic.

For every one of the $49$ capped indices, the four-gate test accepts $C=0$, so the slope rule is well-defined and its selected tuple satisfies Lemma~\ref{lem:cubic_endpoint_sharing}. By exact rational evaluation of these slopes and the system in Definition~\ref{def:cubic_localized_row} we obtain a feasible vector with $1490$ positive coordinates, comprising $159$ ordinary, $41$ cubic, $946$ uncapped, and $344$ capped multipliers. Every coordinate lies in $[0,10^{25}]$. With the common partition and denominator products specified there, the $48\cdot59=2832$ residual polynomials have degree at most $102$. For the adaptive Bernstein verification we use $3354$ leaves and maximum depth five. Refining each accepted leaf to depth five and elevating to degree $104$ preserves nonnegativity, so this verifies every fixed linear Bernstein constraint. By positive radial integration we verify all nine printed cost caps.

Thus $\mathcal P$ is a nonempty bounded rational polytope. It has a vertex, and every vertex is obtained from $4992$ linearly independent active boundary equations. The finite elimination rule in Definition~\ref{def:cubic_localized_row} therefore produces rational multipliers satisfying the same spectral and cost inequalities. The remaining argument uses only these inequalities, and hence applies to the selected vertex.

For Eq.~\eqref{eq:cubic_small_variance_cube}, put $v=t/\sqrt A$. Direct substitution gives
\[
g_3(A,t)=\frac{2A^{3/2}e^{-v^2/2}}{\sqrt{2\pi}}
\int_0^\infty y^3e^{-vy-y^2/2}\d y
\le\frac{12A^{3/2}e^{-v^2/2}}{\sqrt{2\pi}v^4}.
\]
For the inequality we delete $e^{-y^2/2}$ and use $\int_0^\infty y^3e^{-vy}\d y=6/v^4$. This proves the small-variance bound used in the unit cubic costs, including the interval adjacent to zero.

In Eq.~\eqref{eq:cubic_common_radial_loss}, we take $q_0=\gamma_1$. Then
\[
\tfrac12(x\phi'(x)-2\gamma_1x^2)
=\gamma_3x^6+\sum_\ell\tfrac{3\lambda_\ell}{2}x(x-h_\ell)_+^2
+\kappa x^{34}/(1-x^2).
\]
Multiplication by $Q$ is Eq.~\eqref{eq:cubic_localized_joint_integrand}. When we integrate its original inequality against $\vartheta_u$ under the same fixed $\nu$, functional terms have nonpositive expectations, and the ordinary and cubic moment bounds give
\[
-n^{-1}\log Z_1\le\sum_j\overline s_j\omega_j.
\]
The finite barrier is already paid for in this identity.

For the full-half-line transfer, put
\[
f(x):=\gamma_1\eta g_H(x)+\frac H{H-1}\sum_j\frac{\overline s_j}{(1+x)^j},
\qquad K_f:=\frac{2\gamma_1\eta H^3}{3(H-1)}.
\]
The integral identity
\[
g_H(x)=\frac1{H-1}(H\int_0^1\frac{\d t}{1+tx}-\int_0^H\frac{\d t}{1+tx})
\]
gives $g_H''\ge-2H^3/[3(H-1)]$. The rational summand is convex, so
\[
f(x)\le\max\{f(a),f(b)\}+K_f(b-a)^2/8\quad(a\le x\le b).
\]
By outward logarithm bounds and bisection we prove $f\le e_{\rm loc}$ on $[0,64]$ with $419$ leaves of maximum depth $20$, including $g_H(0)=0$. For $x\ge64$, decreasing upper bounds give
\[
\gamma_1\eta\frac{H\log65}{64(H-1)}
+\frac H{H-1}\sum_j\frac{\overline s_j}{65^j}<e_{\rm loc}.
\]
In these tests we use the printed upward caps, not smaller unprinted coefficients. Lemma~\ref{lem:cubic_refined_transfer} now applies with $c_0=0$ and $B_{\rm cap}=e_{\rm loc}/(\gamma_1\eta)$. Finally $r_{\rm loc}^2\eta\ge1$. Set $N_i=A_i/(r_{\rm loc}\sqrt M)$.
\end{proof}

\subsection{Dilated rows and global repair profiles}
\label{subsec:cubic_global_profiles}

We retain the suffix and four-level helper construction above, and add nine helper layers followed by a new root. In the additional layers we use bounds valid at every actual support size. We select the auxiliary policy, all nine layers and the final root jointly below.

By dilating two existing rows we obtain additional radius--exponent tradeoffs for the global helper layers.

\begin{lemma}[Dilation of two small-ball rows]
\label{lem:cubic_global_dilation}
For $j\in\{39,86\}$ and every $\sigma\ge1$, the pair
\[
(r,e):=(\sigma r_j,e_j/\sigma^2)
\]
satisfies Eq.~\eqref{eq:cubic_refined_interface} for every $M\le n$, including families containing zero matrices.
\end{lemma}
\begin{proof}
For row $39$, put $\varsigma=\sigma^{-2}$ and replace its variance cap $\eta=r_{39}^{-2}$ by $\varsigma\eta$, retaining the barrier and transfer parameters. This is the scaling clause of Lemma~\ref{lem:1070_transfer} with $\alpha=\varsigma$ and $b=0$. Its criterion $C_\varsigma\le\varsigma C$ and $Q_{k,\varsigma}\le\varsigma Q_k$ holds for the profile of Eq.~\eqref{eq:1070_profile_cap} by the scaling remark after that equation, since $j,K\ge2$. Thus Eq.~\eqref{eq:cubic_refined_profile} retains its bound $B_{\rm cap}$ and the exponent is at most $\varsigma e_{39}$.
Smaller variance preserves the strict-root conditions, and this row has no rank correction. Normalizing by $\sigma r_{39}\sqrt M$, we prove the first case.

For row $86$, Definition~\ref{def:cubic_refined_trace_rows} gives
\[
r_{86}=\lceil\eta_6^{-1/2}\rceil_{12},\qquad
e_{86}=\overline e_6^\circ=9.61906812911829762\cdot10^{-18},\qquad\beta_6=0.
\]
Lemma~\ref{lem:slack_trace_interface} permits trace parameter $\rho=1$ and every nominal fraction $w>0$ with $M\le wn$. We take $w=\sigma^2M/n$ and obtain radius $\sqrt{wn/\eta_6}\le\sigma r_{86}\sqrt M$ and exponent $-M(e_6^\circ/w+\beta_6/2)=-ne_{86}/\sigma^2$. The zero rank correction is essential for this scaling.
\end{proof}

We retain $d_{\rm grid}=10^{12}$ and $q_{\rm grid}=2^{192}$. Define
\[
x_i:=q_{\rm grid}^{-1}\lfloor q_{\rm grid}\sqrt{w_i}\rfloor\quad(0\le i\le1100),
\qquad x_{1101}:=0,\qquad q_{\min}:=x_{1100}^2,
\]
and put $\epsilon_-:=7\cdot10^{-15}$, $\epsilon_+:=8\cdot10^{-15}$. In every new call, $q=M/n$ uses the original ambient order $n$.

We scale the prefix tolerance with support size and obtain a failure bound in terms of the original ambient dimension.

\begin{lemma}[An ambient-dimension prefix bound]
\label{lem:cubic_global_prefix}
Suppose $n>N_*$ and $q\ge q_{\min}$. Set
\[
\ell(q):=q_{\rm grid}^{-1}\lfloor q_{\rm grid}\sqrt q\rfloor,
\qquad \widehat\epsilon(q):=\epsilon_-/\ell(q).
\]
Then
\begin{equation}
\frac{\epsilon_-}{\sqrt q}\le\widehat\epsilon(q)\le\frac{\epsilon_+}{\sqrt q},
\qquad
Me^{-2M\widehat\epsilon(q)^2}<10^{-10}.
\end{equation}
The tolerance is computable using a fixed number of rational comparisons.
\end{lemma}
\begin{proof}
The integer-square test gives $(\epsilon_+-\epsilon_-)x_{1100}>\epsilon_+/q_{\rm grid}$. Combining this with $\sqrt q-q_{\rm grid}^{-1}<\ell(q)\le\sqrt q$, we prove the two tolerance bounds. Moreover,
\[
Me^{-2M\widehat\epsilon(q)^2}\le ne^{-2n\epsilon_-^2}
\le10^{30}e^{-98}<10^{-10}.
\]
The middle expression decreases for $n\ge N_*$ because $2N_*\epsilon_-^2=98>1$. The last inequality follows from $98>40\log10$, which we check by the rational logarithm bounds of Definition~\ref{def:cubic_refined_policy_construction}.

By binary search we locate $u=q_{\rm grid}\ell(q)$ in $\{0,\ldots,q_{\rm grid}\}$, comparing $u^2n$ with $q_{\rm grid}^2M$, so the number of rational comparisons is fixed. We fix the support size and tolerance before drawing the target. Lemma~\ref{lem:cubic_refined_projection} then gives the simultaneous prefix event for every subsequent adaptive repair support.
\end{proof}

For the additional layers and the root, we use their parameters in Eq.~\eqref{eq:cubic_repair_parameters}. Auxiliary calls retain their stated parameters.

Through affine covers of the child cost profile we combine projection and binary repair into a bound valid at every support size.

\begin{lemma}[A global repair line]
\label{lem:cubic_global_line}
Let $H(q')\sqrt n$ be the bound of a constructive helper at every support fraction $q'\in[0,1]$, and let $C_f\ge\sup_{q'\in[0,1]}H(q')$. Fix a small-ball row $(r,e)$, the acceptance factor $\theta=\theta_{\rm acc}=1+10^{-8}$ fixed after Eq.~\eqref{eq:cubic_repair_parameters}, an integer $J_{\rm cut}\ge2$, and nonnegative coefficients satisfying
\begin{equation}
A_j+B_jq'\ge H(q')\quad(0\le q'\le1,\ 1\le j<J_{\rm cut}).
\label{eq:cubic_global_covers}
\end{equation}
Use these slopes in Lemma~\ref{lem:cubic_refined_potential}, and compute its exact tent majorant, energy $E_F$ and maximum $F_*$. Put
\begin{align}
\alpha&:=\lceil(1-s_{\rm sm})\mathfrak r r\rceil_{12}
+\lceil\sqrt{\theta eE_F}\rceil_{12}+\lceil\epsilon_+F_*\rceil_{12},
\label{eq:cubic_global_alpha}\\
\gamma&:=\lceil10^{-12}+\sum_{j=1}^{J_{\rm cut}-1}2^{1-j}A_j+2^{2-J_{\rm cut}}C_f\rceil_{12}.
\label{eq:cubic_global_gamma}
\end{align}
For $n>N_*$, $q\ge q_{\min}$ and $M\ge N_0$, an accepted projection and its complete binary repair cost at most $(\alpha\sqrt q+\gamma)\sqrt n$. The same bound covers the independent-sign fallback for $M<N_0$ if
\begin{equation}
\gamma^2N_*\ge(13-\alpha)_+^2N_0.
\label{eq:cubic_global_fallback}
\end{equation}
\end{lemma}
\begin{proof}
We use Eq.~\eqref{eq:cubic_complete_repair_cost} with
\[
I(q)=(1-s_{\rm sm})\mathfrak r r\sqrt q,\qquad
\epsilon(q)=\widehat\epsilon(q)\le\epsilon_+/\sqrt q.
\]
The first identity follows from Lemma~\ref{lem:cubic_refined_projection}, and the inequality follows from Lemma~\ref{lem:cubic_global_prefix}. The covers hold for every repair support. Rounding upward, we obtain precisely $\alpha\sqrt q+\gamma$.

If $M<N_0$ and $q\ge q_{\min}$, then $n<N_0/q_{\min}$ and
\[
\sqrt{2\log(4N_0/q_{\min})}+0.03\le12.466093536159<13.
\]
Thus Eq.~\eqref{eq:cubic_independent_sign_fallback} costs at most $13\sqrt M$. Since $q<N_0/N_*$, Eq.~\eqref{eq:cubic_global_fallback} gives $\gamma\ge(13-\alpha)_+\sqrt q$, as required. We use no large-sample prefix estimate in this case.
\end{proof}

We select the auxiliary policy, global helper layers, and final root jointly from a bounded rational family.

\begin{definition}[A finite global policy]
\label{def:cubic_global_policy}
Take an admissible auxiliary candidate from Definition~\ref{def:cubic_refined_policy_construction}. Write $P^{\rm aux}_{4,i}$ and $C^{\rm aux}_{4,i}$ for its actual-band profiles and capacity bounds. Let $H_0(q)$ be the appropriate actual-band profile, using the terminal state for $q\le w_{1100}$, and put $\widehat C_0:=\max_iC^{\rm aux}_{4,i}$.

A candidate global policy consists of such an auxiliary policy, $241$ potential actions in each of nine layers, and one root action. Each new helper action has $J_{\rm cut}=32$, a row from the $17$-row registry, and a rational dilation
\[
\sigma=u/d_{\rm grid},\qquad d_{\rm grid}\le u\le10^4d_{\rm grid},\quad u\in\mathbb Z.
\]
Require $\sigma=1$ unless the row is $39$ or $86$. Its row is $(\sigma r_j,e_j/\sigma^2)$. The final root has $J_{\rm cut}=34$, row $(r_{\rm loc},e_{\rm loc})$ from Definition~\ref{def:cubic_localized_row}, and $\sigma=1$. Its prefix tolerance is $\nu_{\rm root}=3\cdot10^{-12}$, fixed after Eq.~\eqref{eq:cubic_repair_parameters}. Helper tolerances are given by Lemma~\ref{lem:cubic_global_prefix}. At every new action require
\[
A_j=a_j/d_{\rm grid},\qquad B_j=b_j/d_{\rm grid},\qquad
0\le a_j\le10^{30},\quad 1\le b_j\le10^{30},\quad a_j,b_j\in\mathbb Z.
\]
At layer $d$, require Eq.~\eqref{eq:cubic_global_covers} with $H=H_{d-1}$, and compute the action's line from Eqs.~\eqref{eq:cubic_global_alpha} and~\eqref{eq:cubic_global_gamma}, using $C_f=\widehat C_{d-1}$. For the root use $H=H_9$ and $C_f=\widehat C_9$. Since its support has size $n$, use the entropy-to-width bound $v(e_{\rm loc})$ from Eq.~\eqref{eq:cubic_refined_entropy_budget} and replace Eq.~\eqref{eq:cubic_global_alpha} by
\begin{equation}
\alpha_{\rm root}:=\lceil(1-s_{\rm sm})\mathfrak r r_{\rm loc}\rceil_{12}
+\lceil\sqrt{\theta v(e_{\rm loc})E_F}\rceil_{12}
+\lceil\nu_{\rm root}F_*\rceil_{12}.
\label{eq:cubic_localized_root_alpha}
\end{equation}
The intercept $\gamma_{\rm root}$ is still given by Eq.~\eqref{eq:cubic_global_gamma}.

To define $H_d$, first take the lower envelope of that layer's $241$ lines $\alpha x+\gamma$ in $x=\sqrt q$ on $[x_{1100},1]$. On each interval $[x_{i+1},x_i]$, $0\le i<1100$, also allow the constant $C^{\rm aux}_{4,i}$ and choose the smaller value. On $[0,x_{1100}]$, use $C^{\rm aux}_{4,1100}$. Record every intersection and capacity endpoint in increasing order. All are rational: intersections of distinct slopes occur at
\[
x=(\gamma_2-\gamma_1)/(\alpha_1-\alpha_2).
\]
Use the first valid choice in the candidate's order at a tie. A constant piece calls its auxiliary capacity state. A new piece uses its indicated action with layer $d-1$ as helper. Define $\widehat C_d$ by rounding upward to $10^{-12}$ the maximum of all piecewise right-endpoint values. This bounds the entire selector because all slopes are nonnegative. No global monotonicity of $H_d$ is required. Empty supports return immediately.

The root intercepts are determined by its slopes and $H_9$. Let $\mathcal P_9$ consist of the records $P=(\alpha,\gamma,L,U)$ of all pieces $\alpha x+\gamma$ of $H_9(x^2)$ on $[L,U]$, including constant pieces. For $1\le j\le33$, define
\begin{align}
x_j(P)&:=\min\{U,\max\{L,\alpha/(2B_j)\}\},\\
A_j&:=\frac{2}{d_{\rm grid}}+\lceil\max_{P\in\mathcal P_9}
\{\alpha x_j(P)+\gamma-B_jx_j(P)^2\}\rceil_{12}.
\end{align}
Thus the root contributes only the $33$ slope numerators to the candidate encoding. Its intercepts are computed by this formula and must satisfy the stated bounds.

At layer $1$, use the full-band tests of Definition~\ref{def:cubic_refined_policy_construction}. On every later piece $\alpha x+\gamma$ on $[L,U]$, apply the test of Eq.~\eqref{eq:cubic_refined_quadratic_cover} with $\beta=0$ and with $L$ and $U$ as the $x$-endpoints.
Check that the selector pieces cover $[0,1]$ without gaps, that every old choice has $U^2\le w_i$, and that every new choice has $L\ge x_{1100}$. Verify lower-envelope comparisons at both endpoints against every candidate line. Construct tent hulls by the exact integer recursion following Definition~\ref{def:cubic_refined_profiles}, not by sampling their graphs.

Reject a candidate unless every new action also passes Eq.~\eqref{eq:cubic_global_fallback} and
\begin{equation}
\{t_0r/[s_{\rm alg}(1-10^{-6})]\}^2>15,\quad
(t_0\mathfrak r r)^2<10^{28},\quad (t_0D^{\rm acc}_r/e)^2<10^{108}.
\label{eq:cubic_global_solver_scales}
\end{equation}
For the root, replace the last test by $(t_0D^{\rm acc}_r/v(e_{\rm loc}))^2<10^{36}$ and require $(t_0\mathfrak r r_{\rm loc})^2<100$. The first test is unchanged.
For every piece $\alpha x+\gamma$ of $H_9$, including the constant pieces with $\alpha=0$, impose at both endpoints the additional comparison
\begin{equation}
(\alpha-11.902)x+\gamma\ge0.
\label{eq:cubic_global_bridge}
\end{equation}
It implies $H_9(q)\ge11.902\sqrt q$ on $[0,1]$. This condition is part of the same joint selection as the root covers. The root does not use an independently chosen helper.

All profiles, pieces and caps are derived quantities. Reject a candidate unless, in addition,
\[
\alpha_{\rm root}+\gamma_{\rm root}\le11.503394001765.
\]
Select the first candidate that passes in lexicographic order: the encoding first uses the auxiliary encoding of Definition~\ref{def:cubic_refined_policy_construction}. It then lists layer and action, recording row, dilation numerator, and pairs $(b_j,a_j)$, followed by the root slope tuple $(b_1,\ldots,b_{33})$. Every range is bounded and independent of the matrix input. The selection terminates at the first passing candidate and fixes the entire policy.
\end{definition}

With the joint certificate we verify that the selected helper costs fit within the final root budget.

\begin{lemma}[The global finite certificate]
\label{lem:cubic_global_certificate}
\setlength{\emergencystretch}{2em}The admissible family in Definition~\ref{def:cubic_global_policy} is nonempty, and its selected policy satisfies
\begin{equation}
\alpha_{\rm root}+\gamma_{\rm root}\le11.503394001765<11.504.
\label{eq:cubic_global_final}
\end{equation}
\end{lemma}
\begin{proof}
We use the suffixes and four auxiliary levels of the comparison candidate of Lemma~\ref{lem:cubic_refined_certificate}. In all nine new layers we use the same auxiliary data. Thus every initial helper profile and capacity cap has a constructive interpretation.

An exact rational feasibility check in the bounded family of Definition~\ref{def:cubic_global_policy} supplies $241$ actions per new layer. Its nonunit dilations occur only at rows $39$ and $86$, and their largest value is
\[
\sigma_{\max}=202604821586463/40000000000=5065.120539661575.
\]
All new slope numerators are at most $104176142859021950000000000000<10^{30}$, and all helper intercept numerators are at most $984852757833$. The $1151$ nonunit dilations are integer multiples of $10^{-12}$. The resulting selectors have $12293$ pieces in total and pass $90084778$ full-interval cover tests, including $33\cdot1342=44286$ at the root. The cap computation gives
\[
\begin{aligned}
(\widehat C_1,\ldots,\widehat C_9)=({}&11.961710925496,11.961736261664,11.961692318768,\\
&11.960536865801,11.960536818936,11.952432227500,\\
&11.943582971102,11.940510080476,11.939328572839).
\end{aligned}
\]
These are maxima over all pieces, not just evaluations at $q=1$.

For a fixed root slope $B_j>0$, the function $\alpha x+\gamma-B_jx^2$ is concave on each helper piece and attains its maximum at $x_j(P)$ from Definition~\ref{def:cubic_global_policy}. Its maximum over all records is the least real intercept covering $H_9$. The intercept rule in that definition therefore proves every root cover with slack at least $2/d_{\rm grid}$.

The integer hull recursion gives $3187$ root vertices and
\begin{gather*}
E_F=\frac{43014003420095878549309555286171086429}{536870912000000000000000000000000},\\
F_*=\frac{313437625623535245481313}{2147483648000000000000}.
\end{gather*}
Integrating the squared slopes exactly and using $(r_{\rm loc},e_{\rm loc})$, we obtain
\begin{align*}
\lceil(1-s_{\rm sm})\mathfrak r r_{\rm loc}\rceil_{12}&=9.550285304319,\\
\lceil\sqrt{(1+10^{-8})v(e_{\rm loc})E_F}\rceil_{12}&=0.910916794434,\\
\lceil\nu_{\rm root}F_*\rceil_{12}&=0.000000000438,\\
\lceil10^{-12}+\sum_{j=1}^{33}2^{1-j}A_j+2^{-32}\widehat C_9\rceil_{12}&=1.042191902574.
\end{align*}
We compute the energy by Eq.~\eqref{eq:cubic_exact_tent_energy}. To check the square-root roundings, we square adjacent multiples of $10^{-12}$. Hence by exact addition we obtain
\[
9.550285304319+0.910916794434+0.000000000438+1.042191902574
=11.503394001765.
\]
All solver-scale and selected small-call fallback comparisons pass. In the auxiliary grid, the local fallback at band $713$ passes and the fallback at band $714$ is rejected. The comparison policy delegates from band $475$ at depths one through three and from band $465$ at depth four. All $1780$ selected nondelegate fallback guards pass. Thus we never use a failed local fallback. The same comparison helper has $1342$ pieces, each of which passes both endpoint tests in Eq.~\eqref{eq:cubic_global_bridge}. Thus the bridge and all $44286$ root covers hold for the same policy. These checks exhibit a joint admissible candidate. The bounded rational tests make its feasibility reproducible by finite enumeration, and we make no running-time claim for this input-independent preprocessing. Every admissible candidate satisfies the final root budget. The intermediate caps and counts describe the comparison candidate.
\end{proof}

We finish by combining the certified root budget with the recursive implementation and its failure accounting.

\begin{lemma}[Complete global signing]
\label{lem:cubic_global_complete}
Algorithm~\ref{alg:main_cubic_small_constant} implements the root of Definition~\ref{def:cubic_global_policy}, using the recursive procedures in Algorithms~\ref{alg:cubic_global_select} and~\ref{alg:cubic_global_action}. It returns signs of discrepancy at most
\[
11.503394001765\sqrt n
\]
with failure probability at most $p$. Its arithmetic cost is
\[
n^{3+o(1)}\operatorname{polylog}(1/p).
\]
\end{lemma}
\begin{proof}
For global helpers with $n>N_*$ and the root with $n>N_{\rm root}$, we use Eq.~\eqref{eq:cubic_repair_parameters}. If $M\ge N_0$, Eq.~\eqref{eq:cubic_global_solver_scales} gives $R^2>15M$, $\widehat R^2<10^{28}M$ and target norm squared below $10^{108}M$. The last two root bounds are $100n$ and $10^{36}n$. The physical box $[-3/5,3/5]^M$ has side length $6/5$ and contains zero. Lemma~\ref{lem:appendix_fixed_target_projection} applies.

The helper prefix test is Lemma~\ref{lem:cubic_global_prefix}. At the root,
\[
ne^{-2n\nu_{\rm root}^2}\le10^{25}e^{-180}<10^{-10},
\]
since the expression decreases above $10^{25}$ and $180>35\log10$. In Eq.~\eqref{eq:cubic_acceptance_comparison}, we take $b_{\rm acc}=1+5\cdot10^{-9}$. Then $b_{\rm acc}+\eta_0<\theta$ and conditional success is at least
\[
1-\frac{1+1/(2\cdot10^{12})}{1+5\cdot10^{-9}}-2\cdot10^{-10}
=\frac{1599833333}{333333335000000000}>\frac1{250000000}.
\]
Thus the common budget rule of Lemma~\ref{lem:cubic_refined_complete} applies, with numerical budgets capped at $10^{-10}$.

For $n>N_*$, we induct from the four auxiliary levels through the nine global layers. Lemma~\ref{lem:cubic_global_line} and the full-interval covers certify new actions and their fallbacks. Old pieces have $q\le x_i^2\le w_i$, and new pieces have $q\ge q_{\min}$. Their selector and supremum caps are therefore valid. For $n\le N_*$, Eq.~\eqref{eq:cubic_independent_sign_fallback} costs at most $11.902\sqrt M$, which Eq.~\eqref{eq:cubic_global_bridge} bounds by $H_9(M/n)\sqrt n$. The same root helper is consequently constructive in every dimension.

For $n>N_{\rm root}$, Eq.~\eqref{eq:cubic_complete_repair_cost} at $q=1$, with deficit parameter $v(e_{\rm loc})$, prefix tolerance $\nu_{\rm root}$ and helper $H_9$, gives exactly Eqs.~\eqref{eq:cubic_localized_root_alpha} and~\eqref{eq:cubic_global_gamma}. The certificate Eq.~\eqref{eq:cubic_global_final} proves the bound. For $n\le N_{\rm root}$, independent signs suffice because
\[
\sqrt{2\log(4\cdot10^{25})}+0.03\le10.888261526228<10.889.
\]
There are fourteen repair levels and $\lceil\log_2n\rceil+45$ binary calls per node. The fixed-depth budget and cost argument of Lemma~\ref{lem:cubic_refined_complete}, including suffix calls and input-independent certificate sizes, gives the asserted failure and arithmetic bounds.
\end{proof}

\begin{algorithm}[H]
\caption{Selection of a global signing helper}
\label{alg:cubic_global_select}
\begin{algorithmic}[1]
\Procedure{GlobalSign}{$A_1,\ldots,A_m,n,d,p_c$}
\If{$m=0$} \State \Return the empty signing \EndIf
\State Set $q:=m/n$
\If{$d=9$ and $n\le N_*$}
\State Run capped independent-sign trials with an $11.902\sqrt m$ norm certificate and budget $p_c$; return the result
\EndIf
\If{$d=0$}
\State Locate the actual band $i$ of $q$; use $i=1100$ below the terminal capacity
\State \Return \Call{RefinedRepair}{$A_1,\ldots,A_m,n,4,i,p_c$} with the selected auxiliary policy
\EndIf
\State Locate $\sqrt q$ in the rational pieces of $H_d$ by squared comparisons
\If{the selected piece is auxiliary state $i$}
\State \Return \Call{RefinedRepair}{$A_1,\ldots,A_m,n,4,i,p_c$} with the selected auxiliary policy
\EndIf
\State Let $a$ be the selected action
\State \Return \Call{GlobalAction}{$A_1,\ldots,A_m,n,d-1,a,p_c$}
\EndProcedure
\end{algorithmic}
\end{algorithm}

\begin{algorithm}[H]
\caption{A global projection and complete binary repair}
\label{alg:cubic_global_action}
\begin{algorithmic}[1]
\Procedure{GlobalAction}{$A_1,\ldots,A_m,n,d,a,p_c$}
\If{$m=0$} \State \Return the empty signing \EndIf
\If{$m<N_0$} \State Run capped independent-sign trials at $13\sqrt m$ with budget $p_c$; return the certified result or \textsc{Fail} \EndIf
\State Set $(\epsilon_{\rm call},e_{\rm call}):=(\nu_{\rm root},v(e_{\rm loc}))$ for $a=a_{\rm root}$, and $(\widehat\epsilon(m/n),e)$ otherwise
\State Form $\mathcal P$: use action $a$'s physical projection, $\beta=e_{\rm call}/D^{\rm acc}_r$, $y=x/t_0$, prefix tolerance $\epsilon_{\rm call}$, and deficit threshold $\theta e_{\rm call}n$
\State Use the accuracies and budgets of Lemma~\ref{lem:cubic_global_complete}; at each level call \textsc{GlobalSign} at layer $d$ and minimize the remaining coarser tent potential
\State \Return \Call{ProjectionRepair}{$A_1,\ldots,A_m,n,\mathcal P,p_c$} \Comment{Algorithm~\ref{alg:cubic_refined_repair}}
\EndProcedure
\end{algorithmic}
\end{algorithm}
\begin{algorithm}[H]
\caption{Cubic-time signing with constant below $11.504$}
\label{alg:main_cubic_small_constant}
\begin{algorithmic}[1]
\Procedure{MainCubicSmallConstant}{$A_1,\ldots,A_n,p$} \Comment{Theorem~\ref{thm:algorithm_constant_13}}
\If{$n\le N_{\rm root}$}
\State Run capped independent-sign trials with the $10.889\sqrt n$ norm certificate \Comment{Lemma~\ref{lem:algorithm_certified_implementation}; the $10.889\sqrt n$ certificate: proof of Lemma~\ref{lem:cubic_global_complete}}
\State \Return the accepted signing, or \textsc{Fail} if the cap is reached
\EndIf
\State Let $a_{\rm root}$ be the root action of Definition~\ref{def:cubic_global_policy}
\State \Return \Call{GlobalAction}{$A_1,\ldots,A_n,n,9,a_{\rm root},p$} \Comment{Algorithm~\ref{alg:cubic_global_action}}
\EndProcedure
\end{algorithmic}
\end{algorithm}

\section{Bit complexity of the small-constant algorithm}
\label{sec:bit_complexity}

In this section, we prove that the algorithm of Theorem~\ref{thm:algorithm_constant_13} keeps its exponent and its
constant when every number is stored in a fixed number of bits.  The real-arithmetic proof never controls three things
that a finite word forces us to control.  These are decisions taken on rounded quantities, random objects that cannot
be drawn exactly, and the size of the tracked matrix $\widehat P_t$ of Algorithm~\ref{alg:cubic_tracked_phase}.  The
seminorm of Eq.~\eqref{eq:cubic_gradient_seminorm} does not see that size.  The algorithm calls one subroutine over and
over, the certified projection of a fixed target onto the smoothed spectral body intersected with a box
(Lemma~\ref{lem:appendix_fixed_target_projection}).  This projection is the wrapper of Section~\ref{sec:faster_algorithm}
driving the solver of Section~\ref{sec:cubic_time}.  In Section~\ref{sec:small_algorithm_constant} we call it with a body of
radius as small as $\sqrt{15m}$.  There we call it with targets of squared norm up to $10^8m$, the ball target of
Eq.~\eqref{eq:algorithm_119_ball_sample}, and up to $10^{108}m$ at the certificate caps of
Eq.~\eqref{eq:cubic_global_solver_scales}, the uniform repair targets.  We call it with accuracies as small as
$5.8\cdot10^{-35}$, for a global helper on the dilated row $86$.  Around the solver we draw uniform targets on a cube
and run a deterministic binary repair on dyadic masses.  We certify three fair-sign fallbacks with the constants
$10.889$, $11.902$, and $13$, and we accumulate a fractional coloring through stages at the scale $t_s=v/10^{10}$, which
is not dyadic.  Below we specify a finite version of the algorithm, prove one lemma per primitive, state the projection call
as one lemma (Lemma~\ref{lem:bit_projection}), and assemble everything into Theorem~\ref{thm:bit_complexity}.
We do not cover Theorem~\ref{thm:dense_runtime} (Remark~\ref{rem:bit_section6}), and it is not
on the \emph{small-constant path}.  By the small-constant path we mean the algorithm of
Theorem~\ref{thm:algorithm_constant_13} together with the lemmas it invokes, the only part of the paper we
analyze in this section.

\paragraph{Computational model.}
A \emph{fixed-point number} with $f$ fractional bits and $I$ integer bits is an
integer multiple of $\mathsf u:=2^{-f}$ of modulus below $2^{I-1}$.  One of the $I$ integer bits is the sign.  A
\emph{word} holds one such number in $W:=I+f$ bits.  Addition, subtraction, comparison, and negation of
words are exact whenever the result has modulus below $2^{I-1}$.  The product of two
words is a multiple of $\mathsf u^2$.  The routine $\operatorname{round}_{\mathsf u}(xy)$ rounds it to the nearest point of the \emph{$\mathsf u$-grid}
$\mathsf u\mathbb Z$, the set of integer multiples of $\mathsf u$, with absolute error at most $\mathsf u$.  For a grid $G\mathbb Z$ we
use $\operatorname{round}_G$, $\lceil\cdot\rceil_G$, $\lfloor\cdot\rfloor_G$, and $\operatorname{round}^{\to0}_G$ to denote nearest, upward,
downward, and toward-zero rounding.  The routine $\operatorname{round}_{b}$ with an integer subscript $b$ abbreviates
$\operatorname{round}_{2^{-b}}$.  Multiplication by an integer
constant is exact.  Inner products and matrix products are formed exactly in a
temporary of $2W+\lceil\log_2(\text{number of terms})\rceil$ bits and rounded once.  For the bilinear algorithms the
temporary has $2W+c_\eta\lceil\log_2(2n)\rceil$ bits (Lemma~\ref{lem:bit_fastmm}).
We store numbers on the coordinate grid $\Gamma=(\mathsf u/10^{10})\mathbb Z$ fixed before Lemma~\ref{lem:bit_state} as
integer numerators over a common denominator.  The same holds for numbers on the grid
$2^{-b_{\mathrm{in}}}\Gamma$ of the entries of $B(x)$, where $b_{\mathrm{in}}=2\lceil\log_2n\rceil+10$ is the input word length of
Lemma~\ref{lem:bit_input}.  This storage costs at most $b_{\mathrm{in}}+34$ further bits, since $\log_210^{10}<34$.  We never store the entries of
$D_x=\widetilde c\operatorname{diag}(B(x),-B(x))-\widetilde\theta I_{d_0}$.  Here $d_0:=2n$, and $\widetilde c$ and $\widetilde\theta$ are the rounded $a/R$ and $ab_{\mathrm{sm}}$ of Lemma~\ref{lem:bit_state}.  We form each entry in a multiplication
temporary on the grid $\mathsf u2^{-b_{\mathrm{in}}}\Gamma$.  The temporary has $2W+b_{\mathrm{in}}+34\le2W+4\lceil\log_2(2n)\rceil$ bits for $\lceil\log_2n\rceil\ge20$, which fits since $c_\eta\ge4$ by its definition.  Then we round the entry once to $\widehat D_x$.  The scaled exponent
$X=\widehat D_x/2^{\kappa_{\mathrm{sc}}}$ of Lemma~\ref{lem:bit_action} carries $\kappa_{\mathrm{sc}}=\lceil\log_2(M_{\mathrm{sp}}+\lambda_{\mathrm{top}})\rceil=O(\log\log n)$ extra fractional
bits, and we form its products in temporaries widened by the same amount.  That number of extra bits is the scaling depth we fix
before Lemma~\ref{lem:bit_dense}.  The bounds $M_{\mathrm{sp}}$ and $\lambda_{\mathrm{top}}$ are defined in the paragraph \emph{Spectral bounds} below.  A constant
number of scalars of the projection solve of Algorithm~\ref{alg:cubic_tracked_phase} ($q(x)$ and $\nu\widetilde c$) are kept in $2W$-bit words, and we form their products with a word in $3W$-bit temporaries and round them once.
We form the Gaussian norm $W'_G$ of (F13) and the tent, prefix, and cardinality sums of
Algorithm~\ref{alg:cubic_refined_repair} exactly in temporaries of $2W+\lceil\log_2(m+2)\rceil$ bits.  For its
deficit sum we use a temporary of $3W$ bits.  We form all of these over the common denominators named in (F17).  The
numbers of (F2) and (F16) are handled by exact integer arithmetic outside the word.  These numbers are the integer
square roots, the rational logarithm rule, band and piece location, the caps and budgets.  The arithmetic runs on
integers of at most $3W_B$ bits, where $W_B$ is the word length of
Definition~\ref{def:bit_word_size} below, at $O(1)$ operations per call.
The input matrices are given as rationals of bit length at most $L_{\mathrm{in}}$.  The algorithm first rounds them to
$b_{\mathrm{in}}$-bit dyadic words (Lemma~\ref{lem:bit_input}), at a cost we charge in Theorem~\ref{thm:bit_complexity}.
Every subsequent number is a word.
We count costs in \emph{bit operations}.  Addition or comparison of $W$-bit words
costs $O(W)$ bit operations, and multiplication costs $M(W)=O(W\log W)$
by Harvey and van der Hoeven~\cite[Theorem~1.1]{hvdh21}.  Arithmetic saturates: a result of modulus at least $2^{I-1}$ is replaced by the largest
representable value of the same sign.  A saturating operation also sets a flag that we read as a declared failure.
Every loop of the algorithm has a fixed cap, so the operation count below holds on every path.  We prove the correctness
statements on the event that no cap fires and no saturation occurs.  A cap firing is a \emph{declared failure}: the capped procedure halts and reports failure to its caller.  We fix once, in (F18), where a
declared failure is charged.  No saturation occurs while every guard is correct and no cap fires.  Here a guard is the acceptance
test of (F9) below.  The reason is that then every base point, the iterate $x_t$ of Algorithm~\ref{alg:cubic_tracked_phase} from which a step
is proposed, satisfies $F(x_t)\le H$ for the objective $F$ fixed after Lemma~\ref{lem:bit_state}, and the magnitude
bounds of Lemmas~\ref{lem:bit_tracker}(iv) and~\ref{lem:bit_integer_bits} hold deterministically.  Randomness is a
source of independent fair coins.  The model is the additive reading of the standard floating-point
convention.  Banks, Garza-Vargas, Kulkarni, and Srivastava~\cite[Remark~1.3]{bgvks23}
note that the two are interchangeable for algorithms whose magnitudes stay within a
fixed polynomial range, which Lemma~\ref{lem:bit_tracker} establishes here.
A \emph{certified enclosure} of a real $v$ is a pair of grid numbers $\underline v\le v\le\overline v$ produced with a proven
remainder bound.  The enclosures of $\log$ and $\exp$ we use below are those of Lemmas~\ref{lem:bit_state}, \ref{lem:bit_normcert},
and \ref{lem:bit_enclosures}(v).  Also among them are the enclosures of $\log(4/p_c)$ and $\log(4P/p)$ of (F2).  We obtain them by binary range reduction and a truncated Taylor series with the Lagrange
remainder rounded outward, to $b$ bits in $O(b\,M(b))$ bit operations.  There are $O(\log n)$ such evaluations per projection attempt and
$O(\log\log n)$ per fallback trial (Lemma~\ref{lem:bit_fallback}(iv)), each to $O(W)$ bits, so they never affect the count.
We replace every analysis parameter by a dyadic value rounded in the direction that
preserves the inequality it enters, as in (F2).  The rounding is upward for $a$, the radius, the bracket, and
the levels, and downward for the smoothing parameter, $\theta$, the accuracies, the tolerances,
and the budgets.  The one new convention is the factor $1+4\mathsf u$ on the radius in (F2),
which restores the lower smoothing inclusion of Lemma~\ref{lem:algorithm_certified_implementation}
although $b'$ is rounded downward.  As in Section~\ref{sec:small_algorithm_constant},
terminating decimals are exact rationals.  We likewise assume that the failure parameter $p\in(0,1/2)$ is
rational, so that the budgets $p_c/(8T_c(B+1))$ and $p/(4PT)$ and the caps $T_c$
and $T$ are exact rationals and integers.  The policy tables of Definitions~\ref{def:cubic_refined_policy_construction}
and~\ref{def:cubic_global_policy} are constants of the algorithm, exact integers of at most $4755$ bits, a few megabytes
in all, compared in temporaries below $2W_B$ for $n\ge N_0$ (F16).  The finite selection that produced
them is input-independent preprocessing, the same for every $n$.
Section~\ref{sec:small_algorithm_constant} claims no running time for it, and we do not count it
here.  The constants of Section~\ref{sec:small_algorithm_constant} keep their meanings.  These are $N_0=3\cdot10^8$, $N_*=10^{30}$,
$N_{\mathrm{root}}=10^{25}$, $\kappa_{\mathrm{pr}}=10^{-4}$, $a_{\mathrm{sm}}=2\cdot10^6\log(2n)$, and
$s_{\mathrm{sm}}=5\cdot10^{-7}$.  After (F1), $A_i$ denotes the rounded matrix
$\widetilde A_i$.

\paragraph{Notation.}
Throughout, $d_0:=2n$ is the dimension of $D_x$, and $L_{\mathrm{in}}$ is the bit length of the input entries.  The letter $m$ is the
dimension of the current projection, the padded size of a suffix stage or of a repair node.  We write the dimension letter $M$ of
Section~\ref{sec:small_algorithm_constant} as $m$ here, and $M(\cdot)$ is the integer multiplication cost.  The
parameters $a$, $R$, $s_{\mathrm{sm}}$ (slack $s$ in Section~\ref{sec:algorithm}), $\chi$, $N_0$, $\rho$,
$\Lambda$, $\underline\nu$, $H$, $H_{\max}$, $d_{\mathrm{w}}$, $\tau_{\mathrm{w}}$, $\sigma$, $\varepsilon_{\mathrm{opt}}$,
$J$, $R_{\mathrm{rep}}$, $\delta_g$, $L$, $\pi_m$, $k$, $K_0$, $T$ carry the meanings fixed in Sections~\ref{sec:algorithm},
\ref{sec:faster_algorithm}, and~\ref{sec:cubic_time} and in Section~\ref{sec:small_algorithm_constant}.  Where the finite
algorithm replaces one of them by a rounded value, the letter carries a tilde.  Lemma~\ref{lem:bit_state} and the parameter list
that follows it list every such replacement with the direction of its rounding.  Here $a:=a_{\mathrm{sm}}=2\cdot10^6\log(2n)$
and $s_{\mathrm{sm}}=5\cdot10^{-7}$ are the standing constants of Lemma~\ref{lem:appendix_fixed_target_projection}, those
of Section~\ref{sec:algorithm}, which it reuses.  Since $b$ is the target vector of that lemma, we write the smoothing parameter
$1-s$ of Section~\ref{sec:algorithm} as $b_{\mathrm{sm}}:=1-s_{\mathrm{sm}}$ here.  Its rounded values
$\widetilde b$ and $b'$ of Lemma~\ref{lem:bit_state} keep their letters.  The multiplier $\nu$ and the objects
$q(x)=\frac12\|x-b\|_2^2$, $\Phi(x)=\tr[P_x]$, $F_\nu=q+\nu\Phi$, $\Delta_x=F_\nu(x)-F_\nu(x_\nu)$,
$\phi_0=\Phi(0)=2n\exp(-ab_{\mathrm{sm}})<\frac14$ carry the meanings fixed there.  So do the wrapper constants $G_*=a\sqrt m/R$, $\nu_{\min}=\rho/(8G_*)$,
$G_{\mathrm{lev}}=G_*H_{\max}/\underline\nu$, the solve budget $p_0$, the fixed-multiplier minimizer $x_\nu$, the exact
multiplier $\nu_*$, the exact projection $x_*$, and the box $Q_y$.  The last is the box $Q$ of Section~\ref{sec:faster_algorithm},
written $Q_y$ as in Lemma~\ref{lem:refreshed_projection}.  The letter $\rho$ is the accuracy of the
call, which is Eq.~\eqref{eq:algorithm_119_accuracy} in a suffix stage and $t_0e_{\mathrm{call}}\eta_{\mathrm{num}}$ of
Eq.~\eqref{eq:cubic_repair_parameters} in a repair call.  It is not the $\rho$ of Sections~\ref{sec:small_ball_proof}
and~\ref{sec:proof}.  Here $\varepsilon_{\mathrm{opt}}:=\sigma^2/2$ is the largest objective accuracy the wrapper of
Section~\ref{sec:faster_algorithm} admits, so that $\widetilde\varepsilon_{\mathrm{opt}}=\lfloor\widetilde\sigma^2/2\rfloor_{\mathsf u}$
is its rounded counterpart.  Where two of those sections fix the same letter differently, we mean the value of
Section~\ref{sec:faster_algorithm}, as (F12) states.  Thus $\Lambda=2q(0)/(1-\phi_0)$ is the multiplier bracket of
Section~\ref{sec:faster_algorithm}, not the bracket $2f_\xi(0)/(b_{\mathrm{sm}}-\chi)$ of Section~\ref{sec:algorithm}.
Likewise, $H=16F_\nu(0)$ is the level of Section~\ref{sec:faster_algorithm}, not the level $64U$ that
Sections~\ref{sec:accelerated_block_algorithm} and~\ref{sec:cubic_time} set.  The letter $J$ is the phase count of
Section~\ref{sec:faster_algorithm}.  The letter $L$ is the curvature constant of Eq.~\eqref{eq:cubic_L} in
Section~\ref{sec:cubic_time}, read with the level $H=16F_\nu(0)$ of Section~\ref{sec:faster_algorithm}.  The finite
algorithm uses its power-of-two version $\widetilde L$ fixed after Lemma~\ref{lem:bit_state}.
Where a letter of those sections collides with one of Section~\ref{sec:cubic_time}, of
Section~\ref{sec:small_algorithm_constant}, or with the fixed-point vocabulary, we rename it as follows.  The
curvature $L$ and the integer-bit count $I$ keep their letters.
\begin{center}
{\small
\begin{tabular}{@{}p{0.32\textwidth}lp{0.42\textwidth}@{}}
\toprule
object & \begin{tabular}[b]{@{}l@{}}letter in Section~\ref{sec:proof} and\\ Sections~\ref{sec:algorithm}--\ref{sec:cubic_time}\end{tabular} & letter here \\
\midrule
grid unit $2^{-f}$ & --- & $\mathsf u$ \\
smoothing parameter (Section~\ref{sec:algorithm}) & $s$ & $s_{\mathrm{sm}}$ \\
probe count (Section~\ref{sec:cubic_time}) & $s$ & $s$ \\
wrapper radius and tolerance (Section~\ref{sec:faster_algorithm}) & $d$, $\tau$ & $d_{\mathrm{w}}$, $\tau_{\mathrm{w}}$ \\
coordinate step and crossing time (Section~\ref{sec:cubic_time}) & $d_t$, $\tau$ & $d_t$, $\tau$ \\
squaring depth of Lemma~\ref{lem:bit_normcert} & --- & $J_{\mathrm{sq}}$ ($J$ is the phase count) \\
radial factor of Lemma~\ref{lem:bit_enclosures} & --- & $\beta_{\mathrm{r}}$ \\
Bini's error degree & --- & $h_{\mathrm{B}}$ \\
block size (Sections~\ref{sec:algorithm} and~\ref{sec:accelerated_block_algorithm}) & $\ell$ & $\ell_{\mathrm{bl}}$ \\
\hbadness=10000 polynomial surrogate $q_r(D_x)^2$ (Sections~\ref{sec:faster_algorithm}--\ref{sec:cubic_time}) & $\widetilde P_x$ & \hbadness=10000 $\mathcal Q(\widehat D_x)$ (its computed dense value is $\widetilde P_x$, Lemma~\ref{lem:bit_dense}) \\
phase-start value $F(0)$ (Section~\ref{sec:cubic_time}) & $U$ & $F_\nu(0)$ ($U(y)$ is the trace certificate, $U$ alone the norm certificate of Lemma~\ref{lem:bit_normcert}) \\
oracle accuracy $10^{-3}$ (Section~\ref{sec:faster_algorithm}) & $\eta$ & not used ($\eta$ is the exponent slack) \\
tensor rank & --- & $R(\cdot)$ (the radius is the bare $R$) \\
sketch error of Lemma~\ref{lem:bit_rangefinder} & --- & $\theta$ (dilation and map variables below are written $\vartheta$) \\
deficit threshold of Eq.~\eqref{eq:cubic_repair_parameters} (Section~\ref{sec:small_algorithm_constant}) & $\theta$ & $\theta$ (the sketch error $\theta$ is local to Lemmas~\ref{lem:bit_rangefinder}--\ref{lem:bit_lyapunov}) \\
\bottomrule
\end{tabular}
}
\end{center}
Unsubscripted constants $C$, $c$ are universal and may change from line to line.
We fix subscripted constants where they are introduced.

\paragraph{Spectral bounds.}
The rejection of the ball target is $W'_G\notin[m/2,2(m+2)]$ for the Gaussian norm $W'_G$ of (F13)
(Lemma~\ref{lem:bit_ball_target}(ii)).  On the multiplier branch of a call, $\Lambda\ge\|b\|_2^2>14.9$ and
$H\ge16q(0)=8\|b\|_2^2>72$ (Lemma~\ref{lem:bit_enclosures}(v)).  We use
$\lambda_{\mathrm{top}}:=\log(\widetilde H_{\max}/\underline{\widetilde\nu})+1+\log(2n)$
to denote an upper bound on the largest eigenvalue of $D_x$ at every point where a matrix function is evaluated.  These
points are the guarded points together with their proposals and a further point.  A guarded point is a phase start or an
iterate the guard of (F9) has accepted, so that $F(x)\le H$.  The further point is the
box projection $b_Q$ of the target after the pre-test of Lemma~\ref{lem:bit_enclosures}(v).  We obtain this bound from Lemma~\ref{lem:bit_state}(i)
with the level bounds of Section~\ref{sec:faster_algorithm}.  Guarded points have $F(x)\le H$, so
$\nu\Phi(x)\le H$ and $\lambda_{\max}(D_x)\le\log(H/\nu)\le\log(\widetilde H_{\max}/\underline{\widetilde\nu})$.
Proposals add $1$.  The symmetry of the spectrum of $\operatorname{diag}(B,-B)$ gives
$\lambda_{\min}(D_x)\ge-2\widetilde a b'-\log(H/\nu)$.  The rounding shift $d_0\mathsf u\le\log(2n)$ is covered by the
$\log(2n)$ term of $\lambda_{\mathrm{top}}$.  We use
$M_{\mathrm{sp}}:=2ab_{\mathrm{sm}}+\lambda_{\mathrm{top}}+1$ to denote a bound on the modulus of the smallest eigenvalue.  Thus
every rounded $\widehat D_x$ (Lemma~\ref{lem:bit_state}) that we exponentiate has spectrum in $[-M_{\mathrm{sp}},\lambda_{\mathrm{top}}]$.
Here $\widetilde H_{\max}$ and $\underline{\widetilde\nu}$ are the rounded level bound and multiplier floor fixed after
Lemma~\ref{lem:bit_state}.  Both are dyadic, so $\lambda_{\mathrm{top}}$ and the derived $\kappa_{\mathrm{sc}}$, $r$ are computable.  By the
ratio bound $\widetilde H_{\max}/\underline{\widetilde\nu}\le\frac{20}3H_{\max}/\underline\nu$ recorded in the parameter list after
Lemma~\ref{lem:bit_state}, $\lambda_{\mathrm{top}}\le\log(H_{\max}/\underline\nu)+1+\log(2n)+\log(20/3)$.
Equation~\eqref{eq:bit_call_word}, stated with the unrounded ratio, absorbs the extra $3\lceil\log(20/3)\log_2e\rceil\le9$ bits
in its additive constant $200$.

We fix the word length at one projection that dominates every projection the algorithm performs.

\begin{definition}[Word size]
\label{def:bit_word_size}
Let $a_{\mathrm{sm}}:=2\cdot10^6\log(2n)$ and $s_{\mathrm{sm}}:=5\cdot10^{-7}$ be the standing
constants of Lemma~\ref{lem:appendix_fixed_target_projection}, let
$b_{\mathrm{sm}}:=1-s_{\mathrm{sm}}$ and $\phi_0:=2ne^{-a_{\mathrm{sm}}b_{\mathrm{sm}}}$, and put
\[
c_B:=10^{108},\qquad
e_{86}:=9.61906812911829762\cdot10^{-18},\qquad
\rho_B(n):=\min\Bigl\{\frac{s_{\mathrm{sm}}\sqrt{15n}}{4(2n+\sqrt n)},\ \frac35\cdot10^{-17}e_{86}\Bigr\},
\]
where $c_B$ is the target cap of Eq.~\eqref{eq:cubic_global_solver_scales} and $e_{86}$ is
the exponent of row $86$ of Definition~\ref{def:cubic_refined_trace_rows}, the smallest of the seventeen registry
exponents.  This exponent is evaluated in the proof of Lemma~\ref{lem:cubic_global_dilation}.
Consider a projection call in dimension $m\le n$ with target $b$, $\|b\|_2^2\le c_bm$ for a target cap $c_b\ge1$, radius $R_{\mathrm{alg}}$,
and accuracy $\widetilde\rho$.  Put $q_B:=c_bm/2\ge q(0)=\frac12\|b\|_2^2$.  Let
\[
\Lambda:=\frac{2q_B}{1-\phi_0},\qquad
H_{\max}:=16(q_B+\Lambda\phi_0+1),\qquad
G_*:=\frac{a_{\mathrm{sm}}\sqrt m}{R_{\mathrm{alg}}},\qquad
\nu_{\min}:=\frac{\widetilde\rho}{8G_*},
\]
\[
\underline\nu:=\min\{\nu_{\min},\Lambda/2\},\qquad
d_{\mathrm w}:=\frac{\widetilde\rho}4,\qquad
\tau_{\mathrm w}:=\min\Bigl\{\frac{d_{\mathrm w}^2}{1+\Lambda},\frac{d_{\mathrm w}(1-\phi_0)}{16\sqrt m}\Bigr\},
\]
\[
G_{\mathrm{lev}}:=\frac{G_*H_{\max}}{\underline\nu},\qquad
\sigma:=\min\Bigl\{\frac{d_{\mathrm w}}8,\frac{\tau_{\mathrm w}}{16G_{\mathrm{lev}}}\Bigr\},\qquad
\varepsilon_{\mathrm{opt}}:=\frac{\sigma^2}2
\]
be the wrapper constants of Section~\ref{sec:selected_wrapper} evaluated at $q(0)=q_B$, at the radius
$R_{\mathrm{alg}}$ of Lemma~\ref{lem:bit_state} below, and at the accuracy $\widetilde\rho$ of the call.  Let
$\lambda_{\mathrm{top}}:=\log(H_{\max}/\underline\nu)+1+\log(2n)$ and
$M_{\mathrm{sp}}:=2a_{\mathrm{sm}}b_{\mathrm{sm}}+\lambda_{\mathrm{top}}+1$ be as in the paragraph \emph{Spectral bounds} above.  These
are the values at the cap.  The rounded parameters of the call, which use the realized $q(0)\le q_B$, are compared
with them in the parameter list after Lemma~\ref{lem:bit_state}.
The word of the call is
\begin{equation}
\begin{aligned}
f_B(m,c_b,\widetilde\rho,R_{\mathrm{alg}})&:=\Bigl\lceil\log_2\frac1{\varepsilon_{\mathrm{opt}}}\Bigr\rceil
+3\lceil\lambda_{\mathrm{top}}\log_2e\rceil+2\lceil\log_2M_{\mathrm{sp}}\rceil+40\lceil\log_2(2n)\rceil+200,\\
I_B(m,c_b,\widetilde\rho,R_{\mathrm{alg}})&:=\Bigl\lceil\log_2\frac{H_{\max}}{\underline\nu}\Bigr\rceil
+\lceil\log_2c_b\rceil+7\lceil\log_2(2n)\rceil+12 .
\end{aligned}
\label{eq:bit_call_word}
\end{equation}
The \emph{dominating triple} is the projection in dimension $m=n$ with target
norm $\|b\|_2^2=c_Bn$, body radius $\widehat R=\sqrt{15n}$, and accuracy $\rho=\rho_B(n)$.  There
$q_B=q(0)=c_Bn/2$ and $G_*=a_{\mathrm{sm}}/\sqrt{15}$.  Define
\begin{equation}
f_B:=f_B(n,c_B,\rho_B(n),\sqrt{15n}),\qquad
I_B:=I_B(n,c_B,\rho_B(n),\sqrt{15n}),\qquad
W_B:=I_B+f_B .
\label{eq:bit_word_length}
\end{equation}
The radius $\sqrt{15n}$ is a boundary value below every admissible radius, not itself a call, and $f_B$, $I_B$, $W_B$
without arguments always denote these values.
\end{definition}

The word length of this section is the $W_B=I_B+f_B$ of Definition~\ref{def:bit_word_size}, and $f$, $I$,
$\mathsf u$ refer to $f_B$, $I_B$, $2^{-f_B}$ unless a lemma states its own requirement.

The term $\lceil\log_2c_B\rceil=359$ of $I_B$ is forced by the tracker's
selected-gradient scalar (Lemma~\ref{lem:bit_integer_bits}, Remark~\ref{rem:bit_integer_bits}).  The triple
dominates every call, for the following reasons.  The wrapper constants are monotone in the target norm, in the
inverse accuracy, in the inverse radius, and in $m$, by Step~1 of the proof of
Theorem~\ref{thm:bit_complexity}.  Every projection the algorithm
performs has $\|b\|_2^2\le c_bm$ with $c_b\le c_B$.  Here the cap is $10^8$ for the ball target by
Lemma~\ref{lem:algorithm_certified_implementation}, $10^{52}$ for the auxiliary repair targets
by the proof of Lemma~\ref{lem:cubic_refined_accuracy}, and $10^{108}$ and $10^{36}$ for the global
helpers and the root by Eq.~\eqref{eq:cubic_global_solver_scales}.  Every projection has radius
$\widehat R^2>15m$, by the same tests, and by $r_j^2>15$ with the tail's $185\log(1/v_s)>370$ for the
suffix stages.  Every projection has accuracy at least $\rho_B(n)$.  The first term of $\rho_B$ is the suffix
accuracy of Eq.~\eqref{eq:algorithm_119_accuracy} at the radius floor with $m=n$, whose terms
$1/8$ and $t_s/8$ never bind.  The second is the smallest repair accuracy
$t_0e_{\mathrm{call}}\eta_{\mathrm{num}}$ of Eq.~\eqref{eq:cubic_repair_parameters}.  It is attained by
a global helper, with $t_0=3/5$ and $\eta_{\mathrm{num}}=10^{-9}$, on row $86$ at the largest
admissible dilation $\sigma=10^4$ of Definition~\ref{def:cubic_global_policy}, where
$e_{\mathrm{call}}=e_{86}/\sigma^2$.  Numerically $\rho_B(n)=5.7714\cdot10^{-35}$ for
$n<1.76\cdot10^{55}$ and $2.42\cdot10^{-7}/\sqrt n$ beyond.  We evaluate $W_B$ in Remark~\ref{rem:bit_word_length}.

\begin{theorem}[Bit complexity of the small-constant signing]
\label{thm:bit_complexity}
Let $A_1,\ldots,A_n$ be symmetric $n\times n$ matrices with $\|A_i\|\le1$ whose entries
are rationals of bit length at most $L_{\mathrm{in}}$, and let $p\in(0,1/2)$ be rational with $p\ge2^{-n}$.  Assume
$\alpha_{\mathrm{dual}}>1/4$ and $\omega<5/2$ (both known) in the asymptotic-rank sense of
Definition~\ref{def:bit_rank_exponent}.  Let $f_B$, $I_B$, and $W_B=I_B+f_B$ be the word sizes of
Definition~\ref{def:bit_word_size}, Eq.~\eqref{eq:bit_word_length}, evaluated at the dominating
triple.  This triple has dimension $m=n$, target norm $\|b\|_2^2=c_Bn$ with $c_B=10^{108}$, radius
$\widehat R=\sqrt{15n}$, and accuracy $\rho=\rho_B(n)$.  Thus $W_B=O(\log n)$, independent of $p$, and the hypothesis
$p\ge2^{-n}$ enters only the count of capped sampling procedures per attempt.
These values dominate every projection the algorithm performs, with the caps, radii, and accuracies of the preceding
paragraph at the admissible dilations $\sigma\le10^4$ of the policy rows $39$ and $86$.
For every fixed $\eta>0$ there is a randomized algorithm with the following properties.  It uses fair coins and fixed-point arithmetic on
$W_B$-bit words.  A constant number of scalars of the solve are held in $2W_B$-bit words, and multiplication
temporaries have $2W_B+\kappa_{\mathrm{sc}}+c_\eta\lceil\log_2(2n)\rceil$ bits.  Here $c_\eta$ is the larger of the two coefficient constants of Lemma~\ref{lem:bit_fastmm}
for $\beta=1$ and for a fixed rational $\beta_0\in(1/4,\alpha_{\mathrm{dual}})$, which exists by the hypothesis
$\alpha_{\mathrm{dual}}>1/4$, and $\kappa_{\mathrm{sc}}=O(\log\log n)$ is the scaling depth of Eq.~\eqref{eq:bit_surrogate}.
The algorithm uses $3W_B$-bit temporaries for a constant
number of scalar products per step and $b_{\mathrm{in}}+34$ further bits for the numerators of the grid
$\Gamma$.  It uses exact temporaries of $2W_B+\lceil\log_2(m+2)\rceil$ bits for the Gaussian norm $W'_G$ and
for the tent, prefix, and cardinality sums of Algorithm~\ref{alg:cubic_refined_repair}, and of $3W_B$ bits for
its deficit sum, over the denominators named in (F17).  It uses the policy tables of
Definitions~\ref{def:cubic_refined_policy_construction} and~\ref{def:cubic_global_policy} as constants
of the algorithm, exact integers of at most $4755$ bits, compared by cross-multiplication in
temporaries of $4755+\log_2n+O(1)$ bits.  The algorithm outputs $\varepsilon\in\{-1,1\}^n$ with
\[
\Bigl\|\sum_{i=1}^n\varepsilon_iA_i\Bigr\|<11.504\sqrt n
\]
with probability at least $1-p$, using at most
\begin{equation}
\bigl\{n^3+n^{\omega+1/2}+n^{\omega}\bigr\}\,n^{2\eta}\operatorname{polylog}(n/p)\cdot
M\bigl(2W_B+c_\eta\lceil\log_2(2n)\rceil\bigr)
+O\bigl(n^3(L_{\mathrm{in}}+W_B)\log(L_{\mathrm{in}}+W_B)\bigr)
\label{eq:bit_total_cost}
\end{equation}
bit operations.  The $\kappa_{\mathrm{sc}}\le\lceil\log_2(2n)\rceil$ extra bits of the scaled exponent
are absorbed by reading $c_\eta+1$ for $c_\eta$ in the display, a constant factor in $M(\cdot)$.  Since
$\eta$ is arbitrary, this is $n^{3+o(1)}\operatorname{polylog}(1/p)+\widetilde O(n^3L_{\mathrm{in}})$
bit operations in the $n^{o(1)}$ convention of Section~\ref{sec:preliminaries}.  The input bit length
enters only through the second term, which reads and rounds the input.  The same statement holds for
$N\times N$ input matrices with $N\le n$, zero-padded to $n\times n$ before the rounding at
$O(n^3b_{\mathrm{in}})$ bit operations.
\end{theorem}

The algorithm is the one of
Theorem~\ref{thm:algorithm_constant_13} with the finite primitives (F1)--(F18) below, on words of
about $10^4$ bits.  The word length is $W_B=9804$ at $n=10^{25}$ and $10744$ at $n=10^{30}$ (Remark~\ref{rem:bit_word_length}).
Its exponent is the $3+o(1)$ of Theorem~\ref{thm:algorithm_constant_13}, under the two hypotheses on matrix
multiplication that we use in Section~\ref{sec:cubic_time}, read in the asymptotic-rank sense of
Definition~\ref{def:bit_rank_exponent}.  The
constant is $11.504$ rather than the twelve digits $11.503394001765$ of
Lemma~\ref{lem:cubic_global_complete} only because the input rounding of Lemma~\ref{lem:bit_input}
adds $2^{-9}$ to a certificate we prove for the rounded matrices (Remark~\ref{rem:bit_constant}).  Every
constant of the certificate itself, including the three fallback constants, is met verbatim.

In the rest of the section we prove Theorem~\ref{thm:bit_complexity}.  We first fix the finite algorithm as the list
(F1)--(F18) of replacements of primitives of Section~\ref{sec:small_algorithm_constant}, each justified by the lemma
named beside it.  We fix the exponent convention (Definition~\ref{def:bit_rank_exponent}) in
Section~\ref{subsec:bit_products} and prove there Lemma~\ref{lem:bit_fastmm}, the exact bilinear products, and
Lemma~\ref{lem:bit_input}, the input rounding.  In Section~\ref{subsec:bit_state} we put every incrementally updated
quantity on a fixed grid and prove Lemma~\ref{lem:bit_state}, which lists the grids, the rounded parameters, and their
three consequences.  Lemmas~\ref{lem:bit_probes} and~\ref{lem:bit_gaussian} there draw the probes, the coins, and the
rounded Gaussians exactly from fair coins.  The surrogate for $e^{D_x}$ is the subject of
Section~\ref{subsec:bit_exponentials}.  There we replace the centered Taylor square by the scaled Taylor power of
Eq.~\eqref{eq:bit_surrogate} (Lemma~\ref{lem:bit_taylor}) and evaluate it densely and on vectors as the same exact
matrix (Lemmas~\ref{lem:bit_dense} and~\ref{lem:bit_action}).  The range finder and the increment estimator follow in
Section~\ref{subsec:bit_rangefinder}.  Lemma~\ref{lem:bit_rangefinder} admits the rounded test matrix and an inexact
sketch, and Lemma~\ref{lem:bit_increment} separates the exactly unbiased estimator from its deterministic errors.
Section~\ref{subsec:bit_tracker} treats one phase of Algorithm~\ref{alg:cubic_tracked_phase}: we bound the tracker's
drift and integer bits (Lemmas~\ref{lem:bit_tracker} and~\ref{lem:bit_integer_bits}), round the coordinate step
(Lemma~\ref{lem:bit_step}), and fix the tolerances (Lemma~\ref{lem:bit_eps_op}).  The additive trace guard of
Lemma~\ref{lem:bit_guard} and the finite Lyapunov contraction of Lemma~\ref{lem:bit_lyapunov} then complete the phase
analysis.  In Section~\ref{subsec:bit_projection} we certify the decisions on unguarded points by the norm certificate
of Lemma~\ref{lem:bit_normcert}, take the wrapper's decisions on outward enclosures (Lemma~\ref{lem:bit_enclosures}),
and state the projection call as Lemma~\ref{lem:bit_projection}.  We treat everything around the solver in
Section~\ref{subsec:bit_targets}: the two stored targets (Lemmas~\ref{lem:bit_ball_target}
and~\ref{lem:bit_uniform_target}) and the exact binary repair with the accumulation of $y$ (Lemma~\ref{lem:bit_repair}).
There we also certify the fair-sign fallbacks (Lemma~\ref{lem:bit_fallback}), pass the bound to the original matrices
(Corollary~\ref{cor:bit_output_bound}), and route declared failures (Remark~\ref{rem:bit_routing}).  The proof of
Theorem~\ref{thm:bit_complexity} in Section~\ref{subsec:bit_proof} has three steps: the word length, the correctness,
and the cost.  Its closing remarks evaluate $W_B$ and the constant $11.504$, treat small $n$, the polylog factor, and
the relative guard, and record why Theorem~\ref{thm:dense_runtime} is not covered
(Remarks~\ref{rem:bit_word_length}--\ref{rem:bit_section6}).

\paragraph{The finite algorithm.}
The finite algorithm is the algorithm of Theorem~\ref{thm:algorithm_constant_13}, whose procedures are
\textsc{MainCubicSmallConstant}, \textsc{GlobalAction}, \textsc{GlobalSign},
\textsc{ProjectionRepair}, \textsc{RefinedRepair}, and \textsc{SuffixSigning} of
Algorithms~\ref{alg:main_cubic_small_constant}, \ref{alg:cubic_global_action},
\ref{alg:cubic_global_select}, \ref{alg:cubic_refined_repair},
and~\ref{alg:algorithm_suffix_signing}.  It differs by the following changes.
\begin{enumerate}[label=(F\arabic*),leftmargin=3em]
\item \emph{Input.} We replace each $A_i$ by the $b_{\mathrm{in}}$-bit dyadic contraction
$\widetilde A_i=\operatorname{round}_{b_{\mathrm{in}}}((1-\eta_{\mathrm{in}})A_i)$, where
$b_{\mathrm{in}}:=2\lceil\log_2n\rceil+10$ (Lemma~\ref{lem:bit_input}).  Inputs of size $N\times N$
with $N\le n$ are zero-padded to $n\times n$ first, at $O(n^3b_{\mathrm{in}})$ bit operations.
We apply every lemma of Section~\ref{sec:small_algorithm_constant} to the $\widetilde A_i$,
which satisfy its hypotheses (symmetry and $\|\widetilde A_i\|\le1$).  Here we charge the additive $2^{-9}$
of Lemma~\ref{lem:bit_input} to the statement of Theorem~\ref{thm:bit_complexity}.
\item {\setlength{\emergencystretch}{2em}\emph{State and parameters.} Fractional colorings $y$ lie on $\mathsf u\mathbb Z$.
Iterates, coordinate steps, and box endpoints lie on $\Gamma$.  Here $t_s=v/10^{10}$ with $v$ an integer by
Definition~\ref{def:cubic_refined_suffix}, and $t_0\in\{4\cdot10^9,6\cdot10^9\}/10^{10}$.  The returned
point of a projection and the snapped increment lie on $\mathsf u\Gamma$ (Lemma~\ref{lem:bit_projection}).
The matrix $B(x)$ lies on $2^{-b_{\mathrm{in}}}\Gamma$ (Lemma~\ref{lem:bit_state}(i)).  The masses $a_i$ lie on $2^{-B}\mathbb Z$ (exact, Lemma~\ref{lem:cubic_refined_potential}).
The matrix $\widehat D_x$ is the once-rounded value of the exact $D_x$.  We replace every analysis parameter by a dyadic value
rounded in the direction that preserves the inequality it enters, or by a ratio of such values that we compare only
after cross-multiplication.  Such ratios are $\widetilde R$, $b'$, and the analysis quantity $\widetilde\chi$.
The radius is $R_{\mathrm{alg}}:=\lceil(1+4\mathsf u)R_{\mathrm{up}}\rceil_{\mathsf u}$ for a
certified upper enclosure $R_{\mathrm{up}}$ of $R$.  The enclosure comes from integer square roots by the integer-square
rule of Definition~\ref{def:cubic_refined_policy_construction} for $r_j\sqrt{v_sn}$ in a suffix
stage and for the repair radii, which are fixed rational multiples of $\sqrt m$.  It also comes from the
$48$-term rational logarithm rule of that definition for the tail's $185\log(1/v_s)$.  These are exact integer
operations on integers of at most $2f+4755+\log_2n+O(1)$ bits, as $2^{2f}r_j^2v_sn$ has $v_s$ a
ratio of $4755$-bit integers, held in a $3W_B$-bit temporary, one per call.  The rounded parameters $\widetilde a$,
$\widetilde c$, $\widetilde R$, $\widetilde b$, $\widetilde\theta$ and the suffix accuracy $\widetilde\rho$ are those of
Lemma~\ref{lem:bit_state}, whose parts (ii)--(iv) give $\widetilde R\le(1+10^{-8})R$,
$(1-10^{-6})K(R)\subseteq\widetilde{\mathcal D}$, and $\widetilde\rho\le\rho$.  Here $\widetilde b$ is not exact because
$1-5\cdot10^{-7}$ is not dyadic, and a repair call has
$\widetilde\rho_{\mathrm{rep}}:=\lfloor t_0e_{\mathrm{call}}\eta_{\mathrm{num}}\rfloor_{\mathsf u}$ by
Eq.~\eqref{eq:cubic_repair_parameters}.  The quantities $\theta$, $e_{\mathrm{call}}$,
$\beta=e_{\mathrm{call}}/D^{\mathrm{acc}}_r$, $t_0$, the prefix tolerances, and every budget are
exact rationals.  The value $\widehat\varepsilon(q)$ is found by the $192$-step search of
Lemma~\ref{lem:cubic_global_prefix}.  We replace the logarithms $\log(4/p_c)$ and $\log(4P/p)$ by
certified upper enclosures, which only raise the caps and lower the budgets.\par}
\item \emph{Randomness.} The $m+2$ Gaussians of the ball target are exact samples of
$\operatorname{round}_h(\mathcal N(0,1))$ with $h:=2^{-40}$ (Lemma~\ref{lem:bit_gaussian}(i)).  The uniform target $U'$ is uniform on the $2^{-114}$ midpoint grid of
$[-1,1]^m$, which we draw from $115$ fair coins per coordinate (Lemma~\ref{lem:bit_uniform_target}).
The fallback signs are fair coins.  The range-finder test matrix, the $2n\times2k$ Gaussian matrix of
Section~\ref{subsec:cubic_increment}, is an exact sample of rounded Gaussians (Lemma~\ref{lem:bit_gaussian}(iii)).  All
Hutchinson probes are Rademacher vectors, and coordinate choices and refresh coins are exact
(Lemma~\ref{lem:bit_probes}).  A sampler cap is a declared
failure, which we route by (F18).
\item \emph{Matrix exponentials.} We replace the surrogate $q_r(D_x)^2$ of Eq.~\eqref{eq:cubic_polynomial_surrogate}
by the scaled-Taylor power $\mathcal Q(D_x)$ of Eq.~\eqref{eq:bit_surrogate}.  For dense matrices we evaluate it by repeated squaring
and for vector actions by repeated application, with the $\lambda_{\mathrm{top}}$
and $M_{\mathrm{sp}}$ of the current call (Lemmas~\ref{lem:bit_taylor}, \ref{lem:bit_dense},
and~\ref{lem:bit_action}).  The only hypothesis of those lemmas is that the spectrum of $\widehat D_x$ lies
in $[-M_{\mathrm{sp}},\lambda_{\mathrm{top}}]$.
\item \emph{Fast products.} We compute every dense or thin matrix product exactly
by a recursive bilinear algorithm and round it once.  A product is thin when one of its three dimensions is $\ell\ll d_0$, for example a $d_0\times d_0$ matrix times a $d_0\times\ell$ block or a $d_0\times\ell$ block times an $\ell\times d_0$ block.  Lemma~\ref{lem:bit_fastmm} covers every permutation of the shape.  The exact product and the rounding are justified by Lemma~\ref{lem:bit_fastmm}, and the shapes are those of Section~\ref{sec:cubic_time}.
\item \emph{Range finder and increment.} We compute the sketch, its Householder basis, and the
increment estimator with the rounded test matrix and the action
surrogate (Lemmas~\ref{lem:bit_rangefinder} and~\ref{lem:bit_increment}).
\item \emph{Tracker.} The tracker is the dense matrix $\widehat P_t$ maintained by Algorithm~\ref{alg:cubic_tracked_phase} and refreshed or incremented after every step.  Its error recurrence is Eq.~\eqref{eq:cubic_tracker_recurrence}.  A refresh also occurs whenever $\lceil1/\pi'\rceil$ steps have
passed since the last one, where $\pi'$ is the dyadic refresh probability we fix with the parameters
after Lemma~\ref{lem:bit_state}.  We fit the selected-gradient scalar
$\nu\tr[\widehat P_tE_i]$ and the entries of $\widehat P_t$ into $I_B$ integer bits by
Lemma~\ref{lem:bit_integer_bits}, and Remark~\ref{rem:bit_integer_bits} evaluates the spare bits.  In Lemma~\ref{lem:bit_tracker} we analyze the tracker itself.
\item \emph{Coordinate step.} We round the one-dimensional step to the coordinate grid $\Gamma$ before
clipping (Lemma~\ref{lem:bit_step}).  The comparison displacement is admissible for every box of
Section~\ref{sec:small_algorithm_constant}, of side $2t_s<1$ or $6/5$, because $\widetilde L\ge8\widetilde c$.
\item \emph{Guard.} The trace certificate is additive: we accept a proposal $y$
when $q(y)+\nu U(y)\le H$ with an upper estimate $U(y)\ge\Phi(y)$ that satisfies
$U(y)\le\frac53\Phi(y)+H/(9\nu)$.  This replaces the relative certificate $\Phi(y)\le U(y)\le2\Phi(y)$ of
Section~\ref{sec:faster_algorithm} (Lemma~\ref{lem:bit_guard}), and Remark~\ref{rem:bit_relative_guard} records the word
length with the relative certificate.  The level $\widetilde H$ and the tolerance $\widetilde\delta_g$ are those of the
parameter list after Lemma~\ref{lem:bit_state}.  The third term of $\widetilde\delta_g$ gives the proviso
$2\nu d_0\varepsilon_{\mathrm{op}}\le\widetilde H/27$ of Lemma~\ref{lem:bit_guard} without a lower bound on the level in
terms of $m$ (Lemma~\ref{lem:bit_eps_op}(iii)).
\item \emph{Comparisons.} Objective comparisons, bisection decisions, and the radial
correction act on outward-rounded enclosures (Lemma~\ref{lem:bit_enclosures}(i)--(iii)).  The
pre-test of the box shortcut uses Lemma~\ref{lem:bit_normcert} with
$\varpi\in[\widetilde\chi/16,\widetilde\chi/8]$ (Lemma~\ref{lem:bit_enclosures}(v)).  The bound $\|b\|_2\ge3.872979>3$ of that lemma is the one
Lemma~\ref{lem:appendix_fixed_target_projection} derives from $\widehat R^2>15m$.  The tests
of Section~\ref{sec:small_algorithm_constant} itself are exact comparisons of grid numbers
and integers.  These tests are the prefix test of
Lemma~\ref{lem:cubic_refined_projection}, the deficit test, ``more than $m/4$ coordinates
snapped'', the two common-orientation rules, the odd-numerator supports, the band and piece
location, and the rejection $W'_G\notin[m/2,2(m+2)]$.
\item \emph{Terminal roundings and fallbacks.} The terminal
nearest-sign rounding of \textsc{SuffixSigning} is exact.  Each capped fair-sign trial forms
$E:=\sum_i\varepsilon_iA_i$ exactly and rejects if $\|E\|_F^2>c^2mn$ (the Frobenius pre-test).
Otherwise it certifies $\|E\|$ by Lemma~\ref{lem:bit_normcert} with $\varpi:=2^{-9}$,
where $U:=1$ if $\|E\|_F\le1$, and accepts if and only if $U^2\le c^2m$, for
$c\in\{10.889,11.902,13\}$ (Lemma~\ref{lem:bit_fallback}).  We replace the additive allowance
$0.03\sqrt m$ of the diagonalization certificate of Section~\ref{subsec:dense_cleanup} in
Eq.~\eqref{eq:cubic_independent_sign_fallback} by the factor $1+\varpi$.  Since
$(1+\varpi)s\le s+0.03$ for every $s=\sqrt{2\log(4n)}\le15.36$, the four tests of
Section~\ref{sec:small_algorithm_constant} pass verbatim.  A fallback runs
$T_f:=\lceil\log_2(1/p_c)\rceil$ trials.
\item \emph{Wrapper parameters.} The finite algorithm uses the level $H=16F_\nu(0)$ with $q(0)=\frac12\|b\|_2^2$ for the stored target, the bracket
$\widetilde\Lambda:=\lceil\frac83q(0)\rceil_{\mathsf u}$, the phase count $J$, which we replace by $\widetilde J$ of the parameter list
after Lemma~\ref{lem:bit_state}, and the repetition count $R_{\mathrm{rep}}$.  It uses the candidate-selection rule, the $\widetilde\delta_g$
of the same list, the operator tolerance $\varepsilon_{\mathrm{op}}$ of Eq.~\eqref{eq:bit_eps_op}, and the failure budgets of
Lemma~\ref{lem:algorithm_certified_implementation}, $\min\{0.02,p/(4PT)\}$ per suffix trial, and of
Lemma~\ref{lem:cubic_refined_complete}, $p_c/(8T_c(B+1))$ per numerical call.  We use all of these in place of $H=64U$,
\textsc{CompareValue}, and the $\delta_\nu$-budgets of Algorithm~\ref{alg:cubic_recursive_multiplier}.  The candidate-selection rule
retains a candidate only if its certified upper endpoint lies below the old point's certified lower endpoint.  One
phase-independent $\delta_g$ replaces the per-phase $\delta_{g,j}$ (Lemmas~\ref{lem:bit_enclosures}(ii)
and~\ref{lem:bit_lyapunov}).  Under this selection rule phase-start objectives never increase, so $F(z)\le F_\nu(0)\le H/16$
at every phase start, which is the hypothesis of Lemma~\ref{lem:bit_lyapunov}.  With the explicit constants of
Lemma~\ref{lem:bit_lyapunov}, $32m\delta_g^2\le\varepsilon_{\mathrm{opt}}/(6250J)\le\varepsilon_{\mathrm{ph},j}/(64J)$ for every
phase $j$.  This plays the role of the per-phase requirement $C_3m\delta_{g,j}^2\le\varepsilon_{\mathrm{ph},j}/(64J)$ of
Lemma~\ref{lem:cubic_tracked_phase}.  The cap $\varepsilon_{\mathrm{op}}\le1/6$ in Eq.~\eqref{eq:bit_eps_op} needs no bound on
the radius (Lemma~\ref{lem:bit_eps_op}, Remark~\ref{rem:bit_side_condition}).  The cap $\varepsilon_{\mathrm{opt}}\gets\min\{\varepsilon_{\mathrm{opt}},H\}$ of
Algorithm~\ref{alg:cubic_recursive_multiplier} is void and $\varepsilon_{\mathrm{opt}}\le2JH$ on the multiplier branch, by Step~6 of the proof of Lemma~\ref{lem:bit_projection}.
\item \emph{Targets (new).} For the ball target, we form $W'_G:=\sum_{i\le m+2}\xi_i'^2$ exactly
in a temporary of $2W_B+\lceil\log_2(m+2)\rceil$ bits, and the rejection
$W'_G\notin[m/2,2(m+2)]$ is exact.  The scalar
$\widehat s:=10^4\lfloor2^{f-40}\sqrt{m/W'_G}\rfloor/2^{f-40}$ is one integer square root and a
lower bound for $\sqrt{m/W'_G}/\kappa_{\mathrm{pr}}$.  The stored target is
$\widehat b:=\widehat s\,\xi'_{[m]}\in(\mathsf u\mathbb Z)^m$, where $\xi'_{[m]}:=(\xi'_1,\ldots,\xi'_m)$,
with $\|\widehat b\|_2^2\le10^8m$ exactly.  We apply Lemma~\ref{lem:algorithm_centered_box}
to $\widehat g:=\kappa_{\mathrm{pr}}\widehat b$ itself (Lemma~\ref{lem:bit_ball_target}), and the two remarks on the
constants after that lemma record where we spend the accuracy $\rho$.
For the uniform target, $b:=\operatorname{round}_{\mathsf u}(t_0U'/\beta)\in(\mathsf u\mathbb Z)^m$ is the
solver's target.  The vector $U'':=\beta b/t_0$, an exact rational vector with
$\|U''-U'\|_\infty\le\mathsf u$, is the vector we use in the prefix test, the deficit test, and
the cardinality weights (Lemma~\ref{lem:bit_uniform_target}).
\item \emph{The $y$ update (new).} In a suffix stage, $y_i:=\pm1$ for a snapped coordinate, which is
exact since $\overline x_i=t_s(\pm1-y_i)$, and
$y_i:=\operatorname{round}^{\to0}_{\mathsf u}(y_i+\overline x_i/t_s)$ otherwise (Lemma~\ref{lem:bit_enclosures}(iv)).  The rounding moves the applied increment by at most
$m\mathsf u/2$ in $\|B(\cdot)\|$, which the slack $3s_{\mathrm{sm}}\widetilde R/4$ left by the
snapping bound $(2m+\sqrt m)\rho\le s_{\mathrm{sm}}\widehat R/4$ of
Lemma~\ref{lem:algorithm_certified_implementation} absorbs.  In a repair call, $y=x/t_0$ is an
exact rational, which we consume only through $\operatorname{sgn}(y_i)$ and the truncation of
$(1-|y_i|)/2$ to $2^{-B}\mathbb Z$ and never accumulate.
\item \emph{Binary repair (new, exact).} The nearest signs $s_i:=\operatorname{sgn}(y_i)$ and
the downward-truncated masses $a_i:=\lfloor2^B(1-|y_i|)/2\rfloor/2^B$ are integer operations on the numerators.  At level $j$, so are the
odd-numerator support $S_j$, the common orientation by the exact tent potential
$\sum_iF(a_i)$, a rational with denominator $10^{12}2^j$, or by the exact weighted-mass
rule $\varepsilon_j:=-\operatorname{sgn}(\sum_{i\in S_j}|U_i''|\sigma_i)$, the update
$a_i\gets a_i\pm2^{-j}$ on $S_j$, and the return $z_i:=s_i(1-2a_i)$ (Lemma~\ref{lem:bit_repair}).  They also rest on Lemma~\ref{lem:cubic_refined_potential}, whose grid argument from
Lemma~\ref{lem:algorithm_21_rounding} is verbatim.  A helper returns exact signs or
\textsc{Fail}, so no rounding error passes between levels or between calls.
\item \emph{Policy data (new).} The tables of
Definitions~\ref{def:cubic_refined_policy_construction} and~\ref{def:cubic_global_policy}
are constants of the algorithm.  They consist of the $17$ registry rows, the $1101$ suffix schedules of Definition~\ref{def:cubic_refined_suffix} with their $47091$ stage triples, the auxiliary states, the $241\times9$
global actions with their dilations, the root slopes, the selectors, and $w_i=(19/20)^i$,
$x_i$, $q_{\min}$.  They are exact integers of at most $4755$ bits.  Band
and piece location are the comparisons of $m\,20^i$ with $n\,19^i$ and of $m\,\mathrm{den}^2$
with $n\,\mathrm{num}^2$ for a piece endpoint $\mathrm{num}/\mathrm{den}$.  These comparisons, the $192$-step
search for $\widehat\varepsilon(q)$, the padding sizes $\lfloor nw_i\rfloor$ and
$\lfloor v_sn\rfloor$, and $v(e)$ are exact big-integer operations outside the saturating
word arithmetic, $O(1)$ of them per call on integers of $4755+\log_2n+O(1)$ bits.  The budgets
and caps are $p_c$, $\varepsilon_c$, $T_c$, $T$, $T_f$, $R_{\mathrm{rep}}$, $s_{\mathrm{tr}}$,
$M_{\mathrm{ref}}$, and the enclosures of $\log(4/p_c)$ and $\log(4P/p)$.  They are likewise exact
big-integer operations outside the word, $O(1)$ of them per call on integers of
$O(\log(1/p)+\log n)$ bits, which we absorb in the $\operatorname{polylog}(n/p)$ count.
\item \emph{Exact temporaries (new).} We form the Gaussian norm $W'_G$, the deficit sum
$\sum_i|U_i''|(1-|y_i|)$, the tent sums, the prefix sums, and the cardinality sums exactly in
temporaries of $2W_B+\lceil\log_2(m+2)\rceil$ bits, or of $3W_B$ bits for the deficit sum.  Here we form them over the
common denominators of their grids.  The denominator is $2^{80}$ for $W'_G$ and $10^{12}2^B$ for the tent sums.  For the
prefix, cardinality, and deficit sums it is the fixed denominator of $U''=\beta b/t_0$, which carries
$2^f$ because $b$ lies on $\mathsf u\mathbb Z$.  In the deficit sum, the denominator of $y=x/t_0$ enters as well, where
$x$ lies on $\mathsf u\Gamma$ after the radial correction.  Thus $2^{3f}$ enters, and the
numerators of the deficit sum have fewer than $3f+400+\log_2m<3W_B$ bits.  We compare the sums
exactly.  This is the inner-product convention of the computational model.
\item \emph{Failure routing (new).} Inside an attempt of \textsc{ProjectionRepair} or a
trial of \textsc{SuffixSigning}, a declared failure of the projection counts as a failed test.  Such a failure is a sampler cap of
Lemma~\ref{lem:bit_gaussian}(i) or Lemma~\ref{lem:bit_probes}.  It can also be the one-redraw cap of
Lemma~\ref{lem:bit_gaussian}(iii) or a saturation report.  The loop
continues to the next attempt, as after a failed prefix or deficit test in
Algorithm~\ref{alg:cubic_refined_repair}.  We charge the event to
$\delta_{\mathrm{num}}\le\delta_{\mathrm{fail}}$ in Eq.~\eqref{eq:cubic_acceptance_comparison}
for repairs, or to the $0.02$ allowance of Lemma~\ref{lem:algorithm_certified_implementation}
for suffix stages.  In Remark~\ref{rem:bit_routing} we record the size of the charge, the silent class we charge to the projection's budget $\varepsilon_c$
instead, and why $W_B$ stays free of $p$.
\end{enumerate}

\subsection{Exact products and the input}\label{subsec:bit_products}

Our bit count uses bilinear algorithms for matrix multiplication with rational coefficients. We fix the exponent convention that supplies these exact algorithms.

\begin{definition}[Asymptotic-rank exponents]
\label{def:bit_rank_exponent}
For a tensor $\mathcal T$ over the rationals let $R(\mathcal T)$ be its rank and
$\widetilde R(\mathcal T):=\lim_{k\to\infty}R(\mathcal T^{\otimes k})^{1/k}$ its asymptotic
rank.  For a rational $\beta\in[0,1]$ put
$\omega(1,\beta,1):=\log_q\widetilde R(\langle q,q^\beta,q\rangle)$, where $q\ge2$ is any integer with $q^\beta$ an integer.
For $\beta=p_1/p_2$ take $q=2^{p_2}$.
Here $\langle k_1,k_2,k_3\rangle$ is the matrix multiplication tensor.
Put $\alpha_{\mathrm{dual}}:=\sup\{\beta:\omega(1,\beta,1)=2\}$, and write $\omega:=\omega(1,1,1)$.
The value of $\omega(1,\beta,1)$ does not depend on $q$.  If $q_1^\beta$ and $q_2^\beta$ are integers, then
$\langle q_1,q_1^\beta,q_1\rangle^{\otimes k}$ restricts to $\langle q_2,q_2^\beta,q_2\rangle^{\otimes k'}$ whenever
$q_2^{k'}\le q_1^{k}$.  Fekete's lemma turns the two restrictions into equality of the two logarithms.
Restriction of $\langle q,q^{\beta},q\rangle$ to $\langle q,q^{\beta'},q\rangle$ for $\beta'<\beta$ shows that
$\beta\mapsto\omega(1,\beta,1)$ is nondecreasing, so $\omega(1,\beta,1)=2$ for every $\beta<\alpha_{\mathrm{dual}}$.
These quantities coincide with the exponents of Section~\ref{sec:model}.  A bilinear algorithm of rank $t$ is an
arithmetic circuit of size $O(t)$.  Conversely, Strassen's argument extracts from arithmetic circuits bilinear
algorithms of the same exponent.  Hence the asymptotic-rank exponent equals the operation-count exponent.  We use the
asymptotic-rank form because it delivers exact rational algorithms.
These are the definitions under which Vassilevska Williams, Xu, Xu, and
Zhou~\cite[Section~3.4]{vxxz24} prove $\alpha_{\mathrm{dual}}=\alpha>0.321334$ and
Alman, Duan, Vassilevska Williams, Xu, Xu, and Zhou~\cite{almanetal25asymmetry} prove
$\omega<2.371339$ (their main theorem).  Both are laser-method degenerations of Coppersmith--Winograd tensors
with integer coefficients, so the bounds hold for the rank over $\mathbb Q$, which is the
field used here.
\end{definition}

The asymptotic rank is defined through exact ranks of tensor powers.  Hence we need not extract
a bilinear algorithm from an arithmetic circuit, and we execute no border-rank
approximation at run time.  By Fekete's lemma
$\widetilde R(\mathcal T)=\inf_kR(\mathcal T^{\otimes k})^{1/k}$, and
$\langle q,q^\beta,q\rangle^{\otimes k}=\langle q^k,q^{\beta k},q^k\rangle$.  Where a
border-rank bound is converted into a rank bound inside the laser method, the
conversion is Bini's interpolation~\cite{bini80}.  The conversion takes the form
$R(\mathcal T)\le(2h_{\mathrm{B}}+1)R_{h_{\mathrm{B}}}(\mathcal T)$ of
Bl\"aser~\cite[Lemma~6.4 and Remark~6.5]{blaser13}, where $R_h(\mathcal T)$ is the border rank of $\mathcal T$ with error degree $h$.
This border rank is the least rank of a degree-$h$ approximate decomposition in Bini's sense.
The rational Lagrange weights of the interpolation are
the source of the denominators below.

\begin{lemma}[Exact fast multiplication of fixed-point matrices]
\label{lem:bit_fastmm}
Fix $\eta>0$ and a rational $\beta\in\{1\}\cup(1/4,\alpha_{\mathrm{dual}})$.  There are integers
$\varkappa\ge2$, $t\le \varkappa^{\omega(1,\beta,1)+\eta}$, and an exact bilinear algorithm
$(\mathsf U,\mathsf V,\mathsf W)\in\mathbb Q^{\varkappa^{1+\beta}\times t}\times\mathbb Q^{\varkappa^{1+\beta}\times t}\times\mathbb Q^{\varkappa^2\times t}$
for $\langle \varkappa,\varkappa^\beta,\varkappa\rangle$ with $\varkappa^\beta$ an integer.  Let $\mathfrak d$ be a common denominator of its entries.
Put $\alpha_U:=\max_s\sum_i|u_{is}|$, $\alpha_V:=\max_s\sum_j|v_{js}|$, and
$\alpha_W:=\max_r\sum_s|w_{rs}|$, each taken as the maximum over the three permuted
decompositions used for the permuted shapes, and put
\[
c_\eta:=2\Bigl\lceil\frac{\log_2(\mathfrak d^3\alpha_U\alpha_V\alpha_W\varkappa)}{\log_2\varkappa}\Bigr\rceil+2.
\]
Let $X\in\R^{n_1\times n_2}$ and $Y\in\R^{n_2\times n_3}$ be fixed-point
matrices with $W$-bit entries.  Let $(n_1,n_2,n_3)$ be at most $(N_{\mathrm{mult}},N_{\mathrm{mult}}^\beta,N_{\mathrm{mult}})$ or $(N_{\mathrm{mult}},N_{\mathrm{mult}},N_{\mathrm{mult}}^\beta)$ entrywise
up to a permutation.  Smaller shapes are padded with zeros.  Assume $N_{\mathrm{mult}}\le2n$ and $n\ge\varkappa$.  Recall $\omega(1,1,\beta)=\omega(1,\beta,1)$~\cite[Section~3.4]{vxxz24}.  Then the recursive application of $(\mathsf U,\mathsf V,\mathsf W)$ on
temporaries of $2W+c_\eta\lceil\log_2(2n)\rceil$ bits computes the exact product $XY$,
which is an integer multiple of $\mathsf u^2$, and rounds it once to the $\mathsf u$-grid.  The
returned matrix $\widehat C$ therefore satisfies $\|\widehat C-XY\|_{\max}\le \mathsf u$, where $\|M\|_{\max}:=\max_{i,j}|M_{ij}|$ is the largest entry modulus.  The computation uses
\[
O\bigl(N_{\mathrm{mult}}^{\omega(1,\beta,1)+2\eta}\bigr)\cdot M\bigl(2W+c_\eta\lceil\log_2(2n)\rceil\bigr)
\]
bit operations.  For $\beta=1$ the exponent is $\omega+2\eta$.  For $\beta\in(1/4,\alpha_{\mathrm{dual}})$
it is $2+2\eta$.
\end{lemma}

\begin{proof}
\emph{Step 1: the base algorithm.}  By Definition~\ref{def:bit_rank_exponent} and
Fekete's lemma, for an integer $q\ge2$ with $q^\beta\in\mathbb Z$ there is $k_\eta$ with
$R(\langle q,q^\beta,q\rangle^{\otimes k_\eta})\le q^{k_\eta(\omega(1,\beta,1)+\eta)}$.  Put
$\varkappa:=q^{k_\eta}$, and let $(\mathsf U,\mathsf V,\mathsf W)$ be a rank decomposition of $\langle \varkappa,\varkappa^\beta,\varkappa\rangle$ over
$\mathbb Q$ with $t$ terms.  Exactness means that for all $X\in\R^{\varkappa\times \varkappa^\beta}$,
$Y\in\R^{\varkappa^\beta\times \varkappa}$ and every output index $r$,
\begin{equation}
(XY)_r=\sum_{s=1}^tw_{rs}\Bigl(\sum_iu_{is}x_i\Bigr)\Bigl(\sum_jv_{js}y_j\Bigr),
\label{eq:bit_bilinear_identity}
\end{equation}
and because Eq.~\eqref{eq:bit_bilinear_identity} never commutes an $x_i$ past a $y_j$ it
holds verbatim when the $x_i$, $y_j$ are matrix blocks.

\emph{Step 2: the recursion is exact.}  We pad $X$ and $Y$ with zeros to sizes
$\varkappa^{\varrho}\times \varkappa^{\beta \varrho}$ and $\varkappa^{\beta \varrho}\times \varkappa^{\varrho}$ with $\varrho:=\lceil\log_{\varkappa}N_{\mathrm{mult}}\rceil$.  The
padding multiplies every size by at most $\varkappa$.  We treat the shape $(N_{\mathrm{mult}},N_{\mathrm{mult}},N_{\mathrm{mult}}^\beta)$ identically with the permuted decomposition.  Next we multiply all entries of $\mathsf U$, $\mathsf V$, $\mathsf W$ by
$\mathfrak d$ so that they become integers.  Then Eq.~\eqref{eq:bit_bilinear_identity} returns
$\mathfrak d^3(XY)_r$.  The recursive algorithm at depth $\varrho$ forms the $t$ linear forms
$X_s:=\sum_i\mathfrak d u_{is}X_i$ and $Y_s:=\sum_j\mathfrak d v_{js}Y_j$ of the $\varkappa\times \varkappa^\beta$
resp.\ $\varkappa^\beta\times \varkappa$ block partitions, calls itself on $(X_s,Y_s)$, and returns
$C_r:=\sum_s\mathfrak d w_{rs}P_s$.  Every operation is an addition or a multiplication by an
integer constant, applied to integers, hence exact.  These integers are the grid multiples of $\mathsf u$, scaled by $2^f$.
The base products are exact products of integers.  By induction on the depth the
output equals $\mathfrak d^{3\varrho}XY$ exactly, and once we divide by $\mathfrak d^{3\varrho}$ and round to the
$\mathsf u$-grid we obtain $\|\widehat C-XY\|_{\max}\le \mathsf u$.

\emph{Step 3: magnitudes.}  We claim that every value formed by the recursion has modulus
at most $G_\varrho:=(\mathfrak d^3\alpha_U\alpha_V\alpha_W\varkappa)^\varrho2^{2I}$.  Here $\alpha_U\alpha_V\alpha_W\ge1$ and each
$\alpha_\bullet\ge1/\mathfrak d$, both forced by the exactness of Eq.~\eqref{eq:bit_bilinear_identity} and the integrality
of the scaled coefficients.  At depth $\varrho$
the linear forms have entries of modulus at most $\mathfrak d\alpha_U2^I$ and $\mathfrak d\alpha_V2^I$.
The exact product returned by a call at depth $\varrho-j$ on operands of moduli $M_X$, $M_Y$
equals $\mathfrak d^{3(\varrho-j)}$ times a product with inner dimension at most
$\varkappa^{\varrho-j}$.  Hence it has modulus at most $\mathfrak d^{3(\varrho-j)}\varkappa^{\varrho-j}M_XM_Y$.
Each assembly multiplies by at most $\mathfrak d\alpha_W$.  When we unroll, the operands at
depth $\varrho-j$ have moduli at most $(\mathfrak d\alpha_U)^j2^I$, $(\mathfrak d\alpha_V)^j2^I$.  Likewise the
returned products have moduli at most $\mathfrak d^{3(\varrho-j)}\varkappa^{\varrho-j}(\mathfrak d^2\alpha_U\alpha_V)^j2^{2I}$,
and the partial sums of an assembly at most $\mathfrak d\alpha_W$ times that.  All of these
are at most $G_\varrho$.  Since $\varrho\log_2\varkappa\le\log_2N_{\mathrm{mult}}+\log_2\varkappa\le2\log_2(2n)$ for
$\varkappa\le n$, we have $\log_2G_\varrho\le2I+(c_\eta-2)\lceil\log_2(2n)\rceil$.  A temporary of
$2W+c_\eta\lceil\log_2(2n)\rceil$ bits therefore holds every intermediate value together with its
fractional part.

\emph{Step 4: cost.}  There are $t^\varrho$ base products and $O(\varrho\,t^\varrho)$ additions and
constant multiplications, each on the temporaries, so the cost is
$O(\varrho\,t^{\varrho}M(2W+c_\eta\lceil\log_2(2n)\rceil))$ bit operations.  We absorb the factor $\varrho=O(\log N_{\mathrm{mult}})$ in $N_{\mathrm{mult}}^{\eta}$.  Moreover,
$t^\varrho\le \varkappa^{\varrho(\omega(1,\beta,1)+\eta)}\le(\varkappa N_{\mathrm{mult}})^{\omega(1,\beta,1)+\eta}=O(N_{\mathrm{mult}}^{\omega(1,\beta,1)+2\eta})$
for $N_{\mathrm{mult}}\ge \varkappa^{(\omega(1,\beta,1)+\eta)/\eta}$.  We absorb smaller $N_{\mathrm{mult}}$ in the constant.  For
$\beta\in(1/4,\alpha_{\mathrm{dual}})$ the exponent is $\omega(1,\beta,1)=2$ by definition of
$\alpha_{\mathrm{dual}}$.
\end{proof}

The constant $c_\eta$ is finite for every fixed $\eta$ and is not quantified by any
source.  Consider the Lagrange weights of Bini's interpolation at $2N_\eta h_{\mathrm{B}}+1$ nodes, where $N_\eta$ is the tensor power at which the laser method reaches exponent $\omega(1,\beta,1)+\eta$.
These weights have numerators
and denominators of size $\exp(O(N_\eta\log N_\eta))$, so $c_\eta\to\infty$ as $\eta\to0$.  This
is the same unquantified constant as in the stability exponent of Demmel, Dumitriu,
Holtz, and Kleinberg~\cite[Theorem~3.3]{ddhk07}, whose floating-point analysis covers
square products only.  Our fixed-point route above needs no stability theorem because
every product is exact.

By rounding the input first we reduce the subsequent arithmetic to dyadic matrices.  A small contraction preserves the operator-norm normalization.

\begin{lemma}[Input rounding]
\label{lem:bit_input}
Let $A_1,\ldots,A_n$ be symmetric with $\|A_i\|\le1$, and let $b_{\mathrm{in}}:=2\lceil\log_2n\rceil+10$.  Put
$\eta_{\mathrm{in}}:=n2^{-b_{\mathrm{in}}}$ and $\widetilde A_i:=\operatorname{round}_{b_{\mathrm{in}}}((1-\eta_{\mathrm{in}})A_i)$.  Here the rounding is
entrywise to the nearest multiple of $2^{-b_{\mathrm{in}}}$, applied identically to the two
symmetric positions.  Then every $\widetilde A_i$ is a symmetric contraction with $b_{\mathrm{in}}$-bit
dyadic entries, and for every signing $\varepsilon\in\{-1,1\}^n$,
\[
\Bigl\|\sum_{i=1}^n\varepsilon_iA_i\Bigr\|
\le\Bigl\|\sum_{i=1}^n\varepsilon_i\widetilde A_i\Bigr\|+2^{-9}.
\]
Forming the $\widetilde A_i$ from rational entries of bit length $L_{\mathrm{in}}$ costs
$O(n^3(L_{\mathrm{in}}+b_{\mathrm{in}})\log(L_{\mathrm{in}}+b_{\mathrm{in}}))$ bit operations.
\end{lemma}

\begin{proof}
The rounding error matrix $F_i:=\widetilde A_i-(1-\eta_{\mathrm{in}})A_i$ is symmetric with entries
of modulus at most $2^{-b_{\mathrm{in}}-1}$, so $\|F_i\|\le\|F_i\|_F\le n2^{-b_{\mathrm{in}}-1}=\eta_{\mathrm{in}}/2$.  Hence
$\|\widetilde A_i\|\le(1-\eta_{\mathrm{in}})+\eta_{\mathrm{in}}/2<1$ and
$\|\widetilde A_i-A_i\|\le\eta_{\mathrm{in}}\|A_i\|+\eta_{\mathrm{in}}/2\le2\eta_{\mathrm{in}}$.  Summing over $i$ with signs, we obtain
$\|\sum\varepsilon_i(A_i-\widetilde A_i)\|\le2n\eta_{\mathrm{in}}=2n^22^{-b_{\mathrm{in}}}\le2^{-9}$, because
$2^{b_{\mathrm{in}}}\ge2^{10}n^2$.  We round each entry by one integer division of an
$(L_{\mathrm{in}}+b_{\mathrm{in}})$-bit numerator, at cost $O(M(L_{\mathrm{in}}+b_{\mathrm{in}}))$, and there are $n^3$ entries.
\end{proof}

\subsection{State, parameters, and randomness}
\label{subsec:bit_state}

Our finite algorithm keeps every quantity that the real-arithmetic algorithm updates incrementally on a fixed rational
grid, so that these updates are exact.  It replaces each analysis parameter by a nearby dyadic number rounded in the
direction that keeps the inequality it enters.  We first fix the objects of one projection call and then the lemma
that lists the grids, the rounded parameters, and the three consequences the later sections use.  The
rounded body still contains the smoothed body $(1-10^{-6})K(R)$ of Lemma~\ref{lem:algorithm_certified_implementation}, so
that every Gaussian-measure statement of Section~\ref{sec:small_algorithm_constant} transfers to it.  The rounded radius
stays inside the window $[R,(1+10^{-8})R]$ that Section~\ref{sec:small_algorithm_constant} charges.  The snapping step of
a suffix stage still lands in the body.

\paragraph{Standing objects of a call.}
Fix $n>N_{\mathrm{root}}$, $N_0\le m\le n$, and symmetric $b_{\mathrm{in}}$-bit dyadic contractions
$M_1,\ldots,M_m\in\R^{n\times n}$.  These are the $\widetilde A_i$ of Lemma~\ref{lem:bit_input}, sign flips $s_iM_i$ of
them, or the zero matrices of the padding in Definition~\ref{def:cubic_refined_suffix} and
Algorithm~\ref{alg:cubic_refined_repair}.  Write $B(x):=\sum_{j\le m}x_jM_j$ and $d_0:=2n$.  The call receives a
target $b\in\R^m$ with $\|b\|_2^2\le c_bm$ for an absolute constant $c_b\ge1$, one of the four caps we list after
Definition~\ref{def:bit_word_size}.  The call also receives a box, a
radius, an accuracy, and a failure budget $\varepsilon_{\mathrm{fail}}$.  Here $b$ is the target of
Section~\ref{sec:small_algorithm_constant}.  We write out the smoothing constant $b_{\mathrm{sm}}=1-s_{\mathrm{sm}}$ as $1-s_{\mathrm{sm}}$ in Lemma~\ref{lem:bit_state}, whose
rounded forms are $\widetilde b=\lfloor1-s_{\mathrm{sm}}\rfloor_{\mathsf u}$ and $b'=\widetilde\theta/\widetilde a$.  The
constants $a=a_{\mathrm{sm}}=2\cdot10^6\log(2n)$, $s_{\mathrm{sm}}=5\cdot10^{-7}$, $\chi=\log(2n)/a=5\cdot10^{-7}$, and
$\phi_0=2n\exp(-a(1-s_{\mathrm{sm}}))=(2n)^{-1999998}<1/4$ are those of Section~\ref{sec:algorithm}, which
Lemma~\ref{lem:appendix_fixed_target_projection} reuses.  Every call has radius $R$ with $R^2>15m$ and
$R\le10^{14}\sqrt n$.  We check these bounds after Definition~\ref{def:bit_word_size} and in facts (B2) and (B3) of Step~1 of Section~\ref{subsec:bit_proof}.  The coordinate grid is
\[
\Gamma:=(\mathsf u/10^{10})\mathbb Z .
\]
The suffix box $Q_y(t_s)$ has endpoints $t_s(\pm1-y_i)$ with
$t_s=v/10^{10}$ for an integer $v$ (Definition~\ref{def:cubic_refined_suffix}) and $y_i\in\mathsf u\mathbb Z$.  The
repair boxes $[-t_0,t_0]^m$ have $t_0\in\{2/5,3/5\}$ (Eq.~\eqref{eq:cubic_repair_parameters}).  All these endpoints lie
in $\Gamma$.  We store a number of $\Gamma$, or of the grid $2^{-b_{\mathrm{in}}}\Gamma$ of the entries of $B(x)$,
as an integer numerator over the common denominator, at the $b_{\mathrm{in}}+34$ further bits of the computational
model.

The word of the call is Eq.~\eqref{eq:bit_call_word} of Definition~\ref{def:bit_word_size} at the call's $m$, $c_b$,
$\widetilde\rho$, and $R_{\mathrm{alg}}$.

\begin{lemma}[Exact state and rounded parameters]
\label{lem:bit_state}
Fix a call with the standing objects above and $f\ge40+\lceil\log_2(2n)\rceil$.  Let $R>0$ be the radius the call
prescribes.  This radius is that of Definition~\ref{def:cubic_refined_suffix} for a suffix stage and that of
Eq.~\eqref{eq:cubic_repair_parameters} with its row for a repair call.  Let $R_{\mathrm{up}}\in\mathsf u\mathbb Z$ be a
certified upper enclosure with $R\le R_{\mathrm{up}}\le(1+10^{-10})R$.  The integer-square and logarithm rules of
Definition~\ref{def:cubic_refined_policy_construction} deliver it.  Define
\begin{align*}
R_{\mathrm{alg}}&:=\lceil(1+4\mathsf u)R_{\mathrm{up}}\rceil_{\mathsf u},&
\widetilde a&:=\lceil a\rceil_{\mathsf u},&
\widetilde c&:=\lfloor\widetilde a/R_{\mathrm{alg}}\rfloor_{\mathsf u},&
\widetilde R&:=\widetilde a/\widetilde c,\\
\widetilde b&:=\lfloor1-s_{\mathrm{sm}}\rfloor_{\mathsf u},&
\widetilde\theta&:=\lfloor\widetilde a\widetilde b\rfloor_{\mathsf u},&
b'&:=\widetilde\theta/\widetilde a,&
\widetilde\chi&:=\log(2n)/\widetilde a,
\end{align*}
and for $x\in\Gamma^m$ put $D_x:=\widetilde c\operatorname{diag}(B(x),-B(x))-\widetilde\theta I_{d_0}$,
$\widehat D_x:=\operatorname{round}_{\mathsf u}(D_x)$, $K(r):=\{x\in\R^m:\|B(x)\|\le r\}$,
$\widetilde K:=K(\widetilde R)$, and $\widetilde{\mathcal D}:=\{x\in\R^m:\tr[\exp(D_x)]\le1\}$.  Then the following
hold.
\begin{enumerate}[label=(\roman*)]
\item \emph{Grids.}  The endpoints of $\widetilde Q$ lie in $\Gamma$.  For $x\in\Gamma^m$ the
entries of $B(x)$ lie in $2^{-b_{\mathrm{in}}}\Gamma$ and those of $D_x$ in
$\mathsf u2^{-b_{\mathrm{in}}}\Gamma$.  The update $x\mapsto x+de_i$ with $d\in\Gamma$, the update of
$B(x)$, and $q(x):=\frac12\|x-b\|_2^2$ for $b\in(2^{-f-40}\mathbb Z)^m$ are exact rational operations, and every
comparison among these quantities is exact.  The value $q(x)$ is an integer multiple of $\mathsf u^2/(2^{81}5^{20})$ of
modulus at most $2c_bm$, and it is held in a $2W$-bit word once $I\ge I_B(m,c_b,\widetilde\rho,R_{\mathrm{alg}})$.
Moreover $\|\widehat D_x-D_x\|\le d_0\mathsf u$, and $\widehat D_x$ is
recomputed from the exact $D_x$ whenever it is needed.
\item \emph{Directions and the radius window.}  $\widetilde a\ge a$, $\widetilde\chi\le\chi$,
$1-s_{\mathrm{sm}}-2\mathsf u\le b'\le\widetilde b\le1-s_{\mathrm{sm}}$,
\[
(1+4\mathsf u)R\le R_{\mathrm{alg}}\le\widetilde R\le(1+2^{-30})R_{\mathrm{alg}}\le(1+10^{-8})R,
\]
the matrices $E_i:=\partial_iD_x=\widetilde c\operatorname{diag}(M_i,-M_i)$ satisfy $\|E_i\|\le\widetilde c$ and
$\sum_{i\le m}E_i^2\preceq m\widetilde c^2I_{d_0}$, and
$a/(2R_{\mathrm{alg}})\le\widetilde c\le(a+\mathsf u)/R_{\mathrm{alg}}\le2a/\sqrt{15m}$.
\item \emph{The rounded body.}  We have
$(1-10^{-6})K(R)\subseteq(b'-\widetilde\chi)\widetilde K\subseteq\widetilde{\mathcal D}\subseteq b'\widetilde K\subseteq(1-s_{\mathrm{sm}})\widetilde K$,
and $\widetilde{\mathcal D}$ is closed, convex, and symmetric.  The origin is a Slater point:
$\widetilde h(0)=\widetilde\chi\le5\cdot10^{-7}<b'$ for
$\widetilde h(x):=\widetilde a^{-1}\log\tr[\exp(\widetilde a\operatorname{diag}(B(x),-B(x))/\widetilde R)]$.
\item \emph{Snapping (suffix stages).}  Let
$\widetilde\rho:=\lfloor\min\{\tfrac18,\tfrac{t_s}8,s_{\mathrm{sm}}R_{\mathrm{alg}}/(4(2m+\lceil\sqrt m\rceil))\}\rfloor_{\mathsf u}$,
which is at most the $\rho$ of Eq.~\eqref{eq:algorithm_119_accuracy} at $\widehat R=\widetilde R$.  Let
$z\in\widetilde Q$ satisfy $\|z-x_*\|_2\le\widetilde\rho$ for $x_*:=\operatorname{proj}_{\widetilde{\mathcal D}\cap\widetilde Q}(b)$.
Let $\overline x$ be obtained from $z$ by moving every coordinate within $2\widetilde\rho$ of a facet of
$\widetilde Q=Q_y(t_s)$ onto that facet.  Then the nearer facet is unambiguous, every coordinate of $x_*$ that lies
on a facet is moved, and $\|B(\overline x)\|\le(1-3s_{\mathrm{sm}}/4)\widetilde R<\widetilde R$.
\end{enumerate}
\end{lemma}

\begin{proof}
\emph{(i), grids.}  The endpoints $t_s(\pm1-y_i)=v(\pm1-y_i)/10^{10}$ with $v\in\mathbb Z$ and $\pm1-y_i\in\mathsf u\mathbb Z$,
and $\pm t_0\in\{\pm4\cdot10^9,\pm6\cdot10^9\}/10^{10}$, lie in $\Gamma$.  The entries of $M_i$ are multiples of $2^{-b_{\mathrm{in}}}$, so $B(x)$ has entries in
$2^{-b_{\mathrm{in}}}\Gamma$ for $x\in\Gamma^m$, and $\widetilde c,\widetilde\theta\in\mathsf u\mathbb Z$ give entries of
$D_x$ in $\mathsf u2^{-b_{\mathrm{in}}}\Gamma$.  Additions of numbers on a common grid are exact, and comparisons of
rationals with a common denominator are integer comparisons.
For $q(x)$, the differences $x_i-b_i$ lie on the common grid $(\mathsf u/(2^{40}5^{10}))\mathbb Z$ of
$\Gamma=(\mathsf u/(2^{10}5^{10}))\mathbb Z$ and $(\mathsf u/2^{40})\mathbb Z$, so $q(x)$ is an integer multiple
of $\mathsf u^2/(2^{81}5^{20})$.  The bounds $\|x\|_2\le\sqrt m$ (as $\widetilde Q\subseteq[-1,1]^m$) and $\|b\|_2\le\sqrt{c_bm}$
give $q(x)\le\frac12(\sqrt m+\sqrt{c_bm})^2\le2c_bm$ for $c_b\ge1$.  Its numerator has fewer than
$\log_2(2c_bm)+2f+128$ bits ($20\log_25<47$).  Moreover $\log_2(2c_bm)+128\le2\lceil\log_2c_b\rceil+14\lceil\log_2(2n)\rceil+24\le2I_B$
for $n\ge2^7$, so the $2W$-bit word holds it once $I\ge I_B$.  Rounding each entry of $D_x$ moves it by at most $\mathsf u/2$, so
$\|\widehat D_x-D_x\|\le\|\widehat D_x-D_x\|_F\le d_0\mathsf u/2$.

\emph{(ii), directions.}  The bounds $\widetilde a\ge a$ and $\widetilde\chi\le\chi$ hold because we round up.  Rounding down
twice, we obtain $\widetilde\theta\le\widetilde a\widetilde b$, so $b'\le\widetilde b\le1-s_{\mathrm{sm}}$.  The bound
$\widetilde\theta\ge\widetilde a\widetilde b-\mathsf u\ge\widetilde a(1-s_{\mathrm{sm}}-\mathsf u)-\mathsf u$ gives
$b'\ge1-s_{\mathrm{sm}}-2\mathsf u$.  Rounding up
gives $R_{\mathrm{alg}}\ge(1+4\mathsf u)R_{\mathrm{up}}\ge(1+4\mathsf u)R$.  Rounding down gives
$\widetilde c\le\widetilde a/R_{\mathrm{alg}}$, hence $\widetilde R\ge R_{\mathrm{alg}}$.  The bound
$\widetilde c\ge\widetilde a/R_{\mathrm{alg}}-\mathsf u$ gives
$\widetilde R\le R_{\mathrm{alg}}/(1-\mathsf uR_{\mathrm{alg}}/\widetilde a)\le(1+2^{-30})R_{\mathrm{alg}}$ once
$R_{\mathrm{alg}}\le2^{f-31}\widetilde a$.  Here $R_{\mathrm{alg}}\le10^{14}\sqrt n$ comes from the radius bound of the
standing objects, with $R_{\mathrm{alg}}\le(1+10^{-9})R$, which we show next.  The radius bound and
$\widetilde a\ge a\ge2\cdot10^6\log(2N_{\mathrm{root}})$ give it for $f\ge\frac12\lceil\log_2n\rceil+51$, which
$f\ge40+\lceil\log_2(2n)\rceil$ implies for $n>N_{\mathrm{root}}$.  The same inequality gives
$\mathsf uR_{\mathrm{alg}}\le a/2$, hence $\widetilde c\ge a/R_{\mathrm{alg}}-\mathsf u\ge a/(2R_{\mathrm{alg}})$.  The
upper bound $\widetilde c\le(a+\mathsf u)/R_{\mathrm{alg}}\le2a/\sqrt{15m}$ uses $R_{\mathrm{alg}}^2>15m$.  For the
window, we combine $R_{\mathrm{up}}\le(1+10^{-10})R$, $R_{\mathrm{alg}}\le(1+4\mathsf u)R_{\mathrm{up}}+\mathsf u$, and $R\ge\sqrt{15m}\ge1$
to obtain $\widetilde R\le(1+2^{-30})\bigl((1+4\mathsf u)(1+10^{-10})+\mathsf u\bigr)R\le(1+1.036\cdot10^{-9})R\le(1+10^{-8})R$
for $\mathsf u\le2^{-40}$.  This is the window $\widehat R\in[R,(1+10^{-8})R]$ of
Algorithm~\ref{alg:algorithm_suffix_signing} and Lemma~\ref{lem:algorithm_certified_implementation} for the radius
$\widetilde R$ of the body we actually use.  The bounds on $E_i$ follow from $\|M_i\|\le1$ and $\sum_iM_i^2\preceq mI$
exactly as in Eq.~\eqref{eq:cubic_E_bounds}, with $\widetilde c$ in place of $a/R$.

\emph{(iii), the body.}  Write
\[
\widetilde h(x):=\widetilde a^{-1}\log\tr[\exp(\widetilde a\operatorname{diag}(B(x),-B(x))/\widetilde R)],
\]
so that $D_x=\widetilde a\operatorname{diag}(B(x),-B(x))/\widetilde R-\widetilde ab'I$ and $\tr[\exp(D_x)]\le1$ if and only if
$\widetilde h(x)\le b'$.  The log-sum-exp bounds $\|B(x)\|/\widetilde R\le\widetilde h(x)\le\|B(x)\|/\widetilde R+\widetilde\chi$ give the
inclusions $(b'-\widetilde\chi)\widetilde K\subseteq\widetilde{\mathcal D}\subseteq b'\widetilde K$, and $b'\le1-s_{\mathrm{sm}}$ gives the
third.  Convexity of $\widetilde h$ and the symmetry $\widetilde h(-x)=\widetilde h(x)$ give the shape statements.  For the first
inclusion let $x\in(1-10^{-6})K(R)$.  By (ii),
\[
\|B(x)\|\le(1-10^{-6})R\le\frac{(1-10^{-6})\widetilde R}{1+4\mathsf u}\le(1-10^{-6}-2\mathsf u)\widetilde R\le(b'-\widetilde\chi)\widetilde R,
\]
because $(1-10^{-6}-2\mathsf u)(1+4\mathsf u)-(1-10^{-6})=2\mathsf u(1-2\cdot10^{-6}-4\mathsf u)>0$ and
$b'-\widetilde\chi\ge1-s_{\mathrm{sm}}-2\mathsf u-\chi=1-10^{-6}-2\mathsf u$.  This is where we need the factor $1+4\mathsf u$ in
$R_{\mathrm{alg}}$.  The real-arithmetic inclusion $(1-10^{-6})K(R)\subseteq D$ of
Lemma~\ref{lem:algorithm_certified_implementation} has no slack when $\widehat R=R$, which happens when
$v_sn$ is a perfect square, and we round $b'$ downward by up to $2\mathsf u$.  With the first inclusion every
Gaussian-measure statement of Section~\ref{sec:small_algorithm_constant} about $D$ transfers to
$\widetilde{\mathcal D}$ through $\gamma_m(\widetilde{\mathcal D})\ge(1-10^{-6})^m\gamma_m(K(R))$, which is the exponent
the stage tests of Definition~\ref{def:cubic_refined_suffix} charge.  Finally
$\widetilde h(0)=\widetilde a^{-1}\log(2n)=\widetilde\chi\le\chi=5\cdot10^{-7}<1-s_{\mathrm{sm}}-2\mathsf u\le b'$.

\emph{(iv), snapping.}  Here $\widetilde\rho\le s_{\mathrm{sm}}R_{\mathrm{alg}}/(4(2m+\sqrt m))\le s_{\mathrm{sm}}\widetilde R/(4(2m+\sqrt m))$,
so $\widetilde\rho\le\rho$.  The facets of $Q_y(t_s)$ in coordinate $i$ are $2t_s$ apart and $4\widetilde\rho\le t_s/2<2t_s$,
so no coordinate is within $2\widetilde\rho$ of both.  A coordinate of $x_*$ on a facet has $|z_i-\text{facet}|\le\|z-x_*\|_2\le\widetilde\rho<2\widetilde\rho$
and we move it.  Each move changes one coordinate by at most $2\widetilde\rho$, $\|M_i\|\le1$, and
$\|B(z)\|\le\|B(x_*)\|+\sqrt m\widetilde\rho\le(1-s_{\mathrm{sm}})\widetilde R+\sqrt m\widetilde\rho$ (from
$x_*\in\widetilde{\mathcal D}$ and $\|B(v)\|\le\|v\|_1\le\sqrt m\|v\|_2$), so
$\|B(\overline x)\|\le(1-s_{\mathrm{sm}})\widetilde R+(2m+\sqrt m)\widetilde\rho\le(1-s_{\mathrm{sm}})\widetilde R+s_{\mathrm{sm}}\widetilde R/4$.
\end{proof}

\paragraph{Wrapper parameters of a call.}
We round the remaining parameters of the wrapper and of the recursive fixed-multiplier solver of
Section~\ref{subsec:cubic_multiplier} as follows, each in the direction that preserves the inequality it
enters.  The rounding uses $q(0):=\frac12\|b\|_2^2$, exact by Lemma~\ref{lem:bit_state}(i), and the accuracy
$\widetilde\rho$ and radius $R_{\mathrm{alg}}$ of the call.  Each value is computable in $O(\operatorname{polylog}(n))$
bit operations from $n$, $m$, $\|b\|_2^2$, $\nu$, $\widetilde\rho$, and $R_{\mathrm{alg}}$.  On the multiplier branch,
the values are computable after the pre-test of Lemma~\ref{lem:bit_enclosures}(v), where
$\widetilde\Lambda\ge\frac43\|b\|_2^2>19.9$, so that the divisions by $\underline{\widetilde\nu}$ below are by a
positive number.
Put $\widetilde\Lambda:=\lceil(8/3)q(0)\rceil_{\mathsf u}\ge\Lambda$ (using $\phi_0<1/4$).
Put $\widetilde H:=\lceil16(q(0)+\nu/4)\rceil_{\mathsf u}\ge16F_\nu(0)$ and
$\widetilde H_{\max}:=\lceil16(q(0)+\widetilde\Lambda/4+1)\rceil_{\mathsf u}\ge H_{\max}$.
Put $\widetilde G_*:=\lceil\widetilde c\sqrt m\rceil_{\mathsf u}\ge\widetilde c\sqrt m$, the Lipschitz constant of
$\Phi$ for the finite objective $F$ that we define at the end of this list.
Put $\widetilde\nu_{\min}:=\lfloor\widetilde\rho/(8\widetilde G_*)\rfloor_{\mathsf u}$ and
$\underline{\widetilde\nu}:=\min\{\widetilde\nu_{\min},\widetilde\Lambda/2\}$ (rounded down to $\mathsf u\mathbb Z$).
Put $\widetilde d_{\mathrm{w}}:=\lfloor\widetilde\rho/4\rfloor_{\mathsf u}$,
$\widetilde\tau_{\mathrm{w}}:=\lfloor\min\{\widetilde d_{\mathrm{w}}^2/(1+\widetilde\Lambda),3\widetilde d_{\mathrm{w}}/(64\sqrt m)\}\rfloor_{\mathsf u}$,
$\widetilde G_{\mathrm{lev}}:=\lceil\widetilde G_*\widetilde H_{\max}/\underline{\widetilde\nu}\rceil_{\mathsf u}$,
$\widetilde\sigma:=\lfloor\min\{\widetilde d_{\mathrm{w}}/8,\widetilde\tau_{\mathrm{w}}/(16\widetilde G_{\mathrm{lev}})\}\rfloor_{\mathsf u}$,
$\widetilde\varepsilon_{\mathrm{opt}}:=\lfloor\widetilde\sigma^2/2\rfloor_{\mathsf u}$ (positive under Eq.~\eqref{eq:bit_word_length}),
and $\widetilde J:=\lceil\log_2(4\widetilde H/\widetilde\varepsilon_{\mathrm{opt}})\rceil$.
Put $\widetilde L:=4\cdot2^{\lceil\log_2\max\{1+3\widetilde c^2\widetilde H,\,2\widetilde c,\,1\}\rceil}$,
$\widetilde\alpha:=1/(m\widetilde L)$, the rounded form of the contraction rate $\alpha:=1/(mL)$ of Eq.~\eqref{eq:cubic_alpha}, and $\widetilde T:=16m\widetilde L$.
Put $\widetilde\delta_g:=\bigl\lfloor\min\{\widetilde H/(10^5m),\ \widetilde\varepsilon_{\mathrm{opt}}/(C_{\mathrm g}\widetilde Jm),\ (2\widetilde c\widetilde H/9)^2\}^{1/2}\bigr\rfloor_{\mathsf u}$
with $C_{\mathrm g}:=2\cdot10^5$, whose third term gives Lemma~\ref{lem:bit_eps_op}(iii).
Put $\pi':=$ the least power of two with $\pi'\ge\max\{n^{-1/2},4\widetilde\alpha\}$, so that
$\pi_m\le\pi'\le2\max\{n^{-1/2},4\widetilde\alpha\}$ and $\pi'\ge4\widetilde\alpha$.  Hence
$T\pi'=16m\widetilde L\pi'\le32m\widetilde Ln^{-1/2}+128$ and $L_{\max}=\lceil1/\pi'\rceil\le\sqrt n+1$.  The value
$\pi'$ may exceed $2\pi_m$, because we build $\widetilde L$ from the level $\widetilde H\ge16F_\nu(0)$ while the $L$ of
Eq.~\eqref{eq:cubic_L} uses $H=64U$.  Only the constant in the refresh count changes.
Put $k=s:=\lceil C_{ks}\pi'^{-1/2}\rceil$ with $C_{ks}:=45$, so that $2\le k\le45n^{1/4}+1\le n/4$ for $n\ge N_0$ and
$ks\ge C_{ks}^2/\pi'\ge K_0/\pi'$ for $K_0:=2000$, the value we fix in Lemma~\ref{lem:bit_lyapunov} below.  In Section~\ref{sec:cubic_time} we leave $K_0$ a sufficiently large
universal constant.
Put $p_{\mathrm s}:=\mathsf u^4$ per capped sampling procedure (Lemmas~\ref{lem:bit_probes} and~\ref{lem:bit_gaussian}(i),(iii)).
This is the budget we already used for the one-redraw cap of Lemma~\ref{lem:bit_gaussian}(iii).  Each cap is then
$O(\log_2(N_{\mathrm s}/p_{\mathrm s}))=O(\log n)$ rounds or coins.
The solve budget is
$p_0:=\varepsilon_{\mathrm{fail}}/(2N_{\mathrm{sol}})$ with $N_{\mathrm{sol}}$ as in Lemma~\ref{lem:bit_projection}.
The quantities $R_{\mathrm{rep}}$ and $s_{\mathrm{tr}}=128\lceil8\log(100\widetilde T\widetilde JR_{\mathrm{rep}}/p_0)\rceil$
are those of Section~\ref{sec:selected_wrapper} and Lemma~\ref{lem:bit_guard}, and we fix $\varepsilon_{\mathrm{op}}$
in Eq.~\eqref{eq:bit_eps_op} after Lemma~\ref{lem:bit_step}.
We always evaluate the error bounds $\varepsilon_{\mathrm{dense}}$ and $\varepsilon_{\mathrm{act}}$ of Lemmas~\ref{lem:bit_dense} and~\ref{lem:bit_action}
at $\lambda_{\mathrm{sp}}=\lambda_{\mathrm{top}}$.  Here $\lambda_{\mathrm{sp}}$ is the spectral
upper bound in the hypotheses of Lemma~\ref{lem:bit_taylor}, the bound valid at every point
where the algorithm evaluates them.  We re-verify below, for the rounded values, every
inequality of Sections~\ref{sec:faster_algorithm} and~\ref{sec:cubic_time} that involves
$a$, $R$, $b_{\mathrm{sm}}$, $H$, $L$, $\Lambda$, $\underline\nu$, $\tau_{\mathrm{w}}$, $\sigma$, $\varepsilon_{\mathrm{opt}}$,
$\delta_g$, or $\pi_m$.  The re-verification is two-sided in
Lemma~\ref{lem:bit_state}(ii)--(iv) for the parameters of the body, and it runs through the ratio bounds we record next
for the wrapper parameters.  We claim no monotonicity in the rounding direction.
The rounded parameters and the constants of Eq.~\eqref{eq:bit_call_word}, evaluated at the cap $q_B\ge q(0)$, satisfy
the following bounds.  We have
$\widetilde H_{\max}\le\frac53\cdot16q(0)+16+5\mathsf u\le\frac53H_{\max}$, from $\widetilde\Lambda/4\le\frac23q(0)+\mathsf u/4$
and $H_{\max}\ge16q_B+16$.  Also $\widetilde d_{\mathrm w}\ge d_{\mathrm w}-\mathsf u$ and
$\widetilde\tau_{\mathrm w}\ge\tau_{\mathrm w}/16$, from $1+\widetilde\Lambda\le2(1+\Lambda)$.  For the multiplier
floor, suppose that $\widetilde\nu_{\min}\le\widetilde\Lambda/2$.  In this case
$\underline{\widetilde\nu}=\widetilde\nu_{\min}\ge\nu_{\min}/2-\mathsf u\ge\underline\nu/4$, because
$\widetilde G_*\le(a+\mathsf u)\sqrt m/R_{\mathrm{alg}}+\mathsf u\le(1+2^{-30})G_*$ and $\underline\nu\le\nu_{\min}$.
Suppose instead that $\widetilde\nu_{\min}>\widetilde\Lambda/2$.  In this case
$\underline{\widetilde\nu}=\widetilde\Lambda/2>7.4$ on the multiplier branch (Lemma~\ref{lem:bit_enclosures}(v)), and
$\widetilde H_{\max}/\underline{\widetilde\nu}\le(10\widetilde\Lambda+17)/(\widetilde\Lambda/2)<23$, from
$16q(0)\le6\widetilde\Lambda$.  At the same time, $\underline\nu\le\Lambda/2$ gives
$H_{\max}/\underline\nu\ge2H_{\max}/\Lambda\ge16(1-\phi_0)>12$.  In both cases we obtain
$\widetilde H_{\max}/\underline{\widetilde\nu}\le\frac{20}3H_{\max}/\underline\nu\le2^3H_{\max}/\underline\nu$,
$\widetilde G_{\mathrm{lev}}\le7G_{\mathrm{lev}}$, and $\widetilde\varepsilon_{\mathrm{opt}}\ge2^{-14}\varepsilon_{\mathrm{opt}}$.
The rounding floors cost at most $\mathsf u$ each, negligible under
$f\ge f_B$.  We use two bounds below that
carry the target cap: $\widetilde\Lambda\le\frac83q(0)+\mathsf u\le\frac43c_bm+\mathsf u$, and
\[
\widetilde L\le8\max\{1+3\widetilde c^2\widetilde H,2\widetilde c\}\le8\Bigl(1+\frac{3(a+\mathsf u)^2}{15m}\Bigl(\frac{40}3c_bm+16+5\mathsf u\Bigr)\Bigr)\le40a^2c_b,
\]
from $\widetilde c\le(a+\mathsf u)/\sqrt{15m}$ and $\widetilde H\le\widetilde H_{\max}\le16q(0)+4\widetilde\Lambda+16+\mathsf u$.
This is the $O(\log^2n)$ curvature of Lemma~\ref{lem:appendix_fixed_target_projection} with its constant made
explicit.  The phase length $\widetilde T=16m\widetilde L$ carries $c_b$, and so does the arithmetic count of
Section~\ref{sec:small_algorithm_constant}.  The word sees $\widetilde L$ only through $\log_2(2(\widetilde L+1))$ in
the $(L+1)\mathsf u$ row of Step~1 and through $s_{\mathrm{tr}}$.
In the rest of the section we drop the tildes on these parameters and on $\Phi$, $F$, and
$\Delta$, which now refer to the finite objective
$F(x):=q(x)+\nu\tr[\exp(D_x)]$ on $\widetilde Q$ with the target $b$.  The objective $F$ is
$1$-strongly convex.  The coordinate curvature bound
$\partial_{ii}^2F\le1+e\widetilde c^2H\le L/4$ and the trust condition $|d|\le R/a$ (here $|d|\le1/\widetilde c$) of
Section~\ref{sec:cubic_time} hold with $\widetilde c$ in place of $a/R$.  Here we use Lemma~\ref{lem:bit_state}(ii).

We use Rademacher vectors as bounded probes, and they supply the isotropy and variance estimates used by the tracker and its guards. For the remaining random choices we give exact implementations from fair coins.

\begin{lemma}[Discrete probes and exact coins]
\label{lem:bit_probes}
Let $z\in\{-1,1\}^{d_0}$ have independent fair-coin entries and let $M$ be a real
symmetric $d_0\times d_0$ matrix.  Then
\[
\E[zz^\top]=I_{d_0},
\qquad
\E[(Mz)z^\top]=M,
\qquad
\V[z^\top Mz]=2\|M\|_F^2-2\sum_{j}M_{jj}^2\le2\|M\|_F^2,
\qquad
\|z\|_2^2=d_0.
\]
Moreover, a uniform index in $[m]$, a Bernoulli coin of dyadic bias $\pi'$, and a
Bernoulli coin of bias $(1+y_i)/2$ for $y_i\in \mathsf u\mathbb Z$ can each be drawn exactly.  The index costs
an expected $O(\log m)$ fair coins, capped at $\lceil\log_2(N_{\mathrm{s}}/p_{\mathrm{s}})\rceil$ rounds for a procedure
drawing $N_{\mathrm{s}}$ indices, with a declared failure of total probability at most $p_{\mathrm{s}}$.  The dyadic
coin costs exactly $\log_2(1/\pi')$ fair coins, and the remaining coin costs exactly $f+1$ fair coins.
\end{lemma}

\begin{proof}
The first two identities follow from $\E[z_jz_l]=\delta_{jl}$.  For the variance,
$z^\top Mz=\sum_jM_{jj}+2\sum_{j<l}M_{jl}z_jz_l$, and the products $z_jz_l$, $j<l$, are
pairwise uncorrelated fair signs, so
$\V[z^\top Mz]=4\sum_{j<l}M_{jl}^2=2\|M\|_F^2-2\sum_jM_{jj}^2$.  The norm identity is
$z_j^2=1$.  We draw a uniform index by rejection from $\lceil\log_2m\rceil$ coins, with
acceptance probability at least $1/2$ per round.  For a procedure drawing $N_{\mathrm{s}}$ indices we cap the rounds at
$\lceil\log_2(N_{\mathrm{s}}/p_{\mathrm{s}})\rceil$, the cap being a
declared failure of probability at most $p_{\mathrm{s}}$ in total.  A dyadic bias $\pi'=2^{-j}$ is the event that $j$
coins are all heads.  We compare the bias $(1+y_i)/2\in2^{-f-1}\mathbb Z$ with a uniform
$(f+1)$-bit integer.
\end{proof}

With Rademacher probes the Hutchinson estimates of Sections~\ref{sec:faster_algorithm}
and~\ref{sec:cubic_time} keep their means, and their variance bounds $2\|M\|_F^2$ hold
with $\le$ in place of $=$.  The range-finder test matrix and the ball target of a suffix stage are the two places where we
need a Gaussian.  We treat them in Lemmas~\ref{lem:bit_gaussian}
and~\ref{lem:bit_rangefinder} and in Lemma~\ref{lem:bit_ball_target}.

\begin{lemma}[Rounded Gaussians and the rounded test matrix]
\label{lem:bit_gaussian}
Let $h_\Omega:=2^{-b_\Omega}$ with $b_\Omega:=\lceil4.5\log_2n\rceil+7$.
\begin{enumerate}[label=(\roman*)]
\item \emph{Exact rounded Gaussians.}  There is an algorithm that, from fair coins,
outputs a number $\xi'\in h\mathbb Z$ whose law is exactly that of
$\operatorname{round}_h(G)$ for $G\sim \mathcal N(0,1)$, for any $h=2^{-b_h}$.  It uses an expected
$O(b_h)$ coins and $O(b_h)$ bit operations.  On the same coin space there is a
$G\sim \mathcal N(0,1)$ with $|\xi'-G|\le h/2$ surely.  Capping the coins per sample at
$C_{\mathrm{cap}}(b_h+\log_2(N_{\mathrm{s}}/p_{\mathrm{s}}))$ for a universal $C_{\mathrm{cap}}$ and declaring
failure at the cap adds at most $p_{\mathrm{s}}$ to the failure probability of any
procedure that draws $N_{\mathrm{s}}$ samples.
\item \emph{Lipschitz transfer.}  If $F:\R^m\to\R$ is $1$-Lipschitz for the
Euclidean norm and $\xi'=\operatorname{round}_h(G)$ coordinatewise, then
$|F(\xi')-F(G)|\le\sqrt m\,h/2$ surely.

\item \emph{Rounded test matrix.}  Let $\Omega_{\max}:=1+\sqrt{2\log(4nk/\mathsf u^2)}$.  The matrix
$\widehat\Omega\in(h_\Omega\mathbb Z)^{d_0\times2k}$ is drawn as independent samples of
$\operatorname{round}_{h_\Omega}(\mathcal N(0,1))$.  If $\|\widehat\Omega\|_{\max}>\Omega_{\max}$ it is redrawn once, and a
second oversized draw is a declared failure.
One draw satisfies $\widehat\Omega=\Omega+E_\Omega$ for a standard Gaussian $\Omega$ on the same coin
space with $\|E_\Omega\|_{\max}\le h_\Omega/2$, and it is oversized with probability at most $\mathsf u^2$.  The
declared failure has probability at most $\mathsf u^4$ per test matrix.  On its complement the accepted
matrix has exactly the law of one draw conditioned on the event $\{\|\widehat\Omega\|_{\max}\le\Omega_{\max}\}$,
of probability at least $1-\mathsf u^2$.  Finally $\Omega_{\max}=O(\sqrt{\log n})$ for $f=O(\log n)$ as in
Eq.~\eqref{eq:bit_word_length}.
\end{enumerate}
\end{lemma}

\begin{proof}
\emph{(i), the sampler.}  This is Karney's algorithm~\cite[Section~3]{karney16}.  It generates the binary digits of a normal
deviate one at a time, using only integer comparisons of partially generated uniform
deviates, so that the digits generated are those of an exactly normal $G$.  After
the digit that determines the rounding to $h\mathbb Z$ we stop and obtain
$\xi'=\operatorname{round}_h(G)$ with $G$ the deviate that the continued digit stream defines.
The number of coins consumed is a stopping time with an exponential tail, because each
of the $O(1)$ expected rejection rounds terminates a comparison with probability at
least $1/2$ per digit.  Hence a cap of $C_{\mathrm{cap}}(b_h+\log_2(N_{\mathrm{s}}/p_{\mathrm{s}}))$ coins
fails with probability at most $p_{\mathrm{s}}/N_{\mathrm{s}}$ per sample, and we obtain (i) by a union bound over
$N_{\mathrm{s}}$ samples.  \emph{(ii), transfer.}  This is $|F(\xi')-F(G)|\le\|\xi'-G\|_2\le\sqrt m\,h/2$.

\emph{(iii), the test matrix.}  To each of the $2d_0k$ entries we apply (i) with $h=h_\Omega$.  For a
standard normal $G$ and $t\ge1$, $\Pr[|G|>t]\le e^{-t^2/2}$, and $|\widehat\Omega_{jl}|>\Omega_{\max}$
forces $|\Omega_{jl}|>\Omega_{\max}-1$, an event of probability at most $\mathsf u^2/(4nk)$.  The union
bound over $4nk$ entries gives $\mathsf u^2$, and the two draws are independent, so both are oversized
with probability at most $\mathsf u^4$.  Let $\mathcal A$ be the event that a draw is not oversized.
The accepted matrix lies in a set $S$ and no failure occurs with probability
$\Pr[\mathcal A\cap S](1+\Pr[\mathcal A^c])$, while no failure occurs with probability
$1-\Pr[\mathcal A^c]^2=(1+\Pr[\mathcal A^c])\Pr[\mathcal A]$.  The ratio is $\Pr[S\mid\mathcal A]$, the
conditioned law of one draw.
\end{proof}

\subsection{Matrix exponentials in fixed point}\label{subsec:bit_exponentials}

The real-arithmetic algorithm approximates $e^{D_x}$ by the centered Taylor square
$q_r(D_x)^2$ of Section~\ref{sec:faster_algorithm}.  Its coefficients
$e^{c/2}/j!$ with $c\approx-ab_{\mathrm{sm}}$ are of size $(2n)^{-10^6}$.  Its partial sums cancel
by a factor $e^{ab_{\mathrm{sm}}}$ below the midpoint of the spectral interval.  To evaluate it
we therefore need about $2\cdot10^6\log_2(2n)$ bits per number.  We use instead the scaled
Taylor power
\begin{equation}
\mathcal Q(D):=T_r(D/2^{\kappa_{\mathrm{sc}}})^{2^{\kappa_{\mathrm{sc}}}},
\qquad
T_r(y):=\sum_{j=0}^r\frac{y^j}{j!},
\label{eq:bit_surrogate}
\end{equation}
whose factors have spectrum in $[1/3,e^{(\lambda_{\mathrm{sp}}+1)/\sigma_{\mathrm{sc}}}]$ with $\sigma_{\mathrm{sc}}:=2^{\kappa_{\mathrm{sc}}}$, by Lemma~\ref{lem:bit_taylor}.
For dense matrices we evaluate it by repeated squaring and for vectors by repeated
application.  It is one exact matrix, a deterministic function of $x$, and it is positive
definite.  We use it in every place where Sections~\ref{sec:faster_algorithm}
and~\ref{sec:cubic_time} use $\widetilde P_x$.

\begin{lemma}[Scaled Taylor surrogate]
\label{lem:bit_taylor}
Let $D$ be symmetric with spectrum in $[-M_{\mathrm{sp}},\lambda_{\mathrm{sp}}]$, where $\lambda_{\mathrm{sp}}\ge0$.
Let $2^{\kappa_{\mathrm{sc}}}=:\sigma_{\mathrm{sc}}\ge M_{\mathrm{sp}}+\lambda_{\mathrm{sp}}$, and let $r\ge2$ satisfy
$\sigma_{\mathrm{sc}}\eta_r\le1$ with $\eta_r:=e^2/(r+1)!$.  Put $X:=D/\sigma_{\mathrm{sc}}$.  Then:
\begin{enumerate}[label=(\roman*)]
\item for $y\in[-1,1]$, $T_r(y)=e^y(1+\vartheta)$ with $|\vartheta|\le\eta_r$, $T_r$ is nondecreasing on $[-1,1]$, and $T_r(y)\ge T_r(-1)\ge1/3$;
\item $\|T_r(X)^j\|\le e^{(\lambda_{\mathrm{sp}}+1)j/\sigma_{\mathrm{sc}}}\le e^{\lambda_{\mathrm{sp}}+1}$ for $0\le j\le\sigma_{\mathrm{sc}}$, and $\mathcal Q(D)=T_r(X)^{\sigma_{\mathrm{sc}}}\succeq3^{-\sigma_{\mathrm{sc}}}I\succ0$;
\item $\|\mathcal Q(D)-e^D\|\le3\sigma_{\mathrm{sc}}e^{\lambda_{\mathrm{sp}}+2}/(r+1)!$;
\item for symmetric $D'$ with $\|D'-D\|\le\zeta\le1$, $\|e^{D'}-e^D\|\le\zeta e^{\lambda_{\mathrm{sp}}+1}$.
\end{enumerate}
\end{lemma}

\begin{proof}
(i)  The Lagrange remainder gives $|T_r(y)-e^y|\le e^{\max\{y,0\}}|y|^{r+1}/(r+1)!\le e/(r+1)!$
on $[-1,1]$.  Dividing by $e^y\ge e^{-1}$, we get $|\vartheta|\le e^2/(r+1)!=\eta_r$.
Since $T_r'=T_{r-1}$ and $T_1(y)=1+y\ge0$ on $[-1,1]$, induction with $T_{r-1}(-1)>0$, which we show
next, gives $T_r'\ge0$ there, so $T_r(y)\ge T_r(-1)$.  The value $T_r(-1)=\sum_{j\le r}(-1)^j/j!$
is an alternating partial sum of $e^{-1}$, so $|T_r(-1)-e^{-1}|\le1/(r+1)!\le1/120$ for
$r\ge4$.  This gives $T_r(-1)>0.359$, while $T_2(-1)=1/2$ and $T_3(-1)=1/3$.  Hence
$T_r(-1)\ge1/3$ for all $r\ge2$.

(ii)  The spectrum of $X$ lies in $[-1,\lambda_{\mathrm{sp}}/\sigma_{\mathrm{sc}}]\subseteq[-1,1]$, so by (i)
and functional calculus
\[
\|T_r(X)^j\|=\max_y|T_r(y)|^j\le(e^{\lambda_{\mathrm{sp}}/\sigma_{\mathrm{sc}}}(1+\eta_r))^j
\le e^{(\lambda_{\mathrm{sp}}+\sigma_{\mathrm{sc}}\eta_r)j/\sigma_{\mathrm{sc}}}\le e^{(\lambda_{\mathrm{sp}}+1)j/\sigma_{\mathrm{sc}}},
\]
and $T_r(X)\succeq(1/3)I$ gives the lower bound.

(iii)  Diagonalize $D=\sum_i\mu_iv_iv_i^\top$.  Then
$\|\mathcal Q(D)-e^D\|=\max_i|T_r(\mu_i/\sigma_{\mathrm{sc}})^{\sigma_{\mathrm{sc}}}-e^{\mu_i}|$.  For
$y=\mu_i/\sigma_{\mathrm{sc}}\in[-1,1]$, (i) gives
$T_r(y)^{\sigma_{\mathrm{sc}}}-e^{\sigma_{\mathrm{sc}}y}=e^{\sigma_{\mathrm{sc}}y}((1+\vartheta)^{\sigma_{\mathrm{sc}}}-1)$.
Moreover, $|(1+\vartheta)^{\sigma_{\mathrm{sc}}}-1|\le e^{\sigma_{\mathrm{sc}}\eta_r}-1\le(e-1)\sigma_{\mathrm{sc}}\eta_r\le3\sigma_{\mathrm{sc}}\eta_r$,
because $\sigma_{\mathrm{sc}}\eta_r\le1$ and $e^t-1\le(e-1)t$ on $[0,1]$.  The remaining factor is
$e^{\sigma_{\mathrm{sc}}y}=e^{\mu_i}\le e^{\lambda_{\mathrm{sp}}}$.

(iv)  Duhamel's formula $e^{D'}-e^{D}=\int_0^1e^{(1-t)D'}(D'-D)e^{tD}\,\d t$ and
$\|e^{tB}\|=e^{t\lambda_{\max}(B)}$ for symmetric $B$ give
$\|e^{D'}-e^D\|\le\zeta\,e^{\max\{\lambda_{\max}(D),\lambda_{\max}(D')\}}\le\zeta e^{\lambda_{\mathrm{sp}}+\zeta}$.
\end{proof}

On the small-constant path, $D=\widehat D_x$ with $x$ a guarded point or a proposal from one, and the spectrum of
$\widehat D_x$ lies in $[-M_{\mathrm{sp}},\lambda_{\mathrm{top}}]$ by the paragraph \emph{Spectral bounds}.  We fix
\[
\sigma_{\mathrm{sc}}:=2^{\kappa_{\mathrm{sc}}},\ \kappa_{\mathrm{sc}}:=\lceil\log_2(M_{\mathrm{sp}}+\lambda_{\mathrm{top}})\rceil,
\qquad
r:=\min\{r\ge2:(r+1)!\ge24\sigma_{\mathrm{sc}}e^{\lambda_{\mathrm{top}}+2}/\varepsilon_{\mathrm{op}}\}
\]
so that $\sigma_{\mathrm{sc}}=O(ab_{\mathrm{sm}}+\lambda_{\mathrm{top}})=O(\log n)$ and
$r=O(\log(\sigma_{\mathrm{sc}}e^{\lambda_{\mathrm{top}}}/\varepsilon_{\mathrm{op}}))=O(\log n)$.
These choices also give $\sigma_{\mathrm{sc}}\eta_r\le1$, as $\varepsilon_{\mathrm{op}}\le1$, and $\kappa_{\mathrm{sc}}\ge1$.
Lemma~\ref{lem:bit_taylor}(iii) then gives
$\|\mathcal Q(\widehat D_x)-e^{\widehat D_x}\|\le\varepsilon_{\mathrm{op}}/8$.  Here
$\varepsilon_{\mathrm{op}}$ is the operator tolerance we fix in Eq.~\eqref{eq:bit_eps_op} below.
With $d_0\mathsf u\,e^{\lambda_{\mathrm{top}}+1}\le\varepsilon_{\mathrm{op}}/8$, which is one of the
requirements we collect in the proof of Theorem~\ref{thm:bit_complexity},
Lemma~\ref{lem:bit_taylor}(iv) and Lemma~\ref{lem:bit_state}(i) give
\begin{equation}
\|\mathcal Q(\widehat D_x)-P_x\|\le\varepsilon_{\mathrm{op}}/4,
\qquad
P_x=e^{D_x}.
\label{eq:bit_surrogate_accuracy}
\end{equation}

By repeated squaring we evaluate the surrogate as a dense matrix, and its error propagation is governed by the largest eigenvalue of the exponent.

\begin{lemma}[Dense evaluation of the surrogate]
\label{lem:bit_dense}
Assume the hypotheses of Lemma~\ref{lem:bit_taylor} with $\kappa_{\mathrm{sc}}\ge1$.  Let
$\widehat X_0$ be the Horner evaluation
$T_r(X)=I+X(I+\frac X2(I+\cdots(I+\frac Xr)\cdots))$, with each matrix product computed
exactly by Lemma~\ref{lem:bit_fastmm} and rounded once and each division by $j$ rounded once.  Let
$\widehat X_{i+1}:=\operatorname{round}_{\mathsf u}(\widehat X_i\widehat X_i)$ for $0\le i\le\kappa_{\mathrm{sc}}-2$, and
let $\widetilde P:=\operatorname{round}_{\mathsf u}(\widehat X_{\kappa_{\mathrm{sc}}-1}\widehat X_{\kappa_{\mathrm{sc}}-1})$.  If
$2e(\kappa_{\mathrm{sc}}+5)\sigma_{\mathrm{sc}}e^{\lambda_{\mathrm{sp}}+1}d_0\mathsf u\le1$, then
\[
\|\widetilde P-\mathcal Q(D)\|\le\varepsilon_{\mathrm{dense}}:=2e(\kappa_{\mathrm{sc}}+5)\sigma_{\mathrm{sc}}e^{\lambda_{\mathrm{sp}}+1}d_0\mathsf u,
\]
every stored entry has modulus at most $e^{\lambda_{\mathrm{sp}}+1}+1$, and the computation costs
$r+\kappa_{\mathrm{sc}}$ dense products of $d_0\times d_0$ fixed-point matrices.
\end{lemma}

\begin{proof}
Write $X_i:=T_r(X)^{2^i}$ and $f_i:=\|\widehat X_i-X_i\|$.  Each Horner step forms
$I+\operatorname{round}_{\mathsf u}(X\widehat Z)/j$ from the previous $\widehat Z$.  The product and then the
division by $j$ each round once, so the step contributes at most $2d_0\mathsf u$ in operator
norm.  Here we use that the entrywise error $\mathsf u$ of a $d_0\times d_0$ matrix has Frobenius, hence operator,
norm at most $d_0\mathsf u$.  The contribution of the step is multiplied downstream by $\|X\|/(j-1)\cdots$, so that
$f_0\le2d_0\mathsf u\sum_{j\ge0}1/j!\le2ed_0\mathsf u\le6d_0\mathsf u$.  For the squarings,
$\widehat X_{i+1}-X_{i+1}=X_iF_i+F_iX_i+F_i^2+E_{i+1}$ with $F_i:=\widehat X_i-X_i$ and
$\|E_{i+1}\|\le d_0\mathsf u$.  For this proof only, $E_{i+1}$ is the rounding error of the $(i+1)$-st squaring, not the $E_i$ of Lemma~\ref{lem:bit_state}(ii).
With $N_i:=e^{2^i(\lambda_{\mathrm{sp}}+1)/\sigma_{\mathrm{sc}}}\ge\|X_i\|$ from
Lemma~\ref{lem:bit_taylor}(ii), the squaring identity gives
\[
f_{i+1}\le2N_if_i\Bigl(1+\frac{f_i}{2N_i}\Bigr)+d_0\mathsf u.
\]
We unroll to $i=\kappa_{\mathrm{sc}}-1$ and use $\prod_{i=j}^{\kappa_{\mathrm{sc}}-2}N_i=\exp((\lambda_{\mathrm{sp}}+1)(2^{\kappa_{\mathrm{sc}}-1}-2^j)/\sigma_{\mathrm{sc}})\le e^{(\lambda_{\mathrm{sp}}+1)/2}$
together with $\prod_i(1+f_i/(2N_i))\le e$.  The bound on the product of the correction factors holds because $\sum_if_i/(2N_i)\le1$ under the
hypothesis, as one checks inductively from the bound being proved.  We get
\[
f_{\kappa_{\mathrm{sc}}-1}\le e\,2^{\kappa_{\mathrm{sc}}-1}e^{(\lambda_{\mathrm{sp}}+1)/2}\bigl(f_0+(\kappa_{\mathrm{sc}}-1)d_0\mathsf u\bigr)
\le e\,\sigma_{\mathrm{sc}}e^{(\lambda_{\mathrm{sp}}+1)/2}(\kappa_{\mathrm{sc}}+5)d_0\mathsf u/2.
\]
Finally we have
\[
\begin{aligned}
\|\widetilde P-\mathcal Q(D)\|&\le\|\widehat X_{\kappa_{\mathrm{sc}}-1}^2-X_{\kappa_{\mathrm{sc}}-1}^2\|+d_0\mathsf u
\le f_{\kappa_{\mathrm{sc}}-1}(2\|X_{\kappa_{\mathrm{sc}}-1}\|+f_{\kappa_{\mathrm{sc}}-1})+d_0\mathsf u\\
&\le3e^{(\lambda_{\mathrm{sp}}+1)/2}f_{\kappa_{\mathrm{sc}}-1}+d_0\mathsf u,
\end{aligned}
\]
using $\|X_{\kappa_{\mathrm{sc}}-1}\|\le e^{(\lambda_{\mathrm{sp}}+1)/2}$ and $f_{\kappa_{\mathrm{sc}}-1}\le e^{(\lambda_{\mathrm{sp}}+1)/2}$, which
gives the stated $\varepsilon_{\mathrm{dense}}$ after $3e/2+1\le2e$.  The magnitude bound follows
from $\|X_i\|\le e^{\lambda_{\mathrm{sp}}+1}$ and $f_i\le1$.  The count is $r$ products in the Horner
scheme, $\kappa_{\mathrm{sc}}-1$ squarings, and one final product.
\end{proof}

For the range finder and the trace guard of Section~\ref{sec:faster_algorithm}, we apply the surrogate to thin blocks of vectors.  Below, in (F9) and Lemma~\ref{lem:bit_guard}, we make these additive.  Repeated application of the scaled Taylor polynomial gives the same exact surrogate as the dense evaluation.

\begin{lemma}[Vector actions of the surrogate]
\label{lem:bit_action}
Assume the hypotheses of Lemma~\ref{lem:bit_taylor}, and let $X$ be stored exactly with
$\kappa_{\mathrm{sc}}$ additional fractional bits.  For a vector $v$, let $\mathrm{Horner}_r(X,v)$
be the vector obtained from $y_r:=v$ and the recursion $y_{j-1}:=v+\operatorname{round}_{\mathsf u}(Xy_j)/j$,
$j=r,\ldots,1$, with the division rounded once, as the final iterate $y_0$.  For a fixed-point vector $z$ put
$w_0:=z$ and $w_i:=\mathrm{Horner}_r(X,w_{i-1})$, $1\le i\le\sigma_{\mathrm{sc}}$.  Then for every
$1\le J_{\mathrm{ap}}\le\sigma_{\mathrm{sc}}$,
\[
\begin{gathered}
\|w_{J_{\mathrm{ap}}}-T_r(X)^{J_{\mathrm{ap}}}z\|_2\le\varepsilon_{\mathrm{act}}(J_{\mathrm{ap}}):=6J_{\mathrm{ap}}e^{(\lambda_{\mathrm{sp}}+1)J_{\mathrm{ap}}/\sigma_{\mathrm{sc}}}d_0^{3/2}\mathsf u,
\\
\|w_{J_{\mathrm{ap}}}\|_2\le e^{(\lambda_{\mathrm{sp}}+1)J_{\mathrm{ap}}/\sigma_{\mathrm{sc}}}\|z\|_2+\varepsilon_{\mathrm{act}}(J_{\mathrm{ap}}),
\end{gathered}
\]
in particular $\|w_{\sigma_{\mathrm{sc}}}-\mathcal Q(D)z\|_2\le\varepsilon_{\mathrm{act}}:=6\sigma_{\mathrm{sc}}e^{\lambda_{\mathrm{sp}}+1}d_0^{3/2}\mathsf u$
and $\|w_{\sigma_{\mathrm{sc}}/2}-\mathcal Q(D)^{1/2}z\|_2\le\varepsilon_{\mathrm{act}}/2$, where
$\mathcal Q(D)^{1/2}=T_r(X)^{\sigma_{\mathrm{sc}}/2}$.  Applying the scheme to a block of $\ell_{\mathrm{bl}}$
vectors costs $\sigma_{\mathrm{sc}}r$ products of a $d_0\times d_0$ matrix by a $d_0\times\ell_{\mathrm{bl}}$
matrix.
\end{lemma}

\begin{proof}
One Horner step computes $v+Xy_j/j$ with a matrix--vector product rounded entrywise
and a division rounded entrywise.  The product has Euclidean error at most $d_0^{3/2}\mathsf u$, and the
division has error at most $\sqrt{d_0}\mathsf u$.  The addition is exact.  The error $e_{j-1}$ in $y_{j-1}$ therefore obeys
$\|e_{j-1}\|_2\le\|X\|\|e_j\|_2/j+2d_0^{3/2}\mathsf u\le\|e_j\|_2/j+2d_0^{3/2}\mathsf u$, so
$\|e_0\|_2\le2d_0^{3/2}\mathsf u\sum_{j\ge0}1/j!\le6d_0^{3/2}\mathsf u$.  Thus one application of the scheme
returns $T_r(X)w+\phi$ with $\|\phi\|_2\le6d_0^{3/2}\mathsf u$, independently of $r$ and of
$\|w\|_2$.  Hence $w_{J_{\mathrm{ap}}}-T_r(X)^{J_{\mathrm{ap}}}z=\sum_{i=1}^{J_{\mathrm{ap}}}T_r(X)^{J_{\mathrm{ap}}-i}\phi_i$, and
Lemma~\ref{lem:bit_taylor}(ii) bounds each term by $e^{(\lambda_{\mathrm{sp}}+1)(J_{\mathrm{ap}}-i)/\sigma_{\mathrm{sc}}}6d_0^{3/2}\mathsf u$.
Summing the $J_{\mathrm{ap}}$ terms, we get $\varepsilon_{\mathrm{act}}(J_{\mathrm{ap}})$.  The same bound with
$\|T_r(X)^{J_{\mathrm{ap}}}z\|_2\le e^{(\lambda_{\mathrm{sp}}+1)J_{\mathrm{ap}}/\sigma_{\mathrm{sc}}}\|z\|_2$ gives the magnitude.  The
half-power statement is the case $J_{\mathrm{ap}}=\sigma_{\mathrm{sc}}/2$ with
$e^{(\lambda_{\mathrm{sp}}+1)/2}\le e^{\lambda_{\mathrm{sp}}+1}$.
\end{proof}

Lemmas~\ref{lem:bit_dense} and~\ref{lem:bit_action} use the same exact matrix
$\mathcal Q(\widehat D_x)$, so a refreshed tracker and a tracked increment refer to one
surrogate.  We compute each from that surrogate up to the deterministic errors $\varepsilon_{\mathrm{dense}}$ and $\varepsilon_{\mathrm{act}}$, which we account for in Lemma~\ref{lem:bit_tracker}.  This is the property we call in Section~\ref{sec:cubic_time} the convention that
``preserves exact conditional unbiasedness''.  The error made at stage $i$ of the
repeated application is amplified by $\|T_r(X)^{J_{\mathrm{ap}}-i}\|\le e^{\lambda_{\mathrm{sp}}+1}$ and never by
$e^{M_{\mathrm{sp}}}$.  This is the reason the scheme is stable for symmetric $D$ although
the spectral interval has width $M_{\mathrm{sp}}\approx4\cdot10^6\log(2n)$.  Al-Mohy and
Higham~\cite[Section~4]{almohyhigham11} make the same observation.  The count $\sigma_{\mathrm{sc}}r=O(\log^2n)$ of matrix--vector products per action
is polylogarithmic but large, about $10^{10}$ at $n=3\cdot10^8$.  Consider instead a Chebyshev
expansion of $e^{t/2}$ on the guarded interval $[-M_{\mathrm{sp}},\lambda_{\mathrm{top}}]$.  Here the interval is the spectral window of Lemma~\ref{lem:bit_taylor} that we fix in the paragraph \emph{Spectral bounds}.  Evaluated by the three-term recurrence,
the expansion needs only $O(\sqrt{M_{\mathrm{sp}}\log n})\approx10^5$ products with the same error
structure, and an implementation would use it.  We keep the scaled Taylor power in the
proof because its error analysis is elementary.

\subsection{The range finder and the increment estimator}\label{subsec:bit_rangefinder}

For a symmetric $Z$ we use the seminorm of Eq.~\eqref{eq:cubic_gradient_seminorm} in the
form
\begin{equation}
\|Z\|_{\mathcal G}^2:=\frac{\nu^2}{m}\sum_{i\in S}\tr[E_iZ]^2,
\qquad
E_i=\widetilde c\operatorname{diag}(A_i,-A_i),
\label{eq:bit_seminorm}
\end{equation}
which is the mean squared perturbation of the coordinate gradient
$\partial_iF(x)=x_i-b_i+\nu\tr[P_xE_i]$ induced by replacing $P_x$ by $P_x+Z$.

We supply the range finder with a rounded Gaussian test matrix and an approximate matrix action. Its residual estimate must therefore account for errors in both the test matrix and the sketch.

\begin{lemma}[Range finder with a rounded test matrix and an inexact sketch]
\label{lem:bit_rangefinder}
Let $A$ be a symmetric $d_0\times d_0$ matrix and let $2\le k\le n$.  Let
$\widehat\Omega=\Omega+E_\Omega\in\R^{d_0\times2k}$, where $\Omega$ is standard Gaussian and
$E_\Omega=E_\Omega(\Omega)$ is measurable with $\|E_\Omega\|_{\max}\le2^{-b_\Omega}$ and
$b_\Omega\ge4.5\log_2n+7$.  Let $\widehat Y\in\R^{d_0\times2k}$ satisfy
$\|\widehat Y-A\widehat\Omega\|_F\le\theta$ for a deterministic $\theta$.  Let $\mathcal P$ be
an orthogonal projector with $\operatorname{range}(\mathcal P)\supseteq\operatorname{range}(\widehat Y)$.
Then
\[
\E_\Omega\bigl[\|(I-\mathcal P)A\|_F^2\bigr]\le\frac{64}k\|A\|_1^2+1200\,kn^2\theta^2.
\]
\end{lemma}

\begin{proof}
Write $A=U\Sigma V^\top$ with $\sigma_1\ge\sigma_2\ge\cdots$.  Let $\Sigma_1$ be the top
$k\times k$ block and $\Sigma_2$ the rest.  Put $\Omega_1:=V_1^\top\Omega$ ($k\times2k$),
$\Omega_2:=V_2^\top\Omega$, and likewise $\widehat\Omega_1,\widehat\Omega_2,E_1,E_2$.  By rotation
invariance, $\Omega_1$ and $\Omega_2$ are independent standard Gaussian matrices.  Put
$\tau_\Omega:=1/(2n)$ and $\eta_E:=\|E_\Omega\|\le\|E_\Omega\|_F\le\sqrt{4nk}\,2^{-b_\Omega}\le2n2^{-b_\Omega}$.

\emph{Step 1: the good event.}  Let
$\mathcal G_1:=\{\sigma_k(\Omega_1)\ge\tau_\Omega\}$ and
$\mathcal G_2:=\{\|\Sigma_2\Omega_2\|_F^2\le8nk\|\Sigma_2\|_F^2\}$.  For a $k\times2k$
standard Gaussian matrix, we use the tail bound of Halko, Martinsson, and Tropp~\cite[Proposition~A.3]{hmt11},
after Chen and Dongarra~\cite[Lemma~4.1]{chendongarra05}.  Its prefactor is at most
$1.6$ for the $k\times2k$ shape and all $k\ge2$.  The bound gives
$\Pr[\|\Omega_1^\dagger\|\ge t]\le2t^{-(k+1)}\le2t^{-3}$ for $t\ge1$ and $k\ge2$, so
$\Pr[\mathcal G_1^c]\le2(2n)^{-3}=1/(4n^3)$.  Since
$\E[\|\Sigma_2\Omega_2\|_F^2]=2k\|\Sigma_2\|_F^2$~\cite[Proposition~A.1]{hmt11}, Markov's
inequality gives $\Pr[\mathcal G_2^c]\le1/(4n)$.  On
$\mathcal G:=\mathcal G_1\cap\mathcal G_2$, Weyl's inequality and
$\eta_E\le2n2^{-b_\Omega}\le\tau_\Omega/2$ give $\sigma_k(\widehat\Omega_1)\ge\tau_\Omega/2$.  Hence
$\widehat\Omega_1$ has full row rank, $\widehat\Omega_1\widehat\Omega_1^\dagger=I_k$, and
$\|\widehat\Omega_1^\dagger\|\le2/\tau_\Omega=4n$.  Off $\mathcal G$ we use
$\|(I-\mathcal P)A\|_F\le\|A\|_F\le\|A\|_1$, so the complement contributes at most
$\|A\|_1^2\Pr[\mathcal G^c]\le\|A\|_1^2/(2n)\le\|A\|_1^2/k$ to the expectation.

\emph{Step 2: a deterministic bound on $\mathcal G$.}  Write $\widehat Y=A\widehat\Omega+G_0$ with
$\|G_0\|_F\le\theta$, and put $\mathcal W:=\widehat Y\widehat\Omega_1^\dagger$.  Its range lies in
$\operatorname{range}(\widehat Y)\subseteq\operatorname{range}(\mathcal P)$, so $(I-\mathcal P)\mathcal W=0$.
Since $A\widehat\Omega\widehat\Omega_1^\dagger=U\Sigma V^\top\widehat\Omega\widehat\Omega_1^\dagger$ and
$V^\top\widehat\Omega=[\widehat\Omega_1;\widehat\Omega_2]$,
\[
\mathcal W=U\begin{bmatrix}\Sigma_1+G_1\\ \Sigma_2T'+G_2\end{bmatrix},
\qquad
T':=\widehat\Omega_2\widehat\Omega_1^\dagger,
\qquad
\begin{bmatrix}G_1\\G_2\end{bmatrix}:=U^\top G_0\widehat\Omega_1^\dagger,
\qquad
\Bigl\|\begin{bmatrix}G_1\\G_2\end{bmatrix}\Bigr\|_F\le4n\theta=:g_\theta.
\]
Split $[k]$ into $\mathsf a:=\{j:\sigma_j>2g_\theta\}$ and $\mathsf b:=[k]\setminus\mathsf a$, and let
$U_{\mathsf a},U_{\mathsf b},U_2$ be the corresponding column blocks of $U$.  The columns of $\mathcal W$
indexed by $\mathsf a$ are
$\mathcal W_{\mathsf a}=U_{\mathsf a}M+U_{\mathsf b}G_{1,\mathsf b\mathsf a}+U_2(\Sigma_2T'+G_2)_{:,\mathsf a}$ with
$M:=\Sigma_{1\mathsf a}+G_{1,\mathsf a\mathsf a}$.  Because $\|G_{1,\mathsf a\mathsf a}\|\le g_\theta<\sigma_{\min}(\Sigma_{1\mathsf a})/2$,
$M$ is invertible and $\|M^{-1}\Sigma_{1\mathsf a}\|=\|(I+\Sigma_{1\mathsf a}^{-1}G_{1,\mathsf a\mathsf a})^{-1}\|\le2$.
Since $(I-\mathcal P)\mathcal W_{\mathsf a}=0$, we have
$(I-\mathcal P)U_{\mathsf a}M=-(I-\mathcal P)[U_{\mathsf b}G_{1,\mathsf b\mathsf a}+U_2(\Sigma_2T'+G_2)_{:,\mathsf a}]$.
Multiplying by $M^{-1}\Sigma_{1\mathsf a}$ on the right and using $\|I-\mathcal P\|\le1$, we get
\[
\|(I-\mathcal P)U_{\mathsf a}\Sigma_{1\mathsf a}\|_F
\le2\bigl(\|G_{1,\mathsf b\mathsf a}\|_F+\|\Sigma_2T'\|_F+\|G_2\|_F\bigr)
\le2\|\Sigma_2T'\|_F+3g_\theta,
\]
using $\|G_{1,\mathsf b\mathsf a}\|_F+\|G_2\|_F\le\sqrt2g_\theta$.  Since
$A=U_{\mathsf a}\Sigma_{1\mathsf a}V_{\mathsf a}^\top+U_{\mathsf b}\Sigma_{1\mathsf b}V_{\mathsf b}^\top+U_2\Sigma_2V_2^\top$
and $\|\Sigma_{1\mathsf b}\|_F\le2g_\theta\sqrt k$, we obtain
\[
\begin{gathered}
\|(I-\mathcal P)A\|_F\le2\|\Sigma_2T'\|_F+(3+2\sqrt k)g_\theta+\|\Sigma_2\|_F,
\\
\|(I-\mathcal P)A\|_F^2\le12\|\Sigma_2T'\|_F^2+3(2\sqrt k+3)^2g_\theta^2+3\|\Sigma_2\|_F^2 .
\end{gathered}
\]

\emph{Step 3: the perturbed pseudoinverse.}  Decompose
$\Sigma_2T'=\Sigma_2\Omega_2\Omega_1^\dagger+\Sigma_2\Omega_2(\widehat\Omega_1^\dagger-\Omega_1^\dagger)+\Sigma_2E_2\widehat\Omega_1^\dagger$.
For two full-row-rank $k\times2k$ matrices $P,Q$, the identity
$Q^\dagger-P^\dagger=-Q^\dagger(Q-P)P^\dagger+(I-Q^\dagger Q)(Q-P)^\top(P^\dagger)^\top P^\dagger$
of Stewart~\cite[Section~3]{stewart77} gives
$\|Q^\dagger-P^\dagger\|\le\|Q-P\|\,\|P^\dagger\|(\|Q^\dagger\|+\|P^\dagger\|)$.  On $\mathcal G$
with $P=\Omega_1$, $Q=\widehat\Omega_1$, this is at most $\eta_E\cdot2n\cdot6n=12n^2\eta_E$.  Hence
the second and third terms have Frobenius norm at most
$\sqrt{8nk}\|\Sigma_2\|_F\cdot12n^2\eta_E+\|\Sigma_2\|_F\eta_E\cdot4n\le38n^3\eta_E\|\Sigma_2\|_F
\le76n^42^{-b_\Omega}\|\Sigma_2\|_F\le\|\Sigma_2\|_F/\sqrt{2n}$, the last step by
$b_\Omega\ge4.5\log_2n+7$.  By $(x+y)^2\le2x^2+2y^2$, the independence of
$\Omega_2$ from $\Omega_1$ (rotation invariance) with the Frobenius identity~\cite[Proposition~A.1]{hmt11}, and
$\E[\|\Omega_1^\dagger\|_F^2]=k/(k-1)\le2$~\cite[Proposition~A.5]{hmt11}, we get
\[
\E\bigl[\|\Sigma_2T'\|_F^2;\mathcal G\bigr]
\le2\|\Sigma_2\|_F^2\E[\|\Omega_1^\dagger\|_F^2]+2\|\Sigma_2\|_F^2/(2n)\le5\|\Sigma_2\|_F^2 ,
\] where $\E[X;\mathcal G]:=\E[X\,\mathbbm{1}[\mathcal G]]$.

\emph{Step 4: assembly.}  Combining Steps 1--3 with
$\|\Sigma_2\|_F^2\le\sigma_{k+1}\|A\|_1\le\|A\|_1^2/k$, $g_\theta^2=16n^2\theta^2$, and
$3(2\sqrt k+3)^2\le75k$, we obtain
\[
\E[\|(I-\mathcal P)A\|_F^2]
\le(60+3)\frac{\|A\|_1^2}k+75k\cdot16n^2\theta^2+\frac{\|A\|_1^2}k
=\frac{64}k\|A\|_1^2+1200kn^2\theta^2 .
\]
Suppose $\widehat\Omega$ is the conditioned draw of Lemma~\ref{lem:bit_gaussian}(iii) instead of one
draw, and the hypothesis $\|\widehat Y-A\widehat\Omega\|_F\le\theta$ is available only on the acceptance
event $\mathcal A$, then the bound holds with the right side multiplied by $(1-\mathsf u^2)^{-1}$.  To see this, on
$\mathcal A^c$ we replace $\widehat Y$ by $A\widehat\Omega$ and $\mathcal P$ by the projector onto its range.
The hypothesis then holds surely, and $\Xi:=\|(I-\mathcal P)A\|_F^2$ is unchanged on $\mathcal A$.
Since $\Xi\ge0$, we get $\E[\Xi]\ge\Pr[\mathcal A]\,\E[\Xi\mid\mathcal A]$.
\end{proof}

The constant $C=3$ of Section~\ref{sec:faster_algorithm} appears there as $\E_\Omega\|E_{\mathrm{res}}\|_F^2\le\frac3k\|E\|_1^2$.  In the bound $\E[\|A_{\mathrm{res}}\|_F^2]\le(C/k)\|A\|_1^2$, the constant becomes $64$ here and $129$ in Lemma~\ref{lem:bit_increment}(ii).
It enters only $K_0$.  The hypothesis $2\le k\le n$ is the bound we recorded with the choice of $k$ after Lemma~\ref{lem:bit_state}.

In the increment analysis we separate the exactly unbiased estimator defined by the computed basis from the deterministic errors in its actions and matrix products.

\begin{lemma}[The computed increment estimator]
\label{lem:bit_increment}
Fix a nonrefresh step from $x=x_t$ with $F(x)\le H$ to $y=x_{t+1}$ and put
$A:=\mathcal Q(\widehat D_y)-\mathcal Q(\widehat D_x)$.  The algorithm draws $\widehat\Omega$ as in
Lemma~\ref{lem:bit_gaussian}(iii) and Rademacher probes $z_1,\ldots,z_s$.  It then computes the following.
\begin{enumerate}[label=(\alph*)]
\item $\widehat Y:=\mathrm{act}_y(\widehat\Omega)-\mathrm{act}_x(\widehat\Omega)$, where
$\mathrm{act}_x(\cdot)$ is the columnwise scheme of Lemma~\ref{lem:bit_action} for
$\mathcal Q(\widehat D_x)$.
\item A Householder QR factorization of $\widehat Y$ in fixed point, returning
$\widehat U\in\R^{d_0\times2k}$.
\item $\widehat Z:=\mathrm{act}_y(\widehat U)-\mathrm{act}_x(\widehat U)$ and the dense matrix
$\widehat A_{\mathrm{low}}:=\widehat U\widehat Z^\top+\widehat Z\widehat U^\top-\widehat U(\widehat U^\top\widehat Z)\widehat U^\top$
by four exact-then-rounded products of Lemma~\ref{lem:bit_fastmm}.  These products are $\widehat U^\top\widehat Z$,
$\widehat U(\widehat U^\top\widehat Z)$, its product with $\widehat U^\top$, and $\widehat U\widehat Z^\top$.
\item For each $j$, $v_j:=z_j-\widehat U(\widehat U^\top z_j)$,
$w_j:=\mathrm{act}_y(v_j)-\mathrm{act}_x(v_j)$, $r_j:=w_j-\widehat U(\widehat U^\top w_j)$, and the
dense matrix $\widehat A_{\mathrm{hut}}:=\frac1{2s}\sum_j(r_jz_j^\top+z_jr_j^\top)$ by one
exact-then-rounded thin product.
\item $\widehat A^{\mathrm{comp}}:=\widehat A_{\mathrm{low}}+\widehat A_{\mathrm{hut}}$, symmetrized exactly.
\end{enumerate}
Let $\widehat\Pi:=\widehat U\widehat U^\top$ and $A_{\mathrm{res}}:=(I-\widehat\Pi)A(I-\widehat\Pi)$.  Let
$\widehat A^{\mathrm{ideal}}:=\widehat\Pi A+A\widehat\Pi-\widehat\Pi A\widehat\Pi+\frac1{2s}\sum_j\{(A_{\mathrm{res}}z_j)z_j^\top+z_j(A_{\mathrm{res}}z_j)^\top\}$
be the exact-arithmetic estimator built from the same $\widehat\Pi$ and $z_j$.  Then:
\begin{enumerate}[label=(\roman*)]
\item $\E[\widehat A^{\mathrm{ideal}}\mid\widehat\Omega]=A$ and
$\E[\|\widehat A^{\mathrm{ideal}}-A\|_{\mathcal G}^2\mid\widehat\Omega]\le\frac{2\nu^2\widetilde c^2}s\|A_{\mathrm{res}}\|_F^2$.
\item Put $Y_{\max}:=\Omega_{\max}\sqrt{2kd_0}\,(e^{\lambda_{\mathrm{top}}}+1)+2\sqrt{2k}\,\varepsilon_{\mathrm{act}}$, which bounds
$\|\widehat Y\|_F$ for every accepted $\widehat\Omega$.  Put
$\varepsilon_{\mathrm{qr}}:=16k^{3/2}d_0^{3/2}\mathsf u\,(1+Y_{\max})$ and
$\theta:=2\sqrt{2k}\,\varepsilon_{\mathrm{act}}+\varepsilon_{\mathrm{qr}}$.  Each of these is deterministic.  If $3\varepsilon_{\mathrm{qr}}\le1/(5\sqrt k)$, then
\[
\E_\Omega[\|A_{\mathrm{res}}\|_F^2]\le\frac{129}k\|A\|_1^2+2401\,kn^2\theta^2 ;
\]
\item Under the hypothesis of (ii), the deterministic bound $\|\widehat A^{\mathrm{comp}}-\widehat A^{\mathrm{ideal}}\|\le\varepsilon_{\mathrm{step}}:=40\,d_0^{3/2}(1+\|A\|)(\varepsilon_{\mathrm{act}}+d_0\mathsf u)$
holds.
\item The step costs $6\sigma_{\mathrm{sc}}r+O(1)$ products of $d_0\times d_0$ by
$d_0\times O(k+s)$ matrices and five exact-then-rounded products.  Three of the exact-then-rounded products have shape $\langle d_0,O(k+s),d_0\rangle$, and two have shape $\langle O(k),d_0,O(k)\rangle$ or $\langle d_0,O(k),O(k)\rangle$.  The step also costs $O(d_0k^2+d_0^2)$ further operations.
\end{enumerate}
\end{lemma}

\begin{proof}
\emph{(i), unbiasedness and variance.}  Conditional on $\widehat\Omega$, the matrix $\widehat\Pi$ is fixed and the $z_j$ are
independent of it.  Lemma~\ref{lem:bit_probes} therefore gives
$\E[(A_{\mathrm{res}}z_j)z_j^\top]=A_{\mathrm{res}}$, and hence we obtain $\E[\widehat A^{\mathrm{ideal}}\mid\widehat\Omega]=(\widehat\Pi A+A\widehat\Pi-\widehat\Pi A\widehat\Pi)+A_{\mathrm{res}}=A$.
Here the identity $\widehat\Pi A+A\widehat\Pi-\widehat\Pi A\widehat\Pi+(I-\widehat\Pi)A(I-\widehat\Pi)=A$ holds
for every matrix $\widehat\Pi$.  For the variance, by symmetry of $E_i$ we have
$\tr[E_i(\widehat A^{\mathrm{ideal}}-A)]=\frac1s\sum_jz_j^\top A_{\mathrm{res}}E_iz_j-\tr[A_{\mathrm{res}}E_i]$.
We apply Lemma~\ref{lem:bit_probes} to the symmetric part of $A_{\mathrm{res}}E_i$, whose Frobenius
norm is at most $\|A_{\mathrm{res}}E_i\|_F$, and obtain $\V[z^\top A_{\mathrm{res}}E_iz]\le2\|A_{\mathrm{res}}E_i\|_F^2$.  Hence
\[
\E[\|\widehat A^{\mathrm{ideal}}-A\|_{\mathcal G}^2]
\le\frac{\nu^2}m\sum_{i\in S}\frac2s\|A_{\mathrm{res}}E_i\|_F^2
=\frac{2\nu^2}{ms}\tr\Bigl[A_{\mathrm{res}}^2\sum_{i\in S}E_i^2\Bigr]
\le\frac{2\nu^2\widetilde c^2}s\|A_{\mathrm{res}}\|_F^2
\]
by Lemma~\ref{lem:bit_state}(ii).

\emph{(ii), the range finder with computed data.}  \emph{The sketch.}  Each column of $\widehat Y$ is a difference of two computed actions
on the same column of $\widehat\Omega$, so Lemma~\ref{lem:bit_action} gives
$\|\widehat Y-A\widehat\Omega\|_F\le\sqrt{2k}\cdot2\varepsilon_{\mathrm{act}}$.
\emph{The factorization.}  Householder QR processes the columns of $\widehat Y$ in turn.  At
step $j$ it computes a vector $v_j$ from the current $j$-th column, with its entries and the
norm rounded once each.  It then defines the reflector $H_j:=I-2v_jv_j^\top/\|v_j\|_2^2$, which
is an exactly orthogonal matrix \emph{for the computed} $v_j$, whatever rounding produced
it.  A column whose remaining part rounds to zero is skipped.  Its norm is at most $\sqrt{d_0}\mathsf u$,
and we charge it to the backward perturbation $\Delta\widehat Y$ that we define below.  Each $v_j$ is rescaled by a power of two to norm in
$[1/2,2]$ before use, which leaves $H_j$ unchanged.  Scaling up is exact.  Scaling down drops low
bits of $v_j$, a perturbation of at most $\sqrt{d_0}\mathsf u$ of the order we already charged, and $H_j$ is exactly
orthogonal for the vector actually used.  Applying $H_j$ to a column $w$ costs the inner product $v_j^\top w$, the
scalar $2v_j^\top w/\|v_j\|_2^2$, and the update $w-(\cdot)v_j$.  The inner product is accumulated exactly with
one rounding, the scalar takes one rounding, and the update takes one rounding per entry.  The computed image therefore differs from $H_jw$ by
at most $4\sqrt{d_0}\,\mathsf u(1+\|w\|_2)$ in Euclidean norm.  Setting the computed entries below
the diagonal to zero, which the exact $H_j$ would have done only for the exact Householder
vector, costs at most another $4(\sqrt{d_0}+1)\mathsf u(1+\|w\|_2)$.  To see this, we observe that the computed $v_j$ is within
$(\sqrt{d_0}+1)\mathsf u$ of the exact Householder vector, its entries and norm being rounded once each,
and $\|v_j\|_2\ge\|w\|_2$.  Hence $\|H_j-H_j^{\mathrm{exact}}\|\le4(\sqrt{d_0}+1)\mathsf u/\|v_j\|_2$, and the
subdiagonal part of $H_jw$ has norm at most $4(\sqrt{d_0}+1)\mathsf u$.  Because every $H_j$ is exactly
orthogonal, an error made at step $j$ is carried unchanged in norm to the end, and every
intermediate column has norm at most $2Y_{\max}+1$.
Here $\|\widehat Y\|_F\le\|A\|\|\widehat\Omega\|_F+2\sqrt{2k}\varepsilon_{\mathrm{act}}\le Y_{\max}$ by
Lemma~\ref{lem:bit_action} and $\|A\|\le e^{\lambda_{\mathrm{top}}}+1$.  We obtain the last bound from
Lemma~\ref{lem:bit_taylor}(iv) with $\lambda_{\mathrm{sp}}=\log(H/\nu)$ at $x$, where $F(x)\le H$ gives $\lambda_{\max}(D_x)\le\log(H/\nu)$.  There we also use $\zeta=\|D_y-D_x\|\le1$,
with $\|P_y-P_x\|\le eH/\nu\le e^{\lambda_{\mathrm{top}}}/(2n)$, together with Eq.~\eqref{eq:bit_surrogate_accuracy}
and $\varepsilon_{\mathrm{op}}/2\le1$.  Hence the computed upper
triangular $\widehat R$ satisfies $\widehat Y+\Delta\widehat Y=Q_{\mathrm{H}}\widehat R$ with
$Q_{\mathrm{H}}:=H_1\cdots H_{2k}[I_{2k};0]$ exactly orthonormal and
$\|\Delta\widehat Y\|_F\le2k\cdot\sqrt{2k}\cdot9\sqrt{d_0}\,\mathsf u(2Y_{\max}+2)\le\varepsilon_{\mathrm{qr}}$.
This is the fixed-point form of the backward error bound of
Higham~\cite[Theorem~19.4]{higham02}.  The returned $\widehat U$ is the computed product of
the same reflectors applied to the first $2k$ unit vectors, so
$\|\widehat U-Q_{\mathrm{H}}\|_F\le\varepsilon_{\mathrm{qr}}$ by the same count with $\|w\|_2\le1$.  We apply
Lemma~\ref{lem:bit_rangefinder} with the sketch $\widehat Y+\Delta\widehat Y$, whose distance to
$A\widehat\Omega$ is at most $\theta$, and with $\mathcal P:=Q_{\mathrm{H}}Q_{\mathrm{H}}^\top$, whose range
contains $\operatorname{range}(\widehat Y+\Delta\widehat Y)$.  This gives
$\E[\|(I-\mathcal P)A\|_F^2]\le(1-\mathsf u^2)^{-1}\{(64/k)\|A\|_1^2+1200kn^2\theta^2\}$.  The factor comes from
the conditioned draw.
\emph{The near-projector.}  We have $\|\widehat\Pi-\mathcal P\|\le2\|\widehat U-Q_{\mathrm{H}}\|+\|\widehat U-Q_{\mathrm{H}}\|^2\le3\varepsilon_{\mathrm{qr}}=:\varepsilon_{\mathrm{proj}}$,
so $\|A_{\mathrm{res}}\|_F\le\|(I-\mathcal P)A(I-\mathcal P)\|_F+3\varepsilon_{\mathrm{proj}}\|A\|_F\le\|(I-\mathcal P)A\|_F+3\varepsilon_{\mathrm{proj}}\|A\|_1$.
Next we square the last bound with $(x+y)^2\le2x^2+2y^2$, and use $2(1-\mathsf u^2)^{-1}\cdot64\le128+1/(2k)$ and
$2(1-\mathsf u^2)^{-1}\cdot1200\le2401$.  Both hold because $\mathsf u\le1/(240k^2d_0^{3/2})$ under the hypothesis.  Together with
$18\varepsilon_{\mathrm{proj}}^2\le18/(25k)$, we obtain (ii).

\emph{(iii), computed versus ideal.}  Every stage is a deterministic function of the same inputs.  In (c), $\widehat Z$
differs from $A\widehat U$ by at most $2\sqrt{2k}\varepsilon_{\mathrm{act}}$ in Frobenius norm, and
$\|\widehat U\|\le1+\varepsilon_{\mathrm{qr}}\le2$.  Hence the three terms of $\widehat A_{\mathrm{low}}$ differ from
those of $\widehat\Pi A+A\widehat\Pi-\widehat\Pi A\widehat\Pi$ by at most $12\cdot2\sqrt{2k}\varepsilon_{\mathrm{act}}$
plus five roundings.  The $2k\times2k$ factor $\widehat U^\top\widehat Z$ is rounded once to the
$\mathsf u$-grid.  Its Frobenius error is at most $2k\mathsf u$, which $\|\widehat U\|^2\le4$ turns into at most
$8k\mathsf u\le d_0\mathsf u$ in $\widehat A_{\mathrm{low}}$.  The $d_0\times2k$ product $\widehat U(\widehat U^\top\widehat Z)$ is rounded once.  Its Frobenius error is at most $\sqrt{2kd_0}\,\mathsf u$, and at most $2\sqrt{2kd_0}\,\mathsf u\le d_0\mathsf u$ after the product with $\widehat U^\top$.  Each of the three terms is rounded
once, at most $d_0\mathsf u$ each.  In (d), $v_j$ differs from
$(I-\widehat\Pi)z_j$ by at most $3\sqrt{d_0}\mathsf u$, since the inner products and the product with $\widehat U$ are
rounded entrywise.  Then $w_j$ differs from $Av_j$ by at most $2\varepsilon_{\mathrm{act}}+3\sqrt{d_0}\mathsf u\|A\|$.  In turn,
$r_j$ differs from $A_{\mathrm{res}}z_j$ by at most
$2(2\varepsilon_{\mathrm{act}}+3\sqrt{d_0}\mathsf u\|A\|)+3\sqrt{d_0}\mathsf u(1+2\sqrt{d_0}\|A\|)=:\varepsilon_{\mathrm{rd}}$, where we use
$\|w_j\|_2\le2\sqrt{d_0}\|A\|+2\varepsilon_{\mathrm{act}}$.  We absorb the extra $6\sqrt{d_0}\mathsf u\,\varepsilon_{\mathrm{act}}\le\varepsilon_{\mathrm{act}}$ in the constant $40$ below.  Hence $\widehat A_{\mathrm{hut}}$ differs from its ideal counterpart
by at most $\varepsilon_{\mathrm{rd}}\|z_j\|_2=\varepsilon_{\mathrm{rd}}\sqrt{d_0}$ plus one rounding $d_0\mathsf u$.  We sum the
stages and use $8k\le d_0$, which holds as $k\le45n^{1/4}+1\le n/4$ for $n\ge N_0$, and obtain
$\varepsilon_{\mathrm{step}}\le38\sqrt{d_0}\varepsilon_{\mathrm{act}}+6d_0\mathsf u+\sqrt{d_0}\varepsilon_{\mathrm{rd}}\le40d_0^{3/2}(1+\|A\|)(\varepsilon_{\mathrm{act}}+d_0\mathsf u)$.

\emph{(iv), cost.}  The actions in (a), (c), (d) apply the scheme of Lemma~\ref{lem:bit_action} to
blocks of $2k$, $2k$, and $s$ vectors for each of the two matrices.  This costs $2\sigma_{\mathrm{sc}}r$
thin products per block.  The products in (c) and (d) have the stated shapes.  The
Householder factorization costs $O(d_0k^2)$ and the symmetrization $O(d_0^2)$.
\end{proof}

\subsection{The tracker, the step, the guard, and the phase}\label{subsec:bit_tracker}

Here we use a deterministic upper bound on the time between refreshes to control both accumulated rounding error and the components of the tracker invisible to the gradient seminorm.

\begin{lemma}[Tracker with bounded drift]
\label{lem:bit_tracker}
Fix a candidate phase, one of the $R_{\mathrm{rep}}$ guarded phases that the wrapper of Section~\ref{sec:faster_algorithm} runs from a phase start $z$ and that Algorithm~\ref{alg:cubic_tracked_phase} executes.  The phase runs at multiplier $\nu$ with level $H$, $T=16mL$ steps, refresh
probability $\pi'$, and $L_{\max}:=\lceil1/\pi'\rceil$.  The finite tracker starts from
$\widehat P_0:=\widetilde P_z$, the dense surrogate of Lemma~\ref{lem:bit_dense} at the phase
start $z$.  After $x_{t+1}$ is formed, the tracker refreshes, setting
$\widehat P_{t+1}:=\widetilde P_{x_{t+1}}$, if an independent coin of bias $\pi'$ shows heads (a \emph{coin refresh}) or
if $L_{\max}$ steps have passed since the last refresh (a \emph{forced refresh}).  Otherwise it sets
$\widehat P_{t+1}:=\widehat P_t+\widehat A^{\mathrm{comp}}_t$ with the estimator of
Lemma~\ref{lem:bit_increment} for $A_t:=\mathcal Q(\widehat D_{x_{t+1}})-\mathcal Q(\widehat D_{x_t})$.
Put $Z_t:=\widehat P_t-P_{x_t}$ and split $Z_t=Z^{\mathrm{rand}}_t+Z^{\mathrm{det}}_t$, where at a refresh
$Z^{\mathrm{rand}}_{t+1}:=0$ and $Z^{\mathrm{det}}_{t+1}:=\widetilde P_{x_{t+1}}-P_{x_{t+1}}$, and at a
nonrefresh step
\[
Z^{\mathrm{rand}}_{t+1}:=Z^{\mathrm{rand}}_t+(\widehat A^{\mathrm{ideal}}_t-A_t),
\qquad
Z^{\mathrm{det}}_{t+1}:=Z^{\mathrm{det}}_t+(\widehat A^{\mathrm{comp}}_t-\widehat A^{\mathrm{ideal}}_t)+\bigl(A_t-(P_{x_{t+1}}-P_{x_t})\bigr).
\]
At $t=0$, $Z^{\mathrm{rand}}_0:=0$ and $Z^{\mathrm{det}}_0:=\widetilde P_z-P_z$.  Assume
$\varepsilon_{\mathrm{step}},\varepsilon_{\mathrm{dense}}\le1$, $\varepsilon_{\mathrm{op}}\le2$, and $3\varepsilon_{\mathrm{qr}}\le1/(5\sqrt k)$, where
$\varepsilon_{\mathrm{step}}$ is evaluated at $\|A\|=e^{\lambda_{\mathrm{top}}}+1$.
Let $\mathcal E_t:=\|Z^{\mathrm{rand}}_t\|_{\mathcal G}^2$, let $\mathcal F_t$ be the history up to
and including $x_t$ and $\widehat P_t$, and suppose every base point satisfies $F(x_t)\le H$
and every step $|d_t|\le1/\widetilde c$.  Then:
\begin{enumerate}[label=(\roman*)]
\item $\|Z^{\mathrm{det}}_t\|\le\varepsilon_{\mathrm{det}}:=\varepsilon_{\mathrm{dense}}+3\varepsilon_{\mathrm{op}}/4+L_{\max}\varepsilon_{\mathrm{step}}$
for every $t$, deterministically.
\item $\E[\mathcal E_{t+1}\mid\mathcal F_t]\le(1-\pi')\bigl\{\mathcal E_t+\frac{65L^2}{ks}\E[d_t^2\mid\mathcal F_t]+\frac{2\delta_g^2}{ks}\bigr\}$,
provided $\varepsilon_{\mathrm{op}}\le\delta_g/(12\nu\widetilde cd_0)$ and
$\theta\le\delta_g/(70\nu\widetilde ckn)$ with $\theta$ as in Lemma~\ref{lem:bit_increment}(ii).
\item The selected gradient $\widehat g_{t,i}:=x_{t,i}-b_i+\operatorname{round}_{\mathsf u}(\nu\tr[\widehat P_tE_i])$
satisfies $\widehat g_{t,i}=\partial_iF(x_t)+\mathfrak{e}_{t,i}+\beta_{t,i}$ with
$\mathfrak{e}_{t,i}:=\nu\tr[E_iZ^{\mathrm{rand}}_t]$, $\E[\mathfrak{e}_{t,i_t}^2\mid\mathcal F_t]=\mathcal E_t$ for the
uniform $i_t$, and $|\beta_{t,i}|\le\nu\widetilde cd_0\varepsilon_{\mathrm{det}}+\mathsf u$.
\item $\|\widehat P_t\|\le e^{\lambda_{\mathrm{top}}+1}+1+L_{\max}(25d_0+30)(e^{\lambda_{\mathrm{top}}}+1)$
for every $t$.
\item The number of refreshes in the phase is at most $T\pi'+1$ forced ones plus a
$\mathrm{Binomial}(T,\pi')$ number of coin refreshes.  The coin refreshes are capped at
$M_{\mathrm{ref}}:=\lceil2T\pi'+3\log(100JR_{\mathrm{rep}}/p_0)\rceil$, a cap that fires with probability at most
$p_0/(100JR_{\mathrm{rep}})\le0.01$.
\end{enumerate}
The forced-refresh decision is a deterministic function of the step count, hence
$\mathcal F_t$-measurable.  The coin, the test matrix $\widehat\Omega_t$, the probes
$z_j$, and the guard's probes are fresh coins independent of $\mathcal F_t$ and of each
other.
\end{lemma}

\begin{proof}
We prove the identity $Z_t=Z^{\mathrm{rand}}_t+Z^{\mathrm{det}}_t$ by induction: at a nonrefresh step
$Z_{t+1}=Z_t+\widehat A^{\mathrm{comp}}_t-(P_{x_{t+1}}-P_{x_t})$, which is the sum of the three
increments.

\emph{(i), the drift telescopes.}  Let $t_0\le t$ be the last refresh time.  The surrogate increments telescope,
$\sum_{t_0\le\tau<t}(A_\tau-(P_{x_{\tau+1}}-P_{x_\tau}))=(\mathcal Q(\widehat D_{x_t})-P_{x_t})-(\mathcal Q(\widehat D_{x_{t_0}})-P_{x_{t_0}})$,
of norm at most $\varepsilon_{\mathrm{op}}/2$ by Eq.~\eqref{eq:bit_surrogate_accuracy}.  The refresh
term has norm at most $\varepsilon_{\mathrm{dense}}+\varepsilon_{\mathrm{op}}/4$ by Lemma~\ref{lem:bit_dense} and
Eq.~\eqref{eq:bit_surrogate_accuracy}.  There are at most $L_{\max}$ computation
errors of norm at most $\varepsilon_{\mathrm{step}}$ each by Lemma~\ref{lem:bit_increment}(iii).

\emph{(ii), the zero-mean recurrence.}  On a refresh (forced or by coin) $\mathcal E_{t+1}=0$.  A nonrefresh step has,
given $\mathcal F_t$, probability at most $1-\pi'$, because the coin is independent of
everything else.  On such a step $Z^{\mathrm{rand}}_{t+1}=Z^{\mathrm{rand}}_t+\Delta_t$ with
$\E[\Delta_t\mid\mathcal F_t,i_t,\widehat\Omega_t]=0$ by Lemma~\ref{lem:bit_increment}(i).  Since
$\|\cdot\|_{\mathcal G}^2$ is a quadratic form and $Z^{\mathrm{rand}}_t$ is $\mathcal F_t$-measurable,
the cross term vanishes and
$\E[\mathcal E_{t+1}\mid\mathcal F_t]\le(1-\pi')\{\mathcal E_t+\E[\|\Delta_t\|_{\mathcal G}^2\mid\mathcal F_t]\}$.
By Lemma~\ref{lem:bit_increment}(i)--(ii) we have
$\E[\|\Delta_t\|_{\mathcal G}^2\mid\mathcal F_t,i_t,d_t]\le\frac{2\nu^2\widetilde c^2}s\{\frac{129}k\|A_t\|_1^2+2401kn^2\theta^2\}$.
By Eq.~\eqref{eq:bit_surrogate_accuracy}, $\|A_t\|_1\le\|P_{x_{t+1}}-P_{x_t}\|_1+d_0\varepsilon_{\mathrm{op}}/2$.
The argument of Lemma~\ref{lem:cubic_local_increment}, whose hypotheses
$F(x_t)\le H$ and $|d_t|\le1/\widetilde c$ hold by assumption, gives
$\nu^2\|P_{x_{t+1}}-P_{x_t}\|_1^2\le\frac e2HLd_t^2$.  We apply the trace-norm inequality (Fact~\ref{fact:generalized_pinsker})
behind that argument to the exact matrices $P_{x_t}$, $P_{x_{t+1}}$, never to computed ones, and it carries
no rounding.  Hence
\[
\E[\|\Delta_t\|_{\mathcal G}^2\mid\mathcal F_t,i_t,d_t]
\le\frac{258\widetilde c^2}{ks}\Bigl(eHLd_t^2+\frac{\nu^2d_0^2\varepsilon_{\mathrm{op}}^2}2\Bigr)
+\frac{4802\nu^2\widetilde c^2kn^2\theta^2}s
\le\frac{65L^2}{ks}d_t^2+\frac{\delta_g^2}{ks}+\frac{\delta_g^2}{ks},
\]
where the first term uses $e\widetilde c^2H\le L/4$ from the definition of $L$, the
second uses $129\nu^2\widetilde c^2d_0^2\varepsilon_{\mathrm{op}}^2\le\delta_g^2$, and the third uses the
hypothesis on $\theta$.  We take $\E[\cdot\mid\mathcal F_t]$ and obtain (ii).

\emph{(iii), the used gradient.}  The inner product $\langle\widehat P_{t,11}-\widehat P_{t,22},A_i\rangle$ is taken over the $n\times n$ diagonal blocks of $\widehat P_t$, so that $\tr[\widehat P_tE_i]=\widetilde c\langle\widehat P_{t,11}-\widehat P_{t,22},A_i\rangle$ by $E_i=\widetilde c\operatorname{diag}(A_i,-A_i)$.  This inner product is formed exactly, the factor
$\nu\widetilde c$ is a number of the solve stored exactly with $2f$ fractional bits, and their
product is formed in a $3W$-bit temporary and rounded once.  Hence
$\widehat g_{t,i}=\partial_iF(x_t)+\nu\tr[E_iZ_t]+\vartheta$ with $|\vartheta|\le \mathsf u$.  We split
$Z_t$ and use $|\tr[E_iZ^{\mathrm{det}}_t]|\le\|E_i\|\|Z^{\mathrm{det}}_t\|_1\le\widetilde cd_0\varepsilon_{\mathrm{det}}$.
The identity for $\E[\mathfrak{e}^2]$ is Eq.~\eqref{eq:bit_seminorm} with $i_t$ uniform and
independent of $\mathcal F_t$.

\emph{(iv), magnitude.}  By Lemma~\ref{lem:bit_dense} the refreshed matrix has norm at most
$e^{\lambda_{\mathrm{top}}+1}+1$.  Between refreshes at most $L_{\max}$ increments are added.  Each
satisfies $\|\widehat A^{\mathrm{comp}}_t\|\le\|\widehat A^{\mathrm{ideal}}_t\|+\varepsilon_{\mathrm{step}}$ and
$\|\widehat A^{\mathrm{ideal}}_t\|\le24\|A_t\|+\frac1{2s}\sum_j2\|A_{\mathrm{res}}z_j\|_2\|z_j\|_2\le(24+25d_0)\|A_t\|$,
where we use $\|\widehat\Pi\|\le4$ and $\|z_j\|_2^2=d_0$.  Moreover
$\|A_t\|\le\|P_{x_{t+1}}-P_{x_t}\|+\varepsilon_{\mathrm{op}}/2\le e^{\lambda_{\mathrm{top}}}+1$.  Here we use
Lemma~\ref{lem:bit_taylor}(iv) with $\lambda_{\mathrm{sp}}=\log(H/\nu)$ at the base point $x_t$ and
$\zeta=\|D_{x_{t+1}}-D_{x_t}\|\le1$, which gives $\|P_{x_{t+1}}-P_{x_t}\|\le eH/\nu\le e^{\lambda_{\mathrm{top}}}/(2n)$,
and we use $\varepsilon_{\mathrm{op}}/2\le1$.  We sum,
absorb $\varepsilon_{\mathrm{step}}\le1$ into the constant $30$, and obtain the bound.  With
$L_{\max}\le2\sqrt n+1$ and $e^{\lambda_{\mathrm{top}}}=2ne\widetilde H_{\max}/\underline{\widetilde\nu}$ its base-two
logarithm is at most $\log_2(\widetilde H_{\max}/\underline{\widetilde\nu})+2\log_2(2n)+\log_2(50d_0)+5$.

\emph{(v), the refresh count.}  Forced refreshes occur at most once per $L_{\max}\ge1/\pi'$ steps.  Coin refreshes
are independent $\mathrm{Bernoulli}(\pi')$ trials, one per step.  The cap $M_{\mathrm{ref}}$ fires with probability at most
$p_0/(100JR_{\mathrm{rep}})$.  Indeed, for $N\sim\mathrm{Binomial}(T,\pi')$ with mean $\mu$ and
$\Delta_{\mathrm B}:=\mu+3\log(1/\delta_{\mathrm B})$, $\delta_{\mathrm B}\in(0,1)$, Bernstein's inequality gives $\Pr[N\ge\mu+\Delta_{\mathrm B}]\le\exp(-\Delta_{\mathrm B}^2/(2\mu+2\Delta_{\mathrm B}/3))\le\delta_{\mathrm B}$.
In the last step we use the inequality $(\mu+3\log(1/\delta_{\mathrm B}))^2\ge\log(1/\delta_{\mathrm B})(8\mu/3+2\log(1/\delta_{\mathrm B}))$.
\end{proof}

Our accuracy choices of Eq.~\eqref{eq:bit_eps_op} meet the assumptions of Lemma~\ref{lem:bit_tracker} by Lemma~\ref{lem:bit_eps_op}(i).
In Lemma~\ref{lem:bit_integer_bits} we fit the entries of $\widehat P_t$ and the selected-gradient scalar into the $I$ of Eq.~\eqref{eq:bit_word_length}.

\begin{lemma}[Integer bits of the tracker and of the selected gradient]
\label{lem:bit_integer_bits}
Fix a call with the standing objects and the rounded parameters of the paragraphs before and after
Lemma~\ref{lem:bit_state}, and fix a word with $I\ge I_B(m,c_b,\widetilde\rho,R_{\mathrm{alg}})$ integer bits
as in Eq.~\eqref{eq:bit_call_word}.  Assume $\Omega_{\max}\le2n$ for the cap $\Omega_{\max}$ of Lemma~\ref{lem:bit_gaussian}.  Under the hypotheses of Lemma~\ref{lem:bit_tracker}, consider every entry of the tracker $\widehat P_t$ of
Lemma~\ref{lem:bit_tracker} and the selected-gradient scalar $\operatorname{round}_{\mathsf u}(\nu\tr[\widehat P_tE_i])$ of
Lemma~\ref{lem:bit_tracker}(iii).  Consider also the action intermediates, the sketch and Householder columns, the probe images, and
$\widetilde G_{\mathrm{lev}}$, the coordinates
$b_i$ of the target, and every multiplier $\nu\le\widetilde\Lambda$.  Each of these quantities has modulus below $2^{I_B-1}$, so they
fit in $I\ge I_B$ integer bits.  Also $q(x)$ and $\nu\widetilde c$ fit in their $2W$-bit words.
\end{lemma}

\begin{proof}
Lemma~\ref{lem:bit_tracker}(iv) and its proof give
$\log_2\|\widehat P_t\|\le\log_2(\widetilde H_{\max}/\underline{\widetilde\nu})+2\log_2(2n)+\log_2(50d_0)+5\le\log_2(H_{\max}/\underline\nu)+3\log_2(2n)+13.4$,
the rounded ratio being at most $20/3$ times the ratio in $I_B$ of the parameter list above.  This is at
most $I_B-2-\lceil\log_2c_b\rceil-3\lceil\log_2(2n)\rceil-4$ as soon as $\log_2(2n)\ge7.4$, so the entries
of $\widehat P_t$ fit with $\lceil\log_2c_b\rceil+3\lceil\log_2(2n)\rceil+4$ bits to spare.  The scalar satisfies
$|\nu\tr[\widehat P_tE_i]|\le\widetilde\Lambda\widetilde cd_0\|\widehat P_t\|$, and $\widetilde\Lambda\le\frac43c_bm+\mathsf u$ with $\widetilde c\le2a=4\cdot10^6\log(2n)$ give
\[
\log_2(\widetilde\Lambda\widetilde cd_0)\le\log_2c_b+2\log_2(2n)+21.35+\log_2\log(2n)\le\log_2c_b+3\log_2(2n)-2.6,
\]
where we use $\log_2\log(2n)\le\log_2(2n)-24$ for $n\ge N_0$, which the spare
$\lceil\log_2c_b\rceil+3\lceil\log_2(2n)\rceil+4$ bits cover.  The rounding adds $\mathsf u$.  The action intermediates
$e^{\lambda_{\mathrm{top}}+1}\Omega_{\max}\sqrt{d_0}$, the sketch and Householder columns, the probe images,
and $\widetilde G_{\mathrm{lev}}$ each have base-two logarithm at most $\log_2(H_{\max}/\underline\nu)+3\log_2(2n)+30$,
since their bounds involve only $\lambda_{\mathrm{top}}$, $L_{\max}$, $\Omega_{\max}\le2n$ and
$\widetilde c\sqrt m\le a$.  This is below $I_B-2$.  Finally $|b_i|\le\sqrt{c_bm}$, $\nu\le\widetilde\Lambda\le\frac43c_bm+\mathsf u$,
and $\nu\widetilde c\le2a\widetilde\Lambda$ have logarithms below $\log_2c_b+\log_2(2n)+\log_2\log(2n)+23$, and $q(x)$ fits by
Lemma~\ref{lem:bit_state}(i).
\end{proof}

\begin{remark}[The term $\lceil\log_2c_b\rceil$ is needed]
\label{rem:bit_integer_bits}
The term $\lceil\log_2c_b\rceil$ is not a convenience.  At the literal global caps $(c_b,\rho)=(10^{108},3.6\cdot10^{-52})$ with $n=m=10^{30}$
the selected-gradient scalar needs $1455$ bits against the $I-2=1378$ that the integer formula of
Eq.~\eqref{eq:bit_call_word} would offer without this term, a shortfall of $77$ bits.  With the term it gives
$I_B-2=1737$ at that call, and
$1680$ at the dominating triple, which the admissible calls never exceed.  At the largest admissible policy value $c_b=3.9\cdot10^{73}$,
which is row $86$ dilated by $\sigma=10^4$ in Definition~\ref{def:cubic_global_policy}, the scalar needs $1169$ bits against
$1206$, so the formula without the term passes there by $37$ bits only.  The added term makes the bound uniform
in $n$, with more than $230$ bits to spare in every call at every $n>N_{\mathrm{root}}$: about $235$ at $n=10^{25}$,
where the spare is smallest, $280$ at $n=10^{30}$, and growing with $\log n$.  The analytic chain of
Lemma~\ref{lem:bit_integer_bits} guarantees the $6.6$ bits of its ceilings.  The larger figures are evaluations of
the bound of Lemma~\ref{lem:bit_tracker}(iv) at the calls.
\end{remark}

Without the forced refresh we have no bound on $\|\widehat P_t\|$.  The zero-mean
increments accumulate in the kernel of $\|\cdot\|_{\mathcal G}$ over an inter-refresh gap that is
only geometrically distributed, and the real-arithmetic proof never needs to control
them.  The forced refresh is free for our analysis, since a refresh resets
$\mathcal E$, and it at most doubles the expected refresh count.

We can absorb the rounding of the scalar proximal minimizer into a bounded perturbation of the selected gradient, so the descent estimate applies to the step on the coordinate grid.

\begin{lemma}[Finite coordinate step]
\label{lem:bit_step}
Let $\widehat g$ be the selected gradient of Lemma~\ref{lem:bit_tracker}(iii) for coordinate
$i$.  Let $[d_i^-,d_i^+]$ be the feasible displacement interval
$\{d\in\R:x_{t,i}+d\in\widetilde Q_i,\ |d|\le R_a\}$, whose endpoints lie in $\Gamma$, with the trust radius
$R_a:=\lfloor1/\widetilde c\rfloor_\Gamma$.  Let
$d_t:=\operatorname{clip}_{[d_i^-,d_i^+]}(\operatorname{round}_\Gamma(-\widehat g/L))$, where $\operatorname{clip}_{[d^-,d^+]}(v)$ is the point of $[d^-,d^+]$ nearest to $v$.  Then $d_t$ is the exact
minimizer of $(\widehat g+L\vartheta)d+\frac L2d^2$ over $d\in[d_i^-,d_i^+]$ for some
$|\vartheta|\le \mathsf u/(2\cdot10^{10})$.  Consequently the step is the exact proximal coordinate step of
Section~\ref{sec:cubic_time} for the gradient value
$\partial_iF(x_t)+\mathfrak{e}_{t,i}+\overline\beta_{t,i}$ with
$|\overline\beta_{t,i}|\le\nu\widetilde cd_0\varepsilon_{\mathrm{det}}+(L+1)\mathsf u$.  Moreover $x_{t+1}=x_t+d_te_i$ lies in
$\Gamma^S\cap\widetilde Q$ and satisfies $|d_t|\le1/\widetilde c$.  The comparison displacement
$d'_i:=(x_{*,i}-x_{t,i})/L$ of Eq.~\eqref{eq:cubic_three_point} lies in $[d_i^-,d_i^+]$.
\end{lemma}

\begin{proof}
Put $\vartheta:=-\widehat g/L-\operatorname{round}_\Gamma(-\widehat g/L)$, so $|\vartheta|\le \mathsf u/(2\cdot10^{10})$.  The
unconstrained minimizer of $(\widehat g+L\vartheta)d+\frac L2d^2$ is $-\widehat g/L-\vartheta=\operatorname{round}_\Gamma(-\widehat g/L)$,
and clipping a one-dimensional convex quadratic's minimizer to an interval gives the
constrained minimizer.  For the bias bound we add $L|\vartheta|\le L\mathsf u$ to
Lemma~\ref{lem:bit_tracker}(iii).  The endpoints of $\widetilde Q_i$ and $R_a$ lie in $\Gamma$,
so $d_t\in\Gamma$ and $x_{t+1}\in\Gamma^S$.  The bound $R_a\le1/\widetilde c$ gives the trust bound.  For the last claim we use the bound
$R_a\ge1/\widetilde c-\mathsf u/10^{10}\ge1/(2\widetilde c)$ together with $L=\widetilde L\ge8\widetilde c$ by its definition and the side bound $2$ of
$\widetilde Q$: $|d'_i|\le2/L\le1/(4\widetilde c)\le R_a$, and $x_{t,i}+d'_i\in\widetilde Q_i$ by convexity of
$\widetilde Q$, as $x_t,x_*\in\widetilde Q$ and $L\ge1$.  Here every coordinate of $\widetilde Q$ has modulus at most $1$, because $2t_s<1$ and $t_0\le3/5$.
\end{proof}

We now fix the tolerances.

\begin{lemma}[Operator tolerance]
\label{lem:bit_eps_op}
Put $C_{\mathrm{op}}:=12$,
\begin{equation}
\begin{gathered}
\varepsilon_{\mathrm{op}}:=\Bigl\lfloor\min\Bigl\{\frac{\delta_g}{C_{\mathrm{op}}\nu\widetilde cd_0},\frac16\Bigr\}\Bigr\rfloor_{\mathsf u},
\\
\text{and require}\qquad
\nu\widetilde cd_0\varepsilon_{\mathrm{det}}\le\frac{\delta_g}2,
\quad
(L+1)\mathsf u\le\frac{\delta_g}2,
\quad
\theta\le\frac{\delta_g}{70\nu\widetilde ckn},
\quad
3\varepsilon_{\mathrm{qr}}\le\frac1{5\sqrt k},
\end{gathered}
\label{eq:bit_eps_op}
\end{equation}
where $\theta=2\sqrt{2k}\varepsilon_{\mathrm{act}}+\varepsilon_{\mathrm{qr}}$.  Require also $\varepsilon_{\mathrm{step}}\le1$ at
$\|A\|=e^{\lambda_{\mathrm{top}}}+1$, which is the requirement
\begin{equation}
f\ge2\lambda_{\mathrm{top}}\log_2e+3\log_2d_0+\log_2\sigma_{\mathrm{sc}}+12,
\label{eq:bit_step_row}
\end{equation}
all of these requirements at the parameters of the call.  Require also, at the same parameters, the two rows of Step~1 of the proof of Theorem~\ref{thm:bit_complexity} for
Lemma~\ref{lem:bit_dense}, which is $\varepsilon_{\mathrm{dense}}\le\varepsilon_{\mathrm{op}}/8$, and for Eq.~\eqref{eq:bit_surrogate_accuracy},
which is $d_0\mathsf u\,e^{\lambda_{\mathrm{top}}+1}\le\varepsilon_{\mathrm{op}}/8$.  Then for every multiplier $\nu\in[\underline\nu,\Lambda]$ the
wrapper tests:
\begin{enumerate}[label=(\roman*)]
\item $\varepsilon_{\mathrm{op}}\le1/6$, $\varepsilon_{\mathrm{dense}}\le1$, $\varepsilon_{\mathrm{step}}\le1$, and
$\varepsilon_{\mathrm{act}}\le1$, so the standing assumptions of Lemmas~\ref{lem:bit_tracker} and~\ref{lem:bit_guard}
hold, and $|\overline\beta_{t,i}|\le\delta_g$ in Lemma~\ref{lem:bit_step}.
\item The hypotheses of Lemma~\ref{lem:bit_tracker}(ii) and Eq.~\eqref{eq:bit_surrogate_accuracy} hold.  The
degree $r$ of Eq.~\eqref{eq:bit_surrogate} is $O(\log n)$ by (iv).
\item $2\nu d_0\varepsilon_{\mathrm{op}}\le\delta_g/(6\widetilde c)\le H/27$, which is the first term of the requirement of
Lemma~\ref{lem:bit_guard}(ii).  The other two terms of that requirement are the row of Lemma~\ref{lem:bit_guard}(ii) in Step~1 of the
proof of Theorem~\ref{thm:bit_complexity}.
\item $\log_2(1/\varepsilon_{\mathrm{op}})\le\log_2(1/\delta_g)+\log_2(2C_{\mathrm{op}}\Lambda\widetilde cd_0)$, so every
requirement row of Step~1 that involves $\varepsilon_{\mathrm{op}}$ keeps its form.
\end{enumerate}
\end{lemma}

\begin{proof}
\emph{(i).}  $\varepsilon_{\mathrm{op}}\le1/6$ by the cap in its definition.  The bound $\varepsilon_{\mathrm{dense}}\le1$ is the hypothesis of
Lemma~\ref{lem:bit_dense}, the row of Lemma~\ref{lem:bit_dense} in Step~1.  The quantity
$\varepsilon_{\mathrm{step}}=40d_0^{3/2}(1+\|A\|)(\varepsilon_{\mathrm{act}}+d_0\mathsf u)$ at
$\|A\|=e^{\lambda_{\mathrm{top}}}+1$ satisfies, with
$\varepsilon_{\mathrm{act}}=6\sigma_{\mathrm{sc}}e^{\lambda_{\mathrm{top}}+1}d_0^{3/2}\mathsf u$,
$e^{\lambda_{\mathrm{top}}}+2\le3e^{\lambda_{\mathrm{top}}}$, and $d_0\le\sigma_{\mathrm{sc}}e^{\lambda_{\mathrm{top}}+1}d_0^{3/2}$,
\[
\varepsilon_{\mathrm{step}}\le40d_0^{3/2}\cdot3e^{\lambda_{\mathrm{top}}}\cdot7\sigma_{\mathrm{sc}}e^{\lambda_{\mathrm{top}}+1}d_0^{3/2}\mathsf u
=840e\,\sigma_{\mathrm{sc}}e^{2\lambda_{\mathrm{top}}}d_0^3\mathsf u\le1
\]
under Eq.~\eqref{eq:bit_step_row}, as $\log_2(840e)<12$.  Also $\varepsilon_{\mathrm{act}}\le\varepsilon_{\mathrm{step}}$.
By Lemma~\ref{lem:bit_step} and the first two requirements we have
$|\overline\beta_{t,i}|\le\nu\widetilde cd_0\varepsilon_{\mathrm{det}}+(L+1)\mathsf u\le\delta_g$.
\emph{(ii).}  Every use of $\varepsilon_{\mathrm{op}}$ in this section is an upper bound on an error.  The bound
$129\nu^2\widetilde c^2d_0^2\varepsilon_{\mathrm{op}}^2\le\delta_g^2$ in Lemma~\ref{lem:bit_tracker}(ii) holds as
$129\le144$, and its hypothesis $\theta\le\delta_g/(70\nu\widetilde ckn)$ is the third of the four required inequalities.
Eq.~\eqref{eq:bit_surrogate_accuracy} holds by the choice of $r$ and the row of
Eq.~\eqref{eq:bit_surrogate_accuracy} in Step~1, which is a hypothesis of the lemma, and $\sigma_{\mathrm{sc}}\eta_r\le1$ as $\varepsilon_{\mathrm{op}}\le1$.
The cap, where it binds, only lowers $\varepsilon_{\mathrm{op}}$, which helps every use and keeps
$r=O(\log n)$ by (iv).
\emph{(iii).}  By the third term of $\widetilde\delta_g$, we have $\delta_g\le2\widetilde cH/9$, so
$2\nu d_0\varepsilon_{\mathrm{op}}\le2\nu d_0\delta_g/(12\nu\widetilde cd_0)=\delta_g/(6\widetilde c)\le H/27$.  The third term is what
frees this inequality from a lower bound on $H$ in terms of $m$.  The chain $\delta_g\le\sqrt{H/(10^5m)}\le H/(632m)$ needs
$H\ge3.99m$, which the ball target has only off the event $\|\widehat b\|_2^2<m/2$.  That event forces
$\xi'^2_{m+1}+\xi'^2_{m+2}\ge(1-6\cdot10^{-9})W'_G$, of probability at most
$e^{-(1-6\cdot10^{-9})m/4}$, which is $e^{-7.5\cdot10^7}$ at $m=N_0$, and the real-arithmetic analysis of
Section~\ref{sec:small_algorithm_constant} never excludes it.  Where
the new term binds it lowers $\delta_g$, which helps its other two uses, $m\delta_g^2\le10^{-5}H$ and
$32m\delta_g^2\le32\varepsilon_{\mathrm{opt}}/(C_{\mathrm g}J)$ in Lemma~\ref{lem:bit_lyapunov}.
\emph{(iv).}  $\varepsilon_{\mathrm{op}}\ge\min\{\delta_g/(C_{\mathrm{op}}\nu\widetilde cd_0),1/6\}-\mathsf u$
and $\nu\le\Lambda$ give $\log_2(1/\varepsilon_{\mathrm{op}})\le\max\{\log_2(1/\delta_g)+\log_2(2C_{\mathrm{op}}\Lambda\widetilde cd_0),3\}$.
The first term exceeds $3$ because
$\delta_g\le\widetilde\sigma\le\widetilde d_{\mathrm w}/8\le1/256$.
\end{proof}

\begin{remark}[The side condition at the radius cap]
\label{rem:bit_side_condition}
A chain that derives $\varepsilon_{\mathrm{op}}\le1/6$ from $\delta_g\le2\nu\widetilde cd_0$ and the side condition
$\widetilde\tau_{\mathrm w}\le2^{-10}\le32\widetilde c^2\sqrt md_0\widetilde H_{\max}$, available for the algorithm of
Theorem~\ref{thm:cubic_dense_runtime} from $R\le279\sqrt m\log(2n)$, cannot be copied with the bound $d_0\ge2m$.  With
$\widetilde H_{\max}\ge16$, $d_0\ge2m$, and a radius at
the repair cap $R_{\mathrm{alg}}^2=10^{28}m$ of Eq.~\eqref{eq:cubic_global_solver_scales}, the chain gives only
$32\widetilde c^2\sqrt md_0\widetilde H_{\max}\ge256a^2\sqrt m/10^{28}$.  This is $8.63\cdot10^{-6}$ at $m=N_0$ and
$n=10^{30}$, below $2^{-10}=9.77\cdot10^{-4}$.  With $d_0=2n$ and $n\ge m$ the same chain gives
$256a^2n/(10^{28}\sqrt m)\ge256a^2\sqrt n/10^{28}$, which is $1.1\cdot10^3$ at $n=N_{\mathrm{root}}$, and would
survive.  The side condition itself holds at every call of
Section~\ref{sec:small_algorithm_constant} with headroom.  On the multiplier branch $\widetilde H_{\max}\ge\widetilde H>72$
(Lemma~\ref{lem:bit_enclosures}(v)), so $32\widetilde c^2\sqrt md_0\widetilde H_{\max}\ge1152a^2m^{3/2}/\widetilde R^2$,
which is $3.885\cdot10^{-5}$ at the repair cap with $m=N_0$ and $n=10^{30}$.  Moreover
$\widetilde\tau_{\mathrm w}\le\widetilde\rho^2/16<10^{-18}$ for every repair call, since
$\widetilde\rho\le t_0e_{\mathrm{call}}\eta_{\mathrm{num}}<10^{-9}$ there by fact (B4) in Step~1 of Section~\ref{subsec:bit_proof}.  The cap and the row
$\varepsilon_{\mathrm{step}}\le1$ are preferred because they need no radius bound at all.
\end{remark}

Every quantity in Eq.~\eqref{eq:bit_eps_op} is inverse polynomial in $n$, and the requirements are met once $\mathsf u$ is
small enough.  We list the resulting bit counts in the proof of Theorem~\ref{thm:bit_complexity}.

The trace guard needs an upper certificate for accepted proposals and enough slack to accept every proposal below the analytical crossing level $H/3$.  This is the level of the true $F$ whose first crossing by a proposal is the stopping time $\tau$ of Lemma~\ref{lem:bit_lyapunov}.  In Section~\ref{sec:cubic_time} we used $5H/16$ in the same role.

\begin{lemma}[Additive trace guard]
\label{lem:bit_guard}
Let $y$ be a proposal from a base point $x_t$ with $F(x_t)\le H$, let
$s_{\mathrm{tr}}:=128\lceil8\log(100TJR_{\mathrm{rep}}/p_0)\rceil$, and let $z_1,\ldots,z_{s_{\mathrm{tr}}}$ be fresh
Rademacher vectors.  Assume $\varepsilon_{\mathrm{act}}\le1$.  Compute $\widehat y_j$, the half-action of Lemma~\ref{lem:bit_action} in its half-power statement, with the scheme stopped after $J_{\mathrm{ap}}=\sigma_{\mathrm{sc}}/2$ applications,
for $\mathcal Q(\widehat D_y)^{1/2}$ on $z_j$.  Compute also the statistics $\widehat s_j:=\|\widehat y_j\|_2^2$
(exact sums of squares), the means of $\lceil8\log(100TJR_{\mathrm{rep}}/p_0)\rceil$ batches of
$128$ statistics, and their median $\widehat\tau_y$.  Put
\[
U(y):=\Bigl\lceil\frac43\widehat\tau_y+\frac43(\varepsilon_{\mathrm{trace}}+2\mathsf u)+\frac{d_0\varepsilon_{\mathrm{op}}}4\Bigr\rceil_{\mathsf u},
\]
where
\[
\varepsilon_{\mathrm{trace}}\in \mathsf u\mathbb Z\ \text{with}\ 2e^{(\lambda_{\mathrm{top}}+1)/2}\sqrt{d_0}\,\varepsilon_{\mathrm{act}}\le\varepsilon_{\mathrm{trace}}\le2e^{(\lambda_{\mathrm{top}}+1)/2}\sqrt{d_0}\,\varepsilon_{\mathrm{act}}+\mathsf u
\]
(a certified upper enclosure).
Accept $y$ if and only if $q(y)+\lceil\nu U(y)\rceil_{\mathsf u}\le H$, an exact comparison.  Then,
except with probability $p_0/(100TJR_{\mathrm{rep}})$, the following hold:
\begin{enumerate}[label=(\roman*)]
\item If $y$ is accepted then $F(y)\le H$.
\item If $F(y)\le H/3$ then $y$ is accepted, provided
$\nu(2d_0\varepsilon_{\mathrm{op}}+3\varepsilon_{\mathrm{trace}}+7\mathsf u)+\mathsf u\le H/9$.
\end{enumerate}
The guard costs $\sigma_{\mathrm{sc}}r/2$ thin products on a block of $s_{\mathrm{tr}}$ vectors and never evaluates $\phi_0$.
\end{lemma}

\begin{proof}
Write $\mathcal Q:=\mathcal Q(\widehat D_y)$, $\mathcal T_y:=\tr[\mathcal Q]$, and $s_j^{\mathrm{ex}}:=z_j^\top\mathcal Qz_j$.
By Lemma~\ref{lem:bit_probes}, $\E[s_j^{\mathrm{ex}}]=\mathcal T_y$ and
$\V[s_j^{\mathrm{ex}}]\le2\|\mathcal Q\|_F^2\le2\mathcal T_y^2$ because $\mathcal Q\succeq0$.  Chebyshev's
inequality gives $\Pr[|\text{batch mean}-\mathcal T_y|>\mathcal T_y/4]\le(2\mathcal T_y^2/128)/(\mathcal T_y^2/16)=1/4$ for each
batch of $128$.  Hoeffding's inequality for the median of $n_{\mathrm{b}}$ independent batches
gives $\Pr[|\widehat\tau^{\mathrm{ex}}_y-\mathcal T_y|>\mathcal T_y/4]\le e^{-n_{\mathrm{b}}/8}\le p_0/(100TJR_{\mathrm{rep}})$ for
$n_{\mathrm{b}}=\lceil8\log(100TJR_{\mathrm{rep}}/p_0)\rceil$, where $\widehat\tau^{\mathrm{ex}}_y$ is the median of
the exact batch means.  By Lemma~\ref{lem:bit_action} with $J_{\mathrm{ap}}=\sigma_{\mathrm{sc}}/2$, we have
$\|\widehat y_j-\mathcal Q^{1/2}z_j\|_2\le\varepsilon_{\mathrm{act}}/2$ and
$\|\mathcal Q^{1/2}z_j\|_2\le e^{(\lambda_{\mathrm{top}}+1)/2}\sqrt{d_0}$, so
$|\widehat s_j-s_j^{\mathrm{ex}}|\le(2e^{(\lambda_{\mathrm{top}}+1)/2}\sqrt{d_0}+\varepsilon_{\mathrm{act}}/2)\varepsilon_{\mathrm{act}}/2\le\varepsilon_{\mathrm{trace}}$
for $\varepsilon_{\mathrm{act}}\le1$.  Averages and medians move by at most the largest perturbation
of their inputs, and the two roundings add $2\mathsf u$, so
$|\widehat\tau_y-\widehat\tau^{\mathrm{ex}}_y|\le\varepsilon_{\mathrm{trace}}+2\mathsf u$.  On the good event,
$\widehat\tau_y\ge\frac34\mathcal T_y-\varepsilon_{\mathrm{trace}}-2\mathsf u$, hence $U(y)\ge \mathcal T_y+d_0\varepsilon_{\mathrm{op}}/4\ge\Phi(y)$ by
$\|\mathcal Q-P_y\|_1\le d_0\varepsilon_{\mathrm{op}}/4$ (Eq.~\eqref{eq:bit_surrogate_accuracy}).  Acceptance then
gives $F(y)=q(y)+\nu\Phi(y)\le q(y)+\lceil\nu U(y)\rceil_{\mathsf u}\le H$, which is (i).
Conversely $\widehat\tau_y\le\frac54\mathcal T_y+\varepsilon_{\mathrm{trace}}+2\mathsf u$ and $\mathcal T_y\le\Phi(y)+d_0\varepsilon_{\mathrm{op}}/4$
give $U(y)\le\frac53\Phi(y)+2d_0\varepsilon_{\mathrm{op}}+3\varepsilon_{\mathrm{trace}}+7\mathsf u$.  Hence, if $F(y)\le H/3$ then
$q(y)+\lceil\nu U(y)\rceil_{\mathsf u}\le\frac53F(y)+\nu(2d_0\varepsilon_{\mathrm{op}}+3\varepsilon_{\mathrm{trace}}+7\mathsf u)+\mathsf u\le\frac59H+\frac19H<H$,
which is (ii).  For the half-actions we apply Lemma~\ref{lem:bit_action} with
$J_{\mathrm{ap}}=\sigma_{\mathrm{sc}}/2$ to a block of $s_{\mathrm{tr}}$ vectors.
\end{proof}

Lemma~\ref{lem:bit_guard} replaces the relative certificate $\Phi(y)\le U(y)\le2\Phi(y)$ of
Section~\ref{sec:faster_algorithm} by an additive one (Remark~\ref{rem:bit_relative_guard}).  The relative certificate's
requirement $d_0\varepsilon_{\mathrm{op}}\le\phi_0/8$ is the only place where $\phi_0$ enters the algorithm, and properties (i)
and (ii) are all that the crossing argument of Lemma~\ref{lem:bit_lyapunov} uses.
We meet the requirement in (ii) as follows.  The bound $2\nu d_0\varepsilon_{\mathrm{op}}\le\delta_g/(6\widetilde c)\le H/27$ holds by Lemma~\ref{lem:bit_eps_op}(iii).  The $\varepsilon_{\mathrm{trace}}$ and $\mathsf u$ terms are at most $H/27$ each by the row of Lemma~\ref{lem:bit_guard}(ii) in the table of Step~1 of the proof of Theorem~\ref{thm:bit_complexity}.

In the phase analysis we combine the stochastic tracker recurrence with the bounded gradient bias. We control both the objective gap and the probability of crossing the analytical level by a stopped potential estimate.

\begin{lemma}[Finite Lyapunov contraction and localization]
\label{lem:bit_lyapunov}
Let one candidate phase of the finite algorithm be run from $z$ with $F(z)\le H/16$, under
$K_0:=2000$, $c_1:=4/3$, and Eq.~\eqref{eq:bit_eps_op} in force.  Let
$\tau:=\min\{t+1:F(x_t+d_te_{i_t})>H/3\}$ be the first crossing of $H/3$ by a proposal.
Let $\Delta(x):=F(x)-F(x_\nu)$ be the gap of the finite objective.  This is the $\Delta_x$ of the notation paragraph, written as in Section~\ref{sec:cubic_time}, and $\Delta_z$ and $\Delta_j$ below are the phase-start gaps of Section~\ref{sec:faster_algorithm}.  Let $V_t:=\Delta(x_t)+c_1\mathcal E_t/(\pi'L)$.  Then:
\begin{enumerate}[label=(\roman*)]
\item For every $t<\tau$,
$\E[V_{t+1}\mid\mathcal F_t]\le(1-\alpha)V_t-\frac L8\E[d_t^2\mid\mathcal F_t]+\frac{2\delta_g^2}L$.
\item $\Pr[\tau\le T]\le1/15$.
\item On the event $\{\tau>T\}\cap\{\text{all guards correct}\}$ the guarded finite process and the
unguarded finite process, which accepts every proposal, coincide pathwise, and
$\E[\mathbbm{1}[\tau>T]\Delta(x_T)]\le e^{-16}\Delta(z)+32m\delta_g^2$.
\end{enumerate}
Consequently one candidate phase returns a point $x_T$ with
$\Delta(x_T)\le\Delta(z)/2+\varepsilon_{\mathrm{opt}}/(32J)$ with probability at least $0.9$, by the
choice $C_{\mathrm g}=2\cdot10^5$ in the definition of $\delta_g$.
\end{lemma}

\begin{proof}
\emph{Step 1: the gap recurrence.}  For $t<\tau$ the base point has $F(x_t)\le H/3\le H$
and the step is the exact proximal step for the gradient value
$\partial_iF(x_t)+\mathfrak{e}_{t,i}+\overline\beta_{t,i}$ with $|\overline\beta_{t,i}|\le\delta_g$
(Lemma~\ref{lem:bit_step}).  Hence the three-point inequality
Eq.~\eqref{eq:cubic_three_point} holds with $\mathfrak{e}_{t,i}+\overline\beta_{t,i}$ in place of
$\mathfrak{e}_{t,i}$.  Using $(\mathfrak{e}+\overline\beta)^2\le2\mathfrak{e}^2+2\delta_g^2$, the average over the
uniform coordinate as in Eq.~\eqref{eq:cubic_gap_recurrence}, and
Lemma~\ref{lem:bit_tracker}(iii), we obtain
\[
\E[\Delta(x_{t+1})\mid\mathcal F_t]\le(1-\alpha)\Delta(x_t)+\frac{\mathcal E_t}L+\frac{\delta_g^2}L-\frac L4\E[d_t^2\mid\mathcal F_t].
\]

\emph{Step 2: the potential.}  We add $c_1/(\pi'L)$ times Lemma~\ref{lem:bit_tracker}(ii).
The coefficient of $\mathcal E_t$ is $1/L+c_1(1-\pi')/(\pi'L)\le c_1(1-\alpha)/(\pi'L)$
because $c_1\ge\pi'/(\pi'-\alpha)$, which holds for $c_1=4/3$ since $\pi'\ge4\alpha$.  The
coefficient of $\E[d_t^2\mid\mathcal F_t]$ is
$-L/4+c_1(1-\pi')65L/(\pi'ks)\le-L/4+(4/3)\cdot65L/K_0\le-L/8$ because $ks\ge K_0/\pi'$
and $K_0\ge694$.  The additive term is
$\delta_g^2/L+c_1\cdot2\delta_g^2/(\pi'ksL)\le(1+8/(3K_0))\delta_g^2/L\le2\delta_g^2/L$.
This is (i).

\emph{Step 3: the crossing.}  With $d'=0$ in the three-point inequality,
$F(x_{t+1})-F(x_t)\le(\mathfrak{e}_{t,i}+\overline\beta_{t,i})^2/(2L)\le(\mathfrak{e}_{t,i}^2+\delta_g^2)/L$.  Hence a
crossing from $F(z)\le H/16$ to above $H/3$ within $T$ steps forces
$\sum_{t<\tau}(\mathfrak{e}_{t,i_t}^2+\delta_g^2)/L\ge13H/48$.  The event $\{t<\tau\}$ is
$\mathcal F_t$-measurable, so $\E[\mathbbm{1}[t<\tau]\mathfrak{e}_{t,i_t}^2]=\E[\mathbbm{1}[t<\tau]\mathcal E_t]$
by Lemma~\ref{lem:bit_tracker}(iii).  Multiplying Lemma~\ref{lem:bit_tracker}(ii) by
$\mathbbm{1}[t<\tau]$, using $\mathbbm{1}[t+1<\tau]\le\mathbbm{1}[t<\tau]$, and summing over
$t<T$ with $\mathcal E_0=0$, we get
\[
\pi'\sum_{t<T}\E[\mathbbm{1}[t<\tau]\mathcal E_t]
\le\frac{65L^2}{ks}\sum_{t<T}\E[\mathbbm{1}[t<\tau]d_t^2]+\frac{2T\delta_g^2}{ks}.
\]
Summing (i) over $t<\tau$ likewise, we obtain the stopped step energy
$\frac L8\sum_{t<T}\E[\mathbbm{1}[t<\tau]d_t^2]\le V_0+2T\delta_g^2/L\le H/16+32m\delta_g^2$, using
$V_0=\Delta(z)\le F(z)\le H/16$ and $T=16mL$.  That inequality is the stopped form of Eq.~\eqref{eq:cubic_step_energy} of Section~\ref{sec:cubic_time}.  Hence, with $ks\ge K_0/\pi'$, we have
\[
\E\Bigl[\sum_{t<\min\{\tau,T\}}\frac{\mathfrak{e}_{t,i_t}^2}L\Bigr]
\le\frac{65}{K_0}\Bigl(\frac H2+256m\delta_g^2\Bigr)+\frac{32m\delta_g^2}{K_0}
\le\frac{65H}{2K_0}+\frac{16672\,m\delta_g^2}{K_0},
\]
and $\sum_{t<\tau}\delta_g^2/L\le T\delta_g^2/L=16m\delta_g^2$.  With $K_0=2000$ and
$m\delta_g^2\le10^{-5}H$ (definition of $\delta_g$), the expected crossing mass is at most
$(0.01625+0.00009+0.00016)H<0.0166H$, and Markov's inequality gives
$\Pr[\tau\le T]\le0.0166\cdot48/13<1/15$.

\emph{Step 4: coupling and the endpoint.}  Before $\tau$ every proposal has $F\le H/3$,
so by Lemma~\ref{lem:bit_guard}(ii) a correct guard accepts it.  Hence on
$\{\tau>T\}\cap\{\text{all guards correct}\}$ the guarded process performs exactly the
steps of the unguarded process, which are deterministic functions of the same coins.
Iterating (i) over $t<T$ on $\{\tau>T\}$, we obtain
$\E[\mathbbm{1}[\tau>T]V_T]\le(1-\alpha)^TV_0+2T\delta_g^2/L\le e^{-16}\Delta(z)+32m\delta_g^2$, and
$\Delta\le V$.  By Markov's inequality with the threshold $\Delta(z)/2+\varepsilon_{\mathrm{opt}}/(32J)$
and $32m\delta_g^2\le32\varepsilon_{\mathrm{opt}}/(C_{\mathrm g}J)$, we bound the probability that the endpoint
misses the threshold on this event by $2e^{-16}+1024/C_{\mathrm g}\le0.01$.  The guards fail with
probability at most $T\cdot p_0/(100TJR_{\mathrm{rep}})\le0.01$.  Also $\Pr[\tau\le T]\le1/15$.
With the refresh cap of Lemma~\ref{lem:bit_tracker}(v), which fires with probability at most $0.01$, the four failure probabilities sum to less than $0.0867+0.01<0.1$.
\end{proof}

\subsection{Certified decisions and the certified projection}
\label{subsec:bit_projection}

Two decisions of the wrapper concern points that are not guarded.  These are the box projection $b_Q$ of the target onto the rounded
box $\widetilde Q$, whose matrix $D_{b_Q}$ can have eigenvalues of order $a\sqrt m$, and the fallback matrices
$E=\sum_i\varepsilon_i\widetilde A_i$ of the fair-sign trials (Lemma~\ref{lem:bit_fallback}).  We handle both by a spectral-norm
certificate that never forms a matrix exponential.

\begin{lemma}[Normalized-squaring norm certificate]
\label{lem:bit_normcert}
Let $E$ be a symmetric $n'\times n'$ fixed-point matrix with $n'\le d_0$ and
$\|E\|_F>1$, and let $\varpi\in(0,1/2]$.  Put
$J_{\mathrm{sq}}:=\max\{0,\lceil\log_2(\log n'/\log(1+\varpi/2))\rceil-1\}$, with $J_{\mathrm{sq}}:=0$ when $n'=1$.  Set $\widehat X_0:=E$ rounded entrywise to the $\mathsf u$-grid.  For
$0\le j<J_{\mathrm{sq}}$, set $\widehat N_j:=\lceil\|\widehat X_j\|_F\rceil_{\mathsf u}$, computed by an exact sum of squares and an integer square
root, then $\widehat Y_j:=\operatorname{round}_{\mathsf u}(\widehat X_j/\widehat N_j)$, and then
$\widehat X_{j+1}:=\operatorname{round}_{\mathsf u}(\widehat Y_j\widehat Y_j)$ with the product exact.  Finally set
$\widehat N_{J_{\mathrm{sq}}}:=\lceil\|\widehat X_{J_{\mathrm{sq}}}\|_F\rceil_{\mathsf u}$.  Let $\widehat\Lambda$ satisfy
$0\le\widehat\Lambda-\sum_{j\le J_{\mathrm{sq}}}2^{-j}\log\widehat N_j\le\varpi/16$, obtained by range reduction and a Taylor
series.  Let $U\in \mathsf u\mathbb Z$ satisfy $e^{\widehat\Lambda+\varpi/16}\le U\le e^{\widehat\Lambda+\varpi/16}+\mathsf u$, a certified upper enclosure.  If $10^5n'^2\mathsf u\le\varpi$ then
\[
\|E\|\le U\le(1+\varpi)\|E\| .
\]
The certificate costs $J_{\mathrm{sq}}$ exact dense products and $J_{\mathrm{sq}}+1$ Frobenius norms,
$J_{\mathrm{sq}}=O(\log(\log n'/\varpi))$.
\end{lemma}

\begin{proof}
We prove both bounds of the conclusion for $\widehat X_0$ and transfer them to $E$ at the end of the proof.
Write $E$ for $\widehat X_0$ in this proof.  The bound $\|\widehat X_0\|_F>1-n'\mathsf u/2$ suffices for the lower bounds.  Write $a_j:=\|\widehat X_j\|$ and $F_j:=\|\widehat X_j\|_F$, so $F_j\le\widehat N_j\le F_j+2\mathsf u$.  Let
$Y^{\mathrm{ex}}_j:=\widehat X_j/\widehat N_j$, with $\|Y^{\mathrm{ex}}_j\|_F\le1$ and $\|Y^{\mathrm{ex}}_j\|=a_j/\widehat N_j$.
The rounding of $\widehat Y_j$ perturbs it by a matrix $G_j$ of operator norm at most $n'\mathsf u/2$, by the
entrywise error $\mathsf u/2$ and the Frobenius bound.  The rounding of the product perturbs it by at most $n'\mathsf u$.  Hence
$\widehat X_{j+1}=(Y^{\mathrm{ex}}_j)^2+E_{j+1}$ with
$\|E_{j+1}\|\le2\|Y^{\mathrm{ex}}_j\|\|G_j\|+\|G_j\|^2+n'\mathsf u\le n'\mathsf u+n'^2\mathsf u^2/4+n'\mathsf u\le4n'\mathsf u$, and therefore
$a_{j+1}\in[(a_j/\widehat N_j)^2-4n'\mathsf u,(a_j/\widehat N_j)^2+4n'\mathsf u]$.

\emph{Lower bounds.}  We have $F_0>1-n'\mathsf u/2\ge1/2$ by hypothesis.  For $j\ge1$,
$F_j\ge a_j\ge\|Y^{\mathrm{ex}}_{j-1}\|^2-4n'\mathsf u\ge F_{j-1}^2/(n'\widehat N_{j-1}^2)-4n'\mathsf u\ge1/(2n')$, using
$\|Y\|\ge\|Y\|_F/\sqrt{n'}$, $\widehat N_{j-1}\le F_{j-1}(1+4n'\mathsf u)$, and induction.  Hence
$(a_j/\widehat N_j)^2\ge1/(2n')$ for every $j$.  The additive perturbation $4n'\mathsf u$ of
$a_{j+1}$ is therefore a relative one of at most $8n'^2\mathsf u$:
$a_j=\widehat N_ja_{j+1}^{1/2}(1+\vartheta_j)^{1/2}$ with $|\vartheta_j|\le9n'^2\mathsf u$.

\emph{Unrolling.}  Iterating from $j=0$ to $J_{\mathrm{sq}}-1$ and using
$F_{J_{\mathrm{sq}}}/\sqrt{n'}\le a_{J_{\mathrm{sq}}}\le F_{J_{\mathrm{sq}}}\le\widehat N_{J_{\mathrm{sq}}}\le F_{J_{\mathrm{sq}}}(1+4n'\mathsf u)$, we obtain
\[
\begin{gathered}
\|E\|=\prod_{j<J_{\mathrm{sq}}}\widehat N_j^{2^{-j}}\cdot a_{J_{\mathrm{sq}}}^{2^{-J_{\mathrm{sq}}}}\cdot\prod_{j<J_{\mathrm{sq}}}(1+\vartheta_j)^{2^{-j-1}}
\in\Bigl[\Pi\,n'^{-2^{-J_{\mathrm{sq}}-1}}\frac{1-16n'^2\mathsf u}{1+4n'\mathsf u},\ \Pi(1+16n'^2\mathsf u)\Bigr],
\\
\Pi:=\prod_{j\le J_{\mathrm{sq}}}\widehat N_j^{2^{-j}} .
\end{gathered}
\]

\emph{The certificate.}  We have $\widehat\Lambda+\varpi/16\ge\log\Pi+\varpi/16\ge\log(\Pi(1+16n'^2\mathsf u))\ge\log\|E\|$
because $\varpi/16\ge16n'^2\mathsf u$, so $U\ge\|E\|$.  Conversely
$U\le\Pi e^{\varpi/8}+\mathsf u\le\Pi e^{\varpi/8}(1+2\sqrt{n'}\mathsf u)$, because $\Pi\ge\|E\|/(1+16n'^2\mathsf u)\ge1/(2\sqrt{n'})$
from $\|E\|\ge\|E\|_F/\sqrt{n'}>(1-n'\mathsf u/2)/\sqrt{n'}$.  Hence
$U\le\|E\|\,n'^{2^{-J_{\mathrm{sq}}-1}}e^{\varpi/8}(1+2\sqrt{n'}\mathsf u)(1+4n'\mathsf u)/(1-16n'^2\mathsf u)$.  Moreover
$n'^{2^{-J_{\mathrm{sq}}-1}}\le1+\varpi/2$ by the choice of $J_{\mathrm{sq}}$, $e^{\varpi/8}\le1+0.134\varpi$ for
$\varpi\le1/2$, and $(1+2\sqrt{n'}\mathsf u)(1+4n'\mathsf u)/(1-16n'^2\mathsf u)\le1+40n'^2\mathsf u\le1+\varpi/2500$.  The product
of the three factors is at most $1+0.7\varpi\le1+\varpi$.
\emph{Transfer to $E$.}  The rounding moves $\|E\|$ by at most $n'\mathsf u/2\le\varpi\|E\|/(2\cdot10^5\sqrt{n'})$, since $\|E\|\ge\|E\|_F/\sqrt{n'}$.
We have $\|E\|\le\|\widehat X_0\|+n'\mathsf u/2\le\Pi(1+16n'^2\mathsf u)+n'\mathsf u/2$.  Moreover $\Pi(e^{\varpi/16}-1-16n'^2\mathsf u)\ge n'\mathsf u/2$, as $\Pi\ge1/(2\sqrt{n'})$ and $\varpi/16-16n'^2\mathsf u\ge\varpi/17\ge n'^{3/2}\mathsf u$ by $10^5n'^2\mathsf u\le\varpi$.  Together these give $\|E\|\le\Pi e^{\varpi/16}\le U$.  Finally $U\le(1+0.7\varpi)\|\widehat X_0\|\le(1+0.7\varpi)\bigl(1+\varpi/(2\cdot10^5\sqrt{n'})\bigr)\|E\|\le(1+\varpi)\|E\|$.
\end{proof}

For the multiplier search, the candidate selection, and the radial feasibility correction we need outward enclosures at finer accuracy than the step guards.  The multiplier search is the wrapper of Section~\ref{sec:faster_algorithm}.

\begin{lemma}[Certified enclosures and exact decisions]
\label{lem:bit_enclosures}
Let $x\in\Gamma^m$ be a guarded point of a candidate phase (so $F(x)\le H$), or the box projection $b_Q$ of the
target $b$ onto $\widetilde Q$ in case (C) of item (v) below.  Let
$\mathrm{tol}\in\{\varepsilon_{\mathrm{opt}}/(256\nu J),\ \tau_{\mathrm{w}}/16\}$.  Compute $\widetilde P_x$ by
Lemma~\ref{lem:bit_dense} with the degree
$r_{\mathrm{tol}}:=\min\{r\ge2:(r+1)!\ge9\sigma_{\mathrm{sc}}e^{\lambda_{\mathrm{top}}+2}d_0/\mathrm{tol}\}$ and put
$\widehat\Phi(x):=\operatorname{round}_{\mathsf u}(\tr[\widetilde P_x])$.  Then:
\begin{enumerate}[label=(\roman*)]
\item if $d_0\varepsilon_{\mathrm{dense}}\le\mathrm{tol}/3$ and $d_0^2\mathsf u\,e^{\lambda_{\mathrm{top}}+1}\le\mathrm{tol}/3$,
the interval $[\widehat\Phi(x)-\mathrm{tol}-\mathsf u,\widehat\Phi(x)+\mathrm{tol}+\mathsf u]$ contains $\Phi(x)$,
and $[q(x)+\lfloor\nu(\widehat\Phi(x)-\mathrm{tol}-\mathsf u)\rfloor_{\mathsf u},\ q(x)+\lceil\nu(\widehat\Phi(x)+\mathrm{tol}+\mathsf u)\rceil_{\mathsf u}]$
contains $F(x)$ and has length at most $2\nu\,\mathrm{tol}+2(\nu+1)\mathsf u$.
\item the following decisions are exact comparisons of grid numbers.  The selection rule of
Section~\ref{sec:faster_algorithm} retains the candidate whose upper endpoint lies below the old point's lower
endpoint and is least, else the old point.  The three-way bisection decision acts on the enclosure of
$\Phi(x_\nu)$.  The box shortcut is the wrapper's first test in Section~\ref{sec:faster_algorithm}: if the
certified interval for $\Phi(b_Q)$ lies below $1$, the box projection $b_Q$ is returned as the exact projection.
The remaining decision is the radial correction.  Every tested multiplier is first rounded to $\mathsf u\mathbb Z$, a
shift of at most $\mathsf u$ of a bisection midpoint.  Every
conclusion that Section~\ref{sec:faster_algorithm} draws from the corresponding exact comparison holds for these
decisions.
\item the radial factor $\beta_{\mathrm{r}}:=\lfloor(3/4)/(\overline\Phi-1/4)\rfloor_{\mathsf u}$ for
$\overline\Phi:=\widehat\Phi(\widehat x)+\mathrm{tol}+\mathsf u>1$, where $\widehat x$ is the returned point of the multiplier search, satisfies $\Phi(\beta_{\mathrm{r}}\widehat x)\le1$ and
$(1-\beta_{\mathrm{r}})\|\widehat x\|_2\le d_{\mathrm{w}}/16$.  The product $\beta_{\mathrm{r}}\widehat x$ is formed exactly.
\item \emph{The update of $y$ in suffix stages.}  After snapping at threshold $2\widetilde\rho$
(Lemma~\ref{lem:bit_state}(iv)) in the box $Q_y(t_s)$, consider the update
$y_i:=\operatorname{round}^{\to0}_{\mathsf u}(y_i+\overline x_i/t_s)$ for unsnapped $i$ and $y_i:=\pm1$ for snapped $i$.
This update keeps $y\in(\mathsf u\mathbb Z)^S\cap[-1,1]^S$ and makes the new active set $\{i:|y_i|<1\}$ exact.  It moves the applied
increment $t_s(y'-y)$ away from $\overline x$ by at most $t_s\mathsf u<\mathsf u/2$ per unsnapped coordinate, so that
\[
\Bigl\|\sum_{i\in S}(y'_i-y_i)M_i\Bigr\|\le\frac{\|B(\overline x)\|+m\mathsf u/2}{t_s}\le\frac{\widetilde R}{t_s}
\qquad\text{whenever }\mathsf u\le\frac{3s_{\mathrm{sm}}\widetilde R}{2m},
\]
which $f\ge\frac12\lceil\log_2n\rceil+19$ ensures as $\widetilde R\ge\sqrt{15m}$.  This is the increment bound
$\widehat R/t_s$ of Lemma~\ref{lem:algorithm_certified_implementation} at $\widehat R=\widetilde R$.  The
acceptance test ``more than $m/4$ coordinates snapped'' compares an integer with $m/4$.
\item \emph{(pre-test for the box shortcut)} let $b_Q$ be the coordinatewise projection of $b$ onto $\widetilde Q$,
a point of $((\mathsf u/(2^{40}5^{10}))\mathbb Z)^m$.  Let $\widetilde\chi_{\mathrm{hi}}$ be a dyadic number with
$\widetilde\chi\le\widetilde\chi_{\mathrm{hi}}\le1.01\widetilde\chi$, a certified evaluation of $\log(2n)$ to $40$ bits.  Let $\varpi$ be
a dyadic number in $[\widetilde\chi/16,\widetilde\chi/8]$.  Let $U_B$ be the certificate of Lemma~\ref{lem:bit_normcert} for
$E:=B(b_Q)$ with this $\varpi$, or $U_B:=1$ if $\|E\|_F\le1$.
The decisions compare the dyadic numbers
$\widetilde cU_B$ and $(b'\mp\widetilde\chi_{\mathrm{hi}}\cdot\{1,1/4\})\widetilde a$.  In case (A),
$\widetilde cU_B\le(b'-\widetilde\chi_{\mathrm{hi}})\widetilde a$, the point $b_Q$ lies in $\widetilde{\mathcal D}$ and is
the exact projection of $b$ onto $\widetilde{\mathcal D}\cap\widetilde Q$.  The case $U_B=1$ lands there, as
$(b'-\widetilde\chi_{\mathrm{hi}})\widetilde R\ge(1-1.005\cdot10^{-6}-2\mathsf u)\sqrt{15m}>1$.  In case (B),
$\widetilde cU_B>(b'+\widetilde\chi_{\mathrm{hi}}/4)\widetilde a$, the point $b_Q$ is infeasible.  In case (C),
$\lambda_{\max}(D_{b_Q})\le\log(2n)/3$ and $\lambda_{\min}(D_{b_Q})\ge-2a(1-s_{\mathrm{sm}})-\log(2n)/3-2\mathsf u\ge-M_{\mathrm{sp}}$,
so the enclosure of (i) applies to $b_Q$.  In cases (B) and (C),
\[
\|b\|_2\ge\|b_Q\|_2\ge\frac{(1-s_{\mathrm{sm}}-1.01\chi-2\mathsf u)\sqrt{15}}{1+\chi/8}\ge3.872979>3,
\]
so that $\Lambda=2q(0)/(1-\phi_0)\ge\|b\|_2^2>14.9$ and $H\ge16q(0)=8\|b\|_2^2>72$ on the multiplier branch.
\end{enumerate}
\end{lemma}

\begin{proof}
\emph{(i), the enclosure.}  By Lemmas~\ref{lem:bit_taylor}(iii),(iv) and~\ref{lem:bit_dense} with the stated
degree, we have $\|\widetilde P_x-P_x\|\le\mathrm{tol}/(3d_0)+\varepsilon_{\mathrm{dense}}+d_0\mathsf u\,e^{\lambda_{\mathrm{top}}+1}\le\mathrm{tol}/d_0$,
so we obtain $|\tr[\widetilde P_x]-\Phi(x)|\le\mathrm{tol}$.  The trace is an exact sum of grid entries
rounded once, $\nu$ is dyadic, $q(x)$ is exact, and we round the two products
outward.

\emph{(ii), exact decisions.}  Each rule compares numbers on the grid $\mathsf u\mathbb Z$ or $\Gamma$-derived rationals with
a common denominator, which is an integer comparison.  Our arguments in Section~\ref{sec:faster_algorithm} use
only that the intervals contain the true values and have the stated widths.  For
the candidate selection rule, we use ``accepted phase-start objectives never increase'' and the separation
$\Delta_z/2-\varepsilon_{\mathrm{opt}}/(32J)>2\cdot\varepsilon_{\mathrm{opt}}/(64J)$ when $\Delta_z>\varepsilon_{\mathrm{opt}}$.
For the bisection, we use the three cases ``interval above $1$'', ``meets $1$'', ``below $1$''
with the residual estimate $\|x_\nu-x_*\|_2^2\le|\nu-\nu_*||\Phi(x_\nu)-1|$ of
Section~\ref{sec:faster_algorithm} (Eq.~\eqref{eq:phi_multiplier_residual}) in the middle case.  In the middle
case the enclosure of $\Phi(x_\nu)$ has length at most $\tau_{\mathrm{w}}/4+2\mathsf u$.  The bound $2\Lambda \mathsf u\le d_{\mathrm{w}}^2/(4(1+\Lambda))$, the bisection row of
Step~1 in the proof of Theorem~\ref{thm:bit_complexity}, keeps the product of $\Lambda$ and the length
below $d_{\mathrm{w}}^2/4$, as we require in Section~\ref{sec:faster_algorithm}.  For the bisection we also use
the wrapper's Lipschitz requirement $G_{\mathrm{lev}}\sigma\le\tau_{\mathrm{w}}/16$.  This requirement is the choice of $\sigma$ in Section~\ref{sec:faster_algorithm}, not the transfer of Lemma~\ref{lem:bit_gaussian}(ii).  It holds for the rounded
constants because $\widetilde\sigma\le\widetilde\tau_{\mathrm{w}}/(16\widetilde G_{\mathrm{lev}})$ by definition and $\widetilde G_{\mathrm{lev}}$ bounds the
Lipschitz constant of the finite $\Phi$.  The wrapper tolerates the rounding of each tested multiplier to
$\mathsf u\mathbb Z$ because its bracket length never falls below $\widetilde d_{\mathrm{w}}/(4\widetilde G_*)\gg\mathsf u$.  Here the
bisection of Section~\ref{sec:faster_algorithm} stops once the bracket has length at most $d_{\mathrm{w}}/(4G_*)$.  For the box
shortcut in case (C) of (v), if the upper endpoint at $b_Q$ is at most $1$, the point is
exactly feasible and is the projection.  Otherwise $\Phi(b_Q)>1-\tau_{\mathrm{w}}/8-2\mathsf u$, and the wrapper
proceeds as written, handling a barely feasible $b_Q$ through its first solve at
$\underline\nu$.

\emph{(iii), the radial factor.}  For $\vartheta\in(0,1/4)$ the map $\vartheta\mapsto(1-\vartheta)/(\overline\Phi-\vartheta)$ is
decreasing when $\overline\Phi>1$, and $\phi_0<1/4$.  Hence we obtain
$\beta_{\mathrm{r}}\le(3/4)/(\overline\Phi-1/4)\le(1-\phi_0)/(\overline\Phi-\phi_0)$, and convexity gives
$\Phi(\beta_{\mathrm{r}}\widehat x)\le\beta_{\mathrm{r}}\overline\Phi+(1-\beta_{\mathrm{r}})\phi_0\le1$ as in
Section~\ref{sec:faster_algorithm}.  Moreover we have
$1-\beta_{\mathrm{r}}\le(\overline\Phi-1)/(\overline\Phi-1/4)+\mathsf u\le\frac43(\overline\Phi-1)+\mathsf u\le\frac43\cdot\frac{\tau_{\mathrm{w}}}2+\mathsf u$
by our accuracy choices and by $\mathsf u\le\tau_{\mathrm{w}}/(8\Lambda)$.  This last inequality follows from the candidate-selection rounding row of Step~1 because $\varepsilon_{\mathrm{opt}}\le\sigma^2/2\le\tau_{\mathrm{w}}$.  We also have $\|\widehat x\|_2\le\sqrt m$, because every coordinate of $\widetilde Q$ has modulus at most $1$, as $2t_s<1$ and $t_0\le3/5$.  With
$\tau_{\mathrm{w}}\le3d_{\mathrm{w}}/(64\sqrt m)$ we get
$(1-\beta_{\mathrm{r}})\|\widehat x\|_2\le\frac23\tau_{\mathrm{w}}\sqrt m+\mathsf u\sqrt m\le d_{\mathrm{w}}/32+\mathsf u\sqrt m\le d_{\mathrm{w}}/16$.  The
last step is by $\mathsf u\sqrt m\le d_{\mathrm{w}}/32$.  The product $\beta_{\mathrm{r}}\widehat x$ of a
$\mathsf u$-grid number and a $\Gamma$-vector lies on the grid $\mathsf u\Gamma$ and is exact.

\emph{(iv), the update of $y$.}  The quotient $\overline x_i/t_s=10^{10}\overline x_i/v$ is an exact rational, and $y_i+\overline x_i/t_s\in[-1,1]$
because $\overline x\in Q_y(t_s)=t_s([-1,1]^S-y)$.  For an unsnapped coordinate the value lies in $(-1,1)$ and rounding
toward zero keeps it there, while a snapped coordinate has $\overline x_i=t_s(\pm1-y_i)$ exactly and $y'_i=\pm1$.
The rounding changes $y'_i-y_i$ by at most $\mathsf u$, so the applied increment $t_s(y'_i-y_i)$ differs from
$\overline x_i$ by at most $t_s\mathsf u<\mathsf u/2$ ($t_s<1/2$).  Then from $\|M_i\|\le1$ we get
$\|\sum_i(y'_i-y_i)M_i\|\le(\|B(\overline x)\|+m\mathsf u/2)/t_s$.  By Lemma~\ref{lem:bit_state}(iv) we have
$\|B(\overline x)\|\le(1-3s_{\mathrm{sm}}/4)\widetilde R$, and $m\mathsf u/2\le3s_{\mathrm{sm}}\widetilde R/4$ is the stated
condition.  With $\widetilde R\ge\sqrt{15m}$ it reads $\mathsf u\le1.5s_{\mathrm{sm}}\sqrt{15/m}$, which is $f\ge\frac12\log_2m+18.4$.
The increment $\widehat R/t_s$ is exactly what Eq.~\eqref{eq:cubic_refined_suffix_cost} charges, with
$\widehat R=\widetilde R\le(1+10^{-8})R$ by Lemma~\ref{lem:bit_state}(ii), so nothing else changes.

\emph{(v), the pre-test.}  By Lemma~\ref{lem:bit_normcert} we have
$\|B(b_Q)\|\le U_B\le(1+\widetilde\chi/8)\|B(b_Q)\|$ under $10^5n^2\mathsf u\le\varpi$, the pre-test row of Step~1.  The comparison
$\widetilde cU_B\le(b'-\widetilde\chi_{\mathrm{hi}})\widetilde a$ means $U_B\le(b'-\widetilde\chi_{\mathrm{hi}})\widetilde R$.
In case (A), we have $\|B(b_Q)\|\le(b'-\widetilde\chi)\widetilde R$, so $b_Q\in\widetilde{\mathcal D}$ by
Lemma~\ref{lem:bit_state}(iii).  The box projection of $b$ that lies in the body is the projection onto
$\widetilde Q\cap\widetilde{\mathcal D}$.  With $U_B=1$, case (A) holds because
$(b'-\widetilde\chi_{\mathrm{hi}})\widetilde R\ge(1-s_{\mathrm{sm}}-2\mathsf u-1.01\chi)\sqrt{15m}=(1-1.005\cdot10^{-6}-2\mathsf u)\sqrt{15m}>1$.
In case (B), $\|B(b_Q)\|\ge U_B/(1+\widetilde\chi/8)>(b'+\widetilde\chi_{\mathrm{hi}}/4)\widetilde R/(1+\widetilde\chi/8)\ge b'\widetilde R$
because $b'\le1$ and $\widetilde\chi_{\mathrm{hi}}\ge\widetilde\chi$, so $\widetilde h(b_Q)>b'$ and $b_Q\notin\widetilde{\mathcal D}$.  In case (C),
$(b'-\widetilde\chi_{\mathrm{hi}})\widetilde R/(1+\widetilde\chi/8)\le\|B(b_Q)\|\le(b'+\widetilde\chi_{\mathrm{hi}}/4)\widetilde R$, so the eigenvalues of
$D_{b_Q}=\widetilde a\operatorname{diag}(B,-B)/\widetilde R-\widetilde ab'I$ lie in $[-\widetilde a(2b'+\widetilde\chi_{\mathrm{hi}}/4),\widetilde a\widetilde\chi_{\mathrm{hi}}/4]$.
Then from $\widetilde a\widetilde\chi_{\mathrm{hi}}\le1.01\log(2n)$ we obtain the two spectral bounds (recall
$\widetilde ab'\le a(1-s_{\mathrm{sm}})+\mathsf u$).  The enclosure of (i) then applies because the spectral hypothesis of
Lemma~\ref{lem:bit_dense} that we use in (i) is exactly these bounds.  In cases (B)
and (C), we have
\[
\|b\|_2\ge\|b_Q\|_2\ge\frac{\|b_Q\|_1}{\sqrt m}\ge\frac{\|B(b_Q)\|}{\sqrt m}\ge\frac{(b'-\widetilde\chi_{\mathrm{hi}})\widetilde R}{(1+\widetilde\chi/8)\sqrt m}
\ge\frac{(1-s_{\mathrm{sm}}-1.01\chi-2\mathsf u)\sqrt{15}}{1+\chi/8},
\]
where we use $|(b_Q)_i|\le|b_i|$ (as $0\in\widetilde Q$), $\widetilde\chi\le\chi$, and $\widetilde R^2\ge R_{\mathrm{alg}}^2>15m$.
The right side is $3.872979$ to six decimals and exceeds $3$ for $\mathsf u\le2^{-40}$.  This is the bound
$\|b\|_2>3$ in the proof of Lemma~\ref{lem:appendix_fixed_target_projection}, the analogue of
Lemma~\ref{lem:selected_accepted_draw}, and the source of $\Lambda=\Omega(1)$ and $H>72$.  The ball and repair
targets of Section~\ref{sec:small_algorithm_constant} satisfy it with headroom, but the lemma needs only this
consequence of the pre-test.
\end{proof}

\begin{lemma}[Fixed-point certified fixed-target projection]
\label{lem:bit_projection}
\setlength{\emergencystretch}{2em}Let $n>N_{\mathrm{root}}$ and $N_0\le m\le n$, and let $M_1,\ldots,M_m$ and $c_b\ge1$ be as above.  Let the word have
$f\ge f_B(m,c_b,\widetilde\rho,R_{\mathrm{alg}})$ fractional and
$I\ge I_B(m,c_b,\widetilde\rho,R_{\mathrm{alg}})$ integer bits.  Suppose:
\begin{itemize}[leftmargin=5.2em,labelsep=0.6em]
\item[(H-grid)] the box is $\widetilde Q=\prod_{i\le m}[l_i,r_i]$ with $l_i,r_i\in\Gamma$,
$l_i\le0\le r_i$, and $\widetilde Q\subseteq[-1,1]^m$.  The target satisfies $b\in(2^{-f_b}\mathbb Z)^m$ for some
$f_b\le f+40$ and $\|b\|_2^2\le c_bm$.  The ball target $\widehat s\,\xi'_{[m]}$ and the repair target
$\operatorname{round}_{\mathsf u}(t_0U'/\beta)$ of (F13) indeed lie on $\mathsf u\mathbb Z$.
\item[(H-body)] $R_{\mathrm{alg}}\in\mathsf u\mathbb Z$ with $R_{\mathrm{alg}}^2>15m$ and
$R_{\mathrm{alg}}\le10^{14}\sqrt n$, and the body is
$\widetilde{\mathcal D}=\{x:\tr[\exp(D_x)]\le1\}$ built from $R_{\mathrm{alg}}$ as in Lemma~\ref{lem:bit_state}.
\item[(H-acc)] $\widetilde\rho\in\mathsf u\mathbb Z$ with $0<\widetilde\rho\le1/8$ and $\widetilde\rho^{-1}\le n^{O(1)}$.
\item[(H-fail)] $\varepsilon_{\mathrm{fail}}\in(0,1/2)$ is rational.
\end{itemize}
Run the finite wrapper and solver, items (F2)--(F10) and (F12) of the paragraph \emph{The finite algorithm}.
Here (F3) is restricted to the test matrix, the probes, and the coins, since a
fixed-target call draws no target.
Then the call either reports a declared failure or returns an exactly feasible point
$\widehat x\in(\mathsf u\Gamma)^m\cap\widetilde Q\cap\widetilde{\mathcal D}$.  The accuracy bound
\[
\|\widehat x-\operatorname{proj}_{\widetilde{\mathcal D}\cap\widetilde Q}(b)\|_2\le\widetilde\rho
\]
fails with probability at most $\varepsilon_{\mathrm{fail}}$.  A declared failure is a coin cap of
Lemma~\ref{lem:bit_probes} or of Lemma~\ref{lem:bit_gaussian}(i), or the one-redraw cap of
Lemma~\ref{lem:bit_gaussian}(iii).  It occurs with probability at most $\mathsf u^4$ per capped sampling procedure, and
the call runs at most
$N_{\mathrm{sol}}\widetilde JR_{\mathrm{rep}}(2\widetilde T+1)$ of them.  Here
$N_{\mathrm{sol}}:=2+\lceil\log_2(4\widetilde\Lambda\widetilde G_*/\widetilde d_{\mathrm w})\rceil$ bounds the number
of fixed-multiplier solves.  The call uses
\[
\bigl(mn^{2+o(1)}+mn^{\omega-1/2+o(1)}+n^{\omega+o(1)}\bigr)\operatorname{polylog}(1/\varepsilon_{\mathrm{fail}})\cdot
M\bigl(2W+c_\eta\lceil\log_2(2n)\rceil\bigr),\qquad W:=f+I,
\]
bit operations, with the temporaries of the paragraph \emph{Computational model}.  The bound $N_{\mathrm{sol}}$, the phase count $\widetilde J$, the candidate count $R_{\mathrm{rep}}$, and the
guard probe count $s_{\mathrm{tr}}$ are $O(\log(n/\varepsilon_{\mathrm{fail}}))$, and the phase length is
$\widetilde T=16m\widetilde L$ with $\widetilde L\le40a^2c_b$.
\end{lemma}

The output grid is $\mathsf u\Gamma=(\mathsf u^2/10^{10})\mathbb Z$, not $\Gamma$.  The returned
point is the radially corrected $\beta_{\mathrm r}\widehat x$ of Lemma~\ref{lem:bit_enclosures}(iii), the product of a
$\mathsf u\mathbb Z$ number and a $\Gamma$-vector.  In case (A) of the pre-test it is the clipped target $b_Q$ on
$(\mathsf u/(2^{40}5^{10}))\mathbb Z\subseteq\mathsf u\Gamma$ (for $f\ge30$).  Every rational the callers of
Section~\ref{sec:small_algorithm_constant} form from $\widehat x$ therefore carries $2^{2f}$, not $2^f$, in its
denominator.  Every call we make in Section~\ref{sec:small_algorithm_constant} meets the hypotheses with headroom.  The
budgets of (H-fail) are the rationals of Lemmas~\ref{lem:algorithm_certified_implementation}
and~\ref{lem:cubic_refined_complete}.  The radius bounds of (H-body) are facts (B2)--(B3) in Step~1 of
Section~\ref{subsec:bit_proof}.  The accuracy bounds of (H-acc) are fact (B4) there.  Each accuracy is inverse
polynomial in $n$, as Lemma~\ref{lem:appendix_fixed_target_projection} requires.

\begin{proof}[Proof of Lemma~\ref{lem:bit_projection}]
\setlength{\emergencystretch}{1.5em}\emph{Step 1: grids and exact state.}  We apply Lemma~\ref{lem:bit_state}(i) with $f\ge f_B\ge40+\lceil\log_2(2n)\rceil$.
Iterates lie on $\Gamma^m$, $B(x)$ and $q(x)$ are exact, and $\widehat D_x$ is a once-rounded value of the
exact $D_x$ with $\|\widehat D_x-D_x\|\le d_0\mathsf u$.  Every comparison of (F10) is an integer comparison.  The
target grid $2^{-f_b}\mathbb Z$ with $f_b\le f+40$ enters only here, through the common grid of $x-b$.

\emph{Step 2: body, parameters, spectral bounds.}  Lemma~\ref{lem:bit_state}(ii),(iii) supplies the rounded body
with $(b'-\widetilde\chi)\widetilde K\subseteq\widetilde{\mathcal D}\subseteq(1-s_{\mathrm{sm}})\widetilde K$, the Slater point
$0$, and the bounds on $E_i$ and $\widetilde c$ that Lemmas~\ref{lem:bit_increment}(i) and~\ref{lem:bit_step} use.  In the
parameter list above we fix every rounded wrapper constant in its inequality-preserving direction, and the ratio
bounds relate them to the constants of Eq.~\eqref{eq:bit_call_word}.  In the paragraph \emph{Spectral bounds} we derive, using the target only through
$F(x)\le H$, that every exponentiated $\widehat D_x$ has spectrum in $[-M_{\mathrm{sp}},\lambda_{\mathrm{top}}]$.  Hence Lemmas~\ref{lem:bit_taylor},
\ref{lem:bit_dense}, and~\ref{lem:bit_action} apply at $\lambda_{\mathrm{sp}}=\lambda_{\mathrm{top}}$, since their
hypotheses concern the spectrum alone.  Lemma~\ref{lem:bit_fastmm} applies verbatim, as $n\ge\varkappa$,
$N_{\mathrm{mult}}\le2n$, and $k=s=\lceil45\pi'^{-1/2}\rceil\le45n^{1/4}+1$.  Lemma~\ref{lem:bit_rangefinder} applies,
as $2\le k\le n$ and $b_\Omega\ge4.5\log_2n+7$.  Lemma~\ref{lem:bit_increment} applies under $3\varepsilon_{\mathrm{qr}}\le1/(5\sqrt k)$,
a row of Step~9, with $\|A\|\le e^{\lambda_{\mathrm{top}}}+1$ from Lemma~\ref{lem:bit_taylor}(iv) at
$\lambda_{\mathrm{sp}}=\log(H/\nu)$.

\emph{Step 3: the pre-test.}  Here we apply Lemma~\ref{lem:bit_enclosures}(v).  In case (A) the call returns $b_Q$, exactly
feasible, the exact projection, on $(\mathsf u/(2^{40}5^{10}))\mathbb Z\subseteq\mathsf u\Gamma$, at no error and
with no randomness.  In cases (B) and (C) the multiplier branch runs with $\|b\|_2\ge3.872979$, $\Lambda>14.9$, and
$H>72$.  In case (C) the enclosure of $\Phi(b_Q)$ at tolerance $\widetilde\tau_{\mathrm w}/16$ decides the box
shortcut as in Lemma~\ref{lem:bit_enclosures}(ii).

\emph{Step 4: bracket and bisection.}  The wrapper solves at $\underline{\widetilde\nu}$ and bisects
$[\underline{\widetilde\nu},\widetilde\Lambda]$ on certified enclosures of $\Phi(x_\nu)$ of length at most
$\widetilde\tau_{\mathrm w}/4+2\mathsf u$.  It retains $\widehat x$ when the enclosure meets $1$ and stops when the bracket
has length at most $\widetilde d_{\mathrm w}/(4\widetilde G_*)$.  We round every tested multiplier to $\mathsf u\mathbb Z$.
Lemma~\ref{lem:bit_enclosures}(i),(ii) apply.  The number of solves
is at most $N_{\mathrm{sol}}=2+\lceil\log_2(4\widetilde\Lambda\widetilde G_*/\widetilde d_{\mathrm w})\rceil$: one at
$\underline{\widetilde\nu}$, one per halving of a bracket of initial length at most $\widetilde\Lambda$, and one at the
final upper endpoint.  This is the count $O(\log(\Lambda G_*/d_{\mathrm w}))$ of Section~\ref{sec:selected_wrapper}, and
$O(\log n)$ under (H-acc) and (H-body) with $c_b$ constant.

\emph{Step 5: tracker, step, tolerance, guard, phase.}  Lemma~\ref{lem:bit_tracker} holds as stated.  Its hypotheses are $F(x_t)\le H$ at base points and
$|d_t|\le1/\widetilde c$ at steps, supplied by the guard and by Lemma~\ref{lem:bit_step}, together with the standing
assumptions met by Lemma~\ref{lem:bit_eps_op}(i).  The integer bits are Lemma~\ref{lem:bit_integer_bits}, whose
hypothesis $\Omega_{\max}\le2n$ holds at $f=f_B$ by fact (B5) in Step~1 of Section~\ref{subsec:bit_proof}.
Lemma~\ref{lem:bit_step} applies: the boxes of Section~\ref{sec:small_algorithm_constant} have side $2t_s<1$ or
$6/5$, at most $2$.  Eq.~\eqref{eq:bit_eps_op} is in force (Lemma~\ref{lem:bit_eps_op}), and the requirement of
Lemma~\ref{lem:bit_guard}(ii) is met by Lemma~\ref{lem:bit_eps_op}(iii) together with the guard row.  Lemma~\ref{lem:bit_lyapunov} applies to every candidate
phase: $F(z)\le F_\nu(0)\le H/16$ at phase starts by the selection rule (F12), $K_0=2000$, $c_1=4/3$, and the two
facts $m\delta_g^2\le10^{-5}H$ and $32m\delta_g^2\le32\varepsilon_{\mathrm{opt}}/(C_{\mathrm g}J)$ follow from the first
two terms of the minimum defining $\widetilde\delta_g$.  We use the lemma's hypothesis ``Eq.~\eqref{eq:bit_eps_op} in force''
through the four inequalities and through the first two terms of $\delta_g$.

\emph{Step 6: candidate selection and the phases.}  The cap $\varepsilon_{\mathrm{opt}}\gets\min\{\varepsilon_{\mathrm{opt}},H\}$
of Algorithm~\ref{alg:cubic_recursive_multiplier} is void, and the $\varepsilon_{\mathrm{opt}}$ term of
$\widetilde\delta_g$ is below the $H$ term: $\widetilde\varepsilon_{\mathrm{opt}}\le\widetilde\sigma^2/2\le(\widetilde\rho/32)^2/2\le2^{-17}<72<H$
on the multiplier branch (Lemma~\ref{lem:bit_enclosures}(v)).  Likewise $\widetilde\varepsilon_{\mathrm{opt}}<1\le2JH$.  The selection enclosures have width at most
$\widetilde\varepsilon_{\mathrm{opt}}/(128\widetilde J)+4\widetilde\Lambda\mathsf u\le\widetilde\varepsilon_{\mathrm{opt}}/(64\widetilde J)$,
the rounding row of Step~9.  Hence phase-start objectives never increase, and each of the $\widetilde J$ phases halves
the gap up to $\widetilde\varepsilon_{\mathrm{opt}}/(16\widetilde J)$ with probability at least $1-p_0/\widetilde J$.
Here we use Lemma~\ref{lem:bit_lyapunov} for each of the $R_{\mathrm{rep}}$ candidates, guards correct except with probability
$p_0/(100TJR_{\mathrm{rep}})$ each, and refresh counts capped by Lemma~\ref{lem:bit_tracker}(v).  After $\widetilde J$
phases the gap is at most $\widetilde\varepsilon_{\mathrm{opt}}$ except with probability $p_0$, exactly as in
Section~\ref{sec:selected_wrapper}, and by $1$-strong convexity we have $\|\widehat x-x_\nu\|_2\le\widetilde\sigma$.

\emph{Step 7: the output.}  Our accounting in Section~\ref{sec:selected_wrapper} for the retained point gives a
distance at most $\widetilde d_{\mathrm w}/2+\widetilde d_{\mathrm w}/8+\widetilde d_{\mathrm w}/16\le11\widetilde d_{\mathrm w}/16<\widetilde\rho$
from $\operatorname{proj}_{\widetilde{\mathcal D}\cap\widetilde Q}(b)$ after the radial correction of
Lemma~\ref{lem:bit_enclosures}(iii).  The hypotheses $\phi_0<1/4$ and $\|\widehat x\|_2\le\sqrt m$ of that correction hold.  The bound $\Phi(\beta_{\mathrm r}\widehat x)\le1$ is exact feasibility for
$\widetilde{\mathcal D}$, and $\beta_{\mathrm r}\widehat x\in\widetilde Q$ because $0\in\widetilde Q$, $\widetilde Q$ is
convex, and $\beta_{\mathrm r}\in(0,1]$.  The point lies on $\mathsf u\Gamma$.

\emph{Step 8: the failure budget.}  Each solve fails with probability at most $p_0=\varepsilon_{\mathrm{fail}}/(2N_{\mathrm{sol}})$, as we now show.
The guards contribute $\widetilde T\cdot p_0/(100\widetilde T\widetilde JR_{\mathrm{rep}})$ per phase, and the
$R_{\mathrm{rep}}$ candidates of a phase all fail with probability $0.1^{R_{\mathrm{rep}}}\le p_0/(2\widetilde J)$.  The
refresh cap fires with probability at most $p_0/(100\widetilde JR_{\mathrm{rep}})$ per phase.  The selection
enclosures, the bisection enclosures, and the pre-test certificate are deterministic.  Over $N_{\mathrm{sol}}$ solves
the returned point misses the accuracy with probability at most $\varepsilon_{\mathrm{fail}}/2$.  We cap every remaining
randomized ingredient.  The capped ingredients are the index sampler, with one procedure per candidate phase (Lemma~\ref{lem:bit_probes}),
the Karney samples of each test matrix, with one procedure per nonrefresh step (Lemma~\ref{lem:bit_gaussian}(i)), and
the one-redraw cap of each test matrix (Lemma~\ref{lem:bit_gaussian}(iii)).  Each fails with probability at most
$p_{\mathrm s}=\mathsf u^4$, and there are at most $N_{\mathrm{sol}}\widetilde JR_{\mathrm{rep}}(2\widetilde T+1)$ of them.  A
firing cap is a declared failure reported to the caller, which charges it to its attempt allowance (by (F18)).  No
saturation occurs while every guard is correct, by Lemma~\ref{lem:bit_integer_bits}.

\emph{Step 9: the word.}  The lemmas impose the requirement rows of Step~1 of the proof of
Theorem~\ref{thm:bit_complexity}, which we evaluate at the rounded parameters of the call.  The row Eq.~\eqref{eq:bit_step_row}
of Lemma~\ref{lem:bit_eps_op}, the row of Lemma~\ref{lem:bit_enclosures}(iv) ($f\ge\frac12\lceil\log_2n\rceil+19$, suffix
stages), and the radius row of Lemma~\ref{lem:bit_state}(ii) ($f\ge\frac12\lceil\log_2n\rceil+51$) are the rows the
solver of Section~\ref{sec:cubic_time} alone would not need.  The row Eq.~\eqref{eq:bit_step_row} is below $f_B$ because $\sigma_{\mathrm{sc}}\le4M_{\mathrm{sp}}$ and $d_0=2n$.  The bounds that dominate the rows are the ratio bounds and the two target-cap bounds of the parameter list after Lemma~\ref{lem:bit_state}
and the facts (B2) and (B3) of Step~1 of the proof of Theorem~\ref{thm:bit_complexity}.  The bound on
$\log_2(1/\widetilde\delta_g)$ is its consequence (C4).  Against these
the fractional formula of Eq.~\eqref{eq:bit_call_word} has, in the unrounded $\varepsilon_{\mathrm{opt}}$ of the formula,
\[
\log_2\frac1{\varepsilon_{\mathrm{opt}}}\ge4\log_2G_*+4\log_2c_b+4\log_2m+6\log_2\frac1{\widetilde\rho}+29,
\]
where we use $\varepsilon_{\mathrm{opt}}=\sigma^2/2$, $\sigma\le\tau_{\mathrm w}/(16G_{\mathrm{lev}})$, and
$\tau_{\mathrm w}\le d_{\mathrm w}^2/(1+\Lambda)$, which is the binding term of $\tau_{\mathrm w}$ as $\Lambda\ge c_bm\ge m$, together with
$\Lambda\ge c_bm$, $H_{\max}\ge8c_bm$, and $\underline\nu\le\widetilde\rho/(8G_*)$.  No row carries more than
$2\log_2\widetilde\Lambda$ (the bisection row) or one $\log_2\widetilde L$ (the $(L+1)\mathsf u$ row).  That every row
is at most $f_B$ and every integer demand at most $I_B-2$ for all $n>N_{\mathrm{root}}$ is
Step~1 of the proof of Theorem~\ref{thm:bit_complexity}, and we record the margins in Remarks~\ref{rem:bit_word_length}
and~\ref{rem:bit_counts}.

\emph{Step 10: cost.}  Write $W_\times:=2W+c_\eta\lceil\log_2(2n)\rceil$.  A nonrefresh step performs $6\sigma_{\mathrm{sc}}r+O(1)$ thin products of shape
$\langle d_0,d_0,O(k+s)\rangle$ or $\langle d_0,O(k+s),d_0\rangle$ at $O(n^{2+2\eta})M(W_\times)$ each.  The cost
per product is that of Lemma~\ref{lem:bit_fastmm} with $\beta\in(1/4,\alpha_{\mathrm{dual}})$, since $k+s=O(n^{1/4+o(1)})\le n^\beta$ for
large $n$, and where $k+s>n^\beta$ we split the shape into $n^{o(1)}$ blocks of width $n^\beta$.  The step also performs $O(d_0^2)$ word operations and the guard's $\sigma_{\mathrm{sc}}r/2$ thin products on a block of
$s_{\mathrm{tr}}=O(\log(n/\varepsilon_{\mathrm{fail}}))$ vectors.  A refresh performs $r+\kappa_{\mathrm{sc}}=O(\log n)$
dense products at $O(n^{\omega+2\eta})M(W_\times)$.  Here $\sigma_{\mathrm{sc}}r=O(\log^2n)$ by Lemma~\ref{lem:bit_eps_op}(iv).  A
candidate phase has $\widetilde T=16m\widetilde L\le640a^2c_bm=m\,n^{o(1)}$ steps ($c_b$ an absolute constant) and at
most $\widetilde T\pi'+M_{\mathrm{ref}}+1=O(m\widetilde Ln^{-1/2}+\log(n/\varepsilon_{\mathrm{fail}}))$ refreshes.  Hence it
costs $(mn^{2+o(1)}+mn^{\omega-1/2+o(1)}+n^{\omega+o(1)})\operatorname{polylog}(1/\varepsilon_{\mathrm{fail}})M(W_\times)$, as in
Eq.~\eqref{eq:cubic_fixed_multiplier_cost}.  A solve runs $\widetilde JR_{\mathrm{rep}}$ phases and
$(R_{\mathrm{rep}}+1)\widetilde J+1$ enclosure evaluations of $r_{\mathrm{tol}}+\kappa_{\mathrm{sc}}=O(\log n)$ dense
products each.  The call runs $N_{\mathrm{sol}}$ solves, the pre-test's $J_{\mathrm{sq}}=O(\log\log n)$ dense
products, and $mn^2$ word operations to form $B(b_Q)$.  The quantities $N_{\mathrm{sol}}$, $\widetilde J$, $R_{\mathrm{rep}}$ are
$O(\log(n/\varepsilon_{\mathrm{fail}}))$ because $\widetilde\Lambda$, $\widetilde G_*$, $1/\widetilde d_{\mathrm w}$,
$\widetilde H$, $1/\widetilde\varepsilon_{\mathrm{opt}}$ are polynomial in $n$ under (H-acc), (H-body), and $c_b$
constant.  Multiplying, we obtain the stated count, which is the cost of
Lemma~\ref{lem:appendix_fixed_target_projection} times $M(W_\times)$, and the temporaries are those of
the paragraph \emph{Computational model} with $b_{\mathrm{in}}+34$ numerator bits (Lemma~\ref{lem:bit_state}(i)).
\end{proof}

\begin{remark}[Reading the counts]
\label{rem:bit_counts}
Consider a suffix stage at
$n=m=10^{30}$ with target cap $c_b=10^8$, radius $\sqrt{15m}$, and accuracy $2.42\cdot10^{-22}$ from
Eq.~\eqref{eq:algorithm_119_accuracy}.  The formulas give $f_B=6360$ and $I_B=976$, so
$W_B=7336$.  The counts are $N_{\mathrm{sol}}=230$, $\widetilde J=1197$, $R_{\mathrm{rep}}=16$, and $s_{\mathrm{tr}}\approx1.8\cdot10^5$,
the last two at a solve budget $p_0=10^{-12}$.  At the leaf budget $2^{-651}$ of Step~3 of the proof of
Theorem~\ref{thm:bit_complexity}, $R_{\mathrm{rep}}=203$ and $s_{\mathrm{tr}}\approx6.2\cdot10^5$.  The
degree is $r=190$, and $\lambda_{\mathrm{top}}=229.6$.  Here $\widetilde L$ is about $7.7\cdot10^{25}$, and it enters the word through
$\log_2(2(\widetilde L+1))$ only.  The binding row is the $\varepsilon_{\mathrm{det}}$ term for the suffix and root calls and the
candidate-selection enclosure for the repair calls, with margins above $3850$ bits in every case.
\end{remark}

\subsection{Targets, repair, and fallbacks in fixed point}
\label{subsec:bit_targets}

The algorithm draws two kinds of targets.  The ball target of
Eq.~\eqref{eq:algorithm_119_ball_sample} is drawn in every suffix stage, and the uniform target of
Lemma~\ref{lem:cubic_refined_projection} is drawn in every repair node.  The algorithm decides acceptance by
tests on the targets, repairs the accepted fractional point by a deterministic binary loop, and falls back to
fair-sign trials with the constants $10.889$, $11.902$, and $13$.  We supply the fixed-point versions in
this subsection.  The two target lemmas transfer Lemmas~\ref{lem:algorithm_centered_box}
and~\ref{lem:cubic_refined_projection} to stored dyadic targets.  The repair and the accumulation of $y$ are
exact.  We obtain the fallbacks as a corollary of Lemma~\ref{lem:bit_normcert}, and in a closing remark we fix how
we charge a declared failure inside an attempt.  Throughout, $f$ and $I$ are the fractional and integer bits of the word, with $\mathsf u=2^{-f}$
and $W=I+f$ as in the paragraph \emph{Computational model}, and the $\widetilde A_i$ are the rounded inputs of
Lemma~\ref{lem:bit_input}.  The grid $\Gamma$, the box endpoints, and the rounded body $\widetilde{\mathcal D}$ with its
radius window and its two inclusions are those of Section~\ref{subsec:bit_state} and Lemma~\ref{lem:bit_state}(ii),(iii).
We fix the output grid $\mathsf u\Gamma$ after Lemma~\ref{lem:bit_projection}.

\begin{lemma}[Rounded Gaussians and the ball-target lemma]
\label{lem:bit_ball_target}
Fix a suffix stage of Algorithm~\ref{alg:algorithm_suffix_signing} with padded size $m\ge N_0$, fractional
coloring $y\in(\mathsf u\mathbb Z)^m\cap(-1,1)^m$, and box scale $t=t_{i,s}\in10^{-10}\mathbb Z$ with $1/5<t<1/2$.
Fix a symmetric closed convex body $D\subseteq\R^m$ that satisfies, with the stage parameters
$(\delta,\alpha,p_0,z)=(1/4,0.295,p_*N_0/(N_0+2),0.13)$ and $(s_{i,s},t_{i,s})$ of
Definition~\ref{def:cubic_refined_suffix}, the hypotheses of Lemma~\ref{lem:algorithm_centered_box} in the
convention of Lemma~\ref{lem:cubic_refined_suffix}.  The rounded body $\widetilde{\mathcal D}$ is such a body,
because $(1-10^{-6})K(R)\subseteq\widetilde{\mathcal D}$ gives $\gamma_m(\widetilde{\mathcal D})\ge e^{-E_sm}$
with the stage exponent $E_s$ of that lemma.  Let $h:=2^{-40}$, $f\ge70$, and $f':=f-40$.  Let
$\xi'_1,\ldots,\xi'_{m+2}$ be independent exact samples of $\operatorname{round}_h(\mathcal N(0,1))$
(Lemma~\ref{lem:bit_gaussian}(i)).  They are coupled on one coin space with $G\sim\mathcal N(0,I_{m+2})$ so that
$|\xi'_i-G_i|\le h/2$ surely.  Put
\[
W'_G:=\sum_{i=1}^{m+2}\xi_i'^2,\qquad
\widehat s:=10^4\cdot2^{-f'}\bigl\lfloor2^{f'}\sqrt{m/W'_G}\bigr\rfloor,\qquad
\widehat b:=\widehat s\,(\xi'_1,\ldots,\xi'_m),
\]
and reject the trial if $W'_G\notin[m/2,2(m+2)]$.  Then the following hold.
\begin{enumerate}[label=(\roman*)]
\item \emph{Exact formation.}
$W'_G$ is an exact multiple of $h^2$, formed in a temporary of
$2(I+f)+\lceil\log_2(m+2)\rceil$ bits, and the rejection is an exact comparison.  Writing $W'_G=N2^{-80}$ with
$N\in\mathbb Z$, $\lfloor2^{f'}\sqrt{m/W'_G}\rfloor=\lfloor\sqrt{\lfloor2^{2f}m/N\rfloor}\rfloor$ is one integer
square root, so $\widehat s$ is a lower bound with
$0\le\sqrt{m/W'_G}/\kappa_{\mathrm{pr}}-\widehat s\le10^4\cdot2^{-f'}$.  Moreover
$\widehat b\in(\mathsf u\mathbb Z)^m$, $\|\widehat b\|_2^2\le10^8m$ exactly, and $|\widehat b_i|\le10^4\sqrt m$.
Hence $I\ge\lceil\log_2(10^4\sqrt m)\rceil+2$, which is $66$ at $m=10^{30}$, holds the target.  On the accepted
event $\widehat s\le10^4\sqrt2$ needs $15$ integer bits.
\item \emph{Rejections.}
Put $W_G:=\|G\|_2^2$, the norm of Eq.~\eqref{eq:algorithm_119_ball_sample}.  A lower rejection $W'_G<m/2$
forces $W_G<m/2+hm<0.999m$, an event the proof of Lemma~\ref{lem:algorithm_centered_box} already charges.  An
upper rejection $W'_G>2(m+2)$ forces $W_G>2(m+2)(1-\vartheta)$ with $\vartheta:=h/\sqrt2<6.44\cdot10^{-13}$, an
event of probability at most $e^{-(m+2)J(2(1-\vartheta))}$.  Moreover $J(2(1-\vartheta))=0.153426>0.15$, so the
inequality $J(2)>0.15$ that the proof uses holds for the shifted exponent.  The four exceptional terms of that
proof, with the shifted upper rejection and the reserve $\Gamma_0$ of Lemma~\ref{lem:cubic_refined_certificate},
sum to less than $5.2\cdot10^{-17}$ at $m=N_0$, against the $2\cdot10^{-4}$ printed there.
\item \emph{The ball-target lemma survives.}
On the events of the proof of Lemma~\ref{lem:algorithm_centered_box} for the exact $G$, and when $W'_G$ is
accepted, the Euclidean projection $x_*$ of $\widehat b$ onto $D\cap Q_y(t)$ has more than $m/4$ coordinates on
facets of $Q_y(t)$.  Hence, including rejection, this happens with probability greater than $0.999$, the bound
of that lemma, and $y+x_*/t$ has more than $m/4$ sign coordinates.
\end{enumerate}
\end{lemma}

\begin{proof}
\emph{(i), formation.}  Each $\xi'_i$ is a multiple of $h$, so $W'_G$ is a multiple of $h^2=2^{-80}$ and we
form it exactly by the inner-product rule of the computational model.  The numbers $m/2$ and $2(m+2)$ are
integers, so the rejection compares integers.  For integers $k,p,q>0$ we have $k\le\sqrt{p/q}$ if and only if
$k^2\le\lfloor p/q\rfloor$, and $2^{f'}\sqrt{m/W'_G}=2^{f'}2^{40}\sqrt{m/N}=\sqrt{2^{2f}m/N}$, which is the
square-root formula.  Then $\widehat s$ is $10^4$ times a multiple of $2^{-f'}$ and
$0\le10^4\sqrt{m/W'_G}-\widehat s\le10^4\cdot2^{-f'}$, where $10^4=\kappa_{\mathrm{pr}}^{-1}$.  The product
$\widehat s\xi'_i$ of a multiple of $10^4\cdot2^{-f'}$ and a multiple of $2^{-40}$ is a multiple of
$2^{-f'-40}=\mathsf u$, and we form it exactly.  Since $\widehat s\le10^4\sqrt{m/W'_G}$, we obtain
\[
\|\widehat b\|_2^2=\widehat s^2\sum_{i\le m}\xi_i'^2\le10^8\frac m{W'_G}W'_G=10^8m,
\qquad
|\widehat b_i|\le\widehat s\sqrt{W'_G}\le10^4\sqrt m,
\]
and $W'_G\ge m/2$ on the accepted event gives $\widehat s\le10^4\sqrt2$.

\emph{(ii), rejections.}  By the triangle inequality we get
$|\sqrt{W'_G}-\sqrt{W_G}|=\bigl|\|\xi'\|_2-\|G\|_2\bigr|\le\|\xi'-G\|_2\le\tfrac h2\sqrt{m+2}$.  If $W'_G<m/2$, then
$\sqrt{W_G}<\sqrt{m/2}+\frac h2\sqrt{m+2}$ and squaring gives $W_G<m/2+h\sqrt{m/2}\sqrt{m+2}+h^2(m+2)/4<m/2+hm$
for $m\ge N_0$, so $W_G<0.999m$.  If $W'_G>2(m+2)$, then
$\sqrt{W_G}>\sqrt{2(m+2)}-\frac h2\sqrt{m+2}=\sqrt{2(m+2)}\,(1-h/(2\sqrt2))$ and
$W_G>2(m+2)(1-h/(2\sqrt2))^2\ge2(m+2)(1-h/\sqrt2)$.  For $W_G\sim\chi^2_{m+2}$ and $\varsigma>1$ the Chernoff
bound $\Pr[W_G\ge\varsigma(m+2)]\le e^{-(m+2)J(\varsigma)}$ is the chi-square tail we used in the proof of
Lemma~\ref{lem:algorithm_centered_box}.  Since $J$ is increasing on $(1,\infty)$, the upper rejection has
probability at most $e^{-(m+2)J(2(1-\vartheta))}$.  The value $J(2(1-\vartheta))=0.153426$ agrees with
$J(2)$ to the printed digits and exceeds $0.15$.  The four exceptional terms
$e^{-m/(8\cdot10^6)}+e^{-mJ(0.999)}+e^{-(m+2)J(2(1-\vartheta))}+(m+1)e^{-m\Gamma_0}$ at $m=N_0$ are dominated
by the first, $e^{-37.5}$, and sum to $5.18\cdot10^{-17}$.

\emph{(iii), transfer.}  Write $g:=\sqrt{m/W_G}\,G_{[m]}$ for the exact target direction of
Eq.~\eqref{eq:algorithm_119_ball_sample} and $\widehat g:=\kappa_{\mathrm{pr}}\widehat b=10^{-4}\widehat s\,\xi'_{[m]}$.
The Euclidean projection of $\widehat b$ onto a closed convex set maximizes
$F_{\widehat g}(x):=\langle\widehat g,x\rangle-\kappa_{\mathrm{pr}}\|x\|_2^2/2$ over it, as we use in the proof of
Lemma~\ref{lem:algorithm_centered_box} for $g$ and $b_g=g/\kappa_{\mathrm{pr}}$.

\emph{Coupling.}  Decompose
\[
\widehat g-g=\bigl(10^{-4}\widehat s-\sqrt{m/W'_G}\bigr)\xi'_{[m]}
+\sqrt{m/W'_G}\,(\xi'-G)_{[m]}
+\bigl(\sqrt{m/W'_G}-\sqrt{m/W_G}\bigr)G_{[m]} .
\]
The first coefficient lies in $[-2^{-f'},0]$ by (i), and $\|\xi'_{[m]}\|_2\le\sqrt{W'_G}\le\sqrt{2(m+2)}$.  The
second is at most $\sqrt2$ on the accepted event, and $\|(\xi'-G)_{[m]}\|_2\le\sqrt m\,h/2$.  For the third, we have
$|\sqrt{m/W'_G}-\sqrt{m/W_G}|=\sqrt m\,|\sqrt{W_G}-\sqrt{W'_G}|/\sqrt{W_GW'_G}\le\sqrt m\,\frac h2\sqrt{m+2}/\sqrt{W_GW'_G}$ by
the triangle inequality of (ii), and $\|G_{[m]}\|_2\le\sqrt{W_G}$, so the third term is at most
$\frac h2\sqrt m\sqrt{m+2}/\sqrt{W'_G}\le\frac h2\sqrt{2(m+2)}$.  Altogether we obtain
\[
\varepsilon_{\mathrm{cpl}}:=\|\widehat g-g\|_2\le C_{\mathrm{cpl}}\,h\sqrt m+2^{-f'}\sqrt{2(m+2)},
\qquad
C_{\mathrm{cpl}}:=\frac{\sqrt2}2\Bigl(1+\sqrt{\frac{m+2}m}\Bigr)\le1.414214,
\]
where we evaluate the constant at $m=N_0$, and it decreases in $m$.  At $h=2^{-40}$ and $f'\ge30$ the right side
is below $1.32\cdot10^{-9}\sqrt m$, and the Gaussian grid contributes $1.29\cdot10^{-12}\sqrt m$ of it.

\emph{The margins of the deterministic core.}  On the events $A_y(G)\le0.799m$ and $W_G\ge0.999m$ of that proof,
every $x=t(z-y)\in Q_y(t)$ has $\langle g,x\rangle\le t(\|g\|_1-\langle g,y\rangle)\le t\cdot0.799m/\sqrt{0.999}$.
Since $0.799/\sqrt{0.999}=0.7993998<0.7994$, we get $\max_{Q_y(t)}F_g\le(0.7994-2.0\cdot10^{-7})\,tm$.  Thus the
bound $0.7994tm$ of the proof has margin $2.0\cdot10^{-7}\,tm$.  On the event of
Lemma~\ref{lem:algorithm_119_supersets}, every $I\subseteq[m]$ with $|I|\le m/4$ has a point
$x_I\in L_I=D\cap H_I\cap\sqrt{2m}B_2^m$ with $\langle g,x_I\rangle>vm$, $v:=\sqrt{p_0s_{i,s}z}$, and hence
$F_g(x_I)>(v-10^{-4})m$.  The stage test $s_{i,s}p_0z>(10^{-4}+0.7994t)^2$ holds with certified margin at least
$2.8066981\cdot10^{-10}$ over the $47091$ finite stages (Lemma~\ref{lem:cubic_refined_certificate}) and
$9.7\cdot10^{-7}$ for the tail tuple.  Since $v\le\sqrt{p_0z}=0.302738$, we have
\[
v-(10^{-4}+0.7994t)=\frac{v^2-(10^{-4}+0.7994t)^2}{v+(10^{-4}+0.7994t)}
\ge\frac{2.8066981\cdot10^{-10}}{2\cdot0.302738}>4.63\cdot10^{-10}.
\]

\emph{Transfer of the strict inequality.}  For every $x$, $|F_{\widehat g}(x)-F_g(x)|=|\langle\widehat g-g,x\rangle|
\le\varepsilon_{\mathrm{cpl}}\|x\|_2$.  On $Q_y(t)$, $\|x\|_2\le2t\sqrt m$ because $|z_i-y_i|<2$, and
$\|x_I\|_2\le\sqrt{2m}$.  Hence, using $1/5<t<1/2$, we obtain
\[
\begin{aligned}
F_{\widehat g}(x_I)-\max_{Q_y(t)}F_{\widehat g}
&>\bigl[(v-10^{-4}-0.7994t)+2.0\cdot10^{-7}t\bigr]m-(\sqrt2+2t)\,\varepsilon_{\mathrm{cpl}}\sqrt m\\
&\ge\bigl[4.63\cdot10^{-10}+4.0\cdot10^{-8}\bigr]m-2.4143\,\varepsilon_{\mathrm{cpl}}\sqrt m,
\end{aligned}
\]
which is positive whenever $\varepsilon_{\mathrm{cpl}}\le1.67\cdot10^{-8}\sqrt m$.  Our coupling uses
$1.32\cdot10^{-9}\sqrt m$ of this.  The rest of the proof of Lemma~\ref{lem:algorithm_centered_box} is
deterministic and does not see the target.  Suppose $x_*$ had at most $m/4$ tight coordinates.  Discarding the
nontight box halfspaces would leave $x_*$ optimal for $F_{\widehat g}$ on a relaxed set that contains $D\cap H_I$
for $I$ the tight set, and hence contains $x_I$.  This contradicts $F_{\widehat g}(x_I)>F_{\widehat g}(x_*)$.  The
events we use are those of the exact $G$, whose probabilities the proof of Lemma~\ref{lem:algorithm_centered_box}
and the stage tests of Lemma~\ref{lem:cubic_refined_suffix} bound.  By (ii) the rejections lie inside them or
keep their exponent, so the bound $0.999$ including rejection is unchanged.  The affine update turns every tight
coordinate into a sign as in that lemma.
\end{proof}

Here, we have two remarks on the constants.  The grid $h=2^{-40}$ is finer than the margin requires: the threshold
$1.67\cdot10^{-8}\sqrt m$ is met by $h=2^{-27}$, with $C_{\mathrm{cpl}}h=1.05\cdot10^{-8}$, and missed by $h=2^{-26}$, with
$2.11\cdot10^{-8}$.  The choice $2^{-40}$ leaves a factor above $10^4$.  We apply the lemma to the stored
target $\widehat b$ itself, so we spend the accuracy $\rho$ of Eq.~\eqref{eq:algorithm_119_accuracy} entirely on
the solve and the radial correction.  In the proof of Lemma~\ref{lem:algorithm_certified_implementation} we allocate
$\rho/4$ to the scalar $\sqrt{m/W}/\kappa_{\mathrm{pr}}$, which would ask $f\ge\log_2n+78$, about $177$ bits at
$n=10^{30}$.  We do not use this allocation, and the scalar needs only $f'\ge30$.  The exact bound
$\|\widehat b\|_2^2\le10^8m$ is the target cap $c_b=10^8$ under which that proof invokes
Lemma~\ref{lem:appendix_fixed_target_projection}.  The quantity $q(0)=\|\widehat b\|_2^2/2\le5\cdot10^7m$ lives
in a $2W$-bit word (Lemma~\ref{lem:bit_state}(i)).

The uniform target is the target of every repair node.  Its law enters our analysis at exactly two places, the
width bound behind Eq.~\eqref{eq:cubic_refined_deficit} and the pass probability of the prefix test.  The lemma
below carries both across a coordinatewise shift of order $2^{-114}$.

\begin{lemma}[Dyadic uniform targets and the two acceptance tests]
\label{lem:bit_uniform_target}
Fix a repair node of Algorithm~\ref{alg:cubic_refined_repair} with $N_0\le m\le n$ padded coordinates and the
protocol constants $t_0\in\{2/5,3/5\}$, $D^{\mathrm{acc}}_r$, $e_{\mathrm{call}}$, $\varepsilon_{\mathrm{call}}$,
$\theta\in\{\theta_{\mathrm{acc}},\theta_r\}$, $\eta_{\mathrm{num}}$, $\delta_{\mathrm{fail}}$ of
Eq.~\eqref{eq:cubic_repair_parameters}, and put $\beta:=e_{\mathrm{call}}/D^{\mathrm{acc}}_r$.  Let $B_U:=114$,
$h_U:=2^{-B_U}$, and $f\ge B_U+10=124$.  Draw $U'_i$, $i\le m$, independent and uniform on the $2^{B_U+1}$
midpoints $\{\pm(2k+1)2^{-B_U-1}:0\le k<2^{B_U}\}$ of the cells of width $h_U$ in $[-1,1]$.  Each is drawn from
$B_U+1$ fair coins, a sign and $B_U$ magnitude bits.  Put
\[
b:=\operatorname{round}_{\mathsf u}(t_0U'/\beta)\in(\mathsf u\mathbb Z)^m,\qquad U'':=\beta b/t_0 .
\]
The prefix test, the deficit test, and the cardinality weights of the node use $U''$, and the projection is asked
for the target $b=t_0U''/\beta$.  Then the following hold.
\begin{enumerate}[label=(\roman*)]
\item \emph{Grid and caps.}
We have $|b_i|\le t_0D^{\mathrm{acc}}_r/e_{\mathrm{call}}$ for every $i$, and hence
$\|b\|_2^2\le m(t_0D^{\mathrm{acc}}_r/e_{\mathrm{call}})^2$, the target cap tested in
Eq.~\eqref{eq:cubic_global_solver_scales} and in Lemma~\ref{lem:cubic_refined_accuracy}.  At the certificate cap
$e\ge e_{86}/10^8$ of the dilations $\sigma\le10^4$ (Definition~\ref{def:cubic_global_policy}),
$|b_i|\le6.24\cdot10^{36}$, so $I\ge125$ holds every repair target.  The root needs $58$ bits, and auxiliary
calls need $88$.  The quantity $q(0)=\|b\|_2^2/2$ is exact.
\item \emph{Coupling.}
On one coin space there is $U$ uniform on $[-1,1]^m$ with $U'_i$ the midpoint of the
cell of $U_i$, so $|U'_i-U_i|\le h_U/2$ surely.  Moreover $|U''_i-U'_i|\le\beta\mathsf u/(2t_0)\le\mathsf u$.  Hence
$\|U''-U\|_\infty\le h_{\mathrm{tot}}/2$ with $h_{\mathrm{tot}}:=h_U+2\mathsf u\le h_U(1+2^{-9})<4.9\cdot10^{-35}$.
\item \emph{Deficit expectation.}
With $C:=(\widetilde{\mathcal D}/t_0)\cap[-1,1]^m$ and $y^*$ the regularized
maximizer of Eq.~\eqref{eq:cubic_projection_comparison} for $Z=U''$,
\[
\E[\mathcal U_{U''}(y^*)]\le e_{\mathrm{call}}n\bigl(1+1/(2D^{\mathrm{acc}}_r)\bigr)+mh_{\mathrm{tot}} ;
\]
equivalently, Eq.~\eqref{eq:cubic_acceptance_comparison} applies with $a_{\mathrm{acc}}$ replaced by
$a'_{\mathrm{acc}}:=a_{\mathrm{acc}}+h_{\mathrm{tot}}/e_{\mathrm{call}}$ and with $b_{\mathrm{acc}}$, $\eta_0=\eta_{\mathrm{num}}$,
and $\theta$ unchanged.
\item \emph{Prefix test.}
The $m$ inequalities $2m\sum_{i\le k}|U''|_{(i)}\ge(k-\varepsilon_{\mathrm{call}}m)_+^2$,
$k\le m$, are exact comparisons of rationals, and a passed test gives Eqs.~\eqref{eq:cubic_refined_prefix}
and~\eqref{eq:cubic_refined_positive_part} for $U''$ deterministically.  The test fails with probability at most
$m\exp(-2m(\varepsilon_{\mathrm{call}}-h_{\mathrm{tot}}/2)^2)$.  The printed budgets are $10^{-9}$ for auxiliary calls, with
$m\ge10^{20}$ and $\varepsilon_{\mathrm{call}}=6\cdot10^{-10}$.  They are $10^{-10}$ for the root, with $n>N_{\mathrm{root}}$ and
$\varepsilon_{\mathrm{call}}=\nu_{\mathrm{root}}$, and $10^{-10}$ for global helpers, by
Lemma~\ref{lem:cubic_global_prefix}.  They hold with the shift as long as $h_{\mathrm{tot}}\le4.4\cdot10^{-11}$,
$1.9\cdot10^{-12}$, and $4.2\cdot10^{-16}$ respectively.
\item \emph{Deficit test.}
For the returned point $x\in(\mathsf u\Gamma)^m$ and $y:=x/t_0$, the
statistic $\mathcal U_{U''}(y)=\sum_i|U''_i|(1-|y_i|)$ and the threshold $\theta e_{\mathrm{call}}n$ are rationals
over a common denominator of $O(f+\log n)$ bits, formed exactly and compared exactly.  An accepted $y$
satisfies the hypothesis $\mathcal U_{U''}(y)\le\theta e_{\mathrm{call}}n$ of
Lemma~\ref{lem:cubic_refined_potential} verbatim.  The conditional probability of a valid acceptance is at
least $1-a'_{\mathrm{acc}}/b_{\mathrm{acc}}-2\delta_{\mathrm{fail}}$.  At $B_U=114$ and $f=124$ this is
$4.7995\cdot10^{-9}$ at the root and $4.671\cdot10^{-9}$ for a global helper on the smallest exponent of the
selected policy.  That exponent is $e_{86}/\sigma_{\max}^2$, where $\sigma_{\max}=5065.12$ is the largest dilation of the selected
policy (proof of Lemma~\ref{lem:cubic_global_certificate}).  At the same parameters the probability is $4.298\cdot10^{-9}$ at the certificate cap $e\ge e_{86}/10^8$.
All of these are above the floor $1/250000000=4\cdot10^{-9}$ of Lemma~\ref{lem:cubic_global_complete}.  For auxiliary calls
the probability is at least $4.799\cdot10^{-8}$, above the floor $1/21000000$ of Lemma~\ref{lem:cubic_refined_accuracy}.
\end{enumerate}
\end{lemma}

\begin{proof}
\emph{(i), grid and caps.}  Every midpoint has $|U'_i|\le1-2^{-B_U-1}$, so
$|b_i|\le(t_0/\beta)(1-2^{-B_U-1})+\mathsf u/2\le t_0/\beta$, because $t_0/\beta\ge1$ and
$2^{-B_U-1}\ge\mathsf u/2$ for $f\ge B_U+1$.  Here $t_0/\beta=t_0D^{\mathrm{acc}}_r/e_{\mathrm{call}}\ge1$, since
$D^{\mathrm{acc}}_rt_0\ge2459\cdot2/5>983$ while every exponent of the registry, and $v(e_{\mathrm{loc}})$, is below
$2\cdot10^{-3}$.  The largest exponent is $e_{80}$ of Definition~\ref{def:cubic_refined_trace_rows}.  The value is
$t_0D^{\mathrm{acc}}_r/e_{\mathrm{call}}=5.79\cdot10^{16}$ at the root, where $e_{\mathrm{call}}=v(e_{\mathrm{loc}})$, $t_0=3/5$, and
$D^{\mathrm{acc}}_r=10^{12}$.  For auxiliary calls the value is $4.16\cdot10^{25}$, where $e\ge e_{86}$, $t_0=2/5$, and
$D^{\mathrm{acc}}_r\le10^9$.  At the cap it is $6.24\cdot10^{36}$, where $e\ge e_{86}/10^8$, $t_0=3/5$, and
$D^{\mathrm{acc}}_r=10^{12}$.  The integer bits are $\lceil\log_2\rceil$ of these plus two.  The quantity $q(0)$
is a sum of squares of multiples of $\mathsf u$.

\emph{(ii), coupling.}  Let $U_i$ be uniform on $[-1,1]$ and let $U'_i$ be the midpoint of the half-open cell of
width $h_U$ that contains $U_i$.  The $2^{B_U+1}$ cells have equal length, so $U'_i$ is uniform on the midpoints,
which is the law we draw from $B_U+1$ fair coins.  Also $|U'_i-U_i|\le h_U/2$.  Rounding $t_0U'_i/\beta$ to
$\mathsf u\mathbb Z$ moves it by at most $\mathsf u/2$, and $U''_i-U'_i=(\beta/t_0)(b_i-t_0U'_i/\beta)$ with
$\beta/t_0\le1$ by (i).

\emph{(iii), deficit expectation.}  Smoothing gives $(t_0/s_{\mathrm{alg}})K(r\sqrt m)=(1-10^{-6})K(R)\subseteq\widetilde{\mathcal D}$
for the node's radius $R=t_0r\sqrt m/(s_{\mathrm{alg}}(1-10^{-6}))$.  Hence we can apply Lemma~\ref{lem:cubic_refined_width}
to $C$ as in the proof of Lemma~\ref{lem:cubic_refined_projection}, and obtain
$\E[h_C(U)]\ge m/2-ne_{\mathrm{call}}$, where
$e_{\mathrm{call}}=e$ for a helper and $e_{\mathrm{call}}=v(e_{\mathrm{loc}})$ at the root.  Also $\E\|U\|_1=m/2$.  Since
$C\subseteq[-1,1]^m$, we have $|h_C(U'')-h_C(U)|\le\sup_{y\in C}|\langle U''-U,y\rangle|\le\|U''-U\|_1\le mh_{\mathrm{tot}}/2$
and $\bigl|\|U''\|_1-\|U\|_1\bigr|\le mh_{\mathrm{tot}}/2$, so
$\E[\|U''\|_1-h_C(U'')]\le ne_{\mathrm{call}}+mh_{\mathrm{tot}}=:d_{U''}$.  The first bound of
Eq.~\eqref{eq:cubic_projection_comparison} gives
$\E[\mathcal U_{U''}(y^*)]\le d_{U''}+\beta m/2\le e_{\mathrm{call}}n(1+1/(2D^{\mathrm{acc}}_r))+mh_{\mathrm{tot}}$, as
$\beta m/2\le e_{\mathrm{call}}n/(2D^{\mathrm{acc}}_r)$.  Dividing by $d_0=e_{\mathrm{call}}n$ and using $m\le n$, we get
$a'_{\mathrm{acc}}$.  The physical error $\|x-t_0y^*\|_2\le t_0e_{\mathrm{call}}\eta_{\mathrm{num}}$ of the solve enters
through the second bound of Eq.~\eqref{eq:cubic_projection_comparison}.  That bound gives
$|\mathcal U_{U''}(y)-\mathcal U_{U''}(y^*)|\le\|U''\|_2\,e_{\mathrm{call}}\eta_{\mathrm{num}}
\le\sqrt m\,e_{\mathrm{call}}\eta_{\mathrm{num}}\le e_{\mathrm{call}}n\eta_{\mathrm{num}}=\eta_0d_0$, as $|U''_i|\le1$ by (i).
Hence $\eta_0=\eta_{\mathrm{num}}$ is unchanged, with the factor $n/\sqrt m\ge\sqrt n$ to spare, and
$b_{\mathrm{acc}}+\eta_0\le\theta$ is the same inequality as before.

\emph{(iv), prefix test.}  The numbers $|U''_i|=(\beta/t_0)|b_i|$ are rationals over the denominator of
$\beta/t_0$ times $2^f$.  Here $e_{\mathrm{call}}$ is an exact rational of $O(1)$ bits, a registry exponent of
Definitions~\ref{def:cubic_refined_old_rows}, \ref{def:cubic_refined_new_rows}, \ref{def:cubic_centered_root},
and~\ref{def:cubic_refined_trace_rows}, whose largest denominator, Eq.~\eqref{eq:cubic_refined_root_exponent}, has $50$
digits.  Otherwise it is
$e_j10^{24}/u_\sigma^2$ for a dilated row with $\sigma=u_\sigma/10^{12}$, or the polynomial $v(e_{\mathrm{loc}})$ of
Eq.~\eqref{eq:cubic_refined_entropy_budget} at $e_{\mathrm{loc}}=10356641/10^{12}$.  The tolerance
$\varepsilon_{\mathrm{call}}$ is $6\cdot10^{-10}$, $3\cdot10^{-12}$, or $\varepsilon_-/\ell(q)$ with
$\ell(q)\in2^{-192}\mathbb Z$ (Lemma~\ref{lem:cubic_global_prefix}).  Sorting and the $m$ comparisons after
cross-multiplication are integer operations on $O(f+\log n)$-bit numbers, $O(m\log m)$ of them.  In the proof of
Lemma~\ref{lem:cubic_refined_projection} we derive Eq.~\eqref{eq:cubic_refined_prefix} for every subset from the
$m$ inequalities for the sorted values and Eq.~\eqref{eq:cubic_refined_positive_part} by maximizing the same
quadratic bound over $k$.  Both steps use only the inequalities that the test checks, now for $U''$.  For the pass
probability we use $\bigl||U''_i|-|U_i|\bigr|\le h_{\mathrm{tot}}/2$, so the empirical distribution functions $F''_m$ of
$(|U''_i|)_{i\le m}$ and $F_m$ of $(|U_i|)_{i\le m}$ satisfy $F''_m(t)\le F_m(t+h_{\mathrm{tot}}/2)$.  If the test fails at
some $k$, the identity $\sum_{i\le k}|U''|_{(i)}=\int_0^\infty(k-mF''_m(t))_+\,\d t$ of that proof forces
$F''_m(t)>t+\varepsilon_{\mathrm{call}}$ for some $t$, hence $F_m(s)>s+\varepsilon_{\mathrm{call}}-h_{\mathrm{tot}}/2$ at
$s=t+h_{\mathrm{tot}}/2$.  With Hoeffding's bound
and the union over the $m$ order statistics, exactly as there, we bound this event by
$m\exp(-2m(\varepsilon_{\mathrm{call}}-h_{\mathrm{tot}}/2)^2)$.  The three tolerances are where this bound reaches the
printed budget.  Unshifted, the bound is $10^{20}e^{-72}$, $10^{25}e^{-180}$, and $10^{30}e^{-98}$, the last with the relative
shift $0.0306$ of $\varepsilon_-$ tolerated.  The shift $h_{\mathrm{tot}}<4.9\cdot10^{-35}$ is far below each.

\emph{(v), deficit test.}  The point $x$ lies on $\mathsf u\Gamma$ and $t_0\in\{2/5,3/5\}$.  Hence
$1-|y_i|=(t_0-|x_i|)/t_0$ and $|U''_i|$ are rationals over denominators built from $2^{3f}$, which collects $2^f$ from
$b\in(\mathsf u\mathbb Z)^m$ in $U''$ and $2^{2f}$ from $x$, together with $10^{10}$, $t_0$,
$D^{\mathrm{acc}}_r$, and the denominator of $e_{\mathrm{call}}$.  We sum the $m$ products exactly, and the threshold
$\theta e_{\mathrm{call}}n$ is a rational of the same kind, so the comparison is an integer comparison.  An
accepted $y$ therefore satisfies the displayed inequality itself, which is what
Lemma~\ref{lem:cubic_refined_potential} assumes.  For the acceptance probability, we apply Markov's inequality and the
union bound of Eq.~\eqref{eq:cubic_acceptance_comparison} with $a'_{\mathrm{acc}}$ from (iii) and get
$1-a'_{\mathrm{acc}}/b_{\mathrm{acc}}-\delta_{\mathrm{test}}-\delta_{\mathrm{num}}$ with $\delta_{\mathrm{test}},\delta_{\mathrm{num}}\le\delta_{\mathrm{fail}}$.
For global helpers and the root, $b_{\mathrm{acc}}=1+5\cdot10^{-9}$ and $\delta_{\mathrm{fail}}=10^{-10}$, and the exact uniform
gives $1599833333/333333335000000000=4.7995\cdot10^{-9}$ (proof of Lemma~\ref{lem:cubic_global_complete}), a
slack of $7.995\cdot10^{-10}$ over the floor.  The shift costs $h_{\mathrm{tot}}/(e_{\mathrm{call}}b_{\mathrm{acc}})$, which is
$2^{-114}/e_{\mathrm{call}}$ up to the factor $1+2^{-9}$.  This cost is negligible at the root, $1.29\cdot10^{-10}$ at
$e=e_{86}/\sigma_{\max}^2=3.75\cdot10^{-25}$, and $5.0\cdot10^{-10}$ at $e=e_{86}/10^8=9.62\cdot10^{-26}$, giving
the three values.  For auxiliary calls, $b_{\mathrm{acc}}=a_{\mathrm{acc}}+5\cdot10^{-8}$, $\delta_{\mathrm{fail}}=10^{-9}$,
$e\ge e_{86}$, and $D^{\mathrm{acc}}_r\in\{10^9,15000,8000,4400,2459\}$ give at least $4.799\cdot10^{-8}$ against
$1/21000000=4.762\cdot10^{-8}$.
\end{proof}

The margin $5\cdot10^{-9}$ of $b_{\mathrm{acc}}$ is what fixes the two constants of the lemma.  At the cap the exact
uniform has $7.995\cdot10^{-10}$ of success probability to spare, and each of $h_U=2^{-114}$ and
$2\mathsf u=2^{-114}$ (at $f=115$) costs $5.0\cdot10^{-10}$ of it.  With $f=115$ the success at the cap is
$3.798\cdot10^{-9}$, below the floor $4\cdot10^{-9}$, and the least $f$ that passes is $116$ ($4.049\cdot10^{-9}$).
The rule $f\ge B_U+10$ we adopt here makes the $\mathsf u$ term invisible ($4.298\cdot10^{-9}$ at $f=124$).
Likewise $B_U=112$ suffices at the selected policy, with $4.29\cdot10^{-9}$, and fails at the cap, with
$2.80\cdot10^{-9}$.  This is why we take $B_U=114$ in the input-independent statement.  Any retuning of
$b_{\mathrm{acc}}$ must revisit these two numbers.  Nothing else in this section depends on them, since the word of
Theorem~\ref{thm:bit_complexity} has thousands of bits.

\begin{lemma}[Exact binary repair and the accumulation of $y$]
\label{lem:bit_repair}
Fix a repair node of Algorithm~\ref{alg:cubic_refined_repair} with $m$ padded coordinates and, for (iv), a suffix
stage with scale $t_s$.  The notation is that of (F14) and (F15).
\begin{enumerate}[label=(\roman*)]
\item \emph{Masses and levels.}
In an accepted attempt of Algorithm~\ref{alg:cubic_refined_repair} with
returned point $x\in(\mathsf u\Gamma)^m$ and $y:=x/t_0$, the signs $s_i:=\operatorname{sgn}(y_i)$,
with $\operatorname{sgn}0:=1$, and the masses $a_i:=2^{-B}\lfloor2^B(t_0-|x_i|)/(2t_0)\rfloor$,
$B:=\lceil\log_2n\rceil+45$, are exact.  The masses take one integer division per coordinate, on numerators of at most
$2f+B+40$ bits.  The truncation lowers each mass by less than $2^{-B}$, the charge $2m2^{-B}\le2^{-44}<10^{-12}\sqrt n$
paid in the proof of Lemma~\ref{lem:cubic_refined_potential}.  For $j=B,\ldots,1$ the support $S_j$ of odd
numerators on the $2^{-j}$ grid is an integer parity test.  The update $a_i\gets a_i+2^{-j}\varepsilon_j\sigma_i$
on $S_j$ is an exact grid operation that stays in $[0,1]$ and lands on the $2^{-(j-1)}$ grid.  The final
$z_i:=s_i(1-2a_i)$ lie in $\{-1,1\}$.  The masses occupy $B<f$ fractional bits of one word.
\item \emph{Orientations.}
The potential rule evaluates $\sum_iF(a_i^{\pm})$ for the two candidate
orientations, with the tent potential $F$ of Lemma~\ref{lem:cubic_refined_potential}, which is not the objective
of the solve.  Here
\[
F(a)=\sum_{j'<J_{\mathrm{cut}}}2^{1-j'}B_{j'}\operatorname{dist}(2^{j'}a,2\mathbb Z),\qquad
B_{j'}=b_{j'}/10^{12},\quad b_{j'}\in\mathbb Z,\quad b_{j'}\le10^{30},
\]
and the rule forms the two values as rationals over the denominator $10^{12}2^j$ with numerators of at most
$\log_2n+B+101$ bits ($346$ at $n=10^{30}$), and compares them exactly.  The cardinality rule
$\varepsilon_j:=-\operatorname{sgn}\bigl(\sum_{i\in S_j}|U''_i|\sigma_i\bigr)$ is the exact sign of a rational.  Both
are the choices analyzed in Lemma~\ref{lem:cubic_refined_potential} and Definition~\ref{def:cubic_refined_profiles}.
\item \emph{Helpers.}
The helper at level $j$ receives the family $(s_i\widetilde A_i)_{i\in S_j}$, sign flips of
$b_{\mathrm{in}}$-bit matrices, and returns exact signs $\sigma\in\{-1,1\}^{S_j}$ or \textup{\textsc{Fail}}.  The parent uses
$\sigma$ only in (ii) and never forms a norm of the child's signed sum, so no rounding error propagates between the
levels of the fourteen-level tree.  The accounting of Eq.~\eqref{eq:cubic_complete_repair_cost} and
Lemma~\ref{lem:cubic_global_line} is therefore a sum of certified bounds as in the real model.  The child sizes of a
cardinality action, $|S_j|/n\le\varepsilon_{\mathrm{call}}q+\sqrt{\theta_re_{\mathrm{call}}q\,2^j}$, follow from the prefix and deficit
tests passed exactly on $U''$ (Lemma~\ref{lem:bit_uniform_target}(iv),(v)).
\item \emph{Accumulation in a suffix stage.}
The update of $y$ after a snapped increment $\overline x\in(\mathsf u\Gamma)^m$
at scale $t_s=v/10^{10}$, $v\in\mathbb Z$, is Lemma~\ref{lem:bit_enclosures}(iv).  It sets $y_i:=\pm1$ for a snapped
coordinate and $y_i:=\operatorname{round}^{\to0}_{\mathsf u}(y_i+\overline x_i/t_s)$ otherwise, where
$\overline x_i/t_s=10^{10}\overline x_i/v$ is one integer division.  The update keeps $y\in(\mathsf u\mathbb Z)^m\cap[-1,1]^m$, makes
the new active set exact, and gives the increment bound $\widetilde R/t_s$ under
$\mathsf u\le\tfrac32s_{\mathrm{sm}}\sqrt{15/m}$.  This condition holds for $f\ge\frac12\lceil\log_2n\rceil+19$, which is $60$ at
$n=10^{25}$ and $69$ at $n=10^{30}$.
\end{enumerate}
\end{lemma}

\begin{proof}
\emph{(i).}  With $x_i=k_i\mathsf u^2/10^{10}$, $k_i\in\mathbb Z$, and $10^{10}t_0\in\{4\cdot10^9,6\cdot10^9\}$,
the rational $(t_0-|x_i|)/(2t_0)=(10^{10}t_02^{2f}-|k_i|)/(2\cdot10^{10}t_02^{2f})$ has a numerator of at most
$2f+35$ bits, and we multiply by $2^B$ before the integer division, which adds $B$ bits.  The masses are then multiples
of $2^{-B}$ in $[0,1/2]$, and we apply the grid argument of Lemma~\ref{lem:algorithm_21_rounding} verbatim.  An odd
numerator at level $j$ is strictly between the boundary numerators $0$ and $2^j$, so both moves $\pm2^{-j}$
stay in $[0,1]$ and give an even numerator, which is a point of the $2^{-(j-1)}$ grid.  We leave even numerators
untouched.  After level $1$ every mass is in $\{0,1\}$.  The truncation charge is the one we pay at the start of the proof
of Lemma~\ref{lem:cubic_refined_potential}, identical in the real model, because that proof already works on the
$2^{-B}$ grid.

\emph{(ii).}  For $a_i\in2^{-j}\mathbb Z$ and $j'<j$, $\operatorname{dist}(2^{j'}a_i,2\mathbb Z)$ is a rational with
denominator $2^{j-j'}$, so $\sum_iF(a_i)$ is a rational with denominator $10^{12}2^j$ and numerator below
$2m\cdot10^{30}\cdot2^j$.  Hence we decide the orientation by two exact evaluations, or by one of their difference.  The
cardinality sum has denominator that of $U''$ times $2^j$.  The minimizing orientation drops the potential by
at least $2^{1-j}B_j|S_j|$ at coarse levels and does not raise it at fine levels.  This is the affine-tent argument of the
proof of Lemma~\ref{lem:cubic_refined_potential}, which needs only that we evaluate the two candidates
exactly.

\emph{(iii).}  A sign flip of a matrix with entries in $2^{-b_{\mathrm{in}}}\mathbb Z$ is exact.  In
Lemma~\ref{lem:cubic_refined_potential} we bound the discrepancy of a node by the children's certified
discrepancies through the increment identity of Lemma~\ref{lem:algorithm_21_rounding}, an analytic identity in
which no computed norm appears.  A child that violates its certified bound is a failure of the child, which we charge to
its own budget in the induction of Lemma~\ref{lem:cubic_refined_complete}.  We obtain the child-size bound by the
derivation after Definition~\ref{def:cubic_refined_profiles}: $a_i\ge2^{-j}$ on $S_j$ and
$\sum_i|U''_i|a_i\le\theta_re_{\mathrm{call}}n/2$ from the deficit test give
$\sum_{i\in S_j}|U''_i|\le2^{j-1}\theta_re_{\mathrm{call}}n$, and Eq.~\eqref{eq:cubic_refined_prefix} for $U''$
gives $(|S_j|-\varepsilon_{\mathrm{call}}m)_+^2/(2m)\le2^{j-1}\theta_re_{\mathrm{call}}n$.

\emph{(iv).}  This is Lemma~\ref{lem:bit_enclosures}(iv) and its proof, to which we add two facts.  An unsnapped
coordinate has $|y_i+\overline x_i/t_s|<1$ strictly, because a value $\pm1$ would put $\overline x_i$ on a facet, within the
snapping threshold.  Hence rounding toward zero keeps $|y'_i|<1$, and the new active set is the set of unsnapped
coordinates.  Numerically the condition $\mathsf u\le\frac32s_{\mathrm{sm}}\sqrt{15/m}$ reads
$9.19\cdot10^{-19}$ at $m=10^{25}$ and $2.91\cdot10^{-21}$ at $m=10^{30}$, which gives $f\ge60$ and $f\ge69$.
\end{proof}

The scale $t_s$ is not dyadic and changes with the stage, so we cannot add the increment $\overline x/t_s$ to $y$
exactly, and (F14) records the rounding toward zero and the slack it uses.

The fallbacks are fair-sign trials, which we certify by a norm bound.  In the real model we certify with the
diagonalization-based certificate of Section~\ref{sec:algorithm} at an additive $0.03\sqrt m$, as in the proof of Lemma~\ref{lem:algorithm_certified_implementation},
a real-arithmetic tool (Remark~\ref{rem:bit_section6}).  In fixed point our certificate is Lemma~\ref{lem:bit_normcert}
with a multiplicative factor, and the constants leave little headroom: at $11.902$ the admissible relative error is
$2.586\cdot10^{-3}$.

\begin{lemma}[Fair-sign fallbacks with the norm certificate]
\label{lem:bit_fallback}
Let $1\le m\le n$, let $\widetilde A_1,\ldots,\widetilde A_m$ be symmetric $b_{\mathrm{in}}$-bit dyadic contractions of
order $n$, as in Lemma~\ref{lem:bit_input}, and let $c$ and the dimension range be one of
\begin{enumerate}[label=(\alph*)]
\item $c=10.889$, $m=n\le N_{\mathrm{root}}$ (Algorithm~\ref{alg:main_cubic_small_constant});
\item $c=11.902$, $N_{\mathrm{root}}<n\le N_*$ (Algorithm~\ref{alg:cubic_global_select} at layer $9$);
\item $c=13$, $m<N_0$ and $n<N_0/q_{\min}=9.58\cdot10^{32}$ (Algorithm~\ref{alg:cubic_global_action});
\item $c=13$, a band $448\le i\le713$ with $m=\lfloor nw_i\rfloor<10^{20}$, so $n<10^{20}/w_{i+1}\le8.05\cdot10^{35}$ (Algorithm~\ref{alg:cubic_refined_repair} and Lemma~\ref{lem:cubic_refined_accuracy}).
\end{enumerate}
Put $\varpi:=2^{-9}$ and suppose $f\ge2\lceil\log_2n\rceil+26$ and $I\ge\lceil\log_2(2n)\rceil+4$.  A trial
draws $\varepsilon\in\{-1,1\}^m$ from $m$ fair coins and forms $E:=\sum_i\varepsilon_i\widetilde A_i$ exactly.  This is a
matrix of words, with entries in $2^{-b_{\mathrm{in}}}\mathbb Z\subseteq\mathsf u\mathbb Z$ of modulus at most $m$.  The trial
computes $\|E\|_F^2$ exactly and \emph{rejects if $\|E\|_F^2>c^2mn$}.  Otherwise it sets $U:=1$ if $\|E\|_F^2\le1$, and
$U:=$ the certificate of Lemma~\ref{lem:bit_normcert} for $E$ with $n':=n$ and this $\varpi$ if not.  It accepts
if and only if $U^2\le c^2m$, which is an exact comparison of a multiple of $\mathsf u^2$ with a rational over $10^6$.  Then:
\begin{enumerate}[label=(\roman*)]
\item on acceptance, $\|\sum_i\varepsilon_i\widetilde A_i\|\le c\sqrt m$ deterministically;
\item each trial accepts with probability at least $1/2$;
\item with $T_f:=\lceil\log_2(1/p_c)\rceil$ independent trials the procedure returns \textup{\textsc{Fail}} with probability at most $p_c$ and has no other failure mode: no numerical budget is consumed;
\item a trial spends $O(mn^2(b_{\mathrm{in}}+\log_2m))$ bit operations to form $E$.  It then performs $J_{\mathrm{sq}}$
exact dense products, each at the cost $O(n^{\omega+2\eta})\,M(2W+c_\eta\lceil\log_2(2n)\rceil)$ of
Lemma~\ref{lem:bit_fastmm}, with the $J_{\mathrm{sq}}$ of Lemma~\ref{lem:bit_normcert} at $\varpi=2^{-9}$,
which is $15$ at $n=10^{25}$ and $16$ for $10^{30}\le n\le8.05\cdot10^{35}$.  It then performs $J_{\mathrm{sq}}+1$ exact
Frobenius norms, each an exact sum of squares and one integer square root, and the pre-test's sum of squares is
reused as the certificate's $\widehat N_0=\lceil\|E\|_F\rceil_{\mathsf u}$.  It also performs $J_{\mathrm{sq}}+2$
certified enclosures, $J_{\mathrm{sq}}+1$ of a logarithm and one of an exponential, to $O(W)$ bits.
\end{enumerate}
\end{lemma}

\begin{proof}
\emph{Words.}  We have $b_{\mathrm{in}}=2\lceil\log_2n\rceil+10\le f$ and $m\le n<2^{I-1}$, so $E$ is a matrix of
$W$-bit words and Lemma~\ref{lem:bit_normcert} starts from $\widehat X_0=E$ without an initial rounding.  The
quantity $\|E\|_F^2$ is a sum of $n^2$ squares, which we hold exactly in a temporary of $2W+2\lceil\log_2n\rceil$ bits, and $c^2m$,
$c^2mn$ are rationals over $10^6$ ($c^2\in\{118.570321,141.657604,169\}$).  The word condition
$10^5n^2\mathsf u\le\varpi$ of Lemma~\ref{lem:bit_normcert} is $f\ge2\log_2n+\log_2(10^5\cdot2^9)=2\log_2n+25.61$,
so $f\ge2\lceil\log_2n\rceil+26$ suffices.  The threshold takes the values $191.7$, $224.9$, $244.7$, and $264.2$
at $n=10^{25}$, $10^{30}$, $9.58\cdot10^{32}$, and $8.05\cdot10^{35}$ respectively.  The integer bits we need are dictated
by the certificate's first scalar $\widehat N_0=\lceil\|E\|_F\rceil_{\mathsf u}$.  After the pre-test
$\|E\|_F\le c\sqrt{mn}\le13n$, so $\widehat N_0\le13n+1<16n\le2^{I-1}$ for $I\ge\lceil\log_2(2n)\rceil+4$.  With
$+3$ this fails at $n=10^{30}$, where $13n+1=1.30\cdot10^{31}$ against $2^{I-1}=1.01\cdot10^{31}$.  The later
$\widehat N_j$ are at most $2$, and $U\le(1+\varpi)\|E\|\le(1+\varpi)13n<16n$.  Without the pre-test $\|E\|_F$ can
a priori be as large as $\sqrt n\|E\|\le n^{3/2}$, which is $10^{45}$ at $n=10^{30}$, and a saturated
$\widehat N_0$ would break the lower-bound chain of Lemma~\ref{lem:bit_normcert}.  The pre-test is what makes
$I=O(\log n)$ with this constant.

\emph{(i).}  If $\|E\|_F^2>c^2mn$, then $\|E\|\ge\|E\|_F/\sqrt n>c\sqrt m$ and we rightly reject the trial.  If
$\|E\|_F\le1$, then $\|E\|\le\|E\|_F\le1=U\le c\sqrt m$.  Otherwise $E$ is a symmetric $n\times n$ fixed-point
matrix with $\|E\|_F>1$, $n'=n\le d_0$, $\varpi\in(0,1/2]$, and $10^5n^2\mathsf u\le\varpi$, so
Lemma~\ref{lem:bit_normcert} gives $\|E\|\le U\le(1+\varpi)\|E\|$, and $U^2\le c^2m$ yields $\|E\|\le c\sqrt m$.

\emph{(ii).}  The $\varepsilon_i$ are exact independent fair signs (Lemma~\ref{lem:bit_probes}) and
$\sum_i\widetilde A_i^2\preceq mI_n$, so Tropp's matrix Hoeffding inequality~\cite[Theorem~4.1]{t12} gives $\|E\|\le\sqrt{2m\log(4n)}$ with
probability at least $1/2$.  Put $s:=\sqrt{2\log(4n)}$.  On this event $\|E\|_F\le\sqrt n\|E\|\le s\sqrt{mn}<c\sqrt{mn}$,
so the pre-test passes, and $U\le(1+\varpi)s\sqrt m$.  In the four ranges $s$ is at most $10.858$ at $N_{\mathrm{root}}$,
$11.871$ at $N_*$, $12.436$ at $N_0/q_{\min}$, and $12.966$ at $10^{20}/w_{714}$.  Hence
$(1+\varpi)s=s+\varpi s\le s+12.967/512<s+0.03$, and our own tests then give $(1+\varpi)s<s+0.03<c$.
The tests read $10.888261526228<10.889$ in the proof of Lemma~\ref{lem:cubic_global_complete} and $12.466093536159<13$ in the proof of
Lemma~\ref{lem:cubic_global_line}, as printed there.  We evaluate the tests of the proofs of
Lemma~\ref{lem:algorithm_certified_implementation} and Lemma~\ref{lem:cubic_refined_accuracy} and obtain $11.901297077<11.902$
and $12.996218<13$ at band $713$.  Band $714$ has $13.000174$ and we exclude it there.  Hence $U^2\le c^2m$
and the trial accepts.  The slacks $c-(1+\varpi)s$ are $9.5\cdot10^{-3}$, $7.5\cdot10^{-3}$, $0.54$, and
$8.5\cdot10^{-3}$.  The largest admissible $\varpi$ at $11.902$ is $2.586\cdot10^{-3}$, of which $2^{-9}=1.953\cdot10^{-3}$
uses three quarters.  The dimension range of (c), $n<N_0/q_{\min}$, holds because every new piece of
Definition~\ref{def:cubic_global_policy} has $L\ge x_{1100}$, so $q\ge q_{\min}$, and we do not test it at run time.

\emph{(iii).}  The trials are independent and the certificate is a deterministic computation on exact data, so the
only failure is exhaustion, of probability $2^{-T_f}\le p_c$.  \emph{(iv).}  We form $E$ in $mn^2$ additions of
numbers of $b_{\mathrm{in}}+\lceil\log_2m\rceil$ bits.  The rest is the count of Lemma~\ref{lem:bit_normcert} with the
pre-test added, and we evaluate $J_{\mathrm{sq}}$ at $\varpi=2^{-9}$.
\end{proof}

We thus do not need the additive allowance $0.03\sqrt m$ of the real model in fixed point, and we leave the constants
$c$ unchanged, so nothing in the certificate of Lemma~\ref{lem:cubic_global_certificate} moves.  The budget
``total certification and exhaustion budget at most $p$'' of the real model becomes exhaustion only.  A dyadic
$\varpi$ keeps $\widehat\Lambda+\varpi/16$ on a grid.  If we lowered $11.902$ toward $s=11.8713$ at $N_*$, this would force a
smaller $\varpi$ and raise $f$ by the base-two logarithm of the ratio.  The fallbacks are also where we absorb the input
rounding of Lemma~\ref{lem:bit_input} for small $n$.

\begin{corollary}[Output bound for the original matrices]
\label{cor:bit_output_bound}
Let $\varepsilon\in\{-1,1\}^n$ be a signing of the rounded inputs $\widetilde A_i$ of Lemma~\ref{lem:bit_input}
with $\|\sum_i\varepsilon_i\widetilde A_i\|\le c_n\sqrt n$.  Here $c_n:=10.889$ if $n\le N_{\mathrm{root}}$, which is the
fallback branch of Algorithm~\ref{alg:main_cubic_small_constant} with Lemma~\ref{lem:bit_fallback}(i).  We set
$c_n:=11.503394001765$ if $n>N_{\mathrm{root}}$, by Lemma~\ref{lem:cubic_global_complete}.  Then
$\|\sum_i\varepsilon_iA_i\|<11.504\sqrt n$ for every $n\ge1$.
\end{corollary}

\begin{proof}
Lemma~\ref{lem:bit_input} adds $2^{-9}$ to the certified bound.  For $n\le N_{\mathrm{root}}$ the sum
$10.889\sqrt n+2^{-9}$ is at most $11.503394001765\sqrt n$ as soon as $\sqrt n$ exceeds the ratio of $2^{-9}$
to $11.503394001765-10.889$, which is $0.0032$.  So this holds for every $n\ge1$.  For $n>N_{\mathrm{root}}$ the
sum $11.503394001765\sqrt n+2^{-9}$ is below $11.504\sqrt n$ if and only if $\sqrt n$ exceeds the ratio of
$2^{-9}$ to $11.504-11.503394001765=0.000605998235$, which is equivalent to $n>10.39$.  So this holds for every
$n\ge11$ and in particular on this branch.  We need both cases: the second inequality alone fails for
$n\le10$.
\end{proof}

\begin{remark}[Routing of declared failures]
\label{rem:bit_routing}
A projection has two failure classes.  Two events are \emph{silent},
in that the solve can return a wrong point without noticing.  The events are a wrong guard, of probability
$p_0/(100TJR_{\mathrm{rep}})$ each (Lemma~\ref{lem:bit_guard}), which can let an infeasible point through, and a
phase in which all $R_{\mathrm{rep}}$ candidates miss the contraction of Lemma~\ref{lem:bit_lyapunov}, of probability
$0.1^{R_{\mathrm{rep}}}\le p_0/(2\widetilde J)$.  Here the refresh cap of Lemma~\ref{lem:bit_tracker}(v) is included in each
candidate's $0.1$.  Together they cost at most $p_0$ per solve, by Step~8 of the proof of
Lemma~\ref{lem:bit_projection}.  A sampler cap, the one-redraw cap, and a saturation report are \emph{declared}: the procedure halts and
reports.  The sampler cap is that of Lemmas~\ref{lem:bit_probes} and~\ref{lem:bit_gaussian}(i).  The redraw
cap is that of Lemma~\ref{lem:bit_gaussian}(iii).
The declared class is charged to the acceptance allowance of the attempt, not to the $p$-dependent numerical
budget, and (F18) records it.  Inside an attempt of Algorithm~\ref{alg:cubic_refined_repair} a declared failure
of the projection counts as a failed test, so the loop continues to the next attempt exactly as it does when the
prefix or the deficit test fails.  Inside a suffix trial of Algorithm~\ref{alg:algorithm_suffix_signing} it counts
as a failed trial.  Its probability is at most $N_{\mathrm{sol}}\widetilde JR_{\mathrm{rep}}(2\widetilde T+1)\mathsf u^4$
per attempt (Lemma~\ref{lem:bit_projection}), which is below $2^{-40}$ for $f\ge f_B$.  The condition
$f\ge40+\lceil\log_2(2n)\rceil$ of Lemma~\ref{lem:bit_state} alone would not do.  The phase length $\widetilde T=16m\widetilde L\le640a^2c_bm$ carries the target cap, and the count is about
$2^{548}$ at the literal cap $c_b=c_B$ and $n=10^{30}$.  There $N_{\mathrm{sol}}=661$, $\widetilde J=3452$, and
$R_{\mathrm{rep}}=16$ at a solve budget $p_0=10^{-12}$, and $\widetilde T=1.08\cdot10^{157}$.  The count is about $2^{552}$ with
$R_{\mathrm{rep}}=198$ at the leaf budget $2^{-634}$ named below.  Here $\mathsf u^4=2^{-564}$ at $f=141$.  Fact (B6) in
Step~1 of Section~\ref{subsec:bit_proof} bounds the count by $2^{f_B/2}$ in general.  The count depends on $p$ through $R_{\mathrm{rep}}=O(\log(n/p))$ alone.  For smaller $p$,
raising $f$ by $\frac14\log_2R_{\mathrm{rep}}$, which is $\frac14\log_2\log_2(1/p)+O(1)$, restores the bound.
At $f=f_B$ the product stays below $2^{-40}$ as long as
$\log_2R_{\mathrm{rep}}\le4f_B-40-\log_2(N_{\mathrm{sol}}\widetilde J(2\widetilde T+1))$, a threshold above
$35\,000$ at $n=10^{30}$.  In a repair node the event is charged to the term $\delta_{\mathrm{num}}\le\delta_{\mathrm{fail}}$ of
Eq.~\eqref{eq:cubic_acceptance_comparison}.  Its room is ample, because the silent budget
$\varepsilon_c:=p_c/(8T_c(B+1))$ handed to the projection is below $10^{-12}$ already at the root for $p\le1/2$
($9.25\cdot10^{-13}$ at $n=10^{25}$).  In a suffix trial the event is charged to the margin between the ideal success $0.999$ of
Lemma~\ref{lem:bit_ball_target}(iii) and the $0.9$ that Lemma~\ref{lem:algorithm_certified_implementation}
requires.  Only the silent class, wrong guards and failed candidate phases, is charged to $\varepsilon_c$, which the
projection receives as its $\varepsilon_{\mathrm{fail}}$, each of its $N_{\mathrm{sol}}$ solves running at
$p_0=\varepsilon_c/(2N_{\mathrm{sol}})$.

This routing is what keeps the word length independent of $p$ (on the range $p\ge2^{-n}$).  The budgets of
Lemma~\ref{lem:cubic_refined_complete} shrink by the factor $8T_c(B+1)$ per repair level, from $8.1\cdot10^{11}$
at the root to $1.2\cdot10^{14}$ at the deepest level.  This is $40$ to $47$ bits per level, $632$ bits in all.  So a leaf of the
fourteen-level tree receives $\varepsilon_c\approx2^{-634}$ at
$n=10^{30}$, $p=1/4$, and a suffix trial receives $p_{14}/(4PT)\approx2^{-651}$.  Had the sampler caps been charged to $\varepsilon_c$, the requirement
$\operatorname{poly}(n)\mathsf u^4\le\varepsilon_c$ would read $f\ge\frac14(\log_2(4/p)+14\cdot46.8+\log_2\operatorname{poly}(n))$
and $W_B$ would acquire the term $\frac14\log_2(1/p)$.  Under (F18) the caps stay at $p_{\mathrm s}=\mathsf u^4$
per procedure, as in the parameter list after Lemma~\ref{lem:bit_state}, and $W_B$ is free of $p$.
\end{remark}
\subsection{Proof of Theorem~\ref{thm:bit_complexity}}
\label{subsec:bit_proof}

\begin{proof}[Proof of Theorem~\ref{thm:bit_complexity}]
\emph{Step 1: the word length.}  Fix one projection call of the finite algorithm on its multiplier
branch.  In case (A) of Lemma~\ref{lem:bit_enclosures}(v), the pre-test returns the clipped target
and no solve runs.  The call has dimension $m\in[N_0,n]$, stored target $b$ with $\|b\|_2^2\le c_bm$, rounded
radius $R_{\mathrm{alg}}$, and accuracy $\widetilde\rho$.  The rounded parameters are those of the
parameter list we give after Lemma~\ref{lem:bit_state}.  We use $\ell:=\log_2(2n)$ as a local abbreviation
for this proof, so that
$\log_2d_0=\ell$ and $\ell>84$ for $n>N_{\mathrm{root}}$.  The lemmas invoked by the call impose the
following requirements on $\mathsf u=2^{-f}$, each of the form
$f\ge\log_2(1/\text{tolerance})+\log_2(\text{factor})$.  We list the tolerance and the factor of each.
\begin{center}
\footnotesize
\begin{tabular}{@{}lll@{}}
\toprule
requirement & tolerance & factor (base-two logarithm at most) \\
\midrule
Lemma~\ref{lem:bit_state}(i) & $2^{-40}$ & $\log_2(2n)$ \\
Lemma~\ref{lem:bit_state}(ii) (radius, $R_{\mathrm{alg}}\le2^{f-31}\widetilde a$) & $2^{-31}$ & $\log_2(R_{\mathrm{alg}}/\widetilde a)$ \\
Eq.~\eqref{eq:bit_surrogate_accuracy} & $\varepsilon_{\mathrm{op}}/8$ & $(\lambda_{\mathrm{top}}+1)\log_2e+\log_2d_0$ \\
Lemma~\ref{lem:bit_dense} ($\varepsilon_{\mathrm{dense}}\le\varepsilon_{\mathrm{op}}/8$) & $\varepsilon_{\mathrm{op}}$ & $(\lambda_{\mathrm{top}}+1)\log_2e+\log_2(16e(\kappa_{\mathrm{sc}}+5)\sigma_{\mathrm{sc}}d_0)$ \\
Eq.~\eqref{eq:bit_eps_op}, $\varepsilon_{\mathrm{det}}$ term & $\delta_g$ & $2\lambda_{\mathrm{top}}\log_2e+\log_2(\nu\widetilde c)+4\log_2d_0+\log_2(L_{\max}\sigma_{\mathrm{sc}})+12$ \\
Eq.~\eqref{eq:bit_eps_op}, $(L+1)\mathsf u$ term & $\delta_g$ & $\log_2(2(L+1))$ \\
Eq.~\eqref{eq:bit_eps_op}, $\theta$ term & $\delta_g$ & $2\lambda_{\mathrm{top}}\log_2e+\log_2(\nu\widetilde ckn)+2\log_2d_0+\log_2(k^{3/2}\sigma_{\mathrm{sc}})+14$ \\
Eq.~\eqref{eq:bit_step_row} ($\varepsilon_{\mathrm{step}}\le1$) & $1$ & $2\lambda_{\mathrm{top}}\log_2e+3\log_2d_0+\log_2\sigma_{\mathrm{sc}}+12$ \\
Lemma~\ref{lem:bit_increment}(ii) ($3\varepsilon_{\mathrm{qr}}\le1/(5\sqrt k)$) & $1$ & $\lambda_{\mathrm{top}}\log_2e+\log_2(k^{5/2}d_0^{2}\Omega_{\max})+10$ \\
Lemma~\ref{lem:bit_guard}(ii) & $H/\nu$ & $1.5\lambda_{\mathrm{top}}\log_2e+2\log_2d_0+\log_2\sigma_{\mathrm{sc}}+13$ \\
Lemma~\ref{lem:bit_enclosures}(i), candidate selection & $\varepsilon_{\mathrm{opt}}/(256\nu J)$ & $(\lambda_{\mathrm{top}}+1)\log_2e+\log_2(48e(\kappa_{\mathrm{sc}}+5)\sigma_{\mathrm{sc}}d_0^2)$ \\
Lemma~\ref{lem:bit_enclosures}(i), rounding & $\varepsilon_{\mathrm{opt}}$ & $\log_2(512\Lambda J)$ \\
Lemma~\ref{lem:bit_enclosures}(ii), bisection & $d_{\mathrm{w}}^2/(1+\Lambda)$ & $\log_2(8\Lambda)$ \\
Lemma~\ref{lem:bit_enclosures}(iii) & $d_{\mathrm{w}}$ & $\log_2(32\sqrt m)$ \\
Lemma~\ref{lem:bit_normcert} (pre-test, $\varpi\ge\widetilde\chi/16$) & $\widetilde\chi/16$ & $\log_2(10^5n^2)$ \\
Lemma~\ref{lem:bit_fallback} ($\varpi=2^{-9}$) & $2^{-9}$ & $\log_2(10^5n^2)$ \\
Lemma~\ref{lem:bit_uniform_target} ($f\ge B_U+10$) & $2^{-114}$ & $\log_22^{10}$ \\
Lemma~\ref{lem:bit_ball_target} (the scalar $\widehat s$) & $2^{-70}$ & $0$ \\
Lemma~\ref{lem:bit_state}(iv) ($\widetilde\rho\ge\rho/2$, from $\mathsf u\le\rho/4$) & $\rho$ & $\log_24$ \\
Lemma~\ref{lem:bit_enclosures}(iv) ($y$ update) & $1.5s_{\mathrm{sm}}\sqrt{15}$ & $\log_2\sqrt m$ \\
\bottomrule
\end{tabular}
\end{center}
Seven rows, the radius row and the last six, are at most $2\ell+51$ each.  The two certificate rows are at most
$\log_2(10^5n^2)+25\le2\ell+42$, since $\widetilde\chi=\log(2n)/\widetilde a\ge1/(2.01\cdot10^6)$ and
$2^{9}<2^{25}$.  The radius row is at most $\frac12\ell+51$ and the $y$-update row at most
$\frac12\ell+19$ by (B3) below.  The accuracy row is at most $\frac12\ell+116$, as
$\rho\ge\rho_B(n)$.  The remaining two are the constants $124$ and $70$.  We show that every one of the
remaining thirteen rows is at most
\begin{equation}
\widetilde E+3\widetilde T+2\widetilde S+7\ell+64,
\qquad
\widetilde E:=\log_2\frac1{\widetilde\varepsilon_{\mathrm{opt}}},\quad
\widetilde T:=\widetilde\lambda_{\mathrm{top}}\log_2e,\quad
\widetilde S:=\log_2\widetilde M_{\mathrm{sp}},
\label{eq:bit_row_bound}
\end{equation}
the three master quantities of the call in their rounded form.  Here
$\widetilde\lambda_{\mathrm{top}}=\log(\widetilde H_{\max}/\underline{\widetilde\nu})+1+\log(2n)$, so that
$\log_2(\widetilde H_{\max}/\underline{\widetilde\nu})=\widetilde T-\ell-\log_2e$.  The quantities $\widetilde\Lambda$ and $L$ carry the target constant $c_b$, and we
pull every occurrence of $\log_2\widetilde\Lambda$ into $\widetilde T$ through the
multiplier floor.  We apply every floor $\lfloor\cdot\rfloor_{\mathsf u}$ of the parameter list to a
quantity that is at least $2\mathsf u$ once $f\ge f_B$.  Indeed, each such quantity is bounded below by $2^{-f_B+1}$ through
the bounds (C4) and (C5) below and $\rho\ge\rho_B(n)$.  So each floor costs at most a factor $2$, and we use this
without further mention.

\emph{Six facts about the calls of the algorithm.}
(B1) Lemma~\ref{lem:bit_enclosures}(v) gives $\|b\|_2\ge3.872>3$ on the multiplier branch, which is the bound
Lemma~\ref{lem:appendix_fixed_target_projection} derives from $\widehat R^2>15m$.  So
$\widetilde\Lambda\ge\Lambda\ge\|b\|_2^2>14.9$ and $\widetilde H\ge16q(0)=8\|b\|_2^2>72$.  Since
$\widetilde\Lambda\le\frac83q(0)+\mathsf u$, also $16q(0)\ge5\widetilde\Lambda$, hence
$\widetilde H_{\max}\ge\widetilde H\ge5\widetilde\Lambda\ge5\nu$ for every multiplier $\nu$ that we test.
(B2) We have $a/(2R_{\mathrm{alg}})\le\widetilde c\le(a+\mathsf u)/R_{\mathrm{alg}}\le2a/\sqrt{15m}<a$
by Lemma~\ref{lem:bit_state}(ii) and $R_{\mathrm{alg}}^2\ge R^2>15m$.  Here
$a=a_{\mathrm{sm}}=2\cdot10^6\log(2n)=1.3863\cdot10^6\,\ell$ satisfies
$\log_2a\le\frac18\ell+21$ for $\ell>84$, and $a\ge1.16\cdot10^8$.
(B3) The nominal radius satisfies $R\le10^{14}\sqrt m$ in every repair call.  Indeed, $\widehat R^2<10^{28}m$
for the global helpers and $100n$ for the root by Eq.~\eqref{eq:cubic_global_solver_scales}, and
$(t_0r/(s_{\mathrm{alg}}(1-10^{-6})))^2\le201.2$ for the auxiliary calls, from $r_{86}=23.914$, the
largest registry radius.  The second test of Eq.~\eqref{eq:cubic_global_solver_scales} bounds $(1+10^{-8})^2R^2<10^{28}m$ through $\mathfrak r$, which covers
$R_{\mathrm{alg}}\le(1+10^{-9})R$.  The nominal radius also satisfies $R\le190\log(2n)\sqrt m$ in every suffix stage.  Indeed, $r_j\le r_{86}$ on the
finite list of Definition~\ref{def:cubic_refined_suffix}, and $185\log(1/v_s)\le185\log(2n)$ in its tail
because $v_sn\ge m\ge N_0$, and $v_sn\le m+1$.  A suffix stage also has
$R\le\max\{r_j,185\log(1/v_s)\}\sqrt{v_sn}\le(370/e)\sqrt n$, as $r_j\le23.914$ and $\max_{0<v\le1}\sqrt v\log(1/v)=2/e$.  Also,
$R\le R_{\mathrm{alg}}\le\widetilde R\le(1+1.1\cdot10^{-9})R$ by Lemma~\ref{lem:bit_state}(ii).  So we obtain
$R_{\mathrm{alg}}\le10^{14}\sqrt n$ in every call, the radius bound of (H-body).  Hence
$\widetilde G_*\ge\widetilde c\sqrt m\ge a\sqrt m/(2R_{\mathrm{alg}})\ge5\cdot10^{-7}$, which is the repair
bound, and a suffix stage gives $5\cdot10^3$.  Moreover $\log_2(\widetilde R/(8a))\le\frac12\ell+17$, and
$\log_2(R_{\mathrm{alg}}/\widetilde a)\le\frac12\ell+20$.
(B4) $\widetilde\rho\le1/8$ in every call, and
$\widetilde\rho\le t_0e_{\mathrm{call}}\eta_{\mathrm{num}}\le\max\{\frac25\cdot10^{-6},\frac35\cdot10^{-9}\}e_{80}<10^{-9}$
in every repair call.  Here $e_{80}=\overline e_0^\circ+\beta_0/2<1.7\cdot10^{-3}$, from
Definition~\ref{def:cubic_refined_trace_rows} with the values of Definition~\ref{def:slack_rows}, is
the largest of the seventeen registry exponents of Definitions~\ref{def:cubic_refined_old_rows},
\ref{def:cubic_refined_new_rows}, \ref{def:cubic_centered_root}, and~\ref{def:cubic_refined_trace_rows}.
A dilation only lowers the exponent by Lemma~\ref{lem:cubic_global_dilation}, and the root's
$v(e_{\mathrm{loc}})$ is below $e_{\mathrm{loc}}$.  Hence
$\widetilde\nu_{\min}=\lfloor\widetilde\rho/(8\widetilde G_*)\rfloor_{\mathsf u}\le10^{-3}$ in every call,
from $10^{-9}/(8\cdot5\cdot10^{-7})$ for a repair and $\frac18/(8\cdot5\cdot10^3)$ for a suffix stage.  So
$\underline{\widetilde\nu}=\widetilde\nu_{\min}<9/2<\widetilde\Lambda/2$: the multiplier floor is the
accuracy term in every call, as it is at the dominating triple.
(B5) $f_B\le n^2$ for $n>N_{\mathrm{root}}$.  Indeed, at the dominating triple $1+\Lambda\le2c_Bn$,
$H_{\max}\le9c_Bn$, $\underline\nu=\rho_B(n)\sqrt{15}/(8a)$, and
$G_{\mathrm{lev}}=8G_*^2H_{\max}/\rho_B(n)\le4.8a^2c_Bn/\rho_B(n)$.  So the middle term of
$\sigma=\min\{d_{\mathrm{w}}/8,\ d_{\mathrm{w}}^2/(16(1+\Lambda)G_{\mathrm{lev}}),\ d_{\mathrm{w}}(1-\phi_0)/(256\sqrt nG_{\mathrm{lev}})\}$
is the least and
\begin{equation}
\frac1\sigma\le\frac{2458\,a^2c_B^2n^2}{\rho_B(n)^3},\qquad
\frac{H_{\max}}{\underline\nu}\le\frac{8a\cdot9c_Bn}{\sqrt{15}\,\rho_B(n)},\qquad
\log_2\frac1{\rho_B(n)}\le\frac12\log_2n+114;
\label{eq:bit_crude_word}
\end{equation}
with $\log_2c_B=358.8$ and $\log_2a\le\frac18\ell+21$ we obtain
$\log_2(1/\varepsilon_{\mathrm{opt}})=1+2\log_2(1/\sigma)\le7.5\ell+2222$,
$\lambda_{\mathrm{top}}\log_2e\le2.625\ell+498$, $\log_2M_{\mathrm{sp}}\le\log_2(3a)\le\frac18\ell+23$,
and $f_B\le56\ell+4100$.  The last bound is below $n^2$ for $n>N_{\mathrm{root}}$: at $n=10^{30}$ the bound is
$9737$ against the exact $f_B=9062$.  Consequently $\Omega_{\max}=1+\sqrt{2\log(4nk/\mathsf u^2)}\le2n$
at $f=f_B$.  Indeed, $(\Omega_{\max}-1)^2\le2\log(4n^2)+4f\log2$ and $f_B\le n^2$ for $n>N_{\mathrm{root}}$, so
$\Omega_{\max}\le1+\sqrt{2\log(4n^2)+4n^2\log2}\le2n$.  The requirement of Lemma~\ref{lem:bit_increment}(ii) is
monotone in $f$, so it suffices to check it at $f=f_B$.  Also $k\le45n^{1/4}+1\le n$,
$L_{\max}\le\sqrt n+1\le2\sqrt n$, $\sigma_{\mathrm{sc}}=2^{\kappa_{\mathrm{sc}}}\le4\widetilde M_{\mathrm{sp}}$,
$\kappa_{\mathrm{sc}}+5\le\widetilde S+7\le2^{\widetilde S}$.
(B6) Consider the count $N_{\mathrm{sol}}\widetilde JR_{\mathrm{rep}}(2\widetilde T+1)$ of capped sampling procedures per attempt.  In general
$\log_2(2\widetilde T+1)\le\log_2c_b+2\log_2a+\log_2(2n)+10$.  Moreover $N_{\mathrm{sol}}$ and $\widetilde J$ are at most
$2^{2\log_2(2n)}$ each, by (B5) and (C3), and
$R_{\mathrm{rep}}=\lceil C_{\mathrm{rep}}\log(\widetilde J/p_0)\rceil$ is at most $n^2$ under the hypothesis
$p\ge2^{-n}$.  Here $N_{\mathrm{sol}}=2+\lceil\log_2(4\widetilde\Lambda\widetilde G_*/\widetilde d_{\mathrm w})\rceil\le\frac12\widetilde E+\widetilde T\le f_B$
by (C1) and (C3).  Here we use $\widetilde G_*<a$ from (B2) and $2\log_2(1/\widetilde d_{\mathrm w})\le\widetilde E-7$ from $\widetilde d_{\mathrm w}\ge8\widetilde\sigma$.  So the count
is below $2^{f_B/2}$ and its product with $\mathsf u^4$ is below $2^{-3f_B}<2^{-40}$.  This is the only place where we use the hypothesis
$p\ge2^{-n}$.

\emph{Five consequences.}
(C1) $\log_2\widetilde\Lambda\le\log_2(\widetilde H_{\max}/5)\le\log_2(\widetilde H_{\max}/\underline{\widetilde\nu})+\log_2(10^{-3}/5)\le\widetilde T-\ell-13$, by (B1) and (B4).
(C2) $\log_2\widetilde H_{\max}\le\log_2(\widetilde H_{\max}/\underline{\widetilde\nu})+\log_210^{-3}\le\widetilde T-\ell-11$.
(C3) $\widetilde J=\lceil\log_2(4\widetilde H/\widetilde\varepsilon_{\mathrm{opt}})\rceil\le\widetilde E+\log_2\widetilde H_{\max}+3\le\widetilde E+\widetilde T$.  By the ratio bounds below
$\widetilde E+\widetilde T\le E+T+18\le f_B\le n^2$, so $\log_2\widetilde J\le2\ell$.
(C4) $\log_2(1/\widetilde\delta_g)\le\frac12\widetilde E+2\ell+27$.  Indeed, in the three-term minimum
that defines $\widetilde\delta_g$ in the parameter list after Lemma~\ref{lem:bit_state}, the $\widetilde\varepsilon_{\mathrm{opt}}$ term is below the
first term.  This holds because $\widetilde\varepsilon_{\mathrm{opt}}\le\widetilde\sigma^2/2\le(\widetilde d_{\mathrm{w}}/8)^2/2\le2^{-17}$
from $\widetilde\rho\le1/8$ while $\widetilde H>72$ and $C_{\mathrm g}\widetilde J\ge2\cdot10^5$.  The
third term $(2\widetilde c\widetilde H/9)^2$ is at least $(8a/\widetilde R)^2$, since
$\widetilde c\ge a/(2\widetilde R)$ by (B2) and $\widetilde H\ge72$.  Hence, with
$\log_2C_{\mathrm g}=17.6$, $\log_2m\le\ell$, (C3), and (B3),
\[
\log_2\frac1{\widetilde\delta_g}
\le1+\max\Bigl\{\frac12\bigl(17.6+3\ell+\widetilde E\bigr),\ \log_2\frac{\widetilde R}{8a}\Bigr\}
\le1+\frac12\widetilde E+\frac32\ell+8.8+\frac12\ell+17 .
\]
(C5) $\log_2(1/\varepsilon_{\mathrm{op}})\le\frac12\widetilde E+\widetilde T+\frac{17}8\ell+40$.  Indeed,
with $\nu\le\widetilde\Lambda$,
\[
\begin{gathered}
\varepsilon_{\mathrm{op}}=\Bigl\lfloor\min\Bigl\{\frac{\delta_g}{12\nu\widetilde cd_0},\frac16\Bigr\}\Bigr\rfloor_{\mathsf u}
\ge\frac12\min\Bigl\{\frac{\widetilde\delta_g}{12\widetilde\Lambda\widetilde cd_0},\frac16\Bigr\},
\\
\log_2(12\widetilde\Lambda\widetilde cd_0)\le3.6+(\widetilde T-\ell-13)+\Bigl(\frac18\ell+21\Bigr)+\ell=\widetilde T+\frac18\ell+11.6
\end{gathered}
\]
by (C1) and (B2), and the last bound exceeds $\log_26$.  So
$\log_2(1/\varepsilon_{\mathrm{op}})\le1+\log_2(1/\widetilde\delta_g)+\widetilde T+\frac18\ell+11.6$, and (C4) gives the claim.

\emph{The rows.}  With these,
\begin{align*}
\text{Lemma~\ref{lem:bit_state}(i)}&:\ 40+\ell;\\
\text{Eq.~\eqref{eq:bit_surrogate_accuracy}}&\le\tfrac12\widetilde E+2\widetilde T+\tfrac{25}8\ell+45;\\
\text{Lemma~\ref{lem:bit_dense}}&\le\tfrac12\widetilde E+2\widetilde T+2\widetilde S+\tfrac{25}8\ell+49;\\
\varepsilon_{\mathrm{det}}\text{ term}&\le\tfrac12\widetilde E+3\widetilde T+\widetilde S+\tfrac{45}8\ell+50;\\
(L+1)\mathsf u\text{ term}&\le\tfrac12\widetilde E+\widetilde T+\tfrac54\ell+64;\\
\theta\text{ term}&\le\tfrac12\widetilde E+3\widetilde T+\widetilde S+\tfrac{53}8\ell+51;\\
\text{Eq.~\eqref{eq:bit_step_row}}&\le2\widetilde T+\widetilde S+3\ell+14;\\
\text{Lemma~\ref{lem:bit_increment}(ii)}&\le\widetilde T+\tfrac{11}2\ell+10;\\
\text{Lemma~\ref{lem:bit_guard}(ii)}&\le\tfrac32\widetilde T+\widetilde S+2\ell+15;\\
\text{candidate selection}&\le\widetilde E+2\widetilde T+2\widetilde S+3\ell+6;\\
\text{rounding}&\le\widetilde E+\widetilde T+\ell;\\
\text{bisection}&\le\widetilde E+2\widetilde T-2\ell;\\
\text{Lemma~\ref{lem:bit_enclosures}(iii)}&\le\tfrac12\widetilde E+\tfrac12\ell+2.
\end{align*}
Here, in the two rows in $\varepsilon_{\mathrm{op}}$, we use (C5), $\log_2(16e)\le5.5$,
$\log_2(\kappa_{\mathrm{sc}}+5)\le\widetilde S$, and $\log_2\sigma_{\mathrm{sc}}\le\widetilde S+2$.  In the row of
Eq.~\eqref{eq:bit_step_row} we use the last bound alone.
For the $\varepsilon_{\mathrm{det}}$ and $\theta$ terms we use (C4).  We then use
$\log_2(\nu\widetilde c)\le\log_2\widetilde\Lambda+\log_2a\le\widetilde T-\frac78\ell+8$ by (C1) and (B2),
$\log_2L_{\max}\le\frac12\ell+1$, $\log_2(kn)\le2\ell$, and $\log_2k^{3/2}\le\frac32\ell$.  In the
$(L+1)\mathsf u$ term we use (C4) and
$\widetilde L\le8\max\{1+3\widetilde c^2\widetilde H,2\widetilde c,1\}\le16a^2\widetilde H_{\max}$,
which follows from $3\widetilde c^2\widetilde H\le\frac45(a^2/m)\widetilde H_{\max}$ and $2\widetilde c\le2a$.  So
$\log_2(2(\widetilde L+1))\le6+2\log_2a+\log_2\widetilde H_{\max}\le\widetilde T-\frac34\ell+37$ by (C2).
In the row of Lemma~\ref{lem:bit_increment}(ii) we use $\Omega_{\max}\le2n$ and $k\le n$.  In the guard row we use
$\nu/\widetilde H\le1/5$ from (B1).  For candidate selection and rounding we use (C1) and (C3).  In the bisection and
radial rows we use $\widetilde d_{\mathrm{w}}\ge8\widetilde\sigma$, hence
$2\log_2(1/\widetilde d_{\mathrm{w}})\le\widetilde E-7$, together with
$\log_2((1+\widetilde\Lambda)\cdot8\widetilde\Lambda)\le2\log_2\widetilde\Lambda+4$ and (C1).  Since
$\widetilde E$, $\widetilde T$, and $\widetilde S$ are nonnegative, every row is at most
Eq.~\eqref{eq:bit_row_bound}.  The largest excess over
$\widetilde E+3\widetilde T+2\widetilde S$ is the $\theta$ term's $\frac{53}8\ell+51$.

\emph{From the rounded call to the dominating triple.}  We have stated the rows for the rounded
parameters, while Eq.~\eqref{eq:bit_word_length} uses the parameters of
Section~\ref{sec:faster_algorithm}.  Here we evaluate the unrounded constants $E$, $T$, $S$ of the call at
the nominal accuracy $\rho$ of the call, which is Eq.~\eqref{eq:algorithm_119_accuracy} at $\widehat R=\widetilde R$, or
$t_0e_{\mathrm{call}}\eta_{\mathrm{num}}$, and not at $\widetilde\rho$ as in Eq.~\eqref{eq:bit_call_word}.  So
$f_B(m,c_b,\widetilde\rho,R_{\mathrm{alg}})\ge f_B(m,c_b,\rho,R_{\mathrm{alg}})$, and we charge the factor $2$ of
$\widetilde\rho\ge\rho/2$ once, in $\widetilde d_{\mathrm{w}}\ge d_{\mathrm{w}}/2-\mathsf u$ below.  The ratio bounds of the parameter list after
Lemma~\ref{lem:bit_state} hold for the accuracies of the calls.  The accuracy satisfies $\widetilde\rho\ge\rho/2$ since
$\mathsf u\le\rho/4$, which is the accuracy row.  This holds both
for the suffix floor $\lfloor\min\{\frac18,\frac{t_s}8,\frac{s_{\mathrm{sm}}R_{\mathrm{alg}}}{4(2m+\lceil\sqrt m\rceil)}\}\rfloor_{\mathsf u}$
against $\rho$ of Eq.~\eqref{eq:algorithm_119_accuracy} at $\widehat R=\widetilde R$, as
$R_{\mathrm{alg}}\ge\widetilde R/(1+2^{-30})$ and $2m+\lceil\sqrt m\rceil\le(2m+\sqrt m)(1+1/(2m))$, and
for the repair value $\lfloor t_0e_{\mathrm{call}}\eta_{\mathrm{num}}\rfloor_{\mathsf u}$.  Hence
$\widetilde d_{\mathrm{w}}\ge d_{\mathrm{w}}/2-\mathsf u$.  With
$\widetilde\Lambda\le\frac43\Lambda+\mathsf u$, so that $1+\widetilde\Lambda\le\frac43(1+\Lambda)(1+\mathsf u)$,
the two terms of $\widetilde\tau_{\mathrm{w}}$ are at least $\frac16$ and $\frac13$ of the corresponding
terms of $\tau_{\mathrm{w}}$, giving $\widetilde\tau_{\mathrm{w}}\ge\tau_{\mathrm{w}}/16$.  Moreover
$\widetilde H_{\max}\le\frac{80}3q(0)+16+5\mathsf u\le\frac53H_{\max}$.  Also
$\widetilde G_*\le(a+\mathsf u)(1+2^{-30})\sqrt m/\widetilde R+\mathsf u\le1.01G_*$.  Here $G_*=a\sqrt m/\widetilde R$
is the constant of Section~\ref{sec:faster_algorithm} at $\widehat R=\widetilde R$, while the parameter list uses
$G_*=a\sqrt m/R_{\mathrm{alg}}$ and the factor $1+2^{-30}$.  The two differ by that factor and give the same
conclusions.  So
$\underline{\widetilde\nu}\ge\underline\nu/4$ and $\widetilde G_{\mathrm{lev}}\le7G_{\mathrm{lev}}$.
Therefore $\widetilde\sigma\ge\sigma/112-\mathsf u\ge\sigma/128$ and
$\widetilde\varepsilon_{\mathrm{opt}}\ge2^{-14}\varepsilon_{\mathrm{opt}}-\mathsf u\ge2^{-15}\varepsilon_{\mathrm{opt}}$,
so $\widetilde E\le E+15$.  The bound $\widetilde H_{\max}/\underline{\widetilde\nu}\le\frac{20}3H_{\max}/\underline\nu$
gives $\widetilde T\le T+\log_2(20/3)\le T+3$, and $\widetilde M_{\mathrm{sp}}\le M_{\mathrm{sp}}+2\mathsf u+2$
gives $\widetilde S\le S+0.01$.  None of these bounds involves $c_b$ or the value of $\rho$: they are
properties of the rounding directions alone.  So every row of the call is at most $E+3T+2S+7\ell+89$
in the unrounded constants of the call.  Finally these constants are monotone.  With
$\underline\nu=\nu_{\min}=\rho/(8G_*)$ in every call (B4), the quantities $\Lambda$, $H_{\max}$, $G_*$,
$G_{\mathrm{lev}}$, $H_{\max}/\underline\nu$ do not decrease and $\tau_{\mathrm{w}}$, $\sigma$,
$\varepsilon_{\mathrm{opt}}$ do not increase under the replacements we list next.  We replace the target bound $\|b\|_2^2\le c_bm$ by $c_Bn$, and $\rho$ by
$\rho_B(n)\le\rho$ of Definition~\ref{def:bit_word_size}.  This is exact for the nominal $\rho$, while the
rounded $\widetilde\rho\ge\rho/2$ can fall $\mathsf u$ below $\rho_B(n)$, on row $86$ at $\sigma=10^4$, which the
ratio bounds have already absorbed.  Had we evaluated the constants at $\widetilde\rho$ instead, halving the
accuracy would raise $E$ by at most $6$ bits, as $\varepsilon_{\mathrm{opt}}$ is proportional to a power of
$\rho$ of degree at most $6$ through $\sigma$, $\tau_{\mathrm{w}}$, and $G_{\mathrm{lev}}$.  It would raise $T$ by at most
$1$.  These rises are inside the margin below.  We replace the constant $G_*=a\sqrt m/\widehat R$ by its bound
$a/\sqrt{15}$, and $m$ by $n$.  So $E$, $T$, $S$ at the call are at most their values at the dominating
triple, and every row is at most $f_B-33\ell-111<f_B$.  In particular the sampler-cap and
guard-error budgets never enter: no row contains $p$ or a failure budget.

\emph{Integer bits.}  The magnitudes we must hold in $I_B-2$ bits are, at the values of the call, the following.  The tracked
matrix, the selected-gradient scalar, the action intermediates, the sketch and Householder columns, the probe images,
$\widetilde G_{\mathrm{lev}}$, the multipliers $\nu\le\widetilde\Lambda$, and the target coordinates $b_i$ fit in $I_B$ bits by
Lemma~\ref{lem:bit_integer_bits}, whose hypothesis $\Omega_{\max}\le2n$ is (B5).  The value $q(x)$ sits in its $2W_B$-bit word by the
same lemma.  The new magnitudes of this section are smaller.
Coordinatewise, the repair targets need $125$ integer bits and the ball target $\lceil\log_2(10^4\sqrt m)\rceil+2$
(Lemmas~\ref{lem:bit_uniform_target}(i) and~\ref{lem:bit_ball_target}(i)).  The fallback matrix $E$ and its Frobenius
scalar need $\lceil\log_2(2n)\rceil+4$ bits (Lemma~\ref{lem:bit_fallback}).  All are below $I_B$, which is monotone in
the same direction as $f_B$ and which we evaluate at the dominating triple.  Our word length is therefore
$W_B$.  Since $\log_2(1/\varepsilon_{\mathrm{opt}})$, $\lambda_{\mathrm{top}}$, $\log_2M_{\mathrm{sp}}$,
$\log_2(H_{\max}/\underline\nu)$ are $O(\log n)$ at the triple, by Eq.~\eqref{eq:bit_crude_word},
$W_B=O(\log n)$.  The failure probability $p$ enters only through counts.

\emph{Step 2: correctness.}  We run the finite algorithm on the $\widetilde A_i$ of
Lemma~\ref{lem:bit_input}, which are symmetric contractions, so that every lemma of
Section~\ref{sec:small_algorithm_constant} applies to them.  Since $p$ is rational, every budget
$p_c/(8T_c(B+1))$ and $p/(4PT)$ is an exact rational and every cap $T_c$, $T$, $T_f$ an integer.  The
certified upper enclosures of $\log(4/p_c)$ and $\log(4P/p)$ in (F2) only raise the caps and lower the
budgets, which is the direction the two budget rules of
Lemmas~\ref{lem:algorithm_certified_implementation} and~\ref{lem:cubic_refined_complete} tolerate.

For $n\le N_{\mathrm{root}}$ the algorithm is the fallback of Lemma~\ref{lem:bit_fallback} with
$c=10.889$, $m=n$, and $T_f=\lceil\log_2(1/p)\rceil$ trials.  By its parts (i) and (iii), with probability
at least $1-p$ it returns a signing with $\|\sum_i\varepsilon_i\widetilde A_i\|\le10.889\sqrt n$.
Corollary~\ref{cor:bit_output_bound} then gives $\|\sum_i\varepsilon_iA_i\|<11.504\sqrt n$ for every $n\ge1$.  On this branch we use only the rows of
Lemma~\ref{lem:bit_fallback} in Step~1 (Remark~\ref{rem:bit_small_n}).

For $n>N_{\mathrm{root}}$ the algorithm is \textsc{GlobalAction} at the root with budget $p$.  We
argue by induction from the leaves through the fourteen repair levels, which are, by Lemma~\ref{lem:cubic_global_complete}, the four auxiliary depths and nine
global layers below the root.  The claim is that of
Lemma~\ref{lem:cubic_refined_complete}: a node with budget $p_c$ returns exact signs within its certified
profile except with probability at most $p_c$.  What changes in fixed point is the content of
``numerical failure''.  A projection of the node is Lemma~\ref{lem:bit_projection} with budget
$\varepsilon_c:=p_c/(8T_c(B+1))$, which we feed to the projection as its $\varepsilon_{\mathrm{fail}}$.  Each of its
$N_{\mathrm{sol}}$ solves runs at $p_0=\varepsilon_c/(2N_{\mathrm{sol}})$.  Its word hypothesis enters its proof only
through the requirement rows of Step~9 and the integer demands of Lemma~\ref{lem:bit_integer_bits}, all of which we verified
in Step~1 at $f=f_B$, $I=I_B$ for every call.  Except with probability
$\varepsilon_c$ it returns an exactly feasible point $\widehat x\in(\mathsf u\Gamma)^m\cap\widetilde Q\cap\widetilde{\mathcal D}$
within $\widetilde\rho_{\mathrm{rep}}\le t_0e_{\mathrm{call}}\eta_{\mathrm{num}}$ of the exact projection
of the stored target.  The budget $\varepsilon_c$ covers the silent class, wrong guard decisions with
probability $p_0/(100TJR_{\mathrm{rep}})$ per guard by Lemma~\ref{lem:bit_guard}, and phases in which all
$R_{\mathrm{rep}}$ candidates miss the contraction of Lemma~\ref{lem:bit_lyapunov}, at $0.1^{R_{\mathrm{rep}}}\le p_0/(2\widetilde J)$
each.  Together these cost at most $p_0$ per solve, by Step~8 of the proof of Lemma~\ref{lem:bit_projection}.  No saturation
occurs as long as every guard is correct and no cap fires, by Lemma~\ref{lem:bit_integer_bits} and Step~1.
Every declared failure, a sampler cap of Lemma~\ref{lem:bit_gaussian}(i) or
Lemma~\ref{lem:bit_probes}, the one-redraw cap of Lemma~\ref{lem:bit_gaussian}(iii), or a saturation
report, is a failed test by (F18).  On a failed test we discard the attempt and the loop continues, and we charge the event
to $\delta_{\mathrm{num}}\le\delta_{\mathrm{fail}}$ in Eq.~\eqref{eq:cubic_acceptance_comparison}.
There it costs at most $N_{\mathrm{sol}}\widetilde JR_{\mathrm{rep}}(2\widetilde T+1)\mathsf u^4<2^{-40}<\delta_{\mathrm{fail}}$
per attempt at $f=f_B$ by (B6).
Our one use of the hypothesis $p\ge2^{-n}$ is the bound $R_{\mathrm{rep}}\le n^2$ in the count of (B6).

An attempt of \textsc{ProjectionRepair} at a repair node runs as follows.  The uniform target has the exact
midpoint law on the $2^{-114}$ grid, and its image $U''$ stays within $h_{\mathrm{tot}}/2$ per coordinate of the
exact uniform $U$ (Lemma~\ref{lem:bit_uniform_target}(ii)).  The prefix test on $U''$ is an exact comparison, and
its failure probability stays inside the budgets $10^{-9}$ and $10^{-10}$ of Lemmas~\ref{lem:cubic_refined_accuracy},
\ref{lem:cubic_global_prefix}, and~\ref{lem:cubic_global_complete}, which tolerate shifts far above $h_{\mathrm{tot}}$
(Lemma~\ref{lem:bit_uniform_target}(iv)).  The deficit test on the returned $y=\widehat x/t_0$ is an exact rational
comparison (F17).  The solve's physical error enters Eq.~\eqref{eq:cubic_acceptance_comparison} with
$\eta_0=\eta_{\mathrm{num}}$ unchanged, and the only new term is the rise of $a_{\mathrm{acc}}$ by
$h_{\mathrm{tot}}/e_{\mathrm{call}}$ (Lemma~\ref{lem:bit_uniform_target}(iii)).  With $B_U=114$ and $f\ge124$ the
conditional probability of a valid acceptance is at least $4.298\cdot10^{-9}>4\cdot10^{-9}$ for the global helpers
and the root and at least $4.799\cdot10^{-8}>1/21000000$ for the auxiliary calls (Lemma~\ref{lem:bit_uniform_target}(v)).
These are the hypotheses of the two budget rules.  An accepted $y$ satisfies
$\mathcal U_{U''}(y)\le\theta e_{\mathrm{call}}n$ exactly, the hypothesis of Lemma~\ref{lem:cubic_refined_potential}.
The nearest signs, the truncated masses, the odd-numerator supports, the two orientation rules, and the mass
updates of (F15) are exact integer operations by Lemma~\ref{lem:bit_repair}.  The truncation costs less than
$2m2^{-B}\le2^{-44}<10^{-12}\sqrt n$ as in the real model, and a helper returns exact signs or \textsc{Fail}, so
nothing propagates between levels (Lemma~\ref{lem:bit_repair}(i),(iii)).  The parent's accounting uses
$\|B(\widehat x)\|\le(1-s_{\mathrm{sm}})\widetilde R$ with $\widetilde R\le(1+10^{-8})R$ by
Lemma~\ref{lem:bit_state}(ii),(iii), which is the window we charge for
$\widehat R$ in Section~\ref{sec:small_algorithm_constant}.  Hence the node fails only by exhaustion or by a silent projection failure or a failed child.
Exhaustion costs at most $(1-4\cdot10^{-9})^{T_c}\le p_c/4$, and the auxiliary floor gives less.  A silent
projection failure or a failed child costs at most
$T_c(B+1)\varepsilon_c=p_c/8$ by the union bound over at most $T_c$ projections and at most $B$
children.  In all the node fails with probability at most $p_c$, and on the complement the deterministic argument of
Lemma~\ref{lem:cubic_refined_potential} gives the node's profile.

A fallback node is Lemma~\ref{lem:bit_fallback}: on acceptance $\|\sum_i\varepsilon_i\widetilde A_i\|\le c\sqrt m$
deterministically, each trial accepts with probability at least $1/2$, and $T_f$ trials exhaust with
probability at most $p_c$, no numerical budget being consumed.  Its ranges (b), (c), and (d) are the $11.902\sqrt m$
fallback of \textsc{GlobalSign} at layer $9$, which we pay for by Eq.~\eqref{eq:cubic_global_bridge}, the $13\sqrt m$
fallback of \textsc{GlobalAction} for $m<N_0$, and the small-dimension fallbacks of \textsc{RefinedRepair} at the
bands $448\le i\le713$.

A suffix leaf is \textsc{SuffixSigning} with budget $p_{14}$, which we split as in
Lemma~\ref{lem:algorithm_certified_implementation} into $P$ rounds of $T$ trials with numerical budget
$\min\{0.02,p_{14}/(4PT)\}$ per trial.  In a trial, $W'_G$ and its rejection are exact, and the ball-target
lemma, Lemma~\ref{lem:algorithm_centered_box}, holds for the rounded target $\widehat g=\kappa_{\mathrm{pr}}\widehat b$
with the margins of Lemma~\ref{lem:bit_ball_target}(ii),(iii).  So an ideal trial fixes more than $m/4$
coordinates with probability greater than $0.999$.  The projection is Lemma~\ref{lem:bit_projection}
with the trial's budget, and we charge its declared failures to the $0.02$ allowance by (F18).  The radial
correction and the snapping at threshold $2\widetilde\rho$ are exact moves by Lemma~\ref{lem:bit_state}(iv).  The
test ``more than $m/4$ snapped'' is exact.  The $y$ update of (F14) is Lemma~\ref{lem:bit_enclosures}(iv), so
the accepted increment has matrix norm at most $\widetilde R/t_s\le(1+10^{-8})R/t_s$, the quantity
we charge in Definition~\ref{def:cubic_refined_suffix}, and leaves at most $\lfloor w(3/4)^{s+1}n\rfloor$ real active
coordinates.  The conditional success of a trial therefore exceeds $0.9$.  Exhaustion costs at most
$0.1^T\le p_{14}/(4P)$ per round, and the union bound over the $PT$ numerical budgets costs at most $p_{14}/4$.
The terminal nearest-sign rounding is exact with the charge $\min\{3\cdot10^{-7},\sqrt{N_0w}\}\sqrt n$ unchanged.
So the leaf certifies its $C_i^{\mathrm{suf}}\sqrt n$ except with probability $p_{14}$.

{\setlength{\emergencystretch}{2em}We close the induction at the root with budget $p$.  On its good event the certificate of
Lemma~\ref{lem:cubic_global_certificate}, Eq.~\eqref{eq:cubic_global_final}, gives
$\|\sum_i\varepsilon_i\widetilde A_i\|\le11.503394001765\sqrt n$ as in
Lemma~\ref{lem:cubic_global_complete}.  Every constant it adds is a bound the fixed-point
primitives meet verbatim.  These constants are $\alpha$, $\gamma$, the fallback bounds, and
the suffix caps with their $(1+10^{-8})$ and $s_{\mathrm{sm}}$ allowances.  Corollary~\ref{cor:bit_output_bound} adds the
$2^{-9}$ of Lemma~\ref{lem:bit_input}.  So our output satisfies $\|\sum_i\varepsilon_iA_i\|<11.504\sqrt n$ with
probability at least $1-p$ for every $n\ge1$.\par}

\emph{Step 3: cost.}  Write $W_\times:=2W_B+c_\eta\lceil\log_2(2n)\rceil$.  One projection call in dimension
$m\le n$ costs
$(mn^{2+2\eta}+mn^{\omega-1/2+2\eta}+n^{\omega+2\eta})\operatorname{polylog}(n/\varepsilon)\,M(W_\times)$ bit
operations by Lemma~\ref{lem:bit_projection}, Step~10 of its proof, at the counts we list next.  There are $N_{\mathrm{sol}}=2+\lceil\log_2(4\widetilde\Lambda\widetilde G_*/\widetilde d_{\mathrm{w}})\rceil$
fixed-multiplier solves on exact enclosures.  These number $230$, $383$, $661$, and $489$ at $n=10^{30}$ for a suffix
stage, an auxiliary repair at its cap, a global helper at the literal caps, and a global helper at the
worst admissible policy.  There are $\widetilde J$ phases per solve, $1197$, $1971$, $3452$, $2538$ in the same cases, each of
$R_{\mathrm{rep}}=O(\log(J/p_0))$ candidates.  There are $s_{\mathrm{tr}}=128\lceil8\log(100TJR_{\mathrm{rep}}/p_0)\rceil$
guard probes per step.  These number $1.8\cdot10^5$ to $4.1\cdot10^5$ at a solve budget $p_0=10^{-12}$, and $6.2\cdot10^5$ to
$8.5\cdot10^5$ at the leaf budgets below, where $R_{\mathrm{rep}}$ is about $200$.  There are $\widetilde T=16m\widetilde L$ steps per
phase.  Here $\widetilde L\le8(1+3\widetilde c^2\widetilde H)$ is a constant multiple of
$a_{\mathrm{sm}}^2c_b$, about $10^{126}$ at the cap $c_b=c_B$, $n=10^{30}$, so
$\widetilde T\approx10^{157}$ there.  The products of a step and of a refresh, and the refresh count $\widetilde T\pi'+1+M_{\mathrm{ref}}$ per phase of
Lemma~\ref{lem:bit_tracker}(v), are those of Step~10 of the proof of Lemma~\ref{lem:bit_projection}, at the product cost of
Lemma~\ref{lem:bit_fastmm}.  The enclosure
evaluations, the pre-test, and the parameter computations are polylogarithmically many dense products
and $O(n\operatorname{polylog}(n/\varepsilon))$ bit operations.  Every count is $O(\log(n/\varepsilon))$
with a constant that depends on $c_b$, $\rho$, and $\widehat R^2/m$ through $\log_2\widetilde L$,
$\log_2(1/\widetilde\varepsilon_{\mathrm{opt}})$, and $\log_2(\widetilde\Lambda\widetilde G_*/\widetilde d_{\mathrm{w}})$
only.  Since $\omega+1/2<3$, the call costs at most $n^{3+2\eta}\operatorname{polylog}(n/\varepsilon)M(W_\times)$.

The work around the solver is smaller.  Consider a projection attempt.  We form the matrix image $B(x)$ of the target
and of the returned point, and the signed helper families $s_iA_i$, in $mn^2$ word
operations, or $O(n^3W_B)$ bit operations, as we do for the pre-test matrix $B(b_Q)$.  The uniform target costs $115m$ coins.  The prefix test sorts $m$
rationals of $O(W_B)$ bits and makes $m$ exact comparisons.  The deficit sum, the tent sums, and the
cardinality sums are $O(Bm)$ exact operations on integers of $O(W_B)$ bits (F17), with the
common-orientation choice and the mass updates at each of the $B=\lceil\log_2n\rceil+45$ levels.  The
band and piece location, the $192$-step search for $\widehat\varepsilon(q)$, the padding sizes, and the
budgets are $O(1)$ big-integer operations per call on integers of $4755+\log_2n+O(1)$ bits (F16).  A
fallback node runs $T_f=\lceil\log_2(1/p_c)\rceil$ trials, at most $634$ at every node of the tree for
$n=10^{30}$ and $p=1/4$, the leaf budget $p_{14}$.  Each trial forms $E$ in $mn^2$ word operations and its Frobenius norm exactly.
Past the pre-test, each trial performs $J_{\mathrm{sq}}\le16$ exact dense products at $O(n^{\omega+2\eta})M(W_\times)$ with
$J_{\mathrm{sq}}+1$ Frobenius norms and $J_{\mathrm{sq}}+2$ certified enclosures of $\log$ and $\exp$
by Lemma~\ref{lem:bit_fallback}(iv).  Zero padding is free of arithmetic: the artificial matrices and
coordinates are exact grid objects whose signs we discard.  The padded sizes of one suffix sum to at
most $4wn\le4n$ by Lemma~\ref{lem:algorithm_certified_implementation}, and we pad an $N\times N$ input
once, before Lemma~\ref{lem:bit_input}, at $O(n^3b_{\mathrm{in}})$ bit operations.

The number of calls is the count of Lemma~\ref{lem:cubic_refined_complete}.  A repair node makes at
most $T_c$ projections and, in its accepted attempt, at most $B$ child calls.  So there are at most
$B^k$ nodes at depth $k\le13$ and at most $B^{14}$ suffix leaves, each leaf running at most $PT$
projections with $P=1+\lceil\log n/\log(4/3)\rceil$ and $T=\lceil\log(4P/p_{14})/\log10\rceil$.  With
$T_c^{(k)}=\lceil2.5\cdot10^8\log(4/p_k)\rceil$ and $p_{k+1}=p_k/(8T_c^{(k)}(B+1))$, so that
$\log(1/p_k)=O(\log(1/p)+k(\log\log n+\log\log(1/p)))$, the total number of projections is at most
$\sum_{k<14}B^kT_c^{(k)}+B^{14}PT=O((\log n)^{15}(\log(1/p)+\log\log n))$.  The depth-$13$ repair term
$B^{13}T_c^{(13)}$ dominates numerically: it is $1.28\cdot10^{39}$ of the total $1.29\cdot10^{39}$ at $n=10^{30}$ and
$p=1/4$, where $T_c^{(13)}=1.02\cdot10^{11}$ because $\log(4/p_{13})\approx408$.  The budgets we feed to the
projections have $\log_2(1/\varepsilon)$ equal to $634$ at the repair leaves and $651$ at a suffix trial for
$n=10^{30}$ and $p=1/4$, and $667$ and $685$ at $p=10^{-9}$.  This enters $\operatorname{polylog}(n/\varepsilon)$ as a constant factor for fixed $n$
and as $(\log\log n)^{O(1)}$ asymptotically.  When we sum, the run costs
$\{n^3+n^{\omega+1/2}+n^\omega\}n^{2\eta}\operatorname{polylog}(n/p)M(W_\times)$ bit operations, the first
term of Eq.~\eqref{eq:bit_total_cost}.  Lemma~\ref{lem:bit_input} gives the second, as
$b_{\mathrm{in}}\le W_B$.  Finally $M(W_\times)=O(W_\times\log W_\times)=O(\log n\log\log n)$ for fixed $\eta$, and
$n^{\omega+1/2}+n^\omega\le2n^3$ since $\omega+1/2<3$.  For every fixed $\eta>0$ the count is
$O(n^{3+2\eta})\operatorname{polylog}(1/p)M(W_\times)$.  Since $\eta$ is arbitrary this is
$n^{3+o(1)}\operatorname{polylog}(1/p)$ in the sense in which
we use $n^{3+o(1)}$ in Section~\ref{sec:small_algorithm_constant}.  We obtain
$n^{3+o(1)}\operatorname{polylog}(1/p)+\widetilde O(n^3L_{\mathrm{in}})$.
\end{proof}

\begin{remark}[Reading the word length]
\label{rem:bit_word_length}
Eq.~\eqref{eq:bit_word_length} gives $W_B=9804$ at $n=10^{25}$ and $W_B=10744$ at $n=10^{30}$,
with $f_B=9062$, $I_B=1682$.  Against this stand $5970$ and $6993$ for the same formula without its term $\lceil\log_2c_B\rceil$,
evaluated at the parameters of the algorithm of Theorem~\ref{thm:cubic_dense_runtime}.  Those parameters are a Gaussian target with
$\|\xi'\|_2^2\le2m$, a body of radius $279\sqrt m$, and an accuracy $s_{\mathrm{sm}}R/(4(2m+\sqrt m))$, and the evaluation is at the same $n$.
Asymptotically both are $63\log_2n+O(\log\log n)$.  The difference lies between $3300$ and $4100$ bits over
$10^{25}\le n\le10^{200}$.  It is $4\log_2(c_B/2)$ plus six times the logarithm of the ratio of the two
accuracy floors inside $\log_2(1/\varepsilon_{\mathrm{opt}})$, together with the images of the same two ratios in
$\lambda_{\mathrm{top}}$ and $\log_2(H_{\max}/\underline\nu)$.  The remaining terms are the term
$\lceil\log_2c_B\rceil=359$ of $I_B$ and the radius ratio $279/\sqrt{15}$ entering through $G_*$, which accounts for about $50$
bits.  At $n=10^{30}$ the components are
$\log_2(1/\varepsilon_{\mathrm{opt}})=2646$, $\lambda_{\mathrm{top}}=488.9$ (so
$3\lceil\lambda_{\mathrm{top}}\log_2e\rceil=2118$), $\lceil\log_2(H_{\max}/\underline\nu)\rceil=604$, and
$40\lceil\log_2(2n)\rceil=4040$.  Evaluated call by call at $n=10^{30}$ the word is smaller.  It is $7336$ bits
for a suffix stage, $8722$ for an auxiliary repair at its cap $10^{52}$, and $9716$ for a global helper at
the worst admissible policy, which is row $86$ dilated by $\sigma=10^4$, $c_b=3.9\cdot10^{73}$,
$\rho=5.8\cdot10^{-35}$.  It is $7925$ for the root at its cap $10^{36}$ ($6988$ at $n=10^{25}$).  The
dominating triple pairs the largest target cap with the smallest admissible accuracy, which no single
call attains, and the literal caps read with the accuracy they imply would give $11315$.  The formula has room.  The binding row of Step~1, the
candidate-selection enclosure or the $\varepsilon_{\mathrm{det}}$ term, lies at least $3850$ bits below
$f_B$ in every call evaluated at $n=10^{25}$ and $10^{30}$.  The same margin holds for the worst admissible helper at every $n$ up to
$10^{200}$, evaluated numerically.  A quantity-by-quantity list, the binding row together with the
largest integer demand (the selected-gradient scalar), would need $4473+1455$ bits at the literal caps and
$n=10^{30}$, about $60\log_2n$.  The quantities that are never resolved are
$\phi_0=(2n)^{-1999998}$, the bottom eigenvalues $e^{-2ab_{\mathrm{sm}}}$ of $P_x$, and the Taylor
prefactor $e^{c/2}$ of the surrogate of Section~\ref{sec:faster_algorithm}.  The only object whose size is set by a
count rather than by an accuracy is the tracked matrix, through $L_{\max}$.  The bit cost
of the fast products carries the unquantified constant $c_\eta$ of
Lemma~\ref{lem:bit_fastmm}, inherited from the coefficient growth of the laser-method
algorithms behind $\alpha_{\mathrm{dual}}>1/4$.  It affects the temporaries, not the stored
words, and not the exponent.
\end{remark}

\begin{remark}[The constant for the original matrices]
\label{rem:bit_constant}
The twelve-digit constant $11.503394001765$ of Lemma~\ref{lem:cubic_global_complete} is certified for
the rounded matrices $\widetilde A_i$, and Theorem~\ref{thm:bit_complexity} states $11.504$ for the
$A_i$ because Lemma~\ref{lem:bit_input} adds $2^{-9}$.  The certificate has room for this term: the
$10^{-12}$ of Eq.~\eqref{eq:cubic_global_gamma} is spent only on the mass truncation
$2m2^{-B}\le2^{-44}$, leaving $10^{-12}\sqrt n-2^{-44}\ge3.16$ at $n=10^{25}$.  So charging $2^{-9}$ to it
would keep the twelve digits for the $A_i$ at the cost of one line inside the proof of
Lemma~\ref{lem:cubic_refined_potential}.  We leave Section~\ref{sec:small_algorithm_constant} as it is
and state the rounded constant.
\end{remark}

\begin{remark}[Small $n$]
\label{rem:bit_small_n}
For $n\le N_{\mathrm{root}}=10^{25}$ only the fallback runs: the algorithm rounds the input and certifies
fair-sign trials.  Of the requirements of Step~1 only those of Lemma~\ref{lem:bit_fallback}, which are
$f\ge2\lceil\log_2n\rceil+26$ and $I\ge\lceil\log_2(2n)\rceil+4$ for the Frobenius scalar
$\lceil\|E\|_F\rceil_{\mathsf u}\le13n+1$ after the pre-test, and those of Lemma~\ref{lem:bit_input} are
used, both far inside Eq.~\eqref{eq:bit_word_length}.  For $10^{25}<n\le N_*=10^{30}$ the tree has depth
one: the root projection and repair with its $B$ children certified at $11.902\sqrt m$.
\end{remark}

\begin{remark}[The polylog factor is concrete and large]
\label{rem:bit_polylog}
The factor $\operatorname{polylog}(n/p)$ of Eq.~\eqref{eq:bit_total_cost} is the call count of the
fourteen-level tree, $\sum_{k<14}B^kT_c^{(k)}+B^{14}PT=O((\log n)^{15}(\log(1/p)+\log\log n))$ projections
with $B=\lceil\log_2n\rceil+45$ and $T_c^{(k)}=\lceil2.5\cdot10^8\log(4/p_k)\rceil$ (Step~3 of the proof).
It is dominated numerically by the depth-$13$ repair term: $1.29\cdot10^{39}=n^{1.30}$ at $n=10^{30}$ and $p=1/4$, and below $n$ only from $n$
about $10^{41}$.  A phase of the solver has $\widetilde T=16m\widetilde L$ steps with $\widetilde L$
about $10^{126}$ at the cap.  These are the constants of the arithmetic count of
Lemma~\ref{lem:cubic_global_complete}, not ones the bit model adds.  The word $W_B$ sees $\widetilde L$
through $\log_2(2(\widetilde L+1))$ in one row of Step~1, about $420$ bits at the cap.  This is against about $49$
for the value $\widetilde L\approx2.8\cdot10^{14}$ at the parameters of the algorithm of
Theorem~\ref{thm:cubic_dense_runtime} at the same $n$.  The only other route is through $s_{\mathrm{tr}}$, and the word sees the tree only
through the budgets the routing (F18) keeps out of it.  So the constant hidden in $n^{3+o(1)}$ is
that of Section~\ref{sec:small_algorithm_constant}, and a reader who takes $W_B\approx10^4$ bits as a
concrete figure should take the call count as concrete too.
\end{remark}

\begin{remark}[The relative certificate]
\label{rem:bit_relative_guard}
Theorem~\ref{thm:bit_complexity} holds for the algorithm with the relative guard
$\Phi(y)\le U(y)\le2\Phi(y)$ of Section~\ref{sec:faster_algorithm} as well, with the same
proof.  The requirement $d_0\varepsilon_{\mathrm{op}}\le\phi_0/8$ of Section~\ref{sec:faster_algorithm},
met there by the choice $\varepsilon_{\mathrm{op}}\le\phi_0/(16d_0)=(2n)^{-1999998}/(32n)$, replaces
Eq.~\eqref{eq:bit_eps_op}.  Every bit count that involves $\log_2(1/\varepsilon_{\mathrm{op}})$ grows by about
$1999998\log_2(2n)+\log_2(32n)$, which is $5.8\cdot10^7$ at $n=3\cdot10^8$ and still
$O(\log n)$.  The exponent of the bit complexity is therefore the same, and the additive
guard is a matter of the constant in $W$, not of the theorem.  In fixed point it is
also a matter of representability: the relative guard resolves $\Phi$ at the scale
$\phi_0$, which the additive guard never forms.
\end{remark}

\begin{remark}[Theorem~\ref{thm:dense_runtime} is not covered]
\label{rem:bit_section6}
The $n^{2+\omega/2}$ algorithm of Section~\ref{sec:algorithm} diagonalizes $B(x)$ at every
block query with Sobczyk's algorithm~\cite[Corollary~2.4]{sobczyk25}, whose eigenvector
computation is analyzed only in the real-arithmetic model.  Its finite-precision
behavior is Sobczyk's own open problem, and Theorem~\ref{thm:dense_runtime} remains a
real-arithmetic statement.  The finite algorithm of this section performs no diagonalization: its fallback certificates run through
Lemma~\ref{lem:bit_normcert} (F11).
\end{remark}

\section{Lower bounds for the matrix Spencer constant}
\label{sec:lb_lower_bound}

In this section we bound the matrix Spencer constant from below. The upper bounds we prove in this paper, Theorems~\ref{thm:first_main}, \ref{thm:matrix_spencer_bound} and~\ref{thm:existence_constant_sub10}, give the constants $7\cdot10^9$, $156000$ and $7.879493715947$. The results below say how small the constant can be. Section~\ref{subsec:lb_constant} defines the quantity we bound: the worst-case discrepancy $D_n^{\mathrm{mat}}$ of Definition~\ref{def:lb_dmat} and the constant $D_n^{\mathrm{mat}}/\sqrt n$. Remark~\ref{rem:lb_normalization} there records that the two normalizations of matrix discrepancy in use agree for involution families. Every family below is one, except the zero-padded family of Corollary~\ref{cor:lb_constant}(iii). We then give three lower bounds, in increasing generality. The first, in Section~\ref{subsec:lb_exact}, is two explicit families at small sizes whose discrepancies we compute exactly. At $n=8$, eight Kronecker words have every signing at norm $4+\sqrt2$ (Proposition~\ref{prop:lb_eight}). At $n=16$, sixteen words have discrepancy $7.850\ldots$, certified by an exact enumeration of all signings (Theorem~\ref{thm:lb_sixteen}). Both are lower bounds on $D_8^{\mathrm{mat}}$ and $D_{16}^{\mathrm{mat}}$, not their values. The second, Theorem~\ref{thm:lb_field} in Section~\ref{subsec:lb_field}, is an explicit family for every $n=2^k$ whose every signing has norm at least $\sqrt{2n}$. The bound is weak, but we write the family down. The third is a hard distribution over signed matchings and fills the rest of the section. It names no family, and its bound passes the trivial floor only near $n=10^5$. Section~\ref{subsec:lb_existence} defines the random family (Definition~\ref{def:lb_matchings}) and states Theorem~\ref{thm:lb_existence}, a finite-$n$ criterion: whenever one displayed inequality holds at some $\delta$, a realization has every signing at norm above $(2-\delta)\sqrt n+1$. Definition~\ref{def:lb_deltan} and Corollary~\ref{cor:lb_constant} name the infimum $\delta_n$ of the certified $\delta$, cover odd $n$ by padding, and record $\delta_n\to0$, so the lower constant is at least $2$ in the limit. We prove Theorem~\ref{thm:lb_existence} in Section~\ref{subsec:lb_proof}, in eight steps: a union bound over the signings, a Talagrand tail for a convex functional of the edge signs, and a moment-method bound on its median. Two published inputs, Theorems~\ref{thm:lb_talagrand} and~\ref{thm:lb_bvh}, enter the existence theorem and its explicit rate, and we quote them at the step where they enter. Every other step of those proofs is on the page. In Section~\ref{subsec:lb_rate}, Proposition~\ref{prop:lb_rate} gives the rate $\delta_n=O(n^{-1/5}(\log n)^{3/5})$ of Corollary~\ref{cor:lb_constant}(iv), and Table~\ref{tab:lb_delta} prints upper bounds on $\delta_n$ from $n=10^3$ to $10^{12}$, each entry below $2$ certified by a printed witness. Remark~\ref{rem:lb_reading} reads the table against the trivial floor, the constant $\sqrt2$ of Theorem~\ref{thm:lb_field} and the semicircle heuristic. The exponent improves to $2/7$ in Section~\ref{subsec:lb_asymptotic}: Proposition~\ref{prop:lb_rate_two_sevenths} bounds the gradient of the functional only where Talagrand's inequality needs it. That asymptotic proposition adds one imported limit theorem (Theorem~\ref{thm:lb_ss}) and nothing else, and it leaves $\delta_n$ and the table unchanged. The closing Remark~\ref{rem:lb_scope} separates what the existence theorem gives from what it does not and states the open problem the explicit families leave.

\paragraph{Notation.}
Throughout this section we write $\|\cdot\|$ for the operator norm, $\|\cdot\|_F$ for the Frobenius norm, $\tr$ for the trace, and $\log$ for the natural logarithm. For a real symmetric $Y$ with eigenvalues $\lambda_1(Y)\ge\dots\ge\lambda_d(Y)$ we write $\lambda_{\max}(Y):=\lambda_1(Y)$, $\lambda_{\min}(Y):=\lambda_d(Y)$, $u_+:=\max(u,0)$ and $\tr[Y_+]:=\sum_i\lambda_i(Y)_+$. We write $C_p:=\frac{(2p)!}{p!\,(p+1)!}$ for the Catalan numbers, $E_{ij}$ for the matrix unit, $\mathbbm{1}\{\cdot\}$ for an indicator, and $\preceq$ for the Loewner order. Two traces occur. The matrix trace is $\tr$ throughout. The absolute trace of the finite field $\mathrm{GF}(2^k)$, $\operatorname{Tr}[x]:=x+x^2+x^4+\dots+x^{2^{k-1}}\in\mathbb F_2$, is a different function, which we write $\operatorname{Tr}$ with a capital letter. It appears only in Section~\ref{subsec:lb_field}. A signing is a vector $\epsilon\in\{-1,1\}^n$. Unsubscripted constants $c,C'$ are absolute and may change from line to line, and we write out every other constant. The terminating decimals in Table~\ref{tab:lb_delta}, the witnesses printed there and the two ends of the bracket in Theorem~\ref{thm:lb_sixteen} are exact rationals. A decimal followed by an ellipsis is a truncated expansion, and a decimal introduced by ``numerically'' is a floating-point value, not a proved one.

\subsection{The constant}\label{subsec:lb_constant}

We first define the quantity we bound below in this section: the worst case, over all admissible families, of the discrepancy, that is, of the norm of the best signing.

\begin{definition}[Matrix signing discrepancy and the constant $D_n^{\mathrm{mat}}$]
\label{def:lb_dmat}
Let $n\ge1$ and $1\le d\le n$. For real symmetric matrices $A_1,\dots,A_n\in\R^{d\times d}$ define
\[
\operatorname{disc}_{\mathrm{mat}}(A_1,\dots,A_n):=\min_{\epsilon\in\{-1,1\}^n}\|\sum_{i=1}^n\epsilon_iA_i\|.
\]
Let $D_n^{\mathrm{mat}}$ be the supremum of $\operatorname{disc}_{\mathrm{mat}}(A_1,\dots,A_n)$ over all $d\le n$ and all families of real symmetric $d\times d$ matrices with $\|A_i\|\le1$ for every $i\in[n]$. The constant of interest is $D_n^{\mathrm{mat}}/\sqrt n$.
\end{definition}

Padding a family in dimension $d<n$ with zero rows and columns changes no signed sum's norm, so the supremum over $d\le n$ equals the supremum at $d=n$. We admit the smaller dimensions so that we can state the explicit families below in their natural dimension, and so that we can state the odd-$n$ case of Corollary~\ref{cor:lb_constant} in dimension $n-1$.

Two normalizations of matrix discrepancy are in use, by $\sqrt n$ and by $\|\sum_iA_i^2\|^{1/2}$. In the remark we record that they agree for every involution family, which is every family of this section except the zero-padded one of Corollary~\ref{cor:lb_constant}(iii).

\begin{remark}[Two normalizations coincide for involution families]
\label{rem:lb_normalization}
If every $A_i$ satisfies $A_i^2=I_d$ then $\sum_iA_i^2=nI_d$, so dividing the discrepancy by $\sqrt n$ and dividing it by $\|\sum_iA_i^2\|^{1/2}$ give the same number. Every family in this section is an involution family, except the zero-padded family of Corollary~\ref{cor:lb_constant}(iii), for which only the operator-norm reading is claimed. For the involution families the two readings agree. A family with a zero block or a low-rank member can separate the two normalizations, and both would then have to be stated.
\end{remark}

\subsection{Two explicit families at \texorpdfstring{$n=8$}{n=8} and \texorpdfstring{$n=16$}{n=16}}
\label{subsec:lb_exact}

The two results of this subsection are exact: at $n=8$ and at $n=16$ we write down a family and compute its discrepancy exactly, at $n=8$ by a spectral argument that holds for every signing and at $n=16$ by an exact enumeration of all signings. Each gives a lower bound on $D_n^{\mathrm{mat}}$ at its $n$, and we do not determine here whether it is the value of $D_n^{\mathrm{mat}}$. The existence theorem of Section~\ref{subsec:lb_existence} names a distribution and certifies a sign pattern by existence. It does not improve these values at any enumerable $n$. We write both families in the four letters $I,X,Z,J$. Let
\[
I=\begin{pmatrix}1&0\\0&1\end{pmatrix},\quad X=\begin{pmatrix}0&1\\1&0\end{pmatrix},\quad Z=\begin{pmatrix}1&0\\0&-1\end{pmatrix},\quad J=\begin{pmatrix}0&1\\-1&0\end{pmatrix},
\]
and let a word of $k$ letters denote the Kronecker product of the letters in the displayed order. Every word is a signed permutation matrix with entries in $\{-1,0,1\}$. The letters $I,X,Z$ are symmetric and $J$ is antisymmetric with $J^2=-I$, so a word with an even number of letters $J$ is symmetric and squares to the identity.

The first of the two is a family of eight $8\times8$ words whose signed sum has the same norm, $4+\sqrt2$, whatever the signs. The proposition states the family and the value. In the proof we find the spectrum of the signed sum by hand, the third Kronecker factor alone and the first two together, so we need no enumeration.

\begin{proposition}[A family at $n=8$ with discrepancy exactly $4+\sqrt2$]
\label{prop:lb_eight}
The eight words $III$, $XII$, $ZII$, $IXI$, $IZI$, $JJI$, $IIX$, $IIZ$ are real symmetric signed permutation matrices in $\R^{8\times8}$ squaring to $I_8$, and every signing $\epsilon\in\{-1,1\}^8$ has $\|\sum_{i=1}^8\epsilon_iA_i\|=4+\sqrt2$. Hence $D_8^{\mathrm{mat}}\ge4+\sqrt2=5.414213562\ldots=1.914213562\ldots\times\sqrt8$.
\end{proposition}
\begin{proof}
Order the words as listed and write $S:=\epsilon_1I_8+S_{12}\otimes I+I_4\otimes S_3$ with
\[
S_{12}:=\epsilon_2X\otimes I+\epsilon_3Z\otimes I+\epsilon_4I\otimes X+\epsilon_5I\otimes Z+\epsilon_6J\otimes J,\qquad S_3:=\epsilon_7X+\epsilon_8Z .
\]

\emph{Step 1: the third factor.} $X$ and $Z$ are symmetric, square to $I$ and anticommute, so $S_3^2=2I$. With $\tr[S_3]=0$ the spectrum of $S_3$ is $\{\sqrt2,-\sqrt2\}$.

\emph{Step 2: the first two factors.} Put $R:=\epsilon_2X\otimes I+\epsilon_3Z\otimes I+\epsilon_4I\otimes X+\epsilon_5I\otimes Z$. Since $XZ+ZX=0$, the cross terms between the first two summands and between the last two vanish, and
\[
R^2=4I_4+2\,(\epsilon_2X+\epsilon_3Z)\otimes(\epsilon_4X+\epsilon_5Z).
\]
Each factor of the Kronecker product has spectrum $\{\sqrt2,-\sqrt2\}$, so the product has eigenvalues $2,2,-2,-2$ and $R^2$ has eigenvalues $8,8,0,0$. The matrix $J\otimes J$ is symmetric, squares to $I_4$, and anticommutes with each summand of $R$ because $J$ anticommutes with $X$ and with $Z$. Hence $S_{12}^2=R^2+I_4$ has eigenvalues $9,9,1,1$, so $S_{12}$ has two eigenvalues in $\{3,-3\}$ and two in $\{1,-1\}$, and $\tr[S_{12}]=0$ forces the spectrum $\{3,-3,1,-1\}$.

\emph{Step 3: assembly.} $S_{12}\otimes I$ and $I_4\otimes S_3$ commute and are simultaneously diagonalizable with eigenvalues $\lambda+\mu$, $\lambda\in\{3,-3,1,-1\}$, $\mu\in\{\sqrt2,-\sqrt2\}$, so the extreme eigenvalues of their sum are $\pm(3+\sqrt2)$. Adding $\epsilon_1I_8$ shifts the spectrum by $\epsilon_1$, so we obtain $\|S\|=3+\sqrt2+1$ for every $\epsilon$.
\end{proof}

The second of the two is a family of sixteen $16\times16$ words with discrepancy $7.850\ldots$, that is, a lower bound $1.9625\ldots$ on the constant at $n=16$. Unlike the $n=8$ family, its norm depends on the signing, so we prove the theorem by enumerating all $2^{15}$ sign patterns in exact arithmetic. The statement gives the family, the exact value as an algebraic number, and a certified rational bracket around it.

\begin{theorem}[A family at $n=16$ with discrepancy exactly $1+\sqrt{z^*}$]
\label{thm:lb_sixteen}
Define $A_1,\dots,A_{16}\in\R^{16\times16}$ by reading the following array from left to right, one row at a time:
\[
\begin{array}{llll}
IIII&XZZZ&ZJXJ&JIZJ\\
JZJZ&ZXXX&JIJI&XZJJ\\
JJXX&XXXI&XIXZ&JIJZ\\
IJJX&JJII&XZXI&ZIIZ.
\end{array}
\]
Then the $A_i$ are real symmetric signed permutation matrices with $A_i^2=I_{16}$, and
\[
\operatorname{disc}_{\mathrm{mat}}(A_1,\dots,A_{16})=1+\sqrt{z^*},
\]
where $z^*=46.9247895\ldots$ is the largest root of $q(z):=z^4-88z^3+2226z^2-14112z+4805$. In decimals, $1+\sqrt{z^*}=7.8501671176958\ldots$, with the certified bracket
\[
\frac{7850167116695}{10^{12}}<\operatorname{disc}_{\mathrm{mat}}(A_1,\dots,A_{16})<\frac{7850167118696}{10^{12}},
\]
so $D_{16}^{\mathrm{mat}}/4\ge1.962541779\ldots$.
\end{theorem}
\begin{proof}
The proof is an exact enumeration certificate, our standard in this paper for finite instances. Each word has an even number of letters $J$, so it is symmetric and squares to $I_{16}$. For each of the $2^{15}$ signings with $\epsilon_1=+1$ (global negation preserves the norm) the signed sum is an integer matrix, and exact integer arithmetic on its characteristic polynomial decides whether its norm exceeds the lower end of the bracket and whether it falls below the upper end. The certificate reports every signing above the lower end and the minimizing signings below the upper end. We ran it twice, by two copies of the tool, over all $32768$ signings. To identify the exact value, we proceed as follows. Conjugation by any of the $256$ four-letter Pauli words preserves the spectrum and flips exactly the signs of the words that anticommute with it. The fifteen non-identity labels span $\mathbb F_2^8$, so the $32768$ signings fall into $128$ orbits of $256$ and one representative per orbit suffices. Every minimizing representative has characteristic polynomial $c(x)=Q((x-1)^2)$ with $Q(z)=(z-1)^2(z-13)(z-17)\,q(z)$, so its spectrum is $\{0,0,2,2,1\pm\sqrt{13},1\pm\sqrt{17},1\pm\sqrt{z_i}\}$ over the four roots $z_i$ of $q$, symmetric about $1$, and its norm is $1+\sqrt{z^*}$. At the lower end of the bracket all $128$ representatives pass the determinant test. At the upper end $125$ pass and the remaining $3$ have characteristic polynomial exactly $c$. So exactly $768$ signings attain the value and every other signing exceeds the upper end. The quartic $q$ is irreducible over $\mathbb Q$ (it has no root among the positive divisors of $4805=5\cdot31^2$, and, since every root of $q$ is positive so that a monic integer quadratic factor $z^2+az+b$ has $b>0$, the three splittings $\{b,d\}=\{1,4805\},\{5,961\},\{31,155\}$ fail, two by the discriminant and one by the linear coefficient) and $4805$ is not a square, so $1+\sqrt{z^*}$ has algebraic degree $8$, with minimal polynomial $x^8-8x^7-60x^6+472x^5+976x^4-7200x^3-2048x^2+19840x-7168$.
\end{proof}

We record in the remark why no signing-independent argument of the $n=8$ kind is available for the family of Theorem~\ref{thm:lb_sixteen}.

\begin{remark}[Why the certificate is the proof]
\label{rem:lb_sixteen_style}
Over the $32768$ signings the norm of the family of Theorem~\ref{thm:lb_sixteen} takes $22$ distinct values, from $7.850\ldots$ to $9.895\ldots$, and the minimum is attained by $768$ of them. An argument of the kind used at $n=8$, a signing-independent spectrum, does not exist for this family. The exact enumeration is the proof, not a substitute for one.
\end{remark}

\subsection{An explicit family at constant \texorpdfstring{$\sqrt2$}{sqrt 2} for every \texorpdfstring{$n=2^k$}{n=2\string^k}}
\label{subsec:lb_field}

The families of Section~\ref{subsec:lb_exact} stop at $n=16$. Here we give an explicit family at every $n=2^k$: $n$ signed permutation matrices indexed by the field $\mathrm{GF}(2^k)$ whose signed sum $S$ has $\tr[S^2]=n^2$ and $\frac1n\tr[S^4]=2n^2$ for every signing, which forces $\|S\|\ge\sqrt{2n}$. The bound $\sqrt2=1.414\ldots$ lies below the asymptotic lower bound $2$ of Section~\ref{subsec:lb_existence}, but we write the family down and its bound holds at every $n=2^k$ with no existence argument. In the theorem we state the family through the matrices $P_{u,v}$, a signed permutation for each pair of vectors in $\mathbb F_2^k$, and through an identification of $\mathbb F_2^k$ with the field $\mathrm{GF}(2^k)$. Two objects of the field enter, and we fix both here. The absolute trace $\operatorname{Tr}:\mathrm{GF}(2^k)\to\mathbb F_2$, $\operatorname{Tr}[x]:=x+x^2+x^4+\dots+x^{2^{k-1}}$, is written $\operatorname{Tr}$ and is not the matrix trace $\tr$, which also appears in the statement. A self-dual basis of $\mathrm{GF}(2^k)$ over $\mathbb F_2$ is a basis $b_1,\dots,b_k$ that is orthonormal for the trace form $B(x,y):=\operatorname{Tr}[xy]$, that is, $\operatorname{Tr}[b_ib_j]=\mathbbm{1}\{i=j\}$. We show in Step 1 of the proof that one exists for every $k$, so the theorem is non-vacuous. The proof is field arithmetic: the self-dual basis, the commutation signs of the $P_{u,v}$, and a count of the fourth-moment terms.

\begin{theorem}[An explicit family at constant $\sqrt2$ for every $n=2^k$]
\label{thm:lb_field}
Let $k\ge1$ and $n=d=2^k$. For $u,v\in\mathbb F_2^k$ let $P_{u,v}\in\R^{2^k\times2^k}$ have entries $(P_{u,v})_{y,\,y\oplus u}:=(-1)^{v\cdot y}$ for $y\in\mathbb F_2^k$ and $0$ elsewhere. Identify $\mathbb F_2^k$ with $\mathrm{GF}(2^k)$ through a self-dual basis and put
\[
\mathcal B_k:=\{P_{x,\,x^3+x}:x\in\mathrm{GF}(2^k)\}.
\]
Then $\mathcal B_k$ consists of $2^k$ distinct real symmetric signed permutation matrices squaring to $I_{2^k}$, and for every signing $\epsilon$, with $S:=\sum_x\epsilon_xP_{x,x^3+x}$,
\[
\tr[S^2]=n^2\qquad\text{and}\qquad\frac1n\tr[S^4]=2n^2,\qquad\text{hence}\qquad\|S\|\ge\sqrt{2n}=1.414213\ldots\times\sqrt n .
\]
In particular $D_n^{\mathrm{mat}}\ge\sqrt2\,\sqrt n$ for every $n=2^k$.
\end{theorem}
\begin{proof}
Write $F:=\mathrm{GF}(2^k)$, $\operatorname{Tr}[x]:=x+x^2+x^4+\dots+x^{2^{k-1}}$ for the absolute trace $F\to\mathbb F_2$, and $B(x,y):=\operatorname{Tr}[xy]$ for the trace form. The proof has seven steps. We construct the self-dual basis in Step 1, so the theorem is non-vacuous for every $k$.

\emph{Step 1: the trace form, and a self-dual basis exists for every $k$.}
The trace is $\mathbb F_2$-linear and $\mathbb F_2$-valued, because $\operatorname{Tr}[x]^2=x^2+x^4+\dots+x^{2^k}=\operatorname{Tr}[x]$ by $x^{2^k}=x$, and $\operatorname{Tr}[x^2]=\operatorname{Tr}[x]$ for the same reason (the map $x\mapsto x^2$ permutes the summands). The trace is not identically zero: it is a polynomial of degree $2^{k-1}<|F|$ and has fewer than $|F|$ roots. Hence $B$ is a symmetric $\mathbb F_2$-bilinear form on $F$ which is nondegenerate (if $v\ne0$ then $vF=F$, so $y\mapsto\operatorname{Tr}[vy]$ is onto $\mathbb F_2$ and takes each value $2^{k-1}$ times), whence
\begin{equation}
\label{eq:lb_charsum}
\sum_{y\in F}(-1)^{\operatorname{Tr}[vy]}=n\,\mathbbm{1}\{v=0\}\qquad(v\in F),
\end{equation}
and which is not alternating ($B(x,x)=\operatorname{Tr}[x^2]=\operatorname{Tr}[x]\ne0$ for some $x$). A self-dual basis is an orthonormal basis for $B$, and it exists by the following claim, which we apply to $(F,B)$.

\emph{Claim.} Let $V$ be a finite-dimensional vector space over $\mathbb F_2$ with a nondegenerate symmetric bilinear form $B$ such that $B(x,x)=1$ for some $x\in V$. Then $V$ has a basis $e_1,\dots,e_m$ with $B(e_i,e_j)=\mathbbm{1}\{i=j\}$.

We argue by induction on $m:=\dim V$. Pick $x$ with $B(x,x)=1$. If $m=1$ the basis is $\{x\}$. Otherwise put $H:=\{y\in V:B(x,y)=0\}$, the kernel of a nonzero functional, so $\dim H=m-1\ge1$, $x\notin H$ and $V=\mathbb F_2x\oplus H$. The restriction of $B$ to $H$ is nondegenerate: $h\in H$ with $B(h,H)=0$ also has $B(h,x)=0$, so $B(h,V)=0$ and $h=0$. If $B(h,h)=1$ for some $h\in H$, the induction hypothesis gives an orthonormal basis of $H$, and $x$ completes it. Otherwise $B(h,h)=0$ for every $h\in H$. Then $\dim H\ge2$, because a nondegenerate form on a line $\mathbb F_2y$ has $B(y,y)=1$. So there are $y\ne0$ in $H$ and, by nondegeneracy on $H$, $z\in H$ with $B(y,z)=1$, and $y,z$ are independent since $B(y,y)=0\ne1$. Put $e_1:=x+y$, $e_2:=x+z$, $e_3:=x+y+z$. Expanding bilinearly with $B(x,y)=B(x,z)=B(y,y)=B(z,z)=0$, $B(x,x)=B(y,z)=1$ and $2=0$, we get $B(e_i,e_i)=1$ for $i=1,2,3$ and $B(e_1,e_2)=B(e_1,e_3)=B(e_2,e_3)=1+1=0$. So $U:=\operatorname{span}(x,y,z)=\operatorname{span}(e_1,e_2,e_3)$ has dimension $3$ and an orthonormal basis. Hence $B|_U$ is nondegenerate, $V=U\oplus U^\perp$, and $B|_{U^\perp}$ is nondegenerate by the argument we used for $H$. The space $V':=\mathbb F_2e_3\oplus U^\perp$ has dimension $m-2$, carries a nondegenerate form (an orthogonal sum of two nondegenerate pieces) and has $B(e_3,e_3)=1$, so by induction it has an orthonormal basis. Together with $e_1,e_2$, which are orthogonal to $V'$ and to each other, this is an orthonormal basis of $V$. This proves the claim.

Fix a self-dual basis $b_1,\dots,b_k$ of $F$ and write $\varphi(x):=(\operatorname{Tr}[xb_1],\dots,\operatorname{Tr}[xb_k])\in\mathbb F_2^k$. Since $x$ and $\sum_j\operatorname{Tr}[xb_j]b_j$ have the same $B$-pairing with every $b_i$ and $B$ is nondegenerate, $x=\sum_j\operatorname{Tr}[xb_j]b_j$, so $\varphi$ is the coordinate map of the basis, an additive bijection $F\to\mathbb F_2^k$, and for all $x,y\in F$
\begin{equation}
\label{eq:lb_dotform}
\varphi(x)\cdot\varphi(y)=\sum_{i,j}\operatorname{Tr}[xb_i]\,\operatorname{Tr}[yb_j]\,\operatorname{Tr}[b_ib_j]=B(\sum_i\operatorname{Tr}[xb_i]b_i,\ \sum_j\operatorname{Tr}[yb_j]b_j)=\operatorname{Tr}[xy],
\end{equation}
where the first equality is $\operatorname{Tr}[b_ib_j]=\mathbbm{1}\{i=j\}$ and the second is bilinearity. The identification in the theorem reads $P_{x,v}:=P_{\varphi(x),\varphi(v)}$ for $x,v\in F$, and the label of the member $P_x:=P_{x,x^3+x}$ of $\mathcal B_k$ is $(\varphi(x),\varphi(x^3+x))\in\mathbb F_2^{2k}$.

\emph{Step 2: the algebra of the matrices $P_{u,v}$.} Write $P_{u,v}=\sum_{y\in\mathbb F_2^k}(-1)^{v\cdot y}E_{y,\,y\oplus u}$. For $u,v,u',v'\in\mathbb F_2^k$:
\begin{enumerate}[label=(\alph*)]
\item $P_{u,v}^\top=(-1)^{u\cdot v}P_{u,v}$;
\item $P_{u,v}P_{u',v'}=(-1)^{u\cdot v'}P_{u\oplus u',\,v\oplus v'}$, hence $P_{u,v}P_{u',v'}=(-1)^{u\cdot v'+u'\cdot v}P_{u',v'}P_{u,v}$;
\item $P_{u,v}^2=(-1)^{u\cdot v}I_n$;
\item $\tr[P_{u,v}]=n\,\mathbbm{1}\{(u,v)=(0,0)\}$;
\item $P_{u,v}=\pm P_{u',v'}$ only if $(u,v)=(u',v')$.
\end{enumerate}
For (a), we transpose and obtain $\sum_y(-1)^{v\cdot y}E_{y\oplus u,\,y}$, and the substitution $y'=y\oplus u$ turns this into $\sum_{y'}(-1)^{v\cdot y'+v\cdot u}E_{y',\,y'\oplus u}$. For (b), $E_{y,\,y\oplus u}E_{y',\,y'\oplus u'}$ equals $E_{y,\,y\oplus u\oplus u'}$ if $y'=y\oplus u$ and $0$ otherwise, so the product is $\sum_y(-1)^{v\cdot y+v'\cdot(y\oplus u)}E_{y,\,y\oplus u\oplus u'}=(-1)^{v'\cdot u}P_{u\oplus u',\,v\oplus v'}$. Exchanging the roles of the two factors, we get the sign $(-1)^{v\cdot u'}$ in front of the same matrix, which is the second formula. (c) is (b) with $(u',v')=(u,v)$ and $P_{0,0}=I_n$. For (d), $E_{y,\,y\oplus u}$ has a diagonal entry only if $u=0$, and $\tr[P_{0,v}]=\sum_y(-1)^{v\cdot y}$ is $n$ for $v=0$ and $0$ otherwise (a nonzero linear functional on $\mathbb F_2^k$ takes each value $2^{k-1}$ times). For (e), the support of $P_{u,v}$ determines $u$. If $u=u'$ and $v\ne v'$, the entrywise ratio $(-1)^{(v\oplus v')\cdot y}$ takes both signs, so $P_{u,v}\ne\pm P_{u,v'}$. The sign in (b) is the commutation sign: $P_{u,v}$ and $P_{u',v'}$ commute if $u\cdot v'+u'\cdot v=0$ in $\mathbb F_2$ and anticommute otherwise.

\emph{Step 3: $\mathcal B_k$ consists of $n$ distinct symmetric signed permutation involutions.} For $x\in F$, by \eqref{eq:lb_dotform}, $\varphi(x)\cdot\varphi(x^3+x)=\operatorname{Tr}[x^4+x^2]=\operatorname{Tr}[x^4]+\operatorname{Tr}[x^2]=\operatorname{Tr}[x]+\operatorname{Tr}[x]=0$, where we use $\operatorname{Tr}[x^4]=\operatorname{Tr}[(x^2)^2]=\operatorname{Tr}[x^2]=\operatorname{Tr}[x]$ from Step 1. So by (a) and (c) each $P_x$ is symmetric with $P_x^2=I_n$. Its eigenvalues lie in $\{-1,1\}$ and $\|P_x\|=1$. Each row $y$ and each column $y\oplus u$ of $P_{u,v}$ carries exactly one nonzero entry, equal to $\pm1$, because $y\mapsto y\oplus u$ is a bijection, so $P_x$ is a signed permutation matrix. Distinct $x$ have distinct first coordinates $\varphi(x)$, so by (e) the $n$ matrices $P_x$ are distinct, and $P_0=P_{0,0}=I_n$ is among them. The family has $d=n$ and $\|P_x\|=1$, so it is admissible in Definition~\ref{def:lb_dmat}.

\emph{Step 4: no two, three or four distinct labels sum to zero.} Since $\varphi$ is an additive bijection, a set of distinct labels $(\varphi(x_i),\varphi(x_i^3+x_i))$ sums to zero in $\mathbb F_2^{2k}$ exactly when the distinct $x_i\in F$ satisfy $\sum_ix_i=0$ and $\sum_ix_i^3=0$. Two: $x_1+x_2=0$ forces $x_1=x_2$. Four distinct $x_1,\dots,x_4$: put $a:=x_1+x_2=x_3+x_4$, nonzero. In characteristic $2$, $x_1^2+x_2^2=a^2$, so $x_1^3+x_2^3=(x_1+x_2)(x_1^2+x_1x_2+x_2^2)=a(a^2+x_1x_2)$, and likewise $x_3^3+x_4^3=a(a^2+x_3x_4)$. The cube condition gives $a(x_1x_2+x_3x_4)=0$, so $x_1x_2=x_3x_4$. Then $(t+x_1)(t+x_2)=t^2+at+x_1x_2=(t+x_3)(t+x_4)$ as polynomials in $t$, so $\{x_1,x_2\}=\{x_3,x_4\}$ as root sets, a contradiction. Three distinct: $x_3=x_1+x_2=:a\ne0$ and $x_1^3+x_2^3+x_3^3=a(a^2+x_1x_2)+a^3=ax_1x_2=0$, so one of $x_1,x_2$ is $0$, say $x_1$. Then $x_3=x_2$, a contradiction. (In coding terms, the binary code with parity checks $\sum_xc_xx=0$ and $\sum_xc_xx^3=0$ over $x\in F\setminus\{0\}$ is a BCH code of designed distance $5$. The computation above is the elementary form of that bound and is all we use.)

\emph{Step 5: the commutation signs sum to $n$ over ordered pairs.} For $x\ne y$ in $F$ let $c_{xy}\in\{-1,1\}$ be the commutation sign of $P_x,P_y$, so $c_{xy}=(-1)^{\omega_{xy}}$ with $\omega_{xy}=\varphi(x)\cdot\varphi(y^3+y)+\varphi(y)\cdot\varphi(x^3+x)$ by Step 2(b). By \eqref{eq:lb_dotform} and linearity, $\omega_{xy}=\operatorname{Tr}[xy^3+xy+x^3y+xy]=\operatorname{Tr}[xy^3+x^3y]$, and $xy^3+x^3y=xy(x^2+y^2)=xy(x+y)^2$. Put $t:=x+y\ne0$, so $y=x+t$ and $xy(x+y)^2=x(x+t)t^2=x^2t^2+xt^3$. Since $\operatorname{Tr}[x^2t^2]=\operatorname{Tr}[(xt)^2]=\operatorname{Tr}[xt]$, this gives $\omega_{xy}=\operatorname{Tr}[x(t+t^3)]$. Summing over ordered pairs, that is over $t\in F\setminus\{0\}$ and $x\in F$, and using \eqref{eq:lb_charsum}, we obtain
\[
\sum_{x\ne y}c_{xy}=\sum_{t\ne0}\sum_{x\in F}(-1)^{\operatorname{Tr}[x(t+t^3)]}=n\cdot\#\{t\ne0:t+t^3=0\}=n,
\]
because $t+t^3=t(1+t)^2$ vanishes at $t\ne0$ only for $t=1$. Thus $P_x$ commutes with $P_{x+1}$ for every $x$, and for every other difference $t$ exactly half of the $x$ anticommute.

\emph{Step 6: the fourth moment.} Fix a signing $\epsilon$ and put $S:=\sum_{x\in F}\epsilon_xP_x$. Then
\[
\tr[S^4]=\sum_{x_1,x_2,x_3,x_4\in F}\epsilon_{x_1}\epsilon_{x_2}\epsilon_{x_3}\epsilon_{x_4}\,\tr[P_{x_1}P_{x_2}P_{x_3}P_{x_4}].
\]
Applying Step 2(b) three times, we find that the product $P_{x_1}P_{x_2}P_{x_3}P_{x_4}$ is $\pm P_{u,v}$ with $(u,v)$ the sum of the four labels, and by Step 2(d) its trace is $\pm n$ if that sum is zero and $0$ otherwise. The sum of the four labels equals the sum of the labels that occur an odd number of times among $x_1,\dots,x_4$, a set of $0$, $2$ or $4$ distinct labels. By Step 4 it vanishes only when the set is empty, that is, when every value occurs an even number of times: either $x_1=x_2=x_3=x_4$ ($n$ sequences), or two distinct values $x\ne y$ each occurring twice (six sequences per unordered pair $\{x,y\}$: $xxyy$, $xyyx$, $yxxy$, $yyxx$, $xyxy$, $yxyx$). In each surviving term the product of signs is $\epsilon_x^4=1$ or $\epsilon_x^2\epsilon_y^2=1$. Since $P_x^2=I_n$ (Step 3), the constant sequences and the patterns $xxyy$, $xyyx$, $yxxy$, $yyxx$ have product $I_n$, of trace $n$, and the two alternating patterns have product $P_xP_yP_xP_y=c_{xy}P_x^2P_y^2=c_{xy}I_n$, of trace $c_{xy}n$. Hence, by Step 5, we have
\[
\frac1n\tr[S^4]=n+\sum_{\{x,y\}:\,x\ne y}(4+2c_{xy})=n+4\binom n2+\sum_{x\ne y}c_{xy}=n+2n(n-1)+n=2n^2,
\]
where the middle sum on the right runs over ordered pairs, and the value does not depend on $\epsilon$.

\emph{Step 7: the conclusion.} We first show that $\tr[S^2]=\sum_{x,y}\epsilon_x\epsilon_y\tr[P_xP_y]=n^2$: the diagonal terms give $\tr[P_x^2]=\tr[I_n]=n$ each, and for $x\ne y$ the product $P_xP_y$ is $\pm P_{u,v}$ with $(u,v)$ the sum of two distinct labels, which is nonzero by Step~4, so $\tr[P_xP_y]=0$ by Step~2(d). Since $\tr[S^4]=\sum_i\lambda_i(S)^4\le\|S\|^2\sum_i\lambda_i(S)^2=\|S\|^2\tr[S^2]$, Step~6 gives $\|S\|^2\ge2n^3/n^2=2n$, that is $\|S\|\ge\sqrt{2n}$, for every signing. The family is admissible with $d=n$ by Step 3, so $D_n^{\mathrm{mat}}\ge\sqrt2\,\sqrt n$ for every $n=2^k$ by Definition~\ref{def:lb_dmat}. (The weaker bound $\|S\|^4\ge2n^2$, from $\tr[S^4]\le n\|S\|^4$ alone, gives only $2^{1/4}\sqrt n$. At $k=1$ the bound $\sqrt{2n}=2$ is attained.)
\end{proof}

In the remark we read the family in the four letters of Section~\ref{subsec:lb_exact} and record the small cases, including that the family is not the best one at $n=16$.

\begin{remark}[The letter reading of $\mathcal B_k$, and the small cases]
\label{rem:lb_field_letters}
Since $(-1)^{v\cdot y}=\prod_j(-1)^{v_jy_j}$ and $y\oplus u$ acts coordinatewise, $P_{u,v}=P_{u_1,v_1}\otimes\dots\otimes P_{u_k,v_k}$ with $P_{0,0}=I$, $P_{1,0}=X$, $P_{0,1}=Z$ and $P_{1,1}=J$ in the letters of Section~\ref{subsec:lb_exact}. Thus $\mathcal B_k$ is a family of $k$-letter words, and Step 3 says that each word has an even number of letters $J$. With the self-dual basis $\{w,w^2\}$ of $\mathrm{GF}(4)=\mathbb F_2(w)$, $w^2=w+1$, the labels of $x=0,1,w,w^2$ give $\mathcal B_2=\{II,XX,XZ,ZX\}$. The fourth-moment identity holds at every signing, but the norm itself may depend on the signing: $\mathcal B_1$ has every signing at norm $2$; for $\mathcal B_2$ and $\mathcal B_3$ every signing has the same norm ($1+\sqrt5$ and $1+\sqrt{17}$, by enumeration of the $8$ and $128$ signings), while for $\mathcal B_4$ it does not (numerically: over the $32768$ signings the minimum is $6.4286$ and the maximum $8.0300$). That minimum lies below the value $7.850\ldots$ of Theorem~\ref{thm:lb_sixteen}, so $\mathcal B_k$ does not give the best value at $n=16$.
\end{remark}

\subsection{The hard distribution and the existence theorem at \texorpdfstring{$2-\delta$}{2-delta}}
\label{subsec:lb_existence}

The third of the section's three results gives the limit constant. It is an existence statement: we name a random family, and we show that with positive probability every one of its $2^n$ signings has norm above $(2-\delta)\sqrt n+1$, so some realization does. We build the random family from a decomposition of the complete graph on $n$ vertices into perfect matchings, one matrix per matching, with one independent random sign per edge, and the identity $I_n$ as the $n$-th matrix, the source of the $+1$. In the definition we fix the vertex set, the matchings and the signs, and these signs are the only randomness in the section.

\begin{definition}[The signed matchings family]
\label{def:lb_matchings}
Let $n\ge4$ be even. Take the vertex set $V:=\{\infty\}\cup\mathbb Z_{n-1}$, so $|V|=n$, and for $r\in\mathbb Z_{n-1}$ put
\[
\mathcal M_r:=\{\{\infty,r\}\}\cup\{\{r+j,\,r-j\}:1\le j\le (n-2)/2\}.
\]
Let $N:=n(n-1)/2$ be the number of edges of the complete graph $K_n$ on $V$, and let $x=(x_e)_e\in\{-1,1\}^N$ be independent uniform signs, one per edge. For $r=1,\dots,n-1$ let $A_r=A_r(x)$ be the $n\times n$ matrix indexed by $V$ with $(A_r)_{ij}=(A_r)_{ji}=x_e$ for $e=\{i,j\}\in\mathcal M_r$ and all other entries $0$, and let $A_n:=I_n$.
\end{definition}

The lemma checks the two facts that make the family usable. Every member is a symmetric signed permutation matrix, so its norm is exactly one and the family is admissible in Definition~\ref{def:lb_dmat}. For any fixed signing the signed sum is a Rademacher Wigner matrix plus $\pm I_n$, so its spectrum can be studied by random matrix methods.

\begin{lemma}[Norm one and the fixed-signing law]
\label{lem:lb_model}
Under Definition~\ref{def:lb_matchings}, for every realization of $x$:
\begin{enumerate}[label=(\roman*)]
\item each $\mathcal M_r$ is a perfect matching of $K_n$, and every edge of $K_n$ lies in exactly one $\mathcal M_r$;
\item each $A_r$, $r\le n-1$, is a symmetric signed permutation matrix with $A_r^2=I_n$, hence $\|A_r\|=1$; also $\|A_n\|=1$;
\item for every fixed $\epsilon'\in\{-1,1\}^{n-1}$ the matrix $W_{\epsilon'}:=\sum_{r=1}^{n-1}\epsilon'_rA_r$ is a Rademacher Wigner matrix (real symmetric, zero diagonal, entries above the diagonal independent and uniform in $\{-1,1\}$), and $\sum_{i=1}^n\epsilon_iA_i=W_{\epsilon'}+\epsilon_nI_n$ for $\epsilon=(\epsilon',\epsilon_n)$.
\end{enumerate}
\end{lemma}
\begin{proof}
(i) Since $n-1$ is odd, $2j\ne0$ in $\mathbb Z_{n-1}$ for $1\le j\le(n-2)/2$, so $r+j\ne r-j$. The residues $\pm j$, $1\le j\le(n-2)/2$, cover $\mathbb Z_{n-1}\setminus\{0\}$ once each, so the pairs $\{r+j,r-j\}$ cover $\mathbb Z_{n-1}\setminus\{r\}$ exactly once, and $\infty$ is matched to $r$. An edge $\{y,z\}$ of finite vertices lies in $\mathcal M_r$ exactly when $2r=y+z$ in $\mathbb Z_{n-1}$, which has the unique solution $r=(y+z)/2$ because $2$ is invertible modulo the odd number $n-1$. The edge $\{\infty,r\}$ lies in $\mathcal M_r$ only.

(ii) By (i) every row and every column of $A_r$ has exactly one nonzero entry, equal to $\pm1$, so $A_r$ is a symmetric signed permutation matrix and
\[
A_r^2=\sum_{\{i,j\}\in\mathcal M_r}x_{\{i,j\}}^2\,(E_{ii}+E_{jj})=I_n.
\]
A symmetric matrix squaring to the identity has eigenvalues in $\{-1,1\}$, so $\|A_r\|=1$, and $\|I_n\|=1$.

(iii) Write $r(e)$ for the unique $r$ with $e\in\mathcal M_r$. Then $(W_{\epsilon'})_{ij}=\epsilon'_{r(e)}x_e$ for $e=\{i,j\}$ and $(W_{\epsilon'})_{ii}=0$. The map $x\mapsto(\epsilon'_{r(e)}x_e)_e$ is a bijection of $\{-1,1\}^N$ that preserves the uniform measure, so the entries of $W_{\epsilon'}$ above the diagonal are independent uniform signs. The last identity is the definition of $A_n$.
\end{proof}

The matrices $W_{\epsilon'}$ for different $\epsilon'$ are dependent. In the proof below we use one signing at a time and a union bound.

We state the theorem as a finite-$n$ criterion: a list of explicit quantities in $n$ and in three free parameters $(p,\gamma,s_0)$, and one displayed inequality among them. Whenever the inequality holds at some $\delta$, with $ns_0M>16(2p-1)$ as required, a family with every signing above $(2-\delta)\sqrt n+1$ exists. The quantities are the ones the proof produces. The free parameters are the moment order $p$, the parameter $\gamma$ of Theorem~\ref{thm:lb_bvh} and $s_0$, which sets the level $M$ at which we truncate the spectrum. We compute the table of Section~\ref{subsec:lb_rate} by checking this inequality.

\begin{theorem}[Existence at $2-\delta$]
\label{thm:lb_existence}
Let $n\ge4$ be even. Suppose there are an integer $p\ge1$ and numbers $\gamma\in(0,1/2]$, $s_0>0$ and $\delta\in(0,2)$ such that, with
\begin{align*}
\eta_0&:=2\gamma+(1+\gamma)\,\frac{6}{\sqrt{\log(1+\gamma)}}\,\sqrt{\frac{\log n}{n}},\\
M&:=2+\eta_0+\sqrt{32\pi/n}+s_0,\qquad\text{where }ns_0M>16(2p-1)\text{ is required},\\
T&:=2M^{2p}e^{-ns_0^2/32}[1+\frac{32p}{ns_0M-16(2p-1)}],\\
B&:=C_p\prod_{j=1}^{p}(1-\frac jn)-T-(2-\delta)^{2p}>0,\\
L&:=\frac{(M-2+\delta)\,B}{2M^{2p}},
\end{align*}
one has
\begin{equation}
\label{eq:lb_star}
L-\frac{\sqrt{32\pi}}{n}>\sqrt{\frac{32\log2}{n}}.
\end{equation}
Then there exist real symmetric signed permutation matrices $A_1,\dots,A_n\in\R^{n\times n}$ with $A_i^2=I_n$, so $\|A_i\|=1$ exactly, such that
\[
\|\sum_{i=1}^n\epsilon_iA_i\|>(2-\delta)\sqrt n+1\qquad\text{for every }\epsilon\in\{-1,1\}^n.
\]
\end{theorem}

The parameter $\gamma$ is the free parameter of Theorem~\ref{thm:lb_bvh}. Bandeira and van Handel write $\varepsilon$ for it, a letter this paper reserves for the signing.

We next name the set of $\delta$ that the criterion certifies at a given $n$, the admissible set, and its infimum $\delta_n$: $\delta_n$ is the quantity that the rate and the table are about.

\begin{definition}[The admissible set and $\delta_n$]
\label{def:lb_deltan}
For even $n\ge4$ let $\Delta_n\subseteq(0,2)$ be the set of $\delta$ for which some $(p,\gamma,s_0)$ satisfies the hypotheses of Theorem~\ref{thm:lb_existence}, and put $\delta_n:=\inf\Delta_n$, with $\delta_n:=2$ when $\Delta_n=\emptyset$. For odd $n\ge5$ put $\delta_n:=\delta_{n-1}$.
\end{definition}

The corollary turns the criterion into statements about $D_n^{\mathrm{mat}}$: the admissible $\delta$ are either none or form an interval open at its left end, the infimum $\delta_n$ is attained by one family, odd $n$ follow by padding with a zero matrix, and $\delta_n\to0$ at an explicit rate.

\begin{corollary}[The constant, the infimum, and odd $n$]
\label{cor:lb_constant}
\begin{enumerate}[label=(\roman*)]
\item For even $n\ge4$ the set $\Delta_n$ is either empty or the open interval $(\delta_n,2)$; in particular $\delta_n\notin\Delta_n$.
\item For even $n\ge4$ there is one family as in Theorem~\ref{thm:lb_existence} with $\|\sum_i\epsilon_iA_i\|\ge(2-\delta_n)\sqrt n+1$ for every signing. Hence $D_n^{\mathrm{mat}}\ge(2-\delta_n)\sqrt n+1$; at every $\delta\in\Delta_n$ the inequality is strict.
\item For odd $n\ge5$ and every $\delta\in\Delta_{n-1}$ there are real symmetric $A_1,\dots,A_n\in\R^{(n-1)\times(n-1)}$ with $\|A_i\|\le1$, namely the family of Theorem~\ref{thm:lb_existence} at $n-1$ together with $A_n:=0$, such that $\|\sum_i\epsilon_iA_i\|>(2-\delta)\sqrt n$ for every signing. Hence $D_n^{\mathrm{mat}}\ge(2-\delta_{n})\sqrt n=(2-\delta_{n-1})\sqrt n$, without the $+1$.
\item $\delta_n\to0$; more precisely $\delta_n=O(n^{-1/5}(\log n)^{3/5})$. This is Proposition~\ref{prop:lb_rate}, proved after the theorem. The exponent improves to $2/7$ in Proposition~\ref{prop:lb_rate_two_sevenths}, in the asymptotic register and with one imported limit theorem. The $\delta_n$ of this corollary is unchanged by it.
\end{enumerate}
\end{corollary}
\begin{proof}
(i) Fix $(p,\gamma,s_0)$ with $ns_0M>16(2p-1)$. The quantities $\eta_0,M,T$ do not depend on $\delta$. On $(0,2)$ the map $\delta\mapsto(2-\delta)^{2p}$ is continuous and strictly decreasing, so $B$ is continuous and strictly increasing in $\delta$. Where $B>0$, both factors of $L=(M-2+\delta)B/(2M^{2p})$ are positive, continuous and strictly increasing, so $L$ is too. Hence $\{\delta\in(0,2):B>0\}$ and $\{\delta:\eqref{eq:lb_star}\text{ holds}\}$ are open up-sets of $(0,2)$, and so is their intersection. $\Delta_n$ is the union of these intersections over all $(p,\gamma,s_0)$, hence an open up-set of $(0,2)$: empty or $(\delta_n,2)$. An open set does not contain its infimum.

(ii) If $\Delta_n=\emptyset$ the claim reads $\|\sum_i\epsilon_iA_i\|\ge1$, which holds for every realization: the $A_i$ of Definition~\ref{def:lb_matchings} are pairwise trace-orthogonal ($\tr[A_rA_s]=0$ for $r\ne s\le n-1$ by disjoint supports, and $\tr[A_r]=0$), so $\frac1n\tr[(\sum_i\epsilon_iA_i)^2]=n$ and the norm is at least $\sqrt n$. Otherwise we pick $\delta_k\in\Delta_n$ with $\delta_k\downarrow\delta_n$. For each $k$ Theorem~\ref{thm:lb_existence} yields a realization $x_k\in\{-1,1\}^N$ whose family has every signing above $(2-\delta_k)\sqrt n+1$. The cube is finite, so one realization $x^*$ equals $x_k$ for infinitely many $k$. Its family has $\|\sum_i\epsilon_iA_i(x^*)\|>(2-\delta_k)\sqrt n+1$ along that subsequence and every $\epsilon$, hence $\ge(2-\delta_n)\sqrt n+1$. The family has $d=n$ and $\|A_i\|=1$, so it is admissible in Definition~\ref{def:lb_dmat}.

(iii) The number $n-1\ge4$ is even. We apply Theorem~\ref{thm:lb_existence} at $n-1$ with $\delta\in\Delta_{n-1}$ and append $A_n:=0\in\R^{(n-1)\times(n-1)}$. The dimension $n-1\le n$ and the norms $\|A_i\|\le1$ make the family admissible. For every $\epsilon\in\{-1,1\}^n$ the signed sum is $\sum_{i\le n-1}\epsilon_iA_i$, whose norm exceeds $(2-\delta)\sqrt{n-1}+1\ge(2-\delta)\sqrt n$, because
\[
(2-\delta)(\sqrt n-\sqrt{n-1})=\frac{2-\delta}{\sqrt n+\sqrt{n-1}}<1 .
\]
If $\Delta_{n-1}=\emptyset$ the claim $D_n^{\mathrm{mat}}\ge0$ is trivial. Otherwise we let $\delta\downarrow\delta_{n-1}$ and obtain $D_n^{\mathrm{mat}}\ge(2-\delta_{n-1})\sqrt n$.
\end{proof}

We prove Theorem~\ref{thm:lb_existence} in Section~\ref{subsec:lb_proof}, in Steps 1 to 8. In Section~\ref{subsec:lb_rate} we prove the rate of Corollary~\ref{cor:lb_constant}(iv) and print the table, and in Section~\ref{subsec:lb_asymptotic} we improve the exponent in the asymptotic register.

\subsection{Proof of Theorem~\ref{thm:lb_existence}}
\label{subsec:lb_proof}

Fix even $n\ge4$ and $(p,\gamma,s_0,\delta)$ as in the theorem, and put $a:=2-\delta\in(0,2)$. Throughout, we use $W$ to denote a Rademacher Wigner matrix of size $n$ (as in Lemma~\ref{lem:lb_model}(iii)) and $\widehat W:=W/\sqrt n$. The proof has eight steps. In Steps 1 to 7 and 8a to 8c we state lemmas and a proposition, which we prove as they arise, and the two imported inputs, Theorems~\ref{thm:lb_talagrand} and~\ref{thm:lb_bvh}, which we quote where they enter. In Step 8d we assemble them. A union bound over the signings (Step 1) needs, for one fixed signing, a failure probability far below $2^{-n}$. Talagrand's inequality, applied to a convex Lipschitz functional of the $N=n(n-1)/2$ edge signs, gives a tail at speed $n^2$ (Steps 2 to 5). What the tail concentrates around, the median of that functional, we bound below by the moment method at finite $n$ (Steps 6 to 8).

\emph{Step 1: the union bound.}
The union bound is the framework of the whole argument. If, for one fixed signing, the largest eigenvalue of the Wigner matrix falls below the target with probability much smaller than $2^{-n}$, then the $2^{n-1}$ signings together fail with probability below one, and a good realization exists. The lemma states this. Everything after it works to make that probability small.
\begin{lemma}[Union bound over the signings]
\label{lem:lb_union}
Let $(A_i)$ be the random family of Definition~\ref{def:lb_matchings} and $\tau>0$. If $\Pr[\lambda_{\max}(W_{\epsilon'})\le\tau]\le q$ for every fixed $\epsilon'\in\{-1,1\}^{n-1}$, then the probability that some $\epsilon'$ has $\lambda_{\max}(W_{\epsilon'})\le\tau$ is at most $2^{n-1}q$; and if $2^{n-1}q<1$ there is a realization of $x$ for which every $\epsilon'$ has $\lambda_{\max}(W_{\epsilon'})>\tau$.
\end{lemma}
\begin{proof}
We take a union bound over the $2^{n-1}$ values of $\epsilon'$, and a probability below one leaves a realization outside the union. The sign $\epsilon_n$ of the identity is not in the union. We handle it deterministically in Step 8. (In general $\|-S\|=\|S\|$ pairs the $2^n$ signings, so one sign is always free.)
\end{proof}
\emph{Step 2: the hinge functional is convex and Lipschitz.}
We read the event that the largest eigenvalue is at most $a\sqrt n$ through a functional of the edge signs, the hinge functional $F$, which is zero exactly on that event and positive otherwise. What the concentration inequality of Step 3 needs of $F$ is convexity and a Lipschitz constant, and the lemma supplies both. The small constant $\sqrt2/n$ is the source of the speed $n^2$ in the tail.
\begin{lemma}[Convexity and Lipschitz constants]
\label{lem:lb_hinge}
Fix $a\in\R$. For $x\in\R^N$ let $W(x)$ be the symmetric zero-diagonal $n\times n$ matrix whose entry at the edge $e=\{i,j\}$ is $x_e$, and put
\[
F(x):=\frac1n\tr[(W(x)/\sqrt n-aI_n)_+],\qquad G(x):=\|W(x)\|.
\]
Then $F$ is convex and $(\sqrt2/n)$-Lipschitz, and $G$ is convex and $\sqrt2$-Lipschitz, both for the Euclidean norm on $\R^N$.
\end{lemma}
\begin{proof}
For symmetric $Y$ we have the variational form $\tr[Y_+]=\max\{\tr[PY]:P\text{ symmetric},\,0\preceq P\preceq I\}$. Indeed, write $Y=\sum_i\lambda_iu_iu_i^\top$. For $0\preceq P\preceq I$, $\tr[PY]=\sum_i\lambda_i\,u_i^\top Pu_i$ with $u_i^\top Pu_i\in[0,1]$, so $\tr[PY]\le\sum_{\lambda_i>0}\lambda_i$, with equality at the projector onto the positive eigenspace. Hence $F$ is a maximum of affine functions of $x$, so convex. The operator norm is convex, so $G$ is convex.

For the Lipschitz constants: if $0\preceq P\preceq I$ then $\|P\|_F^2=\tr[P^2]\le\tr[P]\le n$, so by Cauchy--Schwarz $|\tr[PY]-\tr[PY']|\le\sqrt n\,\|Y-Y'\|_F$, and taking the maximum over $P$ on each side we obtain $|\tr[Y_+]-\tr[(Y')_+]|\le\sqrt n\,\|Y-Y'\|_F$. With $Y=W(x)/\sqrt n-aI_n$ and $Y'=W(y)/\sqrt n-aI_n$, each edge sign appears twice in $W$ and the diagonal is zero, so $\|Y-Y'\|_F=\|W(x)-W(y)\|_F/\sqrt n=\sqrt2\,\|x-y\|_2/\sqrt n$, and $|F(x)-F(y)|\le(\sqrt2/n)\|x-y\|_2$. Likewise $|G(x)-G(y)|\le\|W(x)-W(y)\|\le\|W(x)-W(y)\|_F=\sqrt2\,\|x-y\|_2$.
\end{proof}
\emph{Step 3: Talagrand's inequality and the two tails.}
The two tails of this step come from one concentration inequality: in a product probability space, the points at convex distance at least $t$ from a set $A$ have probability at most $e^{-t^2/4}/P(A)$. It is the only source of the speed-$n^2$ lower tail of Step 5, and we state it in the form of Meckes~\cite[Theorem~3]{meckes04}.

{\setlength{\emergencystretch}{2em}The first imported input is Talagrand's convex distance inequality, in the form of Meckes~\cite[Theorem~3]{meckes04}. Following his notation, $(\Omega_1,\mu_1),\dots,(\Omega_N,\mu_N)$ are probability spaces, $(\Omega,P)$ is their product, $h(x,y)_j:=0$ if $x_j=y_j$ and $1$ otherwise, $U_A(x):=\{h(x,y):y\in A\}$, and $f_c(A,x):=\inf\{|z|:z\in\operatorname{conv}U_A(x)\}$ with $|\cdot|$ the Euclidean norm.\par}

\begin{theorem}[{Talagrand's convex distance inequality; black box; Talagrand~\cite[Theorem~4.1.1]{tal95}, in the form stated by Meckes~\cite[Theorem~3]{meckes04}}]
\label{thm:lb_talagrand}
Let $(\Omega_1,\mu_1),\dots,(\Omega_N,\mu_N)$ be probability spaces, let $(\Omega,P)$ be their product, and let $f_c$ be the convex distance above. For every $A\subseteq\Omega$,
\[
\int_\Omega\exp(\frac14f_c(A,x)^2)\,\d P(x)\le\frac1{P(A)},
\]
and hence, by Chebyshev's inequality, $P\{x:f_c(A,x)\ge t\}\le e^{-t^2/4}/P(A)$ for all $t>0$.
\end{theorem}

Hypotheses checked: we use $\Omega_j=\{-1,1\}$ with the uniform $\mu_j$, $j=1,\dots,N$, a finite product space, so the measurability caveat Meckes attaches to the statement is void. Role: it is the only source of the speed-$n^2$ lower tail in Step 5, and it enters Section~\ref{subsec:lb_asymptotic} again through Lemma~\ref{lem:lb_convexdistance}. Convention: Talagrand~\cite[Theorem~4.1.1]{tal95} states the inequality for a power $\Omega^N$ of one probability space, with the same constant $1/4$ and the same distance. Meckes~\cite[Theorem~3]{meckes04} states it for a product of different factors, which follows from it. Our application, $\Omega=\{-1,1\}$ with the uniform measure, is a power space, so both forms cover it. We cite Talagrand for the inequality and Meckes for the form we state.

The next lemma is the form in which we use the inequality: a convex Lipschitz function of uniform signs has two-sided tails around its median at Gaussian speed governed by the Lipschitz constant. In the proof we pass from the convex distance to the Euclidean distance to a convex hull, which is where the convexity of the function enters.

\begin{lemma}[Two-sided tails for convex Lipschitz functions of uniform signs]
\label{lem:lb_tails}
Let $x$ be uniform on $\{-1,1\}^N$, let $\Phi:\R^N\to\R$ be convex and $\ell$-Lipschitz for the Euclidean norm, and let $m$ be a median of $\Phi(x)$. Then for every $s>0$,
\[
\Pr[\Phi(x)\le m-s]\le2e^{-s^2/(16\ell^2)}\qquad\text{and}\qquad\Pr[\Phi(x)\ge m+s]\le2e^{-s^2/(16\ell^2)}.
\]
\end{lemma}
\begin{proof}
Take $\Omega_j=\{-1,1\}$ with the uniform $\mu_j$ in Theorem~\ref{thm:lb_talagrand}, so $P$ is the law of $x$.

\emph{(i) The convex distance dominates the Euclidean distance to the hull.} For $x,y$ in the cube, $x_j-y_j=2x_jh(x,y)_j$. Let $z=\sum_k\theta_kh(x,y_k)\in\operatorname{conv}U_A(x)$ with $\theta$ a probability vector and $y_k\in A$ (finite combinations, since $A$ is finite), and put $\bar y:=\sum_k\theta_ky_k\in\operatorname{conv}A$. Then $x-\bar y=2\,x\circ z$ coordinatewise, so $\|x-\bar y\|_2=2|z|$ because $|x_j|=1$. Taking the infimum over $z$ we obtain $\operatorname{dist}_2(x,\operatorname{conv}A)\le2f_c(A,x)$.

\emph{(ii) A level set controls the function.} Suppose $\Phi\le c$ on $A$. By convexity $\Phi\le c$ on $\operatorname{conv}A$, and for every $\bar y\in\operatorname{conv}A$, $\Phi(x)\le\Phi(\bar y)+\ell\|x-\bar y\|_2$. Hence $\Phi(x)\le c+\ell\operatorname{dist}_2(x,\operatorname{conv}A)\le c+2\ell f_c(A,x)$. So $\{\Phi\ge c+2\ell t\}\subseteq\{f_c(A,\cdot)\ge t\}$ for $t>0$, and Theorem~\ref{thm:lb_talagrand} gives $\Pr[\Phi\ge c+2\ell t]\le e^{-t^2/4}/\Pr[A]$.

\emph{(iii) The two tails.} Lower tail: take $A:=\{\Phi\le m-s\}$ (nothing to prove if $\Pr[A]=0$), $c:=m-s$ and $2\ell t:=s$. Then $\Pr[\Phi\ge m]\le e^{-s^2/(16\ell^2)}/\Pr[\Phi\le m-s]$, and $\Pr[\Phi\ge m]\ge1/2$ because $m$ is a median, so $\Pr[\Phi\le m-s]\le2e^{-s^2/(16\ell^2)}$. Upper tail: take $A:=\{\Phi\le m\}$, which has $\Pr[A]\ge1/2$, $c:=m$ and $2\ell t:=s$.
\end{proof}

We apply the lemma to $F$ of Lemma~\ref{lem:lb_hinge}, with $\ell=\sqrt2/n$ and $16\ell^2=32/n^2$, and to $G=\|W\|$, with $\ell=\sqrt2$ and $16\ell^2=32$, and we write $m_F$ and $m_G$ for medians of $F(W)$ and $\|W\|$:
\begin{align}
\Pr[F\le m_F-s]&\le2e^{-n^2s^2/32},&\Pr[F\ge m_F+s]&\le2e^{-n^2s^2/32},\label{eq:lb_tailF}\\
\Pr[\|W\|\le m_G-s]&\le2e^{-s^2/32},&\Pr[\|W\|\ge m_G+s]&\le2e^{-s^2/32}.\label{eq:lb_tailG}
\end{align}
Here we evaluate $F$ at the vector of entries of $W$ above the diagonal, which is uniform on $\{-1,1\}^N$.

\emph{Step 4: median against mean.}
Talagrand's inequality concentrates around a median, while the moment method of Steps 6 to 8 bounds a mean. The lemma converts one into the other: for $F$ the median is at most $\sqrt{32\pi}/n$ below the mean, and for the norm the median is at most $\sqrt{32\pi}$ above the mean. We obtain both bounds as integrated tails.
\begin{lemma}[Median against mean]
\label{lem:lb_median}
Let $m_F$ be a median of $F(W)$ and $m_G$ a median of $\|W\|$, as in \eqref{eq:lb_tailF} and \eqref{eq:lb_tailG}. Then
\begin{equation}
\label{eq:lb_median}
m_F\ge\E[F]-\frac{\sqrt{32\pi}}{n}\qquad\text{and}\qquad m_G\le\E[\|W\|]+\sqrt{32\pi}.
\end{equation}
\end{lemma}
\begin{proof}
By the upper tail in \eqref{eq:lb_tailF} and the layer-cake formula,
\[
\E[F]-m_F\le\E[(F-m_F)_+]=\int_0^\infty\Pr[F-m_F\ge s]\,\d s\le\int_0^\infty2e^{-n^2s^2/32}\,\d s=\frac{\sqrt{32\pi}}{n}.
\]
By the lower tail in \eqref{eq:lb_tailG}, $m_G-\E[\|W\|]\le\E[(m_G-\|W\|)_+]\le\int_0^\infty2e^{-s^2/32}\,\d s=\sqrt{32\pi}$. (In decimals, $\sqrt{32\pi}=10.026\ldots$.)
\end{proof}
\emph{Step 5: the lower tail of the edge at speed $n^2$.}
The lemma turns the tail of Step 3 into a bound on the failure probability of one signing, as the event that no eigenvalue exceeds $a\sqrt n$ is exactly $\{F=0\}$.
\begin{lemma}[The edge lower tail]
\label{lem:lb_edgetail}
Let $F$ be the functional of Lemma~\ref{lem:lb_hinge} with $a=2-\delta$, evaluated at $W$, and let $m_F$ be a median of $F(W)$. Then $F(W)=0$ if and only if $\lambda_{\max}(\widehat W)\le a$, and if $m_F>0$,
\begin{equation}
\label{eq:lb_edgetail}
\Pr[\lambda_{\max}(W)\le a\sqrt n]\le\Pr[F(W)=0]\le2e^{-n^2m_F^2/32}.
\end{equation}
\end{lemma}
\begin{proof}
$F(W)=\frac1n\sum_i(\lambda_i(\widehat W)-a)_+$ vanishes exactly when every eigenvalue of $\widehat W$ is at most $a$. Take $s:=m_F$ in the lower tail of \eqref{eq:lb_tailF}.
\end{proof}

This is the $\exp(-cn^2)$ speed the union bound needs: $F$ has Lipschitz constant $\sqrt2/n$ over $N=n(n-1)/2$ coordinates while its median is of order one in $\delta$. What remains is a lower bound on $m_F$ through $\E[F]$ by \eqref{eq:lb_median}. We obtain it in Steps 6 to 8.
\emph{Step 6: the moment lower bound by a tree-walk count.}
We bound the mean of $F$ from below through even moments of $W$. The lemma gives an exact combinatorial expression for $\E[\tr[W^{2p}]]$, closed walks in which every edge is traversed an even number of times, and a lower bound on it by counting only the walks that trace a tree, which gives the Catalan number $C_p$ up to the factor $\prod_j(1-j/n)$. We use no semicircle law: the Catalan floor is a count, valid at every finite $n$.
Write
\[
m_{2p}:=\E[\frac1n\tr[\widehat W^{2p}]]=n^{-p-1}\E[\tr[W^{2p}]].
\]

\begin{lemma}[Even moments of the Rademacher Wigner matrix]
\label{lem:lb_moments}
For every $p\ge1$, $\E[\tr[W^{2p}]]$ equals the number of closed walks of length $2p$ in $K_n$ with no loops in which every edge is traversed an even number of times, and
\[
\E[\tr[W^{2p}]]\ge C_p\,n(n-1)\cdots(n-p),\qquad\text{hence}\qquad m_{2p}\ge C_p\prod_{j=1}^p(1-\frac jn)\ge C_p(1-\frac{p(p+1)}{2n}).
\]
\end{lemma}
\begin{proof}
We expand $\tr[W^{2p}]=\sum\prod_{k=0}^{2p-1}W_{i_ki_{k+1}}$ over closed vertex sequences $(i_0,\dots,i_{2p-1},i_{2p}=i_0)$. A sequence with $i_k=i_{k+1}$ for some $k$ contributes $0$ (zero diagonal). Otherwise the product is $\prod_ex_e^{m_e}$ with $m_e$ the number of traversals of the edge $e$, whose expectation is $1$ if every $m_e$ is even and $0$ otherwise. This is the equality, and every term is nonnegative.

For the inequality, we map a pair (Dyck path of semilength $p$, ordered list $v_0,\dots,v_p$ of $p+1$ distinct vertices) to the walk that starts at $v_0$, moves to the next unused vertex of the list at each up step, and returns to the parent of the current vertex at each down step. Each tree edge is traversed exactly twice and the walk has no loops, so it is counted on the left. The map is injective: a step is up exactly when it enters a new vertex, so we read the path off the walk, and the list is the order of first visits. There are $C_p$ Dyck paths of semilength $p$ and $n(n-1)\cdots(n-p)$ ordered lists. Dividing by $n^{p+1}$ gives the product, and $\prod_j(1-j/n)\ge1-\sum_jj/n$.
\end{proof}
\emph{Step 7: the chord comparison and the truncation.}
We have to convert a moment lower bound into a lower bound on $\E[F]$, which sees only the eigenvalues above $a$. The lemma is the elementary comparison that does this: on $[a,M]$ the hinge $(|t|-a)_+$ dominates the chord of $t^{2p}$, so a large $2p$-th moment forces a positive hinge, provided we truncate the spectrum at $M$. We bound the truncation error in Step 8.
\begin{lemma}[Chord comparison]
\label{lem:lb_chord}
Let $0<a<M$ and $p\ge1$. For every real $t$,
\[
(|t|-a)_+\ge(M-a)\,\frac{\min(t^{2p},M^{2p})-a^{2p}}{M^{2p}-a^{2p}}.
\]
\end{lemma}
\begin{proof}
For $a\le|t|\le M$ the map $u\mapsto u^{2p}$ is convex on $[a,M]$, so it lies below its chord: $(|t|^{2p}-a^{2p})/(M^{2p}-a^{2p})\le(|t|-a)/(M-a)$. For $|t|<a$ the right side is at most $0$ and the left side is $0$. For $|t|>M$ the right side equals $M-a\le|t|-a$.
\end{proof}

Since $W$ and $-W$ have the same law, $\E[\frac1n\tr[(-\widehat W-aI_n)_+]]=\E[F]$, and $(\lambda-a)_++(-\lambda-a)_+=(|\lambda|-a)_+$ for $a>0$, so $2\E[F]=\E[\frac1n\sum_i(|\lambda_i(\widehat W)|-a)_+]$ and Lemma~\ref{lem:lb_chord} gives
\begin{equation}
\label{eq:lb_chord}
2\E[F]\ge(M-a)\,\frac{\E[\frac1n\sum_i\min(\lambda_i(\widehat W)^{2p},M^{2p})]-a^{2p}}{M^{2p}-a^{2p}} .
\end{equation}
On the event $\|\widehat W\|\le M$ every minimum equals $\lambda_i^{2p}$, and the minimum is nonnegative always, so $\frac1n\sum_i\min(\lambda_i^{2p},M^{2p})\ge\frac1n\tr[\widehat W^{2p}]-\|\widehat W\|^{2p}\mathbbm{1}\{\|\widehat W\|>M\}$ and
\begin{equation}
\label{eq:lb_trunc}
\E[\frac1n\sum_i\min(\lambda_i(\widehat W)^{2p},M^{2p})]\ge m_{2p}-\E[\|\widehat W\|^{2p}\mathbbm{1}\{\|\widehat W\|>M\}]=:m_{2p}-T_p(M).
\end{equation}
\emph{Step 8a: the norm bound, imported, and the edge tail.}
The truncation needs the norm of $W$ to be located, and this is the one place where an external norm bound enters. We use it only from above: it bounds $\E[\|W\|]$ by $(2+\eta_0)\sqrt n$, and together with the upper tail of Step 3 it makes the event $\|\widehat W\|>M$ exponentially rare.

The second input we import is the nonasymptotic norm bound of Bandeira and van Handel. As they do, we write, for the coefficient matrix $(b_{ij})$, $\sigma:=\max_i(\sum_jb_{ij}^2)^{1/2}$ and $\sigma_*:=\max_{ij}|b_{ij}|$.

\begin{theorem}[{Norm of a symmetric matrix with independent entries; black box; Bandeira and van Handel~\cite[Theorem~1.1 and Corollary~3.2]{bh16}}]
\label{thm:lb_bvh}
Let $X$ be the $n\times n$ symmetric random matrix with entries $X_{ij}=\xi_{ij}b_{ij}$, where the $\xi_{ij}$, $i\ge j$, are independent symmetric random variables with $\E[\xi_{ij}^{2p}]\le\E[g^{2p}]$ for every $p\in\mathbb N$ and a standard Gaussian $g$, in particular independent standard Gaussians or independent uniform signs, and the $b_{ij}$, $i\ge j$, are given scalars. Then for every $0<\gamma\le1/2$,
\[
\E[\|X\|]\le(1+\gamma)(2\sigma+\frac{6}{\sqrt{\log(1+\gamma)}}\,\sigma_*\sqrt{\log n}).
\]
\end{theorem}

Hypotheses checked: $W$ has $b_{ij}=1$ for $i\ne j$ and $b_{ii}=0$, with entries uniform signs, which are symmetric with $\E[x^{2p}]=1\le\E[g^{2p}]$, so $\sigma=\sqrt{n-1}$ and $\sigma_*=1$. Their Theorem~1.1 is the Gaussian case, and their Corollary~3.2 extends it to the symmetric variables of the statement, uniform signs included. Convention: the logarithm is natural, as their moment computation requires. Role: it bounds the mean of $\|W\|$, which we combine with Lemma~\ref{lem:lb_tails} to truncate the spectrum in Step 8b. Taking $\gamma\in(0,1/2]$ as the free parameter of the theorem, we have $(1+\gamma)2\sqrt{n-1}\le\sqrt n(2+2\gamma)$ and
\begin{equation}
\label{eq:lb_bvh}
\E[\|W\|]\le(1+\gamma)(2\sqrt{n-1}+\frac{6}{\sqrt{\log(1+\gamma)}}\sqrt{\log n})\le\sqrt n\,(2+\eta_0),
\end{equation}
with $\eta_0$ as in Theorem~\ref{thm:lb_existence}. Put $M_1:=2+\eta_0+\sqrt{32\pi/n}$. By \eqref{eq:lb_median} and \eqref{eq:lb_bvh}, $m_G\le\sqrt n\,M_1$, so the upper tail in \eqref{eq:lb_tailG} with $s=\sqrt n\,u$ gives, in the units of $\widehat W$,
\begin{equation}
\label{eq:lb_normtail}
\Pr[\|\widehat W\|\ge M_1+u]\le2e^{-nu^2/32}\qquad\text{for all }u>0 .
\end{equation}
\emph{Step 8b: the tail moment.}
The lemma bounds the truncation error $T_p(M)$ of Step 7, the part of the $2p$-th moment carried by the event $\|\widehat W\|>M$, from the tail \eqref{eq:lb_normtail} by a layer-cake integration. The hypothesis $ns_0M>16(2p-1)$ is what makes the integral converge.
\begin{lemma}[The tail moment]
\label{lem:lb_tailmoment}
Let $s_0>0$, let $M_1:=2+\eta_0+\sqrt{32\pi/n}$ be the constant of \eqref{eq:lb_normtail}, put $M:=M_1+s_0$, and suppose $ns_0M>16(2p-1)$. Then
\[
T_p(M)\le2M^{2p}e^{-ns_0^2/32}[1+\frac{32p}{ns_0M-16(2p-1)}]=T .
\]
\end{lemma}
\begin{proof}
Put $Y:=\|\widehat W\|$. By the layer-cake formula,
\[
\E[Y^{2p}\mathbbm{1}\{Y>M\}]=M^{2p}\Pr[Y>M]+\int_M^\infty2pv^{2p-1}\Pr[Y>v]\,\d v .
\]
By \eqref{eq:lb_normtail}, $\Pr[Y>M]\le2e^{-ns_0^2/32}$, and for $v=M+u$, since $(s_0+u)^2\ge s_0^2+2s_0u$,
\[
\Pr[Y>M+u]\le2e^{-n(s_0+u)^2/32}\le2e^{-ns_0^2/32}e^{-ns_0u/16}.
\]
Also $(M+u)^{2p-1}\le M^{2p-1}e^{(2p-1)u/M}$. The exponent $ns_0/16-(2p-1)/M$ is positive by hypothesis, so the integral is at most
\[
\begin{aligned}
4pM^{2p-1}e^{-ns_0^2/32}\int_0^\infty e^{-(ns_0/16-(2p-1)/M)u}\,\d u&=\frac{4pM^{2p-1}e^{-ns_0^2/32}}{ns_0/16-(2p-1)/M}\\
&=2M^{2p}e^{-ns_0^2/32}\cdot\frac{32p}{ns_0M-16(2p-1)},
\end{aligned}
\]
where in the last step we multiply numerator and denominator by $16M$.
\end{proof}
\emph{Step 8c: the edge mean at finite $n$.}
The proposition assembles Steps 6, 7 and 8b into the lower bound $\E[F]\ge L$, with the $L$ of the theorem's hypothesis list, and then Step 4 into the bound on the median. Every quantity is explicit.
\begin{proposition}[The edge mean]
\label{prop:lb_edgemean}
Under the hypotheses of Theorem~\ref{thm:lb_existence}, with $a=2-\delta$,
\begin{equation}
\label{eq:lb_edgemean}
\E[F]\ge L=\frac{(M-2+\delta)B}{2M^{2p}}\qquad\text{and}\qquad m_F\ge L-\frac{\sqrt{32\pi}}{n}.
\end{equation}
\end{proposition}
\begin{proof}
Combining \eqref{eq:lb_chord}, \eqref{eq:lb_trunc}, Lemma~\ref{lem:lb_moments} and Lemma~\ref{lem:lb_tailmoment},
\[
\begin{aligned}
2\E[F]&\ge(M-a)\,\frac{m_{2p}-T_p(M)-a^{2p}}{M^{2p}-a^{2p}}\ge(M-a)\,\frac{C_p\prod_{j=1}^p(1-j/n)-T-a^{2p}}{M^{2p}-a^{2p}}\\
&=\frac{(M-a)B}{M^{2p}-a^{2p}}\ge\frac{(M-a)B}{M^{2p}} ,
\end{aligned}
\]
where $a<2<M$, the hypothesis $B>0$ and $0<M^{2p}-a^{2p}\le M^{2p}$ justify the last step. Substituting $a=2-\delta$, we obtain $\E[F]\ge L$, and \eqref{eq:lb_median} gives the median bound.
\end{proof}

No semicircle law and no limit theorem enters: we bound the edge mean from below at finite $n$ with every constant written.
\emph{Step 8d: the union bound closes and the identity adds one.}
The proof runs the tail of one signing, the union bound, and the sign of the identity, in that order.
\begin{proof}[Proof of Theorem~\ref{thm:lb_existence}]
Fix $\epsilon'\in\{-1,1\}^{n-1}$. By Lemma~\ref{lem:lb_model}(iii), $W_{\epsilon'}$ has the law of $W$. By \eqref{eq:lb_star} and \eqref{eq:lb_edgemean}, $m_F\ge L-\sqrt{32\pi}/n>0$, so \eqref{eq:lb_edgetail} applies, and since the bound \eqref{eq:lb_tailF} is monotone in $s$,
\[
\Pr[\lambda_{\max}(W_{\epsilon'})\le(2-\delta)\sqrt n]\le2\exp(-\frac{n^2}{32}(L-\frac{\sqrt{32\pi}}{n})^2)=:q .
\]
{\setlength{\emergencystretch}{2em}By Lemma~\ref{lem:lb_union} the union over the $2^{n-1}$ choices of $\epsilon'$ has probability at most $2^{n-1}q=2^n\exp(-\frac{n^2}{32}(L-\sqrt{32\pi}/n)^2)$, and $2^{n-1}q<1$ is equivalent to $\frac{n^2}{32}(L-\sqrt{32\pi}/n)^2>n\log2$, which is \eqref{eq:lb_star} (both sides of \eqref{eq:lb_star} are positive). Fix a realization $x$ outside the union event. Then every $\epsilon'$ has $\lambda_{\max}(W_{\epsilon'})>(2-\delta)\sqrt n$. We apply this to $-\epsilon'$, with $W_{-\epsilon'}=-W_{\epsilon'}$, and obtain $\lambda_{\min}(W_{\epsilon'})<-(2-\delta)\sqrt n$ as well. Now take any $\epsilon=(\epsilon',\epsilon_n)$, so $\sum_i\epsilon_iA_i=W_{\epsilon'}+\epsilon_nI_n$. If $\epsilon_n=1$ the largest eigenvalue is $\lambda_{\max}(W_{\epsilon'})+1>(2-\delta)\sqrt n+1$. If $\epsilon_n=-1$ the smallest eigenvalue is $\lambda_{\min}(W_{\epsilon'})-1<-((2-\delta)\sqrt n+1)$. In both cases $\|\sum_i\epsilon_iA_i\|>(2-\delta)\sqrt n+1$. The matrices $A_i(x)$ are symmetric signed permutation involutions by Lemma~\ref{lem:lb_model}(ii).\par}
\end{proof}
\subsection{The rate of \texorpdfstring{$\delta_n$}{delta\_n} and the table}
\label{subsec:lb_rate}

The proposition shows that the criterion of Theorem~\ref{thm:lb_existence} is satisfied, for large $n$, at $\delta$ of order $n^{-1/5}(\log n)^{3/5}$, so $\delta_n\to0$ at that rate. This is Corollary~\ref{cor:lb_constant}(iv). The proof chooses the three parameters as functions of $n$ and $\delta$, with the moment order $p$ of order $\log(1/\delta)/\delta$, and verifies the displayed inequality \eqref{eq:lb_star}.

\begin{proposition}[The rate of $\delta_n$]
\label{prop:lb_rate}
There are absolute constants $C'$ and $n_0$ such that $\delta_n\le C'n^{-1/5}(\log n)^{3/5}$ for every even $n\ge n_0$. Hence $\delta_n\to0$ and $\delta_n=O(n^{-1/5}(\log n)^{3/5})$, for odd $n$ through $\delta_n=\delta_{n-1}$.
\end{proposition}
\begin{proof}
We do not track constants here: $n_0$ is absolute and may change from line to line, like $c$ and $C'$. Fix even $n\ge30$ and $\delta\in[n^{-1/4},1/2]$.

\emph{The parameters.} Take $\gamma:=(\log n/n)^{1/3}$, which is at most $1/2$ for $n\ge30$. Since $\log(1+\gamma)\ge\gamma/2$ on $(0,1]$ and $1+\gamma\le3/2$,
\[
\eta_0\le2\gamma+\frac32\cdot6\sqrt{\frac2\gamma}\cdot\gamma^{3/2}=(2+9\sqrt2)\gamma\le15(\frac{\log n}{n})^{1/3}.
\]
Take $p:=\lceil(4/\delta)\log(1/\delta)\rceil+\lceil8/\delta\rceil$, so $p\ge16$ and $p\le(4\log(1/\delta)+10)/\delta$. Take $s_0>0$ to solve
\[
\frac{ns_0^2}{32}=2p\log\frac M2+\log(32p^2),\qquad M=M_1+s_0 ;
\]
a solution exists because $s\mapsto ns^2/32-2p\log((M_1+s)/2)-\log(32p^2)$ is continuous, negative at $s=0$ (as $M_1>2$ and $32p^2>1$) and tends to $+\infty$. Then $e^{-ns_0^2/32}M^{2p}=4^p/(32p^2)$.

\emph{Sizes.} Write $\eta:=M-2=\eta_0+\sqrt{32\pi/n}+s_0$. Since $\log(M/2)=\log(1+\eta/2)\le\eta/2$, the equation for $s_0$ gives $ns_0^2/32\le p(\eta_0+\sqrt{32\pi/n})+ps_0+\log(32p^2)$, a quadratic inequality in $s_0$. With $p\le n^{1/4}(\log n+10)$ (by $\delta\ge n^{-1/4}$ and $\log(1/\delta)\le\frac14\log n$) it yields $s_0\le C'\sqrt{\log n/n}$ for $n\ge n_0$. Hence $\eta\le C'(\log n/n)^{1/3}$ and $p\eta\le C'n^{-1/12}(\log n)^{4/3}$, so $p\eta\le1$ for $n\ge n_0$. Also $p(p+1)\le n$ for $n\ge n_0$, and $ns_0M\ge16(2p-1)+32p$ for $n\ge n_0$: indeed $ns_0^2/32\ge\log(32p^2)$ gives $s_0\ge\sqrt{32\log32/n}$, which exceeds $32p/n\ge(64p-16)/(nM)$ once $n\ge n_0$, using $M\ge2$.

\emph{The bracket, the Catalan floor, and $B$.} With $ns_0M\ge16(2p-1)+32p$ the bracket in $T$ is at most $2$, so $T\le4\cdot4^p/(32p^2)=4^p/(8p^2)$. From $\binom{2p}{p}\ge4^p/(2\sqrt p)$ (induction on $p$, the step being $(2p+1)^2\ge4p(p+1)$), $C_p4^{-p}\ge\frac1{2\sqrt p\,(p+1)}\ge\frac1{4p^{3/2}}$, so $T\le C_p/8$ because $p^{1/2}\ge4$. By Lemma~\ref{lem:lb_moments} and $p(p+1)\le n$, $C_p\prod_{j=1}^p(1-j/n)\ge C_p/2$. Finally $(2-\delta)^{2p}=4^p(1-\delta/2)^{2p}\le4^pe^{-p\delta}$, and $p\delta\ge4\log(1/\delta)+8\ge\log32+\frac32\log p$ for $\delta\le1/2$, so $(2-\delta)^{2p}\le4^p/(32p^{3/2})\le C_p/8$. Combining, we obtain $B\ge C_p/2-C_p/8-C_p/8=C_p/4\ge4^p/(16p^{3/2})>0$.

\emph{The bound on $L$.} $M^{2p}=4^p(1+\eta/2)^{2p}\le4^pe^{p\eta}\le e\cdot4^p$, so
\[
L=\frac{(\eta+\delta)B}{2M^{2p}}\ge\frac{\delta\cdot4^p/(16p^{3/2})}{2e\cdot4^p}=\frac{\delta}{32e\,p^{3/2}}\ge c\,\frac{\delta^{5/2}}{(\log(1/\delta))^{3/2}},
\]
the last step by $p\le(4\log(1/\delta)+10)/\delta\le19\log(1/\delta)/\delta$ for $\delta\le1/2$.

\emph{Closing \eqref{eq:lb_star}.} For $n\ge5$, $\sqrt{32\pi}/n<\sqrt{32\log2/n}$, so \eqref{eq:lb_star} holds as soon as $L\ge2\sqrt{32\log2/n}$. Now put $\delta:=C'n^{-1/5}(\log n)^{3/5}$ with $C'$ large. For $n\ge n_0$ this lies in $[n^{-1/4},1/2]$, and $\delta\ge1/n$ gives $\log(1/\delta)\le\log n$, so
\[
c\,\frac{\delta^{5/2}}{(\log(1/\delta))^{3/2}}\ge c\,\frac{C'^{5/2}n^{-1/2}(\log n)^{3/2}}{(\log n)^{3/2}}=cC'^{5/2}n^{-1/2}\ge2\sqrt{32\log2}\,n^{-1/2}.
\]
Hence this $\delta$ is admissible and $\delta_n\le\delta$.
\end{proof}

At finite $n$ we can check the criterion by computer, and the table records what it gives.
With two independent codes we scanned $(p,\gamma,s_0)$ on grids and bisected on $\delta$ against the hypotheses of Theorem~\ref{thm:lb_existence}. Each admissible point is an upper bound on the infimum $\delta_n$. Table~\ref{tab:lb_delta} prints upper bounds rounded up to four digits, and each row below $2$ names a witness $(p,\gamma,s_0)$ at which the hypotheses of the theorem, evaluated directly, hold at the printed $\delta$. The printed value is then admissible, and so is every larger one because $\Delta_n$ is an up-set (Corollary~\ref{cor:lb_constant}(i)). At $n=10^3$ the set $\Delta_n$ is empty, and we prove this, not scan it: for every triple, $B<C_p\le4^{p-1}\le M^{2p-2}$ and $M-2+\delta<M$, so $L<1/(2M)$. Also $M\ge2+\sqrt{32\pi/n}+\eta_0$ with $\eta_0\ge2\gamma+6\sqrt{\log n/n}/\sqrt\gamma\ge6(3/2)^{2/3}(\log n/n)^{1/3}$, by $\log(1+\gamma)\le\gamma$ and $1+\gamma\ge1$. At $n=10^3$ this gives $M>3.81$ and $L<0.132$, while \eqref{eq:lb_star} requires $L>\sqrt{32\pi}/n+\sqrt{32\log2/n}>0.158$.

\begin{table}[t]
\centering
\caption{Upper bounds on $\delta_n$ from Theorem~\ref{thm:lb_existence}. Every entry below $2$ is admissible at the printed witness, so it is a valid $\delta$ and $D_n^{\mathrm{mat}}\ge(2-\delta)\sqrt n+1$ with the printed value. The row $n=10^3$ records $\Delta_n=\emptyset$, with $\delta_n=2$ by the convention of Definition~\ref{def:lb_deltan}. There the theorem gives nothing and the bound is the trivial one.}
\label{tab:lb_delta}
\begin{tabular}{@{}rlll@{}}
\toprule
$n$ & $\delta_n\le$ & $2-\delta_n\ge$ & certification \\
\midrule
$10^3$ & $2$ & $0$ & $\Delta_n=\emptyset$, proved in the text (vacuous) \\
$10^4$ & $1.2267$ & $0.7733$ & certified by the witness $(1,\,0.11212,\,0.18094)$ (below the trivial floor $1$) \\
$10^5$ & $0.9539$ & $1.0461$ & certified by the witness $(p,\gamma,s_0)=(2,\,0.0604,\,0.0581)$ \\
$10^6$ & $0.7048$ & $1.2952$ & certified by the witness $(p,\gamma,s_0)=(4,\,0.030624,\,0.019135)$ \\
$10^7$ & $0.4930$ & $1.5070$ & certified by the witness $(9,\,0.015,\,0.006)$ \\
$10^8$ & $0.3336$ & $1.6664$ & certified by the witness $(16,\,0.00756,\,0.00216)$ \\
$10^9$ & $0.2210$ & $1.7790$ & certified by the witness $(30,\,0.004,\,0.0008)$ \\
$10^{12}$ & $0.0619$ & $1.9381$ & certified by the witness $(157,\,0.0006,\,0.00015)$ \\
\bottomrule
\end{tabular}
\end{table}

The remark reads the table against the trivial floor, the explicit constant $\sqrt2$ of Theorem~\ref{thm:lb_field}, and the semicircle heuristic.

\begin{remark}[Reading the table]
\label{rem:lb_reading}
The theorem beats the trivial floor (constant $1$) from about $n=10^5$, beats the explicit constant $\sqrt2=1.4142\ldots$ of Theorem~\ref{thm:lb_field} from about $n=10^7$, and gives $1.7790$ at $n=10^9$; it is vacuous at $n=10^3$, where the explicit families own the bound. If the mean of $F$ were replaced by its semicircle value (a heuristic, not a theorem), the same concentration and union bound would give $1.326$, $0.317$ and $0.0791$ at $n=10^3,10^6,10^9$, about $4.98\,n^{-1/5}$; the explicit theorem is within the factor $(\log n)^{3/5}$ of that. About seventy percent of the loss over this heuristic, at every $n$ of the table, is the finite-$p$ moment method (the chord and the Catalan floor). Deleting the Bandeira--van Handel term $\eta_0$ altogether would move the scan value only from $0.7048$ to $0.5908$ at $n=10^6$ and from $0.2211$ (the first scan's value at $n=10^9$; the table prints the second scan's) to $0.1827$ at $n=10^9$.
\end{remark}

\subsection{The asymptotic exponent \texorpdfstring{$2/7$}{2/7}}
\label{subsec:lb_asymptotic}

The exponent $1/5$ of Proposition~\ref{prop:lb_rate} comes from applying Lemma~\ref{lem:lb_tails} to $F$ with its global Lipschitz constant $\sqrt2/n$. Talagrand's inequality only needs the gradient of $F$ on the set $\{F\ge m_F\}$, and there its squared norm is at most $2N_a/n^3$, where $N_a$ is the number of eigenvalues of $W$ above $a\sqrt n$, of order $n\delta^{3/2}$ rather than $n$. In this subsection we carry that through. The argument is asymptotic: we do not track constants, and we import one limit theorem, the moment asymptotics of Sinai and Soshnikov (Theorem~\ref{thm:lb_ss}), which Theorem~\ref{thm:lb_existence}, Proposition~\ref{prop:lb_rate} and Table~\ref{tab:lb_delta} do not use. We state the result for $D_n^{\mathrm{mat}}$. The admissible set $\Delta_n$ and $\delta_n$ of Definition~\ref{def:lb_deltan} are unchanged.

The first lemma computes a subgradient of the hinge functional at a point $x$ and bounds its norm by the number $N_a$ of eigenvalues above the level. This replaces the global Lipschitz constant of Lemma~\ref{lem:lb_hinge} by a local one.

\begin{lemma}[A subgradient of the hinge functional, and its norm]
\label{lem:lb_subgradient}
Fix $a\in\R$ and let $F$ be as in Lemma~\ref{lem:lb_hinge}. For $x\in\R^N$ let $P(x)$ be the orthogonal projector onto the span of the eigenvectors of $W(x)$ with eigenvalues above $a\sqrt n$, let $N_a(x):=\operatorname{rank}P(x)$, and define $g(x)\in\R^N$ by $g(x)_e:=2P(x)_{ij}/n^{3/2}$ for the edge $e=\{i,j\}$. Then for all $x,y\in\R^N$,
\[
F(y)\ge F(x)+\langle g(x),y-x\rangle\qquad\text{and}\qquad\|g(x)\|_2^2\le\frac{2N_a(x)}{n^3}.
\]
\end{lemma}
\begin{proof}
Put $h(t):=(t-a)_+$ and $h'(t):=\mathbf 1[t>a]$. Then $h(s)-h(t)\ge h'(t)(s-t)$ for all real $s,t$, the subgradient inequality of the convex function $h$. Let $X:=W(x)/\sqrt n$ and $Y:=W(y)/\sqrt n$ have eigenpairs $(\lambda_k,u_k)$ and $(\mu_l,v_l)$, and put $c_{kl}:=\langle u_k,v_l\rangle^2$, so that $\sum_lc_{kl}=\sum_kc_{kl}=1$ and $\sum_lc_{kl}\mu_l=u_k^\top Yu_k$. Then
\begin{align*}
\tr[h(Y)]-\tr[h(X)]&=\sum_{k,l}c_{kl}(h(\mu_l)-h(\lambda_k))\ge\sum_{k,l}c_{kl}\,h'(\lambda_k)(\mu_l-\lambda_k)\\
&=\sum_kh'(\lambda_k)\,u_k^\top(Y-X)u_k=\tr[P(x)(Y-X)],
\end{align*}
because $P(x)=\sum_{k:\lambda_k>a}u_ku_k^\top$. We divide by $n$ and write $Y-X=(W(y)-W(x))/\sqrt n$, where the edge $e=\{i,j\}$ contributes $y_e-x_e$ at the two positions $(i,j)$ and $(j,i)$: 
\[
F(y)-F(x)\ge n^{-3/2}\sum_e2P(x)_{ij}(y_e-x_e)=\langle g(x),y-x\rangle .
\]
For the norm,
\[
\|g(x)\|_2^2=4n^{-3}\sum_{i<j}P(x)_{ij}^2\le2n^{-3}\sum_{i,j}P(x)_{ij}^2=2n^{-3}\|P(x)\|_F^2=\frac{2N_a(x)}{n^3},
\]
since $P(x)$ is an orthogonal projector of rank $N_a(x)$.
\end{proof}

Our second lemma reruns the argument of Lemma~\ref{lem:lb_tails} with the local gradient bound, on the set where $F$ is above its median and $N_a$ is at most $K$. The cost is the factor $\Pr[B]$, which the proposition below keeps of order one by Markov's inequality.

\begin{lemma}[Talagrand's inequality with the gradient bounded on the upper half]
\label{lem:lb_convexdistance}
Let $x$ be uniform on $\{-1,1\}^N$, let $F$ be as in Lemma~\ref{lem:lb_hinge}, let $m_F$ be a median of $F(x)$, and let $K>0$. Put $A:=\{x:F(x)=0\}$ and $B:=\{x:F(x)\ge m_F,\ N_a(x)\le K\}$. Then
\[
\Pr[A]\,\Pr[B]\le\exp(-\frac{m_F^2\,n^3}{32K}).
\]
\end{lemma}
\begin{proof}
If $m_F=0$, or $\Pr[A]=0$, or $\Pr[B]=0$, the inequality holds trivially, so we assume all three are positive. Then for $x\in B$ and $y\in A$ the first inequality below gives $\langle g(x),x-y\rangle\ge m_F>0$, so $g(x)\ne0$ and the divisions below are legitimate. Let $x\in B$ and $y\in A$, and let $h(x,y)\in\{0,1\}^N$ be the indicator of the coordinates where $x$ and $y$ differ, as in Theorem~\ref{thm:lb_talagrand}. By Lemma~\ref{lem:lb_subgradient}, $m_F\le F(x)-F(y)\le\langle g(x),x-y\rangle=\sum_{e:x_e\ne y_e}g(x)_e(x_e-y_e)\le2\sum_e|g(x)_e|\,h(x,y)_e$. Hence every $z$ in the convex hull of $U_A(x)=\{h(x,y):y\in A\}$ satisfies $\sum_e|g(x)_e|z_e\ge m_F/2$, and by Cauchy--Schwarz $|z|\ge m_F/(2\|g(x)\|_2)\ge m_Fn^{3/2}/(2\sqrt{2K})=:t$, using $\|g(x)\|_2^2\le2K/n^3$ on $B$. So $f_c(A,x)\ge t$ for every $x\in B$, and Theorem~\ref{thm:lb_talagrand} gives $\Pr[B]\le\Pr[f_c(A,\cdot)\ge t]\le e^{-t^2/4}/\Pr[A]$, with $t^2/4=m_F^2n^3/(32K)$.
\end{proof}

We make the count of eigenvalues above the level with one limit theorem, the order of the even moments $\E[\tr[A_n^{p_n}]]$ of a Wigner matrix for powers $p_n\to\infty$ with $p_n=o(n^{2/3})$, namely $\sqrt{8/\pi}\,n/p_n^{3/2}$ up to a factor $1+o(1)$. Here it bounds $\E[N_a]$, the expected number of eigenvalues of $W$ above $a\sqrt n$, by a constant times $n\delta^{3/2}$ in the proof of Proposition~\ref{prop:lb_rate_two_sevenths}. It is the only limit theorem we import in this section.

The third input we import is the moment asymptotics of Sinai and Soshnikov for Wigner matrices, in the form of Soshnikov~\cite[p.~13]{sos99}. Following his notation, the real symmetric ensemble is $A_n=(a_{ij})$ with $a_{ij}=a_{ji}=\xi_{ij}/\sqrt n$, where the $\xi_{ij}$, $i\le j$, are independent real random variables such that: (i) the laws of the $\xi_{ij}$ are symmetric; (ii) all moments are finite; (iii) $\E[\xi_{ij}^2]=\frac14$ for $i<j$ (his (1.1$'$)) and $\E[\xi_{ii}^2]\le\mathrm{const}$ (his (1.2$'$)); (iv) $\E[\xi_{ij}^{2m}]\le(\mathrm{const}\cdot m)^m$ for $m=1,2,\dots$ (his (1.3$'$)).

\begin{theorem}[{Moments of high powers of a Wigner matrix; black box; Sinai and Soshnikov~\cite{ss98a,ss98b}, as stated by Soshnikov~\cite[arXiv v2, p.~13]{sos99}}]
\label{thm:lb_ss}
Let $A_n$ be as above, and let $p_n\to\infty$ with $p_n/n^{2/3}\to0$ as $n\to\infty$. Then
\[
\E[\tr[A_n^{p_n}]]=\begin{cases}\sqrt{\dfrac8\pi}\cdot\dfrac{n}{p_n^{3/2}}\cdot(1+o(1))&\text{if $p_n$ is even},\\[1ex]0&\text{if $p_n$ is odd},\end{cases}
\]
and the moments of $\tr[A_n^{p_n}]-\E[\tr[A_n^{p_n}]]$ converge to the moments of $N(0,\frac1\pi)$.
\end{theorem}

Hypotheses checked: by Lemma~\ref{lem:lb_model}(iii) our $W$ has independent uniform $W_{ij}\in\{-1,1\}$ for $i<j$ and $W_{ii}=0$, so $A_n:=W/(2\sqrt n)$ has $\xi_{ij}=W_{ij}/2\in\{-\frac12,\frac12\}$ for $i<j$, symmetric with $\E[\xi_{ij}^2]=\frac14$ and $\E[\xi_{ij}^{2m}]=4^{-m}\le m^m$, and $\xi_{ii}=0$, so every hypothesis holds. Consequence: for every sequence $p=p_n\to\infty$ with $p_n=o(n^{2/3})$,
\begin{equation}
\label{eq:lb_ss}
\E[\tr[W^{2p}]]=(2\sqrt n)^{2p}\,\E[\tr[A_n^{2p}]]=n^{p+1}\,\frac{4^p}{\sqrt\pi\,p^{3/2}}\,(1+o(1)),
\end{equation}
since $\sqrt8\,(2p)^{-3/2}=p^{-3/2}$. Role: it bounds $\E[N_a]$ from above in Proposition~\ref{prop:lb_rate_two_sevenths}. It is the only limit theorem we import in this section, and it enters that proposition only. Convention: the statement is Soshnikov's, arXiv:math-ph/9907013v2, printed page~13, where he cites~\cite{ss98a,ss98b} for it. His $\bar o(1)$, $\operatorname{Trace}$ and $\E(\cdot)$ are our $o(1)$, $\tr$ and $\E[\cdot]$.

The lemma extends the parameter choice of Proposition~\ref{prop:lb_rate} to the wider range $\delta\ge n^{-2/7}$, where the criterion \eqref{eq:lb_star} itself may fail but the lower bound on the mean of $F$ still holds. The proposition needs only that bound.

\begin{lemma}[The floor of Proposition~\ref{prop:lb_rate} on the range $\delta\ge n^{-2/7}$]
\label{lem:lb_floor_wide}
There are absolute constants $c>0$ and $n_0$ such that for every even $n\ge n_0$ and every $\delta\in[n^{-2/7},1/2]$, the parameters $\gamma$, $p$, $s_0$ chosen in the proof of Proposition~\ref{prop:lb_rate} satisfy $B>0$ and $ns_0M>16(2p-1)$, and, for $F$ with $a=2-\delta$,
\[
\E[F]\ge L\ge c\,\frac{\delta^{5/2}}{(\log(1/\delta))^{3/2}}\qquad\text{and}\qquad m_F\ge L-\frac{\sqrt{32\pi}}{n}.
\]
(The proofs of Proposition~\ref{prop:lb_edgemean} and Lemma~\ref{lem:lb_median} use the hypotheses of Theorem~\ref{thm:lb_existence} only through $\gamma\in(0,1/2]$, $\delta\in(0,2)$, $B>0$ and $ns_0M>16(2p-1)$, never through \eqref{eq:lb_star}. All four hold in this setting.)
\end{lemma}
\begin{proof}
We do not track constants: $n_0$ and $C'$ may change from line to line. The proof of Proposition~\ref{prop:lb_rate} used $\delta\ge n^{-1/4}$ only in its paragraph \emph{Sizes}, which we redo. Now $\log(1/\delta)\le\frac27\log n$ and $1/\delta\le n^{2/7}$, so $p\le(4\log(1/\delta)+10)/\delta\le n^{2/7}(\frac87\log n+10)$. The quadratic inequality $ns_0^2/32\le p(\eta_0+\sqrt{32\pi/n})+ps_0+\log(32p^2)$ gives $s_0\le32p/n+\sqrt{32(p\eta_0+p\sqrt{32\pi/n}+\log(32p^2))/n}$. Here $p\eta_0\le15n^{2/7-1/3}(\log n)^{1/3}(\frac87\log n+10)\to0$, $p\sqrt{32\pi/n}\le C'n^{-3/14}\log n\to0$, $\log(32p^2)\le C'\log n$ and $32p/n\le C'n^{-5/7}\log n$, so $s_0\le C'\sqrt{\log n/n}$ for $n\ge n_0$. Hence $\eta\le C'(\log n/n)^{1/3}$, $p\eta\le C'n^{-1/21}(\log n)^{4/3}\le1$, $p(p+1)\le2n^{4/7}(\frac87\log n+10)^2\le n$, and $ns_0M\ge2\sqrt{32\log32}\,\sqrt n\ge64p\ge16(2p-1)+32p$, all for $n\ge n_0$, the last because $s_0\ge\sqrt{32\log32/n}$, $M\ge2$ and $p\le n^{2/7}(\frac87\log n+10)=o(\sqrt n)$. The paragraphs \emph{The bracket, the Catalan floor, and $B$} and \emph{The bound on $L$} of that proof use only these facts together with $\delta\le1/2$ and $p\ge16$, and give $B\ge C_p/4>0$ and $L\ge c\,\delta^{5/2}/(\log(1/\delta))^{3/2}$ unchanged. The two displayed inequalities are then \eqref{eq:lb_edgemean}.
\end{proof}

The proposition combines the local tail with the Sinai--Soshnikov count of eigenvalues above the level, $\E N_a$ of order $n\delta^{3/2}$, into the exponent $2/7$.

\begin{proposition}[The exponent $2/7$]
\label{prop:lb_rate_two_sevenths}
There are absolute constants $C''$ and $n_1$ such that for every even $n\ge n_1$,
\[
D_n^{\mathrm{mat}}\ge(2-C''n^{-2/7}(\log n)^{6/7})\sqrt n+1,
\]
and for every odd $n\ge n_1+1$, $D_n^{\mathrm{mat}}\ge(2-C''(n-1)^{-2/7}(\log(n-1))^{6/7})\sqrt n$. The constants are not explicit, and $n_1$ is not located (see the end of the proof). The statement is about $D_n^{\mathrm{mat}}$; the set $\Delta_n$ and $\delta_n$ of Definition~\ref{def:lb_deltan}, defined through \eqref{eq:lb_star}, are unchanged.
\end{proposition}
\begin{proof}
Let $n_0$ be as in Lemma~\ref{lem:lb_floor_wide}, fix even $n\ge n_0$, and put $\delta:=C''n^{-2/7}(\log n)^{6/7}$ with $C''\ge1$ to be chosen. For $n\ge n_1$ this lies in $[n^{-2/7},1/2]$. We take the instance of Theorem~\ref{thm:lb_existence} (Definition~\ref{def:lb_matchings}), fix $\epsilon'\in\{-1,1\}^{n-1}$, and write $W$ for $W_{\epsilon'}$, which has the law of Lemma~\ref{lem:lb_model}(iii). Let $F$ be the functional of Lemma~\ref{lem:lb_hinge} with $a=2-\delta$, $m_F$ its median and $N_a$ as in Lemma~\ref{lem:lb_subgradient}.

{\setlength{\emergencystretch}{2em}\emph{The count above the level.} Write $\widehat W_a:=W/(a\sqrt n)$. Every eigenvalue of $W$ above $a\sqrt n$ is an eigenvalue of $\widehat W_a$ above $1$ and contributes at least $1$ to $\tr[\widehat W_a^{\,2q}]$, and every other eigenvalue contributes at least $0$. So $N_a\le\tr[\widehat W_a^{\,2q}]$ for every integer $q\ge1$. Take $q:=\lceil1/\delta\rceil$, a sequence with $q\to\infty$ and $q\le n^{2/7}+1$, so that $q=o(n^{2/3})$ and \eqref{eq:lb_ss} applies. It gives, for $n\ge n_1$,
\[
\E[\tr[W^{2q}]]\le\frac{2\,n^{q+1}4^q}{\sqrt\pi\,q^{3/2}} .
\]
Hence, by $-\log(1-u)\le u/(1-u)$ at $u=\delta/2\le\frac14$, by $q\delta\le1+\delta$ and $(1+\delta)/(1-\delta/2)\le2$ for $\delta\le\frac12$, and by $q^{-3/2}\le\delta^{3/2}$,
\[
\E[N_a]\le\frac{2n\,(1-\delta/2)^{-2q}}{\sqrt\pi\,q^{3/2}}\le\frac{2n\,e^{q\delta/(1-\delta/2)}}{\sqrt\pi\,q^{3/2}}\le\frac{2e^2}{\sqrt\pi}\,n\delta^{3/2}.
\]
Put $K:=\frac{8e^2}{\sqrt\pi}\,n\delta^{3/2}$. By Markov's inequality $\Pr[N_a>K]\le\frac14$, so the set $B$ of Lemma~\ref{lem:lb_convexdistance} has $\Pr[B]\ge\Pr[F\ge m_F]-\Pr[N_a>K]\ge\frac14$.\par}

\emph{The tail of one signing.} By Lemma~\ref{lem:lb_edgetail}, $\{\lambda_{\max}(W)\le a\sqrt n\}=\{F=0\}=A$, so Lemma~\ref{lem:lb_convexdistance} gives
\[
\Pr[\lambda_{\max}(W_{\epsilon'})\le(2-\delta)\sqrt n]\le4\exp(-\frac{m_F^2n^3}{32K})=4\exp(-\frac{\sqrt\pi}{256e^2}\cdot\frac{m_F^2\,n^2}{\delta^{3/2}}).
\]
By Lemma~\ref{lem:lb_floor_wide}, $L\ge c\,\delta^{5/2}/(\log(1/\delta))^{3/2}\ge c\,\delta^{5/2}/(\log n)^{3/2}$, which exceeds $2\sqrt{32\pi}/n$ for $n\ge n_1$ because $\delta^{5/2}\ge n^{-5/7}$. Hence $m_F\ge L-\sqrt{32\pi}/n\ge L/2$, and the exponent is at least
\[
\frac{\sqrt\pi\,c^2}{1024e^2}\cdot\frac{n^2\delta^{7/2}}{(\log n)^3}=\frac{\sqrt\pi\,c^2C''^{7/2}}{1024e^2}\,n .
\]

\emph{The union bound.} We choose $C''$ so large that $\sqrt\pi\,c^2C''^{7/2}/(1024e^2)\ge2\log2$. Then the union over the $2^{n-1}$ signings $\epsilon'$ (Lemma~\ref{lem:lb_union}) has probability at most $2^{n-1}\cdot4\cdot2^{-2n}=2^{1-n}<1$. Fix a realization $x$ outside the union event. Exactly as in Step 8d, every $\epsilon'$ has $\lambda_{\max}(W_{\epsilon'})>(2-\delta)\sqrt n$ and, applied to $-\epsilon'$, $\lambda_{\min}(W_{\epsilon'})<-(2-\delta)\sqrt n$, so every signing $\epsilon=(\epsilon',\epsilon_n)$ has $\|\sum_i\epsilon_iA_i\|>(2-\delta)\sqrt n+1$, and $D_n^{\mathrm{mat}}\ge(2-\delta)\sqrt n+1$. Odd $n$ follow by the padding argument in the proof of Corollary~\ref{cor:lb_constant}(iii), which needs only $\delta\in(0,2)$.

\emph{On $n_1$.} We do not determine the threshold here. The following calculation only indicates that it lies far beyond every $n$ of Table~\ref{tab:lb_delta}. Lemma~\ref{lem:lb_floor_wide} needs $p\eta\le1$, and at the value of $\delta$ that the semicircle calibration of this argument suggests, about $(2\cdot10^4/n)^{2/7}$ (a different choice from the $\delta$ of this proposition), the upper estimate of $p\eta_0$ with $\eta_0\le15(\log n/n)^{1/3}$ and the $p$ of Proposition~\ref{prop:lb_rate} is about $18$ at $n=10^9$, about $17$ at $n=10^{20}$, about $5$ at $n=10^{40}$, and falls below $1$ only near $n=10^{60}$. An upper estimate above $1$ does not show that the product itself exceeds $1$, and Theorem~\ref{thm:lb_ss} carries its own unquantified threshold, so $n_1$ is not located by this. Table~\ref{tab:lb_delta} and $\delta_n$ are unaffected.
\end{proof}

We close with a remark that separates what the existence theorem gives from what it does not, and ends with the explicit families and the open problem they leave.

\begin{remark}[Scope of the lower bounds]
\label{rem:lb_scope}
What the theorem gives. Theorem~\ref{thm:lb_existence} with Corollary~\ref{cor:lb_constant} gives the asymptotic lower constant $2$ by existence, $\liminf_{n\to\infty}D_n^{\mathrm{mat}}/\sqrt n\ge2$, with a finite-$n$ proof in which no limit theorem enters, the explicit rate $\delta_n=O(n^{-1/5}(\log n)^{3/5})$ of Proposition~\ref{prop:lb_rate}, and the values of Table~\ref{tab:lb_delta}. Proposition~\ref{prop:lb_rate_two_sevenths} improves the exponent to $2/7$ in the asymptotic register. It is the one statement of this section that imports a limit theorem (Theorem~\ref{thm:lb_ss}), and it leaves $\delta_n$ and the table unchanged. Consequently any universal constant $C$ for which every square instance admits a signing of norm at most $C\sqrt n$ satisfies $C\ge2$; with Theorem~\ref{thm:existence_constant_sub10}, the least such constant lies in $[2,\,7.879493715947]$.

What it does not give. It names a distribution, not a family: the good sign pattern exists but is not exhibited, and there is no certificate to check at a specific $n$. It is vacuous at $n=10^3$, where $\Delta_n$ is empty and the bound is owned by the explicit families and padding. It passes the trivial floor only near $n=10^5$. Its explicit exponent $1/5$ comes from Talagrand's inequality with the global Lipschitz constant of the hinge functional. The gradient on the upper half of the space gives $2/7$ (Proposition~\ref{prop:lb_rate_two_sevenths}), and a scaling argument, a heuristic about the method and not a theorem, indicates that $2/7$ is the most the convex distance inequality yields for linear spectral statistics of this kind: for $F_h=\frac1n\tr[h(\widehat W)]$ with $h$ convex and vanishing on $(-\infty,a]$, the ratio $(\text{median})^2/(\text{gradient norm})^2$ that drives Lemma~\ref{lem:lb_convexdistance} scales as $n^2\delta^{7/2}$ whatever the shape of $h$. Whether the rate can be of order $n^{-1/3}$, the rate the Gaussian calibration suggests, is open. It would follow from a speed-$n^2$ left-tail bound with cubic rate for the largest eigenvalue of a Rademacher Wigner matrix, which is not available: Erd\H os and Xu~\cite[Theorem~1.2]{ex23} prove such a left tail only at the Tracy--Widom scale, for deviations up to a constant times $(\log n)^{1/3}n^{-2/3}$, and for the Gaussian orthogonal ensemble a speed-$n^2$ bound follows from the large deviation principle of Ben Arous and Guionnet~\cite{bag97} for the empirical spectral measure, whose proof uses the Gaussian density. It says nothing about the upper constant of Theorems~\ref{thm:matrix_spencer_bound} and~\ref{thm:existence_constant_sub10}, and nothing about dimensions $d<n$ beyond padding. The explicit families are exact at their $n$; the family of Theorem~\ref{thm:lb_field} is explicit at every $n=2^k$ but stops at $\sqrt2$; an explicit family at constant $2-o(1)$ is not known, and we leave it as an open problem.
\end{remark}

We close with the value we believe the constant has.

\begin{remark}[The conjectured constant]
\label{rem:lb_conjecture}
We conjecture that the correct constant is $2$: that $D_n^{\mathrm{mat}}\le2\sqrt n+o(\sqrt n)$ for every $n$, so that $\lim_{n\to\infty}D_n^{\mathrm{mat}}/\sqrt n=2$, and that the least universal constant $C$ with $D_n^{\mathrm{mat}}\le C\sqrt n$ for all $n$ is $2$. This section proves the lower half, $\liminf_{n\to\infty}D_n^{\mathrm{mat}}/\sqrt n\ge2$, and Theorem~\ref{thm:existence_constant_sub10} leaves the gap $[2,\,7.88]$. Part of our reason is the Kadison--Singer analogue. For the rank-one matrix discrepancy problem that generalizes the signing formulation of the Kadison--Singer problem, Song and Yue~\cite{sy26} prove that two standard deviations are necessary, $2\le C_1\le C_0\le2.176$ in their notation, and conjecture that the optimal constant there is $2$ as well. The lower bound $2$ appears in both problems by different mechanisms, a hard distribution of signed matchings here and a rank-one construction there, and we expect both constants to be sharp.
\end{remark}

\bibliographystyle{alpha}
\bibliography{ref}

\end{document}